\documentclass[a4paper,twoside,12pt]{book}
\usepackage{graphicx}
\usepackage{braket}
\usepackage{algorithm}
\usepackage{algorithmicx}
\usepackage{algpseudocode}
\usepackage{subfig}
\usepackage{tabularx}
\usepackage{array}
\usepackage{amsmath, amssymb, amsthm}
\usepackage{bbm}
\usepackage{thmtools} 
\usepackage{thm-restate}
\usepackage[nottoc]{tocbibind}
\usepackage{color}
\usepackage{multirow}
\usepackage[hidelinks]{hyperref}
\usepackage{setspace}
\usepackage[a4paper, margin=2cm, marginparwidth=0cm, marginparsep=0cm, outer=0cm, headheight=11pt, marginpar=0pt, top=30mm, bottom=20mm, bindingoffset=20mm, inner=20mm, outer=20mm]{geometry}
\usepackage{cite}

\usepackage{pdflscape}

\usepackage{booktabs}

\usepackage{graphics}
\usepackage{amssymb}
\usepackage{amsmath, framed}
\usepackage{braket}
\usepackage{epsfig}
\usepackage{amsmath}
\usepackage{amsfonts}
\usepackage{amssymb}
\usepackage{enumerate}

\usepackage{color,soul}

\usepackage{ifpdf}
\usepackage{	qcircuit}
\usepackage{pifont}
\usepackage{graphicx}
\usepackage{braket}
\usepackage{pifont}
\usepackage{array}
\usepackage{longtable}
\usepackage{bm}
\usepackage{hyperref}
\usepackage{color}
\usepackage{makeidx}
\usepackage{cite}
\usepackage{mathtools}
\usepackage{bbm}
\usepackage{wrapfig}
\usepackage[skip=0pt]{caption}
\usepackage{titlesec}

\usepackage[font=footnotesize]{caption}
\usepackage[table]{xcolor}
\definecolor{matteRed}{RGB}{222,115,116} % Defining an Excel matte red color
\definecolor{matteYellow}{RGB}{255,176,59}
\definecolor{matteGreen}{RGB}{198,224,180} % Approximate Excel matte green color
\definecolor{matteBlue}{RGB}{141,180,226}  % Approximate Excel matte blue color

\usepackage{adjustbox}

\usepackage{graphics}
\usepackage{amssymb}
\usepackage{amsmath, framed}
\usepackage{braket}
\usepackage{epsfig}
\usepackage{amsmath}
\usepackage{amsfonts}
\usepackage{amssymb}
\usepackage{enumerate}
\usepackage{blkarray}

\usepackage{color,soul}

\usepackage{ifpdf}
\usepackage{	qcircuit}
\usepackage{pifont}
\usepackage{graphicx}
\usepackage{braket}
\usepackage{pifont}
\usepackage{array}
\usepackage{longtable}
\usepackage{bm}
\usepackage{hyperref}
\usepackage{color}
\usepackage{makeidx}
\usepackage{cite}
\usepackage{mathtools}
\usepackage{bbm}
\usepackage{wrapfig}

\usepackage{tikz}
\usepackage{circuitikz}
\usepackage{tikz-cd}
\usetikzlibrary{positioning,shapes.geometric, arrows.meta, decorations.pathmorphing}

\usepackage{verbatim}

\usepackage{listofitems}
\usetikzlibrary{arrows.meta} % for arrow size
\usepackage[outline]{contour} % glow around text
\contourlength{1.4pt}
\colorlet{myred}{red!80!black}
\colorlet{myblue}{blue!80!black}
\colorlet{mygreen}{green!60!black}
\colorlet{myorange}{orange!70!red!60!black}
\colorlet{mydarkred}{red!30!black}
\colorlet{mydarkblue}{blue!40!black}
\colorlet{mydarkgreen}{green!30!black}

\tikzset{
  >=latex, % for default LaTeX arrow head
  node/.style={thick,circle,draw=myblue,minimum size=22,inner sep=0.5,outer sep=0.6},
  node in/.style={node,green!20!black,draw=mygreen!30!black,fill=mygreen!25},
  node hidden/.style={node,blue!20!black,draw=myblue!30!black,fill=myblue!20},
  node convol/.style={node,orange!20!black,draw=myorange!30!black,fill=myorange!20},
  node out/.style={node,red!20!black,draw=myred!30!black,fill=myred!20},
  connect/.style={thick,mydarkblue}, %,line cap=round
  connect arrow/.style={-{Latex[length=4,width=3.5]},thick,mydarkblue,shorten <=0.5,shorten >=1},
  node 1/.style={node in}, % node styles, numbered for easy mapping with \nstyle
  node 2/.style={node hidden},
  node 3/.style={node out}
}
\usepackage{cancel}

\newcommand{\innerproduct}[2]{\langle #1, #2 \rangle}
\newcommand{\Real}{\mathbb{R}}
\newcommand{\Complex}{\mathbb{C}}

\newcommand{\ketpsi}{\ket{\psi}}
\newcommand{\kz}{\ket{0}}
\newcommand{\ko}{\ket{1}}
\newcommand{\bz}{\bra{0}}
\newcommand{\bo}{\bra{1}}

\newcommand{\pmat}[4]{\begin{pmatrix} #1 & #2 \\ #3 & #4\end{pmatrix}}
\newcommand{\bellzz}{\frac{\ket{00} + \ket{11}}{\sqrt{2}}}

\newcommand{\ketbra}[2]{| #1\rangle\!\langle #2 |}

\newcommand{\Tr}{\mathrm{Tr}}

\newcommand{\spn}{\text{span}}

\newcommand{\T}{\textsf{\scriptsize T}}

\newcommand{\inv}{{\,\text{-}\hspace{-1pt}1}}
\newcommand{\ad}{\mathrm{ad}}
\newcommand{\Ad}{\mathrm{Ad}}

\newcommand{\K}{\mathbb{K}}

\newcommand{\M}{\mathcal{M}}

\newcommand{\X}{\mathcal{X}}
\newcommand{\Y}{\mathcal{Y}}
\newcommand{\N}{\mathcal{N}}
\newcommand{\Hilb}{\mathcal{H}}

\newcommand{\Z}{\mathbb{Z}}
\newcommand{\R}{\mathbb{R}}
\newcommand{\C}{\mathbb{C}}
\newcommand{\Km}{\mathbb{K}}

\renewcommand{\u}{\mathfrak u}
\newcommand{\su}{\mathfrak{su}}
\renewcommand{\sl}{\mathfrak{sl}}

\newcommand{\g}{\mathfrak g}
\newcommand{\h}{\mathfrak h}
\renewcommand{\t}{\mathfrak t}
\renewcommand{\a}{\mathfrak a}
\renewcommand{\k}{\mathfrak k}
\newcommand{\p}{\mathfrak p}
\newcommand{\m}{\mathfrak m}

\newcommand{\BH}{\mathcal{B}(\Hilb)}

\newcommand{\sutwon}{\text{SU}(2^n)}
\newcommand{\sutwo}{\text{SU}(2)}

\newcommand{\liesutwo}{\frak{su}(2)}

\newcommand{\liesun}{\frak{su}(n)}

\newcommand{\projdelta}{\text{proj}_\Delta}

\newcommand{\trace}{\text{Tr}}

\newcommand{\tpn}{T_p\M}
\newcommand{\tpndual}{T_p^*\M}
\newcommand{\tpnn}{T_p\N}

\newcommand{\cinfm}{C^\infty(\M)}
\newcommand{\epsint}{(-\epsilon,\epsilon)}
\newcommand{\pardivx}{\frac{\partial}{\partial x^\mu}}
\newcommand{\pardivxdual}{dx^\mu}
    \newcommand{\pardiv}{\left(\frac{\partial}{\partial x^\mu}\right)_p}

\DeclareMathOperator{\tr}{tr}

\DeclareMathOperator{\Prb}{\mathbb{P}}

\newtheorem{theorem}{Theorem}[section]
\newtheorem{axiom}{Axiom}[section]

\newtheorem{proposition}{Proposition}[section]
\newtheorem{definition}[theorem]{Definition} % Definition numbering same as theorem
\newtheoremstyle{mystyle}
  {} % Space above
  {} % Space below
  {\normalfont} % Body font
  {} % Indent amount
  {\bfseries} % Theorem head font
  {.} % Punctuation after theorem head
  {.5em} % Space after theorem head
  {} % Theorem head spec (can be left empty, meaning `normal`)
  
\theoremstyle{mystyle}

\numberwithin{equation}{section}

\usepackage[toc,acronym,nonumberlist,nopostdot,style=super]{glossaries}

\makenoidxglossaries
\newacronym{qip}{QIP}{Quantum Information Processing}
\newacronym{nisq}{NISQ}{Noisy Intermediate-Scale Quantum}
\newacronym{cad}{CAD}{Computer-Aided Design}

\hypersetup{pdfpagelayout=TwoPageRight}

\frontmatter
\begin{document}
\pdfbookmark[0]{Cover}{cover}
\begin{titlepage}
\begin{center}
\vspace{1cm}
{\Large{\textbf{Quantum Geometric Machine Learning}\par}}
\vspace{1cm}
by
\textbf{Elija Perrier} \\
\vspace{0.5cm}

Thesis submitted in fulfillment of the requirements of the degree of\\
\vspace{0.5cm}
\textbf{Doctor of Philosophy}\\
\vspace{0.5cm}
under the supervision of \\
\vspace{0.5cm}
\textbf{A/Prof. Dr. Christopher Ferrie} \\
\vspace{0.5cm}
\textbf{Prof. Dr. Dacheng Tao} \\
\vspace{0.5cm} 
% \textbf{A/Prof. Dr.supervisor3} \\
% \vspace{0.5cm}
University of Technology Sydney \\
Faculty of Engineering and Information Technology \\
\vspace{0.5cm}
May, 2024\\
\end{center}
\end{titlepage}
%%%%%%%%%%%%%%%%%%%%%%%
\newpage\null\thispagestyle{empty}\newpage
% \begin{table}
% \begin{tabular}{|p{14cm}|}
% \hline
% CERTIFICATE OF ORIGINAL AUTHORSHIP \\
% \\
% I, Elija Timothy Perrier declare that this thesis, is submitted in fulfillment of the requirements for the award of Doctor of Philosophy, in the Faculty of Engineering and Information Technology at the University of Technology Sydney. \\
% \\
% This thesis is wholly my own work unless otherwise referenced or acknowledged. In addition, I certify that all information sources and literature used are indicated in the thesis. \\
% \\
% This document has not been submitted for qualifications at any other academic institution. \\
% \\
% This research is supported by the Australian Government Research Training Program. \\
% \\
% Signature: \\
% \\
% Date:\\
% \hline
% \end{tabular}
% \end{table}
\begin{figure}
    \centering
    \includegraphics{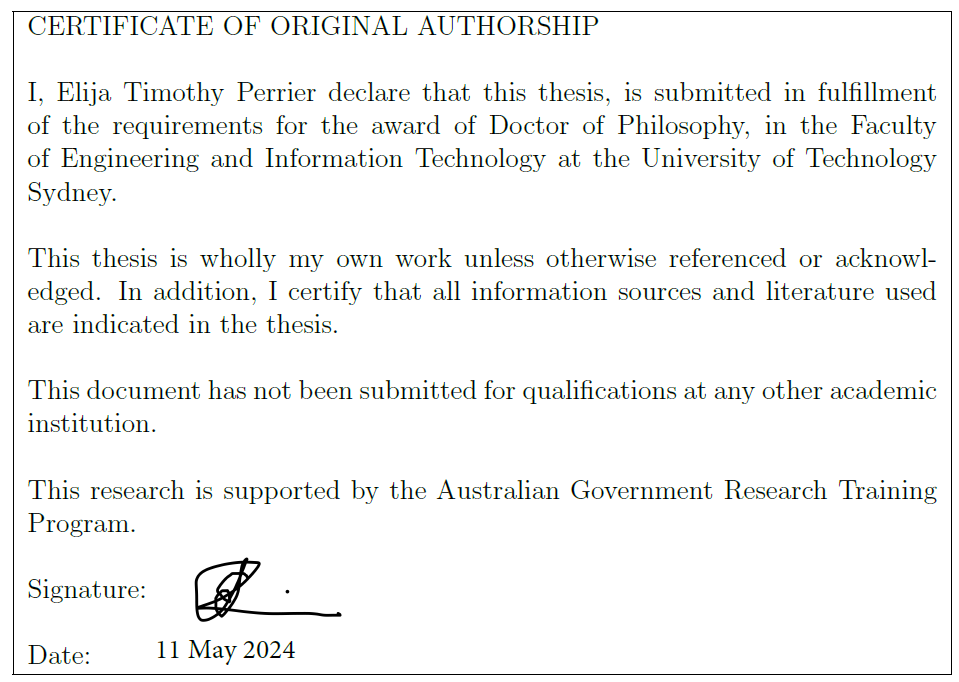}
    \label{fig:enter-label}
\end{figure}
%%%%%%%%%%%%%%%%%%%%%
\pagenumbering{gobble}
\newpage\null\thispagestyle{empty}\newpage
\begin{titlepage}
\begin{center}\Large{\textbf{Acknowledgments}}\end{center}
I would like to thank my principal supervisor Associate Professor Dr. Christopher Ferrie for his patient, instructive and illuminating supervision throughout the tenure of my research. In addition, I would like to thank A/P Ferrie for the opportunity to participate in an AUSMURI collaborative research project while working with senior and experienced researchers across quantum science disciplines. I would like to acknowledge and thank my co-supervisor, Professor Dr. Dacheng Tao, for his counsel and informative advice, especially relating to information theory, machine learning and other computational science. I would like to express particular gratitude to Dr. Christopher Jackson of the University of Waterloo. Dr Jackson's insights and mentorship have had a significant positive impact on both my research and academic career. I would like to acknowledge the significant assistance of the iHPC training facility at UTS, whose resources and expertise were important elements in running large-scale quantum simulations over several months and years. I would like to acknowledge the generous financial support provided by the Australian Government via the Australian Research Training Program scholarship and by the UTS Faculty of Engineering and Information Technology. I would like to thank staff, faculty and students (past and present) at the Centre for Quantum Software and Information, UTS. In particular, I acknowledge the support and discussions with colleagues Professor Dr. Michael Bremner, Associate Professor Dr. Simon Devitt and Professor Dr. Min-Hsiu Hsieh, together with Dr. Akram Youssry, Dr. Arinta Auza, Dr. Maria Quadeer and Lirnad\"e Pira. I would like to also thank my academic colleagues at the Australian National University whose engagement on deep learning, artificial intelligence and other technical matters was highly beneficial, especially Professor Dr. Seth Lazar of the ANU, Professor Dr. Tiberio Caetano of the Gradient Institute and Professor Dr. Kimberlee Weatherill of the University of Sydney. I would like to also thank colleagues at Stanford University, including Mauritz Kop and encouragement from Professor Dr. Mateo Aboy at Cambridge University. 

Finally I would like to thank my family including my mother, Janice Perrier, whose help in caring for our young children (including a newborn) over many months was a lifesaver, and my aunt, Alexsis Starcevich, whose support made that help possible. Also thank you to my father Chris Perrier and his wife Cecilia Basile, and my sister, Mirabai Perrier, who in the final weeks gave her all. 

Most of all, I would like to express unending gratitude to my partner, Paige Newman and my children, Violet Perrier-Newman and Scarlett Perrier-Newman. Their sacrifices in time, space and attention to accommodate and support my research were essential, without which I could not have completed this work, especially during the challenging period of the COVID-19 pandemic through which this thesis was formulated.

\end{titlepage}
\newpage\null\thispagestyle{empty}\newpage
%%%%%%%%%%%%%%%%%%%%%%%
\begin{titlepage}
\vspace*{\fill}
\begin{center}
This is a THESIS BY COMPILATION.	Parts of this thesis have been already published. The content have been edited to suit the formatting of the thesis and to maintain its coherence.
\end{center}
\vfill 
\end{titlepage}
%%%%%%%%%%%%%%%%%%%%%%%%%
\newpage\null\thispagestyle{empty}\newpage
\begin{titlepage}
\begin{center} DECLARATION OF PUBLICATIONS INCLUDED IN THE THESIS \end{center}

\begin{figure}[h!]
    \centering
    \includegraphics{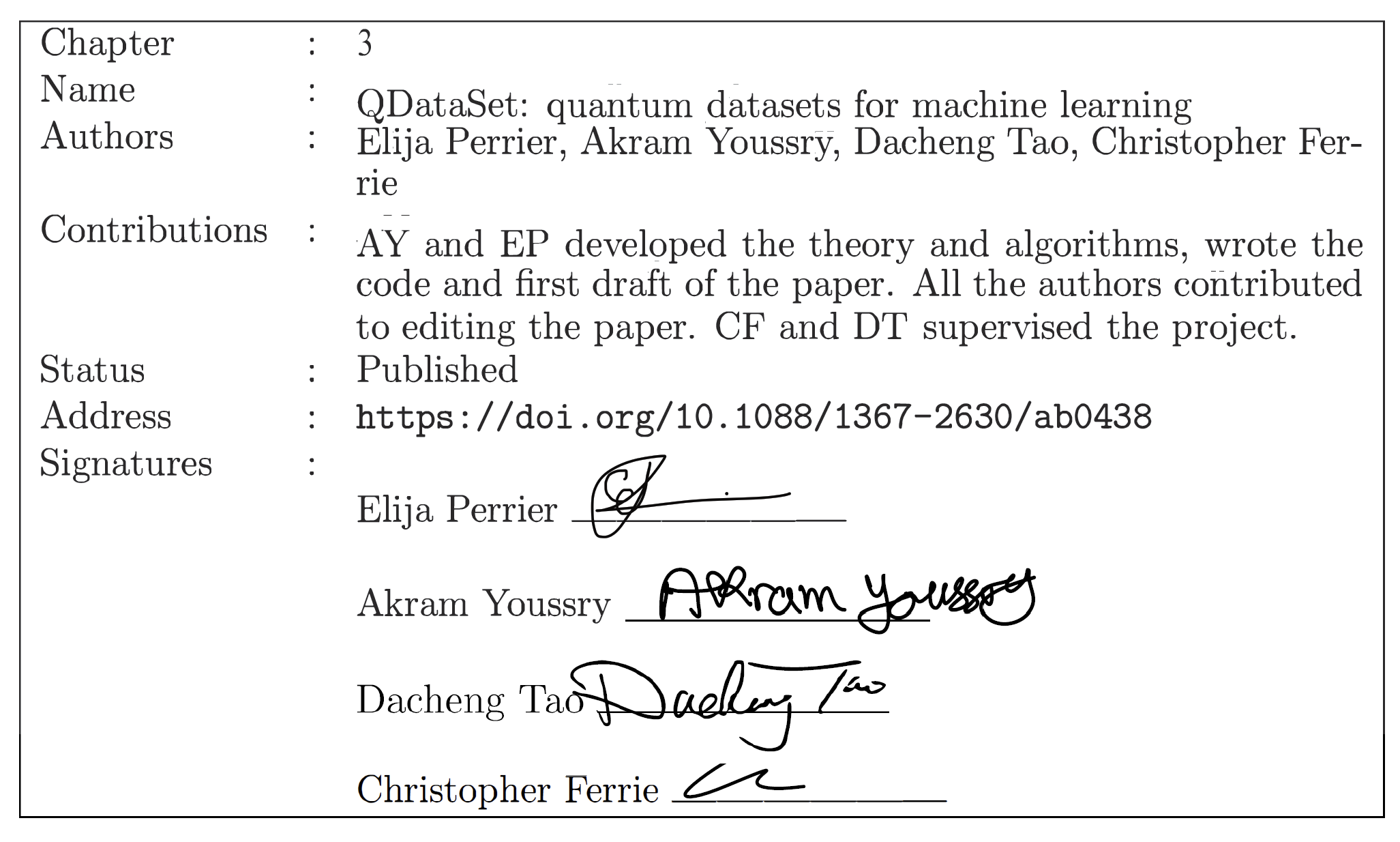}
    % \caption{Caption}
    % \label{fig:enter-label}
\end{figure}

\begin{figure}[h!]
    \centering
    \includegraphics{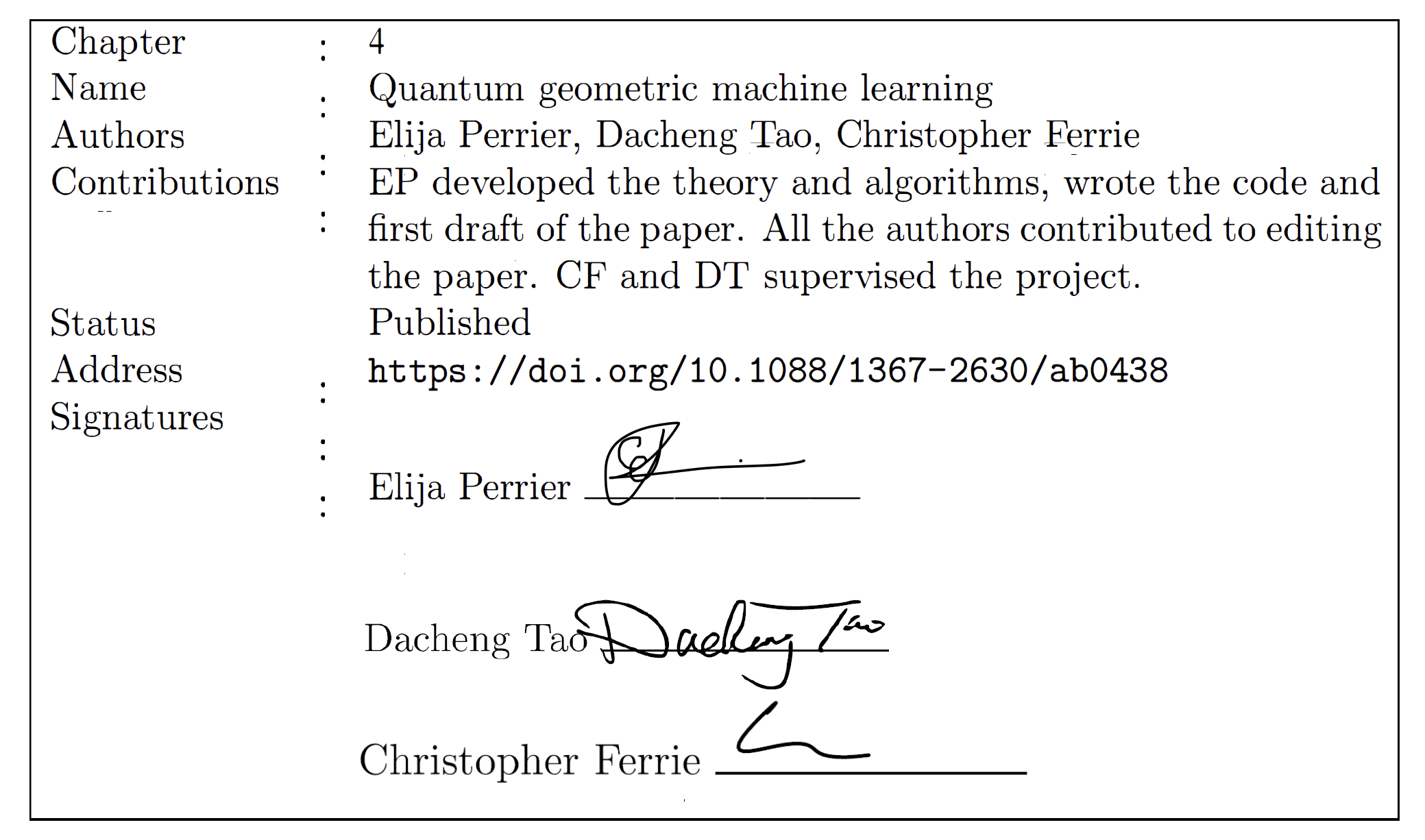}
    % \caption{Caption}
    % \label{fig:enter-label}
\end{figure}

\newpage

\begin{figure}[h!]
    \centering
    \includegraphics{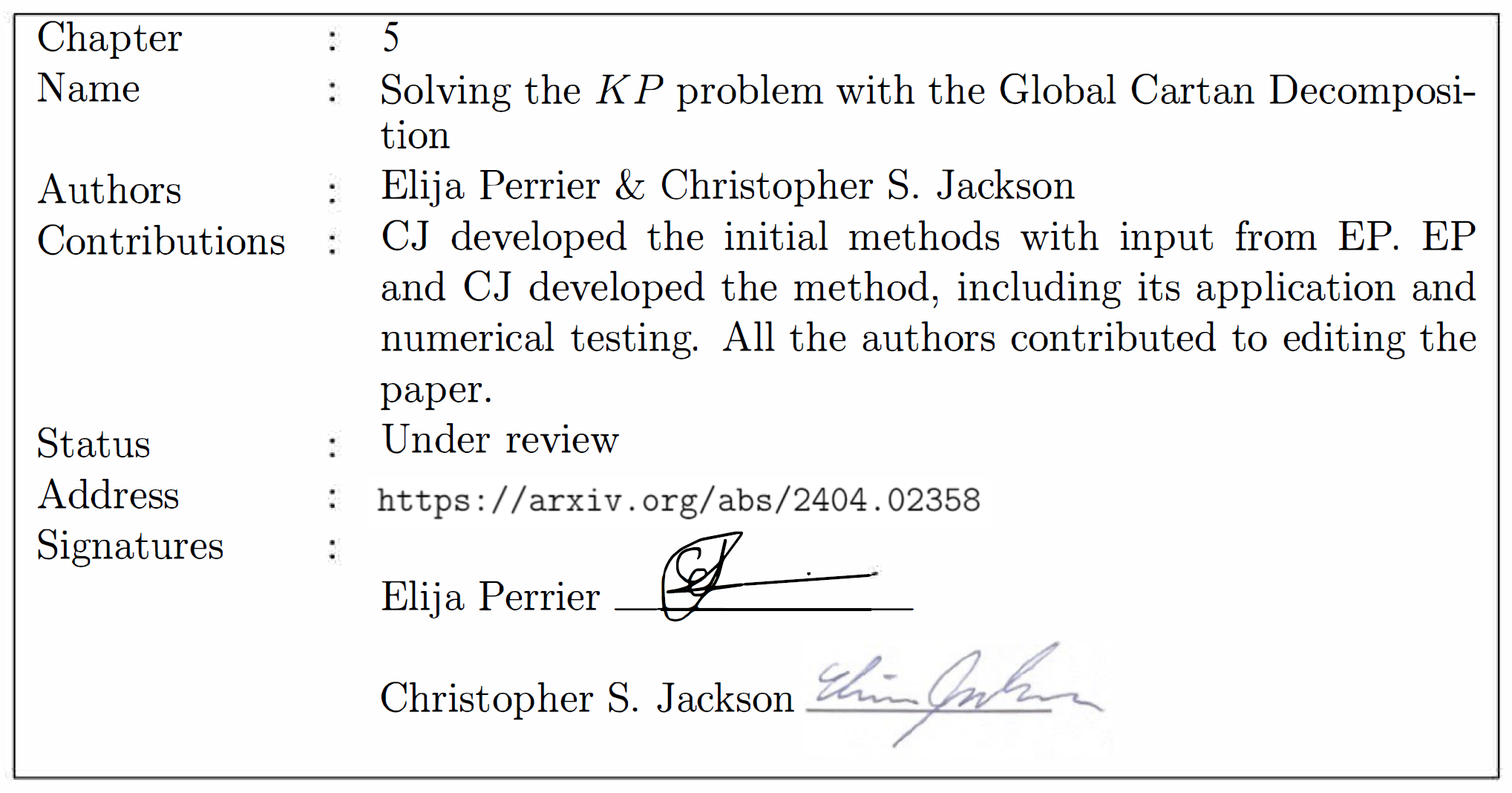}
    % \caption{Caption}
    % \label{fig:enter-label}
\end{figure}

\end{titlepage}
%%%%%%%%%%%%%%%%%%%%%%
\newpage\null\thispagestyle{empty}\newpage
\pagenumbering{roman}
\setcounter{page}{1}
\setcounter{secnumdepth}{4}
\setcounter{tocdepth}{4}
\pdfbookmark{\contentsname}{Contents}
\tableofcontents
\listoffigures
\listoftables
\printnoidxglossary[type=\acronymtype,sort=letter,title=List of Abbreviations]
% \tableofcontents
\mainmatter
% \Chapter*{Abstract}
% \addcontentsline{toc}{chapter}{Introduction}

% \newpage\null\thispagestyle{empty}\newpage
% %%%%%%%%%%%%%%%%%%%%%%%%%%%%%%%%%%%%%%%
% \mainmatter
\chapter{Introduction}

\section*{Abstract}
The use of geometric and symmetry techniques in quantum and classical information processing has a long tradition across the physical sciences as a means of theoretical discovery and applied problem solving. In the modern era, the emergent combination of such geometric and symmetry-based methods with quantum machine learning (QML) has provided a rich opportunity to contribute to solving a number of persistent challenges in fields such as QML parametrisation, quantum control, quantum unitary synthesis and quantum proof generation. In this thesis, we combine state-of-the-art machine learning methods with techniques from differential geometry and topology to address these challenges. We present a large-scale simulated dataset of open quantum systems to facilitate the development of quantum machine learning as a field. We demonstrate the use of deep learning greybox machine learning techniques for estimating approximate time-optimal unitary sequences as geodesics on subRiemannian symmetric space manifolds. Finally, we present novel techniques utilising Cartan decompositions and variational methods for analytically solving quantum control problems for certain classes of Riemannian symmetric space.
\newpage

% \section{Introduction}
\begin{quote}
    Geometry will draw the soul toward truth and create the spirit of philosophy (\textit{Plato})
\end{quote}
\begin{quote}
    There is geometry in the humming of the strings, there is music in the spacing of the spheres (\textit{Pythagoras})
\end{quote}
\begin{quote}
    Spacetime tells matter how to move; matter tells spacetime how to curve (\textit{John Wheeler})
\end{quote}
\begin{quote}
    One geometry cannot be more true than another; it can only be more convenient (\textit{\'Elie Cartan})
\end{quote}

\section{Overview}
This thesis introduces \textit{quantum geometric machine learning} (QGML), presenting results in quantum control and quantum machine learning using techniques from differential geometry and Lie theory. Our work, as we discuss below and throughout this thesis, represents a synthesis of four distinct but related disciplines: quantum information science (focusing specifically on quantum computing and quantum control), abstract algebra and representation theory, differential geometry and quantum machine learning. By combining insights and methods from these fields, we have sought to leverage theory and architectures to explore how hybrid quantum-classical systems can simulate quantum systems, learn underlying geometric symmetries and be used as tools for optimisation problems. Throughout this thesis, we have endeavoured to connect the relevance of these important concepts to quantum information processing, control and machine learning in a way that is accessible to a cross-disciplinary audience. Synthesising techniques across disciplines can be a challenging and at times unwieldy task as drawing precise maps between concepts faces a thicket of jargon, terms of art and disciplinary convention that characterise any academic field. To add clarity to the process, we therefore foreshadow below how each discipline relates to the overall contributions of this thesis.
\begin{enumerate}[(i)]
    \item \textit{Quantum information processing}. The primary focus of this thesis is to address problems in quantum information processing, specifically problems related to simulating quantum systems and the control of quantum systems. Quantum information processing is a vast discipline, incorporating quantum computing, quantum communication and quantum sensing. Our focus is on the first of these, quantum computing, but we retain a general information-theoretic framing given the overlap with information sciences through quantum machine learning theory. The targets and objective of the work in this thesis are the construction of operators and computations represented by unitary operations $U(t)$ acting on quantum states belonging to a Hilbert space $\ket{\psi(t)} \in \Hilb$ that evolve the state over times $t=0,...,T$ to a target end-state at time $T$ given by $\ket{\psi(T)}=U(T)\ket{\psi(0)}$ from some initial state $\ket{\psi(0)}$ and which meet certain optimality criteria, such as minimal time or energy. The particular states of interest are mostly single- and multi-qubit systems (such that $G=SU(2)$ or a tensor product thereof), but our work also encompasses higher-order dimensions of quantum system (such as qudits related to $SU(3)$).
    \item \textit{Algebra}. Algebra, by which we primarily mean the theory of continuous Lie groups and representation theory, informs this thesis in many ways but its main relevance is because, at least for closed quantum systems, unitary operations the discovery of which is the primary objective above form a Lie group $G$ whose corresponding Lie algebra $\g$ is the basis for constructing the quantum mechanical Hamiltonians governing the evolution of quantum systems. To this end, we explore the deep connections between algebra and symmetry reflected in the work of Cartan and others via concepts such as Cartan decompositions, abstract root systems and diagrammatic tools such as Dynkin diagrams.
    \item \textit{Geometry}. Geometry enters the thesis as a way to frame problems of interest such that we can leverage specific results in the theory of classical (and quantum) geometric control theory. In geometric terms, the unfolding of a quantum computation via the representation of unitary evolution can be construed as the tracing out of a curve on a differentiable manifold $\M$ corresponding to the Lie group $G$. The evolution of the curve is generated by vectors in the tangent space $T\M$ which can be equated with the said Lie algebra $\g$ group generators above. Time or energy-optimal (minimal) curves correspond to geodesics, so the search for time-optimal quantum circuits or evolution becomes a question of using the machinery of differential geometry to find geodesics with minimal arc-length $\ell$. For certain classes of Lie group $G$, the symmetry properties of those groups allow the corresponding manifold $G/K$ (for a chosen isometry group $K$) to be construed as a Riemannian (or subRiemannian) symmetric space admitting a Cartan decomposition $\g = \k \oplus \p$. In such cases, results from geometric control theory show that general optimality (Pontryagin Maximum Principle) criteria can be met under assumptions about the Lie group (and Lie algebra) being fully reachable under the operation of the Lie derivative (commutator). Moreover, the symmetry properties of such groups (specifically their partition into horizontal $H\M$ and vertical $V\M$ subspaces corresponding to the subalgebras $\p$ and $\k$ respectively) together with the Lie triple property $[\p,[\p,\p]] \subseteq \p$ mean that we can make simplifying assumptions about (a) the existence of curves $\gamma(t)$ between $U(0)$ and $U(T)$ on $\M$ and (b) the uniqueness (via being minimal length) of such curves, such as when those curves generated by Hamiltonians drawn from $\p$ only. In this case, the optimisation problem is considerably simplified, becoming a question of finding the minimal (optimal) (in terms of energy or time) control functions $u(t)$ which apply to such generators to evolve curves. The final substantive Chapter \ref{chapter:Time optimal quantum geodesics using Cartan decompositions} provides a new method for calculating such time-optimal curves for use in quantum information.
    \item \textit{Quantum machine learning}. Quantum and classical machine learning enter the thesis as a way to then solve for this specific problem of finding optimal control functions $u(t)$ by training hybrid classical-quantum algorithms on data such as data about quantum geodesics. The quantum element of machine learning enters via a hybrid quantum-classical stack whereby the rules of quantum evolution (such as via Schr\"odinger's equation and via Hamiltonians comprising Lie algebra elements $\g$) are encoded in the neural network stack. The neural network architecture then seeks to leverage classical machine learning to learn control functions $u(t)$ to achieve objectives of synthesising $U(T)$ (such as via maximising fidelity among the model estimate $\hat U(T)$ and $U(T)$ itself). We call this hybrid model a \textit{greybox} model of quantum machine learning as it incorporates a whitebox of both known facts (such as the laws of quantum mechanical state evolution) and a blackbox in terms of unknown non-linear parameters. We show the utility of this approach in Chapter \ref{chapter:QDataSet and Quantum Greybox Learning}, where we evidence its benefits in constructing a large-scale quantum computing dataset of specific relevance to researchers in both closed- and open-quantum systems research and in Chapter \ref{chapter:Quantum Geometric Machine Learning}, where we show how for certain classes of problem, the greybox method is more effective at learning the geometric and geodesic structure of time-optimal unitaries from data than other methods.
\end{enumerate}

Our contribution to the cross-disciplinary pollination among these disciplines is necessarily specific to our problems of interest. Nevertheless, we hope that this thesis makes a modest and useful contribution to joining the dots as it were between disciplines and serves as a motivation for researchers within such disciplines to explore the rich tapestries of complementary methods available to them across the academic aisles.

\subsection*{Historical background}\label{sec:intro:Historical background}
Geometric methods have a long lineage in quantum mechanics and quantum information. The historical development of late 19th and early 20th century physics was informed in tandem with and significantly by geometric techniques, from general relativity, to quantum mechanics. Geometry informed much of 19th century mathematical physics, including Hamilton's reformulation of Newtonian mechanics, Riemann's work on differential geometry and Klein's \textit{Erlangen} program. As we explore below, the work of Poincar\'{e}, Cartan, Weyl and others was significantly influenced by revolutionary geometric ideas. In modern quantum field theory and other disciplines, geometry has and continues to play a significant role. The synthesis of geometry with other fields of mathematics, including group theory, analysis and number theory has in turn opened up rich seams of insight via cross-disciplinary pollination. What is perhaps not often emphasised in modern formalism is how the early pioneers of quantum mechanics were profoundly influenced by the emerging tools of non-Euclidean geometry and its relationship with symmetry (see \cite{hawkins_emergence_2012} for a detailed history). For context, we explore a bit more historical background below, before proceeding to a summary of contributions of this thesis and a summary of each Chapter and the supplementary Appendices. 

Understanding the historical development of geometric methods is a useful \textit{aid\'e-m\'emoire} for situating the modern adoption of differential geometry within quantum information and machine learning contexts. The development of geometry and algebra, two of the four pillars of modern mathematics \cite{gowers_princeton_2010}, has been a tandem one as each discipline influenced the other via through methodologies and conceptual abstraction. Philosophically, the relationship between algebra and geometry has exhibited a symbiotic waxing and waning throughout the development of mathematics' early formalism via the then-canonical ancient distinctions of the science of magnitude (geometry) and the science of multitude (arithmetic) \cite{detlefsen_formalism_2007}. Traditionally, geometric and algebraic (or what today would be regarded as proto-algebraic) proofs or demonstration took relatively distinct forms. For example, Euclid's geometric proofs in the \textit{Elements} \cite{heath_thirteen_1956} were characterised by geometric demonstrative visual methods, while alternative proofs, such as the celebrated proof of the infinity of primes, relied upon explicit logical inference and deduction.
Over successive centuries, the distinctions between geometric and algebraic methods began to blur as scholars uncovered significant connections between the fields. An example includes Galois's infamous relation of geometric constructions (in terms of canonical ruler/compass structures) to symmetry properties of polynomials and groups. Similarly, significant contributions by Gauss, Riemann, Cayley and others helped set the scene for the emergence of later seminal results at the end of the 19th and start of the 20th centuries and for the interplay of algebraic and geometric techniques drawn upon by modern physics and information science.
%
% \subsubsection{Erlangen programme}
Of particular note, the late nineteenth-century \textit{Erlangen Program} of Felix Klein \cite{gray_felix_2005} mentioned above put forward a systematic means of describing geometries in terms of their symmetry properties, as described by the mathematics of groups. Klein's innovative approach unified diverse geometric concepts under a common framework that significantly informed the subsequent development of mathematical fields and physical sciences. In particular, the unification of geometry and group theory via the identification of geometries with specific transformation groups originally developed by Sophus Lie has had a profound impact on the development of not just mathematics, but also classical and quantum physics. Lie's \textit{id\'ee fixe} in describing the transformation of geometric objects in terms of underlying symmetries formed a primary motivation for his later development of the theory of continuous transformation groups (Lie groups). Together with other developments, Lie groups thus became an indispensable fulcrum through which the mathematical pillars of geometry, analysis (especially the study of continuity) and geometry were tied together with spectacular results across the mathematical sciences. \\
\\
An example of an important discovery across disciplines \cite{sharpe_differential_2000} is Klein's insight that non-Euclidean geometries could be recognised as examples of coset spaces $G/H$ (for Lie groups $G$ and subgroups $H$) seemed incompatible with the other central trend of nineteenth century geometry, the development of Riemannian geometry (owing to a focus on discrete algebraic structures in the former case and reliance upon smoothly varying metrics and continuous curvature in the latter case). Klein's results promoted a synthesis of algebraic, Lie theoretic, and geometric methods within the then-emerging field of representation theory, providing a powerful framework for analysing structural symmetries of mathematical and physical systems. As Knapp \cite{knapp_lie_1996} notes, the development of representation theory and Lie theoretic-methods has taken diverse approaches within the literature. Some tomes have emphasised primarily the algebraic and group-theoretic basis of Cartan decompositions and optimisation. Others, such as the work of Helgason \cite{helgason_differential_1962,helgason_differential_1979}, have followed more closely Cartan's genesis, an approach that elucidates the concurrent development of differential-geometric principles and Lie theory together. Of particular note for this thesis is the oeuvre of Cartan, specifically the unification of geometry, symmetry and algebra via Cartan's seminal classification of Riemannian symmetric spaces according to the taxonomy of semi-simple Lie groups is regarded as one of the 20th century's most elegant mathematical achievements. Moreover, the emerging synthesis of geometry and algebra had profound impacts on physics itself where geometric and Lie algebraic principles were usefully adopted across the field, reaching an apotheosis of sorts in the classical theory of general relativity. Similarly, non-commutative algebra and geometry played a significant role in the development of quantum mechanics in the early part of the 20th century.

\subsection*{The Nature of Learning}
The search for symmetry has also shaped computational science in profound ways since its inception, including via techniques to optimise network flows, analyse data, solve cryptographic problems and facilitate error correction. More recently, the advent of quantum information sciences \cite{nielsen_quantum_2011}, specifically quantum computing, has breathed new life into the use of symmetry reduction techniques drawn from these fields as researchers seek new methods to overcome challenges posed by computational intractability or resource constraints. The emergence of sophisticated machine learning methods and technologies has also brought about an impetus to explore how well-established mathematics of symmetry can be leveraged together with learning protocols in order to solve optimisation problems in quantum information science in general and quantum control specifically. The nature of learning in classical and quantum information science is a rich and widely studied problem \cite{hastie_elements_2013,vapnik_nature_1995}. At the heart of theories of learning are methods, protocols or even unknown phenomena that in some way increase the knowledge of a system about the world or some ground truth. In this sense quantitative sciences of information inhere a certain philosophical stance or claim about what learning is. In the abstract, learnability of systems is (or can be) construed using stateful representation, in which the knowledge or information of a system represents an epistemic state, such as one encoded within the parameters of a neural network \cite{goodfellow_deep_2016} or quantum algorithm, as a means of in effect quantifying (together with, for example, rules of inference) the knowledge or information of a system. Yet information is not the sole measure relevant to learning: knowledge or information \textit{about} the state of the world is key as assessed by figures of merit such as prediction metrics (such as loss functions), which most often stand in as proxies for a system's epistemic state. 

Thus a system that randomly assimilates large amounts of information with poor predictive capacity (that one which fails to generalise or one which overfits) is considered one with less capacity to learn (or that has learnt less) than one with less information but greater generalisability, in both classical and quantum contexts. To learn thus involves complex interactions between the amount of information a system can represent on the one hand to enable its versatile generalisability and the extent to which that system develops sufficiently accurate representations about the world in order to make accurate predictions, something manifest in practice in bias-variance trade and studied formally within statistical learning theory. A learning protocol is thus one which increases this epistemic state and enables the learning of some structure and predictive improvement. Thus learning can be construed, and in computational science usually is construed, inherently as an optimisation task given objectives and data about the world. Machine learning, classically, involves the search for and design of models which satisfy such epistemic objectives to optimise this learning task using classical data and algorithms. Quantum machine learning has the same overall aim, yet is complicated by the unique characteristics of quantum information (as manifest in properties of quantum data and quantum algorithms), imposing constraints upon how learning occurs in a quantum context.

\section{Related work}
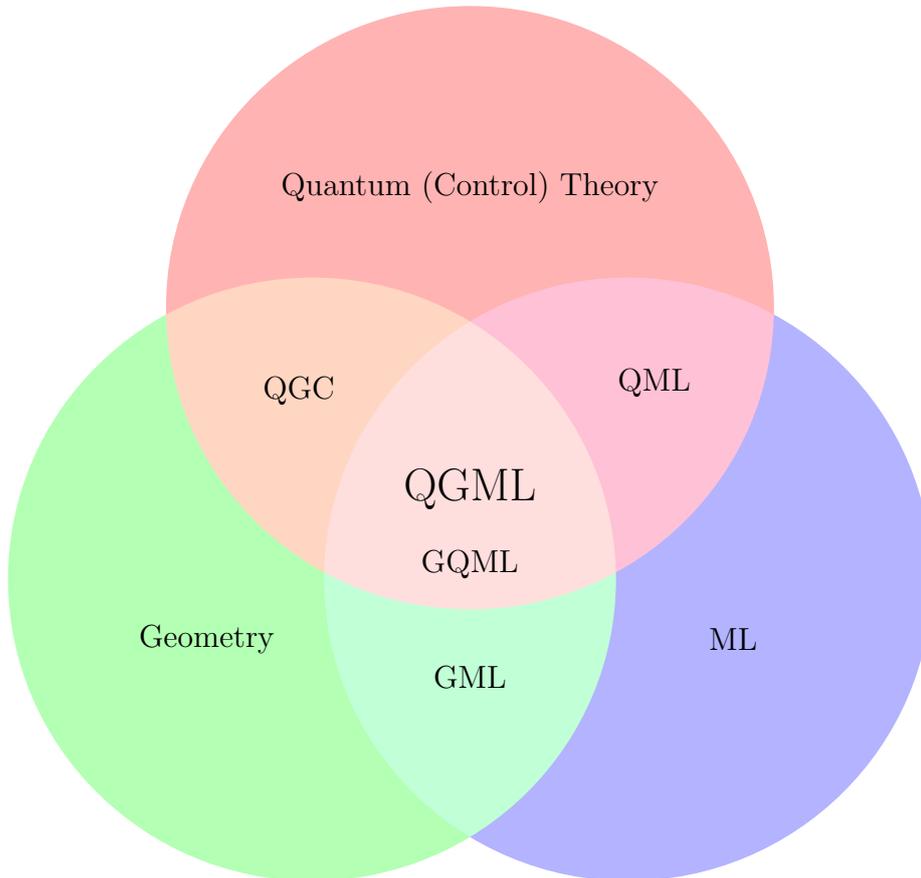
\begin{figure}[h!]
    \centering
    \begin{tikzpicture}[scale=2]
  \begin{scope}[blend group = soft light]
    \fill[red!30!white]   ( 90:1.2) circle (2);
    \fill[green!30!white] (210:1.2) circle (2);
    \fill[blue!30!white]  (330:1.2) circle (2);
  \end{scope}
  \node at ( 90:2)    {Quantum (Control) Theory};
  \node at ( 150:1.30)    {QGC};
  \node at ( 90:-1.25)    {GML};
  \node at ( 210:2)   {Geometry};
  \node at ( 330:2)   {ML};
  \node at ( 210:-1.40)   {QML};
  \node [font=\Large] {QGML};
  \node at (90:-0.5) {GQML};
\end{tikzpicture}
    \caption{Venn diagram describing overlap of quantum information processing and control theory (Quantum Control Theory), Geometry and Machine Learning (ML). QGML sits at the intersection of quantum computing and control theory, geometry and machine learning. Related fields, including quantum geometric control (QGC), geometric machine learning (GML) and quantum machine learning (QML) can be situated accordingly. The most overlap between existing literature and the present work on QGML is geometric quantum machine learning (GQML) which seeks to encode symmetries across entire quantum machine learning networks.}
    \label{fig:intro-venndiagram}
\end{figure}

As foreshadowed above, symmetry and geometric methods have been an integral part of the development of modern mathematics and physics since ancient times, thru to the development of classical physics and the revolution in quantum physics of last century. The thesis draws upon this lineage, intersecting with quantum information, Lie theory and differential geometry to problems in quantum control and statistical learning. For contextualisation, we situate QGML within the intersection of these existing disciplines as set out in the Venn diagram in Figure \ref{fig:intro-venndiagram}. Our work primarily is focused on quantum information theory with a focus on quantum control theory (QCT), both in terms of simulating open quantum systems (not only for control) and time-optimal control for optimal unitary synthesis. In QCT (see Appendix \ref{chapter:Background: Quantum Information Processing} and section \ref{sec:quant:Quantum Control}), we are primarily interested in how to apply controls in order to obtain quantum states belonging to Hilbert spaces $\ket{\psi(t)} \in \Hilb$ ($\rho(t) \in B(\Hilb)$). For convenience, we rely upon the unitary operator (channel) representation of states $\rho(t) = U(t)\rho(0)U(t)^\dagger$, hence the quantum state of interest is represented via the unitary operator applied to the initial quantum state $\rho(0)$ at time $t=0$. Quantum state evolution is given by Schr\"odinger's equation $dU/U = -iHdt$ the solutions to which belong to unitary or special unitary groups $G$. We rely (as is often the case in QCT literature) upon time-independent approximate solutions to Schr\"odinger's equation. In QCT, the generic quantum control problem is described via a Hamiltonian $H$ comprising a drift part $H_d$ and a control Hamiltonian $H_c$ part:
\begin{align}
    H = H_d + \sum_k H_k u_k \qquad \dot U =-i \left(H_d + \sum_k H_k u_k \right)U
\end{align}
where initial conditions are chosen depending on the problem at hand (where time dependence is understood). The QCT problem becomes how to select controls $u_k \in \R$ (bounded) to reach some target $U(T)$ at time $t=T$ ideally in a time or energy-optimal fashion \cite{dalessandro_introduction_2007,dalessandro_dynamical_2021,dalessandro_k-p_2019,albertini_controllability_2018,jurdjevic_controllability_1978}. The algebraic and Lie theoretic features of this problem enter quantum information-based control theory because quantum computation in general must be unitary (or represented by unitary channels) in order for quantum state and quantum data (and measure) to be preserved (as distinct from collapsing to classical data). This is represented via the computations of interest being described as unitary group elements $U(t) \in G$ where $G$ is a unitary or special unitary group (see definitions \ref{defn:quant:Unitary Operator}, \ref{defn:alg:Unitary matrix and unitary group} and sections \ref{sec:quant:Quantum evolution} generally). As a result, this necessitates that Hamiltonians $H(t)$ in Schr\"odinger's equation (definition \ref{eqn:quant:schroddensitycontrol1}) (whose solutions are such unitaries $U(t)$) must (in the closed form case) be comprised of elements from the corresponding Lie algebra $\g$ which generate those group elements (note we discuss the impact of noise in sections below). Control theory is then characterised by two primary objectives: (a) determining whether the system is \textit{controllable}, that is, whether the desired or target computation or end point can be reached given the application of a control system; and (b) strategies for determining the optimal control for satisfying an objective such as energy or time minimisation.

The theory of quantum geometric control (QGC), such as early work in quantum control leveraging Cartan decompositions in NMR contexts \cite{khaneja_cartan_2000,khaneja_time_2001,khaneja_sub-riemannian_2002}, derives from a mixture of algebraic techniques (such as Cartan decompositions) and classical geometric control theory \cite{jurdjevic_geometric_1997,boscain_k_2002,jurdjevic_non-euclidean_1995,jurdjevic_control_1981}. The fulcrum concept here is the seminal correspondence between Lie groups $G$ and Lie algebras $\g$ on the one hand and differential manifolds $\M$ and tangent spaces $T\M$ on the other. To this end, this work fits within an emerging literature on the application of symmetry reduction and symmetry-based optimisation methods applied to quantum machine learning. To solve this optimisation problem (and indeed to simulate open and closed quantum systems for quantum control as per Chapter \ref{chapter:QDataSet and Quantum Greybox Learning}), we leverage techniques in control theory, geometry and quantum machine learning. Our work involves leveraging machine learning (ML) techniques via a hybrid quantum-classical approach adopting quantum machine learning (QML) principles. It focuses on using parametrised quantum circuits (see section \ref{sec:ml:Variational quantum circuits}) in neural network architecture which encodes structural features (such as unitarity of outputs) while enabling machine learning optimisation. 

Other related fields to our work include geometric machine learning (GML) \cite{bronstein_geometric_2021} where machine learning problems, such as parameter manifolds and gradient descent, are studied from an information-theory perspective, bearing similarity with geometric information science \cite{amari_information_2016} and Lie group machine learning \cite{li_lie_2019}. Probably the field with the most overlap with QGML is that of geometric QML (GQML) \cite{larocca_group-invariant_2022,larocca_diagnosing_2021,larocca_theory_2021,larocca_exploiting_2020,nguyen_theory_2022,cerezo_challenges_2022,cerezo_variational_2020,ragone_representation_2023,caro_generalization_2022} a recently-developed area which studies how symmetry techniques can be used in the design of or embedded within quantum machine learning architectures (usually parametrised or variational quantum circuits) in ways such that those networks respect to varying degrees those underlying symmetries. GQML techniques have also been shown to have import in addressing aspects of the barren plateau problem in QML. GQML is probably the most similar field to the work herein due to its use of dynamical Lie algebras in QML network design.  Most of the literature in the field is concerned with constructing quantum machine learning architectures (e.g. variational quantum circuits) which respect compositionally certain symmetries by encoding Lie algebraic generators within VQC layers, whereas the work in Chapter \ref{chapter:Quantum Geometric Machine Learning} for example does not seek an entire network that respects symmetry properties, but rather only a sub-network of layers whose output (e.g. in the form of unitary operators $U(t)$) respects symmetries inherent in unitary groups. Our work also takes on a more differentially geometric flavour than GQML and also emphasises the relationship with formal control theory to some degree more than GQML literature usually does. Moreover, our work in that Chapter is concerned with design choices to enable the network to learn geodesic approximations via simplifying the problem down to one of learning optimal control functions.

\section{Contributions}
In Chapters \ref{chapter:QDataSet and Quantum Greybox Learning}, \ref{chapter:Quantum Geometric Machine Learning} and \ref{chapter:Time optimal quantum geodesics using Cartan decompositions}, we present three distinct but connected contributions to the literature on geometric techniques in quantum machine learning. The first is the presentation of a large-scale synthetic dataset of open quantum systems evolving in the presence of noise, which we denote the QDataSet (Chapter \ref{chapter:QDataSet and Quantum Greybox Learning}). While quantum machine learning has evolved for more than a decade, the field has been characterised by a sparsity of large-scale datasets from which to benchmark and develop algorithms (as is the case classically). To remedy this gap, the QDataSet was produced. It presents a novel machine learning architecture that combines both blackbox and whitebox techniques into a so-called greybox architecture, the underlying rationale being that learning tasks are more efficient when algorithms are encoded with prior information. 
The next novel contribution of this thesis, in Chapter \ref{chapter:Quantum Geometric Machine Learning}, is the demonstration of the use of quantum machine learning techniques for learning approximate time-optimal unitary sequences in the form of geodesics over multi-qubit manifolds $SU(2^n)$. The results in that Chapter evidence the learnability, in certain cases, of sequences of unitary operators $(U_j) \in G$ (where $G \simeq SU(2)$) and multi-qubit variations which constitute geodesics along the implicit Lie group manifold $G$. As a result, they may be interpreted as time-optimal sequences for generating target unitaries $U_T \in G$. Furthermore, we validate the utility of an experimental `greybox' approach to hybrid quantum-classical machine learning architecture for solving the learning problem, demonstrating its benefits over blackbox methods. Code for the algorithms (and benchmarking examples) is available via a repository in line with principles of open access and reproducibility.

The final contribution of this thesis is in Chapter \ref{chapter:Time optimal quantum geodesics using Cartan decompositions}, where we present a novel analytical method for determining Hamiltonians that generate time-optimal unitary sequences for certain classes symmetric spaces. These results draw upon deep connections between algebra and geometry as pioneered by the work of Cartan and subsequently developed by others across geometric control theory and quantum control. The results show that under reasonable simplifying assumptions related to the Cartan decomposition of the Lie group manifold $G$, Hamiltonians for generating certain time-optimal unitaries $U_T \in G$ can be analytically determined using variational means. The novel result (where $G/K$ (for an isometry subgroup $K$) can be construed as a globally Riemann symmetric space) provides a means to tackle one of the key barriers to learning time-optimal unitary sequences, namely the exponentially hard problem of globally minimising over all geodesic paths.

\section{Chapter Synopses}
In this section, we set out a synopsis of each Chapter below. The first five Chapters are self-contained and should be regarded as the thesis proper. Note that we have included both a background theory Chapter (see Chapter \ref{chapter:Background: Background Theory}) and extensive supplementary background information in the form of Appendices. The rationale for this is that the subject matter of this work is inherently multi-disciplinary in nature, synthesising concepts from quantum information processing with those in algebra, geometry and machine learning. Each supplementary Appendix is tailored to provide additional background material in a relatively contained way for readers whom may be familiar with some, but not all, of these diverse scientific disciplines. The Appendices reproduce or paraphrase standard results in the literature with source material identified at the beginning of each Appendix. Proofs are omitted for brevity but can be found in the cited sources and other standard texts. The substantive Chapters \ref{chapter:QDataSet and Quantum Greybox Learning}, \ref{chapter:Quantum Geometric Machine Learning} and \ref{chapter:Time optimal quantum geodesics using Cartan decompositions} have been tailored to cross-refer to the Appendices' sections, definitions, theorems and equations in order to assist readers who may wish to delve deeper.

Supplemental Appendix \ref{chapter:background:Differential Geometry} (Differential Geometry) goes into slightly greater detail than other Chapters in order to lay out the theoretical underpinnings of geodesics, measuring distance and ultimately optimisation of unitary sequences via minimisation of arc-length. One of the observations throughout the years during which this thesis was produced was that while researchers in each discipline, such as quantum information processing or machine learning, generally found the key concepts of the other accessible, this was much less so for geometry and representation theory. Geometry, especially coordinate-free non-Euclidean geometry, in particular is a thicket of concepts wrapped in idiosyncratic terminology such as bundles, pullbacks, pushforwards, forms and so on. While this is not unique to geometry as a discipline, it was observed that this generally added to its opacity and difficulty in being understood or usefully adopted by researchers outside the confines of mathematical research. This opacity also, in the author's opinion at least, somewhat contributes to the lack of appreciation of how intertwined the development of geometry and quantum mechanics was in the late 19th and early 20th centuries, where, in the modern era, the connection of geometry to familiar tools such as commutation (or structure constants) is obscured, but becomes apparent for example when considering Cartan's structural equations (see theorem \ref{thm:geo:Cartanstructuralequations}) or the Riemannian curvature tensor (definition \ref{defn:geo:Riemann curvature tensor}).

\subsection*{Chapter \ref{chapter:Background: Background Theory} (Background Theory)}
This Chapter provides an abridged summary of background theory across quantum information processing, Lie theory and representation theory, differential geometry and classical / quantum machine learning relevant to the substantive results in Chapters \ref{chapter:QDataSet and Quantum Greybox Learning},\ref{chapter:Quantum Geometric Machine Learning} and \ref{chapter:Time optimal quantum geodesics using Cartan decompositions}. The Chapter begins by surveying quantum information processing, covering elementary axioms of quantum mechanics as they apply to quantum computing. It covers operator formalism, evolution of quantum states, measurement, quantum control and open quantum systems. In addition it introduces concepts of quantum registers and quantum channels. It then covers algebra and Lie theory, including Lie groups, representation theory, Cartan algebras, root systems and Dynkin diagrams. Following this, it summarises basic principles of differential geometry, including differential manifolds, vector fields, tensor fields, tangent planes, fibre bundles, geodesics, Riemannian and subRiemannian manifolds and symmetric spaces. It also covers elementary theory of geometric control. The final section summarises principles of machine learning from classical and quantum perspectives, including statistical learning theory and variational quantum circuits.

\subsection*{Chapter \ref{chapter:QDataSet and Quantum Greybox Learning} (QDataSet and Quantum Greybox Learning)}
This Chapter presents material published in relation to the \textit{QDataSet}, a large-scale synthetic dataset of open quantum systems set out in \cite{perrier_qdataset_2022}. The Chapter explores the importance of large-scale datasets in machine learning for the development of novel algorithms relevant to theoretical and applied problems in the field. It sets out theory and implementation of open quantum systems models used to develop the dataset, together with background information on open quantum systems noise protocols and the theory of open quantum systems. It presents examples of how to use the dataset (along with code) related to common tasks such as quantum tomography and quantum control. The related repository contains code for generating the dataset, together with example Python code and worked examples for using the dataset for benchmarking in a variety of use cases. As at the date of this thesis, the paper and dataset has been referenced in over 20 publications.

\subsection*{Chapter \ref{chapter:Quantum Geometric Machine Learning} (Quantum Geometric Machine Learning)}
This Chapter presents results related to machine learning architectures for learning time-optimal subRiemannian unitary geodesics $U_T \in SU(2^n)$ set out in \cite{perrier_quantum_2020}. The work surveys geometric methods in quantum computing, including the work of Nielsen and others applying variational techniques to circuit synthesis. It reviews the underlying theory canvassed in earlier Chapters as applicable to the case of multi-qubit circuit synthesis. It presents machine learning architectures and experimental protocols for learning geodesic structure from underlying training data constructed from horizontal distributions belonging to relevant Lie algebras $\Delta \in \su(2^n)$. A raft of machine learning architectures are presented in order to demonstrate the utility of `greybox' algorithm design (which incorporates known information into stochastic machine learning architecture). Results of experiments are presented along with algorithm pseudocode and links to source-code material. 

\subsection*{Chapter \ref{chapter:Time optimal quantum geodesics using Cartan decompositions} (Global Cartan Decompositions for $KP$ Problems)}
The final Chapter presents novel results set out in \cite{perrier2024solving} related to the use of Cartan decompositions for solving certain unitary sequence optimisation $KP$ quantum control problems for targets $U_T \in G=K \oplus P$ (for symmetric subgroup $K$ and antisymmetric subgroup $P$), where $G/K$ is a symmetric space and $G$ is equipped with a Cartan decomposition $G=KAK$. The Chapter develops background theory presented in earlier Chapters related to symmetric spaces and semi-simple Lie groups. It connects this theory to the $KAK$ representation of unitary targets and unitary evolution. It posits a constant-$\theta$ ansatz where $\theta$ parametrises generators drawn from the subset $\a \subset \p$ from a maximally non-compact Cartan subalgebra $\h \subset \g$. It provides a general proof and method that uses Cayley transformations and variational (Lagrangian) methods to show how the global optimisation problem can, for certain classes of symmetric space, be analytically solved. It also presents results related to root systems and quantum control. 

\subsection*{Appendix \ref{chapter:Background: Quantum Information Processing} (Quantum Information Processing)} This supplemental Appendix contains the background theory and context for quantum information processing and quantum control. The first part presents a short recapitulation of primary theorems and definitions in quantum information theory covering background information on state spaces, operators and system evolution, density operators, composite systems and measurement. The second part provides background detail on quantum control formalism and open quantum systems theory relevant to later Chapters.

\subsection*{Appendix \ref{chapter:Background: Geometry, Lie Algebras and Representation Theory} (Algebra)}
This supplemental Appendix includes background information on Lie theory and representation theory relevant to geometric concepts applied in our results Chapters. The summary begins with contextualisation of the role of geometry in the development of algebraic and quantum techniques. It then surveys essential results from the theory of Lie groups and Lie algebras which are of central importance to the results we present further on. It covers aspects of representation theory relevant to the work, including root system derivation, Cartan matrices and Dynkin formalism. It focuses in particular on Cartan decompositions from an algebraic standpoint. 

\subsection*{Appendix \ref{chapter:background:Differential Geometry} (Differential Geometry)}
This supplemental Appendix provides a summary and discussion of background theory related to differential geometry, such as the theory of manifolds, fibre and tangent bundles and tensor formalism. It begins with a coverage of elementary geometric concepts, such as differentiable manifolds, charts, tangent and cotangent (dual) spaces and tangent planes. It then proceeds to summarise key concepts such as fibre bundles (and vertical/horizontal subspaces), connections on manifolds, parallel transport and horizontal lifts and covariant differentiation. With this background in train, it focuses on key concepts for optimisation of geodesics and measuring arc-length. Its particular focus is Riemannian manifolds, drawing in connections with Cartan's classification of symmetric spaces in terms of semi-simple Lie groups. It then surveys concepts in subRiemannian geometry (and the theory of distributions) relevant to quantum control relevant to Chapters \ref{chapter:Quantum Geometric Machine Learning} and \ref{chapter:Time optimal quantum geodesics using Cartan decompositions}. It concludes with a focus on geometric control theory relevant to quantum control, with an emphasis on the application of Cartan methods that enable simplification (or symmetry reduction).

\subsection*{Appendix \ref{chapter:Background: Classical, Quantum and Geometric Machine Learning} (Quantum Machine Learning)} This supplemental Appendix surveys key background literature in classical and quantum machine learning. It begins with an overview of classical statistical learning theory, following which common methods in machine learning are examined. It sets out methods and theory relevant in particular to machine learning architecture used in later Chapters, such as deep neural networks. It also examines these results from the perspective of geometric, algebraic and functional analysis theory touched on in previous Chapters. The Chapter then proceeds to a summary of key elements of quantum machine learning, especially variational and other methods used in areas such as quantum control and algorithm synthesis. It examines different algorithmic design approaches from `black-box' methods to more engineered `whitebox' methods where existing assumptions and prior knowledge (such as with respect to quantum theory) are encoded in algorithms for efficiency purposes. It briefly summarises the use and role of geometric and Lie algebraic techniques in classical machine learning and control. It concludes with a synopsis of a few recent emergent trends combining Lie algebraic theory and quantum machine learning.

\section{A Note on Notation}
Consistency of notation is notoriously often honoured in the breach in many fields where different symbolism or conventions for equivalent concepts can confuse readers unfamiliar with content. This can be compounded when seeking to build connections across disciplines, such as between quantum information notation and algebraic or geometric nomenclature. Geometry in particular can often be riddled with subtle yet distinct ways of presenting concepts, especially when grappling with coordinate versus coordinate-free means of expressing results. We mention a few examples below primarily for readers unfamiliar with differential geometry notation. This can be skipped for those familiar with geometry without loss of generality as it is more a foreshadowing for readers from other disciplines. 

In general we have aimed for notational consistency across the Chapters below, such as opting for representing Lie group manifold elements as $\gamma(t) \in \M$ where $\gamma(t)$ is often the symbolism for curves along manifolds (see section \ref{defn:geo:Curves on manifold} for example). We then connect $\gamma(t)$ as the expression of curves traced out on $G$ via unitary group elements $U(t) \in G$.

For convenience, we usually interchange $G$ and $\M$ to emphasise the differentiable manifold properties of $G$ and leave the more formal relation with principal $G$-bundles understood while pointing readers towards relevant literature. In other cases, we endeavour for consistency across geometric, algebraic and control theory notation. For example, in traditional control theory, $x(t)$ is used for state variables when discussing the Pontryagin Maximum Principle (e.g. see section \ref{sec:geo:Pontryagin Maximum Principle}), whereas we use $\gamma(t)$ again as a means of describing how state evolution (in this case represented by the unitary channel $U(t)$) can be geometrically represented in terms of curves on manifolds. Another example is the important relationship between Lie algebras $\g$ and tangent planes (see section \ref{sec:geo:Tangent planes and Lie algebras}) which allows us to (under certain conditions) equate the geometric concept of tangent vectors describing evolution of $\gamma(t)$ in terms of generators from the Lie algebra $\g$ generating evolutions $U(t)$ via the Hamiltonian $H(t)$. Here the Hamiltonian is in effect constructed from Lie algebra elements and the correspondence with tangent planes and $\g$ allows us to then think of $H(t)$ as constructed from tangent bundle elements (which is not unrelated to how state evolution is presented in quantum field theory for example).

%==========
%==========
%===BACKGROUND
\chapter{Background Theory} \label{chapter:Background: Background Theory}
\section{Overview}
Quantum geometric machine learning intersects multiple mathematical and scientific disciplines. In this Chapter, we provide a brief synopsis of background theory relevant to our contributory Chapters \ref{chapter:QDataSet and Quantum Greybox Learning}, \ref{chapter:Quantum Geometric Machine Learning} and \ref{chapter:Time optimal quantum geodesics using Cartan decompositions}. In addition to the material below, we have linked to extensive supplemental material set out in the Appendices. The rationale for this is primarily the interdisciplinary nature of the subject matter that spans quantum information processing, algebra and representation theory, differential geometry and machine learning (both classical and quantum).  

\section{Quantum Information Processing}
Our starting point in this work is the field of quantum information processing. This section covers key elements of quantum information processing theory relevant to quantum control and quantum simulation results in Chapters \ref{chapter:QDataSet and Quantum Greybox Learning}, \ref{chapter:Quantum Geometric Machine Learning} and \ref{chapter:Time optimal quantum geodesics using Cartan decompositions}, summarising the more expansive treatment of background quantum information processing theory set out in Appendix \ref{chapter:Background: Quantum Information Processing}.  We begin with the concept of the state of a quantum system represented as a unit vector $\psi$ over a complex Hilbert space $\Hilb=\Hilb(\C)$. Two unit vectors $\psi_1, \psi_2\in\mathcal{H}$, where $\psi_2 = c\psi_1$ for $c \in \mathbb{C}$, correspond to the same physical state (axiom \ref{axiom:quant:quantumstates}). In density operator formalism, the quantum system is described by the bounded linear operator $\rho \in \mathcal{B}(\Hilb)$ acting on $\Hilb$. For bounded $A \in \mathcal{B}(\Hilb)$, the expectation value of $A$ is $\Tr(\rho A)$. Composite quantum systems are represented by tensor products $\psi_k \otimes \psi_j$ (or equivalently in operator formalism, $\rho_k \otimes \rho_j$). More generally, $\Hilb$ is a state space $V(\mathbb{K})$ construed as a vector space over a field $\mathbbm{K}$, typically $\mathbbm{C}$. Elements $\psi \in V$ represent possible states of a quantum system (definition \ref{defn:quant:State space}). Quantum information processing synthesises such theory with foundational concepts of information via the concepts of classical registers and states (definition \ref{defn:quant:Classical registers and states}) where a classical register $X$ is either (a) a simple register, being an alphabet $\Sigma$ (describing a state); or (b) a compound register, being an n-tuple of registers $X=(Y_k)$. \textit{Quantum registers} and states are then defined as an element of a complex (Euclidean) space $\C^\Sigma$ for a classical state set $\Sigma$ satisfying specifications imposed by axioms of quantum mechanics (definition \ref{defn:quant:Quantum registers and states}). Quantum registers (and by extension quantum states $\psi$) are assumed to encode complete information about the quantum system (such that even open quantum systems (section \ref{sec:quant:Open quantum systems}) can be reframed as complete closed systems encompassing system and environment dynamics - see below). 

Hilbert spaces are equipped with important structural features (we describe in more detail below). First is the inner product (definition \ref{defn:quant:Inner Product}) \( \langle \cdot, \cdot \rangle: V \times V \rightarrow \C, (\psi,\phi) \mapsto \langle \psi,\phi \rangle \) for $\psi,\phi \in V, c \in \C$. Together with the Cauchy-Schwartz inequality (definition \ref{defn:quant:Cauchy-Schwarz Inequality}) we then define a norm (definition \ref{defn:quant:Norm} on $V(\C)$, given by \( \lVert . \rVert: V \to \mathbb{R} \), \( \psi \mapsto \lVert \psi \rVert \) for $\psi,\phi \in V$, which in turn imposes a distance (metric) function $d$ on $V$. 
Normed vector spaces are Banach spaces satisfying certain convergence and boundedness properties and allowing definition of an operator norm (definition \ref{defn:quant:Operator norm}) ensuring for example that operations (evolutions) remain within $V(\C)$. 
This structure allows us to define in a quantum informational context the concept of the \textit{dual space}, a particularly important concept when we come to geometric and algebraic representations of quantum states and evolutions. The dual space (definition \ref{defn:quant:Dual space}) is defined as the set of all bounded linear functionals is denoted the dual space $V^*(\mathbb{K})$ to $V(\mathbbm{K})$ such that $\chi: V \to \mathbbm{K}, \psi \mapsto a||\psi||$ for some (scaling) $a \in \mathbbm{K}$.

A Hilbert space is then formally defined as a vector space $\mathcal{H}(\mathbbm{C})$ with an inner product \( \langle \cdot, \cdot \rangle \) complete in the norm defined by the inner product $||\psi|| = \sqrt{\langle \psi, \psi \rangle}$ (definition \ref{defn:quant:Hilbert Space}). Hilbert spaces can be composites of other Hilbert spaces, such as in the case of direct-sum (spaces) $\Hilb = \Hilb_i \oplus \Hilb_j$ (relevant to state-space decomposition) (definition \ref{prop:quant:Hilbert Space Direct Sum}). Moreover, a Hilbert space admits orthonormal and orthogonal basis (definitions \ref{defn:quant:Orthogonality and orthonormality} allowing us to work with $\Hilb$ using an orthonormal basis \ref{defn:quant:Orthonormal Basis}), which is of fundamental importance to the formalism of quantum computation.

In many cases (such as those the subject of Chapters \ref{chapter:QDataSet and Quantum Greybox Learning} and \ref{chapter:Quantum Geometric Machine Learning}) we are interested in two-level quantum systems denoted \textit{qubit} systems (equation \ref{eqn:quant:qubit}) along with multi-qubit systems, hence we are interested in \textit{tensor products} of Hilbert spaces $\Hilb_i \otimes \Hilb_j$ (definition \ref{defn:quant:Tensor Product}). Tensor products, which also have a representation as a bilinear map $T:\Hilb_i \times \Hilb_j \to \C$ are framed geometrically later on forming an important part of differential geometric formalism (such as relating to contraction operations and measurement). They exhibit certain universal and convergence properties required for representations of quantum states and their evolutions (see section \ref{sec:quant:Tensor product states}). Using the formalism above, one often uses bra-ket notation where states $\psi \in \Hilb$ are represented via ket notation $\ket{\psi}$ with their duals $\psi^\dagger$ represented via bra notation $\bra{\psi}$. Connecting this notation, ubiquitous in quantum information, to the formalism of vector spaces, duals and mappings provides a useful bridge to geometric framings of quantum information (we expand upon this further on). 
%========quantum evolution
\subsection{Operator formalism}
Quantum control, quantum machine learning and unitary synthesis is fundamentally concerned with the directed evolution of quantum systems to obtain some final quantum state. The rules governing quantum state dynamics present fundamental constraints upon how such systems evolve. We characterise quantum state evolution and interaction via operator formalism and measurement axioms such that every measureable physical observable quantity is defined by a real-valued function $M$ defined on classical phase space, there exists a self-adjoint (Hermitian) linear measurement operator $M$ acting on $\mathcal{H}$ (axiom \ref{axiom:quant:observableoperators}). The operators in question are drawn from the set of (norm) bounded endomorphic linear maps $A:\Hilb \to \Hilb$ which we denote $\mathcal{B}(\Hilb)$ (definition \ref{defn:quant:Bounded linear operators}). For any operator $A \in \mathcal{B}(\Hilb)$ we define the adjoint $A^\dagger: \Hilb \to \Hilb$ (definition \ref{defn:quant:Adjoint of an Operator}) as the unique operator such that $\langle \phi, A\psi \rangle = \langle A^\dagger\phi, \psi \rangle$ for all $\phi,\psi \in \Hilb$. Adjoint operators satisfy a number of important properties, including being Hermitian operators with real eigenvalues if $A^\dagger=A$. In quantum information, we are particularly focused on positive semi-definite operators $A = B^\dagger B \in \mathcal{B}(\Hilb)$ such that the norm is non-negative $||B \psi ||^2 \geq 0$ (definition \ref{defn:quant:Positive Semidefinite Operators}). Doing so allows us to define \textit{projection operators} $P$ such that $P^2 = P$ and, for some closed $V \in \Hilb$, $P(V)=I$ and $P(V^\perp)=0$, which are essential for quantum measurement formalism. We can also define an important class of positive semi-definite \textit{isometry} operators (definition \ref{defn:quant:isometries}) for $A \in \mathcal{B}(\X,\Y)$ which preserve the norm $||Av||=||v||, v \in \Hilb$. This in turn allows us to define normal operators ($A^\dagger A = A A^\dagger)$ and in turn the crucial set of isometry operators denoted \textit{unitary operators} $U \in \mathcal{B}(\Hilb)$ which preserve inner products $\langle U\phi, U\psi \rangle = \langle \phi, \psi \rangle$. From this relation we deduce that $||U||=I$ and thus that $U^\dagger=U^\inv$ such that $UU^\dagger = U^\dagger U = I$ which is often the form in which they are first introduced in quantum contexts. Importantly, the norm (inner-product) preserving homomorphic characteristic of unitary operators gives rise to the preservation of Haar measure (see definition \ref{defn:quant:Haarmeasure}) in turn relating to conservation of probability. The unitary group $U(n)$ is a central Lie group in quantum information processing and each Chapter in this work and is a concept we revisit in several guises.

Operator formalism above allows the assertion of the \textit{spectral theorem} which states that each normal operator $X \in \mathcal{B}(\Hilb)$ can be decomposed as the linear combination of projection operators $\Pi_k$ (changing notation to avoid confusion with $P$ subspaces below) onto their corresponding eigenspace with eigenvalues $\lambda_k$, that is $X = \sum_k^m \lambda_k \Pi_k$ (see definition \ref{thm:quant:spectral}). In quantum measurement formalism (section \ref{sec:quant:Measurement}), the spectral theorem manifests via measurement operators being framed as projection operators $M_m$ whose eigenvalues $m$ are observed with probability $\braket{\psi|M^\dagger_m M_m |\psi}$ (see equation \ref{eqn:quant:measurementprobinnerproduct}). Other key operations include the trace operator (definition \ref{defn:quant:Trace}) such that for $A \in \mathcal{B}(\Hilb)$ we have $\trace (A) = \sum_j \braket{e_j, A e_j}$ for an orthonormal basis $\{ e_j \}$ of $\Hilb$. The trace can be construed as a tensorial contraction (see section \ref{defn:geo:Tensor contractions}) with a range of important properties discussed below related to both measurement and system-environment interactions in open quantum systems (section \ref{sec:quant:Open quantum systems} and Chapter \ref{chapter:QDataSet and Quantum Greybox Learning}) via the partial trace (definition \ref{defn:quant:partialtrace}).

An important operator representation in quantum information contexts and one used throughout this work is that of \textit{density operators} which function as in effect representations of distributions over quantum states. First, we define a Hilbert-Schmidt operator $A \in \mathcal{B}(\Hilb)$ as one such that $\trace(A^\dagger A) < \infty$. A density operator is then a positive semi-definite operator $\rho \in \mathcal{B}(\Hilb)$ that is self-adjoint with non-negative trace $\trace(\rho)=1$. The set of density operators denoted $D(\Hilb)$ is a convex set (important for transitions among quantum states represented by $\rho$). Moreover, density operators can be considered as a statistical description of quantum states, analogous to probability distributions over pure states $\rho_a$ (indexed by the set $\Gamma$) where $\rho = \sum_{a\in\Gamma}p(a) \rho_a$. Here $p(a)$ describes the probability of observing (observations that characterise the system in state) $\rho_a$ and so in this sense represents a probability distribution over (pure) states.
Density operator formalism thus allows quantum states and operations upon them to be represented via operator formalism. In particular, we then arrive at two crucial distinctions between \textit{pure} quantum states (those with $\trace(\rho^2)=1$) and mixed (ensembles) of quantum states $\{\rho_i\}$ such that $\trace(\rho) < 1$ (see also other measures such as quantum relative entropy (definition \ref{defn:quant:quantumrelativeentropy}) for discerning pure versus mixed states). Pure states represent the most specific (and complete) information that can be known about a quantum system (where uncertainty is inherently ontological), while mixed states represent epistemic uncertainty about the quantum state \cite{harrigan_einstein_2010} relevant in particular to open quantum systems. Density matrix formalism allows state similarity to be more easily framed, such as via \textit{trace distance} (definition \ref{defn:quant:Tracedistance}). The formalism also allows the definition of coherent and incoherent superpositions (see definition \ref{defn:quant:coherent_superposition}) which is of particular relevance to understanding decohering phenomena in noise-models of open quantum systems (such as those we discuss in later Chapters).

Another set of concepts addressed in this work drawn from information theory is that of quantum channels, which are maps between the space of linear operators i.e. $\Phi: \mathcal{B}(\Hilb_1) \to \mathcal{B}(\Hilb_2)$. Such channels are completely positive trace-preserving (CPTP) maps which play an important role in quantum information theory where we often are interested in operations upon (or maps between) operator representations. The set of such channels is denoted $C(\Hilb_1,\Hilb_2)$. An example is framing unitary evolution in terms of \textit{unitary channels} (definition \ref{defn:quant:Unitary Channel}) where the operation of $U \in \mathcal{B}(\Hilb)$ is represented as $\Phi(X) = U X U^\dagger$ for $X \in \mathcal{B}(\Hilb)$. Furthermore, we can also then define \textit{quantum-classical channels} (definition \ref{defn:quant:Quantum-classical channel}) $\Phi \in C(\Hilb_1,\Hilb_2)$ which transform a quantum state $\rho_1 \in \mathcal{B}(\Hilb)$ (with possible off-diagonal terms) into a classical distribution of states represented by a diagonal density matrix $\rho_2 \in \BH$. Quantum-classical channels are a means of representing general measurement in quantum information (see generally \cite{watrous_theory_2018}).

\subsection{Evolution}
The evolution of quantum systems is framed by the postulate (axiom \ref{axiom:quant:evolution}) that quantum states evolve according to the Schr\"odinger equation $i d\psi = H(t) \psi dt$ (see definition \ref{eqn:quant:schrodingersequation}) where $\psi \in \Hilb$ and the self-adjoint operator that generates evolutions in time, $H(t) \in \BH$, is denoted the Hamiltonian (definition \ref{defn:quant:Hamiltonian}). The Hamiltonian (which represents the set of Hamilton's equations \cite{goldstein_classical_2002} for quantised contexts \cite{hall_quantum_2013})) of a system is the primary means of mathematically characterising the dynamics of quantum systems, describing state evolution and control architecture. The equation has equivalent representations in density operator form as $d\rho = -i[H,\rho]dt$ (equation (\ref{eqn:quant:schrodingerdensityform})) and unitary form as $dU U^\inv = -iH dt$ (equation (\ref{eqn:quant:schrodingerdUUinv})). The utility of each representation depends on the problem at hand. We focus primarily on the unitary form due to the apparent relationship with unitary Lie groups and geometry. Time-dependent solutions to Schr\"odinger's equation take the form of unitaries (equation (\ref{eqn:quant:unitarysolutiontimedependentschrod})) $U = \mathcal{T}_+\exp\left(-i\int_0^{\Delta t} H(t) dt\right)$ (see section \ref{sec:chapter1:evolution:Hamiltonian formalism}) the set of which as we shall see form a Lie group. In practice solving for the time-dependent form of $U(t)$ can be difficult or intractable. Instead a time-independent approximation $U(t) \approx U(t-1),,,U(0)$ (equation (\ref{eqn:quant:timeindependentschrodunitary})) is adopted by assuming that Hamiltonians $H(t)$ (and thus $U(t)$) can be regarded as constant over sufficiently small time scales $\Delta t$. Doing so allows a target unitary $U(T)$ to be decomposed as a sequence $U(T)=U(t_{n})...U(t_0)$. As discussed below, for full controllability such that arbitrary $U(t)$ are reachable (within a reachable set $\mathcal{R}$), we rely upon the Baker-Campbell-Hausdorff theorem (definition \ref{defn:alg:Baker-Campbell-Hausdorff}) to ensure the subspace $\p \subset \g$ of Hamiltonian generators are bracket-generating (definition \ref{defn:geo:bracketgenerating}) such that $\g$ is generated under the operation of the Lie derivative (\ref{defn:alg:Lie algebraliederivative}). We explain these terms in more detail below.   

\subsection{Measurement}
Measurement formalism is fundamentally central to quantum information processing. We can parse the measurement postulate of quantum mechanics into two parts. The first (axiom \ref{axiom:quant:probabilitymeasurement}) provides that given $\psi = \sum_j^\infty a_j e_j, a_j \in \C$ where $\psi \in \Hilb$, the probability of measuring an observable (using operator $M$) with value $m$ is given by $|a_m|^2$ (with $e_m$ as eigenstates of the measurement operator). This is expressed as $p(m) = \braket{\psi | M_m^\dagger M_m |\psi}  = \text{tr}(M_m^\dagger M_m \rho)$ (equation (\ref{eqn:quant:measurementprobinnerproduct})). The effect of measurement (usually rolled into one axiom) by $M$ on state $\rho$ in axiom (\ref{axiom:quant:effectofmeasurement}) measurement leads to a post-measurement state $\rho'$ given by (equation (\ref{eqn:quant:postmeasurementstatedensity})):
\begin{align*}
    \rho'=\frac{M_m\rho M_m^\dagger}{\braket{M_m^\dagger M_m,\rho}} 
\end{align*}
due to the (partial) trace operation of measurement $\text{tr}(M_m^\dagger M_m \rho)$. This is sometimes describe as the Copenhagen interpretation or collapse of the wave function view of quantum mechanics. In information theory terms, each measurement outcome $m$ is associated with a positive semi-definite operator (equation (\ref{eqn:quant:measuretoposdef})). In more advanced treatments, we are interested in measurement statistics drawn from positive operator-valued measures (POVMs), a set of positive operators $\{  E_m \}=\{M^\dagger_m M_m\}$ satisfying $\sum_m E_m=\mathbb{I}$ in a way that gives us a complete set of positive operators with which to characterise $\rho$ (see sections \ref{sec:quant:POVMs and Kraus Operators} and \ref{sec:quant:Informational completeness and POVMs}). Because of this collapse, quantum measurement must be repeated using identical initial conditions (state preparations) in order to generate sufficient measurement statistics from which quantum data, such as quantum states $\rho$ or operators such as $U(t)$ may be reconstructed or inferred (e.g. via state or process tomography). Chapter \ref{chapter:QDataSet and Quantum Greybox Learning} of this work explores the role of measurement in quantum simulations. In Chapters \ref{chapter:Quantum Geometric Machine Learning} and \ref{chapter:Time optimal quantum geodesics using Cartan decompositions}, we assume access to data about $U(t)$ and estimates $\hat U(t)$ assume the existence of a measurement process that provides statistics that enable characterisation of operators and states and in order to calculate loss for machine learning optimisation protocols. As flagged above, measurement can be framed in terms of quantum-to-classical channels (equation \ref{eqn:quant:measurementchannel1}) and permits composite (e.g. multi-qubit) system measurement (see section \ref{sec:quant:Composite system measurement}). Moreover, as we discuss in section \ref{sec:ml:Quantum measurement and machine learning}, the distinct characteristics of quantum versus classical measurement have implications for quantum control and quantum machine learning (such as that typical online instantaneous feedback based control is unavailable due to the collapse of $\rho$ under measurement). 

Quantum measurements then form a key factor in being able to distinguish quantum states and operators according to \textit{quantum metrics} (section \ref{sec:quant:quantummetrics}). The choice of appropriate metric on $\Hilb$ or $\mathcal{B}(\Hilb)$ is central to quantum algorithm design and machine learning where, for example, one must choose an appropriate distance measure for comparing function estimators $\hat f$ to labels drawn from $\Y$ (see section \ref{sec:ml:Quantum Machine Learning}). State discrimination (section \ref{sec:quant:State discrimination}) for both quantum and probabilistic classical states requires incorporation of stochasticity (the probabilities) together with a similarity measure. For example, the Holevo-Helstrom theorem (\ref{thm:quant:Holevo-Helstrom theorem}) quantifies the probability of distinguishing between two quantum states given a single measurement $\mu$. A range of metrics such as Hamming distance, quantum relative entropy (definition \ref{defn:quant:quantumrelativeentropy}) and fidelity exist for comparing states and operators. In this work, we focus upon the fidelity function (see section \ref{sec:quant:Fidelity function}) allowing state and operator discrimination $F(\rho,\sigma) = \big|\big|\sqrt{\rho}\sqrt{\sigma}  \big|\big|_1 =  \Tr\left(\sqrt{\sqrt{\sigma}\rho\sqrt{\sigma}}\right)$ (equation \ref{eqn:quant:fidelityfunction}) which is related to the trace distance (definition \ref{defn:quant:Tracedistance}). The fidelity function is chosen as the loss function for our empirical risk measure in Chapters \ref{chapter:QDataSet and Quantum Greybox Learning} and \ref{chapter:Quantum Geometric Machine Learning} via the cost functional given in equation \ref{eqn:qgml:batchfidelity}.

\subsection{Quantum control}
Quantum control problems consider how to exert control over quantum systems in order to obtain certain computational results, or steer quantum systems in ways corresponding to computations represented by $U(t)$. Quantum control is at the centre of each of Chapters \ref{chapter:QDataSet and Quantum Greybox Learning} (as a use-case for the QDataSet and via learning noise-cancellation operators $V_0$), \ref{chapter:Quantum Geometric Machine Learning} for finding time-optimal geodesics using machine learning and \ref{chapter:Time optimal quantum geodesics using Cartan decompositions} for finding time-optimal Hamiltonians using global Cartan decompositions. For a closed quantum system (noiseless and isolated with no environmental interactions) $\rho \in \mathcal{B}(\Hilb)$ the Hamiltonian can be partitioned into a \textit{drift} $H_d(t)$ and \textit{control} $H_c(t)$ part $H_0(t) =  H_d(t) + H_{\text{ctrl}}(t)$ such that Schr\"odinger's equation (equation (\ref{eqn:quant:schroddensitycontrol2})) becomes $i d\rho = [ H_d(t) + H_{\text{ctrl}}(t), \rho(t)]dt$ (see definition \ref{defn:quant:controlequation}). The drift term represents the uncontrollable evolution of the quantum system, while the control term represents the controllable part. As discussed in sections \ref{sec:quant:controlsystem} and \ref{sec:geo:Time optimal control}, quantum control problems usually follow the Pontryagin Maximum Principle \cite{dalessandro_introduction_2007,jurdjevic_geometric_1997} (see section \ref{sec:geo:PMP and quantum control}) where the state of a system $x$ is described according to a (first order) differential equation $\dot x = f(t,x,u)$ where $x$ represents the system state, $f$ is a vector field while $u=u(t)$ are real-valued time varying (or constant over small $\Delta t$ interval) control functions. In quantum settings, equation (\ref{eqn:quant:controlsystem1}) is described by the Schr\"odinger equation. In unitary form this is equivalent to (equation (\ref{eqn:quant:control-state-eqn})):
\begin{align}
    \dot x \sim  \dot U \qquad f(x,t,u) \sim -iH(u(t)) U(t) \label{eqn:quant:control-state-eqn}
\end{align}
where $iH(u(t))$ is a Hamiltonian comprising controls and generators (for us, drawn from a Lie algebra $\g$). The general Pontryagin Maximum Principle (PMP) control problem then incorporates adjoint (costate) variables and other variational assumptions in order to obtain the general form of control problem, the quantum versions of which are set out in section \ref{sec:geo:Time optimal control} (and see generally \cite{dalessandro_introduction_2007,jurdjevic_geometric_1997,sachkov_control_2009}). The drift Hamiltonian $H_d$ and control Hamiltonian $H_c$ combine as (equations (\ref{eqn:quant:control-math-form}) and (\ref{eqn:quant:control-unitary form})):
\begin{align}
    H(u) = H_d + \sum_k H_k u_k \qquad \dot U =-i \left(H_d + \sum_k H_k u_k \right)U
\end{align}
where initial conditions are chosen depending on the problem at hand (often $U(0)=I$) or, as per our global Cartan decomposition method in Chapter \ref{chapter:Time optimal quantum geodesics using Cartan decompositions} in order to streamline solving the optimal control problem. The objective of time-optimal control is to assume a target $U_T \in \mathcal{\R}^m$ (a reachable set given applicable controls) which is reachable by applying control functions $u_j(t) \in U \subset \R^m$ to corresponding generators in $\g$ or a control subset $\p \subset \g$, subject to a bounded norm on the controls $||u_j(t)|| \leq L$. The targets are quantum states $\rho(T)$ at final time $T$ represented via CPTP unitary operators (channels) $U(T)$ acting on initial conditions (also represented in terms of unitaries). As such, for the quantum case, we are interested in targets as elements of Lie groups of interest to quantum computation, such as unitary groups and special unitary groups (see definition (\ref{defn:alg:Unitary matrix and unitary group})). Given theoretical assumptions as to the existence of length minimising geodesics on Riemannian or subRiemannian manifolds (see sections \ref{sec:geo:Riemannian Manifolds and Metrics} and \ref{sec:geo:SubRiemannian geometry}), time optimisation becomes equivalent to length (equation (\ref{eqn:geo:arclength})) or energy (equation (\ref{eqn:geo:energyhorizontalcurve})) minimisation (see section \ref{sec:geo:Time optimal control}). 

Lie groups $G$ are equipped with the structure of a differentiable manifold $\M$ (see section \ref{sec:geo:Manifolds and charts} and definition \ref{defn:geo:Differentiable manifold}). Thus we assume suitable structure on the underlying manifold $\M$ such as the existence of a metric (e.g. Riemannian or subRiemannian, section \ref{sec:geo:Riemannian Manifolds and Metrics} and \ref{sec:geo:SubRiemannian control and symmetric spaces}) which for Lie groups and their corresponding Lie algebras is the metric induced by the Killing form (definition \ref{defn:alg:Killing form}). 
 
In practice the minimisation problem proceeds using variational methods (see sections \ref{sec:geo:variational methods} and \ref{sec:cartan:generalmethod}) and so in practice for controls $u(t)$ means minimising arc-length via minimising control pulses $\ell(t) = \min_{u(t)} \int_0^T ||u(t)||dt$ (see equations (\ref{eqn:geo:arclengthparamby})) having regard to manifold curvature (which enters the minimisation problem via the metric). 

\subsection{Open quantum systems}
The role of external noise and environmental interactions in quantum information processing is central to solving problems in quantum control for open quantum systems. While Chapters \ref{chapter:Quantum Geometric Machine Learning} and \ref{chapter:Time optimal quantum geodesics using Cartan decompositions} assume a closed system, Chapter \ref{chapter:QDataSet and Quantum Greybox Learning} explicitly models its quantum simulations and formalism on the basis of open quantum systems (section \ref{sec:quant:Open quantum systems}). Open quantum systems formalism involves constructing an overall quantum system comprising the \textit{system} (or closed quantum evolution) and the \textit{environment} (i.e. interactions with the environment). Such open systems can be modelled (equation (\ref{eqn:quant:openquantHamiltonianSE})) in a simple case via introducing environmental dynamics into the Hamiltonian:
    \begin{align}
        H(t) = H_0(t) + H_1(t) \doteq \underbrace{H_d(t) + H_{\text{ctrl}}(t)}_{H_0(t)} + \underbrace{H_{SE}(t) + H_E(t)}_{H_1(t)} \label{eqn:quant:openquantHamiltonianSE}
    \end{align}
 where $H_0(t)$ is defined as above encompassing drift and control parts of the Hamiltonian and where $H_1(t)$ consists of two terms: (i) a system-environment interaction part $H_{SE}(t)$ and (ii) a free evolution part of the environment $H_E(t)$.

 To model the effect of noise on quantum evolution, we utilise the concept of a \textit{superoperator} (definition \ref{defn:quant:superoperators manifold}) $\mathcal{S}: \mathcal{B}(\mathcal{H}) \rightarrow \mathcal{B}(\mathcal{H})$ which is trace preserving or decreasing, exhibits convex linearity and complete positivity such that it is a CP map (equation (\ref{eqn:quant:superoperatorkraus})). Superoperators then allow us to represent the effect of environmental interactions (noise) as quantum or classical channels via the effect on density operator evolution. A common method of representing such open-state evolution is via the \textit{Lindblad master equation} (section \ref{sec:quant:Noise and decoherence}) which formalises non-unitary evolution arising as a result of interaction with noise sources, such as baths. Here mixed unitary and non-unitary evolution of the quantum system is given by (equation (\ref{eqn:quant:lindbladmasterequation})):
  \begin{align}
     \frac{d\rho}{dt} = -i[H, \rho] + \sum_{k} \gamma_k\left( L_k \rho L_k^\dagger - \frac{1}{2} \left\{ L_k^\dagger L_k, \rho \right\} \right)
 \end{align}   
with $\rho$ being the state representation, $H(t)$ the closed system Hamiltonian, $\sum_{k}$ a summation over all noise channels with dephasing (decoherence) rates $\gamma_k$. The Lindblad operators $L_k$ acting on $\rho$ encode how the environment affects the closed quantum state. For modelling of noise in Chapter \ref{chapter:QDataSet and Quantum Greybox Learning}, we impose noise $\beta(t)$ along various axes of qubits (see section \ref{sec:qdata:Noise profile details}), an example being a single qubit Hamiltonian with noise $\beta_z(t)$ imposed along the $z$-axis:
\begin{align*}
    H = \frac{1}{2}\left(\Omega + \beta_z(t)\right)\sigma_z + \frac{1}{2}f_x(t) \sigma_x.
\end{align*}
where the type of noise is characterised according to regimes set-out therein. In general the effect of noise on controls $u(t)$ is modelled via the power spectral density for which section \ref{sec:qdata:Noise profile details} provides a discussion.

\section{Algebra and Lie Theory}
\subsection{Lie groups and Lie algebras}
This work, especially Chapters \ref{chapter:QDataSet and Quantum Greybox Learning}, \ref{chapter:Quantum Geometric Machine Learning} and \ref{chapter:Time optimal quantum geodesics using Cartan decompositions}, focuses on quantum control problems where target computations are represented as elements of unitary Lie groups $U_T \in G$. Synthesising unitaries to reach $U_T$ in a time-optimal fashion involves Hamiltonians composed using generators of the corresponding Lie algebra $\g$ and associated control functions. In particular, we are interested in leveraging the symmetry and differential geometric structure of Lie groups to model time-optimal unitary synthesis as a problem of learning minimal geodesics on Riemannian (section \ref{sec:geo:Riemannian manifolds}) and subRiemannian (section \ref{sec:geo:SubRiemannian geometry}) manifolds. This is especially the case in Chapter \ref{chapter:Time optimal quantum geodesics using Cartan decompositions} where we leverage symmetry structure of semi-simple Lie groups via global Cartan decompositions to solve for time-optimal Hamiltonians analytically. In this section, we cover underlying principles of Lie theory, such as Lie groups, Lie algebras, Killing forms and approximation theorems. We summarise Cartan decompositions and important root-space decompositions. 

The starting point for our analysis is the concept of a \textit{Lie group} $G$ being a Hausdorff topological group equipped with a smooth manifold structure (definition \ref{defn:alg:liegroups}). Lie groups are \textit{continuous} groups owing to their parametrisation via the continuous parameter $t \in \R$ and thus enable the study of infinitesimal transformations and associated symmetry properties. We are primarily interested in matrix Lie groups (section \ref{sec:alg:Matrix Lie groups}) which are closed subgroups of the general linear group of all invertible n-dimensional complex matrices $GL(n;\C) = \{ A \in M_n(\C) | \det(A) \neq 0 \}$. In particular, the existence of Lie group homomorphisms (definition \ref{defn:alg:Lie group homomorphism}) between Lie groups $\varphi: G_1 \to G_2$ allows for various groups of interest in quantum computing to be studied by way of representations of $GL(n;\C)$. Unitary operators (and channels) described above then have a representation as a unitary group (definition \ref{defn:alg:Unitary matrix and unitary group}). Importantly unitary groups are \textit{connected} Lie groups (definition \ref{defn:alg:connected_matrix_lie_group}) such that for all $A, B \in G$ there exists a continuous path $\gamma : [0,1] \rightarrow M_n(\mathbb{C})$ such that $\gamma(0) = A$ and $\gamma(1) = B$ with $\gamma(t)\in G$ for all $t$. This is important for establishing the existence of geodesics $\gamma(t) \in G$ (as time-optimal paths in control theory). For qubits, our target unitaries are drawn from the special unitary group $SU(2)$ (equation \ref{eqn:alg:su2group}).

Associated with each Lie group $G$ is a \textit{Lie algebra} $\g$ (definition \ref{defn:alg:Lie algebraliederivative}) which is of central importance as the algebras from which Hamiltonians governing quantum evolution are composed. Lie algebras are vector spaces over a field $\R$ or $\C$ equipped with a bilinear form (product) $[X,Y]$ for $X,Y \in \frak{g}$ satisfying (a) $[X,Y]=-[Y,X]$ and (b) the Jacobi identity. The bracket $[X,Y]$ we identify as the commutator in quantum contexts or the \textit{Lie derivative} (owing to its status in terms of derivations) satisfying a number of algebraic properties (proposition \ref{prop:alg:Lie bracket properties}). As we discuss in the next section, Lie algebras have a natural interpretation and nice geometric intuition in terms of tangent bundles over Lie group manifolds $\M$. Importantly (particularly for our exegesis in Chapter \ref{chapter:Time optimal quantum geodesics using Cartan decompositions}) the Lie derivative can be identified with the endomorphic adjoint action of a Lie algebra upon itself (definition \ref{defn:alg:Adjoint action}) $\ad: \frak{g}\to \text{End}_\Km (\frak{g})$ such that $\ad_X(Y) = [X,Y]$. As with Lie groups, Lie algebra homomorphisms (definition \ref{defn:alg: Lie algebra homomorphism}) allow mappings between Lie algebras which is important in geometric (fibre bundle) formulations of quantum evolution over manifolds and corresponding tangent bundles (see Appendix \ref{chapter:background:Differential Geometry}). Semi-simple Lie algebras and Lie groups admit a classification system based upon concepts of ideals (definition \ref{defn:alg:Simple and semisimple}) which is of fundamental importance to Cartan's classification of symmetric spaces. Establishing time optimality using variational techniques requires metrics (inner products) on Hamiltonians composed from $\g$ which in turn requires a way of combining generators in $\g$. This is given by the \textit{Killing form} (definition \ref{defn:alg:Killing form}) which is a bilinear mapping over $\g$ given by $B(X,Y) = \Tr(\ad_X, \ad_Y)$ for $X,Y \in \g$. Additionally, Cartan's criterion for semisimplicity is that the Killing form is non-degenerate, that is $\g$ is semisimple if and only if the Killing form for $\g$ is non-degenerate, namely $B(X,Y) \neq 0$ for all $X,Y \in \g$.

Quantum control in terms of Lie groups and Lie algebras relies upon the important bridge between the two provided by the \textit{exponential map} between $G$ and $\g$. In essence unitaries $U(t)$ represent exponentiated Lie algebra $\g$ elements $G \sim \exp(\g)$. This relationship as we discuss is algebraic but connects nicely (not focused on here) with wave function representations of solutions to Schr\"odinger's equation in terms of exponentials (or their sine and cosine expansions). In this context, the \textit{matrix exponential} is an important bridging concept. It is defined (definition \ref{defn:alg:Matrix exponential}) via the power series $e^X = \sum_{n=0}^\infty \frac{X^n}{n!}$ for $X \in \g$ and satisfies a number of properties we rely on throughout (set out in theorem (\ref{thm:alg:matrix_exponential_properties})). Where the exponential map allows one to transition from $\g$ to $G$, the derivative of the matrix exponential at $t=0$ allows one to transition from $g \in G$ to $X \in \g$ via $ \left. \frac{d}{dt} e^{tX} \right|_{t=0} = X$ (equation (\ref{eqn:alg:matrixexpderivativeatzero})). Moreover, the mapping gives rise to an important unique homomorphism between $G$ and $\g$ at the identity, allowing symmetries and properties of $G$ to be explored (and, in a control sense, manipulated) by way of $\g$ itself. Formally, the Lie algebra of a matrix group $G \subseteq GL(n; \mathbb{C})$ is then given by (definition \ref{defn:alg:lie_algebra_of_matrix_lie_group}) $ \mathfrak{g} = \{ X \in M_n(\mathbb{C}) \mid e^{tX} \in G \text{ for all } t \in \mathbb{R} \}$. Moreover, this relationship allows us to map between the adjoint action of $\g$ and that of $G$ itself (equation (\ref{eqn:alg:liealgderivatidentity})) $[X, Y] := XY - YX = \frac{d}{dt}e^{tX}Ye^{-tX} \big|_{t=0}$ which we rely upon for example in our final Chapter.

An important connection between algebraic and geometric methods in quantum information and machine learning arises from the correspondence between Lie algebras $\g$ and the tangent plane (or bundle) of a manifold $T\M$. The correspondence (theorem \ref{thm:alg:Lie algebra tangent space correspondence}) provides that each Lie algebra $\g$ of $G$ is equivalent to the tangent space to $G$ at the identity element of $G$. This equivalence means that $\g$ has a representation as $X \in M_n(\C)$. The vectors $X$ represent derivatives at $t=0$ of smooth curves $\gamma:\R \to G$ with $\gamma(0)=I,\gamma'(0)=X$, allowing characterisation of sequences of unitaries $(U_j(t)) \in G$ (here $j$ indexes the sequence) as paths $\gamma_j(t)$, i.e. we are equating curves with sequences of unitaries. Note that $G$ can also carry a representation in $M_n(\C)$. Furthermore, the Lie algebra / group homomorphism (section \ref{sec:alg:Homomorphisms}) $\Phi:G\to H, \varphi:\g \to \h$ such that $d\Phi = \varphi$. The homomorphism is fundamental to important theorems such as the Baker-Campbell-Hausdorff theorem (section \ref{sec:alg:Baker-Campbell-Hausdorff theorem}):
\begin{align*}
    \exp(X)\exp(Y) = \exp\left(X + Y + \frac{1}{2}[X,Y] + \frac{1}{12}[X,[X,Y]] +...  \right)
\end{align*}
for $X,Y,Z \in \g$ which facilitates approximations for quantum control. Moreover it importantly allows us to understand how evolutions on $\M$ (represented via diffeomorphic exponential group elements) are shaped by not just Lie algebra elements individually, but their conjugation (adjoint action). For example in our final Chapter \ref{chapter:Time optimal quantum geodesics using Cartan decompositions} with Cartan decompositions (section \ref{sec:cartan:Cartan decomposition}) $\g =\p \oplus \k$, it is the fact that $[\p,\p]\subseteq \k$ (together with the other relations constitutive of the Lie triple property of such symmetric spaces) which fundamentally allows elements in $K \subset G$ to be reachable via application of controls to $\p$ and for such curves $\gamma(t)$ to have a representation in the homogeneous space $G/K$. 

\subsection{Representation theory}
Representation theory is an important feature of the application of Lie groups (and non-commutative algebra) in quantum control and machine learning settings. It is of particular relevance to our final Chapter, but also others below. Denoting $GL(n; \C)$ as $G(V)$ we define a \textit{representation} of $G$ as a finite-dimensional continuous homomorphism of $\pi:G \to GL(V)$ such that $\pi([X,Y]) = [\pi(X),\pi(Y)]$ (definition \ref{defn:alg:Finite-Dimensional Representation of a Lie Group}). Lie algebras may also carry representations into $GL(V)$. Representations satisfy a number of properties set out in section \ref{sec:alg:Representations} which allow groups and algebras to be studied by way of such representations. An important such representation is the \textit{adjoint representation}. Given a Lie group $G$ and Lie algebra $\g$ with smooth isomorphism $\Ad_g(h) = g h g^\inv$ where $g,h \in G$, there exists a corresponding Lie algebra isomorphism $\ad_X: \g \to \g$ such that $\exp((\ad_\g)(X)) = g(\exp(X))g^\inv$ where $\exp(\ad_\g(X))=\Ad_G(\exp(X))$. Importantly, the adjoint representation maps Lie algebra elements to automorphisms of the Lie algebra itself. The benefit of adopting the adjoint representation is also to some degree one of computational efficiency where symmetry structure can be more easily diagnosed or embedded (such as in the case of equivariant neural networks \cite{ragone_representation_2023,nguyen_theory_2022,larocca_group-invariant_2022}). Adjoint expansions (section \ref{sec:alg:Adjoint expansions}) are an important tool in later Chapters (particularly in Chapter \ref{chapter:Time optimal quantum geodesics using Cartan decompositions}). Recalling group adjoint action as $\text{Ad}_h(g) = hgh^\inv$ with the related Lie algebra adjoint action $\ad_X(Y) = [X,Y]$, we note that such conjugation can be expanded in sine and cosine terms (equation (\ref{eqn:alg:econjsinhcosh})):
\begin{align}
    e^{i\Theta}Xe^{-i\Theta} = e^{i\ad_\Theta}(X) = \cos\ad_\Theta(X)+i\sin\ad_\Theta(X)
\end{align}
This relation together with the sine and cosine (equation (\ref{eqn:alg:cosadthetaXexpansion})) expansions of the adjoint action are central to the close-form results for calculation of minimum time (equation (\ref{eqn:cartan:OmegaT-minsinadtheta})) time-optimal Hamiltonians for quantum control involving Riemannian symmetric spaces (see section \ref{sec:cartan:generalmethod}). 

Lie algebras are defined over $\R$ (real) and $\C$ (complex) fields. Complex Lie algebras $\g$ (over $\C$) and real Lie algebras $\g_0$ (over $\R$) are related via complexification and the related concept of the real form of a complex Lie algebra. A real vector space $V(\R)$ is the real form of a complex vector space when the complexification of $V$ is given by $W^\R = V \oplus i V$ (definition \ref{defn:alg:realforms}). In Lie algebraic terms this is $\g = \g_0^\C = \g_0 + i\g_0$ (equation (\ref{eqn:alg:liealgebracomplexification})). In particular, this allows us to write $A \in M_n(\C)$ as a real-valued matrix only (equation (\ref{eqn:alg:complextorealmx})) which we leverage in Chapter \ref{chapter:Quantum Geometric Machine Learning}.

\subsection{Cartan algebras and Root-systems}
Cartan decompositions of symmetric space Lie groups provide a technique for solving certain classes of time-optimal quantum control problems. Cartan algebraic formalism (see section \ref{sec:alg:Cartan algebras and Root-systems} for an expanded summary) is crucial to the symmetry-reduction techniques examined in this work. Moreover, Cartan's pioneering methods are at the heart of establishing connections between algebraic (Lie-theoretic) formalism and differential geometric framing. Here we define Cartan subalgebras and connect them with root systems and accompanying diagrammatic and geometric interpretations (summarising primarily standard literature from \cite{knapp_lie_1996,hall_beyond_2007,helgason_differential_1962,helgason_differential_1979}).

Root systems are primarily used in the classification and structure theory of Lie algebras, while weight systems are used in the study and classification of their representations. In particular, in both cases we effectively probe the structure of $\g$ using the adjoint action in the adjoint representation to reveal its symmetry structure used in time-optimal synthesis. This is the case for example in Chapter \ref{chapter:Time optimal quantum geodesics using Cartan decompositions} where the commutation table, Table \ref{tab:su3commutationHIII}, evidences the Cartan commutation relations (section \ref{sec:alg:Cartan decompositions}) via the adjoint action. For a subalgebra $\h \subset \g$ with a (diagonal) basis $H$ and a dual space $\h^*$ with elements $e_j \in \h^*$ (a dual space of functionals $e_j: V \to \C$), we can construct a basis of $\g$ given by $\{\h,E_{ij}\}$ where $E_{ij}$ is 1 in the $(i,j)$ location and zero elsewhere (section \ref{sec:alg:Roots}). We study the adjoint action of elements of $H \in \h$ on each such eigenvector:
\begin{align*}
    \ad_H(E_{ij}) = [H,E_{ij}] = (e_i(H)-e_j(H))E_{ij} = \alpha E_{ij}
\end{align*}
Here $\alpha = e_i - e_j$ is a linear functional on $\h$ such that $\alpha:\h \to \C$. Such functionals $\alpha$ are denoted roots. The Lie algebra $\g$ can then be decomposed as (equation (\ref{eqn:alg:rootdecompositionLiealgebra})):
\begin{align*}
    \g = \h \oplus_{i\neq j} \C E_{ij} = \h \oplus_{\alpha \in \Delta} \g_\alpha \qquad \g_\alpha = \{X \in \g | \ad_H(X) = \alpha X, H \in \h \}
\end{align*}
which satisfies certain commutation relations given in equation (\ref{eqn:alg:rootcommutationrelations}). Roots may then be placed in a positive or negative ordered sequence. Each root is a non-generalised weight of $\ad_\h(\g)$ (a root of $\g$ with respect to $\h$). The set of roots is denoted $\Delta(\g,\h)$. The \textit{weights} (definition \ref{defn:alg:weights}) for a given representation $\pi: \h \to V(\C)$ can then be constructed as a generalised weight space $V_\alpha = \{  v \in V | (\pi(H)-\alpha(H)1)^nv=0, \forall H \in \h \}$ for $V_\alpha\neq 0$ ($v \in V_\alpha$ are generalised weight vectors with $\alpha$ the weights). Here $\pi$ is the adjoint action of $\h$ on $X \in \g$. Weights allow $\g$ to be decomposed as $\g = \h \oplus_{\alpha \in \Delta} \g_\alpha$ (\ref{eqn:alg:rootspace decompositiongalpha}). Note that $\g_0$ is then the weight space subalgebra associated with the zero weight under this adjoint action. This leads to the important definition of a \textit{Cartan subalgebra} (definition \ref{defn:alg:cartansubalgebra}) $\h \subset \g$ as a nilpotent Lie algebra of a finite-dimensional complex Lie algebra $\g$ such that $\h = \g_0$. Cartan subalgebras are maximally abelian subalgebras of $\mathfrak{g}$. There may be multiple Cartan subalgebras in $\g$ and each is conjugate via an automorphism of $\g$ (which for Cartan decompositions means under conjugation by an element of the isometry group $K$). As we discuss in Chapter \ref{chapter:Time optimal quantum geodesics using Cartan decompositions}, the choice of Cartan subalgebra is of pivotal significance to the application of global Cartan decompositions for time-optimal control. 

Cartan subalgebras allow the generalisation of the concept of roots to a \textit{root system}. In this formulation (definition \ref{defn:alg:roots}), roots are the non-zero generalised weights of $\ad_\h(\g)$ with respect to the Cartan subalgebra. Recall we denote the set of roots $\alpha$ is $\Delta(\g,\h)$. Cartan subalgebras $\h$ are the maximally abelian subalgebras such that the elements of $\ad_\g(\h)$ are simultaneously diagonalisable given that $[H_k,H_j]=0$ for $H_k,H_j \in \h$. Roots act as functionals such that for $\alpha \in \Delta$, we have that $\alpha(H) = B(H,H_\alpha), \forall H \in \h$ where $B(\cdot,\cdot)$ denotes the Killing form on $\g$ and $\h$ (see section \ref{sec:alg:Root system properties} for more detail). From this concept of roots, we can define \textit{abstract root systems} (section \ref{sec:alg:Abstract root systems}) that satisfy a range of properties set out in definition \ref{defn:alg:Root System}. These in turn allow roots to be framed in geometric terms related to the Weyl group and concepts of angles between roots (see section \ref{sec:alg:Reduced abstract root systems}) which are expressed via Dynkin diagrams (see below), together with an ordering of roots (see section \ref{sec:alg:Ordering of root systems}). Chapter \ref{chapter:Time optimal quantum geodesics using Cartan decompositions} includes derivation of a root system, for example, in relation to $SU(3)$ connected with the introduction of our method for time-optimal synthesis. Cartan algebras can be used to construct a \textit{Cartan-Weyl basis} which is a basis of the Lie algebra $\g$ comprising the Cartan subalgebra $\h$ together with root vectors, where each root can be thought of as a symmetry transformation (see equation (\ref{eqn:alg:rootdecompositionLiealgebra})).

Abstract root systems moreover allow the construction of a \textit{Cartan matrix} (definition \ref{defn:alg:Cartan matrix}) where given $\Delta \in V$ and simple root system $\Pi = {\alpha_k}$, $k=1,...,n=\dim V$, the Cartan matrix of $\Pi$ an $n \times n$ matrix $A = (A_{ij})$ whose entries are given by $A_{ij} = 2\braket{\alpha_i,\alpha_j}/|\alpha_i|^2$. Cartan matrices are then used to construct \textit{Dynkin diagram} (definition \ref{defn:alg:dynkindiagram}) as set out in Figure \ref{fig:alg:dynkin_diagram_an_expanded}. Dynkin diagrams allow for visual representation of root system, encoding angles and lengths between roots. From a quantum control perspective, roots can be related to the transition frequencies between different energy levels of a quantum system. Importantly, the Cartan matrix (definition \ref{defn:alg:Cartan matrix}), derived from the inner products of simple roots, tells us about the relative strengths and coupling between these transitions (see equation (\ref{eqn:cartan:Hamiltonianroots}) for a generic example). 

\subsection{Cartan decompositions}
We conclude this section with a discussion of the key concept of Cartan decompositions (section \ref{sec:alg:Cartan decompositions}) of semisimple Lie groups and their associated algebras.  For a given a complexified semisimple Lie algebra $\g = \g_0^\C$, a corresponding Cartan involution $\theta$ is associated with the decomposition $\g = \k \oplus \p$. In turn, this allows a global decomposition of $G=KAK$ where $K = e^\k$ and $A = e^\a$ where $\a \subset \p$ (a maximal abelian subspace of $\p$), which is a generalisation of polar decompositions of matrices. For classical semisimple groups, the real matrix Lie algebra $\g_0$ over $\R,\C$ is closed under conjugate transposition, in which case $\g_0$ is the direct sum of its skew-symmetric $\k_0$ and symmetric $\p_0$ members. It can be shown \cite{knapp_lie_1996} that for semi-simple Lie algebras this admits a decomposition $\g_0 = \k_0 \oplus \p_0$ which in turn can be complexified as $\g = \k \oplus \p$. To obtain this decomposition, we consider the fundamental involutive symmetry automorphism on $\g$ denoted the \textit{Cartan involution}. Using the Killing form $B$, $X,Y\in\g$ and an involution $\theta$, a Cartan involution (definition \ref{defn:alg:cartaninvolution}) is a positive definite symmetric bilinear form such that $B_\theta(X,Y) = -B(X,\theta Y)$. From the existence of this involution, we infer the existence of the relevant \textit{Cartan decomposition} given by $\g = \k \oplus \p$ with $\k$ the +1 symmetric eigenspace $\theta(\k)=\k$ and $\p$ the -1 anti(skew)symmetric eigenspace $\theta(\p)=-\p$ satisfying the following commutation relations from equation (\ref{eqn:alg:cartancommutationrelationsmain}):
\begin{align*}
    &[\k,\k] \subseteq \k &[\k,\p] &\subseteq \p & [\p,\p] \subseteq \k
\end{align*}
The subspaces $\k,\p$ are orthogonal with respect to the Killing form in that $B_\g(X,Y)=0=B_\theta(X,Y)$ i.e. for $X \in \k$ and $Y \in \p$ we have $B(X,Y) = 0$. In a quantum control context, this aligns with the decomposition of $\Hilb$ corresponding to states generated by $\p$ and $\k$. The remarkable and important feature of the commutations above is that under the operation of the Lie bracket, Hamiltonians comprised solely of elements in $\p$ can in fact reach targets in $\k$ despite the fibre bundle structure partitioning the space (see the following section). The corresponding Lie group decomposition is given by $G=KAK$  where $K = \exp(\k), A=\exp(\a)$ (definition \ref{defn:alg:KAKdecomposition}) for $\a \subset \h$. Every element $g \in G$ thus has a decomposition as $g=k_1 a k_2$ where $k_1,k_2 \in K, a \in A$. This allows unitary channels to be decomposed as $U = kac$ and $U=pk$ (see \cite{dalessandro_introduction_2007} for the latter) for $k,c \in \exp(\k) = K$ and $p \in \exp(\p)$. We note also, as discussed in Chapter \ref{chapter:Time optimal quantum geodesics using Cartan decompositions}, that for our global decomposition method, we seek $\a \subset \p$ which means that the Cartan subalgebra must have a compact and non-compact subset i.e. $\h \cap \k \neq 0$ and $\h \cap \p \neq 0$, where we seek the maximally non-compact (most overlap with $\p$) subalgebra. Because $\h$ may lie entirely within $\k$, it may be that a Cayley transform is required which in essence involves conjugation of elements of $\h$ with a root vector (see Knapp \cite{knapp_lie_1996} and section \ref{sec:alg:Cayley transforms} for detail). An example of Cayley transforms for $SU(3)$ is set out in section \ref{sec:cartan:cayley transforms and Dynkin diagrams}. We now progress to cover important concepts from differential geometry applicable to methods used in this work, especially those from geometric control theory.

%=========DIFFERENTIAL GEOMETRY
\section{Differential geometry}
\subsection{Manifolds and tangents}
In this section we provide a synopsis of key concepts from geometry related to quantum information processing and machine learning. A more extensive exegesis is set out in Appendix \ref{chapter:background:Differential Geometry}. This section and the Appendix summarise standard material from  \cite{isham_modern_1999,frankel_geometry_2011,montgomery_tour_2002,do_carmo_differential_2016,knapp_lie_1996,helgason_differential_1979,goldstein_classical_2002,kobayashi_foundations_1963}. We begin with an outline of basic concepts such as manifolds, tangent spaces, tensor fields (including vector fields, integral curves and local flows), together with cotangent spaces, metrics and tangent planes. We then explore fibre bundle theory and its central use in geometric contexts via partitioning bundles (tangent bundles for our purposes) into horizontal and vertical subspaces. We develop the theory of connections on a bundle as a lead into definitions of geodesics and parallel transport. We relate these concepts to those of curvature and the definitions of Riemannian (and subRiemannian) manifolds and symmetric spaces (of particular importance as we model our quantum control problems as symmetric space control problems). We then discuss subRiemannian geometry before discussing geometric control theory and variational methods such as the Pontryagin Maximum Principle. 

The study of geometric methods usually begins with the definition and study of the properties of differentiable manifolds. One begins with the simplest concept of a set of elements, characterised only by their distinctiveness (non-identity), akin to pure spacetime points $\M$. The question becomes how to impose structure upon $\M$ in a useful or illuminating manner for explaining or modelling phenomena. A differentiable manifold (definition \ref{defn:geo:Differentiable manifold}) is a Hausdorff topological space $\M$ together with a global differentiable structure. The pure manifold $\M$ lacks sufficient structure for analytic operations such as differentiability to be well-defined. For this, we associate to points in $\M$ an analytic structure equipped with sufficient structure for differential operations. This is via equipping $\M$ with \textit{coordinate charts} (definition \ref{defn:geo:Coordinate charts and atlases}), each of which is a pair $(U,\phi)$ on $\M$ where $U \subset \M$ (an open subset) and $\phi:U \to \Km^m$ where $p \mapsto \phi(p)$ (usually taking $\Km = \R $ or $\C$) is a homeomorphism. The coordinates of a point $p$ are maps from $p\in U \subset  \mathcal{M}$ to the Euclidean space $\K^m$ i.e. $(\phi^1(p),...,\phi^m(p))$. The \textit{coordinate functions} are each of these individual functions $\phi^\mu: U \to \K$. Lie groups (proposition \ref{prop:geo:Lie Group (Manifold)}) can then be construed as a group $G$ equipped with a differentiable structure is a smooth differentiable manifold via the smoothness of group operations (arising from the continuity properties of $G$ i.e. being a topological group). We are often interested in functions defined on $p \in \M$. Functions that preserve such structure are played by the role of $C^r$ functions, those which are differentiable up to order $r$ where differentiability requires $r \geq 1$ while smoothness requires $r=\infty$ i.e. $f \in \cinfm$ (see section \ref{sec:geo:Manifolds and charts} for more detail).

Tangent spaces $T\M$ (section \ref{sec:geo:Tangent spaces}) of a manifold are a crucial concept for geometric characterisation of quantum control and machine learning. We start with the concept of a \textit{curve} (definition \ref{defn:geo:Curves on manifold}) on the manifold $\gamma(t) \in \M$ being a smooth (i.e. $C^\infty$) map $\gamma: \R \supset (-\epsilon,\epsilon) \to \mathcal{M}, t \mapsto \gamma(t)$ from an interval in $\R$ crossing 0 into $\mathcal{M}$, where $\gamma(0)=p$ (the `start' of the curve). With this definition, we then define tangents (definition \ref{defn:geo:Tangent}) in terms of the derivatives of curves in a local coordinate system $(x^1,...,x^m)$ such that two curves are tangent if their derivatives in $\R^m$ at $\gamma(t) = p$ align, that is:
        \begin{align*}
            \left.\frac{dx^i}{dt}(\gamma_1(t))\right|_0 = \left.\frac{dx^i}{dt}(\gamma_2(t))\right|_0 \qquad i=1,...,m.
        \end{align*}
Thus the intuitive notion of a vector `sticking out' from a surface or along a curve is replaced with analytic (derivation) properties of $\M$ via its representation in local charts. The equivalence class of curves satisfying this condition at $p \in \M$ is sometimes denoted $[\gamma]$, hence we can regard tangents as vectors. The \textit{tangent space} $T_p\M$ is defined as the set of all tangents at $p \in \M$, while the tangent bundle $T\M$ is defined as the union of all $T_p\M$. We also define a projection map from the tangent bundle to $\M$ via $\pi: T \mathcal{M} \to \mathcal{M}$ which, as we shall see, is related in Lie theoretic contexts to the exponential map from $\g$ to $G$. Tangent vector $v \in T_p\mathcal{M}$ can be used to define the \textit{directional derivative} on $f \in \cinfm$ via considering $v$ as operators on $f$ via $v(f) = \frac{df(\gamma(t))}{dt}\big|_0$. The directional derivative points in the direction of steepest ascent and is central to defining the all important \textit{gradient} (definition \ref{defn:geo:gradient}) central also to machine learning, quantum information processing. 
For $f: \mathcal{M} \rightarrow \mathbb{R}$ on $\mathcal{M}$, the \textit{gradient} of $f$ at a point $p \in \mathcal{M}$, denoted $\nabla f(p)$ is the unique vector $v \in T_p\M$ satisfying $v(f) = \langle \nabla f(p), v \rangle$. Here $v(f)$ denotes the directional derivative of $f$ in the direction of $v$ (as above), and $\langle \cdot , \cdot \rangle$ denotes the applicable metric. We discuss gradients at some length in later sections especially in the context of backpropagation and stochastic gradient descent.

While intuitively one often regards a problem as confined to a single $\M$ and bundle $T\M$, each $T_p\M$ in the general case is not necessarily identical. For a map between manifolds $h: \M \to \N$, then we can define a corresponding \textit{pushforward} (definition \ref{defn:geo:Push-forward}) as taking a vector in the tangent space associated with $p \in \mathcal{M}$, namely $v \in T_p \mathcal{M}$ and represented as $h_*(v)$ to the tangent space associated with the element of $\mathcal{N}$ i.e. $h_*: T\M \to T\N$. As discussed in section \ref{sec:geo:Tensor fields and tangent spaces}, tangents can be characterised operators or linear maps acting on functions $f \in \cinfm$ such that $T\M$ is a derivation. Each tangent vector can be written as $v = \sum_\mu^m v^\mu (\partial_\mu)$ where $(\partial_\mu)$ represent a basis of partial derivatives of $T\M$ with $v^\mu \in \R$ the component functions (equation (\ref{eqn:geo:tangentpartialderivsum})). 

\subsection{Vector fields}
We construct the notion of a \textit{vector field} as an assignment of $T_p\M$ to each $p \in \M$ satisfying certain linear properties (see definition \ref{defn:geo:vectorfield}) and closure under the commutator (Lie derivative) $[X,Y]$ (see section \ref{sec:geo:Vector fields and commutators} for detail). Vector fields, having a basis in differential operators, can be thought of as operators on $f\in \cinfm$ which are themselves defined on $\M$. We denote the set of vector fields $\mathfrak{X}(\M)$. Thus we can map between Lie algebras and vector fields in specific cases of interest in quantum control. Vector fields are importantly regarded as generators of infinitesimal diffeomorphisms (curves $\gamma(t)$) on $\M$ in the form $\delta(x^\mu) = \epsilon X^\mu(x)$. For optimal control where all $U \in G$ (or $p \in \M$) are reachable, we require $\M$ to be filled with a family of curves $\gamma(t)$ such that the tangent vector to any such curve at $\gamma(p)$ is the vector field $X_p$. Such curves are denoted \textit{integral curves} defined as a curve $\gamma: (-\epsilon,\epsilon) \to \M, t \mapsto \gamma(t) $ such that (at the identity in $\R$) we have $\gamma(0) = p$ and the push-forward $\gamma_{*}$ equates to the vector field at $p$, that is $\gamma_{*}(d/dt) = X_p$ (definition \ref{defn:geo:Integral curves}). The vector field can then be regarded as the generator of infinitesimal translations on $\M$ constituted by a group of one-parameter (e.g. parametrised by $t \in I \subset \R)$ diffeomorphisms, such that vector fields are equated with diffeomorphisms in an equivalent way to Lie algebras with Lie groups via the exponential map. Thus we can connect Lie theory and differential geometric formulations of curves to map generators of curves in $\g$ to vector fields, facilitating later discussion of Hamiltonians generating (geodesic) curves $\gamma(t)$ which are integral curves. More particularly, we have local flow (section \ref{sec:geo:Local flows}) which provides a way to construct integral curves from the vector field in a way parametrised by $t$.

In earlier sections we introduced notions of dual spaces in the context of quantum information processing and algebra. We relate these to important concepts of one-forms, $n$-forms and tensor products. From dual spaces we define \textit{cotangent vectors} as a map from the tangent space to $\R$ i.e. $k:T_p\M \to \R$ (definition \ref{defn:geo:Cotangent vectors}). This can be thought of as selecting, for $v \in T_p\M$ the appropriate dual $k$ such that $\braket{k,v}$ is in $\R$. The \textit{cotangent space} is the space of all cotangent vectors i.e. at $p \in \M$ it is the set $T^*_p\M$ of all such linear maps constituting cotangent vectors. If $T_p\M$ has a basis $\{e_1,...,e_n\}$, then the \textit{dual basis} for $T^*_p\M$ is a collection of vectors $\{f^1,...,f^n \}$ is uniquely specified by the criterion that $\braket{f^i, e_j} = \delta^i_j \in \R$ i.e. they contract to unity (equation (\ref{eqn:geo:dualdelta})). Understanding how vectors and dual basis elements essentially contract each other down to scalars is a fundamental point of tensorial contractions. The \textit{cotangent bundle} $T^*\M$ is the collection (a vector bundle) of all such cotangent for all $p \in \M$ i.e. $T^*\M = \bigcup_{p\in\M} T^*_p\M$.

From cotangent vectors, we then construct the definition of a \textit{one-form} $\omega$ on $\M$ as the assignment of a cotangent vector (being a smooth linear map) $\omega_p$ to every point $p \in \M$ (definition \ref{defn:geo:One-form}) i.e. $\omega:X \to \R, X \in \mathfrak{X}(\M)$. We discuss this in more detail in section \ref{sec:geo:One-Forms}, including the relationship of one forms to inner products and metrics. As noted in the literature, given $h: \M \to \N$ while we cannot in general obtain a differential form to act as a pushforward $h_*: T_p\M \to T_{h(p)}\N$ between tangent spaces, we can always obtain a \textit{pullback} that maps in the other direction $ h^*:T_{h(p)}^*\N \to T_{p}^*\M$ (definition \ref{defn:geo:Pull-back}). Pullbacks provide a means of transforming differential forms, tensors and other objects between manifolds in order to understand how they transform. In Lie derivative contexts (section \ref{sec:geo:Lie derivatives and pullbacks}), for a vector field $X$ on $\M$ with flow $\phi_t^X$, the pullback $\phi_t^{X*}\omega$ denotes how $\omega$ transforms under each diffeomorphism $\phi_t$. The Lie derivative of $\omega$ with respect to $X$, $L_X\omega$ is then the rate of change of $\phi_t^*\omega$ as it `travels backwards' to 0 (equation \ref{eqn:geo:liederivativepullback}), describing how $\omega$ instantaneously changes along the flow of $X$. In the general case, we move from the Lie derivative to the \textit{exterior derivative} to understand how differential forms $df$ (for any $f \in \cinfm$) change locally on manifolds $\M$ (see definition \ref{defn:geo:exterior_derivative}).

\subsection{Tensors and metrics}
With the formalism of vectors and forms, we can then express tensor products in terms of tensor products of tangent and cotangents. A \textit{tensor of type} $(r,s) \in T_p^{r,s}\M$ at a point $p\in\M$ belongs to the tensor product space (equation (\ref{eqn:geo:Trstensortype})):
\begin{align}
        T_p^{r,s}\M := \Bigg[\otimes^r T_p\M \Bigg] \otimes \Bigg[\otimes^s T^*_p \M\Bigg]
    \end{align}
 i.e. $r$ tensor products of the tangent space with itself tensor-producted with $s$ tensor products of the dual cotangent space. Moreover, tensors can be regarded as mappings taking elements from $T_p\M$ and the dual space $T_p^*\M$ and contracting them down to scalars (equation (\ref{eqn:geo:tensorasmapping})). This formalism of contraction is fundamental to calculating metrics (via metric tensors), curvature (via the Riemann curvature tensor) and other operations. Recalling the contraction to unity of basis elements in equation (\ref{eqn:geo:dualdelta}), we can see that contraction of arbitrary tensors $\braket{a f^i, b e_j} = ab\braket{f^i, e_j} = ab\delta^i_j$ effectively multiplies the remaining tensors (vectors and/or covectors) by the scalar product of coefficients (see section \ref{sec:geo:General tensors and n-forms}). This can be seen in particular for tensor contractions (definition \ref{defn:geo:Tensor contractions}). In quantum information and control contexts, the quantum measurement (via the partial or full trace) can be considered in effect a tensorial contraction in this way. Thus tensors play a central role in quantum theory and dynamics. 

 A tensor of central importance to calculation of time-optimal unitary sequences, measurement and machine learning with quantum systems is the \textit{metric tensor} (section \ref{sec:geo:Metric tensor}) defined as a [0,2]-form tensor field mapping $g_p: T_p\M \times T_p\M \to \R$ given by $g := g_{ij} dx^i \otimes dx^j$ with metric components $g_{ij} = \braket{e_i,e_j}$ and inverse components $g^{ij}=\braket{dx^i,dx^j}$. As we discuss below, it is the existence of a metric tensor on Riemannian and subRiemannian (symmetric) spaces of interest that allows for calculation of time-optimal metrics, such as arc-length (minimal time) or energy. The general principles above can be related to $n$-forms (section \ref{sec:geo:n-forms and exterior products}) in order to build up the concept of an \textit{exterior product} of forms $\omega_1 \wedge \omega_2$ (definition \ref{defn:geo:Exterior product}) and \textit{exterior derivative} (definition \ref{defn:geo:exterior_derivative}). The former is related to Cartan's structural equations (theorem \ref{thm:geo:Cartanstructuralequations}) and the Maurer Cartan form which can be related to measures of curvature for example.
 
 \subsection{Tangent planes and Lie algebras}
 A fundamental aspect of Lie group theory is the isomorphism between the set of all left-invariant vector fields on a Lie group $G$, denoted $L(G)$, and the tangent space $T_eG$ at the identity element $e$ of $G$. This isomorphism implies that one can understand the behaviour of left-invariant vector fields by examining transformations within $T_eG$ (section \ref{sec:geo:Tangent planes and Lie algebras}). The concept of left- and right-invariance of a group action begins with elementary concepts of left translation $l_g: G \to G, g' \mapsto gg'$ and equivalent right-translation. We then say that a vector field $X$ on $G$ is \textit{left-invariant} if $l_{g^*}X = X, \forall g \in G$ i.e. that the left $g$-action on $X$ (represented by the pushforward $l_{g^*}$ - recall we are mapping algebraic and geometric concepts here) keeps $X$ constant. Left invariance also applies to the Lie derivative (equation (\ref{sec:geo:leftinvariancecommutator})). Thus we can construe $\g$ in terms of invariant properties of vector fields $X$. Exponential maps can be framed in terms of integral curves, namely as the unique integral curve satisfying $t \to \gamma^{L^A}(t),       A=\gamma_*^{L^A} \left( \frac{d}{dt} \right)_0$ (section \ref{sec:geo:Exponentials, integral curves and tangent spaces}) of the left invariant vector field $L^A$ associated with the identity in $\M$, that is $\gamma^{L^A}(0)=e$ and which is defined for all $t \in R$. The notation $\gamma^{L^A}$ refers to the integral curve generated by the left-invariant vector field $L^A$ originating from the identity element $e \in G$, reflective of the idea that the evolution of curves $\gamma(t)$ (including those with which we are concerned in relation to time-optimality) can be studied in terms of the canonical Lie algebra associated with $T_eG$. 
 In more advanced treatments, which we leverage in Chapter \ref{chapter:Time optimal quantum geodesics using Cartan decompositions}, the \textit{Maurer-Cartan form} (definition \ref{defn:geo:Maurer-Cartan Form}) provides a useful way identifying the geometric encoding of symmetries of $G$ and $\g$ (see section \ref{sec:geo:Maurer-Cartan Form}), in particular by more explicitly showing how underlying structural features of $\g$ are related to geometric properties such as curvature and indeed Hamiltonian evolution. The Maurer-Cartan form is a $\mathfrak{g}$-valued one-form $\omega$ on $G$ given by $\omega_g(v) = l_{g^{-1}_*}(v)$. The Maurer-Cartan equation then describes the differential form via $d\omega^\alpha + \frac{1}{2}\sum_{\beta,\gamma=1}^n C_{\beta \gamma}^\alpha \omega^\beta \wedge \omega^\gamma=0$ (see equation (\ref{eqn:geo:Maurer-Cartan equation}) and definition \ref{defn:geo:Maurer-Cartan Form}). In Chapter \ref{chapter:Time optimal quantum geodesics using Cartan decompositions} we relate the Schr\"odinger equation to the Maurer-Cartan form as a means of connecting to Lie group-related forms (see section \ref{sec:cartan:generalmethod} and equation (\ref{eqn:cartan:general:maurercartan})).

 \subsection{Fibre bundles and Connections}
 Geometric concepts of fibre bundles and connections are integral to geometric control theory and geometric techniques in quantum information processing and machine learning. The represent central concepts leveraged in Chapter \ref{chapter:Quantum Geometric Machine Learning} and \ref{chapter:Time optimal quantum geodesics using Cartan decompositions}. They play a fundamental structural role in the abstract underpinning of how vectors transform across manifolds $\M$, curvature and parallel transport. Usually in quantum computing contexts, one assumes the existence of a single Hilbert space $\Hilb$ within which vectors are transported.  In geometric framings, instead usually one begins with the type of differentiable manifold $\M$ equipped with a topology of interest. One then associates to points in $\M$ additional abstract structure to enable actions like differentiation and mappings of interest to be well-defined, such as a real or complex valued vector space, but it may be some other type of structure. In order to obtain the type of structural consistency quantum information researchers are used to (i.e. the ease of dealing with a single $\Hilb$), one needs to impose additional geometric structure, including bundles and connections. To see this, note that in geometric framings, the evolution of a system from one state to another is represented as a transformation from $p \to p'$ for $p,p' \in \M$. Each $p \in \M$ has its own abstract space e.g. its own vector space. One must therefore define how vectors in one space e.g. $T_p\M$ transforms to $T_{p'}\M$. 
 
 To handle this abstract formulation, geometry adopts the concept of a \textit{bundle} is defined as a triple $(E, \pi, \mathcal{M})$, where $E$ (denoted the \textit{base space}) and $\mathcal{M}$ are linked via a continuous \textit{projection map} $\pi: E \to \mathcal{M}$ with a corresponding inverse from $\pi^\inv:\M \to E$. The idea of a bundle encapsulates the means of assigning abstract structures such as vector spaces to points $p \in \M$. These abstract structures are the \textit{fibres} associated with $p$. The base space is then the union of all fibres. Thus we define a \textit{base (bundle) space} as $E = \bigcup_{p \in \mathcal{M}} F_p$, with $F_p$ being fibres which remain abstract at this stage. Formally (definition \ref{defn:geo:fibre}), a fibre over $p$ is the inverse image of $p$ under $\pi$. It arises via the map $\pi^{-1}:\mathcal{M} \to T\mathcal{M}$, an example of a fibre bundle associating $p \in \mathcal{M}$ with the tangent space $T_p \mathcal{M}$. The projection $\pi$ associates each fibre $F_p$ with a point $p \in \mathcal{M}$, where $F_p = \pi^{-1}(\{p\})$ defines $F_p$ as the preimage $\{p\}$ of $p$ under $\pi$. Certain bundles have the special property that the fibres $\pi^{-1}(\{p\}), p \in \M$ are all \textit{homeomorphic} (\textit{diffeomorphic} for manifolds) to $F$. In such cases, $F$ is known as the \textit{fibre} of the bundle and the bundle is said to be a \textit{fibre bundle}. For vectors, this is the set of all vectors that are tangent to the manifold at the point $p$. The fibre bundle is sometimes visualised in diagrammatic form (see Isham \cite{isham_modern_1999} $\S$5.1.2):
    \[ \begin{tikzcd}
F \arrow{r} & E \arrow{d}{\pi} \\%
& \M
\end{tikzcd}
\]
An important canonical fibre bundle is the principal fibre bundle (section \ref{sec:geo:Principal fibre bundles}) whose fibre acts as a Lie group. A principal fibre bundle has a typical fibre that is a Lie group $G$, and the action of $G$ on the fibres is by right multiplication, which is free and transitive. The fibres of the bundle are the orbits of the $G$-action on $E$ and hence are not generally homeomorphic to each other. When $G$ is the fibre itself and the action on $G$ is both free and transitive. When dealing with principal fibre bundles, often the notation $P=\bigcup_{p \in \mathcal{M}} F_p$ is used to denote the total space (i.e. $E$) in order to emphasise that each fibre is isomorphic to $G$ itself and that all fibres in the bundle are homogeneous i.e. they are all structurally isomorphic (so we can utilise a single representation for each). This is not necessarily the case in general for base spaces $E$.

To obtain the more specific form of a vector bundle or where fibres are the Lie algebras $\g$ themselves (which intuitively connects with $T\M \sim \g$ correspondence), we rely on the definition of an \textit{associated bundle} (definition \ref{defn:geo:Associated bundle}) which effectively allows fibres bundles that are vector bundles (definition \ref{defn:geo:Vector bundle}), being a type of fibre bundle in which each fibre exhibits the structure of an $n$-dimensional vector space. Vector bundles usefully allow for the definition of \textit{connections} (see below) which describe how fibres (or vector spaces) are connected over different points in $\M$. Intuitively the idea of a connection is a means of associating vectors between infinitesimally adjacent tangent planes $T_p\M \to T_{p+dp}\M$ as one progresses from $\gamma(t)=p$ to $\gamma(t+dt)=p+dt$ on $\M$.

Connections (section \ref{sec:geo:Connections}) are of fundamental importance to results in our final Chapter and also to definitions of vertical and horizontal subspaces in subRiemannian control problems further on. A connection on a principal bundle defines a notion of horizontal and vertical subspaces within the tangent space of the total space. Hence they are a fundamental means of distinguishing between Riemannian and subRiemannian manifolds via the decomposition of fibres (and Lie algebras) into horizontal and vertical subspaces. This distinction is crucial for defining parallel transport and curvature, concepts that are central to understanding the dynamics and control of systems with symmetry. A connection on $P$ (which we treat in terms of a vector bundle) provides a smooth splitting of the tangent space $T_pP$ at each point $p \in P$ into \textit{vertical} and \textit{horizontal} subspaces, $T_pP = V_pP \oplus H_pP$, where $V_pP$ is tangent to the fibre (recall we measure tangency here in terms of inner products) and $H_pP$ is isomorphic to $T_{\pi(p)}\mathcal{M}$. The vertical subspace (definition \ref{defn:geo:Vertical subspace}) $V_pP$ at $p \in P$ is defined as the kernel of the differential of the projection map $\pi: P \rightarrow \mathcal{M}$. $V_pP$ consists of tangent vectors to $P$ at $p$ that are `vertical' in the sense that they point along the fibre $\pi^{-1}(\pi(p))$. These vectors represent infinitesimal movements within the fibre itself, without leading to any displacement in the base manifold $\mathcal{M}$. The \textit{horizontal subspace} $H_pP$ (definition \ref{defn:geo:Horizontal subspace}) at $p \in P$ consists of vectors that are `horizontal' in the sense that they correspond to displacements that lead to movement in the base manifold $\mathcal{M}$ when considered under parallel transport defined by the connection. A \textit{connection} (definition \ref{defn:geo:Connection}) (on a principal $G$-bundle but we focus here on vector bundles) $G \to E \to \M$ is a smooth assignment of horizontal subspaces $H_pP \subset T_pP$, to each point $p \in \M$. Essentially they allow the partitioning of the vector bundle and importantly specify the horizontal subspace which in turn is crucial for generating geodesics along $\M$. For example, in Chapter \ref{chapter:Quantum Geometric Machine Learning} and \cite{swaddle_subriemannian_2017,swaddle_generating_2017}, generators chosen from $H_p\M$ (the distribution $\Delta$) are relied upon in order to generate geodesic training data. 

In Chapter \ref{chapter:Time optimal quantum geodesics using Cartan decompositions}, our control subset $\p \subset \g$ can be construed as a horizontal subspace (see section \ref{sec:geo:Relation to Cartan decompositions} for more discussion).  The decomposition into vertical and horizontal subspaces discussed above is particularly relevant to time optimal control problems. In essence, given the Cartan decomposition (\ref{defn:alg:cartandecomposition}) $\g = \k \oplus \p$ we associate $\k$ as the vertical and $\p$ as the horizontal subspace. We can then understand the symmetry relations expressed by the commutators related to this vertical and horizontal sense of directionality: i.e. given $[\k,\k]\subset \k, [\k,\p]\subset \p$ and $[\p,\p]\subset \k$ we can see that the horizontal generators under the adjoint action shift from $p \in G/K$ to $p' \in G/K$, while the vertical generators in k do not translate those points in $G/K$.

\subsection{Geodesics and parallel transport}
We now define concepts of parallel transport, covariant differentiation and geodesics, each of which are essential to the geometric time-optimal control formulation applied in later Chapters. We are interested in \textit{horizontal vector fields} whose flow lines move from one fibre to another, intuitively constituting translation across a manifold via generators in the horizontal subspace.  Parallel transport shifts vector fields vectors along integral curves such that they are parallel according to a specified connection. The concept of \textit{horizontal lifts} (definition \ref{defn:geo:Horizontal lift}) is central to this process by enabling (i) preservation of a concept of direction and (ii) preserving `horizontality' such that the notion of straightness or parallelism of Euclidean space can be extended to curved manifolds, essential for the study of geodesics. For a smooth curve $\gamma$ in $\mathcal{M}$, a horizontal lift $\gamma^\uparrow: [a,b] \to P$ is a curve whose tangent vectors are in the horizontal subspaces. The vector field $(X^{\uparrow}_p)$ is identified as the \textit{horizontal lift} of $X$ as it `lifts' up the vector field $X$ on $\M$ into the \textit{horizontal subspace} of $TP$. The requirement of $V_p(X^{\uparrow}_p) = 0$, indicates that that $X^{\uparrow}$ lies entirely in the horizontal subspace $H_p\M$, encapsulating the essence of parallel transport as maintaining the direction of $X$ through the fibres of $P$.

Given a curve $\gamma(t) \in \M$, the vector field $X$ is parallel along $\gamma(t)$ if the covariant derivative vanishes along the curve. We denote such fields as \textit{parallel vector fields} (definition \ref{def:geo:parallelvectorfields}). The notion of \textit{parallel transport} along $\gamma$ from $\gamma(a)$ to $\gamma(b)$ is defined as a map $\tau: T_{\gamma(a)}\M \to T_{\gamma(b)}\M$ satisfying $\tau: \pi^{-1}(\{ \gamma(a) \}) \to \pi^{-1}(\{ \gamma(b) \})$ (definition \ref{def:geo:Paralleltransport}). In more familiar language this can be shown to be equivalent to $\nabla_{\dot \gamma(t)}X = 0$ by maintaining vectors in the horizontal subspace $H_p\M$. Note sometimes we define for $\dot \gamma(t) \in T_p\M$ the terms $\nabla_{\dot \gamma(t)}:= \nabla_{\gamma(t)}$ where $\gamma(t)$ is any curve that belongs to the equivalence class of $[\dot \gamma(t)]$ (Isham $\S 6.7$). The \textit{covariant derivative} (definition \ref{defn:geo:covariantderivative}) $\nabla$ can then be defined in terms of parallel transport as:
\begin{align*}
    \nabla_{\gamma}X := \lim_{t \to 0} \frac{\tau_{t}^{-1} X(\gamma(t)) - X(\gamma(0))}{t}
\end{align*}
where $X \in \mathfrak{X}$. In this work we generally use the notation $\nabla_{\dot \gamma(t)}$ to emphasise the covariant derivative is with respect to $\dot \gamma(t)$. The notation in geometric settings (section \ref{sec:geo:Geodesics and parallelism}) can get quite dense, but we can see in the definition above that $\tau_{t}^{-1}$ transports vectors back along $\gamma(t) \to \gamma(0)=p_0$ while preserving parallelism. Here $X(p_0)$ can be thought of as the initial vector field at $\gamma(0)=p_0$ (the start of our curve) and $X(\gamma(t))$ the vector field at some later point $\gamma(t)$ with $\tau_{t}^{-1}X(\gamma(t))$ the vector at a later time transported back. The equation above is then zero in such case as $\tau_{t}^{-1} X(\gamma(t)) - X(\gamma(0))$ which requires identity between the vectors, hence they are `parallel'.  The $\nabla_X$ operator is also linear in $\mathfrak{X}(\M)$ (which can also be regarded as a module over $\cinfm$). These properties are expressed by considering $\nabla_X$ as an \textit{affine connection} an operator $\nabla: \mathfrak{X}(\M) \times \mathfrak{X}(\M) \to \mathfrak{X}(\M)$ which associates with $X \in \mathfrak{X}(\M)$ a linear mapping $\nabla_X$ of $\mathfrak{X}(\M)$ satisfying affine conditions (see definition \ref{defn:geo:Affine connection}). As we mention below, Riemannian manifolds area equipped with an important unique affine connection denoted the Levi-Civita connection. The Levi-Civita connection is unique in that it is torsion-free (so characterised by curvature only) and by virtue of its compatibility with the Riemannian metric (i.e. it preserves the inner product of tangent vectors under parallel transport, in turn preserving angles and lengths along curves).

We can now define a geodesic (definition \ref{defn:geo:Geodesic}) using such notions of parallelism and the covariant derivative. Generally, a geodesic is a curve that locally minimises distance and is a solution to the geodesic equation derived from a chosen connection. Denote $\gamma: I \to \M, t \mapsto \gamma(t)$ for an interval $I \subset \Real$ (which we generally without loss of generality specify as $I=[0,1]$) with an associated tangent vector $\dot\gamma(t)$. Here $\gamma$ is regular. Two vector fields $X,Y$ are parallel along $\gamma(t)$ if $\nabla_XY = 0, \forall t \in I$. A curve $\gamma: I \to \M, t \mapsto \gamma(t)$in $\M$ is denoted a \textit{geodesic} if the set of tangent vectors $\{\dot\gamma(t)\} = T_{\gamma(t)}\M$ is parallel with respect to $\gamma$, corresponding to the condition that $\nabla_{\gamma}\dot\gamma=0$, which we denote the geodesic equation. In a coordinate frame the geodesic equation is expressed in its somewhat more familiar form as (equation \ref{eqn:geo:geodesiccoordinateframe}):
\begin{align*}
    \frac{d^2u^\gamma }{ds^2} +   \Gamma^\gamma_{\alpha \beta} \frac{du^\alpha }{ds} \frac{du^\beta }{ds} = 0 
\end{align*}
where:
\begin{align*}
\Gamma^\gamma_{\alpha \beta} = \frac{1}{2} g^{\gamma \mu} \left( \frac{\partial g_{\mu \alpha}}{\partial u^\beta} + \frac{\partial g_{\mu \beta}}{\partial u^\alpha} - \frac{\partial g_{\alpha \beta}}{\partial u^\mu} \right)
\end{align*}
are the Christoffel symbols of the second kind (essentially connection coefficients) with $g^{\gamma \mu}$ the inverse of the metric tensor and $ds$ usually indicates parametrisation by arc length. Solutions to this equation are geodesic curves $\gamma(t)$. For Lie group manifolds, all geodesics are generated by generators from the horizontal subspace $H_p\M$, but not all curves generated from the horizontal subspace are geodesics.

\subsection{Riemannian and subRiemannian manifolds}
Equipped with definitions of geodesics and parallelism we can define Riemannian and subRiemannian manifolds (section \ref{sec:geo:Riemannian manifolds}). A Riemannian manifold (see definition \ref{defn:geo:Riemannian Manifold}) is the tuple $(\M,g)$ (i.e a manifold $\M$ with a metric $g$) where to each $p \in \M$ is assigned a positive definite map $g_p: T_p\M \times T_p\M \to \R$ (described usually in terms of being an inner product) and an associated norm $||X_p||: T_p\M \to \R$. Usually we first define a \textit{Riemannian structure} (definition \ref{defn:geo:(Pseudo)-Riemannian structure}) as a type-(0,2) tensor field $g$ such that (a) $g(X,Y) = g(Y,X)$ (symmetric) and (b) for $p \in \M$, $g$ is a non-degenerate bilinear form $g_p: T_p\M \times T_p\M \to \R$. A Riemannian manifold is then formally defined as a connected $\cinfm$ manifold with a Riemannian structure such that there exists a unique affine connection satisfying (a) $\nabla_X Y - \nabla_Y X = [X,Y]$ (zero torsion) and (b) $\nabla_Z g = 0$ (parallel transport preserving inner products where $g$ is the Riemannian metric). For a richer discussion of the importance of curvature and torsion to such definitions, see definitions of the Riemann curvature tensor (definition \ref{defn:geo:Riemann curvature tensor}) and Cartan's structural equations in theorem \ref{thm:geo:Cartanstructuralequations}. We can then define the important concept of a \textit{Riemannian metric} (definition \ref{defn:geo:Riemannian metric}) as an assignment to each $p \in \M$ of a positive-definite inner product $g_p: T_p\M \times T_p\M \to \R$ with an induced norm $||\cdot ||_p: T_p\M \to \Real, (v,w)\mapsto \sqrt{g_p(v,w)}$. Note that the metric is a $(0,2)$-form thus a tensor, aligning with definition (\ref{def:geo:metric_tensor}).

With the metric in hand, we can now posit a definition of \textit{arc length} (definition \ref{defn:geo:arclength}) which is central to the concept of measuring distance on manifolds and, by extension, time-optimal paths. Given a curve $\gamma(t) \in \M$ with $t \in [\alpha,\beta]$ and metric $g$, the arc length of the curve from $\gamma(0)$ to $\gamma(T)$ is given by:
    \begin{align}
    \ell(\gamma) = \int_0^T \left( g(\dot\gamma(t),\dot\gamma(t)) \right)^{1/2} dt 
\end{align}
Assuming $\M$ is simply connected, then all $p,q \in \M$ can be joined via a curve segment. We then define the metric of \textit{distance} (equation (\ref{eqn:geo:riemannmetric})) between $p,q \in \M$ as the infimum of the shortest curve measured according to the equation (\ref{eqn:geo:arclength}) above, $d(p,q) = \inf_\gamma \ell(\gamma)$. The preservation of distances can then be understood in terms of total geodesicity of manifolds. 
A sub-manifold $S$ of a Riemannian manifold $\M$ is \textit{geodesic at} $p$ if each geodesic tangent to $S$ at $p$ is also a curve in $S$. The submanifold $S$ is \textit{totally geodesic} if it is geodesic for all $p \in S$. It can then be shown that if $S$ is totally geodesic, then parallel translation along $\gamma \in S$ always transports tangents to tangents, that is $\tau: S_p\M \to S_{p'}\M$. Further background detail relevant to the differential geometric theory underpinning the core idea of being able to calculate (and compare) distances on manifolds $\M$ of interest (such as key concepts of first and second fundamental forms and Gauss's \textit{Theorema Egregium} which provides that Gaussian curvature $K$ is invariant under local isometries) can be found in section \ref{sec:geo:Fundamental forms}. 

\subsection{Symmetric spaces}
Chapters \ref{chapter:Quantum Geometric Machine Learning} and \ref{chapter:Time optimal quantum geodesics using Cartan decompositions} focus on quantum control problems where $G$ corresponds to Lie groups admitting Cartan decompositions. As manifolds $\M$ these groups constitute symmetric spaces (section \ref{sec:geo:Symmetric spaces}) and may be classified according to Cartan's regimen. Symmetric spaces were originally defined as Riemannian manifolds whose curvature tensor is invariant under all parallel translation. Locally, they resemble Riemannian manifolds of the form $\R^n \times G/K$ where $G$ is a semi-simple lie group (definition \ref{defn:alg:Simple and semisimple}) with an involutive automorphism whose fixed point set is the compact group $K$ while $G/K$, as a homogeneous space, is provided by a $G$-invariant structure (see \cite{knapp_lie_1996}). Cartan thus allowed the classification of all symmetric spaces in terms of classical and exceptional semi-simple Lie groups.  

The study of symmetric spaces then becomes a question of studying specific involutive automorphisms of semi-simple Lie algebras, thus connecting to the classification of semi-simple Lie groups. 
\textit{Geodesic symmetry} is defined in terms of diffeomorphisms $\varphi$ of $\M$ which fix $p \in \M$ and reverse geodesics through that point i.e. when acting the map $f: \gamma(1) = q \to \gamma(-1)=q'$, such that $d\varphi_p: T_p\M \to T_p\M, X \mapsto d\varphi(X) = -X$ (i.e. $d\varphi_p \equiv -I$) where $I$ denotes the identity in $T_p\M$. Here $d\varphi$ can be thought of as the effect of $\varphi$ on the tangent space e.g. in Lie algebraic terms $\varphi(\gamma(t))=\varphi(\exp(X))=\exp(d\varphi(X))=\exp(-X)$ for $X \in \g$. This means manifold $\M$ is locally symmetric around $p \in \M$. The manifold $\M$ is then \textit{locally (Riemannian) symmetric} (definition \ref{defn:geo:Riemannian local symmetric space}) if each geodesic symmetry is isometric that is, if there is at least one geodesic symmetry about $p$ which is an isometry (and globally so if this applies for all $p \in \M$). As noted in the literature, this is equivalent to the vanishing of the covariant derivative of the curvature tensor along the geodesic. A globally symmetric space is one where geodesic symmetries are isometries for all $\M$.
A manifold being \textit{affine locally symmetric} is one which, given $\nabla$, the torsion tensor $T$ and curvature tensor $R$, $T=0$ and $\nabla_Z R=0$ for all $Z \in T\M$. A manifold is \textit{Riemannian globally symmetric} space if all $p \in \M$ are fixed points of the Cartan involutive symmetry $\theta$ (see definition \ref{defn:alg:cartaninvolution}) such that $\theta^2(p) = \theta$. It can be shown then that the Riemann curvature tensor for the symmetric space (homogeneous) $G/K$ with a Riemannian metric allows for curvature, via $R$:
% , to be related to the Lie triple property, namely:
%
\begin{align}
    R_p(X,Y)Z = -[[X,Y],Z] \qquad X,Y,Z \in \p
\end{align}
 
Curvature plays an important role in the classification of symmetric spaces. Three classes of symmetric space can be classified according to their sectional curvature as follows. Given a Lie algebra $\g$ equipped with an involutive automorphism $\theta^2=I$, with corresponding group $G$ with $G/K$ as above, we have a $G$-invariant structure on the Riemannian metric i.e. $g(X,Y) = g(hX,hY)$ for $h\in G/K$. Then the three types of symmetric space are (i) $G/K$ compact, then $K(X,Y) > 0$, (ii) $G/K$ non-compact, then $K(X,Y) < 0$ and (iii) $G/K$ Euclidean, then $K(X,Y)=0$. Curvature can be related in an important way to commutators as per the presence of commutator terms in the Riemannian curvature tensor (section \ref{defn:geo:Riemann curvature tensor}) and the second of the Cartan structural equations (theorem \ref{thm:geo:Cartanstructuralequations}). The classification table for symmetric spaces is given in section \ref{sec:geo:Classification of symmetric spaces}. Our focus in later Chapters is on $SU(2)$ (and tensor products thereof) together with $SU(3)$. 

\subsection{SubRiemannian Geometry}
With the understanding of geometric concepts above, we can now progress to key concepts in subRiemannian geometry and geometric control which are applied in later Chapters. SubRiemannian geometry is relevant to quantum control problems where only a subspace $\p \subset \g$ is available for Hamiltonian control. Intuitively, this means that the manifold exhibits differences with a Riemannian manifold where the entirety of $\g$ is available, including in relation to geodesic length. In this sense subRiemannian geometry is a more generalised form of geometry with Riemannian geometry sitting within it conceptually. We detail a few key features of subRiemannian theory below before moving onto geometric control.  

SubRiemannian geometry involves a manifold $\M$ together with a distribution $\Delta$ upon which an inner product is defined.  Distribution in this context refers to a linear sub-bundle of the tangent (vector) bundle of $\M$ and corresponds to the horizontal subspace of $T\M$ discussed above and where the vertical subspace is non-null. Formally it is defined (definition \ref{defn:geo:SubRiemannian manifold}) as consisting of a distribution $\Delta$, being a vector sub-bundle $H_p\M \subset T\M$ together with a fibre inner product on $H_p\M$. The sub-bundle corresponds to the \textit{horizontal distribution}, having the meaning ascribed to horizontal subspace $H_p\M$ above. In the language of Lie algebra, where for a decomposition $\g = \k \oplus \p$, we have for our accessible (or control) subspace $\p \subset \g$ rather than $\p = \g$. Thus quantum control scenarios where only a subspace of $\g$ is available for controls can be framed as subRiemannian problem (under typical assumptions). Geometrically, this means that the generators or vector fields $\mathfrak{X}(\M)$ are constrained to limited directions. A transformation not in the horizontal distribution may be possible where the space exhibits certain symmetry structure such as for symmetric spaces equipped with Lie triple property (as we discuss for certain subspaces where $[\p,\p]\subseteq \k$), but in a sense these are indirect such that the geodesic paths connecting the start and end points will be longer than for a Riemannian geometry on $\M$. A curve on $\M$ is a \textit{horizontal curve} if it is tangent to $H_p\M$. \textit{SubRiemannian length} $\ell = \ell(\gamma)$ (for $\gamma$ smooth and horizontal) is then defined in the same way as Riemannian via $\ell(\gamma) = \int || \dot \gamma(t) || dt$. SubRiemannian distance $d_s$ is similarly defined (noting that the subRiemannian distance may be less than or equal to the Riemannian distance as is the subRiemannian metric denoted a \textit{cometric} (definition \ref{defn:geo:SubRiemannian (co)metric})). From the subRiemannian metric we specify a system of Hamilton-Jacobi equations on $T^*\M$ the solution to which is a subRiemannian geodesic. This subRiemannian Hamiltonian formalism can be used to define a \textit{subRiemannian Hamiltonian} given by $H(p,\alpha) = \frac{1}{2}(\alpha,\alpha)_p$ where $\alpha \in T^*\M$ and $(\cdot,\cdot)_p$ is the cometric. SubRiemannian geometry has its own form of Hamiltonian differential equations (see \cite{montgomery_tour_2002}  and Appendix \ref{chapter:background:Differential Geometry}) given by $\dot x^i = \partial H\partial p_i, \dot\partial p_i = -\partial H/\partial x^i$, denoted the \textit{normal geodesic equations} (theorem \ref{thm:geo:Normal subRiemannian geodesics}). The coordinates $x^i$ are the position and $p_i$ and momenta functions for coordinate vector fields. Under these conditions and the assumption that the distribution $\Delta$ is bracket generating (definition \ref{defn:geo:bracketgenerating}) in the same way a Lie subalgebra may be, the theory (Chow and Raschevskii theorem, see \cite{montgomery_tour_2002}) guarantees the existence of geodesics that minimise subRiemannian distance 

\section{Geometric control}
\subsection{Overview}
We conclude this section by bringing the above concepts together through the fulcrum of geometric control theory (section \ref{sec:geo:Geometric control theory}) relative to quantum control problems of interest. The primary problem we are concerned with in our final two chapters is solving time optimal control problems for certain classes of Riemannian symmetric space. The two overarching principles for optimal control are (a) establishing the \textit{existence} of controllable trajectories (paths) and thus reachable states; and (b) to show that the chosen path (or equivalence class of paths) is \textit{unique} by showing it meets a minimisation (thus optimisation) criteria. The Pontryagin Maximum Principle provides a framework and conditionality for satisfying these existence and uniqueness conditions. As we discuss below, it can, however, often be difficult or infeasible to find solutions to an optimisation problem expressed in terms of the PMP. However, for certain classes of problem (with which we are concerned) involving target states in Lie groups $G$ with generators in $\g$, results from algebra and geometry can be applied to satisfy these two requirements. Thus, under appropriate circumstances, where targets are in Lie groups $G$ with the Hamiltonian constructed from Lie algebra element $\g$, so long as our distribution $\Delta \subset \g$ is bracket-generating (so we can recover all of $\g$ via nested commutators), then $G$ is in principle reachable. This satisfies the existence requirement. To satisfy uniqueness requirement, we then apply results from differential geometry and algebra regarding constructing Hamiltonians to ensure paths generated are geodesics, thus optimal by virtue of being minimal time trajectories.\\

In our case, time optimal control is equivalent to finding the time-minimising subRiemannian geodesics on a manifold $\M$ corresponding to the homogeneous symmetric space $G/K$. Our particular focus is the $KP$ problem, where $G$ admits a Cartan $KAK$ decomposition where $\g = \k \oplus \p$, with the control subset (Hamiltonian) comprised of generators in $\p$. In particular such spaces exhibit the Lie triple property $[[\p,\p],\p] \subseteq \k$ given  $[\p,\p]\subseteq \k$. In such cases $\g$ remains in principle reachable, but where minimal time paths constitute subRiemannian geodesics. Such methods rely upon \textit{symmetry reduction} \cite{sheller_symmetry_nodate}. As D'Alessandro notes \cite{dalessandro_introduction_2007}, the primary problem in quantum control involving Lie groups and their Lie algebras is whether the set of reachable states $\mathcal{R}$ (defined below) for a system is the connected Lie group $G$ generated by $\mathcal{L} = \spn\{-H(u(t))\}$ for $H \in \g$ (or some subalgebra $\h \subset \g$) and $u \in U$ (our control set, see below). This is manifest then in the requirement that $\mathcal{R} = \exp(\mathcal{L})$. In control theory $\mathcal{L}$ is designated the \textit{dynamical Lie algebra} and is generated by the Lie bracket (derivative) operation among generators in $H$. The dynamical adage is a reference to the time-varying nature of control functions $u(t)$.
We explicate a few key concepts below. The first is the general requirement that tangents $\dot \gamma(t)$ be essentially bounded so that $\braket{H(t),H(t)}_S \leq N$ for all $t \in [0,T]$ for some constant $N$ (definition \ref{defn:geo:Essentially bounded}). For time-optimal synthesis, we seek geodesic (or approximately geodesic) curves $\gamma(t) \in \M$. For this, we draw upon the theory of horizontal subspaces described above manifest via the concept of a \textit{horizontal control curve} (definition \ref{defn:geo:Horizontal control curves}). Given $\gamma(t) \in \M$ with $\dot\gamma(t) \in \Delta_{\gamma} \subseteq H_p\M$, we can define horizontal control curves as (equation (\ref{eqn:geo:horizontalcurves})):
\begin{align*}
        \dot\gamma(t) = \sum_j^m u_j(t) X_j(\gamma(t))
\label{eqn:geo:horizontalcurves}
    \end{align*}
where $u_j$ are the control functions given by $ u_j(t) = \braket{X_j(\gamma(t)),\dot\gamma(t)} $. The length of a horizontal curve, which is essentially what we want to minimise in optimisation problems, is given by (equation (\ref{eqn:geo:arclengthparamby})):
\begin{align}
    \ell(\gamma) = \int_0^T ||\dot \gamma(t)|| dt =  \int_0^T \sqrt{\braket{\dot\gamma(t),\dot\gamma(t)}} dt = \int_0^T \sqrt{\sum_j^m u^2_j(t)} dt
\end{align}
Note that when $[0,T]$ is normalised this is equivalent to parametrisation by arc length.

\subsection{Time optimisation}
For subRiemannian and Riemannian manifolds, the problem addressed in our final two chapters in particular is, for a given target unitary $U_T \in G$, how to identify the minimal time for and Hamiltonian to generate a minimal geodesic curve such that $\gamma(0)=U_0$ and $\gamma(T)=U_T$ for $\gamma:[0,T] \to \M$. This is described \cite{albertini_symmetries_2018} as the \textit{minimum time optimal control problem} \cite{jurdjevic_geometric_1997}. In principle, the problem is based upon the two equivalent statements (a) $\gamma$ is a minimal subRiemannian (normal) geodesic between $U_0$ and $U_T$ parametrised by constant speed (arc length); and (b) $\gamma$ is a minimum time trajectory subject to $||\vec u|| \leq L$ almost everywhere (where $u$ stands in for the set of controls via control vector $\vec u$). SubRiemannian minimising geodesics starting from $q_0$ to $q_1$ (our $U_T$) subject to bounded speed $L$ describe \textit{optimal synthesis} on $\M$. There are also subtleties relating to whether loci of geodesics are in critical loci or cut loci (see section \ref{sec:geo:Time optimal control}), something explored in the geometric control literature.  The critical locus $CR(\M)$ identifies $t$ (and thus $\gamma(t)$) for which a geodesic ceases to be time optimal beyond a marginal extension $\epsilon$ of the parameter interval $I$, effectively bounding the reachable set (see below) for time-optimality to within this interval. A cut locus $CL(\M)$ indicates the set of $p \in \M$ where multiple minimal geodesics intersect, providing a measure of redundancy or flexibility given controls (multiple options for time-optimality) in control problems. The concept of reachability is framed in terms of \textit{reachable sets} - put simply, for any control problem to be well-formed, we require that the target or end-point ($U_T \in G$ in our case) be `reachable' by application of controls to generators. Formally (definition \ref{defn:geo:Reachable set}), the set of all points $p \in \M$ such that, for $\gamma(t): [0,T] \to \M$ there exists a bounded function $\vec u, ||\vec u||\leq L$ where $\gamma(0)=q_0$ and $\gamma(T) = p$ is called the reachable set and is denoted $\mathcal{R}(T)$.

\subsection{Variational methods}
Of fundamental importance to optimal control solutions (in geometric and other cases) is the application of variational calculus in terms of Euler-Lagrange, Hamiltonian and control optimisation formalism. Sections \ref{sec:geo:variational methods} and \ref{sec:geo:KP Problems} set out in some detail the formalism of geometric control leveraged in this work, supplemented by discussion of generic quantum control principles in section \ref{sec:quant:Quantum Control}.

As noted in our discussion of quantum control above, the Pontryagin Maximum Principle (PMP) (section \ref{sec:geo:Pontryagin Maximum Principle}) is one of the fundamental principles governing optimal control theory. The principle sets out the necessary and sufficient conditions for optimal control of classical and quantum systems, where system evolution is represented geometrically by time (or energy) minimal paths $\gamma(t)$ on $\mathcal{M}$. 
Assume a differentiable manifold $\M$ on which we define $\cinfm$ functions $(f_1(\gamma,u),...,f_m(\gamma,u))$ where our coordinate charts $\phi \in S \subset \R^n$ parametrise curves $\gamma(t) \in \M$. We have as our control variable $u \in U \subset \R^m$ where $U$ is our control set. Both are parametrised as $u=u(t),\gamma=\gamma(t)$ for $t \in I = [t_0,t_1]$ (assuming boundedness and measurability). The evolution of $\gamma \in \M$ is determined by the state and the control, according to the differential \textit{state equation} $\dot\gamma_i(t) = f_i(\gamma(t),u(t))$ (equation (\ref{eqn:geo:pontrydotgamma}))
for almost all $t \in I$. Solution curves $\gamma(t)$ to the state equation are typically unique under these conditions. Importantly, the PMP provides a framework to understand how small variations in the initial conditions affect the system's evolution, necessary for determining optimal trajectories that satisfy constraints and minimise the cost functional (see equation (\ref{eqn:geo:costfunctional})). The PMP encodes dynamical constraints via  \textit{adjoint (costate) equations} where each state variable $\gamma_i(t)$ has a corresponding adjoint variable or co-state $p_i(t)$. The dynamics of these adjoint variables are described by the adjoint equations  $\frac{dp_i}{dt} = -\frac{\partial H}{\partial \gamma_i}(\gamma(t), p(t), u(t))$ (equations (\ref{eqn:geo:pontryadjointsystem})) where $H$ is the Hamiltonian (equation (\ref{eqn:geo:PMPHamiltonian})). The adjoint variables $p(t)$ effectively serve a role similar to Lagrange multipliers, providing a mechanism to incorporate the state dynamics and constraints directly into the Hamiltonian. The PMP then provides (definition \ref{defn:geo:Pontryagin maximum principle}) a maximum principle for solving the optimal control problem by requiring that for a trajectory $(\gamma(t),u(t))$ evolving from $a \to b$ over interval $t \in I=[0,T]$, there exists a non-zero absolutely continuous curve $p(t)$ on $I$ satisfying constraints that (a) the set of states, costates and controls $(\gamma(t),p(t),u(t))$ is a solution curve to the Hamiltonian equations; (b) that the Hamiltonian is maximal $H(x(t), p(t), u(t)) = H_M(x(t), p(t))$ for almost all $t \in [0, T]$ and (c) $p_0(T) \leq 0$ and $H_M(x(T), p(T)) = 0$.

In practice the PMP can be difficult or complicated to solve, thus in control theory one often seeks ways to simplify problems. The primary relevance of the PMP in our work relates to guarantees regarding the existence of time-optimal solutions which we seek to learning Chapter \ref{chapter:Quantum Geometric Machine Learning} using machine learning and provide an alternative method for deducing in Chapter \ref{chapter:Time optimal quantum geodesics using Cartan decompositions}. This is especially the case where target states $\gamma(T)$ belong to Lie groups $G$ with associated Lie algebras $\g$, certain symmetry reduction techniques can make the problem tractable, which often means reducing the task to a simpler minimisation problem. A class of such problems is where $G/K$ is a Riemannian symmetric space and $KP$ problems which are the subject of our last two chapters. 

\subsection{KP problems}
A particular type of subRiemannian optimal control problem with which we are concerned in the final Chapter is the $KP$ problem. In control theory settings, the problem was articulated in particular via Jurdjevic's extensive work on geometric control \cite{jurdjevic_abstract_1970,jurdjevic_control_1972,jurdjevic_control_1981,jurdjevic_controllability_1978}, drawing on the work of \cite{brockett_sub-riemannian_nodate} as particularly set out in \cite{jurdjevic_geometric_1997} and \cite{jurdjevic_hamiltonian_2001}. Later work building on Jurdjevic's contribution includes that of Boscain \cite{boscain_introduction_2021,boscain_k_2002,boscain_invariant_2008}, D'Alessandro \cite{dalessandro_introduction_2007,dalessandro_k-p_2019} and others. $KP$ problems are also the focus of a range of classical and control problems focused on the application of Cartan decompositions of target manifolds $G=K \oplus P$ where elements $U \in G$ can be written in terms of $U=KP$ or $U=KAK$ (see \cite{dalessandro_introduction_2007}). In this formulation, the Lie group and Lie algebra can be decomposed according to a Cartan decomposition (definition \ref{defn:alg:cartandecomposition}) $\g = \k \oplus \p$ (and associated Cartan commutation relations). The space is equipped with a Killing form (definition \ref{defn:alg:Killing form}) which defines an implicit positive definite bilinear form $(X,Y)$ which in turn allows us to define a Riemannian metric restricted to $G/K$ in terms of the Killing form.
Such control problems posit controls in $\p$ with targets in $\k$ and are a form of \textit{subRiemannian control} problem.

Assume our target groups are connected matrix Lie groups (definition \ref{defn:alg:connected_matrix_lie_group}). Recall equation (\ref{eqn:geo:horizontalcurves}) can be expressed as:
\begin{align}
    \dot\gamma(t) = \sum_j X_j(\gamma) u_j(t)
\end{align}
where $X_j \in \Delta = \p$, our control subset. For the $KP$ problem, we can situate $\gamma(0) = 1 \in \M$ (at the identity) $||\hat u|| \leq L$, in turn specifying a reachable set $\mathcal{R}(T)$.  As D'Alessandro et al. \cite{albertini_symmetries_2018,dalessandro_time-optimal_2020} note, reachable sets for $KP$ problems admit reachable sets for a larger class of problems. Connecting with the language of control, we can frame equation (\ref{eqn:geo:horizontalcurves}) in terms of \textit{drift} and \textit{control} parts with:
\begin{align}
    \dot\gamma(t) = A \gamma(t) + \sum_j X_j(\gamma_j) u_j(t) \label{eqn:geo:horizontalcontroSchrod}
\end{align}
where $A\gamma(t)$ represents a drift term for $A \in \k$. Our unitary target in $G$ can be expressed as:
\begin{align}
    \dot \gamma(t) = \sum_j \exp(-At)X_j\exp(A t) \gamma(t) u_j
    \label{eqn:geo:dotU=sumjKAKdal}
\end{align}
for bounded $||A_p|| = L$. For the $KP$ problem, the PMP equations are integrable. One of Jurdjevic's many contributions was to show that in such $KP$ problem contexts, optimal control for $\vec u$ is related to the fact that there exists $A_k \in \k$ and $A_p \in \p$ such that:
\begin{align}
    \sum_j^m X_j u_j(t) = \exp(At)X_j\exp(-A t)
\end{align}

Following Jurdjevic's solution  \cite{jurdjevic_hamiltonian_2001} (see also \cite{albertini_symmetries_2018}), optimal pathways are given by:
\begin{align}
    \dot\gamma(t) &= \exp(A_kt)A_p\exp(-A_k t) \gamma(t) \qquad \gamma(0)=1 \\
    \gamma(t) &= \exp(-A_k t) \exp((A_k + A_p)t)
\end{align}
resulting in analytic curves. Our final Chapter \ref{chapter:Time optimal quantum geodesics using Cartan decompositions} sets out an alternative method for obtaining equivalent time-optimal control solutions for certain classes of problem.

%========Quantum machine learning
\section{Quantum machine learning}
Quantum machine learning (QML) adapts concepts from modern machine learning (and statistical learning) theory to develop learning protocols for quantum data and quantum algorithms. This section summarises key concepts relevant to the use of machine learning in Chapters \ref{chapter:QDataSet and Quantum Greybox Learning} and \ref{chapter:Quantum Geometric Machine Learning}. A more expansive discussion of background concepts is set out in Appendix \ref{chapter:Background: Classical, Quantum and Geometric Machine Learning} from which this section draws. The landscape of QML is already vast thus for the purposes of our work it will assist to situate our results within the useful schema set out in \cite{schuld_machine_2021} below in Table (\ref{table:quantumclassical}) from Appendix \ref{chapter:Background: Classical, Quantum and Geometric Machine Learning} which we reproduce below. Our results in Chapters \ref{chapter:QDataSet and Quantum Greybox Learning} and \ref{chapter:Quantum Geometric Machine Learning} fall somewhere between the second (classical machine learning using quantum data) and fourth (quantum machine learning for quantum data) categories, whereby we leverage classical machine learning and an adapted bespoke architecture equivalent to a parametrised quantum circuit \cite{benedetti_parameterized_2019} (discussed below). 

%===QUANTUM CLASSICAL TABLE
\begin{center}
\begin{table}[h!]
\begin{tabular}{ |p{3cm}||p{3cm}|p{3cm}|p{3cm}|  }
 \hline
 \multicolumn{4}{|c|}{QML Taxonomy} \\
 \hline
 \textbf{QML Division}  & \textbf{Inputs} & \textbf{Outputs} & \textbf{Process} \\
 \hline
 Classical ML   & Classical    & Classical &   Classical \\
 \hline
 Applied classical ML &   Quantum (Classical)  & Classical (Quantum)   & Classical\\
 \hline
 Quantum algorithms for classical problems &Classical & Classical&  Quantum\\
 \hline
 Quantum algorithms for quantum problems    &Quantum & Quantum&  Quantum\\
 \hline
\end{tabular}
\caption{Quantum and classical machine learning table. Quantum machine learning covers four quadrants (listed in `QML Division') which differ depending on whether the inputs, outputs or process is classical or quantum.}
\label{table:quantumclassical2}
\end{table}
\end{center}

\subsection{Statistical learning theory}
The formal theory of learning in computational science is classical statistical learning theory \cite{vapnik_nature_1995,james_introduction_2013,vapnik_overview_1999,hastie_elements_2013} which sets out theoretical conditions for function estimation from data. In the most general sense, one is interested in estimating $Y$ from as some function of $X$, or $Y = f(X) + \epsilon$ (where $\epsilon$ indicates random error independent of $X$). Statistical learning problems are typically framed as \textit{function estimation models} \cite{vapnik_nature_1995} involving procedures for finding optimal functions $f: \mathcal{X} \to \mathcal{Y}, x \mapsto f(x) = y$ (where $\X,\Y$ are typically vector spaces) where it is usually assumed that $X,Y \sim \Prb_{XY}$ i.e. random variables drawn from a joint distribution with realisations $x,y$ respectively. Formally, this model of learning involves (a) a generator of the random variable $X \sim \Prb_X$, (a) a supervisor function that assigns $Y_i$ for each $X_i$ such that $(X,Y) \sim \Prb_{XY}$ and (c) an algorithm for learning a set of functions (the hypothesis space) $f(X,\Theta)$ where $\theta \in \Theta$ (where $\theta \in \mathbb{K}^m$ parametrises such functions.

The two primary types of learning problems are supervised and unsupervised learning (defined below) where the aim is to learn a \textit{model family} $\{f\}$ or distribution $\Prb$. \textit{Supervised learning} (definition \ref{defn:ml:Supervised learning}) problems are framed in terms of a given sample $D_n = \{ X_i, Y_i\}$ comprising i.i.d. tuples $(X_i, Y_i) \in \mathcal{X} \times \mathcal{Y}$ with where $X,Y \sim \Prb_{XY}$. The supervised learning task consists of learning the mapping $f: \X \to \Y$ for both in-sample and out-of-sample $(X_i,Y_i)$ (it is denoted `supervised' owing to the joint distribution of features $X$ with labels $Y$). \textit{Unsupervised learning} by contrast is defined for $D = \{X_i\}, X \sim \Pr_X$ without output labels $Y_i$ for each input $X_i$. Rather the labels, such as in the form of classifications or clusters, are learnt (usually via some form of distance metric applied to $D$ and protocol optimising for statistical properties of clustering). The next two Chapters deal with supervised learning problems. 

Finding optimal functions intuitively means minimising the error $\epsilon = |f(x)-y|$ for a given function. This is described generically as \textit{risk minimisation} where, as we discuss below, there are various measures of risk (or uncertainty) which typically are expectations (averages) of loss. One assumes the existence of a well-defined (or chosen) \textit{loss function} $L:Y \times y \to R, (f(x),y) \mapsto L(f(x),y)$ and a joint probability distribution over $X \times Y$ denoted by $(X,Y) \sim P_{XY}$ and associated density $p_{XY}$. The \textit{statistical (true) risk} is then given by the expectation $R(f) = E[L(f(X),Y) | f]$. The \textit{minimum risk (Bayes risk)} is the infimum of $R(f)$ over the set $\mathcal{F}^*$ all such measureable functions (learning rules) $f$ denoted $R^* = \underset{f \in \mathcal{F}^*}{\inf} R(f)$ (note that $\mathcal{F}^*$ is usually larger than $\mathcal{F}$). The objective of supervised learning then becomes to use data generate functional estimators conditioned on the data samples $\hat{f}_n(x) = f(x; D_n)$ in order to minimise the estimate $E[R(\hat{f}_n)]$. The overall task is then one of \textit{generalisation}, which is intuitively the minimisation of risk across both in-sample and out-of-sample data.  

Formally, the task is then defined as one of \textit{empirical risk minimisation} where \textit{empirical risk} is $\hat{R}_n(f) = \frac{1}{n}\sum_i^n L(F(X_i),Y_i)$ (definition \ref{defn:ml:empiricalrisk}). It assumes the existence of a sample $D_n = \{X_i,Y_i\}^n$, loss function $L$ and family of functions (e.g. classifiers) $F$. The objective then becomes learning an algorithm (rule) that minimises empirical risk, thereby obtaining a best estimator across sampled and out-of-sample data, that is $\hat{f}_n = \arg \underset{f \in \mathcal{F}}{\min} \hat{R}_n(f)$. 
Usually $f$ is a \textit{parameterised} function $f = f(\theta)$ such that the requisite analytic structure (parametric smoothness) for learning protocols such as stochastic gradient descent is provided for by the parametrisation, typically where parameters $\theta \in \mathbb{R}_\theta$. The analyticity of the loss function $L$ means that $\hat{R}_n(f) = \frac{1}{n}\sum_i^n L(F(X_i(\theta)),Y_i)$ is smooth in $\theta$, which implies the existence of a gradient $\nabla_\theta \hat{R}_n(f(\theta))$. The general form of \textit{parameter update rule} is then a transition rule on $\mathbb{R}_\theta$ that maps at each iteration (epoch) $\theta_{i+1} = \theta_i - \gamma(n) \nabla_\theta \hat{R}_n(f(\theta))$ (see discussion of gradient descent below).

\subsection{Loss functions and model complexity}
A crucial choice in any machine learning architecture - and one we justify in detail in later Chapters - is that of the loss function. Two popular choices across statistics and also machine learning (both classical and quantum) are (a) \textit{mean-squared error} (MSE) and (b) root-mean squared error (RMSE). The MSE (equation (\ref{eqn:ml:MSE})) for a function $f$ parameterized by $\theta$ over a dataset $D_n$ is given by $ \text{MSE}(f_\theta) = \frac{1}{n}\sum_{i=1}^n \left(\hat f_\theta(X_i(\theta)) - Y_i\right)^2$. Other common loss functions include (i) cross-entropy loss e.g. Kullback-Leibler Divergence (see \cite{hastie_elements_2013} $\S14)$ for classification tasks and comparing distributions (see section \ref{sec:quant:quantummetrics} for quantum analogues), (ii) mean absolute error loss and (iii) hinge loss. The choice of loss functions has statistical implications regarding model performance and complexity, including bias-variance trade-offs.

As we discuss below, there is a trade-off between the size of $\mathcal{F}$ and empirical risk performance in and out of sample. We can minimise $\hat{R}_n(f)$ by specifying a larger class of estimation rules. At one extreme, setting $f(x) = Y_i$ when $x=X_i$ and zero otherwise (in effect, $F$ containing a trivial mapping of $X_i$) sends $\hat{R}_n(f) \to 0$, but performs poorly on out of sample data. At the other extreme, one could set $f(x)$ to capture all $Y_i$, akin to a scatter-gun approach, yet this would inflate $\hat{R}_n(f)$. The relation between prediction rule $\mathcal{F}$ size and complexity illustrates a tradeoff between approximation and estimation (known as `bias-variance' tradeoff for squared loss functions). The tradeoff is denoted \textit{excess risk}, defined as the difference between expected empirical risk and Bayes risk (equation (\ref{eqn:ml:excessrisk})):
\begin{align*}
    E[R_n(\hat{f}_n)] - R^* = \underbrace{E[\hat{R}_n(\hat{f}_n)] - \underset{f \in \mathcal{F}}{\inf} R(f)}_{\text{estimation error}} + \underbrace{\underset{f \in \mathcal{F}}{\inf} R(f) - R^*}_{\text{approximation error}} 
\end{align*}
Here estimation error reflects how well $\hat{f}$ compares with other candidates in $\mathcal{F}$, while approximation error indicates performance deterioration by restricting $\mathcal{F}$. To reduce empirical risk (and thus excess risk), two common strategies are (a) limiting the size of $\mathcal{F}$ and thus estimation error and (b) \textit{regularisation} techniques, constituting inclusion of a penalty metric that penalises increases in variance (overfitting) of models expressed via $\hat{f}_n = \arg\min_{f \in \mathcal{F}} \{ \hat{R}_n(f) + C(f)   \}$ (equation (\ref{eqn:ml:penaltycostregularisation})). While such strategies are commonplace and deployed in our machine learning architectures in later Chapters, we note the existence of no free lunch theorems which state that no algorithm (or choice of $\hat f$) can universally minimise statistical risk across all distributions, placing effective limits on learnability in terms of restrictions on generalisability (see section \ref{sec:ml:No-Free Lunch Theorems}). In section \ref{sec:ml:Statistical performance measures} we also set out a few key performance measures that are often used in classical and quantum machine learning literature. These measures, such as binary classification loss, accuracy, AUC/ROCR scores and F1-scores all seek to assess model performance. The AUC (Area Under the Curve) score represents the area under the ROC (Receiver Operating Characteristic) curve, the latter of which is a plot of the true positive rate (sensitivity) against the false positive rate (1-specificity) different thresholds. Intuitively, a higher AUC score represents a higher ratio over such thresholds between true positives and false positives, thus providing a measure of how well the model performs (see \cite{hastie_elements_2013}).

\subsection{Deep learning}
The machine learning architectures focused on in Chapters \ref{chapter:QDataSet and Quantum Greybox Learning} and \ref{chapter:Quantum Geometric Machine Learning} are deep learning neural network architectures. In this section we briefly note a few key features of deep learning, focusing on its technical relationship with generalised linear model theory and on architectural principles. Classical neural networks derive from \textit{generalised linear models}. The basic form of such model (definition \ref{defn:ml:Linear models}) starts with i.i.d. samples $X \in \R^m$ and $Y \in \R$ where samples $(X,Y) \sim \Prb_{XY}$ and task of estimating $Y$ according to:
\begin{align*}
        \hat Y = \hat \beta_0 + \sum_j^m X_j \hat \beta_j + \epsilon.
    \end{align*}
Here $\hat \beta_0 \in \R$  is the estimate of bias, $\beta \in \R^m$ is a vector of coefficients and $\epsilon$ uncorrelated (random) errors. The adaptation of generalised linear models to neural networks effectively arises from the introduction of non-linear functions of the linear components of the model (e.g. project pursuit regression as discussed below) \cite{hastie_elements_2013}. A simple example of regularisation in such models is given by ridge regression $\hat{\beta}_{\text{ridge}} = \arg\min_{\beta} \left\{ L(\mathbf{y}, \mathbf{X}\beta) + \lambda \|\beta\|_2^2 \right\}$ (equation  (\ref{eqn:ml:ridgeregression})). Here $\lambda \in \R$ is a penalty term which inflates the loss more if the parameters $\beta$ are too large in a process known as regularisation. Moreover, from such formalism we obtain \textit{ridge function} $f(X) = g\braket{X, a}$ for $X \in \R^m$, $g: \R \to \R$ being a univariate function (a function on the loss function in effect), $a \in \R^m$ with the inner product. Ridge functions essentially wrap the linear models of statistics in a non-linear kernel sometimes denoted project pursuit regression $f(X) = \sum_n^N g_n(\omega_m^T X)$ (equation (\ref{eqn:ml:projectpursuitregression})).

\subsection{Neural networks}
The ridge functions $n_m(\omega_n^T X)$ vary only in the directions defined by $\omega_n$ where the feature $V_m = \omega_m^T X$ can be regarded as the projection of $X$ onto the unit vector $\omega_m$. For sufficiently large parameter space, such functions can be regarded as universal approximators (for arbitrary functions) and form the basis of neural networks. The architecture of neural networks in quantum and classical machine learning involves a number of characteristics such as network topology, constraints (e.g. regularisation strategies or dropout), initial condition choices and transition functions. Neural network architectures are accordingly modelled using a Fiesler framework such that where they are formally defined as a nested 4-tuple $NN=(T,C,S(0),\Phi)$ where $T$ is network topology, $C$ is connectivity, $S(0)$ denotes initial conditions and $\Phi$ denotes transition functions (definition \ref{def:ml:neuralnetworkgeneral}). This framework has since influenced modern descriptions of neural networks and their architecture. 
Neural networks can abstractly be regarded as extensions of non-linearised linear models (as per project pursuit regression above) constituted via functional composition via layers. Each layer takes generally speaking data as an input (from previous layers or initial inputs) which become functions in linear models, which in turn are arguments in a non-linear function denoted an \textit{activation function} which represents the output of the layer. Formally we define an activation function such that for a vector $X \in \R^n$, weight $\omega \in \R^{m \times n}$ and bias term $\beta_0 \in \R^m$, we have an affine (linear) transformation $z = \omega X + \beta_0 \in \R^m$. The activation function (definition \ref{defn:ml:Activation function}) is then the function $\sigma:\R^m \to \R^m$ with $\sigma(z) = \sigma(\omega X + \beta_0) = (\sigma(z_1), \sigma(z_2),\ldots, \sigma(z_m))^T$. 

The function-compositional nature of neural networks is usefully elucidated by considering the basic \textit{feed-forward (multi-layer perceptron) neural network (definition \ref{defn:ml:Feed-forward neural network}}). Such a network comprises multiple layers $a^l_i$ of neurons such that each neuron in each layer (other than the first input layer) is a compositional function of neurons in the preceding layer. A fully-connected network is one where each layer's neurons are functions of each neuron in the previous layer. We represent each layer $l$ as (equation \ref{eqn:ml:feedforwardactivation-al}):
    \begin{align*}
    a_i^{(l)} = \sigma^{l}_i\left(\sum_{j=1}^{n_{l-1}} w_{ij}^{(l)} a_j^{(l-1)} + \beta_i^{(l)}\right). 
\end{align*}
The typical neural network schema (section \ref{sec:ml:neuralnetworkschema}) then involves (a) \textit{input layers} where (feature) data $X$ is input, (b) \textit{hidden layers} of the form above followed by (c) an \textit{output layer} $a^L$ that outputs according to the problem at hand (e.g. for classification or regression problems). Sometimes the final layer is considered in addition to the network itself. There is a constellation of sophisticated and complex neural network architectures emerging constantly within the computer science literature. 

\subsection{Optimisation methods}
The \textit{learning} component of machine learning is an optimisation procedure designed to reduce excess risk via minimising empirical risk. 
 In general, the hidden layers and output layers of the network referred to above comprise parameters (weights) $\theta$ which are dynamically updated over training runs. Most optimisation algorithms used across machine learning involve some sort of \textit{stochastic gradient descent} whereby parameters $\theta$ of estimated functions $\hat f(\theta)$ are updated according to a generalised directional derivative. The most common method for calculating the gradient used in this update rule is \textit{backpropagation}, which seeks to update parameters via propagating errors $\delta \sim |\hat f_\theta - Y|$ across the network. Backpropagation consists of two phases: (a) a \textit{forward pass}, this involves calculating, for each unit in the layer $a_i^{(l)}$, the layer's activation function $\sigma_i^{(l)}$ (see equation (\ref{eqn:ml:feedforwardactivation-al})); and (b) a \textit{backward pass} where the backpropagation updates are calculated. The most common method of updating parameters in the backward pass is stochastic \textit{gradient descent} (definition \ref{defn:ml:Gradient descent optimisation}) which is defined via the mappings:
    \begin{align*}
\omega_{ij}^{(l+1)} &= \omega_{ij}^{(l)} - \gamma_l \sum_{k=1}^{N} \frac{\partial \hat R_k}{\partial \omega_{ij}^{(l)}} = \omega_{ij}^{(l)} - \gamma_l \sum_{k=1}^{N} \nabla_{\omega_{ij}^{(l)}} \hat R_k.    
\end{align*}
Here $\omega_{ij}^{(l)}$ is the weight vector for neuron $i$ in layer $l$ that weights neuron $j$ in layer $l-1$, $a^{(l)}$ is layer $l$, $a_i^{(l)}$ is the $ith$ neuron in layer $l$ and $n_l$ is the number of neurons (units) in layer $l$. The error quantities used for the updating are given by the \textit{backpropagation formula} (equation (\ref{eqn:ml:backpropfinal})):
\begin{align*}
    \delta^{(l)}_i &= \sigma'^{l}_i(z_i^{(l)}) \sum_{\mu=1}^{n_{l+1}} \omega_{i\mu}^{(l+1)}  \delta_\mu^{(l+1)}.    
\end{align*}
Here $\delta^{(l)}_i$ represents the error term for layer $l$ and neuron $i$ which can be seen to be dependent upon the error term $\delta_\mu^{(l+1)}$ of the subsequent $l+1$th layer (see section \ref{sec:ml:Backpropagation} for a full derivation). The backpropagation equations thus allow computation of gradient descent:
\begin{align}
    \frac{\partial \hat R_i}{\partial \omega_{ij}^{(l)}} &= \nabla_{\omega_{ij}^{(l)}} \hat R_i = \sigma'^{l}_i(z_i^{(l)}) a^{(l-1)}_j \sum_{\mu=1}^{n_{l+1}} \omega_{i\mu}^{(l+1)}  \delta_\mu^{(l+1)}. 
\end{align}
In this way, errors are `back-propagated' through the network in a manner that updates weights $\theta$ in the direction of minimising loss (i.e. optimising), thus steering the model $f$ towards the objective. Note our discussion in section \ref{sec:ml:Backpropagation} on differences in the quantum case, primarily arising from quantum state properties and the effects of quantum measurement in collapsing quantum states (see section \ref{sec:quant:Measurement}). In general, there are a number of considerations regarding how gradient descent is calculated and how hyperparameters are tuned. Backpropagation equations (or variants thereof) are as a practical matter encoded in software such as TensorFlow \cite{abadi_tensorflow_2016} (used in this work) where one can either rely on the standard packages inherent in the program, or tailor a customised version. In our approach in the following Chapters, we adopted a range of such methods as detailed below. See sections \ref{sec:ml:Natural gradients} and \ref{sec:ml:Regularisation and Hyperparameter tuning} for more detail. 

\subsection{Machine learning and quantum computing}
The preceding sections on machine learning have focused on primarily classical machine learning principles. Quantum machine learning represents an extension of such principles where either the relevant data is quantum data or the information processing regimen is quantum in nature. Quantum machine learning algorithms include a diverse array of architectures, including quantum neural networks, parametrised (variational) quantum circuits, quantum support vector machines and other techniques \cite{cerezo_challenges_2022,dunjko_machine_2018}. Deep learning and neural network analogues have also been a feature of quantum machine learning literature for some time, with a variety of designs for quantum neural networks within the literature (see \cite{schuld_machine_2021,beer_quantum_2022} for an overview). In more recent years,  quantum deep learning architectures (including quantum analogues of graph neural networks, convolutional neural networks and others \cite{verdon_learning_2019,verdon_universal_2018,verdon_quantum_2019-2,verdon_quantum_2017}) have continued to emerge and shape the discipline. As we discuss below, the quantum nature of data or information processing inherent in these approaches gives rise to a number of differences and challenges distinct from its classical counterpart. We itemise a few of these key differences below.
\begin{enumerate}
    \item \textit{Dissipative versus unitary}. As noted in the literature \cite{schuld_machine_2021} and discussed in section \ref{sec:ml:Neural networks and quantum systems}, classical neural network architectures are fundamentally based upon non-linear and dissipative dynamics \cite{hopfield_neural_1982}, contrary to the linear unitary dynamics of quantum computing. Thus leveraging the functional approximation benefits of neural networks while respecting quantum information constraints requires specific design choices (see \cite{beer_quantum_2022,killoran_continuous-variable_2019,franken_explorations_2020} for examples). One of the motivations for the greybox architecture that characterises our approach in Chapters \ref{chapter:QDataSet and Quantum Greybox Learning} and \ref{chapter:Quantum Geometric Machine Learning} is to design machine learning systems that explicitly overcome this challenge by embedding unitarity constraints within the network itself.
    \item \textit{Quantum statistical learning}. As with its classical counterpart, quantum machine learning has its own analogue of statistical learning theory, sometimes denoted quantum (statistical) learning theory which explores bounds and results on expressibility, complexity and boundedness of quantum machine learning architectures. Quantum machine learning also faces challenges with regard to entanglement (see section \ref{sec:quant:Quantum entanglement}) growth as systems scale affecting the performance of algorithms.
    \item \textit{Quantum measurement}. The unique nature of quantum measurement (section \ref{sec:quant:Measurement}) gives rise to challenges from the fact that measurement causes quantum states $\rho$ to collapse onto eigenstates of the measurement operator $M$ via decohering quantum-classical measurement channels (definition \ref{defn:quant:Quantum-classical channel}). This causes a loss of quantum information in contrast to the classical case. 
    \item \textit{Barren plateaus}. Barren plateaus \cite{mcclean_barren_2018} bear some similarity to the classical vanishing gradient problem (albeit with specific differences as noted in \cite{larocca_thanasilp_cerezo_2024}) where gradient expectation decreases exponentially with the number of qubits (see section \ref{sec:ml:Barren Plateaus}), interfering with the ability of the learning protocol as a result. Proposals exist in the literature (such as weight initialisation and quantum circuit design) to address or ameliorate their effects, but barren plateaus remain another quantum-specific phenomenon that must be addressed in quantum circuit design.
    \item \textit{Data encoding}. Data encoding strategies (section \ref{sec:ml:Encoding data in quantum systems}) also differ in the quantum case, with data being encoded usually via either state representations such as via binary encoding 0 for $\ket{0}$ and 1 for $\ket{1}$ or, for continuous data, phase encoding e.g. via relative phases $\exp(i\eta)$ where $\eta \in (-\pi,\pi)$. Encoding strategies thus differ from their classical counterparts. They are important for quantum and classical data processing as they enable leveraging the potential advantages that motivate the use of quantum algorithms in the first place. 
\end{enumerate}

\subsection{Parametrised variational quantum circuits}
In Chapters \ref{chapter:QDataSet and Quantum Greybox Learning} and \ref{chapter:Quantum Geometric Machine Learning} we adapt parametrised variational quantum circuits \cite{benedetti_parameterized_2019,cerezo_variational_2022} for solving specific problems in quantum simulation and quantum control. Variational quantum circuits (see section \ref{sec:ml:Variational quantum circuits}) can be defined as a unitary operator $U(\theta(t)) \in \mathcal{B}(\Hilb)$ parametrised by the set of parameters $\theta \in \R^m$. A parametrised quantum circuit is a sequence of unitary operations $U(\theta,t) = \mathcal{T}_+ \exp\left(-i\int_0^T H(\theta,t')dt'\right)$ (equation (\ref{eqn:ml:parametrised circuit-timedependent})) and for the time-independent approximation \\$U(\theta,t)\big|_{t=T} = U_{T-1}(\theta_{T-1})(\Delta t)...U_1(\Delta t)(\theta) $. The optimisation problem is then one of minimising the cost functional cost functional on the parameter space given by $C(\theta): \R^\m \to \R$ by learning $\theta_* = \arg\min_\theta C(\theta)$. We let $f_\theta:\X \to \Y$ and denote $U(X,\theta)$ (a quantum circuit) for initial state $X \in \X$ and parameters $\theta \in \R^m$. Let $\{M\}$ represent the set of (Hermitian) observable (measurement) operators $f_\theta(X) = \trace(M \rho(X,\theta))$ (equation (\ref{eqn:ml:qml-VQC-ftheta})) for $\rho(X,\theta)=U(X,\theta)\rho_0 U(X,\theta)^\dagger$. Such parametrised circuits are denoted \textit{variational} because variational techniques are used to solve the minimisation problem of finding $\theta_*$. For our purposes in Chapters \ref{chapter:QDataSet and Quantum Greybox Learning} and \ref{chapter:Quantum Geometric Machine Learning} we parametrised unitaries by control functions $u_j(t)$. It is these control functions that are actually parametrised such that $u(t) = u(\theta(t))$ where a range of deep learning neural networks are used to learn optimal $u(t)$ which is then applied to a Hamiltonian to generate $U=U(\theta(t))$.

The optimisation procedure adopted is based on the fidelity function as central to the loss function (cost functional) according to which the classical register $\theta$ is updated. The loss function based on the fidelity metric (definition \ref{defn:quant:Fidelity}) adopted in equation (\ref{eqn:ml:batchfidelityMSE}) in Chapter \ref{chapter:Quantum Geometric Machine Learning} (\textit{batch fidelity}) takes the mean squared error (MSE, see equation (\ref{eqn:ml:MSE}) above) of the loss (difference) between fidelity of the estimated unitary and target as the measure of empirical risk (section \ref{sec:ml:Statistical Learning Theory}) using the notation for cost functionals $C$:
\begin{align*}
    C(F,1) = \frac{1}{n}\sum_{j=1}^n (1 - F(\hat{U}_j,U_j))^2 \label{eqn:ml:batchfidelityMSE}
\end{align*}
In doing so, we assume the existence of measurement protocols that adequately specify $\hat U_j$ and $U_j$. In our case, parameters $\theta$ are equivalent to a classical register $\Sigma$ which is updated according to classical gradient descent (section \ref{sec:ml:Optimisation methods}). In section \ref{sec:ml:QML Optimisation} we discuss a number of hybrid quantum-classical and primarily quantum mechanical means of effectively calculating gradient descent in a QML context, including parameter shift rules, quantum natural gradients, quantum backpropagation and back-action based methods.

In section \ref{sec:ml:Symmetry-based QML} we briefly cover a range of these areas, both classical and quantum as follows. These include (i) the use of machine learning in optimal (geometric) control theory, essentially as an optimisation method for PMP-based protocols, (ii) \textit{geometric information theory} where infamous relationships between Fisher-information metrics and Riemannian metrics have seen the application of differential geometric techniques for optimisation of information-theoretic (or register-based) problems, (iii) \textit{Lie group machine learning} which was a relatively early application of some symmetry techniques in Lie theory (familiar to control and geometric theory) to problems in machine learning (such as learning symmetries in datasets reflective of group symmetries) and (iv)\textit{geometric quantum machine learning} which leverages Lie theory and particularly dynamical Lie algebras in the construction and design of compositional neural networks and to address issues around barren plateaus. Of these, geometric quantum machine learning (GQML), a field which has emerged relatively recently, is the area within which this work is most well-situated. See section \ref{sec:ml:Geometric QML} for more detail.

\subsection{Greybox machine learning}
Finally, we briefly outline the key hybrid classical-quantum greybox machine learning (see section \ref{sec:ml:Greybox machine learning}) which forms the basis of our design choices in Chapters \ref{chapter:QDataSet and Quantum Greybox Learning} and \ref{chapter:Quantum Geometric Machine Learning}.  By encoding priori information within a classical neural network stack, we obviate the need for the network to learn such rules (such as requirements regarding unitarity or Hamiltonian construction) and thereby affording guarantees (such as unitarity of outputs) in a way that blackbox learning only at best approximates. Variational quantum circuits and other approaches adopt implicitly similar techniques when, for example, seeking to generate unitaries using Pauli generators from $\su(2)$. We focus on synthesising the use of Lie algebras with time-optimal methods in geometric control as encoded within a hybrid quantum-classical variational (parametrised) circuit network. 

In our case, especially in Chapter \ref{chapter:Quantum Geometric Machine Learning}, our targets are $U_T \in G \simeq \M$ (for Lie group $G$) generated (via Hamiltonian flow) by Hamiltonians $H$ from the corresponding Lie algebra $\g$. We focus on an architecture that learns optimal control from training data comprising geodesic unitary sequences in order to generate geodesics (or approximate them) on $G$. As discussed above, learning optimal control functions $u_j(t)$ (for directions $j$) where $j = 1,...,\dim \g$ is a form of optimisation in a machine learning context and  $u_j(\theta(t))$  parametrise the system consistent and are learnt from training data that satisfies PMP constraints. Control functions are adjusted to minimise our loss functions (explicated below in terms of fidelity $F(\hat U_T,U_T)$). Geometrically, this is represented as transformations in one or more of the directions of elements of $H_p\M$ (our distribution $\Delta$). A sketch of the architecture is below:
\begin{enumerate}[(i)]
    \item \textit{Objective}. In both Chapters \ref{chapter:QDataSet and Quantum Greybox Learning} and \ref{chapter:Quantum Geometric Machine Learning}, our aim is to provide a sequence of controls for synthesising $U_T \in G$. In Chapter \ref{chapter:QDataSet and Quantum Greybox Learning}, this takes the form of a hybrid quantum-classical circuit which learns control pulses that steer the Hamiltonians comprising Lie algebra generators from $\su(2)$ towards generating candidate unitaries $\hat U_T$ which minimise fidelity error with our labelled data $U_t$. In Chapter \ref{chapter:Quantum Geometric Machine Learning}, we treat the sequence of unitaries (denoting the sequence $(U_j)$) as training data that is generated according to PMP and geometric principles in order to be geodesic. We embed a control function layer which is parameterised by $\theta$ as per above, details of which can be extracted once fidelity of the candidate geodesic sequence $(\hat U_j)$ reaches an acceptable threshold.
    \item \textit{Input layers}. Input layers $a^{(0)}$ thus vary but essentially in the case of Chapter \ref{chapter:Quantum Geometric Machine Learning} comprise target unitaries $U_T$ which are then fed into subsequent layers $a^{(l)}$.
    \item \textit{Feed-forward layers}. The feed-forward layers (definition \ref{defn:ml:Feed-forward neural network}) then comprise typical linear neurons with an activation function $\sigma$ given by $a^{(1)}=\sigma(W^T a^{(0)} + b)$.
    \item \textit{Control pulse layers}. The feed-forward layers then feed into bounded control pulse layers, there being one parametrised control function for each generator in the Hamiltonian. 
    \item \textit{Hamiltonian and unitary layers}. The control functions are then combined into Hamiltonian layers which are then fed into a layer comprising unitary activation functions. In Chapter \ref{chapter:Quantum Geometric Machine Learning} this enables generation of the candidate sequence $(\hat U_j)$.
    \item \textit{Optimisation strategies}. We utilise batch fidelity via an empirical risk measure that is the MSE of 1 minus fidelity of $U_j$ and $\hat U_j$ or otherwise $U_T, \hat U_T$ (see equation \ref{eqn:qgml:fidelity}). We also experimented with a variety of other hyperparameters (see section \ref{sec:ml:Regularisation and Hyperparameter tuning}) including the use of dropout \cite{srivastava_dropout_2014} which effectively prunes neurons in order to deal with overfitting.
\end{enumerate}
Pseudo-code describing algorithms is set out in the relevant chapter. As noted in the subsequent Chapters, code for the relevant models is available on repositories.
%=========================
%=========================
%=========================
%=====QDataSet and Quantum Greybox Learning
%=========================
%=========================
%=========================

\chapter{QDataSet and Quantum Greybox Learning}
\label{chapter:QDataSet and Quantum Greybox Learning}
\section{Abstract}
The availability of large-scale datasets on which to train, benchmark and test algorithms has been central to the rapid development of machine learning as a discipline. Despite considerable advancements, the field of quantum machine learning has thus far lacked a set of comprehensive large-scale datasets upon which to benchmark the development of algorithms for use in applied and theoretical quantum settings. In this Chapter, we introduce such a dataset, the QDataSet, a quantum dataset designed specifically to facilitate the training and development of quantum machine learning algorithms. The QDataSet comprises 52 high-quality publicly available datasets derived from simulations of one- and two-qubit systems evolving in the presence and/or absence of noise. The datasets are structured to provide a wealth of information to enable machine learning practitioners to use the QDataSet to solve problems in applied quantum computation, such as quantum control, quantum spectroscopy and tomography. Accompanying the datasets on the associated GitHub repository are a set of workbooks demonstrating the use of the QDataSet in a range of optimisation contexts.

The QDataSet is constructed to mimic conditions in laboratories and experiments where inputs and outputs to quantum systems are classical, such as via classically characterised controls (pulses, voltages) and measurement outcomes in the form of a classical probability distribution over observable outcomes of measurement (see section \ref{sec:quant:Measurement}). Actual quantum states, coherences and other characteristically quantum features of the system, while considered ontologically extant, are, epistemologically speaking,  reconstructions conditioned upon classical input and output data. In a machine learning context, this means that the encoding of quantum states, quantum processes (such as unitary evolution) represents the encoding of constraints upon how computation may evolve. To this end, we follow the data generation protocols set out in \cite{youssry_characterization_2020} which we explicate below. 

\section{Introduction}
Quantum machine learning (QML) is an emergent multi-disciplinary field combining techniques from quantum information processing, machine learning and optimisation to solve problems relevant to quantum computation \cite{verdon_universal_2018,amin_quantum_2018,biamonte_quantum_2017,schuld_evaluating_2019}. The last decade in particular has seen an acceleration and diversification of QML across a rich variety of domains. As a discipline at the interface of classical and quantum computing, subdisciplines of QML can usefully be characterised by where they lie on the classical-quantum spectrum of computation \cite{aimeur_machine_2006}, ranging from quantum-native (using only quantum information processing) and classical (using only classical information processing) to hybrid quantum-classical (a combination of both quantum and classical). At the conceptual core of QML is the nature of how quantum or hybrid classical-quantum systems can \textit{learn} in order to solve or improve results in constrained optimisation problems. The type of machine learning of relevance to QML algorithms very much depends on the specific architectures adopted. This is particularly the case for the use of QML to solve important problems in quantum control, quantum tomography and quantum noise mitigation. Thus QML combines concepts and techniques from quantum computation and classical machine learning, while also exploring novel quantum learning architectures.

 While quantum-native QML is a burgeoning and important field, the more commonplace synthesis of machine learning concepts with quantum systems arises in classical-quantum hybrid architectures \cite{schuld_evaluating_2018,verdon_quantum_2019-1,zhou_machine_2017,vidick_quantum_2016,schuld_machine_2021}. Such architectures are typically characterised by a classical parametrisation of quantum systems or degrees of freedom (measurement distributions or expectation values) whose classical parameters are updated according to classical optimisation routine (such as variational quantum circuits discussed in section \ref{sec:ml:Variational quantum circuits}). In applied laboratory and experimental settings, hybrid quantum-classical architectures remain the norm primarily due the fact that most quantum systems rely upon classical controls \cite{dong_quantum_2010,viola_dynamical_1999}. To this end, hybrid classical-quantum QML architectures which are able to optimise classical controls or inputs for quantum systems have wider, more near-term applicability for both experiments and NISQ \cite{preskill_quantum_2018, bharti_noisy_2021} devices. Recent literature on hybrid classical-quantum algorithms for quantum control \cite{youssry_characterization_2020,perrier_quantum_2020}, noisy control \cite{youssry_modeling_2020} and noise characterisation \cite{youssry_characterization_2020} present examples of this approach. Other recent approaches include the hybrid use of quantum algorithms and classical objective functions for natural language processing \cite{lorenz_qnlp_2021}. Thus the search for optimising classical-quantum QML architectures is well-motivated from a theoretical and applied perspective.
 
 Despite the increasing maturity of hybrid classical-quantum QML as a discipline, the field lacks many of the characteristics that have been core to the extraordinary successes of classical machine learning in general and deep learning in particular. Classical machine learning has been driven to a considerable extent by the availability of large-scale, high-quality accessible datasets against which algorithms can be developed and tested for accuracy, reliability and scalability. The availability of such datasets as MNIST \cite{lecun_mnist_1998}, ImageNet \cite{russakovsky_imagenet_2015}, Netflix \cite{zhou_large-scale_2008} and other large scale corpora has acted as a catalyst for not just innovations within the machine learning community, but also for the development of benchmarks and protocols that have helped guide the field. Such datasets have also fostered important cross-collaborations among disciplines in ways that have advanced classical machine learning. By contrast, QML as a discipline lacks a similarly standardised set of canonical large-scale datasets against which machine learning researchers (along the quantum-classical spectrum) may benchmark their algorithms and upon which to base innovation. Moreover, the absence of such large-scale standardised datasets arguably holds back important opportunities for cross-collaboration among quantum physicists, computer science and other fields.
 
 In this Chapter, we seek to address this gap in QML research by presenting a comprehensive dataset for use in QML tailored to solving problems in quantum control, quantum tomography and noise mitigation. The QDataSet is a dedicated resource designed for researchers across classical and quantum computation to develop and train hybrid classical-quantum algorithms for use in theoretical and applied settings relating to these subfields of QML related to control, tomography and noise characterisation of quantum systems. We name this dataset \textit{QDataSet} and our project the \textit{QML Dataset Project}. The motivation behind the QML Dataset Project is to map out a similar data architecture for the training and development of QML as exists for classical machine learning. The focus of the QDataSet on quantum control, tomography and noise problems means that it is of most relevance to these areas as distinct from other areas of QML, such as the development of quantum-only machine learning architectures per se. The contributions of our paper are as follows:
 \begin{enumerate}
     \item Presentation of QDataSet for quantum machine learning, comprising multiple rich large-scale datasets for use in training classical machine learning algorithms for a variety of quantum information processing tasks including quantum control, quantum tomography, quantum noise spectroscopy and quantum characterisation;
     \item Presentation of desiderata of QML datasets in order to facilitate their use by theoretical and, in particular, applied researchers; and
     \item Demonstration of using the QDataSet for benchmarking classical and hybrid classical-quantum algorithms for quantum control.
 \end{enumerate}
We set out below principles of large-scale datasets that are desirable in a quantum machine learning context and which were adopted in the preparation of the QDataSet. More detail on QML is set out in Appendix \ref{chapter:Background: Classical, Quantum and Geometric Machine Learning}.

\section{Overview of QML}
Quantum machine learning, a cross-disciplinary field that explores techniques and synergies between quantum information processing and machine learning, has emerged as a cutting-edge research field at the frontiers of quantum information and computational science. In just a few short years, QML has expanded into a diverse multitude of sub-disciplines, covering such topics as quantum algorithm design, algorithmic optimisation, error correction, quantum control, tomography and topological state classification \cite{biamonte_quantum_2017,schuld_introduction_2015,caro_generalization_2022,ciliberto_quantum_2018,dunjko_machine_2018,schuld_machine_2021,bilkis_semi-agnostic_2021,dunjko_quantum-enhanced_2016}. More broadly, QML intersects classical and quantum sub-disciplines both in its domain, namely its application and utilisation of classical and quantum data, and functionality/methodology, in its utilisation of either classical or quantum (or a hybrid combination of both) information processing methods. To this end, QML can be usefully segmented according to a classical-quantum typology \cite{aimeur_machine_2006,schuld_introduction_2015} depending on whether quantum or classical data or information processing is utilised:
\begin{enumerate}[(i)]
    \item \textit{Classical-classical} (CC): classical data is processed via classical information processing methods, typically this characterises the vast majority of machine learning disciplines;
    \item \textit{Quantum-Classical} (QC): quantum data is processed using classical information processing;
    \item \textit{Classical-Quantum} (CQ): classical data is processed using quantum information processing techniques, often characterised as classical information via quantum channels \cite{watrous_theory_2018,nielsen_quantum_2011}; and
    \item \textit{Quantum-Quantum} (QQ): quantum data is processed using quantum information processing, covering typical prospective use-cases of quantum computing devices.
\end{enumerate}
The typography above is useful for classifying the various optimisation strategies adopted across QML and quantum computing more widely. We expand on it in a bit more detail below in order to situate the QDataSet within the literature. Optimisation strategies across QML and quantum information processing vary considerably depending upon use-cases, research programmes and objectives. Understanding the differences between both classical and quantum data and between classical and quantum information processing, is integral to any research programme involving QML. In particular, while classical techniques and datasets often taken as a standard reference against which the various strands of QML are benchmarked, the distinct natures of classical and quantum data and information processing mean that in many cases direct analogies between quantum and classical information strategies are unavailable. Indeed the non-equivalence of quantum and classical information processing is precisely the underlying motivation behind the pursuit of quantum computation itself. It should be noted, however, that while `data' versus `processing' distinction is a useful heuristic, in reality the nature of quantum and classical data is inseparable from the information processing associated with quantum and classical physics. Furthermore, and more generally, classical information itself is considered an emergent limit of underlying quantum information processing \cite{joos_decoherence_2013}.

%==================

\section{QML Objectives}
\subsection{Overview}
Cross-disciplinary programmes focused on building quantum datasets for machine learning will benefit from a framework to categorise and classify the particular objectives of QML architectures and articulation of number of design principles relevant to the taxonomy of QML datasets. 
Designing large-scale datasets for QML requires an understanding of the objectives for which QML research is undertaken and the extent to which those objectives involve classical and/or quantum information processing. 
 Following \cite{aimeur_machine_2006}, the application of machine learning techniques to quantum information processing can be usefully parsed into a simple input / output and process taxonomy on the basis of whether information and computational processes are classical or quantum in nature. Here a process, input or output being `quantum in nature' refers to the phenomenon by which the input or output data was generated, or by which the computational process occurs, is itself quantum in nature given that measurement outcomes are represented as classical datasets from which the existence of quantum states or processes is inferred. Quantum data encoded in logical qubits, for example in quantum states (superpositions or entangled), is different from classical data, in practical terms information about such quantum data arises by inference on measurement statistics whose outcomes are classical (see section \ref{sec:ml:Encoding data in quantum systems} for a discussion of data encoding). This taxonomy can be usefully partitioned into four quadrants depending on the objectives of the QML task (to solve classical or quantum problems) and the techniques adopted (classical or quantum computational methods). Table (\ref{table:quantumclassical}) lists the various classical and quantum inputs according to this taxonomy.

 \begin{enumerate}
     \item  \textit{Classical machine learning for classical data}. The first quadrant covers the application of classical computational (machine learning) methods to solve classical problems, that is, problems not involving data or processes of a quantum character. 
\item \textit{Classical machine learning for quantum data}. 
The second quadrant covers the application of classical computational and machine learning techniques to solving problems of a quantum character. Specifically, this subdivision of QML covers using  standard machine learning techniques to solving problems specific to the theoretical or applied aspects of quantum computing, including optimal circuit synthesis \cite{youssry_characterization_2020, perrier_quantum_2020, riviello_searching_2015}, design of circuit architectures and so on. Either input or output data are quantum in nature, while the computational process by which optimisation, for example, occurs is itself classical.  \item \textit{Quantum algorithms for classical optimisation}.
The third quadrant covers the application of quantum algorithmic techniques to solving classical problems. In this subdivision, algorithms are designed leveraging the unique characteristics of quantum computation, in ways that assist in optimising classical problems or solve certain classes of problems which may be intractable on a classical computer. Quantum algorithms are designed with machine learning characteristics, potentially utilising certain computational resources or processes unavailable when constrained to classical computation. Examples of such algorithms include variational quantum eigensolvers \cite{peruzzo_variational_2014, yang_optimizing_2017, wecker_progress_2015, mcclean_theory_2016}, quantum analogues of classical machine learning techniques (e.g. quantum PAC learning \cite{lloyd_quantum_2014}) and hybrid quantum analogues of deep learning architectures (see \cite{verdon_universal_2018, vidick_quantum_2016, schuld_supervised_2018} for background).  
\item \textit{Quantum algorithms for quantum information processing}.
The fourth quadrant covers the application of quantum algorithms to solve quantum problems, that is, problems whose input or output data is itself quantum in nature. This division covers the extensive field of quantum algorithm design, including the famous Grover and Shor algorithms \cite{grover_quantum_1997, brassard_exact_1997, shor_polynomial-time_1997}.  
\end{enumerate}
The QDataSet fits within the second subdivision of QML, its primary use being envisaged as assisting in the development of classical algorithms for optimisation problems of engineered quantum systems. Our focus on classical techniques applied to quantum data is deliberate: while advancements in quantum algorithms are both exciting and promising, the unavailability of a scalable fault-tolerant quantum computing system and limitations in hybrid NISQ devices mean that for the vast majority of experimental and laboratory use cases, the application of machine learning is confined to the classical case. Secondly, as a major motivation of this work is to provide an accessible basis for classical machine learning practitioners to enter the QML field, it makes sense to focus primarily on applying techniques from the classical domain to quantum data.

%=====characteristics of datasets

\subsection{Large-Scale Data and Machine Learning}
Classical machine learning has become one of the most rapidly advancing scientific disciplines globally with immense impact across applied and theoretical domains. The advancement and diversification of machine learning over the last two decades has been facilitated by the availability of large-scale datasets for use in the research and applied sciences. Large-scale datasets \cite{lecun_mnist_1998, deng_imagenet_2009, goodfellow_deep_2016} have emerged in tandem with increasing computational power that has seen the velocity, volume and veracity of data increase \cite{kitchin_what_2016, hastie_elements_2013}. Such datasets have both been a catalyst for machine learning advancements and a consequence or outcome of increasing scope and intensity of data generation. The availability of large-scale datasets led to the evolution of data mining, applied engineering and even theoretical results in high energy physics \cite{albertsson_machine_2018}. 

An important lesson for QML is that developments within these fields have been facilitated using such datasets in a number of ways. Large-scale datasets improve the trainability of machine learning algorithms by enabling finer-grained optimisations via commonplace methods such a backpropagation (discussed in Appendix \ref{chapter:Background: Classical, Quantum and Geometric Machine Learning}). This has particularly been true within the field of deep learning and neural networks \cite{goodfellow_deep_2016}, where large-scale datasets have enabled the development of deeper and richer algorithmic architectures able to model complex non-linearities and functional forms, in turn leading to drastic improvements and breakthroughs across domains such as image classification, natural language processing \cite{devlin_bert_2019, brown_language_2020} and time series analysis. With an abundance of data on which to train algorithms, new techniques such as regularisation and dimensionality reduction to address problems arising from large-scale datasets, including overfitting and complexity considerations, have emerged, in turn spurring novel innovations that have contributed to the advancement of the field. Large-scale datasets have also served a standardising function by providing a common basis upon which algorithmic performance may be benchmarked and standardised. By providing standard benchmarks, large-scale datasets have enabled researchers to focus on important features of algorithmic architecture in the design of improvements to training regimes. Such datasets have also enabled the fostering of the field via competitive platforms such as Kaggle, where researchers compete to improve upon state of the art results.

\subsection{Taxonomy of large-scale datasets}
Large-scale classical machine learning datasets share common structural and architectural characteristics designed to facilitate the objectives for which the datasets were compiled. There are a range of considerations when generating these types of datasets, including the specific objectives, the types of data to be stored, the degree of structuring of data (including whether highly structured or unstructured), the dimensionality of datasets, the extent of preprocessing of datasets required, data quality issues (such as missing, uncertain or incorrect data - an issue for example in quantum information processing contexts given sources of error and uncertainty), data imputation for missing datasets and whether data is visible or hidden data (e.g. whether data is direct or a feature constructed from other data), the number of data points, format, default tasks of the datasets, data temporality (how contemporaneous data is), control of datasets and access to data. Datasets are also structured depending on the machine learning algorithms for which they were developed, taking into account the types of objectives, loss functions, optimisers, development environment and programming languages of interest to researchers. Table (\ref{table:characteristics}) sets out a range of issues and desiderata in this regard.

%======large scale dataset characteristics

\begin{center}
\begin{table}
\begin{tabular}{ |p{3cm}||p{12cm}|  }
 \hline
 \multicolumn{2}{|c|}{Data Set Characteristics} \\
 \hline
 \textbf{Item}  & \textbf{Description}  \\
 \hline
 Objectives  & Specification of the objectives for which the dataset was both created and to be used \\
 \hline
 Description  & Sufficient description of data, representation and theoretical description\\
 \hline
  Training/test  & Identification of training (in-sample) and test (out-of-sample) subsets of data\\
 \hline
 Data Types  & Specification of the types of data and formats to be used   \\
 \hline
 Structuring  & Degree to which data is structured or unstructured    \\
  \hline
 Dimensionality  & The dimension of the datasets, dimensional reduction or kernel methods needed   \\
  \hline
 Preprocessing  & Extent to which preprocessing is required in dataset, covering transformations of data such as  sparsification or  decomposition  \\
  \hline
 Data quality, consistency and completeness & Extent to which data is missing, uncertain or incorrect or noisy along with any necessary imputation methods  \\
  \hline
 Visible v. hidden  & Extent to which data points are `direct' or `indirect' (inferred) and whether imputations necessary   \\
 \hline
\end{tabular}
 \caption{Taxonomy of large-scale datasets which can guide the generation of QML datasets and development of QML dataset taxonomies.}
 \label{table:characteristics}
\end{table}
\end{center}
Large-scale dataset characteristics affect the utility of the datasets in applied contexts. Such characteristics are relevant to the design of quantum datasets. Below we set out a number of principles used in the design of the QDataSet which we believe provide a useful taxonomy for the QML community to consider when generating data for use in machine learning-based problems. The aim of the proposed taxonomy for quantum datasets is to facilitate their interoperability across machine learning platforms (classical and quantum) and for use in optimisation for experimentalists and engineered quantum systems. While taxonomies and specific architectures will differ across domains, our proposed QDataSet taxonomy we believe will assist the QML and classical ML community to guide large-scale data generation towards principles of interoperability summarised in Table (\ref{table:characteristics}) and explained below:
\begin{enumerate}
    \item \textit{Objectives}. Quantum datasets, as with classical datasets, benefit from being constructed with particular objectives in mind. Most major classical large-scale datasets are compiled for specific objectives such as, for example, classification or regression tasks. In a quantum setting, such objectives include quantum algorithm design, circuit synthesis, quantum control, tomography or measurement-based objectives (such as sampling). The QDataSet's objectives are to provide training data for use in the development of machine algorithms for controlled experimental and engineered quantum systems. This objective has informed the feature selection and structural design, such as inclusion of measurement statistics, Hamiltonian and unitary sequences and the various types of noise and distortion. 
    \item \textit{Description}.  Sufficiently describing datasets, efficiently representing the data and providing theoretical context for how and why the datasets are so represented, enhances their utility. Representation of data (its form, structure, data types and so on) affects their ease of use and uptake. For machine learning, optimal data representation is an important aspect of feature learning, representation learning \cite{hamilton_representation_2017}. In this work, we go to some lengths to describe the various structural aspects of the QDataSet in order to facilitate its uptake by researchers in designing algorithms. We especially have set-out background information for machine learning practitioners who may be unfamiliar with quantum data in an effort to reduce barriers facing cross-disciplinary collaboration. 
    \item \textit{Training and test sets}. Applied datasets for machine learning require training, validation and test sets in order to adequately train algorithms for objectives, such as quantum control. The design of quantum datasets in general, and the QDataSet in particular, has been informed by desirable properties of training sets. These include, for example: (i) \textit{interoperability}, ensuring training set data can be adequately formatted for use in various programming languages (for example storing QDataSet data via matrices, vectors and tensors in Python); (ii) \textit{generalisability}, preprocessing of datasets to improve generalisability of algorithmic results, especially to test or out-of-sample data \cite{hastie_elements_2013} (in the QDataSet, we do this via providing a variety of noise-affected datasets) and minimise measures such as empirical risk (equation (\ref{defn:ml:empiricalrisk})) (see in general Appendix \ref{chapter:Background: Classical, Quantum and Geometric Machine Learning}); (iii) \textit{feature smoothing} trained algorithms can often focus on information-rich yet small subspaces of data which, while informative for in-sample prediction, can lead to decreased generalisability across the majority of in- and out-of-sample data lacking such features. Feature smoothing is a technique to coarse-grain features so that less weight is put on rarer though information rich features in order to improve generalisation. In a quantum context, this may involve an iterative process of trial and error that trains datasets and seeks to identify relevant features; alternatively, it may involve using techniques from quantum information theory to classify regions of high and low information content e.g. via entropy measures. 
    \item \textit{Data precision and type}. Data precision and data typing is an important consideration for quantum datasets, primarily to facilitate ease of interoperability between software and applied/hardware in the quantum space. Others considerations can include the degree of precision with which data should be presented. Ideally, quantum data for use in developing algorithms for application in experimental quantum control or measurement scenarios should allow flexibility of data precision to match instruments used in laboratories on a case-by-case basis. For example, the QDataSet choices regarding noise degrees of freedom (such as amplitude, mean and standard deviation) have been informed by collaborations with experimental groups. 
    \item \textit{Structuring}. Data structuring, the degree to which data is structured according to taxonomies, is an important characteristic of classical datasets that affects their use and functionality. For quantum datasets, structuring encompasses the types of information that would be included and how that information is categorised. In the selection of real-world applicable datasets, researchers will have a range of choices of salient information to includes, including: theoretical details of the candidate Hamiltonians, details of the physical laboratory setting such as controls, exogenous parameters such as temperature (a significant environmental variable affecting quantum systems), noise or other disturbances; the characteristics of measurement devices and so on. Spectroscopic information, including details of the spectroscopy used, may also be included. What to include and not include will depend upon the particular uses cases and generality (or specificity) of the datasets. In each case, it makes sense for quantum datasets to contain as much useful information as possible such as about parameters, say exogenous environment parameters, or distortion information which may affect measurement devices. Doing so enables algorithms trained on quantum datasets to improve their performance and generalise better. Examples of such information in the QDataSet include details we have included regarding noise profiles and distortion simulations.
    \item \textit{Dimensionality}. The dimensionality of datasets is an important consideration. Large-dimensional datasets often require adaptation in order to facilitate algorithmic learning. This is especially in order to address the ubiquitous curse of dimensionality \cite{bellman_dynamic_1956}, where, as dimensions of datasets increase, algorithms may fail to converge, gradients may vanish or become stuck in barren plateaus. This may occur during the preprocessing stage, in-processing or during post-processing. Techniques such as principal component analysis \cite{hastie_elements_2013}, matrix factorisation \cite{ciliberto_quantum_2018}, feature extraction together with algebraic techniques such as singular value decompositions are all motivated primarily to reduce the dimensionality and complexity of datasets, thereby minimising the hypothesis search space. Moreover, learning algorithms which can efficiently solve problems with sparse datasets often have computational advantages.  Quantum data by its nature rapidly becomes higher-dimensional as the number of qubit or computations resources increases. Such vast search spaces present challenges for QML, such as the barren plateaus problems \cite{mcclean_barren_2018}, the quantum analogue of the vanishing gradient problem in classical machine learning (albeit with differences arising due to exponentially-large search spaces) (see section \ref{sec:ml:Barren Plateaus}).
    \item \textit{Preprocessing}. Datasets often require or benefit from preprocessing in order to ameliorate problems during training, such as vanishing gradients, bias or problems with convergence. Preprocessing data can include techniques such as sparsification \cite{ravishankar_online_2015} or smoothing or other strategies. For example, for quantum circuit synthesis, ensuring training data samples are drawn from across the Hilbert space of interest rather than limited to subspaces can assist with generalisation (see \cite{perrier_quantum_2020} for a geometric example). In such cases, quantum dataset preparation may benefit from preprocessing to address sparsity concerns (see \cite{marrero_entanglement_2020} for examples and for classical analogues of vanishing gradients \cite{pascanu_difficulty_2013}). 
    \item \textit{Visibility}. Classical machine learning is often concerned with extraction - or development - of features. In many forms of classical machine learning, such as those using kernel methods, or deep learning, features of importance to optimal performance of an algorithm may need to be inferred from the data. Quantum datasets in many ways face such challenges from the outset as quantum data can never be directly observed, rather it must be inferred from measurement statistics. When constructing quantum datasets, the extent to which such inferred (as distinct from directly observed) data will be an important choice. In the QDataSet, we have chosen to include a range of such `hidden' or `inferred' data to assist practitioners with use of the dataset, including the intermediate forms of Hamiltonian, unitaries and other data that is not itself directly accessible but is a by-product of our simulation (accessible via intermediate layers).
\end{enumerate}
Studying features of particular datasets and their use in classical contexts assists in extracting desirable features for large-scale quantum datasets. ImageNet is one of the leading research projects and database architectures for images \cite{deng_imagenet_2009, russakovsky_imagenet_2015, goodfellow_deep_2016}. The dataset is one of the most widely cited and important datasets in the development of machine learning, especially for image-classification algorithms using convolutional, hierarchical and other deep-learning based neural networks. The evolution of ImageNet and its use within machine learning disciplines provides a useful guide and comparison for the development of QML datasets in general. ImageNet comprises two main components: (i) a public semi-structured dataset of images together with (ii) an annual competition and workshop. The dataset provides a `ground truth' standardised set of images against which to train categorical image classification algorithms. The competition and workshop provided and continue to provide an important institutional practice driving machine learning development. While beyond the scope of this work, the development of QML would arguably be considerably assisted by the availability of QML-focused competitions akin to those commonplace within the classical machine learning community. Such competitive frameworks would motivate and drive the development of scalable and generalisable QML algorithms. As is also evident from classical machine learning, competitive formats are also a useful way for laboratories, companies or other projects to leverage the expertise of the diverse machine learning community.

Another example from the machine learning community which can inform the development of QML is Kaggle, a leading online platform for machine learning-based competitions. Kaggle runs competitions where competitors are provided with prediction tasks, training sets and constraints upon the type of algorithm design (such as resource use and so on). Competitors then develop models aiming to optimise a measure of success, such as a standard machine learning metric of accuracy, AUC or some other measure \cite{bojer_kaggle_2021}. The open competitive nature of Kaggle is designed to crowd-source solutions and expertise to problems in machine learning and data science. A `quantum Kaggle' would be of considerable benefit to the QML community by providing a platform through which to spur collaborative and competitive development of quantum algorithms.

%====characteristics

\subsection{QML Datasets}
While quantum datasets for machine learning (quantum and classical) are neither as prevalent nor as ubiquitous as those in the classical realm, there are some examples in the literature of quantum or hybrid quantum-classical datasets generated for use in machine learning contexts. 
QML datasets can be categorised into: (1) \textit{general quantum datasets} produced for purposes other than QML, such as quantum datasets in quantum chemistry of other fields, which can be preprocessed or used as training data in QML contexts. Such datasets are not specifically produced for the purposes of QML per se; (2) dedicated \textit{QML-specific quantum datasets}, generated and structured for the purposes of QML. This second category mainly consists of quantum datasets for use in classical or hybrid machine learning contexts. Quantum datasets currently available tend towards one or other of these classifications, though there is overlap, for example, with quantum datasets designed for use in machine learning which are nevertheless highly domain-specific. Examples include quantum chemistry datasets for use in deep tensor neural networks \cite{schutt_quantum-chemical_2017}, datasets for learning spectral properties of molecular systems \cite{chmiela_machine_2017, ramakrishnan_electronic_2015, rupp_fast_2012, bartok_machine-learning_2013} and for solid-state physics \cite{sutton_crowd-sourcing_2019, nyshadham_machine-learned_2019, szlachta_accuracy_2014,faber_machine_2016}.

A recent example is provided by the dedicated quantum chemistry datasets known as the \textit{QM7-X dataset} \cite{hoja_qm7-x_2021}, an expansive dataset of physiochemical properties for several million equilibrium and non-equilibrium structures of small organic molecules. The QM7-X dataset spans regions of chemical compound space and was generated to provide a basis for machine-learning assisted design of molecules with specific properties. The dataset builds upon previous iterations of QM-series and related quantum chemistry datasets \cite{rupp_fast_2012, ruddigkeit_enumeration_2012, smith_ani-1ccx_2020}. Structurally, the dataset combines global (molecular) properties and local (atomic) information, including ground state quantities (spectra and moments) along with response quantities (related to polarisation and dispersion). The dataset is highly domain-specific and represents a salient example of a dataset designed to spur machine-learning driven research within a particular field. More recent examples include entangled datasets for use in QML research \cite{schatzki_entangled_2021}.
\subsection{QML and QC platforms}
The use of quantum datasets in machine learning has been facilitated over the last several years by a surge in quantum programming and languages and platforms for both QML and quantum computing generally. While such platforms are dynamic and changing, it is important that quantum datasets for machine learning be constructed to be as interoperable with platforms in the quantum and classical machine learning community. Generators of large-scale quantum datasets should be cognisant of how their data can be (more easily) used in such platforms below and also how their datasets can be designed in ways that facilitates their ease of use within common machine learning languages, such as TensorFlow, PyTorch and across languages, such as Python, C\#, Julia and others.  The QDataSet has been specifically designed in relatively elementary Python packages such as Numpy in order to facilitate its use across the machine learning community, but also in a way that we hope makes it useable and clearly understandable by the quantum engineering community. We deliberately selected Python as the language of choice within which to build the QDataSet simulation given its status as the leading classical programming language of choice for machine learning. It also is a language adopted across many of the quantum platforms above. We built the QDataSet using Numpy to produce datasets as transferable as possible (rather than, for example, in Qutip). A familiarity with the emerging quantum programming and QML ecosystem is useful for the design of quantum datasets. We set out a few examples of leading quantum programming platforms below.   

\textit{Qutip} \cite{johansson_qutip_2013} is a leading quantum simulation and algorithmic package in Python used for open quantum systems' simulation. The package, while not developed specifically for QML, is widely used for hybrid quantum-classical systems' research. Inputs to Qutip algorithms are Numpy-based vectors, tensors and matrices used to represent density matrices, quantum states and operators. Qutip permits a wide range of simulations to be run and data to be generated, including for state preparation, control and drift Hamiltonians, pulse sequences and noise modelling. As discussed below, the QDataSet, which was built in Python using primarily the Numpy package, but was verified using Qutip. \textit{Q\#} is Microsoft's primary open-source programming language for quantum algorithm design and execution. The platform comprises a number of libraries, simulators and a software development kit (QDK). \textit{Quantum Tensorflow} (QTF) \cite{broughton_tensorflow_2020} is a hybrid quantum-classical version of Google's leading open-source machine learning Tensorflow platform. QTF is constructed to enable the synthesis of classical and quantum algorithmic machine learning, for example classically parameterised quantum circuits, variational quantum circuits and eigensolvers, quantum convolutional neural networks and other quantum analogues of classical machine learning architectures. QTF follows Tensorflow's overall machine learning structure and data taxonomy. Input data is usually in the form of tensors. QTF's in-platform datasets vary depending on use case, but the platform primarily draws upon classical datasets for hybrid use cases (quantum computation applied to solving classical optimisation tasks).  For simulated quantum-native data, QTF draws upon \textit{Cirq}, Google's open source framework for programming quantum computers \cite{cirq_developers_cirq_2021}. Cirq is focused on providing a software library for research into and simulations of quantum circuits, the idea being to develop quantum algorithms in Cirq that can be run on quantum computers and simulators. 

\textit{Strawberry Fields} \cite{killoran_strawberry_2019} is an open-source QML and quantum algorithm programming platform developed by Xanadu for photonic quantum computing. \textit{Qiskit} is another open source software development kit for quantum circuits, control and applications on quantum and hybrid computers  \cite{aleksandrowicz_qiskit_2019}. Qiskit is based on an open source quantum assembly language (QASM) standardised abstraction of quantum circuits. Other platforms enabling the integration of quantum datasets and QML algorithms (either quantum or classical) include those available via IBM's Quantum Experience. The QDataSet has been designed to for interoperability across most of these platforms. Practically speaking, this means that researchers can select dataset features of interest, such as tensors of quantum states, Hamiltonians, unitary operators (gates) or even noise information and integrate as datasets for use in algorithms designed using platforms above. Similarly, machine learning researchers should find the form of data relatively familiar to typical datasets in machine learning, where information is encoded in tensors, lists or matrices. Examples of using similar datasets in customised TensorFlow machine learning models can be found in various sources \cite{youssry_characterization_2020, perrier_quantum_2020}.
\subsection{Quantum data in datasets}
An important aspect of quantum dataset design is the decision regarding what quantum information to include in the dataset. In this section, we list types of quantum data which may be included in large-scale quantum datasets. By \textit{quantum data}, we refer to data generated or characterising quantum systems or processes. More specifically, by quantum data we refer to quantum registers (see definition \ref{defn:quant:Quantum registers and states}) discussed in Appendix \ref{chapter:Background: Quantum Information Processing} together with an assumed state preparation procedure (see \cite{watrous_theory_2018,wiseman_quantum_2010} on state preparation generally and the Appendices for more discussion) which sufficiently encode information into quantum states that subsist and transform according to the protocols of quantum information processing. 
Quantum data may comprise a range of different properties, features or characteristics of quantum systems, the environment around quantum systems. Such data is described using a classical alphabet $\Sigma$ and encoded in quantum states in ways that constitute a quantum register of that information. It may comprise data and information abstracted into a particular representation or form, such as circuit gates, algebraic formulations, codes etc or more physical forms, such as statistics or read-outs from measurement devices. For QML datasets, it is useful to ensure that quantum data is sufficient for classical machine learning researchers to understand and for integrating quantum data into their algorithmic techniques. For example, a classically parameterised quantum circuit (section \ref{sec:ml:Variational quantum circuits}), as common throughout the QML literature, would typically include data or tensors of the relevant parameters, the operators related to such parameters (such as generators) and the quantum states (vectors or density operators) upon which the circuit acts.

%========Dataset preliminaries
\subsection{QDataSet Methodological Overview} \label{sec:qdata:QDataSet Methodological Overview}
In this section, we provide an overview of the methods according to which the QDataSet was developed. Our exposition includes detail of (i) the Hamiltonians used to generate the dataset, (ii) characteristics of control mechanisms used to control the evolution of the quantum simulations and (ii) measurement procedures by which information is extracted from the evolved quantum simulations. This section is stand-alone but can also be read in tandem with the Appendices. We aim to equip classical machine learning practitioners with a minimum working knowledge of our methodology to enable them to understand both how the datasets were generated and the ways in which quantum data and quantum computing differ from their classical counterparts relevant to solving problems in quantum control, tomography and noise spectroscopy. Our focus, as throughout this work, is on the application of classical and hybrid classical-quantum machine learning algorithms and techniques to solve constrained optimisation problems using quantum data (as distinct from the application of purely quantum algorithms). 

A synopsis of quantum postulates and quantum information processing set out in Appendix \ref{chapter:Background: Quantum Information Processing} aims also to provide a mapping between ways in which data and computation is characterised in quantum contexts and their analogues in classical machine learning. For example, the description of dataset characteristics, dimensionality, input features, training data to what constitutes labels, the types of loss functions applicable are all important considerations in classical contexts. By providing a straightforward way to translate quantum dataset characteristics into typical classical taxonomies used in machine learning, we aim to help lower barriers to more machine learning practitioners becoming involved in the field of QML. A high-level summary of some of the types of quantum features that QML-dedicated datasets ideally would contain is set out in the background Appendices. Our explication of the QDataSet below provides considerable detail on each quantum data feature contained within the datasets including parameters, assumptions behind our choice of quantum system specifications, measurement protocols and noise context. A dictionary of specifications for each example in the QDataSet is set out in Tables (\ref{table:datasetproperties1}) and (\ref{table:datasetproperties2}). 

\subsubsection{Formalism}
Recall in density operator formalism (definition \ref{defn:quant:Density operator}), a quantum system can be described via a Hermitian positive semi-definite density operator $\rho \in \mathcal{B}(\Hilb)$ with trace unity acting on the state space of the system (such that if the system is in state $\rho_i$ with probability $p_i$ then $\rho = \sum_i p_i \rho_i$ as per equation (\ref{eqn:quant:rhomixedstate})). Density operators are generalised operator-representations of probability distributions over quantum states with particular properties: all their eigenvalues have to be real, non-negative, sum to unity, inheriting the necessary constraints of a probability distribution. States $\ketpsi$ in the Hilbert space may be composite systems, described as the tensor product of states (see \ref{sec:quant:Tensor product states}) spaces of the component physical systems, that is $\ketpsi = \otimes_i \ket{\psi_i}$. We also mention here (as discussed in section \ref{sec:quant:Open quantum systems}) the importance of open quantum systems sect where a total system Hamiltonian $H$ can be decomposed as $H = H_S + H_E + H_I$ (equation (\ref{eqn:quant:openquantHamiltonianSE})), comprising a closed quantum system Hamiltonian $H_S$, an environment Hamiltonian $H_E$ and an interaction Hamiltonian term $H_I$, which is typically how noise is modelled in quantum contexts. Open quantum systems are typically approximated using master equations to capture the dissipative effects of system/environment interaction. The dissipative nature of open quantum systems has parallels with the dissipative characteristics of neural networks (see \cite{schuld_supervised_2018}). Simulated data of open quantum systems can be generated in a number of packages, such as Qutip. For the QDataSet, we made a design decision to directly simulate the effects of coupling to dissipative environmental baths using more elementary Monte Carlo methods. The reason for this is that master equation formalism both requires a number of assumptions on the system (see \cite{wiseman_quantum_2010, johansson_qutip_2013}) which may be difficult to apply to experimental setups. We also chose to manually engineer the effect of baths in order to minimise the theoretical barriers for classical machine learning practitioners using the QDataSet. 

In this work, we focus on both closed and open quantum system simulation. Closed quantum systems evolve over time $\Delta t= t_1 - t_0$ according to the Schr\"odinger equation (\ref{eqn:quant:schrodingersequation}) whose solutions take the form of unitary channels. As discussed below, given the difficulties in solving for time dependency, unitaries are typically approximated by time-independent sequences of unitaries (see section \ref{sec:quant:Time-independent approximations}). The evolution of quantum states is characterised by such unitaries operating upon vectors that transform (transition) to other states (section \ref{sec:quant:Evolution, Hamiltonians and control}). Intermediate quantum states $\ket{\psi'}$ may be represented as the result of applying unitary operators to initial states $\ketpsi$ such that $\ket{\psi'} = U(t) \ket{\psi_0}$ (or $\rho' = U (t) \rho_0 U(t)^\dagger$). Given initial quantum states, quantum state evolution can be represented entirely via operator dynamics and representations or more information-theoretically in terms of quantum channels acting to transition among quantum registers. In standard programming languages, operators will take the form of matrices or tensors. 
It is worth noting that the operator acting on a quantum state $\rho$ is a unitary $U(t) \in \mathcal{B}(\Hilb)$ which is itself (in a closed quantum system) a function (or representation) of the Hamiltonian $H(t)$ governing the system dynamics. In practice, unitaries are formed by exponentiating time-independent approximations of Hamiltonians. These unitaries represent solutions to the time-dependent Schr{\"o}dinger equation governing evolution for a closed system. 

The evolutionary dynamics of the quantum system are completely described by the Hamiltonian operator acting on the state $\ketpsi$ such that $\ket{\psi(t)} = U(t)\ket{\psi(t=0)}$ (section \ref{sec:quant:Quantum evolution}). In density operator notation, such evolution is represented as $\rho(t) = U(t) \rho(t_0) U(t)^\dagger$. Typically solving the continuous form of the Schr{\"o}dinger equation is intractable or infeasible, so a discretised approximation as a discrete quantum circuit (where each gate $U_i$ is run for a sufficiently small time-step $\Delta$t) is used (e.g. via Trotter-Suzuki decompositions). Open quantum systems are described by the formalism of Lindbladian master equations (equation \ref{eqn:quant:masterequation}), representing the effect of noise sources upon the evolution of quantum states.

\subsubsection{Additional methodological concepts} \label{sec:qdata:Additional methodological concepts}
There are a number of other important concepts for classical machine learning practitioners to be aware of when using quantum datasets such as the QDataSet. We set out some of these: (a) \textit{relative phase} (see section \ref{sec:quant:Density operators and multi-state systems}), for a qubit system, amplitudes $a$ and $b$ differ by a relative phase if $a = \exp(i\theta)b, \theta \in \Real$ (relative phase is an important concept as the relative phase encodes quantum coherences and is, together with basis encoding, a primary means of encoding data in quantum systems); (b) \textit{entanglement} (see section \ref{sec:quant:Quantum entanglement}), composite states (known as EPR or Bell states), may be entangled, meaning that their measurement outcomes are necessarily correlated. For a two-qubit state: 
\begin{align}
    \ketpsi = \bellzz
\end{align}
a measurement outcome of $0$ on the first qubit necessarily means that a measurement of the second qubit will result in the post-measurement state $\kz$ also i.e. the measurement statistics of each qubit correlate. States that are entangled cannot be written as tensor products of component states i.e. $\ketpsi \neq \ket{\psi_1}\ket{\psi_2}$ (see section \ref{sec:quant:Quantum entanglement}); (c) \textit{expectation}, expectation values of an operator $A$ (e.g. a measurement) can be written as $E(A) = \text{tr}(\rho A)$ (see equation (\ref{eqn:quant:averagevalue})); (d) \textit{commutativity}, multiple measurements are performed on a system, the outcome will be order-dependent if the measurement operators corresponding to those measurements do not commute i.e if $[A,B]\neq 0$ (see definition \ref{defn:alg:Lie algebraliederivative}). This is distinct from the classical case; and (e) \textit{no cloning}, quantum data cannot be copied, fundamentally distinguishing it from classical data (see theorem \ref{thm:quant:No-Cloning Theorem}). 

There are a range of other aspects of quantum systems that are relevant to the use of machine learning algorithms for solving optimisation problems (see section \ref{sec:ml:Optimisation methods}) which we pass over but which are relevant to research programmes using such algorithms, including the role of error-correcting codes (designed to limit or self-correct errors to achieve fault-tolerant quantum computing). While not the focus of the QDataSet, it is worth noting for machine learning practitioners that a distinction is usually drawn in the quantum information literature between \textit{physical} and \textit{logical} qubits. A physical qubit is a two-level physical quantum system. A logical qubit is itself an abstraction from a collection of physical qubits which in aggregate behave according to qubit parameters \cite{shaw_encoding_2008}. 
% The QDataSet is generated on the assumption that each qubit is a logical qubit (which may or not equate to a single corresponding physical qubit). 
For more complex treatments (involving a multitude of physical qubits) in quantum control or quantum error correction, the underlying simulation code may be adapted accordingly.

\subsubsection{One- and two-qubit Hamiltonians}\label{sec:qdata:One- and two-qubit Hamiltonians}
We now describe the QDataSet Hamiltonians which are integral to understanding the method by which the datasets were generated. First we describe the single-qubit Hamiltonian and then move to an exposition of the two-qubit case. Each Hamiltonian has been designed consistent with solving use-cases below. In particular, we have adopted quantum control formalism for Hamiltonians (as discussed in Appendices \ref{chapter:Background: Quantum Information Processing} and \ref{chapter:Background: Geometry, Lie Algebras and Representation Theory}) in terms of drift and control components (see section \ref{sec:quant:Quantum Control}). For the single-qubit system, the drift Hamiltonian $H_d(t)$ is fixed in the form:
\begin{align}
    H_d(t) = H_d =  \frac{1}{2} \Omega \sigma_z.
\end{align}
Here $\sigma_z$ is the Pauli generator (see equation (\ref{eqn:qdata:paulimatrices})) for $z$-axis rotations. The $\Omega$ term represents the energy gap of the quantum system (the difference in energy between, for example, the ground and excited state of the qubit, recalling qubits are characterised by having two distinct quantum states). The single-qubit drift Hamiltonian for the QDataSet is time-independent for simplicity, though in realistic cases it will contain a time-dependent component. For the single-qubit control and noise Hamiltonians we have two cases based upon the concept of which axes controls and noise are applied. Recall we can represent a single qubit system on a Bloch sphere, with axes corresponding to the expectations of each Pauli operator and where operations of each Pauli operator constitute rotations about the respective axes. Our controls are control functions (equation (\ref{eqn:quant:controlsystem1})), mostly time-dependent, that apply to each Pauli operator (generator). They act to affect the amplitude over time of rotations about the respective Pauli axes. More detailed treatments of noise in quantum systems and quantum control contexts can be found in \cite{wiseman_quantum_2010,dalessandro_introduction_2007}.

As discussed above, the functional form of the control functions $f_\alpha(t)$ varies (see section \ref{sec:quant:Quantum Control} for background on control functions). We select both square pulse and Gaussian pulse wavefunctions as the form (see below). Each different noise function $\beta_\alpha(t)$ is parameterised differently depending on various assumptions that are more specifically detailed in \cite{youssry_characterization_2020} and summarised below.  Noise and control functions are applied to different qubit axes in the single-qubit and two-qubit cases. For a single qubit, we first have single-axis control along $x$-direction:
\begin{align}
    H_{\text{ctrl}}(t) &= \frac{1}{2} f_x(t) \sigma_x 
\end{align}
with the noise (interaction) Hamiltonian $H_1(t)$ along $z$-direction (the quantification axis):
\begin{align}
    H_1(t) &= \frac{1}{2} \beta_z(t) \sigma_z
\end{align}
Here the function $\beta_z(t)$ (a classical noise function $\beta(t)$ applied along the $z$-axis) may take a variety of forms depending on how the noise was generated (see below for a discussion of noise profiles e.g. N1-N6). See section \ref{sec:quant:Noise and quantum evolution}. It should be noted that noise rarely has a functional form and is itself difficult to characterise (so $\beta(t)$ should not be thought of as a simple function). For the second case, we implement multi-axis control along $x$- and $y$- directions and noise along $x$- and $z$-directions in the form:
\begin{align}
    H_{\text{ctrl}}(t) &= \frac{1}{2} f_x(t) \sigma_x + \frac{1}{2} f_y(t) \sigma_y \\
    H_1(t) &= \frac{1}{2} \beta_x(t) \sigma_x + \frac{1}{2} \beta_z(t) \sigma_z.
\end{align}
Noiseless evolution may be recovered by choosing $H_1(t)=0$. Note given the choice of the $z$-axis as the quantification axis, the application of $x$-axis noise may give rise to decohering effects.
For the two-qubit system, we chose the drift Hamiltonian in the form:
\begin{align}
    H_d(t) = \frac{1}{2} \Omega_1 \sigma_z \otimes \sigma_0 + \frac{1}{2} \Omega_2 \sigma_0 \otimes \sigma_z.
\end{align}
For the control Hamiltonians, we also have two cases. The first one is local control along the $x$-axis of each individual qubit, akin to the single-qubit case each. In the notation, $f_{1\alpha}(t)$ indicates that the control function is applied to, in this case, the second qubit, while the first qubit remains unaffected (denoted by the `1' in the subscript and by the identity operator $\sigma_0$). We also introduce an \textit{interacting control}. This is a control that acts simultaneously on the $x$-axis of each qubit, denoted by $f_{xx}(t)$:  
\begin{align}
     H_{\text{ctrl}}(t) = \frac{1}{2} f_{x1}(t) \sigma_x \otimes \sigma_0 + \frac{1}{2} f_{1x} \sigma_0 \otimes \sigma_x + f_{xx}(t) \sigma_x \otimes \sigma_x.
\end{align}
The second two-qubit case is for local-control along the $x-$axis of each qubit only and is represented as:
\begin{align}
     H_{\text{ctrl}}(t) = \frac{1}{2} f_{x1}(t) \sigma_x \otimes \sigma_0 + \frac{1}{2} f_{1x} \sigma_0 \otimes \sigma_x,
\end{align}
For the noise, we fix the Hamiltonian to be along the $z$-axis of both qubits, in the form: 
\begin{align}
    H_{1}(t) &= \frac{1}{2} \beta_{z1}(t) \sigma_z \otimes \sigma_0 + \frac{1}{2} \beta_{1z} \sigma_0 \otimes \sigma_z.
\end{align}
Notice, that for the case of local-only control and noiseless evolution, this will correspond to two completely-independent qubits and thus we do not include this case, as it is already covered by the single-qubit datasets. We also note that not all interaction terms (such as $\sigma_z \otimes \sigma_z$) need be included in the Hamiltonian. The reason for this is that to achieve universal control equivalent to including all generators, one only need include one-local control for each qubit together with interacting (entangling) terms (though we note recent results regarding constraints on 2-local operations in the presence of certain symmetries \cite{marvian_restrictions_2022}). Assuming one has a minimal (bracket-generating, see definition \ref{defn:geo:bracketgenerating}) set of Pauli generators in $\su(2)$ in the Hamiltonian, one may synthesise any Pauli gate of interest for the one- or two-qubit systems (i.e. given two Pauli gates, one can synthesise the third) making the set of targets reachable (definition \ref{defn:geo:Reachable set}), thus achieve effective universal control (see section \ref{sec:geo:Geometric control theory} for bracket-generating subalgebras and equation \ref{eqn:alg:Baker-Campbell-Hausdorff} for discussion of the BCH formula).
\\
To summarise, the QDataSet includes four categories for the datasets set-out in Table \ref{tab:cats}. The first two categories are for 1-qubit systems, the first is single axis control and noise, while the second is multi-axis control and noise. The third and fourth categories are 2-qubit systems with local-only control or with an additional interacting control together with noise. 

\subsection{Hamiltonians: drift, control, noise} \label{sec:qdata:Hamiltonians: drift, control, noise}
Recapping above, the QDataSet comprises datasets for one- and two- qubits systems evolving in the presence and absence of noise. The canonical forms of Hamiltonian discussed above from which those in the QDataSet are developed are given in \cite{youssry_characterization_2020}. In that work, a limited set of single-qubit systems subject to external environmental noise (baths) was used as input training data for a novel greybox machine learning alternative method for characterising noise. In this work, the underlying simulation was modified to generate a greater variety of qubit-noise examples for the single qubit case. The simulation was then extended beyond that in \cite{youssry_characterization_2020} to generate examples for the two-qubit case (in the presence or absence of noise). As discussed above, the evolution of closed and open quantum systems is described by Hamiltonian dynamics, which encode time-dependent functions into operators which are the Lie algebra generators (see section \ref{sec:alg:Lie algebras}) of time-translations (operators) acting on quantum states. 
% The general form of the Hamiltonian for the QDataSet is given by:
% \begin{align}
%     H(t)&= H_{\text{drift}}(t) + H_{\text{ctrl}}(t) + H_1(t)
% \end{align}
To summarise: the Hamiltonian comprises three elements: (i) a drift Hamiltonian $H_d(t)$, encoding the endogenous evolution of the quantum system; (ii)  a control Hamiltonian $H_{\text{ctrl}}(t)$, encoding evolution due to the application of classical controls which may be applied to the quantum system to steer its evolution; and (iii) and an interaction (noise) Hamiltonian $H_1(t)$, encoding the effects of coupling of the quantum system to its environment, such as a decohering noise source (a bath). The Hamiltonians are composed using Pauli generators (see section \ref{sec:qdata:Pauli matrices} and equations (\ref{eqn:qdata:paulimatrices})), representing elements of the Lie algebra $\su(2)$. We express the Hamiltonians in the Pauli operator basis $\{ \sigma_i \}$ which forms a complete basis for our one- and two-qubit systems. Our control functions are represented as $f_\alpha(t)$ for $\alpha = x,y,z$ where the subscript indicates which generator the control applies to. Concretely, for example, $\sigma_z$ control is denoted $f_z(t) \sigma_z$. In general, continuous time-dependent control formulations are difficult - or infeasible - to solve analytically, where solving here means finding a suitable representation of the control unitary (equation (\ref{eqn:quant:unitarysolutiontimedependentschrod})): 
\begin{align}
    U_{\text{ctrl}} = \mathcal{T}_{+} e^{-i\int_0^T f_\alpha \sigma_\alpha/2 dt}
\end{align}
Most controls are usually classical i.e. $f_\alpha(t) \in \Real$. The simplest control functional form is fixed amplitude control \cite{schattler_geometric_2012} or what is also described as a \textit{square pulse}, where constant energy (expressed as amplitudes) is expressed for a discrete time-step $\Delta t$. Other control waveforms include Gaussian. Moreover, quantum control in the QDataSet context has two primary imperatives. The first is the use of control in closed noise-free idealised quantum systems where the objective is the use of controls to steer the quantum system to some desired objective state. This is equivalent to the synthesis of quantum circuits (sequences of quantum gates) from the identity $I$ to a target unitary $U_T$. The second is the use of controls in the presence of noise, where the quantum system is coupled to an environment that potentially decoheres the system. In this second case, ideally the controls are tailored to neutralise the effect of noise on the evolution of the quantum system - a process typically described by dynamic decoupling \cite{gupta_machine_2018, mavadia_prediction_2017} (see for example Hahn echo or other examples). Crafting suitable controls to mitigate noise effects is challenging. One must properly time and select appropriate amplitudes for the application of controls to counteract decohering interference, modelled as a superoperator term in master equations, for example (see equation (\ref{eqn:quant:lindbladmasterequation})). Typically, it requires information about the noise spectrum itself, obtained using techniques from quantum noise spectroscopy \cite{wiseman_quantum_2010}. It also requires an understanding of how control and noise signals convolve in the context of quantum evolution. Dealing with noise is a central imperative of quantum information processing and the engineering of quantum systems where the aim is to preserve and extend coherence times of quantum information and to correct for errors that arise during the evolution of quantum computational systems. To this end, a major stream of research in quantum information processing concerns quantum error-correcting codes (QEC) as means of encoding quantum information in a way that renders it robust to noise-induced errors and/or enables `self-correction' of errors as a quantum system evolves.

\subsubsection{QDataSet Control} \label{sec:qdata:QDataSet Control}
Recall that the control pulse sequences in the QDataSet consist of two types of waveforms. The first is a train of Gaussian pulses, and the other is a train of square pulses, both of which are very common in actual experiments. Square pulses are the simplest waveforms, consisting of a constant amplitude $A_k$ applied for a specific time interval $\Delta t_k$:
\begin{align}
    f(\Delta t_k) = f = A_k \Delta t_k
\end{align}
where $k$ runs over the total number of time-steps in the sequence. The three parameters of such square pulses are the amplitude $A_k$, the position in the sequence $k$ and the time duration over which the pulse is applied $\Delta t$. In the QDataSet, the pulse parameters are stored in a sequence of vectors $\{a_n \}$. Each vector $a_n$ is of dimension $r$ parameters of each pulse (e.g. the Gaussian pulse vectors store the amplitude, mean and variance, the square pulse vectors store pulse position, time interval and amplitude), enabling reconstruction of each pulse from those parameters if desired. For simplicity, we assume constant time intervals such that $\Delta t_k = \Delta t$. The Gaussian waveform can be expressed as: 
\begin{align}
    f(t) = \sum_{k=1}^{n} A_k e^{-(t-\mu_k)^2 /2 \sigma_k^2}.
\end{align}
where $n$ is the number of Gaussian pulses in the sequence. The parameters of the Gaussian pulses differ somewhat from those of the square pulses. Each of the $n$ pulses in the sequence is characterised by a set of $3$ parameters: (i) the amplitude $A_k$ (as with the square pulses), (ii) the mean $\mu_k$ and (iii) the variance $\sigma_k$ of the pulse sequence. Thus in total, the sequence is characterised by $3n$ parameters.
The amplitudes for both Gaussian and square pulses are chosen uniformly at random from the interval $[A_{\text{min}}, A_{\text{max}}]$, the standard deviation for all Gaussian pulses in the train is fixed to $\sigma_k = \sigma$, and the means are chosen randomly such that there is minimal amplitude in areas of overlapping Gaussian waveform for the pulses in the sequence. The pulse sequences can be represented in the time or frequency domains \cite{bandrauk_quantum_2003}. The QDataSet pulse sequences are represented using the time-domain as it has been found to be more efficient feature for machine learning algorithms \cite{youssry_characterization_2020}.

As discussed in \cite{youssry_characterization_2020}, the choice of structure and characteristics of quantum datasets depends upon the particular objectives and uses cases in question, the laboratory quantum control parameters and experimental limitations. Training datasets in machine learning should ideally be structured so as to enhance the generalisability. In the language of statistical learning theory, datasets should be chosen so as to minimise the empirical risk (equation \ref{eqn:ml:empiricalrisk}) associated with candidate sets of classifiers \cite{vapnik_nature_1995, hastie_elements_2013}. In a quantum control context, this will include understanding for example the types of controls available to researchers or in experiments, often voltage or (microwave) pulse-based \cite{hollenberg_charge-based_2004}. The temporal spacing and amplitude of each pulse in a sequence of controls applied during an experiment may vary by design or to some degree uncontrollably. Pulse waveforms can also differ. For example, the simplest pulse waveform is a constant-amplitude pulse applied for some time $\Delta t$ \cite{khaneja_optimal_2005}. Such pulses are characterised by for example a single parameter, being the amplitude of the waveform applied to the quantum system (this manifests as we discuss below as an amplitude applied to the algebraic generators of unitary evolution (see  \cite{dalessandro_introduction_2007, perrier_quantum_2020} for an example)). Other models of pulses (such as Gaussian) are more complex and require more sophisticated parametrisation and encoding with machine learning architectures in order to simulate. More detail on such considerations and the particular pulse characteristics in the QDataSet are set-out in Tables (\ref{table:datasetproperties1}) and (\ref{table:datasetproperties2}).

\subsubsection{QDataSet Metrics}\label{sec:qdata:QDataSet Metrics}
For most machine learning practitioners using the QDataSet, the entry point will be the application of known classical machine learning metrics. More advanced uses of the QDataSet may utilise quantum-specific metrics directly, for example, via reconstruction of quantum states from measurement statistics. Some use cases combine the use of classical and quantum metrics. For example, \cite{perrier_quantum_2020, youssry_characterization_2020} combine average operator fidelity (equation \ref{eqn:quant:fidelityfunction}) with standard mean-squared error (MSE) (equation (\ref{eqn:ml:MSE}))  into a measure of empirical risk denoted as `batch fidelity' as per equation (\ref{eqn:ml:batchfidelityMSE}). 
In those examples, the objective in question is to train a greybox algorithm (section \ref{sec:ml:Greybox machine learning}) to model certain control pulses needed to synthesise target unitaries. The algorithms learn the particular control pulses which are applied to generators. While it is the extraction of control pulses which are of interest to experimentalists, the final output of the algorithm is a sequence of fidelities where the fidelity of generated (synthesised) unitaries is compared against the target (label) unitaries $U_T$. This sequence of fidelities is then compared against a vector of ones with the loss function set to minimise the MSE (distance) between the fidelity sequence and the label vector. In doing so, the algorithms are effectively trained to maximise fidelity (as fidelities $\approx 1$ are desirable) yet do so using a hybrid approach. The QDataSet has been generated such that a combination of classical, quantum and hybrid metrics of divergence measures may be used in the training process.

\section{Experimental Methods}\label{sec:qdata:Experimental Methods}
The QDataSet comprises 52 datasets based on simulations of one- and two-qubit systems evolving in the presence and/or absence of noise subject to a variety of controls. It has been developed to provide a large-scale set of datasets for the training, benchmarking and competitive development of classical and quantum algorithms for common tasks in quantum sciences, including quantum control, quantum tomography and noise spectroscopy. It has been generated using customised code drawing upon base-level Python packages and TensorFlow in order to facilitate interoperability and portability across common machine learning and quantum programming platforms. Each dataset consists of 10,000 samples which in turn comprise a range of data relevant to the training of machine learning algorithms for solving optimisation problems. The data includes a range of information (stored in list, matrix or tensor format) regarding quantum systems and their evolution, such as: quantum state vectors, drift and control Hamiltonians and unitaries, Pauli measurement distributions, time series data, pulse sequence data for square and Gaussian pulses and noise and distortion data. 
\\
\\
The total compressed size of the QDataSet (using Pickle and zip formats) is around 14TB (uncompressed, several petabytes). Researchers can use the QDataSet in a variety of ways to design algorithms for solving problems in quantum control, quantum tomography and quantum circuit synthesis, together with algorithms focused on classifying or simulating such data. We also provide working examples of how to use the QDataSet in practice and its use in benchmarking certain algorithms. Each part below provides in-depth detail on the QDataSet for researchers who may be unfamiliar with quantum computing, together with specifications for domain experts within quantum engineering, quantum computation and quantum machine learning. Previous chapters Appendix \ref{chapter:Background: Quantum Information Processing} contains extensive background material for researchers unfamiliar with quantum information processing. The Appendices also contain discussions of relevant quantum control definitions and concepts relevant to the construction and use of the QDataSet. We also set out further below examples applications of the QDataSet together with links to corresponding Jupyter notebooks. The notebooks are designed to illustrate basic problem solving in tomography, quantum control and quantum noise spectroscopy using the QDataSet. They are designed to enable machine learning researchers to input their algorithms into the relevant section of the code for testing and experimentation. Machine learning uptake often occurs via adapting example code and so we regard these examples as an important demonstration of the use-case for the QDataSet.
%
%
%
%=========quantum methods background
\subsection{QDataSet and Scalability} \label{sec:qdata:QDataSet and Scalability}
The QDataSet was based on simulations of one and two-qubit systems only. The rationale for this choice was primarily one of computational feasibility. The QDataSet was generated over a six-month period using the University of Technology, Sydney's High Performance Computing (HPC) cluster, with some computations taking several weeks alone. In principle larger datasets based on our simulation code can be prepared, however we note the computational resources of doing so may be considerable. To generate the datasets, we wrote bespoke code in TensorFlow which enabled us to leverage GPU resources in a more efficient manner. As we discuss below, we were interested in developing a dataset that simulated noise affecting a quantum system. This required performing Monte Carlo simulations (see section \ref{sec:qdata:Monte Carlo measurements}) and solving Schr{\"o}dinger's equation several times for each dataset. While existing packages, such as Qutip (see below) are available to model the effect of noise on quantum systems, we chose not to rely upon such systems. The reason was that Qutip relies upon Lindblad master equations to simulate system/noise interactions which in turn rely upon the validity of certain assumptions and approximations. Chief among these is that noise is Markovian. In our datasets, we included coloured noise with a power-spectrum density which is non-Markovian. Furthermore, Qutip's methods assumes a model of a quantum system interacting with a fermionic or bosonic bath which was not applicable in our case given we were modelling the imposition of classical noise using Monte Carlo methods.  

The resource cost for simulating the various qubit systems depended upon whether we sought to simulate noise or distortion. We found, however, that simulating the two-qubit systems took a significant amount of time, nearly four-weeks of runtime for a single two-qubit system. While multiple nodes of the HPC cluster were utilised, even on the largest node on the cluster (with at least 50-100 cores and two GPUs), the simulation time was extensive, even using GPU-equipped clusters. We estimate that more efficient speedup could be obtained by directly simulating in lower-order languages, such as C++. For this reason, we restricted the QDataSet to simulations of at most two-qubit systems. Such a choice obviously limits direct real-world applications of algorithms trained on the QDataSet to one- and two-qubit systems generally. While this may appear a notable limitation given the growing abundance of higher-order multi-qubit NISQ systems, it remains the case that many experimental laboratories remain limited to small numbers of qubits. We expect in most situations that one and two-qubit gates are all that are available. Engineering more than two-body interactions is an incredible challenge and only available in certain architectures. 

NISQ devices offer promising next steps, but it is primarily one- and two-qubit systems that have demonstrated the type of long coherence times, fast gate execution and fault-tolerant operation needed for truly scalable quantum computation \cite{chatterjee_semiconductor_2021,srinivas_high-fidelity_2021,preskill_quantum_2018}. As a result, the QDataSet can be considered relevant to the state of the art. Additionally, simulating more than two qubits would have exceeded computational capacity constraints related to our specific simulated code which includes interactions and iterations over different noise profiles. Moreover, developing algorithms on the basis of small qubit systems is a commonplace way of forming a basis for algorithms for larger multi-qubit systems: training classical machine learning algorithms on lower-order qubit systems has the benefit of enabling researchers to consider how such algorithms can or may learn multi-qubit relations which in turn can assist in algorithm design when applied to higher-order systems. Doing so will be an important step in building scalable databases for applying machine learning to problems in quantum computing.

\subsubsection{QDataSet and Error-Correction} \label{sec:qdata:QDataSet and Error-Correction}
The QDataSet was generated for non-QEC encoded data. The reasoning behind this was that (i) specific error-correcting encodings differ considerably from case to case, whereas unencoded quantum information is more prevalent in the experimental/laboratory setting; and (ii) quantum computational and NISQ devices are yet to reach the scale and prevalence necessary for practical testing of QEC at scale. The simulations in the QDataSet are based upon an alternative technique for quantum control in the presence of a variety of noise \cite{youssry_characterization_2020}, where a greybox neural network (see section \ref{sec:ml:Greybox machine learning}) is used to learn only those characteristics of the noise spectrum relevant to the application of controls (a comparatively simpler problem than seeking to determine the entire spectrum). In this context, the objective of the QDataSet is to enable algorithms to natively learn optimal error correction regimes from the data itself (rather than by encoding in a QEC) via inferring the types of countervailing controls (e.g. control pulses) that should be applied to minimise errors. In principle, the same type of machine-learning control architecture could also apply to QEC encoded data: the machine learning algorithms would in effect learn optimal quantum control conditioned on the data being encoded in an error-correcting code. Moreover, there is an emergent literature on using machine learning for QEC discovery itself. For machine learning practitioners, the QDataSet thus provides a useful way to seek to apply advanced classical machine learning techniques to challenging but important problems. 

%=======noise

\subsection{QDataSet Noise Methodology} \label{sec:qdata:QDataSet Noise Methodology}

\subsubsection{Noise characteristics} \label{sec:qdata:Noise characteristics}
The QDataSet was developed using methods that aimed to simulate realistic noise profiles in experimental contexts. Noise applicable to quantum systems is generally classified as either \textit{classical} or \textit{quantum} \cite{paz-silva_extending_2019}. Classical noise is represented typically as a stochastic process \cite{wiseman_quantum_2010} and can include, for example (i) slow noise which is pseudo-static and not varying much over the characteristic time scale of the quantum system and (ii) fast or `white' noise with a high frequency relative to the characteristic frequencies (energy scales) of the system \cite{wardrop_exchange-based_2014}. The effect of quantum noise on quantum systems is usually classified in two forms. The first is dephasing ($T_2$) noise, which characteristically causes quantum systems to decohere, thus destroying or degrading quantum information encoded within qubits (see section \ref{sec:quant:Noise and decoherence}). Such noise is usually characterised as an operator acting transverse to the quantisation axis of chosen angular momentum. The second type of noise $(T_1)$ can shift the energy state of the system (e.g. shifting the system from a ground to an excited state).

What this means in practice for the use of the QDataSet is usefully construed as follows using a Bloch sphere. Once an orientation ($x,y,z$-axes) is chosen, one is effectively choosing a choice of basis i.e. the basis of a typical qubit $\ketpsi = a\kz + b \ko$ is the basis of eigenstates of the $\sigma_z$ operator. When noise acts along the $z$-axis (i.e. is associated to the $\sigma_z$ operator), then it has the potential to (if the energy of the noise is sufficient) shift the energy state in which the quantum system is in, represented by a `flip' in the basis from $\kz$ to $\ko$ for example. This type of noise is $T_1$ noise. By contrast, noise may act along $x$- and $y$-axes of a qubit, which is represented as being associated with the $\sigma_x$ and $\sigma_y$ operators. These axes are `transverse' to the quantisation axis. Noise along these axis has the effect of dephasing a qubit, thus affecting the coherences encoded in the relative phases of the qubit. Such noise is denoted $T_2$ noise. Open quantum systems' research (section \ref{sec:quant:Open quantum systems}) and understanding noise in quantum systems is a vast and highly specialised topic. As we describe below, the QDataSet adopts the novel approach outlined in \cite{youssry_characterization_2020} where, rather than seeking to fully characterise noise spectra, the only the information about noise relevant to the application of controls (to dampen noise) is sought. Such information is encoded in the $V_O$ operator, which is an expectation that encodes the influence of noise on the quantum system (see subsection \ref{sec:qdata:QDataSet noise profiles} below). In a quantum control problem using the QDataSet samples containing noise, for example, the objective would then be to select controls that neutralise such effects.

\subsection{QDataSet noise profiles}\label{sec:qdata:QDataSet noise profiles}
In developing the QDataSet, we chose sets of noise profiles with different statistical properties. The selected noise profiles have been chosen to emulate commonplace types of noise in experimental settings. Doing so improves the utility of algorithms trained using the QDataSet for application in actual experimental and laboratory settings. While engineers and experimentalists across quantum disciplines will usually be familiar with theoretical and practical aspects of noise in quantum systems, many machine learning and other researchers to whom the QDataSet is directed will not. To assist machine learning practitioners whom may not be familiar with elementary features of noise, it is useful to understand a number of conceptual classifications related to noise used in the QDataSet as follows: (i) power spectral density (which describes the distribution of the noise signal over frequency); (ii) white noise (usually high-frequency noise with a flat frequency); (iii) coloured noise, a stochastic process where values are correlated spatially or temporally; (iv) autocorrelated stochasticity, which describes where the noise waveform characteristics are biased by tending to be short (blue) or long (red) as distinct from unautocorrelated noise, where waveforms are relatively uniformly distributed across wavelengths; and (v) stationary noise (a waveform with  a constant time period) and non-stationary noise (a waveform with a varying time period). See literature on noise in signal processing for more detail (such as \cite{orfanidis_introduction_1995} for an introduction or \cite{wiseman_quantum_2010} for a more advanced quantum treatment). The noise realizations are generated in time domain following one of these profile listed as follows (see  \cite{youssry_characterization_2020} for specific functional forms):
\begin{itemize}
\item N0: this is the noiseless case (indicated in the QDataSet parameters as set out in Tables (\ref{table:datasetproperties1}) and (\ref{table:datasetproperties2}));
    \item N1: the noise $\beta(t)$ is described by its power spectral density (PSD) $S_1(f)$, a form of $1/f$ noise with a Gaussian bump;
    \item N2: here $\beta(t)$ is stationary Gaussian coloured noise described by its autocorrelation matrix; chosen such that it is coloured, Gaussian and stationary (typically lower frequency) and is produced via convolving Gaussian white noise with a deterministic signal;
    \item N3: here the noise $\beta(t)$ is non-stationary Gaussian coloured noise, again described by its autocorrelation matrix which is chosen such that it is coloured, Gaussian and non-stationary. The noise is simulated via multiplication of a deterministic time-domain signal with stationary noise;
    \item N4: in this case, the noise $\beta(t)$ is described by its autocorrelation matrix chosen such that it is coloured, non-Gaussian and non-stationary. The non-Gaussianity of the noise is achieved via squaring the Gaussian noise so as to achieve requisite non-linearities;
    \item N5: a noise described by its power spectral density (PSD) $S_5(f)$, differing from N1 only via the location of the Gaussian bump; and
    \item N6: this profile is to model a noise source that is correlated to one of the other five sources (N1 - N5) through a squaring operation. If the $\beta(t)$ is the realization of one of the five profiles, N6 will have realisations of the form $\beta^2(t)$. This profile is used for multi-axis and multi-qubit systems. 
\end{itemize}
The N1 and N5 profiles can be generated following the method described in  \cite{youssry_characterization_2020} (see the section entitled ``Implementation'' onwards). Regarding the other profiles, any standard numerical package can generate white Gaussian stationary noise. The QDataSet noise realisations were encoded using the Numpy package of Python. We deliberately did so in order to avoid various assumptions used in common quantum programming packages, such as Qutip. To add colouring, we convolve the time-domain samples of the noise with some signal. To generate non-stationarity, we multiply the time-domain samples by a some signal. Finally, to generate non-Gaussianity, we start with a Gaussian noise and apply non-linear transformation such as squaring. The last noise profile is used to model the case of two noise sources that are correlated with each other. In this case we generate the first one using any of the profiles N1-N5, and the other source is completely determined. 

\subsubsection{Noise profile details} \label{sec:qdata:Noise profile details}
Following on from discussion of noise spectral density and open quantum systems (see section \ref{sec:quant:Noise and spectral density}), we specify three primary noise profiles for the construction of the QDataSet below (see \cite{youssry_beyond_2020}). 
\begin{enumerate}
    \item \textit{Single-axis noise; orthogonal pulse}. For a qubit with single-axis noise and orthogonal control pulses, the Hamiltonian is given by
\begin{align}
    H = \frac{1}{2}\left(\Omega + \beta_z(t)\right)\sigma_z + \frac{1}{2}f_x(t) \sigma_x.
\end{align}
With $z$-axis noise PSD:
\begin{align}
    S_Z(\omega) = \begin{cases}
    \frac{1}{\omega+1}+0.8e^{-\frac{(\omega-20)^2}{10}}             & 0 < \omega \le 50 \\
    0.25+0.8e^{-\frac{(\omega-20)^2}{10}}                      & f > 50
    \end{cases}
\end{align}
\item \textit{Multi-axis noise; two orthogonal pulses}. For single-qubit multi-axis noise and two orthogonal control pulses, the Hamiltonian is: 
\begin{align}
    H = \frac{1}{2} \left(\Omega + \beta_z(t)\right) \sigma_z + \frac{1}{2} \left( f_x(t) + \beta_x(t) \right) \sigma_x + \frac{1}{2} f_y(t) \sigma_y
\end{align}
with $x$-axis noise PSD:
\begin{align}
    S_X(\omega) = \begin{cases}
    \frac{1}{(\omega+1)^{1.5}}+0.5e^{-\frac{(\omega-15)^2}{10}}          & 0 < \omega \le 20 \\
    (5/48)+0.5e^{-\frac{(\omega-15)^2}{10}}                        & \omega > 20.
    \end{cases}
\end{align}
Here $z$-axis noise is as per the first case.
\item For the noiseless qubit with two orthogonal pulses, the Hamiltonian is:
\begin{align}
    H = \frac{1}{2} \Omega \sigma_z + \frac{1}{2} f_x(t) \sigma_x + \frac{1}{2} f_y(t) \sigma_y.
\end{align}
\end{enumerate}

%==============connection to master equations
We examine the relationship to master equation formalism in section \ref{sec:quant:Noise and quantum evolution} and the Lindblad master equation (\ref{eqn:quant:lindbladmasterequation}). Consider a single qubit with single-axis noise profile $\beta_z(t)$ along the $z$-axis and orthogonal control pulses $f_x(t)$ applied along the $x$-axis. The Hamiltonian is:
\begin{align}
    H = \frac{1}{2}\left(\Omega + \beta_z(t)\right)\sigma_z + \frac{1}{2}f_x(t) \sigma_x.
\end{align}
To derive the Lindblad master equation, first we look for factors affecting system evolution in terms of unitary and non-unitary dynamics: (a) unitary evolution driven by the deterministic part of the Hamiltonian and (b) non-unitary evolution resulting from the interaction with the environment. Thus $\beta_z(t)$ is the environmental interaction term which is ultimately related to the decoherence rate $\gamma_z$ via the power spectral density $S_Z(\omega)$ (see equation (\ref{eqn:quant:gammakdecoherencerateSomega}) and section \ref{sec:quant:Noise and decoherence} for more detail). To identify the Lindblad operators, we note that the noise is along the $z$-axis such that $L_k = \sqrt{\gamma_z} \sigma_z$, where $\gamma_z$ is the decoherence rate, related as per above related to the noise PSD. The Lindblad master equation is then:
\begin{align}
    \frac{d\rho}{dt} = -i\left[\frac{1}{2}\left(\Omega \sigma_z + f_x(t) \sigma_x\right), \rho\right] + \left( \sqrt{\gamma_z} \sigma_z \rho \sqrt{\gamma_z} \sigma_z - \frac{1}{2} \{ \gamma_z \sigma_z^2, \rho \} \right).
\end{align}
Noting $\sigma_z^2 = I$ (the identity), the dissipative term simplifies to:
\begin{align}
    \sqrt{\gamma_z} \sigma_z \rho \sqrt{\gamma_z} \sigma_z - \frac{1}{2} \gamma_z \{ I, \rho \} = \gamma_z (\sigma_z \rho \sigma_z - \rho).
\end{align}
such that the Lindblad master equation for our system under $z$-axis noise is:
\begin{align}
    \frac{d\rho}{dt} = -i\left[\frac{1}{2}\left(\Omega \sigma_z + f_x(t) \sigma_x\right), \rho\right] + \gamma_z (\sigma_z \rho \sigma_z - \rho).
\end{align}
Thus we can see the link between the measurable characteristics of environmental noise and the theoretical description of its effects on quantum systems via equation (\ref{eqn:quant:lindbladmasterequation}).

%==============connection to master equations

\subsubsection{Distortion}\label{sec:qdata:Distortion}
In physical experiments, the control pulses are physical signals (such as microwave pulses), which propagate along cables and get processed by different devices. This introduces distortions which cannot be avoided in any real devices. However, by properly engineering the systems, the effects of these distortions can be minimized and/or compensated for in the design. In this work, we used a linear-time invariant system to model distortions of the control pulses, and the same filter is used for all datasets. We chose a Chebychev analogue filter \cite{popescu_chebyshev_2016} with an undistorted control signal is the input and distorted the filter output signal. Table (\ref{tab:params}) sets out a summary of key parameters.

\subsubsection{QDataSet noise operators} \label{sec:qdata:QDataSet noise operators}
In developing the QDataSet, we have assumed that the environment affecting the qubit is classical and stochastic, namely that $H_1(t)$ will be a stochastic term that acts directly on the system. The stochasticity of $H_1(t)$ means that the expectation of any observable measured experimentally will be given as:
\begin{align}
    \braket{O}_c = \braket{\Tr{(\rho(T) O)}}_c,    
\end{align}
where $O$ represents the measurement operator corresponding to the observable of interest (e.g. $O \dot{=} M_m$ in notation above) and the $\braket{\cdot}_c$ is a classical expectation over the distribution of the noise realizations. It can then be shown (see \cite{youssry_characterization_2020}) that this can be expressed in term of the initial state $\rho(0)$, and the evolution fixed over the time interval $[0,T]$ as:
\begin{align}
    \braket{O(T)} = \Tr{\left(V_O(T) U_0^{\dagger}(T) \rho(0) U_0(T) O\right)}
    \label{eqn:voeqn}
\end{align}
where $U_0(T) = \mathcal{T}_{+} e^{-i\int_0^T H_0(t) dt}$ is the evolution matrix in the absence of noise and: 
\begin{align}
    V_O(T) &= \braket{W_O(T)}_c\\
        &\doteq \braket{O^{-1}\tilde{U}_I^{\dagger}(T) O \tilde{U}_I(T)}_c.
\end{align}
is a novel noise operator introduced in \cite{youssry_characterization_2020} which characterises the expectation of noise relevant to synthesising counteracting control pulses. We encapsulate the full quantum evolution via the operator $W_O(T)$. Note that $V_O$ is formed via the partial tracing out of the effect of the noise and its interaction with the control pulses, so encodes only those features of importance or relevance to impact of noise (not the full noise spectrum). Importantly, the use of the $V_O$ operator is designed to allow information about noise and observables to be separated (a hallmark of dynamic decoupling approaches). Thus of particular importance is the assumption (see section \ref{sec:quant:Noise and spectral density}) that noise is bounded by an upper frequency limit i.e. $|\omega| \leq \Omega_0$. 
The modified interaction unitary $\tilde{U}_I(T)$ is defined such that:
\begin{align}
    U(T) = \tilde{U}_I(T) U_0(T),
\end{align}
where $U(T)= \mathcal{T}_{+} e^{-i\int_0^T H(t) dt} $ is the full evolution matrix. This contrasts the conventional definition of the interaction unitary which takes the form $U(T) = U_0(T) U_I(T) $. The $V_O$ operator is used in the simulation code for the QDataSet to characterise the effect of noise on such values. Ideally, in a noise-free scenario, those expectations should tend to zero (representative of the absence of noise). The idea for including such noise operators is that this data can then be input into machine learning models to assist the algorithms to learn appropriate, for example, pulse sequences or controls that send $V_O \to I$ (neutralising the noise). 

A detailed explanation and example of the use of the $V_O$ operator is provided in \cite{youssry_characterization_2020}. For machine learning practitioners, the operator $V_O$ may, for example, be used in an algorithm that seeks to negate the effect of $V_O$. The utility of this approach is that full noise spectroscopy is not required.

\subsection*{QDataSet Measurement Methodology} \label{sec:qdata:QDataSet Measurement Methodology}

\subsubsection{QDataSet Measurements} \label{sec:qdata:QDataSet Measurements}
The simulated quantum measurements of the QDataSet are inherently probabilistic. As set out in section \ref{sec:quant:Measurement}, measurement of quantum systems yields an underlying probability distribution over the possible measurement outcomes (observables) $m$ of the system which are in turn determined by the state of the system and the measurement process.  There are several ways to describe quantum measurements mathematically. The most common method (which we adopted) involves projective measurements. In this case, an observable $O$ is described mathematically by a Hermitian operator. The eigendecomposition of the operator can be expressed in the form $O= \sum_{m} m P_m$, where $m$ are the eigenvalues, and $P_m$ (sometimes denoted $M_m$ to emphasise their role in measurement) are the associated projectors (definition \ref{defn:quant:projection_operators}) into the corresponding eigenspace. The projectors $P_m$ must satisfy that $P_m^2 = P_m$, and that $\sum_m P_m = I$ (the identity operator), to ensure we get a correct distribution for the outcomes. In more sophisticated treatments, $O$ may belong to POVM described above which partition the Hilbert space $\mathcal{H}$ into distinct projective subspaces $\mathcal{H}_m$ associated with each POVM operator $O$ (see definition \ref{defn:quant:POVM} and section \ref{sec:quant:POVMs and Kraus Operators}). The probability of measuring an observable is given by:
\begin{align}
    \Pr(m) = \Tr(\rho P_m),
\end{align}
for a system in the state $\rho$ (equation (\ref{eqn:quant:measurementdensitymatrixtrace})). The expectation value of the observable is given by equation:
\begin{align}
    \braket{O}= \Tr{(\rho O)} = \Tr{\left(\rho \sum_m m P_m\right)} = \sum_m m \Pr(m).
\end{align}
As detailed below, the QDataSet contains measurement outcomes for a variety of noiseless and noisy systems. The measurement operators chosen are the Pauli operators described below and the QDataSet contains the expectation values for each Pauli measurement operator. In a classical machine learning context, these measurement statistics form training data labels in optimisation problems, such as designing algorithms that can efficiently sequence control pulses in order to efficiently (time-minimally) synthesise a target state or unitary (and thus undertake a quantum computation) of interest.

\subsubsection{Pauli matrices} \label{sec:qdata:Pauli matrices}
The set of measurement operators for the QDataSet is the set of Pauli operators which are important operators in quantum information processing involving qubit systems. This is in part because such qubit systems can be usefully decomposed into a Pauli operator basis via the Pauli matrices:
\begin{align}
    \sigma_x = \begin{pmatrix}0 & 1 \\ 1 & 0\end{pmatrix}, \sigma_y = \begin{pmatrix}0 & -i \\ i & 0\end{pmatrix}, \sigma_z &= \begin{pmatrix}1 & 0 \\ 0 & -1\end{pmatrix} \label{eqn:qdata:paulimatrices}
\end{align}
together with the identity (denoted $\sigma_0$). Pauli operators are Hermitian (with eigenvalues $+1$ and $-1$), traceless and satisfy that $\sigma_i^2 = I$. Together with the identity matrix (which is sometime denoted by $\sigma_0$), they form an orthonormal basis (with respect to the Hilbert-Schmidt product defined as $\braket{A,B} = \Tr{(A^{\dagger}B)}$) for any $2 \times 2$ Hermitian matrix. QDataSet qubit states can then be expressed in this basis via the density matrix:  
\begin{align}
    \rho = \frac{1}{2}\left(I + \mathbf{r}\cdot \sigma \right)
\end{align}
where the vector $\mathbf{r}=(r_x, r_y, r_z)$ is a unit vector called the Bloch vector, and the vector $\sigma=(\sigma_x, \sigma_y, \sigma_z)$. The dot product of these two vectors is just a shorthand notation for the expression $\mathbf{r}\cdot \sigma= r_x\sigma_x + r_y\sigma_y+ r_z \sigma_z$. As such, a time-dependent Hamiltonian of a qubit can be expressed as 
\begin{align}
    H(t) = \sum_{i\in\{x,y,z\}} {\alpha_i(t) \sigma_i}
\end{align}
with the time-dependence absorbed in the coefficients $\alpha_i(t)$. 

\subsubsection{Pauli measurements} \label{sec:qdata:Pauli measurements}
The measurements simulated in the QDataSet are Pauli measurements. These are formed by taking the expectation value of each Pauli matrix e.g. $\Tr{(\rho \sigma_i)}$ for $i \in \{x,y,z\}$ (the identity is omitted). The resultant measurement distributions will typically form labelled data in a machine learning context. Measurement distributions are ultimately how various properties of the quantum system are inferred (i.e. via reconstructive inference), such as the characteristics of quantum circuits, evolutionary paths and tomographical quantum state description. As we describe below, measurements in the QDataSet comprise measurements on each eigenstate (six in total) of each Pauli operator by all Pauli operators. Hermitian operators have a spectral decomposition in terms of eigenvalues and their corresponding projectors
\begin{align}
    P_0 &= \ketbra{0}{0} = \frac{1}{2}(I + \sigma_z) = \begin{pmatrix} 
    1 & 0 \\
    0 & 0
    \end{pmatrix}\\
    P_1 &= \ketbra{1}{1} = \frac{1}{2}(I - \sigma_z) = \begin{pmatrix} 
    0 & 0 \\
    0 & 1
    \end{pmatrix}
\end{align}
thus we can write:
\begin{align}
\sigma_z = 1 \times \begin{pmatrix} 
    1 & 0 \\
    0 & 0
    \end{pmatrix}  - 1 \times \begin{pmatrix} 
    0 & 0 \\
    0 & 1
    \end{pmatrix}  
\end{align}
For example, a Pauli measurement on a qubit in the $-1$ eigenstate with respect to the $\sigma_z$ operator:
\begin{align}
    \Tr( \rho \sigma_z^\dagger) =   \Tr \left( \begin{pmatrix} 
    0 & 0 \\
    0 & 1
    \end{pmatrix}
    \begin{pmatrix} 
    1 & 0 \\
    0 & -1
    \end{pmatrix} \right) = \Tr \begin{pmatrix} 
    0 & 0 \\
    0 & -1
    \end{pmatrix} = -1
\end{align}
which is as expected. The probability of observing $\lambda=-1$ in this state we should expect to be unity (given the state is in the eigenstate):
\begin{align}
   Pr(m=-1) = \Tr (P_1^2 \rho) = 1
\end{align}
For $n$-qubit systems (such as two-qubit systems in the QDataSet), Pauli measurements are represented by tensor-products of Pauli operators. For example, a $\sigma_z$ measurement on the first qubit and $\sigma_x$ on the second is represented as:
\begin{align}
    \sigma_z \otimes \sigma_x
\end{align}
In programming matrix notation, this becomes represented as a $4 \times 4$ matrix (tensor):
\begin{align}
    \sigma_z \otimes \sigma_X &=\begin{pmatrix} 
    1 & 0 \\
    0 & -1
    \end{pmatrix} \otimes \begin{pmatrix} 
    0 & 1 \\
    1 & 0
    \end{pmatrix} \\
    &=\begin{pmatrix} 
    1 \times \begin{pmatrix} 
    0 & 1 \\
    1 & 0
    \end{pmatrix}  & 0 \\
    0 & -1 \times \begin{pmatrix} 
    0 & 1 \\
    1 & 0
    \end{pmatrix} 
    \end{pmatrix} \\
    &= \begin{pmatrix} 
    0 & 1 & 0 & 0\\
    1 & 0 & 0 & 0\\
    0 & 0 & 0 & -1\\
    0 & 0 & -1 & 0
    \end{pmatrix} 
\end{align}

\begin{figure}[ht]
\centering
\includegraphics[width=0.5\linewidth]{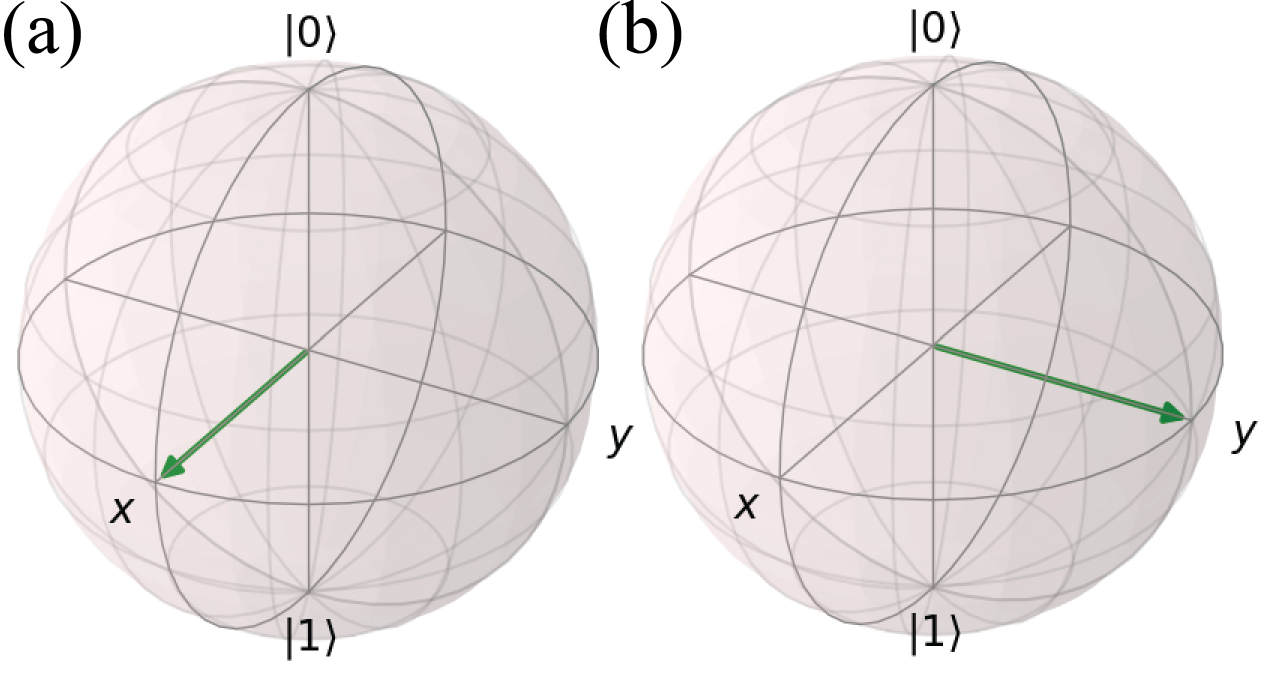}
\caption{An example of a quantum state rotation on the Bloch sphere. The $\kz,\ko$ indicates the $\sigma_z$-axis, the $X$ and $Y$ the $\sigma_x$ and $\sigma_y$ axes respectively. In (a), the vector is residing in a $+1$ $\sigma_x$ eigenstate. By rotating about the $\sigma_z$ axis by $\pi/4$, the vector is rotated to the right, to the $+1$ $\sigma_y$ eigenstate. A rotation about the $\sigma_Z$ axis by angle $\theta$ is equivalent to the application of the unitary $U(\theta) = \exp(-i \theta_z \sigma_z/2)$.}
\label{fig:blochrotation}
\end{figure}

The Pauli representation of qubits used in the QDataSet can be usefully visualised via the Bloch sphere as per Figure \ref{fig:blochrotation}. The axes of the Bloch sphere are the expectation values of the Pauli $\sigma_x, \sigma_y$ and $\sigma_z$ operators respectively. As each Pauli operator has eigenvalues $1$ and $-1$, the eigenvalues can be plotted along axes of the 2-sphere. For a pure (non-decohered) quantum state $\rho$, $|\rho|=\sqrt{r_x^2 + r_y^2 + r_z^2}=1$ (as we require $\Tr{\rho^2=1}$), thus $\rho$ is represented on the Bloch 2-sphere as a vector originating at the origin and lying on the surface of the Bloch 2-sphere. The evolution of the qubit i.e. a computation according to unitary evolution can then be represented as rotations of $\rho$ across the Bloch sphere. In noisy contexts, decohered $\rho$ are represented whereby $|\rho|<1$ i.e. the norm of $\rho$ shrinks and $\rho$ no longer resides on the surface.

For machine learning practitioners, it is useful to appreciate the operation of the QDataSet Pauli operators $\sigma_x, \sigma_y, \sigma_z$ as the generators of rotations about the respective axes of the Bloch sphere. Represented on a Bloch sphere, the application of $\sigma_z$ to a qubit is equivalent to rotating the quantum state vector $\rho$ about the $z$-axis (see Figure \ref{fig:blochrotation}). Conceptually, a qubit is in a $z$-eigenstate if it is lying directly on either the north ($+1$) or south ($-1$) pole. Rotating about the $z$-axis then is akin to rotating the vector on the spot, thus no change in the quantum states (or eigenvalues) for $\sigma_z$ occurs because the system exhibits symmetry under such transformations. This is similarly the case for $\sigma_x, \sigma_y$ generators with respect to their eigenvalues and eigenvectors. However, rotations by $\sigma_\alpha$ will affect the eigenvalues/vectors in the $\sigma_\beta$ basis where $\alpha \neq \beta$ e.g. a $\sigma_x$ rotation will affect the component of the qubit lying along the $\sigma_z$ axis. Similarly, a $\sigma_z$ rotation of a qubit in a $\sigma_x$ eigenstate will alter that state (shown in (a) and (b) of Figure \ref{fig:blochrotation}). An understanding of Pauli operators and conceptualisation of qubit axes is important to the understanding of the simulated QDataSet. An understanding of symmetries of relevance to qubit evolution (and quantum algorithms) is also beneficial. As we describe below, controls or noise are structured to be applied along particular axes of a qubit and thus can be thought of as a way to control or distortions upon the axial rotations of a qubit effected by the corresponding Pauli generator.

There exist higher dimensional generalization to the Pauli matrices that allow forming orthonormal basis to represent operators in these dimensions. In particular if we have a system of $N$ qubits, then one simple generalization is to form the set $\{\sigma_{i_1}^{(1)} \otimes \sigma_{i_2}^{(2)} \otimes \cdots \sigma_{i_N}^{(N)} \}_{i_j \in \{0,x,y,z\}}$. In other words we take tensor products of the Pauli's which gives a set of size $4^N$. For example, for a two-qubit system we can form the $16-$ element set $\{\sigma_0 \otimes \sigma_0, \sigma_0 \otimes \sigma_x, \sigma_0 \otimes \sigma_y, \sigma_0 \otimes \sigma_z, \sigma_x \otimes \sigma_0, \sigma_x \otimes \sigma_x, \sigma_x \otimes \sigma_y, \sigma_x \otimes \sigma_z, \sigma_y \otimes \sigma_0, \sigma_y \otimes \sigma_x, \sigma_y \otimes \sigma_y, \sigma_y \otimes \sigma_z, \sigma_z \otimes \sigma_0, \sigma_z \otimes \sigma_x, \sigma_z \otimes \sigma_y, \sigma_z \otimes \sigma_z\}$. Moreover, for many use cases, we are interested in the \textit{minimal} number of operators, such as Pauli operators, required to achieve a requisite level of control, such as universal quantum computation.  

For the single qubit system, initial states are the two eigenstates of each Pauli operator. As noted above, the quantum state can be decomposed in the Pauli basis as $\rho_j =\frac{1}{2}(I\pm \sigma_j)$, for $j=1,2,3$. This gives a total of 6 states. We perform the three Pauli measurements on each of these states, resulting in a total of 18 possible combinations. These 18 measurements are important to characterize a qubit system. For two-qubits, it will be similar but now we initialize every individual qubit into the 6 possible eigenstates, and we measure all 15 Pauli observables (we exclude identity). This gives a total of 540 possible combinations. 

\subsubsection{Monte Carlo measurements} \label{sec:qdata:Monte Carlo measurements}

Measurements of the one- and two-qubit systems for the QDataSet are undertaken using Monte Carlo techniques. This means that a random Pauli measurement is undertaken multiple times, with the measurement results averaged in order to provide the resultant measurement distribution for each of the operators. The measurement of the quantum systems is contingent on the noise realisations for each system. For the noiseless case, the Pauli measurements are simply the Monte Carlo averages (expectations) of the Pauli operators. Systems with noise will have one or more noise realisations (applications of noise) applied to them. To account for this, we include two separate sets of measurement distribution. The first the expectation value of the three Pauli operators over all possible initial states for each different noise realisation. These statistics are given by the set $\{ V_O \}$ in the QDataSet. Thus for each type of noise, there will be a set of measurement statistics. The second is a set of measurement statistics where we average over all noise realisations for the dataset. This second set of measurements is given by the set $\{E_O \}$. Including both sets of measurements enables algorithms trained using the QDataSet to be more fine-grained in their treatment of noise: in some contexts, while noise profiles may be uncertain, it is clear that the noise is of a certain type, so the first set of measurement statistics may be more applicable. For other cases, there is almost no information about noise profiles or their sources, in which case the average over all noise realisations may be more appropriate. 

\subsubsection{Monte Carlo Simulator} \label{sec:qdata:Monte Carlo Simulator}
For the benefit of researchers using the QDataSet, we briefly set out a bit more detail of how the datasets were generated. The simulator comprises three main components. The first approximates time-ordered unitary evolution. The second component generates realisations of noise given random parametrisations of the power spectral density (PSD) of the noise. The third component simulates quantum measurement. The simulations are based upon Monte Carlo methods whereby $K$ randomised pulse sequences give rise to noise realisations. The quantum systems are then measured to determine the extent to which the noise realisations affect the expectation values. Trial and error indicated a stabilisation of measurement statistics at around $K=500$, thus $K \geq1000$ was chosen for the final simulation run to generate the QDataSet. The final Pauli measurements are then averages over such noise realisations. The parameter $K$ is included for each dataset and example (as described below). For more detail, including useful pseudocode that sets out the relationship between noise realisations, $\beta(t)$ and measurement, see supplementary material in \cite{youssry_characterization_2020} (which we set out for completeness in section \ref{sec:qdata:Monte Carlo algorithm}).

\section{Simulation Results} \label{sec:qdata:Simulation Results}

\subsection{QDataSet form} \label{sec:qdata:QDataSet form}
Quantum information in the QDataSet is stored following the Tensorflow convention of interpreting multidimensional arrays. For example the noise Hamiltonian for one example is stored as a $(1, M , K , 2 , 2)$ array, where the first dimension is the batch, the second is time assuming $M$ steps, then whatever comes next is related to the object itself. In this case the third dimension denotes the noise realization assuming a maximum of $K$ realizations, and the last two dimensions ensure we have a square matrix of size 2. The simulation of the datasets is based on a Monte Carlo method, where a number of evolution trajectories are simulated and then averaged to calculate the observables. The exact details can be found in \cite{youssry_characterization_2020} (see ``Implementation'' section) which we reproduce and expand upon in this work for completeness. 

\subsection{QDataSet parameters} \label{sec:qdata:QDataSet parameters}
Further detail regarding the 52 datasets (including code for simulations) that we present in this work for use solving engineering applications discussed in section \ref{sec:qdata:Usage Notes} using classical machine learning can be found on the repository for the QDataSet \cite{perrier_qdataset_2022}. Table (\ref{table:datasetcharacteristics}) sets out the taxonomy of each of the 52 different datasets. Each dataset comprises 10,000 examples that are compressed into a Pickle file which is in turn compressed into a zip file. The \textit{Item} field indicates the dictionary key and the \textit{Description} field indicates the dictionary value.

\subsection{Greybox Algorithms} \label{sec:qdata:Greybox Algorithms}
The \texttt{simulation.py} module is responsible for simulating a quantum system in the presence of noise. This module is constructed using the TensorFlow framework and consists of the following classes (background in relation to which is set out in section \ref{sec:ml:Greybox machine learning}) above. We set out a brief description of key components of the code below:

\begin{itemize}
    \item \textit{Noise\_Layer}: An internal class dedicated to the generation of noise within the simulation as set out in the Monte Carlo method in \cite{youssry_characterization_2020}.
    
    \item \textit{HamiltonianConstruction}: An internal Python class designed for constructing the Hamiltonians necessary for the quantum system's dynamics.
    
    \item \textit{QuantumCell}: This internal Python class is essential for realizing the time-ordered evolution of the quantum system. 
    
    \item \textit{QuantumEvolution}: This Python class implements the time-ordered quantum evolution, ensuring the system evolves in accordance with the specified Hamiltonians.
    
    \item \textit{Quantum Measurement}: An internal Python class that models the effects of coupling losses at the output of the quantum system, simulating the process of quantum measurements.
    
    \item \textit{$V_O$ Layer}: An internal Python class used to calculate the Vo operator (see section \ref{sec:qdata:QDataSet noise operators} above).
    
    \item \textit{quantumTFsim}: This is the principal class in the Python module. It defines the machine learning model for the qubit, seamlessly integrating quantum simulation with the power of TensorFlow.
\end{itemize}

\subsection{Datasets and naming convention}\label{sec:qdata:Datasets and naming convention}
Each dataset can be categorised according to the number of qubits in the system and the noise profile to which the system was subject. Table (\ref{tab:cats}) sets out a summary of such categories. While other types of profiles or combinations could have been utilised, our aim was to select categories which reflect the types of noise and categorisations relevant to experimental laboratories working on problems such as quantum computation. For category 1 of the datasets, we created datasets with noise profiles N1, N2, N3, N4, together with the noiseless case. This gives a total of 5 datasets. For category 2,  the noise profiles for the X and Z axes respectively are chosen to be (N1,N5), (N1,N6), (N3,N6). Together with the noiseless case, this gives a total of 4 datasets. For category 3 (two-qubit system), we chose only the 1Z (identity on the first qubit, noise along the $z-$axis for the second) and Z1 (noise along the $z-$axis for the first qubit, identity along the second) noise to follow the (N1,N6) profile. This category simulates two individual qubits with correlated noise sources. For category 4, we generate the noiseless, (N1,N5), and (N1,N6) for the 1Z and Z1 noise. This gives 3 datasets. Therefore, the total number of datasets at this point is 13. Including the two types of control waveforms, this gives a total of 26. If we also include the cases of distortion and non-distorted control, then this gives a total of 52 datasets. Comprehensive detail on the noise profiles used to generate the datasets is set-out above.

We chose a convention for the naming of the dataset to try delivering as much information as possible about the chosen parameters for this particular dataset. The name is partitioned into 6 parts, separated by an underscore sign ``\_''. We explicate each part below: 
\begin{enumerate}
    \item The first part is either the letter ``G'' or ``S'' to denote whether the control waveform is Gaussian or square.
    \item The second part is either "1q" or ``2q'' to denote the dimensionality of the system (i.e. the number of qubits).
    \item The third part denotes the control Hamiltonian. It is formed by listing down the Pauli operators we are using for the control for each qubit, and we separate between qubit by a hyphen ``-''. For example, category 1 datasets will have ``X'', while category 4 with have ``IX-XI-XX''. 
    \item The fourth part and fifth parts indicate (i) the axis along which noise is applied (fourth part) and (ii) the type of noise along each axis (fifth part). So ``G\_2q\_IX-XI\_IZ-ZI\_N1-N6'' represents two qubits with control along the $x$ axis of each qubit, while the noise is applied along the $z$-axis of each. In this case, N1 noise is applied along the $z$-axis of the first qubit and N6 noise is applied along the $z$-axis of the second qubit. For datasets where no noise is applied, these two parts are omitted.
    \item Finally, the sixth part denotes the presence of control distortions by the letter ``D'', otherwise it is not included.
\end{enumerate}

For example, the dataset ``G\_2q\_IX-XI-XX\_IZ-ZI\_N1-N6'' is two qubit, Gaussian pulses with no distortions, local X control on each qubit and an interacting XX control along with local noise on each qubit with profile N1 on the first qubit $z$-axis and N6 on the second qubit $z$-axis. Another example the dataset ``S\_1q\_XY\_D", is a single-qubit system with square distorted control pulses along X and Y axis, and there is no noise.

%========technical validation

\section{Technical Validation} \label{sec:qdata:Technical Validation}

Technical validation of the QDataSet was undertaken by comparing the QDataSet data against leading quantum simulation toolkit \textit{Qutip}, an open-source software for simulating the dynamics of open quantum systems \cite{johansson_qutip_2013}. For each of the 26 different simulated systems (each comprising a noisy and noise-free case), the procedure set out below was adopted. A Jupyter notebook containing the code used for technical validation and verification of the datasets is available on the QDataSet repository.

\subsection{Distortion analysis} \label{sec:qdata:Distortion analysis}
\begin{figure}[ht]
\centering
\includegraphics[width=0.5\linewidth]{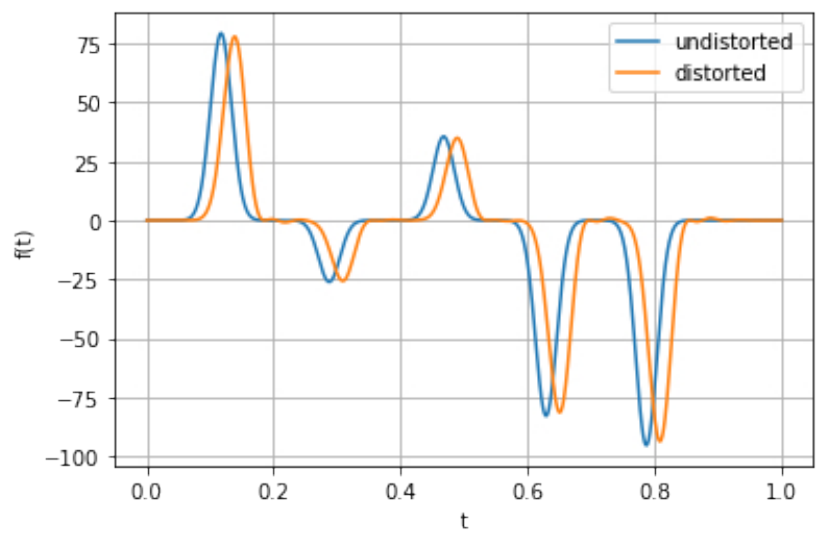}
\caption{Plot of an undistorted (orange) pulse sequence against a related distorted (blue) pulse sequence for the single-qubit Gaussian pulse dataset with $x$-axis control (`G\_1q\_X') over the course of the experimental runtime. Here $f(t)$ is the functional (Gaussian) form of the pulse sequence for time-steps $t$. These plots were used in the first step of the verification process for QDataSet. The shift in pulse sequence is consistent with expected effects of distortion filters. The pulse sequences for each dataset can be found in \textit{simulation\_parameters} $\implies$ \textit{dynamic\_operators} $\implies$ \textit{pulses} (undistorted) or \textit{distorted\_pulses} for the distorted case (see Table (\ref{table:datasetcharacteristics}) for a description of the dataset characteristics).}
\label{fig:pulses}
\end{figure}

\begin{figure}[ht]
\centering
\includegraphics[width=1\linewidth]{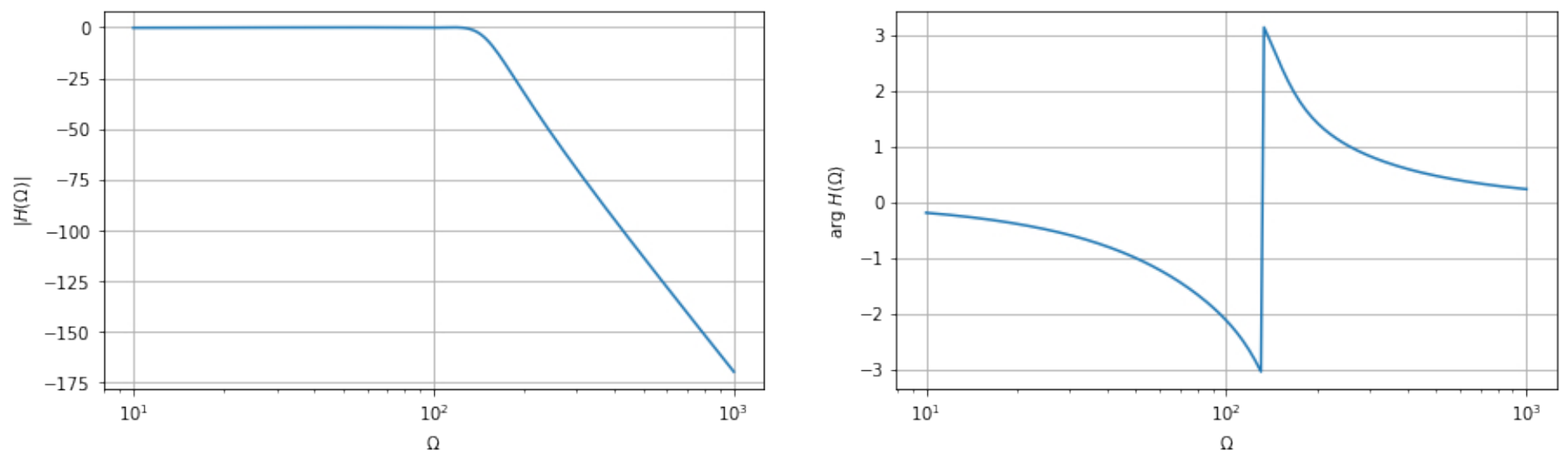}
\caption{The frequency response (left) and the phase response (right) of the filter that is used to simulate distortions of the control pulses. The frequency is in units of Hz, and the phase response is in units of rad.}
\label{fig:distortion}
\end{figure}

Firstly, for each of the one- and two-qubit datasets, the distorted and undistorted pulse sequences were compared for a sample of examples from each dataset in order to assess the effect of relevant distortion filters. A plot of the comparison for the single-qubit case with Gaussian control along the $x$-axis is presented in Figure (\ref{fig:pulses}). The plot compares the distorted sequence of control Gaussian control pulses for the undistorted case (blue) and distorted case (orange). The expectation was for a shift in the Gaussian pulse curves as a result of the distortion filters. An error with the datasets would have seen the distorted pulses not resemble such the undistorted pulses significantly and, moreover, would have likely seen non-Gaussian pulse forms. As can be seen from Figure (\ref{fig:pulses}), the distorted and undistorted pulses appear almost identical but for a shift and a minor amplitude ($f(t)$) reduction in the distorted case , which was seen for each example. This provided us with assurance that simulation was appropriately modelling distortion of the pulses. 
As a second assurance, we plotted the effect of the distortion filter (when applied) and evaluate the frequency response of the filter. The aim of this process was to identify visually whether the form frequency response $H(\Omega)$ and phase response $\Omega$ exhibit the appropriate form (i.e. no obvious errors). The verification plot is shown in Figure (\ref{fig:distortion}).

\subsection{Comparison with Qutip} \label{sec:qdata:Comparison with Qutip}
The second and primary technical validation of the QDataSet was undertaken by comparing mean expectation values of observables for subsamples of each of the datasets against the equivalent expectations for simulations and measurements undertaken in using Qutip \cite{johansson_qutip_2013}. To generate the Qutip equivalents, the equivalent parameters (e.g. Hamiltonian parameters, pulse parameters) were input into Qutip to generate the relevant outputs. For each dataset in the QDataSet, the verification procedure was run on varying samples. To undertake this process, we adopted two validation strategies.
\begin{itemize}
    \item \textit{Mean expectation of all observables over all noise realisations}. In this case, for a sample of examples from each dataset in the QDataSet, the mean expectation over all noise realisations for all observables (i.e. measurements) was compared against the same mean measurements for the equivalent simulation generated in Qutip. This was done for the noiseless and noisy case. The two means were then compared. On average the error (difference between the means) of the order $10^{-06}$, demonstrating the equivalence of the QDataSet simulation code with that from Qutip. 
    \item \textit{Mean expectation of single observables over separate noise realisations}. In the second case, the mean expectation over all noise realisations for each separate observable was compared against the same mean measurements for the equivalent simulation generated in Qutip. Again, this was done for the noiseless and noisy case. Comparison of the two means showed that on average the error (difference between the means) of the order $10^{-07}$, in turn demonstrating the equivalence of the QDataSet simulation code with that from Qutip.  
\end{itemize}

\section{Usage Notes}\label{sec:qdata:Usage Notes}

\subsection*{Overview}
In this section, we include further usage notes related to the 52 QDataSet datasets based on simulations of one- and two-qubit systems evolving in the presence and/or absence of noise subject to a variety of controls. Recall that the QDataSet has been developed primarily for use in training, benchmarking and competitive development of classical and quantum algorithms for common tasks in quantum control, quantum tomography and noise spectroscopy. It has been generated using customised code drawing upon base-level Python packages in order to facilitate interoperability and portability across common machine learning and quantum programming platforms. Each dataset consists of 10,000 samples which in turn comprise a range of data relevant to the training of machine learning algorithms for solving optimisation problems. The data includes a range of information (stored in list, matrix or tensor format) regarding quantum systems and their evolution, such as: quantum state vectors, drift and control Hamiltonians and unitaries, Pauli measurement distributions, time series data, pulse sequence data for square and Gaussian pulses and noise and distortion data.

Researchers can use the QDataSet in a variety of ways to design algorithms for solving problems in quantum control, quantum tomography and quantum circuit synthesis, together with algorithms focused on classifying or simulating such data. We also provide working examples of how to use the QDataSet in practice and its use in benchmarking certain algorithms. Each part below provides in-depth detail on the QDataSet for researchers who may be unfamiliar with quantum computing, together with specifications for domain experts within quantum engineering, quantum computation and quantum machine learning.
 
The aim of generating the datasets is threefold: (a) simulating typical quantum engineering systems, dynamics and controls used in laboratories; (b) using such datasets as a basis to train machine learning algorithms to solve certain problems or achieve certain objectives, such as attainment of a quantum state $\rho$, quantum circuit $U$ or quantum control problem generally (among others); and (c) enable optimisation of algorithms and spur development of optimised algorithms for solving problems in quantum information, analogously with the role of large datasets in the classical setting. We explain these use cases in more detail below:
\begin{enumerate}
    \item \textit{Datasets as simulations.} Firstly, we have aimed to generate datasets which abstractly simulate the types of data, characteristics and features which would be commonly used in laboratories and experimental setups. Each dataset is an abstractions (say of particular Hamiltonians, or noise profiles) which can have any number of physical \textit{realisations} depending on the experimental design. So different experiments can ultimately realise, in the abstract the same or a sufficiently similar structure as that provided by the data. This is an important design choice relating to how the QDataSet is intended to be used. For example, the implementation of the particular Hamiltonians or state preparation may be done using trapped-ion setups, quantum dot or transmon-based qubits \cite{dewes_characterization_2012}, doped systems or otherwise. We assume the availability of a mapping between the dataset features, such as the controls pulses, and particular control devices (such as voltage or microwave-based controls), for example, in the laboratory. 
    \item \textit{Training algorithms using datasets.} The second use case for the QDataSet is related but distinct from the first. The aim is that training models using the datasets has applicability to experimental setups. Thus, for example, a machine learning model trained using the datasets in theory should provide, for example, the optimal set of pulses or interventions needed to solve (and, indeed, optimise) for some objective. It is intended that the output of the machine learning model is an abstraction which can then be realised via the specific experimental setup. The aim then is that the abstraction of each experiments setup allows the application of a variety of machine learning models for optimising in a way that is directly applicable to experimental setups, rather than relying upon experimentalists to then work-out how to translate the model's output into their particular experimental context. Requiring conformity of outputs within these abstract criteria thus facilitates a greater, practical, synthesis between machine learning and the implementation of solutions and procedures in experiments.
    \item \textit{Benchmarking, development and testing.} The third primary use of the datasets is to provide a basis for benchmarking, development and testing of existing and new algorithms in quantum machine learning for quantum control, tomography and related to noise mitigation. As discussed above, classical machine learning has historically been characterised by the availability of large-scale datasets with which to train and develop algorithms. The role of these large datasets is multifaceted: (i) they provide a means of \textit{benchmarking} algorithms (see above), such that a common set of problem parameters, constraints and objectives allows comparison among different models; (ii) their size often means they provide a richer source of overt and latent (or constructible) features which machine learning models may draw upon, improving the versatility and diversity of models which may be usefully trained. The aim of the QDataSet is then that it can be used in tandem by researchers as benchmarking tool for algorithms which they may wish to apply to their own data or experiments.
\end{enumerate}

\subsection{QDataSet Control Sets} \label{sec:qdata:QDataSet Control Sets}
For some machine learning applications, we are interested in minimising the number of controls that must be applied to a quantum system (thus minimising the resources required to control the system). In such cases, we may seek a minimal control algebra or gate set. For example, the minimal number of Pauli operators required to achieve a complete control set of generators for synthesising an arbitrary unitary acting on $n$-qubits in a $2^n$ dimensional Hilbert space is given by a bracket-generating set (distribution) $\Delta \subseteq \mathfrak{su}(2^n)$  \cite{montgomery_tour_2002, perrier_quantum_2020} (see discussion in Appendix \ref{chapter:Background: Geometry, Lie Algebras and Representation Theory}) which can be understood in more complex treatments in the context of Lie algebras and representation theory (the subject of Chapter \ref{chapter:Time optimal quantum geodesics using Cartan decompositions} in particular). Here $\mathfrak{su}(2^n)$ represents the Lie algebra corresponding to the Pauli group SU$(2^n)$, the complete set of generators required to span the $n$ dimensional Hilbert space in the Pauli basis. Given $\Delta$, we can reconstruct the full Pauli basis via the operation of the Lie bracket \cite{swaddle_generating_2017} (though noting this may require additional algorithmic design choices involving ancilla for 2-local operations in the presence of unitary symmetries as per \cite{marvian_restrictions_2022}). The QDataSet generators for one- and two-qubit systems are simply one and two (tensor-product) sets of Paulis respectively (i.e. not the minimal set $\Delta$). For higher-dimensional problems, whether to restrict generators to those within $\Delta$ becomes a consideration for machine learning architectures (see literature on time-optimal quantum control such as \cite{dalessandro_introduction_2007}). These generators will typically be used as the tensors or matrices to which classical controls are applied within machine learning architectures. We explore these questions in other Chapters, especially in the geometric and algebraic context of subRiemannian quantum control and also classical Pontryagin-based control theory.

%========using dataset

\section{Machine learning using the QDataSet} \label{sec:qdata:Machine learning using the QDataSet}

There are many problems related to the characterization and control of quantum systems that can be solved using machine learning techniques. In this section, we give an overview of a number of such problems and how to approach them using machine learning. We provide a brief overview of the different types of problems in quantum computing, engineered quantum systems and quantum control for which the QDataSet and algorithms trained using it may be useful. The list is necessarily non-exhaustive and is intended to provide some direction mainly to machine learning researchers unfamiliar with key problems in applied quantum science.

\subsection{Benchmarking} \label{sec:qdata:Benchmarking}
 Benchmarking algorithms using standardised datasets is an important developmental characteristics of classical machine learning. Benchmarks provide standardised datasets, preprocessing protocols, metrics, architectural features (such as optimisers, loss functions and regularisation techniques) which ultimately enable research communities to precisify their research contributions and improve upon state of the art results.  Results in classical machine learning are typically presented by comparison with known benchmarks in the field and adjudged by the extent to which they outperform the current state of the art benchmarks. Results are presented in tabular format with standardised metrics for comparison, such as accuracy, F1-score or AUC/ROCR statistics. The QDataSet has been designed with these metrics in mind. For example, a range of classical or quantum statistics (e.g. fidelity) can be used to benchmark the performance of algorithms that use the datasets in training. The role of benchmarking is important in classical contexts. Firstly, it enables a basis for researchers across machine learning subdisciplines to gauge the extent to which their results correlate to algorithmic design as distinct from unique features of training data or use cases. Secondly, it provides a basis for better assessing the algorithmic state of the art within subfields.
 Given its relative nascency, QML literature tends to focus on providing proof-of-concept examples as to how classical, hybrid or quantum-native algorithms can be used for classification or regression tasks. There is little in the way of systematic benchmarking of QML algorithms against their classical counterparts in terms of performance of specifically machine learning algorithms.
 
 Recent examples in a QML setting of benchmarking include comparisons of using different error distributions relevant to quantum chemistry (and how these affect performance) \cite{pernot_impact_2020}, benchmarking machine learning algorithms for adaptive phase estimation \cite{costa_benchmarking_2021} and generative machine learning with tensor networks \cite{wall_generative_2021}. In quantum information science more broadly, comparison with classical algorithms is often driven from computational complexity considerations and the search for quantum supremacy or outperformance, namely whether there exists a classical algorithm which can achieve results with equivalent efficiency of the quantum algorithm. Users of the QDataSet for QML research into quantum control, tomography or noise mitigation would benefit from adopting (and adapting) practices common in classical machine learning when reporting results, especially the inclusion of benchmarks against leading state of the art algorithms for particular use-cases, such as classification or regression tasks. Selecting the appropriate benchmarking algorithms itself tends to benefit from domain expertise. The QDataSet has been designed in order to be benchmarked against both classical and quantum algorithms. 
 
 %==
 
 \subsection{Benchmarking by learning protocol} \label{sec:qdata:Benchmarking by learning protocol}
 As discussed in Appendix \ref{chapter:Background: Classical, Quantum and Geometric Machine Learning}, typically machine learning algorithm classification is based firstly on whether the learning protocols are \textit{supervised}, \textit{unsupervised} or semi-supervised \cite{hastie_elements_2013,goodfellow_deep_2016}. \textit{Supervised learning} uses known input and output (label) data to train algorithms to estimate label data. Algorithmic models are updated according to an optimisation protocol, typically gradient descent, in order to achieve some objective, such as minimisation of a loss function that compares the similarity of estimates to label data. \textit{Unsupervised learning}, by contrast, is a learning protocol where label or classification of data is unknown and must be estimated via grouping or clustering together in order to ascertain identifying features. Common techniques include clustering, dimensionality reduction techniques or graph-based methods. \textit{Semi-supervised} learning is an intermediate algorithmic classification drawing on aspects of both supervised and unsupervised learning protocols. Usually known label data, say where only part of a dataset is labelled or classified, is included in architectures in order to learn the classifications which in turn can be used in supervised contexts. The QDataSet can be used in a variety of supervised, unsupervised or semi-supervised contexts. For example, training an algorithm for optimal quantum control can be undertaken in a supervised context (using pulse data, measurement statistics or Hamiltonian sequences) as label data and modelling estimates accordingly. Alternatively, semi-supervised or unsupervised protocols for tomographic classification can be trained using the QDataSet. In any case, an understanding of standard and state of the art algorithms in each category can provide QML researchers using the QDataSet with a basis for benchmarking their own algorithms and inform the design of especially hybrid approaches (see \cite{schuld_supervised_2018} for an overview and for quantum examples of the above).
 
\subsection{Benchmarking by objectives and architecture} \label{sec:qdata:Benchmarking by objectives and architecture}
The choice of benchmarking algorithms will also be informed by the objectives and architecture. Classically, algorithms can be parsed into various categories. Typically they are either \textit{regression}-based algorithms (see discussion of linear models in section \ref{sec:ml:Linear models}), used where the objective is to estimate (and minimise error in relation to) continuous data or \textit{classification}-based algorithms, where the objective is to classify data into discrete categories. \textit{Regression algorithms} are algorithms that seek to model relationships between input variables and outputs iteratively by updating models (and estimates) in order to minimise error between estimates and label data. Typical regression algorithms usually fall within broader families of generalised linear models (GLMs) \cite{gelman_data_2007} and include algorithms such as ordinary least squares, linear and logistic regression, logit and probit models, multivariate models and other models depending on link functions of interest. GLMs are also characterised by regularisation techniques that seek to optimise via penalising higher complexity, outlier weights or high variance. GLMs offer more flexibility for use in QML and for using the QDataSet in particular as they are not confined to assuming errors are normally distributed. Other approaches using Bayesian methods, such as naive Bayes, Gaussian Bayes, Bayesian networks and averaged one-dependence estimators provide yet further avenues for benchmarking algorithms trained on the QDataSet for classification or regression tasks. 

\textit{Classification} models aim to solve decision problems via classification. They typically compare new data to existing datasets using a metric or distance measure. Examples include clustering algorithms such as k-nearest neighbour, support vector machines, learning vector quantisation, decision-trees, locally weighted learning, or graphical models using spatial filtering. Most of the algorithms mentioned thus far fall within traditional machine learning.

Over the last several decades or so, \textit{neural network} architectures have emerged as a driving force of machine learning globally. Quantum analogues and hybrid neural network architecture has itself a relatively long lineage, including quantum analogues of perceptrons, quantum neural networks, quantum Hopfield networks (see \cite{schuld_supervised_2018, dunjko_quantum-enhanced_2016}) through to modern deep learning architectures (such as convolutional, recurrent, graphical and hierarchical neural networks, generative models \cite{goodfellow_deep_2016}) and transformer-based models \cite{vaswani_attention_2017}. See section \ref{sec:ml:Neural networks} for detailed discussion of neural network components and architecture.  
One feature of algorithmic development that is particularly important is dealing with the curse of dimensionality - and in a quantum context, barren plateaus \cite{mcclean_barren_2018} (see section \ref{sec:ml:Barren Plateaus}). Common techniques to address such problems include dimensionality reduction techniques or symmetry-based (for example, tensor network) techniques whose ultimate goal is to reduce datasets down to their most informative structures while maintaining computational feasibility. While the QDataSet only extends to two-qubit simulations,  the size and complexity of the data suggests the utility of dimensionality-reduction techniques for particular problems, such as tomographic state characterisation. To this end, algorithms developed using the QDataSet can benefit from benchmarking and adapting classical dimensionality-reduction techniques, such as principal component analysis, partial regression, singular value decompositions, matrix factorisation and other techniques \cite{hastie_elements_2013}. It is also important to mention that there has been considerable work in QML generally toward the development of quantum and hybrid analogues of such techniques. These too should be considered when seeking benchmarks.

Finally, it is worth mentioning the use (and importance) of ensemble methods in classical machine learning. Ensemble methods tend to combine what are known as `weak learner' algorithms into an ensemble which, in aggregate, outperforms any individual instance of the algorithm. Each weak learner's performance is updated by reference to a subset of the population of weak learners. Such techniques would be suitable for use when training algorithms on the QDataSet. Popular examples of such algorithms are gradient-boosting algorithms, such as xGboost \cite{chen_xgboost_2016}.

%======examples

\section{Example applications of the QDataSet}\label{sec:qdata:Example applications of the QDataSet}
In this section, we outline a number of applications for which the QDataSet can be used. These include training machine learning algorithms for use in quantum state (or process) tomography, quantum noise spectroscopy and quantum control. The QDataSet repository contains a number of example Jupyter notebooks corresponding to the examples below. The idea behind these datasets is that machine learning practitioners can input their own algorithms into the code to run experiments and test how well their algorithms perform.

\subsection{Quantum state tomography} \label{sec:qdata:Quantum state tomography}
Quantum state tomography involves reconstructing an estimate $\hat{\rho}$ of the state of a quantum system given a set of measured observables. Here we summarise the use of the QDataSet for tomography tasks. Discussion of measurement protocols is expanded upon in more detail in section \ref{sec:quant:Measurement}.  The quantum state of interest may be in either a mixed or pure state and the task is to uniquely identify the state among a range of potential states. Tomography requires that measurements be \textit{tomographically complete} (and therefore informationally complete, see definition \ref{defn:quant:Informationally-complete measurements}), which means that the set of measurement operators form a basis for the Hilbert space of interest. That is, a set of measurement operators \( \{M_m\} \) is tomographically complete if for every operator \( A \in \mathcal{B}(\mathcal{H}) \), there exists a representation of \( A \) in terms of \( \{M_m\} \).

Abstractly, the problem involves stipulating a set of operators $\{O_i\}_i$ as input, and the corresponding desired target outputs $\{\braket{O}_i\}_i$. 
The objective is to find the best model that fits this data. We know that the relation between these is given by $\braket{O_i}= \Tr(\rho O_i)$ and we can use this fact to find the estimate of the state. Tomography requires repeatedly undertaking different measurements upon quantum states described by identical density matrices which in turn gives rise to a measurement distribution from which probabilities of observables can be inferred. Such inferred probabilities are used to in turn construct a density matrix consistent with observed measurement distributions thus characterising the state. More formally, assuming an informationally complete positive-operator valued measure (POVM) $\{O_i \}$ spanning the Hilbert-Schmidt space $\mathcal{B}(\mathcal{H})$ of operators on $\mathcal{H}$, we can write the probability of an observation $i$ given density matrix $\rho$ using the Hilbert-Schmidt norm above i.e:
\begin{align}
        p(i|\rho) = \braket{O_i} = \Tr(\rho O_i )
    \end{align}
Data are gathered from a discrete set of experiments, where each experiment is a process of initial state preparation, by applying a sequence of gates $\{ G_j \}$ and measuring. This experimental process and measurement is repeated $N$ times leading to a frequency count $n_i$ of a particular observable $i$. The probability of that observable is then estimated as:
 \begin{align*}
            p(i|\rho) \approx \frac{n_i}{N} = \hat{p}_i
        \end{align*}
% We then have:
% \begin{align*}
%     \braket{O_i} = \hat{p}_i
% \end{align*}
from which we reconstruct the density matrix $\rho$. For a detailed exposition of tomography formalism and conditions, see D'Alessandro \cite{dalessandro_introduction_2007}. Quantum process tomography is a related but distinct type of tomography. In this case, we also have a set of test states $\{ \rho_j \}$ which span $\mathcal{B}(\mathcal{H})$. To undertake process tomography, an unknown gate sequence $G_k$ comprising $K$ gates is applied to the states such that:
 \begin{align}
            p(i|G,\rho_j) \approx \frac{n_i}{N} = \hat{p}_{j,i}
        \end{align}
Connecting with discussion of measurement in Appendix \ref{chapter:Background: Quantum Information Processing}, we observe that a set of measurement operators $\{M_m\}$ is tomographically complete if for every operator $A \in \mathcal{B}(\mathcal{H})$, there exists a representation of $A$ in terms of $\{M_m\}$, i.e.,
\begin{align*}
    A = \sum_m a_m M_m,
\end{align*}
where $a_m \in \C$. Given a POVM $\{E_m\}$ and a series of measurement outcomes $\{p(m)\}$ for a state $\rho$, $\rho$ can be reconstructed via solving the following set of linear equations:
\begin{align*}
    p(m) = \text{Tr}(\rho E_m), \quad \forall m \in \Sigma,
\end{align*}
subject to the constraints that $\rho$ is positive semi-definite.

The QDataSet can be used to train algorithms for machine learning algorithms for tomography. Quantum state and process tomography is particularly challenging. One must ensure that the estimate we get is physical, i.e. positive semi-definite with unit trace. Furthermore, the number of measurements $N$ required for sufficient precision to completely characterise $\rho$ scales rapidly. Each of the $K$ gates in a sequence $G_k$ requires $d^2(d-1)$ (where $d = \dim |\mathcal{B}(\mathcal{H})|$) experiments (measurements) to sufficiently characterise the quantum process is $Kd^4-(K-2)d^2 - 1$ (see \cite{greenbaum_introduction_2015} for more detail). Beyond a small number of qubits, it becomes computationally infeasible to completely characterise states by direct measurement, thus parametrised or incomplete tomography must be relied upon. Machine learning techniques naturally offer potential to assist with such optimisation problems in tomography, especially neural network approaches where inherent non-linearities may enable sufficient approximations that traditional tomographic techniques may not. Examples of the use of classical machine learning include demonstration of improvements due to neural network-based (non-linear) classifiers over linear classifiers for tomography tasks \cite{gao_experimental_2018} and classical convolutional neural networks to assess whether a set of measurements is informationally complete \cite{teo_benchmarking_2021}.

The objective of an algorithm trained using the QDataSet may be, for example, be to predict (within tolerances determined by the use case) the tomographic description of a final quantum state from a limited set of measurement statistics (to avoid having to undertake $N$ such experiments for large $N$). Each of the one- and two-qubit datasets is informationally complete with respect to the Pauli operators (and identity) i.e. can be decomposed into a one- and two-dimensional Pauli basis. There are a variety of objectives and techniques which may be adopted. Each of the 10,000 examples for each profile constitutes an experiment comprising initial state preparation, state evolution and measurement. One approach using the QDataSet would be to try to produce an estimate $\hat{\rho}(T)$ of the final state $\rho(T)$ (which can be reconstructed by application of the unitaries in the QDataSet to the initial states) using the set of Pauli measurements $\{ E_m \}$. To train an algorithm for tomography without a full set of $N$ measurements being undertaken, one can stipulate the aim of the machine learning algorithm as being to take a subset of those Pauli measurements as input and try to generate a final state $\hat{\rho}(T)$ that as closely approximates the known final state $\rho(T)$ provided by the QDataSet. 

A variety of techniques can be used to draw from the measurement distributions and iteratively update the estimate $\hat{\rho}(T)$, for example gradient-based updating of such estimates \cite{youssry_efficient_2019}. The distance measure could be any number of the quantum metrics described in the background chapters above, including state or operator fidelity, trace distance of quantum relative entropy. Classical loss functions, such as MSE or RMSE can then be used (as is familiar to machine learning practitioners) to construct an appropriate loss function for minimisation. A related, but alternative, approach is to use batch fidelity where the loss function is to minimise the error between a vector of ones and fidelities, the vector being the size of the relevant batch. Similar techniques may also be used to develop tools for use in gate set tomography, where the sequence of gates $G_k$ is given by the sequence of unitaries $U_0$ in the QDataSet. In that case, the objective would be to train algorithms to estimate $G_k$ given the set of measurements, either in the presence of absence of noise. Table (\ref{table:quantumtomography}) sets out an example summary for using the QDataSet for tomography.

%========quantum noise spectroscopy

\subsection{Quantum noise spectroscopy} \label{sec:qdata:Quantum noise spectroscopy}
The QDataSet can be used to develop and test machine algorithms to assist with noise spectroscopy. In this problem, we are interested in finding models of the noise affecting a quantum system given experimental measurements. More background on noise and quantum measurement is set out in section \ref{sec:quant:Noise and decoherence}. In terms of the $V_O$ operators discussed earlier, we would like to find an estimate of $V_O$ given a set of control pulse sequences, and the corresponding observables. The QDataSet provides a sequence of $V_O$ operators encoding the average effect of noise on measurement operators. This set of data can be used to train algorithms to estimate $V_O$ from noisy quantum data, such as noisy measurements or Hamiltonians that include noise terms. An example approach includes as follows and proceeds from the principle that we have known information about quantum systems that can be input into the algorithmic architecture (initial states, controls, even measurements) and we are trying to estimate unknown quantities (the noise profile). Intermediate inputs would include the system and noise Hamiltonians $H_0,H_1$ and/or the system and noise unitaries $U_0, U_1$. Alternatively, inputs could also include details of the various noise realisations. The type of inputs will depend on the type of applied use case, such as how much information may be known about noise sources. Label data could be the set of measurements $\{ E_O \}$ (expectations of the observables). Given the inputs (control pulses) and outputs, the problem becomes estimating the mapping $\{ V_O \}$, such that inputs are mapped to outputs via equation (\ref{eqn:voeqn}). Note that details about noise realisations or distributions are never accessible experimentally.

Alternatively, architectures may take known information about the system such as Pauli measurements as inputs or adopt a similar architecture to that in \cite{youssry_characterization_2020,youssry_efficient_2019} and construct a multi-layered architecture that replicates the simulation, where the $\{ \hat{V}_O \}$ are extracted from intermediate or custom layers in the architecture. Such greybox approaches may combine traditional and deep-learning methods and have the benefit of providing finer-grained control over algorithmic structure by allowing, for example, the encoding of `whitebox' or known processes from quantum physics (thereby eliminating the need for the algorithm to learn these processes). Table (\ref{table:quantumspectroscopy}) sets out one example approach that may be adopted.

\subsection{Quantum control and circuit synthesis} \label{sec:qdata:Quantum control and circuit synthesis}
The QDataSet has been designed in particular to facilitate algorithmic design for quantum control. As discussed in our sections on quantum control (sections \ref{sec:quant:Quantum Control} and \ref{sec:geo:Geometric control theory}), we wish to compare different (hybrid and classical) machine learning algorithms to optimise a typical problem in quantum control, namely describing the optimal sequence of pulses in order to synthesise a target unitary $U_T$ of interest. Here the datasets form the basis of training, validation and test sets used to train and verify each algorithm. The target (label) data for quantum control problems can vary. Typically, the objective of quantum control is to achieve a reachable state $\rho(T)$ via the application of control functions to generators, such as Pauli operators. Achieving the objective means developing an algorithm that outputs a sequence of control functions which in turn describe the sequence of experimental controls $f_\alpha(t)$. A typical machine learning approach to quantum control takes $\rho(T)$ as an input together with intermediate inputs, such as the applicable generators (e.g. Pauli operators encoded in the system Hamiltonian $H_0(t)$ of the QDataSet). The algorithm must learn the appropriate time-series distribution of $f_\alpha(t)$ (the set of control pulses included in the QDataSet, their amplitude and sequence in which they should be applied) in order to synthesise the estimate $\hat{\rho}(T)$. Some quantum control problems are agnostic as to the quantum circuit pathway (sequence of unitaries) taken to reach $\hat{\rho}(T)$, though usually the requirement is that the circuit be resource optimal in some sense, such as time-optimal (shortest time) or energy-optimal (least energy). 

One approach is to treat $f_\alpha(t)$ as the label data and $\rho(T)$ as input data to try to learn a mapping between them. A naive blackbox approach is unlikely to efficiently solve this problem as it would require learning from scratch solutions to the Schr{\"o}dinger equation. A more efficient approach may be to encode known information, such as the laws governing Hamiltonian evolution etc into machine learning architecture, such as greybox approaches described above. In this case, the target $f_\alpha(t)$ must be included as an intermediate input into the system Hamiltonians governing the evolution of $\rho(t)$, yet remains the output of interest. In such approaches, the input data would be the initial states of the QDataSet with the label data being $\rho(T)$ (and label estimate $\hat{\rho}(T)$). Applicable loss functions then seek to minimise the (metric) distance between $\rho(T)$ and $\hat{\rho}(T)$, such as fidelity $F(\rho(T),\hat{\rho}(T))$. To recover the sought after sequence $f_\alpha(t)$, the architecture then requires a way to access the intermediate state of parameters representing $f_\alpha(t)$ within the machine learning architecture.

If path specificity is not important for a use case, then  trained algorithms may synthesise any pathway to achieve $\hat{\rho}(T)$, subject to the optimisation constraints. The trained algorithm need not replicate the pathways taken to reach $\rho(T)$ in the training data. If path specificity is desirable, then the QDataSet intermediate operators $U_0(t)$ and $U_1(t)$ can be used to reconstruct the intermediate states i.e. to recover the time-independent approximation:
\begin{align}
    U(t)^\dagger \rho U(t) \approx (U_n...U_1) \rho (U_1...U_n)
\end{align}
An example of such an approach is in \cite{perrier_quantum_2020} where time-optimal quantum circuit data, representing geodesics on Lie group manifolds, is used to train algorithms for generating time-optimal circuits. Table (\ref{table:quantumcontrol}) sets out schemata for using the QDataSet in a quantum control context.

\section{Discussion} \label{sec:qdata:Discussion}
In this work, we have presented the QDataSet, a large-scale quantum dataset available for the development and benchmarking of quantum machine learning algorithms. The 52 datasets in the QDataSet comprise simulations of one- and two-qubit datasets in a variety of noise-free and noisy contexts together with a number of scenarios for exercising control. Large-scale datasets play an important role in classical machine learning development, often being designed and assembled precisely for the purpose of algorithm innovation. Despite its burgeoning status, QML lacks such datasets designed specifically to facilitate QML algorithm development. The QDataSet has been designed to address this need in the context of quantum control, tomography and noise spectroscopy, by providing a resource for cross-collaboration among machine learning practitioners, quantum information researchers and experimentalists working on applied quantum systems. In this work we have also ventured a number of principles which we hope will assist producing large-scale datasets for QML, including specification of objectives, quantum data features, structuring, preprocessing. We set-out a number of key desiderata for quantum datasets in general. We also have aimed to provide sufficient background context across quantum theory, machine learning and noise spectroscopy for machine learning practitioners to treat the QDataSet as a point of entry into the field of QML. The QDataSet is sufficiently versatile to enable machine learning researchers to deploy their own domain expertise to design algorithms of direct use to experimental laboratories.

While designed specifically for problems in quantum control, tomography and noise mitigation, the scope for the application of the QDataSet in QML research is expansive. QML is an emerging cross-disciplinary field whose progression will benefit from the establishment of taxonomies and standardised practices to guide algorithm development. In this vein, we sketch below a number of proposals for the future use of the QDataSet, building upon principles upon which the QDataSet was designed, in order to foster the development of QML datasets and research practices.
\begin{enumerate}
    \item \textit{Algorithm development}. The primary research programme flowing from the QDataSet involves its use in the development of algorithms with direct applicability to quantum experimental and laboratory setups. As discussed above, the QDataSet has been designed to be versatile and of use across a range of use cases, such as quantum control, tomography, noise spectroscopy. In addition, its design enables machine learning practitioners to benchmark their algorithms. Future research involving the QDataSet could cover a systematic benchmarking of common types of classical machine learning algorithms for supervised and unsupervised learning. We also anticipate research programmes expanding upon greybox and hybrid models, using the QDataSet as a way to benchmark state of the art QML models.
    \item \textit{Quantum taxonomies}. While taxonomies within and across disciplines will differ and evolve, there is considerable scope for research programmes examining optimal taxonomic structuring of quantum datasets for QML. In this work, we have outlined a proposed skeleton taxonomy that datasets for QML may wish to adopt or adapt, covering specification of objectives, ways in which data is described, identification of training (in-sample) and test (out-of-sample) data, data typing, structuring, completeness and visibility. Further research in these directions could include expanding taxonomic classifications of QML in ways that connect with classical machine learning taxonomies, taking the QDataSet as an example. Doing so would facilitate greater cross-collaboration among computer scientists and quantum researchers by allowing researchers to easily transfer their domain expertise.
    \item \textit{Experimental interoperability}. An important factor in expanding the reach and impact of QML is the extent to which QML algorithms are directly applicable to solving problems in applied engineering settings. Ideally, QML results and architecture should be `platform agnostic' - able to be applied to a wide variety of experimental systems, such as superconductor, transmon, trapped ion, photonic or quantum dot-based setups. Achieving interoperability across dynamically evolving technological landscapes is challenging for any discipline. For QML, the more that simulations within common platforms (such as those mentioned above) can easily integrate into each other and usefully simulate applied experiments, the greater the reach of algorithms trained using them. To the extent that the QDataSet can demonstrably be used across various platforms, algorithm design using it can assist these research imperatives.
\end{enumerate}
We encourage participants in the quantum community to advance the development of dedicated quantum datasets for the benefit of QML and expect such efforts to contribute significantly to the advancement of the field and cross-disciplinary collaboration.

%=======code availability
\section{Code availability} \label{sec:qdata:Code availability}
The datasets are stored in an online repository and are accessible via links on the site. The largest of the datasets is over 500GB (compressed), the smallest being around 1.4GB (compressed). The QDataSet is provided subject to open-access MIT/CC licensing for researchers globally.  The code used to generate the QDataSet is contained in the associated repository (see below), together with instructions for reproduction of the dataset. The QDataSet code requires Tensorflow $>$ 2.0 along with a current Anaconda installation of Python 3. The code used to simulate the QDataSet is available via the Github repository \cite{perrier_qdataset_2021} (\hyperlink{https://github.com/eperrier/QDataSet}{https://github.com/eperrier/QDataSet}). A Jupyter notebook
containing the code used for technical validation and verification of the datasets is available on this QDataSet Github repository. Please note that at the time of compilation of this thesis, the QDataSet was in the process of being updated to another data repository.

\section{Figures \& Tables}\label{sec:qdata:Figures and  Tables}

%======Longtable example

\begin{center}
\begin{table}
\footnotesize
\begin{tabular}{|p{2.5cm}||p{12.5cm}|}
\hline
 \textbf{Item}  & \textbf{Description}   \\
 \hline
 
 \textit{simulation\_ \:parameters}  & \textit{name}: name of the dataset;\\
     & \textit{dim}: the dimension $2^n$ of the Hilbert space for $n$ qubits (dimension 2 for single qubit, 4 for two qubits);\\
     & $\Omega$: the spectral energy gap;
     \\ &\textit{static\_operators}: a list of matrices representing the time-independent parts of the Hamiltonian (i.e. drift components);
     \\& \textit{dynamic\_operators}: a list of matrices representing the time-dependent parts of the Hamiltonian (i.e. control components), without the pulses. So, if we have a term $f(t) \sigma_x + g(t) \sigma_y$, this list will be $[\sigma_x, \sigma_y]$. This dynamic operators are further distinguished (and labelled) according to being (i) undistorted pulses (labelled \textit{pluses}) or (ii) distorted pulses (labelled \textit{distorted});
     \\& \textit{noise\_operators}: a list of time-dependent parts of the Hamiltonian that are stochastic (i.e. noise components). so if we have terms like $\beta_1(t) \sigma_z + \beta_2(t) \sigma_y$, the list will be $[\sigma_z, \sigma_y]$;
     \\& \textit{measurement\_operators}: Pauli operators (including identity) ($I,\sigma_x,\sigma_y, \sigma_z$)
     \\& \textit{initial\_states}: the six eigenstates of the Pauli operators;
     \\& \textit{T}: total time (normalised to unity);
     \\& \textit{num\_ex}: number of examples, set to 10,000;
     \\& \textit{batch\_size}: size of batch used in data generation (default is 50);
     \\& $K$: number of randomised pulse sequences in Monte Carlo simulation of noise (set to $K = 2000$);
     \\& \textit{noise\_profile}: N0 to N6 (see above);
     \\& \textit{pulse\_shape}: Gaussian or Square;
     \\& \textit{num\_pulses}: number of pulses per interval;
     \\& \textit{elapsed\_time}: time taken to generate the datasets.
 \\
 \hline
 \textit{pulse\_parameters}  & The control pulse sequence parameters for the example:
 \\& Square pulses: $A_k$ amplitude at time $t_k$;
     \\& Gaussian pulses: $A_k$ (amplitude), $\mu$ (mean) and $\sigma$ (standard deviation).
 \\
 \hline
  \textit{time\_range}  & A sequence of time intervals $\Delta t_j$ with $j = 1,...,M$;\\
 \hline
\textit{pulses}  & Time-domain waveform of the control pulse sequence.\\
 \hline
 \textit{distorted\_pulses}  & Time-domain waveform of the distorted control pulse sequence (if there are no distortions, the waveform will be identical to the undistorted pulses).\\
  \hline
\textit{expectations}  & The Pauli expectation values 18 or 52 depending on whether one or two qubits (see above). For each state, the order of measurement is: $\sigma_x, \sigma_y, \sigma_z $ applied to the evolved initial states. As the quantum state is evolving in time, the expectations will range within the interval $[1,-1]$. \\
 \hline
 \textit{$V_O$ operator} & The $V_O$ operators corresponding to the three Pauli observables, obtained by averaging the operators $W_O$ over all noise realizations.\\
 
 \hline
  \textit{noise} & Time domain realisations of the relevant noise.\\
   \hline
  \textit{$H_0$}  & The system Hamiltonian $H_0(t)$ for time-step $j$.\\
   \hline
  \textit{$H1$}& The noise Hamiltonian $H_1(t)$ for each noise realization at time-step $j$.\\
  \hline
 \textit{$U_0$}  & The system evolution matrix $U_0(t)$ in the absence of noise at time-step $j$.\\
 \hline
 \textit{$U_I$} & The interaction unitary $U_I(t)$ for each noise realization at time-step $j$.\\
 \hline
 \textit{$V_O$}  & Set of $3 \times 2000$ expectation values (measurements) of the three Pauli observables for all possible states for each noise realization. For each state, the order of measurement is: $\sigma_x, \sigma_y, \sigma_z $ applied to the evolved initial states.

\\
 \hline
 \textit{$E_O$}  & The expectations values (measurements) of the three Pauli observables for all possible states averaged over all noise realizations. For each state, the order of measurement is: $\sigma_x, \sigma_y, \sigma_z $ applied to the evolved initial states.
 
 % \\
 % \hline 
 % See next page & Continued
 \\
 \hline
 \end{tabular}
 \caption{QDataSet characteristics. The left column identifies each item in the respective QDataSet examples (expressed as keys in the relevant Python dictionary) while the description column describes each item.}
\label{table:datasetcharacteristics}
 \end{table}
\end{center}

\newpage

%========tomography table
\begin{center}
\begin{table}[!ht]
\begin{tabular}{ |p{3cm}||p{13cm}|  }
 \hline
  \textbf{Item}  & \textbf{Description}   \\
 \hline
 Objective  & Algorithm to learn characterisation of state $\rho$ given measurements $\{ E_O\}$.  \\
 \hline
 Inputs  & Set of Pauli measurements $\{ E_O\}$, one for each of the $M$ experiments (in the QDataSet, this is  \\
 \hline
 Label  & Final state $\rho(T)$\\
 \hline
  Intermediate inputs  & Hamiltonians, Unitary operators, Initial states $\rho_0$
  
  \\

 \hline
 
Output  & Estimate of final state $\hat{\rho}(T)$\\
 \hline
 Metric  & State fidelity $F(\rho,\hat{\rho})$, Quantum relative entropy

 \\
%   \hline
% Pseudocode  & [*] \\
  
 \hline
\end{tabular}
\caption{QDataSet features for quantum state tomography. The left columns lists typical categories in a machine learning architecture. The right column describes the corresponding feature(s) of the QDataSet that would fall into such categories for the use of the QDataSet in training quantum tomography algorithms. }
\label{table:quantumtomography}
\end{table}
\end{center}

%=========table for noise spectroscopy

\begin{center}
\begin{table}[!ht]
\begin{tabular}{ |p{3cm}||p{13cm}|  }
 \hline
 \textbf{Item}  & \textbf{Description}   \\
 \hline
 Objective  & Algorithm to estimate noise operators $\{ V_O\}$, thereby characterising relevant features of noise affecting quantum system.  \\
 \hline
 Inputs  & Pulse sequence, reconstructed from the \textit{pulse\_parameters} feature in the dataset.  \\
 \hline
 Label  & Set of measurements $\{ E_O\}$\\
 \hline
  Intermediate inputs  & Hamiltonians, Unitary operators, Initial states $\rho_0$

  \\
 \hline
Output  & Estimate of measurements $\{ \hat{E}_O\}$\\
 \hline
 Metric  &  MSE (between estimates and label data) $MSE(E_O,\hat{E}_O)$
 \\
%   \hline
% Pseudocode  & [*] \\ 
 \hline
\end{tabular}
\caption{QDataSet features for quantum noise spectroscopy. The left columns lists typical categories in a machine learning architecture. The right column describes the corresponding feature(s) of the QDataSet that would fall into such categories for the use of the QDataSet in training quantum tomography algorithms. }
\label{table:quantumspectroscopy}
\end{table}
\end{center}

%=========table for quantum control
% \subsection*{Quantum Control}
\begin{center}
\begin{table}[!ht]
\begin{tabular}{ |p{3cm}||p{12.5cm}|  }
 \hline
 \textbf{Item}  & \textbf{Description}   \\
  \hline
 Objective  & Algorithm to learn optimal sequence of controls to reach final state $\rho(T)$ or (equivalently) synthesise target unitary $U_T$.  \\
 \hline
 Inputs  &   Hamiltonians containing Pauli generators $H_0(t)$ \\
 \hline
 Label  & Final state $\rho(T)$ and (possibly) intermediate states $\rho(t_j)$ for each time-interval $t_j$.\\
 \hline
  Intermediate fixed inputs  & Sequence of unitary operators $U_0(t),U_1(t)$, Initial states $\rho_0$

  \\
 \hline
 Intermediate weights  & Sequence of pulses $f_\alpha(t)$ including parameters depending on whether square or Gaussian (for example)
  
  \\
 \hline

Output  & Estimate of final state $\hat{\rho}(T)$ and intermediate states $\hat{\rho}(t_j)$\\
 \hline
 Metric  & Average operator fidelity $F(\rho,\hat{\rho})$
 
 \\
%   \hline
% Pseudocode  & [*] \\
  
 \hline
\end{tabular}
\caption{QDataSet features for quantum control. The left columns lists typical categories in a machine learning architecture. The right column describes the corresponding feature(s) of the QDataSet that would fall into such categories for the use of the QDataSet in training quantum control algorithms. The specifications are just one of a set of possible ways of framing quantum control problems using machine learning.}
\label{table:quantumcontrol}
\end{table}
\end{center}

\begin{table}[h]
    \centering
    \begin{tabular}{|c|c|c|c|c|}
    \hline
         Category &  Qubits & Drift & Control & Noise \\
         \hline
         1 & 1 & $(z)$ & $(x)$ & $(z)$ \\
         \hline
         2 & 1 & $(z)$ & $(x,y)$ & $(x,z)$ \\
         \hline
         3 & 2 & $(z1, 1z)$ & $(x1, 1x)$ & $(z1, 1z)$ \\
         \hline
         4 & 2 & $(z1, 1z)$ & $(x1, 1x, xx)$ & $(z1, 1z)$ \\
         \hline
    \end{tabular}
    \caption{The general categorization of the provided datasets. The QDataSet examples were generated from simulations of either one or two qubit systems. For each one or two qubit simulation, the drift component of the Hamiltonian was along a particular axis (the $z$-axis) for the single-qubit case and the $z$-axis of the first qubit for the two-qubit case (but not the second qubit) or vice versa. Controls were applied along different axes, such as $x$- or $y$- axes. Finally, noise was similarly added to different axes: the $z$-axis (and in some cases the $x$-axis) of the single qubit case and the $z$-axis case of the first or second qubit for the two-qubit case.}
    \label{tab:cats}
\end{table}

\begin{table}[h]
    \centering
    \begin{tabular}{|c|c|}
    \hline
        Parameter & Value \\
        \hline
        $T$ & 1 \\
        \hline
        $M$ & 1024 \\
        \hline 
        $K$ & 2000 \\
        \hline
        $\Omega$ & 12 \\
        \hline
        $\Omega_1$ & 12 \\
        \hline
        $\Omega_2$ & 10 \\
        \hline
        $n$ & 5 \\
        \hline
        $A_{\text{min}}$ & -100 \\
        \hline
        $A_{\text{max}}$ & 100 \\
        \hline
        $\sigma$ & T/(12M) \\
        \hline
    \end{tabular}
    \caption{Dataset Parameters:  $T$: total time, set to unity for standardisation; $M$: the number of time-steps (discretisations); $K$: the number of noise realisations; $\Omega$: the energy gap for the single qubit case (where subscripts 1 and 2 represent the energy gap for each qubit in the single qubit case); $n$: number of control pulses; $A_{\text{max}},A_{\text{min}}$: maximum and minimum amplitude; $\sigma$: standard deviation of pulse spacing (for Gaussian pulses).}
    \label{tab:params}
\end{table}

%======QDataSet filename table
% \begin{landscape}
\begin{center}
\begin{table}
% {\tiny
% \begin{tabular}{|p{4cm}||p{13cm}|}
\scriptsize
\begin{tabularx}{\linewidth}{|l|X|} % The table will fit to the width of the page

\hline

 \textbf{Dataset}  & \textbf{Description}   \\
 \hline
 %=====Gaussian pulse
 G\_1q\_X  & (i) Qubits: one; (ii) Control: $x$-axis, Gaussian; (iii) Noise: none; (iv) No distortion. 
 \\
 \hline
 G\_1q\_X\_D  & (i) Qubits: one; (ii) Control: $x$-axis, Gaussian; (iii) Noise: none; (iv) Distortion. 
 \\
 \hline
 
 G\_1q\_XY  & (i) Qubits: one; (ii) Control: $x$-axis and $y$-axis, Gaussian; (iii) Noise: none; (iv) No distortion. 
 \\
 \hline
 G\_1q\_XY\_D  & (i) Qubits: one; (ii) Control: $x$-axis and $y$-axis, Gaussian; (iii) Noise: none; (iv) Distortion. 
 \\
 \hline
 
  G\_1q\_XY\_XZ\_N1N5  & (i) Qubits: one; (ii) Control: $x$-axis and $y$-axis, Gaussian; (iii) Noise: N1 on $x$-axis, N5 on $z$-axis; (iv) No distortion. 
 \\
 \hline
 G\_1q\_XY\_XZ\_N1N5\_D  & (i) Qubits: one; (ii) Control: $x$-axis and $y$-axis, Gaussian; (iii) Noise: N1 on $x$-axis, N5 on $z$-axis; (iv) No distortion. 
 \\
 \hline
 
 G\_1q\_XY\_XZ\_N1N6  & (i) Qubits: one; (ii) Control: $x$-axis and $y$-axis, Gaussian; (iii) Noise: N1 on $x$-axis, N6 on $z$-axis; (iv) Distortion. 
 \\
 \hline
 G\_1q\_XY\_XZ\_N1N6\_D  & (i) Qubits: one; (ii) Control: $x$-axis and $y$-axis, Gaussian; (iii) Noise: N1 on $x$-axis, N6 on $z$-axis; (iv) No distortion. 
 \\
 \hline
 
  G\_1q\_XY\_XZ\_N3N6  & (i) Qubits: one; (ii) Control: $x$-axis and $y$-axis, Gaussian; (iii) Noise: N1 on $x$-axis, N6 on $z$-axis; (iv) Distortion. 
 \\
 \hline
 G\_1q\_XY\_XZ\_N3N6\_D  & (i) Qubits: one; (ii) Control: $x$-axis and $y$-axis, Gaussian; (iii) Noise: N1 on $x$-axis, N6 on $z$-axis; (iv) No distortion. 
 \\
 \hline
 G\_1q\_X\_Z\_N1  & (i) Qubits: one; (ii) Control: $x$-axis, Gaussian; (iii) Noise: N1 on $z$-axis; (iv) No distortion. 
 \\
 \hline
 G\_1q\_X\_Z\_N1\_D  & (i) Qubits: one; (ii) Control: $x$-axis, Gaussian; (iii) Noise: N1 on $z$-axis; (iv) Distortion. 
 \\
 \hline
 G\_1q\_X\_Z\_N2  & (i) Qubits: one; (ii) Control: $x$-axis, Gaussian; (iii) Noise: N2 on $z$-axis; (iv) No distortion. 
 \\
 \hline
 G\_1q\_X\_Z\_N2\_D  & (i) Qubits: one; (ii) Control: $x$-axis, Gaussian; (iii) Noise: N2 on $z$-axis; (iv) Distortion. 
 \\
 \hline
 G\_1q\_X\_Z\_N3  & (i) Qubits: one; (ii) Control: $x$-axis, Gaussian; (iii) Noise: N3 on $z$-axis; (iv) No distortion. 
 \\
 \hline
 G\_1q\_X\_Z\_N3\_D  & (i) Qubits: one; (ii) Control: $x$-axis, Gaussian; (iii) Noise: N3 on $z$-axis; (iv) Distortion. 
 \\
 \hline
 G\_1q\_X\_Z\_N4  & (i) Qubits: one; (ii) Control: $x$-axis, Gaussian; (iii) Noise: N4 on $z$-axis; (iv) No distortion. 
 \\
 \hline
 G\_1q\_X\_Z\_N4\_D  & (i) Qubits: one; (ii) Control: $x$-axis, Gaussian; (iii) Noise: N4 on $z$-axis; (iv) Distortion. 
 \\
 \hline
 G\_2q\_IX-XI\_IZ-ZI\_N1-N6  & (i) Qubits: two; (ii) Control: $x$-axis on both qubits, Gaussian; (iii)
      Noise: N1 and N6 $z$-axis on each qubit; (iv) No distortion. 
 \\
 \hline
 G\_2q\_IX-XI\_IZ-ZI\_N1-N6\_D  & (i) Qubits: two; (ii) Control: $x$-axis on both qubits, Gaussian; (iii) Noise: N1 and N6 $z$-axis on each qubit; (iv) Distortion. 
 \\
 \hline
 G\_2q\_IX-XI-XX  & (i) Qubits: two; (ii) Control: single $x$-axis control on both qubits and $x$-axis interacting control, Gaussian; (iii) Noise: none; (iv) No distortion. 
 \\
 \hline
 G\_2q\_IX-XI-XX\_D  & (i) Qubits: two; (ii) Control: single $x$-axis control on both qubits and $x$-axis interacting control, Gaussian; (iii) Noise: none; (iv) Distortion. 
 \\
 \hline
 G\_2q\_IX-XI-XX\_IZ-ZI\_N1-N5  & (i) Qubits: two; (ii) Control: single $x$-axis control on both qubits and $x$-axis interacting control, Gaussian; (iii) Noise: N1 and N5 on $z$-axis noise on each qubit; (iv) No distortion. 
 \\
 \hline
 G\_2q\_IX-XI-XX\_IZ-ZI\_N1-N5  & (i) Qubits: two; (ii) Control: single $x$-axis control on both qubits and $x$-axis interacting control, Gaussian; (iii) Noise: N1 and N5 on $z$-axis noise on each qubit; (iv) Distortion.

 \\ 
 % \hline
 % See next page & Continued\\
 \hline

 % \end{tabular}
 \end{tabularx}
 % }
 \caption{QDataSet File Description (Gaussian). The left column identifies each dataset in the respective QDataSet examples while the description column describes the profile of the Gaussian pulse datasets in terms of (i) number of qubits, (ii) axis of control and pulse wave-form (iii) axis and type of noise and (iv) whether distortion is present or absent. }
\label{table:datasetproperties1}

\end{table}
\end{center}
% \end{landscape}

\newpage
%===part 2

%===part2

\begin{center}
\begin{table}
\scriptsize
\begin{tabularx}{\linewidth}{|l|X|}
% \begin{tabular}{|p{4cm}||p{13cm}|}

\hline

 \textbf{Dataset}  & \textbf{Description } \\
 \hline
 %=========square pulse
 
 S\_1q\_X  & (i) Qubits: one; (ii) Control: $x$-axis, square; (iii) Noise: none; (iv) No distortion. 
 \\
 \hline
 S\_1q\_X\_D  & (i) Qubits: one; (ii) Control: $x$-axis, square; (iii) Noise: none; (iv) Distortion. 
 \\
 \hline
 
 S\_1q\_XY  & (i) Qubits: one; (ii) Control: $x$-axis and $y$-axis, square; (iii) Noise: none; (iv) No distortion. 
 \\
 \hline
 S\_1q\_XY\_D  & (i) Qubits: one; (ii) Control: $x$-axis and $y$-axis, square; (iii) Noise: none; (iv) Distortion. 
 \\
 \hline
 
  S\_1q\_XY\_XZ\_N1N5  & (i) Qubits: one; (ii) Control: $x$-axis and $y$-axis, square; (iii) Noise: N1 on $x$-axis, N5 on $z$-axis; (iv) No distortion. 

\\
 \hline
 S\_1q\_XY\_XZ\_N1N5\_D  & (i) Qubits: one; (ii) Control: $x$-axis and $y$-axis, Gaussian; (iii) Noise: N1 on $x$-axis, N5 on $z$-axis; (iv) No distortion. 
 \\
 \hline
 
 S\_1q\_XY\_XZ\_N1N6  & (i) Qubits: one; (ii) Control: $x$-axis and $y$-axis, square; (iii) Noise: N1 on $x$-axis, N6 on $z$-axis; (iv) Distortion. 
 \\
 \hline
 S\_1q\_XY\_XZ\_N1N6\_D  & (i) Qubits: one; (ii) Control: $x$-axis and $y$-axis, square; (iii) Noise: N1 on $x$-axis, N6 on $z$-axis; (iv) No distortion.
 
 \\
 \hline
 
  S\_1q\_XY\_XZ\_N3N6  & (i) Qubits: one; (ii) Control: $x$-axis and $y$-axis, square; (iii) Noise: N1 on $x$-axis, N6 on $z$-axis; (iv) Distortion. 
 \\
 \hline
 S\_1q\_XY\_XZ\_N3N6\_D  & (i) Qubits: one; (ii) Control: $x$-axis and $y$-axis, square; (iii) Noise: N1 on $x$-axis, N6 on $z$-axis; (iv) No distortion.
 
 \\

 \hline

 %=====square pulse

 S\_1q\_X\_Z\_N1  & (i) Qubits: one; (ii) Control: $x$-axis, square; (iii) Noise: N1 on $z$-axis; (iv) No distortion. 
 \\
 \hline
 S\_1q\_X\_Z\_N1\_D  & (i) Qubits: one; (ii) Control: $x$-axis, square; (iii) Noise: N1 on $z$-axis; (iv) Distortion. 
 \\
 \hline
 S\_1q\_X\_Z\_N2  & (i) Qubits: one; (ii) Control: $x$-axis, square; (iii) Noise: N2 on $z$-axis; (iv) No distortion. 
 \\
 \hline
 S\_1q\_X\_Z\_N2\_D  & (i) Qubits: one; (ii) Control: $x$-axis, Gaussian; (iii) Noise: N2 on $z$-axis; (iv) Distortion. 
 \\
 \hline
 S\_1q\_X\_Z\_N3  & (i) Qubits: one; (ii) Control: $x$-axis, square; (iii) Noise: N3 on $z$-axis; (iv) No distortion. 
 \\
 \hline
 S\_1q\_X\_Z\_N3\_D  & (i) Qubits: one; (ii) Control: $x$-axis, square; (iii) Noise: N3 on $z$-axis; (iv) Distortion. 
 \\
 \hline
 S\_1q\_X\_Z\_N4  & (i) Qubits: one; (ii) Control: $x$-axis, square; (iii) Noise: N4 on $z$-axis; (iv) No distortion. 
 \\
 \hline
 S\_1q\_X\_Z\_N4\_D  & (i) Qubits: one; (ii) Control: $x$-axis, square; (iii) Noise: N4 on $z$-axis; (iv) Distortion. 
 \\
 \hline
 S\_2q\_IX-XI\_IZ-ZI\_N1-N6  & (i) Qubits: two; (ii) Control: $x$-axis on both qubits, square; (iii) Noise: N1 and N6 $z$-axis on each qubit; (iv) No distortion. 
 \\
 \hline
 S\_2q\_IX-XI\_IZ-ZI\_N1-N6\_D  & (i) Qubits: two; (ii) Control: $x$-axis on both qubits, square; (iii) Noise: N1 and N6 $z$-axis on each qubit; (iv) Distortion. 
 \\
 \hline
 S\_2q\_IX-XI-XX  & (i) Qubits: two; (ii) Control: single $x$-axis control on both qubits and $x$-axis interacting control, square; (iii) Noise: none; (iv) No distortion. 
 \\
 \hline
 S\_2q\_IX-XI-XX\_D  & (i) Qubits: two; (ii) Control: single $x$-axis control on both qubits and $x$-axis interacting control, square; (iii) Noise: none; (iv) Distortion. 
 \\
 \hline
 
 S\_2q\_IX-XI-XX\_IZ-ZI\_N1-N5  & (i) Qubits: two; (ii) Control: $x$-axis on both qubits and $x$-axis interacting control, square; (iii) Noise: N1 and N5 $z$-axis on each qubit; (iv) No distortion. 
 \\
 \hline
 S\_2q\_IX-XI-XX\_IZ-ZI\_N1-N5\_D  & (i) Qubits: two; (ii) Control: $x$-axis on both qubits and $x$-axis interacting control, square; (iii) Noise: N1 and N5 $z$-axis on each qubit; (iv) Distortion. 
 \\
 \hline
 
 S\_2q\_IX-XI-XX\_IZ-ZI\_N1-N6  & (i) Qubits: two; (ii) Control: $x$-axis on both qubits and $x$-axis interacting control, square; (iii) Noise: N1 and N6 $z$-axis on each qubit; (iv) No distortion. 
 \\
 \hline
 S\_2q\_IX-XI-XX\_IZ-ZI\_N1-N6\_D  & (i) Qubits: two; (ii) Control: $x$-axis on both qubits and $x$-axis interacting control, square; (iii) Noise: N1 and N6 $z$-axis on each qubit; (iv) Distortion. 
 \\ 
 \hline

 % \end{tabular}
 \end{tabularx}
 \caption{QDataSet File Description (Square). The left column identifies each dataset in the respective QDataSet examples while the description column describes the profile of the square pulse datasets in terms of (i) number of qubits, (ii) axis of control and pulse wave-form (iii) axis and type of noise and (iv) whether distortion is present or absent. }
\label{table:datasetproperties2}
\end{table}
\end{center}

\begin{center}
\begin{table}[h]
\begin{tabular}{ |p{3cm}||p{12cm}|  }
 \hline

 \textbf{Item}  & \textbf{Description}   \\
 
  \hline
 Quantum states  & Description of states in computational basis, usually represented as vector or matrix (for $\rho$). May include initial and evolved (intermediate or final) states \\
 \hline
 Measurement operators & Measurement operators used to generate measurements, description of POVM. \\
 \hline
 Measurement distribution  & Distribution of measurement outcome of measurement operators, either the individual measurement outcomes or some average (the QDataSet is an average over noise realisations). \\
 \hline
 Hamiltonians & Description of Hamiltonians, which may include system, drift, environment etc Hamiltonians. Hamiltonians should also include relevant control functions (if applicable). \\
 \hline
 Gates and operators  & Descriptions of gate sequences (circuits) in terms of unitaries (or other operators). The representation of circuits will vary depending on the datasets and use case, but ideally quantum circuits should be represented in a way easily translatable across common quantum programming languages and integrable into common machine learning platforms (e.g. TensorFlow, PyTorch).\\
 \hline
 Noise  & Description of noise, either via measurement statistics, known features of noise, device specifications. \\
 \hline
 Controls  & Specification and description of the controls available to act on the quantum system.\\
 \hline
\end{tabular}
\caption{An example of the types of quantum data features which may be included in a dedicated large-scale dataset for QML. The choice of such features will depend on the particular objectives in question. We include a range of quantum data in the QDataSet, including information about quantum states, measurement operators and measurement statistics, Hamiltonians and their corresponding gates, details of environmental noise and controls.}
\label{table:quantumdatatable}
\end{table}
\end{center}

%=========figures

\begin{figure}[ht]
\centering
\includegraphics[width=0.5\linewidth]{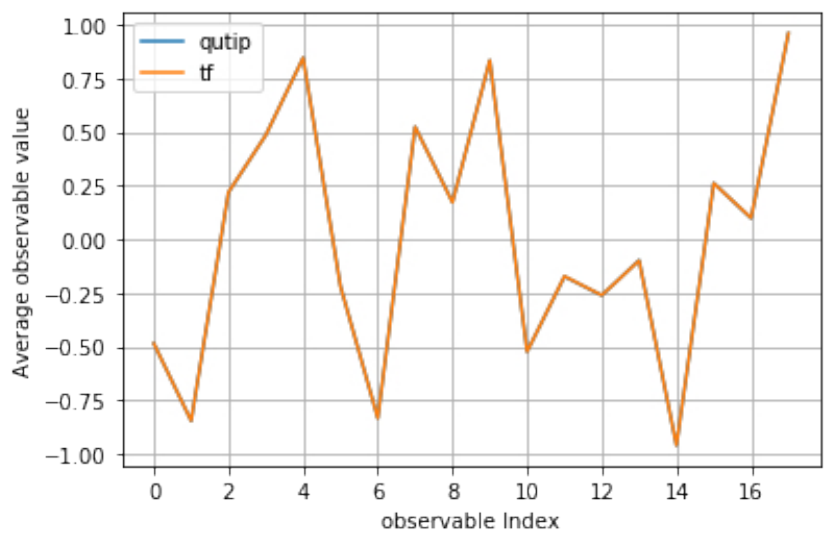}
\caption{Plot of average observable (measurement) value for all observables (index indicates each observable in order of Pauli measurements) for all measurement outcomes for samples drawn from dataset G\_1q\_X (using TensorFlow `tf', orange line) against the same mean for equivalent simulations in Qutip (blue line - not shown due to identical overlap) for a single dataset. Each dataset was sampled and comparison against Qutip was undertaken with equivalent results. The error between means was of order $10^{-6}$ i.e. they were effectively identical (so the blue line is not shown).  }
\label{fig:qutipcheck}
\end{figure}

\newpage
\subsection{Monte Carlo algorithm}\label{sec:qdata:Monte Carlo algorithm}
We set out below the Monte Carlo pseudocode referred to in section \ref{sec:qdata:Monte Carlo Simulator}. This pseudocode is reproduced from \cite{youssry_beyond_2020} (see Algorithm \ref{alg:simulation} (Monte Carlo simulation of a noisy qubit) below).
\begin{algorithm}[ht!]
	\caption{Monte Carlo simulation of a noisy qubit}
	\label{alg:simulation}
	\begin{algorithmic}
	\Function{Evolve}{$H$, $\delta$}
	    \State $U \gets I$
	    \For{$t \gets 0, M-1$}
	        \State $U_t \gets e^{-iH_t\delta}$
	        \State $U \gets U_t U$
        \EndFor
        \State \Return $U$
	\EndFunction
	\Function{GenerateNoise}{$S$, $T$, $M$}
	    \State $N \gets \frac{M}{2}$
        \For{$j \gets 0, N-1$}
            \State $\phi \gets \text{Random}(0,1)$  
	        \State $P_j \gets \frac{M}{\sqrt{T}}\sqrt{S_j}e^{2\pi i\phi}$
	        \State $Q_{N-j} \gets \bar{P}$ 
	   \EndFor
	   \State $P \gets$ \Call{Concatenate}{$P$, $Q$}
	   \State $\beta \gets \text{Re}\{\text{ifft}(P)\}$
	   \State \Return $P$
	\EndFunction
	\Function{simulate}{$\rho$, $O$, $T$, $M$, $f_x$, $f_y$, $f_x$, $S_X$, $S_Y$, $S_Z$ }
	    \State $\delta \gets \frac{T}{M}$
	    \State $E \gets 0$
	    \For{$k \gets 0, K-1$} 
	        \State $\beta_x \gets$ \Call{GenerateNoise}{$S_X$, $T$, $M$}
	        \State $\beta_y \gets$ \Call{GenerateNoise}{$S_Y$, $T$, $M$}
	        \State $\beta_z \gets$ \Call{GenerateNoise}{$S_Z$, $T$, $M$}
	        \For{$j \gets 0, M-1$}
	            \State $t \gets (0.5+j)\delta $ 
	            \State $H_j \gets \frac{1}{2}\left(\Omega + \beta_z(t)\right)\sigma_z + \frac{1}{2}\left(f_x(t) + \beta_x(t)\right) \sigma_x + \frac{1}{2}\left(f_y(t) + \beta_y(t)\right) \sigma_y$
	        \EndFor
	        \State $U \gets$ \Call{Evolve}{$H$, $\delta$}
	        \State $E \gets E + \tr{\left(U \rho U^{\dagger} O\right)}$
	    \EndFor
	    \State $E \gets \frac{E}{K}$
        \State \Return $E$
	\EndFunction
	\end{algorithmic}
\end{algorithm}

%=================================
%=======================================
%CHAPTER: QUANTUM GEOMETRIC MACHINE LEARNING

\chapter{Quantum Geometric Machine Learning} \label{chapter:Quantum Geometric Machine Learning}
\section{Abstract}
The application of machine learning techniques to solve problems in quantum control together with established geometric methods for solving optimisation problems leads naturally to an exploration of how machine learning approaches can be used to enhance geometric approaches for solving problems in quantum information processing. In this Chapter, we review and extend the application of deep learning to quantum geometric control problems. Specifically, we demonstrate enhancements in time-optimal control in the context of quantum circuit synthesis problems by applying novel deep learning algorithms in order to approximate geodesics (and thus minimal circuits) along Lie group manifolds relevant to low-dimensional multi-qubit systems, such as $SU(2)$, $SU(4)$ and $SU(8)$. We demonstrate the superior performance of greybox models, which combine traditional blackbox algorithms with whitebox models (which encode prior domain knowledge of quantum mechanics), as means of learning underlying quantum circuit distributions of interest. Our results demonstrate how geometric control techniques can be used to both (a) verify the extent to which geometrically synthesised quantum circuits lie along geodesic, and thus time-optimal, routes and (b) synthesise those circuits. Our results are of interest to researchers in quantum control and quantum information theory seeking to combine machine learning and geometric techniques for time-optimal control problems.
\section{Introduction} \label{sec:qgml:Introduction}
Machine learning-based approaches to solving theoretical and applied problems in quantum control have gained considerable traction over recent years as researchers leverage access to enhanced computational resources in order to solve numerical optimisation problems \cite{youssry_characterization_2020,batra_recommender_2022,chen_fidelity-based_2014,flynn_quantum_2021}. Concurrently, geometric control techniques (section \ref{sec:geo:Geometric control theory}) in which the tools of differential geometry and topology are applied to problems in quantum information processing, have been applied in a variety of quantum control programmes \cite{khaneja_cartan_2001,khaneja_model_2009,khaneja_optimal_2005,ekert_geometric_2000}. 
The synthesis of geometry and quantum information has also recently emerged of interest to researchers in complexity geometry \cite{brown_complexity_2019,lin_complexity_2020}. 
It is natural therefore that the intersection between geometric and machine learning techniques in quantum control emerge as a cross-disciplinary research direction. Understanding such synergies between techniques within geometric control, quantum information processing and machine learning opens up promising pathways within theoretical and applied quantum computational research, with potential application across other research domains.
In this Chapter, we extend previous research seeking to combine techniques from geometric control, quantum information processing and machine learning in order to synthesise time-optimal quantum circuits for multi-qubit quantum systems. The development of techniques for improving the time-optimality of quantum circuit synthesis is of interest to researchers across the spectrum of theoretical \cite{farhi_quantum_2014} 
% \cite{FarhiQAOA} 
and applied quantum information science given the difficulties and challenges of synthesising quantum circuits for desired computations, let alone time-optimal ones. We approach this ubiquitous problem by extending geometric methods for generating approximate normal subRiemannian geodesic (and thus time-optimal) paths along certain Lie group manifolds of interest to quantum information processing (such as SU$(2^n)$) with tailored deep learning-based machine learning techniques. 
Our results consist of: (1) an evaluation of certain existing approaches for approximating geodesics along Lie manifolds via discrete sequences of unitary propagators; (2) determination of the optimal set of controls for generating discrete approximations to certain geodesic sequences of unitaries in SU$(2^n)$ for application in multi-qubit systems; and (3) demonstration of the utility of adopting so-called `greybox' machine learning architectures \cite{youssry_beyond_2020} which combine `whitebox' architectures, i.e. prior information (such as known laws of quantum mechanics) with `blackbox' architectures, such as various neural network architectures, into synthesising quantum circuits.

\section{Preliminaries}\label{sec:qgml:Preliminaries}
\subsection{Problem description} \label{sec:qgml:Problem description}
The focus of this Chapter is on the development of novel machine learning architectures that leverage results from subRiemannian geometry in a quantum control setting. Such techniques are of relevance to practitioners within quantum control for a variety of reasons. First, as we discuss below in our explication of subRiemannian geometry in quantum control settings, subRiemannian control problems are a generalisation of standard Riemannian control problems in that they represent a more general form of Riemannian geometry (see sections \ref{sec:geo:SubRiemannian geometry} and \ref{sec:geo:Geometric control theory}). 

Second, subRiemannian quantum control problems arise where only a subset of the full Lie algebra (of generators) is itself directly accessible as a control subset. This is of direct relevance to many quantum control scenarios which may be envisioned for quantum computing devices in which one does not have access to the full set of underlying generators, say for arbitrary multi-qubit (qudit) systems with a limited gate set. Many quantum control problems are in fact, when characterised geometrically, subRiemannian quantum control problems. 

Third, is the result that synthesis of quantum circuits (i.e. sequences of unitary propagators) in a time-optimal fashion using geometric techniques (in which time-optimality is equated with generating discretised approximations to minimal distance geodesics on underlying Lie group manifolds) may call for subRiemannian rather than Riemannian geometric characterisation. 
The reason for this is that in order to generate such geodesic approximations, it may be necessary to restrict the underlying control subset of generators to a subset of the full Lie algebra (section \ref{sec:alg:Lie algebras}). For many multi-qubit systems, quantum circuits may be constructed in order to approximate geodesics (and thus be characterised as time optimal) where the generating Lie algebra is restricted to what are known as one- and two-body Pauli operators (tensor products of at most two standard Pauli operators), rather than the full Lie algebra. These three issues - the prevalence of subRiemannian geometric features in quantum control problems, the restricted availability of generators when undertaking control and the need to synthesise circuits in a time-optimal fashion - motivate the use of geometric techniques applied in this Chapter. 

\subsection{New contributions} \label{sec:qgml:New contributions}
In this Chapter, we report a number of experimental results based upon simulations of machine learning models for quantum circuit synthesis.

First, we report improved machine learning architectures for quantum circuit synthesis. We demonstrate in-sample improvements according to standard machine learning metrics (see section \ref{sec:ml:Statistical performance measures}) including MSE and average operator fidelity (equation (\ref{eqn:ml:batchfidelityMSE})) for training, validation and generalisation data by several orders of magnitude compared with relevant state of the art methods. We demonstrate that customised deep learning architectures which utilise a combination of standard and bespoke neural network layers, together with customised objective functions (such as fidelity measures) of relevance to quantum information processing, achieve superior results. This approach is denoted as `greybox' machine learning (section \ref{sec:ml:Greybox machine learning}), is characterised by models that combine known prior assumptions about quantum information processing with machine learning architectures. We demonstrate enhanced performance of greybox over blackbox models.

Second, we report an improvement on previous work combining subRiemannian geometric training with data and deep learning \cite{swaddle_generating_2017} to synthesise quantum circuits. We show that optimal sets of controls may be obtained using a feed-forward fully-connected, Gated Recurrent Unit (GRU) Recurrent Neural Network (RNN) and custom geometric machine learning models. However, we also report on difficulties in usefully adapting such approaches for generalisation. 
Third, we demonstrate that machine learning protocols to learn discretised geodesic approximations in $SU(2^n)$ are particularly sensitive to hyperparameter tuning, including time for application of generators and coverage of training geodesics over manifolds of interest. We show that selection of small time-steps for discretised unitary evolution will result in geodesic approximations highly proximal to the identity in $SU(2^n)$ (as was the case in \cite{swaddle_generating_2017}), resulting in a deterioration in the ability to (in-sample and out of sample) learn geodesic approximations to target unitaries further away (by whatever relevant distance metric or norm is adopted) along the manifolds. Improving model performance to generalise beyond the proximity of the identity is shown to require small evolutionary timescales but also an increased number of segments of the geodesic approximation i.e. a correct balance of timescale and segmentation.

\subsection{Structure} \label{sec:qgml:Structure}
The structure of this Chapter is as follows. Part \ref{sec:qgml:quantcontrolgeom} provides an overview of key quantum control concepts and literature relevant to our experiments. It draws upon material explicated in more detail in supplementary Appendices below, particularly sections \ref{sec:quant:Quantum Control}, \ref{sec:alg:Lie theory} and \ref{sec:geo:Geometric control theory} and examines the formulation of quantum control problems geometrically in terms of Lie groups and differential geometry. It also explores seminal expositions from Nielsen et al. \cite{nielsen_geometric_2006} in which time-optimal quantum circuit synthesis problems are framed in terms of generating approximate geodesics along relevant group manifolds. 
 
Part \ref{sec:qgml:subRiemsection} details the application of subRiemannian geometric theory to quantum circuit synthesis, drawing upon the exegesis in section \ref{sec:geo:SubRiemannian control and symmetric spaces}. Part \ref{sec:qgml:expdesign} lays out the design principles behind the series of experiments undertaken to develop improved machine learning architectures for quantum circuit synthesis via approximate discretised geodesics. This section provides a detailed implementation of greybox variational quantum circuits for machine learning discuss in sections \ref{sec:ml:Variational quantum circuits} and \ref{sec:ml:Greybox machine learning}. Readers interested only in the technical details of the architectures should skip to this section. Part \ref{sec:qgml:results} details the results of the various experiments, with discussion set-out in Part \ref{sec:qgml:discussion}. Future work and directions emerging from this research are then discussed in Part \ref{sec:qgml:conclusion}. Code for the experiments may be found at  GitHub\footnote{Codebase: \url{https://github.com/eperrier/quant-geom-machine-learning}.}.

\section{Quantum control and geometry} \label{sec:qgml:quantcontrolgeom}
\subsection{Overview} 
The necessity of quantum control for various quantum information and computation programmes globally has seen the emergent application of classical geometric control in an effort to solve threshold problems such as how to synthesise time optimal circuits \cite{jurdjevic_geometric_1997,brockett_sub-riemannian_nodate,boscain_k_2002}. Nearly two decades ago, developments in applied quantum control \cite{khaneja_time_2001,khaneja_cartan_2001,khaneja_sub-riemannian_2002} spurned the use of geometric tools to assist in solving optimisation problems in quantum information processing contexts such as applied NMR \cite{khaneja_optimal_2005,khaneja_time_2001,khaneja_cartan_2001,khaneja_sub-riemannian_2002}. Related work also explored the use of Lie theoretic, geometric and analytic techniques for controllability of spin particles\cite{dalessandro_constructive_2001}. Since that time, the connections between geometry and quantum control/information processing across cross-disciplinary fields, via the explication of transformations that enable problems in one field, in this case quantum control optimisation objectives (such as minimising controls for synthesis or reachable targets) to be translated into another, namely the language of differential geometry. 
Of particular note, Nielsen et al. \cite{nielsen_optimal_2006} demonstrated that calculating quantum gate complexity could be framed in terms of a distance-minimisation problem in the context of Riemannian manifolds. In that work, upper and lower bounds on quantum gate complexity, relating to the optimal control cost in synthesising an arbitrary unitary $U_T \in SU(2^n)$, were shown to be equivalent to the geometric challenge of finding minimal distances on certain Riemannian manifolds (section \ref{sec:geo:Riemannian manifolds}), subRiemannian (section \ref{sec:geo:SubRiemannian geometry}) and Finslerian manifolds. Subsequently, geometric techniques were utilised \cite{dowling_geometry_2008, gu_quantum_2008} to find a lower bound on the minimal number of unitary gates required to exactly synthesise $U_T$, thereby specifying a lower bound on the number of gates required to implement a target unitary channel. 

Research into quantum control \cite{dalessandro_k-p_2019, dalessandro_lie_2008} and geometric circuit synthesis \cite{gu_quantum_2008,leifer_quantum_2008,li_geometry_2013}  has built upon results regarding the use of geometric techniques in quantum control settings. Of interest to researchers at the intersection of geometric and machine learning approaches for quantum circuit synthesis, and the focus of this Chapter, is a technique developed in \cite{swaddle_generating_2017, swaddle_subriemannian_2017} that combines subRiemannian geometric techniques with deep learning in order to approximate normal subRiemannian geodesics (definition \ref{thm:geo:Normal subRiemannian geodesics}) for synthesis of time-optimal or nearly-time optimal quantum circuits. Our results present improved machine learning architectures tailored to learning such approximate geodesics.

%=========CHECKPOINT 1

\subsection{Quantum control formalism} \label{sec:qgml:Quantum control formalism}
\subsubsection{Control formulations} \label{sec:qgml:Control formulations}
The affinity between quantum control methods and geometric control and non-control methods arises from many sources within the literature. One fundamental reason is the intimate connection between Lie algebraic formulations of control problems, in classical and quantum settings, and the differential /geometric formulations of Lie theories on the other (as covered in sections \ref{sec:geo:Geometric control theory} and \ref{sec:quant:quantummetrics}). In typical Lie theoretic approaches to quantum control problems \cite{dalessandro_introduction_2007,jurdjevic_geometric_1997,sachkov_control_2009,boscain_introduction_2021} such as synthesis of quantum circuits, the quantum unitary of interest $U$ is drawn from a Lie group $G$ (definition \ref{defn:alg:liegroups}). A feature of Lie groups is that they are mathematical structures that are at once \textit{groups} but also \textit{differentiable manifolds} (definition \ref{defn:geo:Differentiable manifold}), topological structures equipped with sufficient geometric and analytical structure to enable analytic machinery, such as the tools of differential geometry, to be applied to their study \cite{do_carmo_differential_2016}.

A typical formulation of control problems in such Lie theoretic terms takes a target unitary $U_T$ to be an element of a Lie group, such as $SU(2^n)$, represented as a manifold. Associated with the underlying Lie group $G$ is an Lie algebra $\frak{g}$ (definition \ref{defn:alg:Lie algebraliederivative}), say $\frak{su}(2^n)$, comprising the generators of the underlying Lie group of interest. The Lie algebra $\g$ exhibits a homomorphism with the Lie group $G$ such that it is both the generator of group action $G$ and allows the symmetry properties of $G$ to be studied by querying the algebra $\g$ itself (see definition \ref{defn:alg: Lie algebra homomorphism}). Quantum control objectives can then be characterised as attempts to synthesise a target unitary propagator \cite{khaneja_time_2001} belonging to such a Lie group $G$ via application of generators belonging to $\frak{g}$ in a controlled manner. In the simplest (noise-free) non-relativistic settings, computation is effected via evolution from $U(0)=I$ to $U_T$ according the time-dependent Schr{\"o}dinger equation (definition \ref{eqn:quant:schrodingersequation}): 
\begin{align}
    U(t)&=\mathcal{T}_+ \exp\left(-i\int_{0}^t H(s) ds\right).\label{eqn:scrhodtime}
\end{align}
where $\mathcal{T}_+$ represents the time-ordering operator to ensure appropriate causal ordering of operators  and we have set $\hbar=1$ for convenience. The above formulation may also be expressed in terms (discussed in more detail below) of time dependent drift $H_d(t)$ and control $H_c(t)$ Hamiltonians characteristic of quantum control settings (section \ref{sec:quant:Quantum Control}):
\begin{align}
    \dot{U}(t) &= -i(H_d(t) + H_c(t))U(t). \label{eqn:schrodunitary}
\end{align}
The drift part of the Hamiltonian represents the (directly) `uncontrollable' aspect of evolution (and is discussed in more detail below), while the control Hamiltonians represent evolution generated by those elements (generators) of the quantum system which are controllable (see  definition \ref{defn:quant:controlequation}), namely the generators of a Lie algebra of interest, such as, in the case of qubit systems, generalised Pauli operators (equation (\ref{eqn:qdata:paulimatrices})). 

The $H_c$ terms represent the control Hamiltonians :
\begin{align}
    H_c(t) = \sum_{k=1}^m v_k(t) \tau_k.
\end{align}
parametrised by control functions $v_k(t)$ which are continuous functions of time. These Hamiltonians are composed from (usually linear) functions of the generators $\tau_k \in \frak{g}$ (where $\dim \frak{g}=m$ and $k$ indexes the generators) belonging to the corresponding Lie algebra (such as generalised (tensor products) of Pauli operators for $SU(2^n)$. The time-dependence of the Hamiltonians is encoded in these time-dependent control functions as the generators themselves are not time-dependent. While often linear, the functional time dependence can and does often assume non-linear and complicated functional forms, especially in the presence of noise.

Analytically solving for the form of the control function is difficult and usually intractable for higher-order qudit systems or open quantum systems in the presence of noise, with numerical methods usually adopted instead \cite{khaneja_optimal_2005}. This is due to the limited circumstances in which purely analytic solutions are possible (see \cite{khaneja_cartan_2001, dalessandro_introduction_2007} for examples). It is also because, more practically, such analytic methods are insufficient for solving quantum control problems in open quantum systems where noise is present unless the full noise spectrum is known, which is rare (see discussion of noise in section \ref{sec:quant:Noise and quantum evolution}). In both cases, solutions to control problems usually rely upon well-established numerical techniques, such as dynamical decoupling \cite{viola_dynamical_1999}. One of the motivations for the use of machine learning in quantum control problems is precisely their potential utility in learning a sufficient approximation of control functions needed to achieve quantum control objectives, including in the presence of noise \cite{youssry_modeling_2020}.

It is common (as discussed below), in appropriate circumstances, to simplify the typical time-dependent Schr{\"o}dinger equation with its time-independent discretised approximation (definition \ref{eqn:quant:timeindependentschrodunitary}). In this case, evolution towards a target unitary propagator $U_T$ is approximated via a sequence of successive unitaries generated by time-independent Hamiltonians $H_j$ applied at time $t_j$ for duration $\Delta t_j$: 
\begin{align}
    U(t)&=\mathcal{T}_+ \exp\left(-i \int_{0}^t H(s) ds\right)\\
    &=\lim_{N\to\infty} \prod_{j=N}^0 \exp(-iH_j(t_j)\Delta t)\\
    &\approx \prod_{j=N}^0 \exp(-iH_j(t_j)\Delta t)\\
    &=\prod_{j=N}^0 U_j = U_N...U_j...U_0
    \label{eqn:qgml:schrodind}
\end{align}
where $\Delta t_j = \Delta t = T/N$. That is, the unitary propagator at time $t$ (from the identity) is the cumulative reverse product (forward-solved cumulant) of a sequence of $U_j$. This approximation is considered appropriate where $\Delta t$ is small by comparison to total evolution time $T$ (or equivalently total energy) (and resulting cumulative errors from the product of such unitaries are sufficiently small) and is an approximation adopted in our experiments detailed below. 
%=========image

\begin{figure}
\begin{center}
\includegraphics[width=0.75\linewidth]{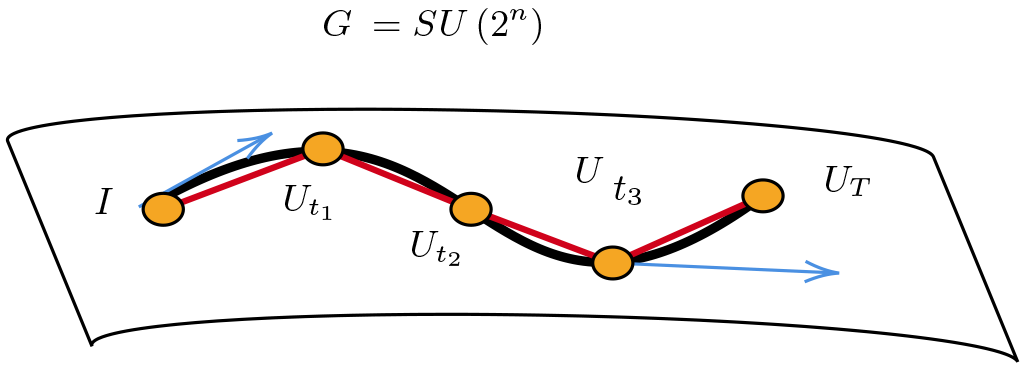}
\caption{Sketch of geodesic path. The evolution of quantum states is represented by the evolution according to Schr{\"o}dinger's equation of unitary propagators $U$ as curves (black line) on a manifold $U\in G$ generated by generators (tangent vectors) (blue) in the time-dependent case (\ref{eqn:schrodunitary}). For the time-independent case, the geodesic is approximated by evolution of discrete unitaries for time $\Delta t$, represented by red curves (shown as linear for ease of comprehension). Here $U_{t_i}$ represents the evolved unitary at time $t_i$.}
\label{schema:geodesic}
\end{center}
\end{figure}

%=========image

Adopting this approximation allows  (\ref{eqn:schrodunitary}) to be expressed as:
\begin{align}
    \dot{U}_j &= -i(H_{d,j} + H_{c,j})U_j\\
    &= -i(H_{d,j} + \sum_k^m v_{k,j} \tau_k)U_j. \label{eqn:qgml:control}
\end{align}
Here, $H_{d,j} = H_d(t_j)$ designates the drift (or internal) part of the Hamiltonian at time-step $t_j$ and similarly for $H_{c,j}$. In the discretised approximation, the control functions $v_{k,j}$ above now represent the amplitude (energy) to be applied at time-step $t_j$ for time $\Delta t$ (duration) and typically correspond, for example, to the application of certain voltages or magnetic fields for a certain period of time. The functional form of the controls $v_{k,j}$ can vary, with common (idealised) representations including Gaussian or `square' pulses.  

The objective of time optimal control is then to select the set of controls to be applied (when using a discretised approximation) at time $t_j$ for time $\Delta t_j$ in order to synthesise $U_T$ in the shortest amount of total time. Such geometric approaches involve reparametrisation of quantum circuits, which are discrete, as approximations to geodesics on Lie group manifolds of interest to quantum information processing \cite{nielsen_optimal_2006, dowling_geometry_2008, gu_quantum_2008}. A schema illustrating the application of discretised time-independent unitaries in order to approximate the geodesic quantum circuits is presented above in Figure \ref{schema:geodesic}. It is the desire to solve this optimisation problem that motivates geometric recharacterisation of problems in quantum information, such as determining and solving geodesic equations of motion. It should be noted that in practice the properties or characteristics of $U_T$ may be known with greater certainty than each $U_j$. In this work, we assume the existence of a measurement (and indeed tomographic) process by which $U_j$ may be sufficiently reconstructed such that knowledge about $U_j$ is accessible.

\subsubsection{Path-length and Lie groups} \label{sec:qgml:Path-length and Lie groups}
The adaptation of geometric methods and variational methods for solving optimisation problems in quantum information processing is characterised in terms of minimising distance of curves along Lie group manifolds $G$. In geometric contexts, equation (\ref{eqn:qgml:control}) above can be expressed as horizontal curves (see definition \ref{defn:geo:Horizontal control curves}) using equation (\ref{eqn:geo:horizontalcontroSchrod}) where $U(t) \sim \gamma(t)$ i.e. as a point in the Lie group manifold $G \sim \M$. 

Doing so requires selection of a (subRiemannian or Riemannian) metric (definition \ref{defn:geo:Riemannian metric}) (for use in a cost functional) that intuitively measures the distance (or arc length i.e. using equation (\ref{eqn:geo:arclength})) between elements in the associated Lie algebra $\g$ which, in geometric terms, are represented by tangent vectors belonging to the associated tangent space $TG$ (see section \ref{sec:geo:Tangent planes and Lie algebras} for discussion of tangent bundle and Lie algebra correspondence). Cost-functionals (see equation (\ref{eqn:geo:costfunctional})) are essentially analogous to variational functional equations such that:
\begin{align}
    C = \int_a^b g_{\alpha\beta} \frac{dx^\alpha}{dt}\frac{dx^\beta}{dt}
\end{align}
where $g_{\alpha\beta}$ represents the (in the most general case, not necessarily constant) metric tensor (definition \ref{defn:geo:Riemannian metric}), $dx/dt$ represents the differential Lie group elements $x \in G$ with respect to the unique single-parametrisation (i.e. time). Solving the optimisation problem of interest, such as synthesising a circuit in minimal time or with minimal energy, becomes a question of minimising the cost function according to the Pontryagin Maximum Principle (see section \ref{sec:geo:PMP and quantum control}). 
Variational methods in this approach set $\delta C=0$ and consequently use standard techniques from variational calculus to derive respective equations of motion, differential equations (see section \ref{sec:geo:variational methods}). The solutions (usually) take the form of exponentiated Lie algebraic elements i.e. unitary propagators, which ultimately minimise the cost functional and solve the underlying optimisation problem.

It is worth explicating the form of cost functionals for quantum information practitioners who may be less familiar with geometric methods. In the discretised case, we essentially replace the integral with a sum over the various Hamiltonians such that we have:
\begin{align}
    C_f = \int_a^b f(H(t)) dt
\end{align}
for the continuous case and
\begin{align}
    C_f = \sum_{j=a}^b f(H_j(t_j))
\end{align}
for the discrete case, where $f$ represents the control function(s) applicable to the Hamiltonian $H(t)$. By selecting the appropriate parametrisation of curves on the manifold (such as a typical parametrisation by arc-length e.g. equation (\ref{eqn:geo:arclengthparamby})), distance along a curve (representing evolution from one unitary, such as the identity, to another) can be equated to minimal time required to evolve (and synthesise) a target unitary $U_T \in G$ of interest. In cases where there are multiple curves between two points, then one must select the minimal path over all such paths \cite{nielsen_geometric_2006} consistent with existence theorems regarding subRiemannian geodesics (see discussion of Chow's theorem in section \ref{sec:geo:SubRiemannian Geodesics}).  Because minimising the cost functional depends itself upon solutions (unitaries) which are themselves generated by Lie algebraic elements subject to control functions, the optimisation problem of quantum control thus becomes a problem of identifying the optimal set (sequence) of control functions to be applied over time in order to minimise the cost functional. Note that the cost functional above in terms of arc-length is different from the cost functional using fidelity (equation (\ref{eqn:qgml:batchfidelity})) as part of our machine learning architecture below.

Applying standard techniques from the calculus of variations (e.g. the Pontryagin Maximum Principle \cite{earp_constructive_2005} (section \ref{sec:geo:PMP and quantum control})) with respect to the cost functional results in the geodesic equation of motion \cite{nielsen_geometric_2006, dowling_geometry_2008} (definition \ref{defn:geo:Geodesic}) which specifies the path that minimises the action and which is typically (for constant metric Riemannian manifolds) is given by equation (\ref{eqn:geo:geodesiccoordinateframe}):
\begin{equation}
    \frac{d^2x^j}{dt^2} +\Gamma_{kl}^j \frac{dx^k}{dt}\frac{dx^l}{dt}=0 \label{eqn:geodesicconst}
\end{equation}
where $x=x(t) \in G$ are the unitary group elements while $dx/dt \in \frak{g}$ represent the differential operators (tangent vectors/generators) of the associated Lie algebra. Also in (\ref{eqn:geodesicconst}) it is implied that the form of applicable geodesic $g_{\alpha\beta}$ is itself identical across the manifold (which may not always be the case). $\Gamma_{kl}^j$ represent Christoffel terms obtained by variation with respect to the metric.  Given a small arc along a geodesic on a Riemannian manifold, the remainder of the geodesic path is completely determined by the geodesic equation. Solutions to the geodesic equation are, in the continuous case, horizontal curves (definition \ref{defn:geo:Horizontal control curves}) and in the discrete case approximations to curves, on the manifold of interest. Such curves are interpretable in terms of quantum circuits which are time-optimal when such geodesics also represent the minimal distance curve linking two unitary group elements on a manifold. In this way, variational methods leveraging geometric techniques and characterisation may be utilised for synthesising quantum circuits.

\subsubsection{Accessible controls and drift Hamiltonians} \label{sec:qgml:Accessible controls and drift Hamiltonians}
Minimising cost functionals in the way described above involves understanding what in classical control theory is described as the set of \textit{accessible} controls available. Those unitaries which may be synthesised via application of the controls are termed reachable targets, that is reachable via application of the controls given the generators (see definition \ref{defn:geo:Reachable set}). Designing appropriate machine learning algorithms using geometric methods or otherwise thus requires information on the form of control function and generators that are available to reach a desired target, such as a target unitary or quantum state. For a given Lie group $G$, access to the entire set of generators $\frak{g}$ renders any element $U \in G$ reachable. In quantum control settings, access to the full Lie algebraic array of generators occasionally render the problem of unitary synthesis, i.e. the sequence of generators and control pulses, analytically or trivially obtainable using geometric means, such as Euler decompositions where $G=SU(2)$ \cite{boozer_time-optimal_2012}. In certain cases (such as those explored below), we are constrained or seek to synthesise target unitaries $U_T$ using only a subset of the relevant Lie algebra, a subset named the \textit{control set} (or \textit{control subset}) $\frak{p} \subset \frak{g}$ with a decomposition $\g = \p \oplus\k$ (note we denote the control subset as a set rather than subalgebra as where such a decomposition is a Cartan decomposition, the non-compactness of $\p$ i.e. that $[\p,\k] \subseteq \k$ results in $\p$ not being a subalgebra of $\g$). In such cases, the full set of generators is not directly accessible. However, one may still be able to reach the target unitary of interest if the elements of $\frak{p}$ may be combined (by operation of the Lie bracket or Lie derivative, as discussed below) in order to generate the remaining generators belonging $\frak{g}$, thus providing access to the $\frak{g}$ in its entirety. This will be the case when the control subset $\p$ satisfies the Lie triple property (equation (\ref{eqn:geo:lietripleproperty})) $[[\p,\p],\p] \subseteq \p$. We distinguish such cases by denoting the first case as a case of \textit{directly} accessible controls, while the second case represents \textit{indirectly} accessible controls. 

Returning to the quantum control paradigm (equation \ref{eqn:qgml:control}), the drift Hamiltonian $H_d$ represents the evolution of a quantum system which cannot be \textit{directly} controlled. It may represent a noise term or the interaction of a system with an environment in open quantum systems' formulations. Where a control subset $\frak{p} \subset \frak{g}$ represents only a subset of the relevant Lie algebra, we can think of the complement $\frak{k} = \frak{p}^\bot$ (where $\frak{g} = \frak{k}\oplus \frak{p}$) as generators from which the drift term $H_d$, or at least elements of it (noting that, for example, in open quantum systems or non-unitary evolutions, generators are not necessarily Lie algebraic in character), above is composed, i.e. that $H_d \in \frak{k}$. 

The interaction between the drift $H_d$ (named due to its origins in fluid dynamics) and control $H_j$ Hamiltonians depends on the set of such accessible controls available to solve the quantum control problem of interest. The application of control Hamiltonians in this case represents, in effect, an attempt to `steer' a system evolving according to $H_d$ towards a desired target via the adjoint action of Lie group elements generated by $\frak{p}$ (see \cite{khaneja_time_2001} for a discussion). 

Understanding the nature of relevant control algebras and the composition of drift Hamiltonians is an important consideration when designing and implementing machine learning architectures for geometric quantum control, including recent novel approaches applying machine learning for modelling and control of a reconfigurable photonic circuit \cite{youssry_modeling_2020} and to learn characteristics of $H_d$ via quantum feature engineering \cite{youssry_beyond_2020}.
One of the motivations of the present work is to demonstrate the utility of being able to encode prior information about the relevant control subset into machine learning protocols whose objective is the output of a time-optimal sequence of control pulses, a design choice that requires information about precisely what generators are accessible.

\subsubsection{Geometric optimisation} \label{sec:qgml:Geometric optimisation}
Selecting the specific control subset and set of control amplitudes in order to generate time-optimal quantum circuits is a difficult task. Solving this optimisation problem in quantum control and quantum circuit literature using geometric techniques follows two broad directions which synthesise results from geometric control theory (see section \ref{sec:geo:Geometric control theory} and \cite{jurdjevic_geometric_1997} for a comprehensive review). One such approach uses symmetric space formalism and Cartan decompositions \cite{khaneja_cartan_2001, khaneja_time_2001,khaneja_cartan_2000} to decompose the Lie algebra $\frak{g}$ associated with a given Lie group $G$ into symmetric and antisymmetric subspaces such that $\frak{g} = \frak{p} \oplus \frak{k}$ (see section \ref{sec:alg:Cartan decompositions} and \cite{knapp_lie_1996} generally). Here $\frak{p}$ is the control subset (containing accessible generators) and $\frak{k}$ is the subalgebra generating the non-directly controllable evolution of the system. If a suitable partition can be found satisfying certain Levi (Cartan) commutation relations (see \cite{dalessandro_lie_2008, graaf_lie_2000}) set out in definition \ref{defn:alg:cartandecomposition}, then the Lie group can be decomposed into a Cartan decomposition $G = KAK$. By doing so, the problem of selecting the appropriate set of generators $\tau \in \frak{p}$ and control amplitudes is simplified (see section  (\ref{sec:qgml:onetwobodyoperators}) for a discussion and \cite{khaneja_cartan_2001, dalessandro_lie_2008} in particular). A drawback of such methods as currently applied to problems in quantum control is their limited scope of application, namely that such methods apply only to limited symmetric space manifolds for which the methods were developed. Furthermore, the particular methods in \cite{khaneja_cartan_2001} used to determine the appropriate generators are limited in their generality. Chapter \ref{chapter:Time optimal quantum geodesics using Cartan decompositions} presents novel results that seek to, for certain classes of symmetric space control, address some of the challenges in using such methods.

An alternative, but ultimately related, method explored by Nielsen et al. in a range of papers \cite{nielsen_geometric_2006, nielsen_optimal_2006, dowling_geometry_2008, gu_quantum_2008} approaches the problem of finding optimal generators and controls via modifying metrics applicable to cost functionals. In \cite{nielsen_geometric_2006}, geometric techniques are applied to determine the minimal size circuit to exactly implement a specific $n$-qubit unitary operation combining variational and geometric techniques from Riemannian geometry (we discuss in detail in section \ref{sec:geo:Riemannian Manifolds and Metrics}), with the paper detailing a method for determining the lower bound of circuit complexity and circuit size by reference to the length of the local minimal geodesic between $U_T$ and $I$ (where length is determined via a Finsler metric on $\su(2^n)$). In later work \cite{nielsen_optimal_2006, dowling_geometry_2008, gu_quantum_2008}, particular metrics with penalty terms are chosen that add higher-weights to higher order Pauli operators in order to steer the generating set towards one- and two-body operators which are assessed as being optimal for geodesic synthesis (see Appendix (\ref{sec:qgml:Nielsen's approach}) for a discussion and section \ref{sec:ml:Reducing empirical risk} for a discussion of penalty metrics and regularisation generally). It is shown that in limiting cases applying the variational techniques and penalty metric of Nielsen et al., the optimal set of generators are one- and two-body terms \cite{wang_quantum_2015}. 

Such variational and penalty metric-based approaches have their drawbacks, however: there are limited convergence guarantees due to, for example, the existence of exponentially Pauli geodesics (many unitaries have minimal Pauli geodesics of exponential length (see \cite{huang_explicit_2007})), reliance upon complicated boundary conditions, or the difficulty in discovering homotopic maps with which to deform known geodesics into other geodesic paths \cite{swaddle_subriemannian_2017, gu_quantum_2008, wang_quantum_2015}. The approach in the work of Nielsen et al. is also less general in that it assumes the entire distribution is the Lie algebra $\frak{su}(2^n)$.

A common characteristic of both approaches in the case of $SU(2^n)$ is a preference for control subspaces comprising only one- and two-body Pauli operators (operators that are tensor products of at most one or two Pauli operators) \cite{nielsen_optimal_2006}. In \cite{khaneja_cartan_2000, khaneja_cartan_2001}, the rationale is that higher-order (more than two-body) generators introduce coupling terms which increase evolution time. In \cite{nielsen_geometric_2006, dowling_geometry_2008, gu_quantum_2008}, this rationale manifests in the imposition of penalty metrics upon higher-order terms in cost functionals. This approach penalises higher-order generators by assigning to them a higher weighting in the metric, thereby penalising higher-order terms in the cost function which seeks to minimise the metric of interest (in the case of \cite{dowling_geometry_2008}, often Finslerian metrics $F$).  

Thus there are strong motivations for preferencing one- and two-body generator control subsets when devising strategies for quantum circuit synthesis. It can be shown that for higher-order SU$(2^n)$ systems, three or more body generators can themselves be composed via one- and two-body generators \cite{swaddle_subriemannian_2017} when they form a bracket-generating set \cite{montgomery_tour_2002}. These combined results motivate the selection of a (minimal) set of one- and two-body generators that can generate the entire Lie algebra $\frak{su}(2^n)$. This is a characteristic of the \textit{bracket-generating} set (or \textit{distribution}) $\Delta$ adopted in \cite{swaddle_generating_2017} (see definition \ref{defn:geo:bracketgenerating}), where instead of imposing penalty metrics or relying on decompositions to obtain optimal control subsets, the control subsets are selected initially to comprise only one- and two-body terms. Such reasoning does not guarantee the utility of one- and two-body terms per se (see \cite{swaddle_subriemannian_2017} for technical examples) but can provide in certain cases a basis for potentially preferring such generators when designing optimisation protocols, such as via machine learning, to approximate geodesics. We also note recent important results which may affect these methods from Marvian \cite{marvian_restrictions_2022} regarding constraints on universality of control from 2-local operations in absence of certain correctives being applied.

%=============Swaddle and circuit synthesis

\section{SubRiemannian quantum circuit synthesis} \label{sec:qgml:subRiemsection}
\subsection{Overview}
The difficulties of synthesising geodesics are well-known throughout geometric and control literature \cite{noakes_global_1998} (see section \ref{sec:geo:Geometric control theory}). The geodesically-driven control methods articulated above face considerable challenges in terms of the complexities of the relevant boundary-value problem when adopting certain `penalty' metrics designed to enforce the geodesic constraints on Finslerian manifolds. Though analytic or numerical (including machine learning) architectures are unlikely to provide means of systematically synthesising approximate geodesics and time-optimal unitary synthesis for arbitrary propagators or higher-dimensional Lie groups, they have potential utility for lower-order qudit systems. 

In \cite{swaddle_generating_2017, swaddle_subriemannian_2017}, an approach leveraging subRiemannian, rather than Riemannian, geometry is adopted in order to overcome some of these barriers to quantum circuit synthesis using geodesic approximations. SubRiemannian geometry \cite{montgomery_tour_2002, nielsen_optimal_2006,shizume_quantum_2012,dalessandro_k-p_2019,albertini_symmetries_2018} is a generalised form of Riemannian geometry that is well-developed in classical control contexts. In its simplest description, it covers typical geometries where only a subset of the full Lie algebra $\frak{g}$ is directly accessible. We discuss subRiemannian geometry in section \ref{sec:geo:SubRiemannian geometry} and follow-on sections.

For the purposes of quantum control, it is helpful to characterise subRiemannian manifolds in Lie theoretic terms (see \cite{montgomery_tour_2002} for a more formal treatment). The concepts below, especially the important correspondence between Lie algebras and tangent bundles are covered extensively in Appendices \ref{chapter:Background: Geometry, Lie Algebras and Representation Theory} and \ref{chapter:background:Differential Geometry}. For a given Lie group manifold $G$, the Lie algebra $\frak{g}$ comprises generators which also form (i.e. correspond to) a basis of the tangent space $T\M$ (section \ref{sec:geo:Tangent planes and Lie algebras}). Curves $\gamma(t)$ along $G$ are those generated by generators $\tau \in \frak{g}$ such that the generators may be thought of as tangent vectors tangent to the curves they generate. The curves which may be generated on a manifold in many ways characterise the manifold. A distinguishing feature of Riemannian and subRiemannian manifolds is the set of accessible generators. Riemannian manifolds are characterised by full direct access to $\frak{g}$, that is, all generators in $\frak{g}$ may generate curves on $G$. In more formal language, the directions a curve may evolve along (or subalgebra of its generators) is characterised by certain subsets $\Delta$ of the tangent bundle $T\M$ for a manifold $G$. The distribution $\Delta$ in this context (see definition \ref{defn:geo:SubRiemannian manifold}) is also denoted the \textit{horizontal} tangent space $H_p\M$, which intuitively refers to tangent vectors being `tangent' and along the manifold but more formally refers to the fact that the covariant derivative (definition \ref{defn:geo:covariantderivative}) of those (generating) tangent vectors $X$ (along the vector field $X_f$) along the curve is zero, that is
\begin{align*}
    \nabla_{\gamma(t)} X = 0.
\end{align*}
The vanishing of the covariant derivative is the characteristic definition of parallel transport (section \ref{sec:geo:Parallel transport and horizontal lifts}). By contrast, it may be the case that only a subspace $\frak{p} \subset \frak{g}$ where $\frak{g} = \frak{p} \oplus \frak{k}$ is accessible for generation of curves on $G$. In this case, evolution of curves tangent to certain directions of tangent vectors in $\frak{k}$ (that is, along the fibres (section \ref{sec:geo:Fibre bundles})) is not directly possible. This set of directly inaccessible generators $\k$ is orthogonal to the set $\frak{p}$ of horizontal tangent vectors and is described as \textit{vertical} (see definition \ref{defn:geo:Vertical subspace}). More formally, the vertical subspace $V_p\M$ of $T\M$ comprises vectors $X$ whose evolution along the curve $\gamma(t)$ is such that $\nabla_{\gamma(t)} X \neq 0$ (having some component not tangent to the manifold). In this second case, the manifold is characterisable as subRiemannian rather than Riemannian. Elements of the vertical subspace $\frak{k}$ may still affect the evolution of curves, but only indirectly to the extent the generators in $\frak{k}$ are able to be generated by the application of the Lie bracket via the BCH formula (see equation (\ref{eqn:qgml:bch}) below and definition \ref{defn:alg:Baker-Campbell-Hausdorff})) i.e. if the distribution is bracket-generating. A number of theorems of subRiemannian geometry \cite{montgomery_tour_2002} then guarantee the existence and uniqueness of certain normal subRiemannian geodesics on $G$ which are both unique and minimal in length. 

Thus, for generating circuits on $G=SU(2^n)$, by constructing a distribution $\Delta$ that is bracket-generating and comprising only one- and two-body generators, it can be shown \cite{swaddle_subriemannian_2017} that normal subRiemannian geodesics may be generated which are minimal and unique, thus approximating the minimal circuits between $I$ and $U_T$. In the next section, we detail the approach in \cite{swaddle_generating_2017} that leverages such subRiemannian geometric insights. We do so in order to provide insight into the subRiemannian machine learning detailed below.

\subsection{SubRiemannian Normal Geodesics}\label{sec:qgml:SubRiemannian Normal Geodesics}
The motivation behind the approach in \cite{swaddle_generating_2017} is to solve the problem of finding time-optimal sequences of gates via approximating subRiemannian normal geodesics on Lie group manifolds \cite{dowling_geometry_2008, brandt_jacobi_2010,brandt_riemannian_2010,brandt_aspects_2012,brandt_tools_2012} in order to synthesise target unitary propagators $U_T \in SU(2^n)$.  The basis of the approach is to firstly adopt the time-independent approximation (\ref{eqn:qgml:schrodind}) and express $U_T$ as an approximate product of exponentials:
\begin{align}
U_T \approx U_n...U_1 &\approx E(c) = \prod_j^n \underbrace{\exp\left(  \sum_k^m \left(h c^k_j \tau_k\right) \right)}_{U_j}\\
U_j(\Delta t) &= \exp\left(  \sum_k^m \left(c^k_j \tau_k\right) \Delta t \right) = \exp(H_j \Delta t)
\label{eqn:qgml:swaddleunitaryseq}
\end{align}
where we have absorbed the imaginary unit $-i$ into the generators. Here $U_j$ are referred to herein as (right-multiplicative or right acting) \textit{subunitaries} for convenience, again justifiable in the large $m$, small $h$ limit where $h=\Delta t$, the evolution time of each $U_j$. The terms $c^k_j:=c^k_j(t)$ represent the amplitudes of the $c_j^k$ (square) control pulses applied to $k$ generators at time interval $t_j$ for duration $\Delta t_j = h$ to generate unitary $U_j$ (i.e. $j$ indexes the segment, $k$ indexes the control amplitude $c^k$ paired with the generators $\tau_k$). For notational clarity, in sections \ref{sec:quant:Quantum Control} and \ref{sec:geo:Geometric control theory} the set of $c_j(t)$ are denoted $u_j(t) := (u^k_j(t)) \in U \subset \R^m, j=1,...,m$ (as per section \ref{sec:geo:Pontryagin Maximum Principle}). The method in \cite{swaddle_generating_2017} in effect becomes a `bang-bang control' problem \cite{schattler_geometric_2012,lasalle_bang-bang_1960} in which the time-dependent Schr\"odinger equation is approximated by a sequence of time-independent solutions $U_j$ where control Hamiltonians $H_j$ are applied via the application of a constant amplitude $c^k_j$ for discrete time interval $\Delta t = h=T/N$ (with $N$ the number of segments).
The term $E(c)$ represents an embedding function that maps controls from $\C^{n \times m}$ into the Lie group manifold: 
% \begin{eqnarray}
%     E: \C^{n \times m} \to SU(2^n) \\ c =(c^1_1,...,c_1^m,...,c^1_N,...,c^m_N) \mapsto \prod_j^n\prod_k^m \exp(h c^k_j \tau_k)
% \end{eqnarray}
% \hl{correct I think wrong not product over generators should be?}
\begin{eqnarray}
    E: \C^{n \times m} \to SU(2^n) \\ c =(c^1_1,...,c_1^m,...,c^1_N,...,c^m_N) \mapsto \prod_j^n \exp\left(\sum_k^mh c^k_j \tau_k\right)
\end{eqnarray}
with the set of coefficients $c \in \C$. The generators $\tau_i$ form a basis for the bracket generating subset $\Delta \in \frak{su}(2^n)$ of dimension  $m$. The Hamiltonian that generates each subunitary $U_j$ is the linear sum of $m$ controls  applied to $m$ generators.  By comparison with the conventional control setting described above (\ref{eqn:qgml:control}), the coefficients $c_j^k$ correspond to $v_{k,j}$.

Because $\Delta$ constitutes the set of generators of the entire Lie algebra $\frak{su}(2^n)$ which in turn acts as the generator of its associated Lie group $\sutwon$, an arbitrary unitary $U \in \sutwon$ can in principle be obtained to arbitrary precision with sufficiently-many products of exponentials. This results from the application of the Baker-Hausdorff-Campbell (BCH) theorem (see definition \ref{defn:alg:Baker-Campbell-Hausdorff} and \cite{nielsen_geometric_2006} for a generalised explication), namely that:
\begin{equation}
    \exp(A)\exp(B) = \exp(A + B + \frac{1}{2}[A,B]+...).
    \label{eqn:qgml:bch}
\end{equation}
The approach in \cite{swaddle_generating_2017, swaddle_subriemannian_2017} is to constrain application to cases where $U$ may be synthesised as a product of a polynomial in $n$ terms, meaning the number of exponentials (subunitaries) required to synthesise $U$ is at most a polynomial function of the number of sub-unitaries $n$. We discuss the effect for machine learning algorithms of increasing $n$ on outcomes such as fidelity measures below.

In the control setting discussed above (in which each $U_j$ is decomposed into its BCH product with coefficients $c_j^k$) each $c^k_j$ sought to be found constitutes some optimal application of the generator $\tau_k$. This is consistent with the result in \cite{khaneja_cartan_2001} (see Appendix (\ref{sec:qgml:onetwobodyoperators})), in which the minimum time for synthesising the target unitary propagator is given by the smallest summation of the coefficients (controls) of the generators $\sum_{i=1}^n |\alpha_i|$ which, in our notation, would sum over all control coefficients for all subunitaries i.e. $\sum_{j=1}^N\sum_{k=1}^m |c_j^k|$. 

It is worth noting that the assumption in \cite{khaneja_cartan_2001} and even \cite{dowling_geometry_2008} and other analytic results in control is that in effect the controls can be applied `instantaneously' such that the minimum time for evolution of a unitary (via the adjoint action of control generators on drift Hamiltonians) is lower-bounded by the evolution driven by the drift Hamiltonian $H_d$. That is, many such control regimes assume that control amplitudes can be applied without energy constraints, which is equivalent to being applicable within infinitesimal time. Often this assumption is justified by the fact that a control voltage may be many orders of magnitude greater than the energy scales of the quantum systems to be controlled. In cases where control amplitudes (for example, voltages) are, in any significant sense, upper-bounded say by energy constraints, then time for optimal synthesis of circuits will of course increase as the assumption of instantaneity will not hold. For our purposes, in a bang bang control scenario and assuming evolution according to any drift Hamiltonian sets a lower-bound on evolution time, we consider the control amplitudes $c_j^k$ as applied for time $h$ rather than instantaneously.

\subsection{One- and two-body terms}  \label{sec:qgml:One- and two-body terms}

 As discussed above, the approach in \cite{swaddle_generating_2017, swaddle_subriemannian_2017} is to in essence circumvent the need for elaborate penalty terms in bespoke metrics to penalise higher-order generalised Pauli geodesic generators (akin to regularisation) by instead simply constraining the control subset, the distribution $\Delta$, to be the Kronecker product of one- and two-body Pauli operators (here $\iota$ ranges over the standard index $0,1,2,3$ of the Paulis including identity): 
\begin{equation}
    \triangle = \text{span}\left\{\frac{i}{\sqrt{2^n}} \sigma_\iota^j, \frac{i}{\sqrt{2^n}} \sigma_\iota^k \sigma_\iota^l \right\}
\end{equation}
where $\sigma_\iota^j$ indicates the n-fold Kronecker product/tensor product of Pauli operators at position $\{1,...,j,...,m\}$ with the two-dimensional identity operator at other indices. 

The underlying approach of the geodesic approximation method in these papers is to seek to learn the inverse map: 
\begin{align}
E^{-1}: \sutwon \to \C^{m \times n} \label{eqn:qgml:inversemap}    
\end{align}
and thus, by doing so, learn the appropriate sequence of control pulses necessary to generate time optimal evolution of unitaries and, consequently, time optimal quantum circuits. The method involves generating training data in the form of normal sub-Riemannian geodesics on $\sutwon$ from $I$ to $U_T$. The exponential product (equation (\ref{eqn:qgml:schrodind})) represents a path $\gamma(t)$ along the $\sutwon$ manifold, however there may be an infinity of paths between $I$ and $U_T$ such that the map $E$ is not injective (or unique, thus minimal), meaning $E^{-1}$ is not well-defined.

\subsection{Generating geodesics}  \label{sec:qgml:Generating geodesics}
To solve this uniqueness problem, \cite{swaddle_generating_2017, swaddle_subriemannian_2017} propose to synthesise paths that approximate minimal normal sub-Riemannian geodesics described above. To generate normal subRiemannian geodesics in $\sutwon$, \cite{swaddle_generating_2017} limit the norm of boundary conditions (a computational efficiency choice) and apply a generalised form of the Pontryagin Maximum Principle \cite{schattler_geometric_2012}. They follow well-established variational approaches in \cite{sachkov_control_2009} where subRiemannian geodesics may be found by minimising the energy (cost) functional: 
\begin{equation}
    \mathcal{E}[\gamma] = \int_0^1 dt \langle \dot{\gamma}(t), \dot{\gamma}(t)\rangle. \label{eqn:qgml:energy}
\end{equation}
Specifically, $\braket{\cdot,\cdot}$ is the restriction of the bi-invariant norm (induced by the inner product on the tangent bundle) to $\Delta \in \frak{su}(2^n)$. For $\su(2)$, this arises from the Killing form which induces a metric on the manifold (see section \ref{sec:geo:KP Problems}). The fibre bundle structure allows partitioning of $T\M$ into horizontal and vertical subspaces (albeit if $\p = \g$ then $V\M = \emptyset$). Equation (\ref{eqn:qgml:energy}) is the energy equation for a horizontal curve specified in equation (\ref{eqn:geo:energyhorizontalcurve}). Here the curve $\gamma(t) \in \M$ (path) varies over $t \in [0,1]$ with tangent vectors to the curve (i.e. along the vector field) given by $\dot{\gamma}(t) \in T\M$. Hence we can see how minimising path length equates to minimisation of energy. This approach uses variational methods (section \ref{sec:geo:variational methods}) to minimise the path length. To contextualise this formulation in Lie theoretic terms, $\gamma(t)$ represent unitaries $U(t) \in SU(2^n)$ and $\dot{\gamma}$ the corresponding tangent (Lie algebraic) vectors. Distance along a path $\gamma(t)$ generated by the tangent vectors (generators) $\dot{\gamma}(t)$ is measured in by subRiemannian (or in the general case, Riemannian) metrics applied to the tangent space (see general exposition in section \ref{sec:geo:Riemannian Manifolds and Metrics}). The other key assumption behind this method is that the applicable metric $g_{\alpha\beta}$ is constant.

The normal subRiemannian geodesic equations arising from minimising the energy functional above can be written in differential form \cite{sachkov_control_2009} as:
\begin{align}
\dot{\gamma}(t) & = u \gamma(t) \label{eqn:qgml:gammadot}\\
\dot{\Lambda} &= [\Lambda, u] \label{eqn:qgml:costate} \\
u &= \text{proj}_\Delta(\Lambda) \label{eqn:qgml:projsimp}
\end{align}
It is worth unpacking each of these terms in order to connect the equations above to the control and geometric formalism above and because they are integrated into the subRiemannian machine learning model detailed below. In control theory formalism (section \ref{sec:geo:Pontryagin Maximum Principle}), equation (\ref{eqn:qgml:gammadot}) is the state equation for state variable $\gamma(t)$, corresponding to equations (\ref{eqn:geo:pontrydotgamma}) and (\ref{eqn:quant:controlsystem1}). In quantum contexts it corresponds to the Schr\"odinger equation, hence we can identify $\gamma(t) \equiv U(t)$ (see equation (\ref{eqn:quant:control-state-eqn})). The $u$ term represents an element of the Lie algebra $u \in \Delta \subset \frak{su}(2^n)$ parameterised by $t \in [0,1]$, i.e. $u: [0,1]\to \frak{su}(2^n)$ with $t \mapsto u(t)$. As such, it represents the generator of evolutions on the underlying manifold $\sutwon$. For each value $t$, the curve $\gamma(t)$ represents an element of the Lie group i.e. $\sutwon$, again parametrised by $t \in [0,1]$. Equation (\ref{eqn:qgml:costate}) is the costate (adjoint) equation with costate (adjoint) variables $\Lambda$ taking values in Lie algebra $\liesutwo$. 
$\Lambda$ represents the costate (adjoint) variable (akin when integrated to a Lagrange multiplier) encoding control constraints of the Pontryagin Maximum Principle. They differ from $u$ in that while $u$ are direct elements of the distribution $\Delta$, $\Lambda(t)$ are elements of the overall Lie algebra $\frak{su}(2^n)$ that are generated by the Lie-bracket between other $\Lambda$ and $u$, hence $\Lambda: [0,1] \to \frak{su}(2^n)$. The relationship with the Lie bracket is instructive in that the Lie bracket also has an interpretation as the Lie derivative (definition \ref{defn:alg:Lie algebraliederivative}), the adjoint equation represents its dynamics.

The time-derivative $\dot{\Lambda}$ refers to how the Lie bracket commutator indicates the change in a vector field along the path $\gamma(t)$. In a control setting, the Lie derivative (definition \ref{defn:alg:Lie algebraliederivative}) tells us how $\Lambda$ changes as it is evolved along curves $\gamma(t)$ generated by elements $u$ of the control subset. For parallel transport along geodesics (section \ref{sec:geo:Parallel transport and horizontal lifts}), as mentioned above, we require this change to be such that the covariant derivative (definition \ref{defn:geo:covariantderivative}) of $\Lambda_0$ as it is parallel transported along the curve is zero, that is:
\begin{align}
    \nabla_{\gamma(t)} \Lambda_0 = 0. \label{eqn:covariant}
\end{align}

The last term (\ref{eqn:qgml:projsimp}) indicates that $u$ resides in the distribution $\Delta$ by virtue of the projection of $\Lambda$ onto the distribution $\Delta$: 
\begin{equation}
    \text{proj}_\Delta(x) = \sum_i \trace( x^\dagger \tau_i) \tau_i \in \Delta. \label{eqn:qgml:projection}
\end{equation}
This projection function is important in that it ensures that the generators of $U_j$ remain within $\Delta$, facilitating the parallel transport of $\Lambda_0$ and that $U_j$ are therefore able to be synthesised from the control subset in our machine learning protocols. 
Here $\gamma(t) \sim U(t)$ and $\dot{\gamma}(t) \sim \Lambda(t)$. The geodesic curves $\gamma(t)$ depend on the initial condition $\Lambda(0)$ (the `momentum' term) with the initial `position' in the manifold being the identity unitary. In the geometric control setting over Lie group manifolds, such as unitary groups, selecting an initial generalised coordinate (akin to `position' in the manifold) and generalised momentum, which in turn amounts to selecting an initial `starting' unitary from the Lie group for the evolution at $t=0$, usually the identity $U(0) \in \sutwon$  along with a starting momentum $\Lambda(0)$ drawn from the associated Lie algebra $\frak{su}(2^n)$. Given these initial operators, the geodesic equations then allow determination of tuples of unitaries and generators (positions in the Lie group manifold, momenta in the Lie algebra) for any particular time  value $t \in [0,1]$. That is, they provide a formula for determining $U(t)$ and $\Lambda(t)$. The distribution (control subset) determines the types of geodesics that may be evolved along. Because the distribution is bracket generating, in principle any curve along $SU(2^n)$ may be synthesised in this way (though not necessarily directly).

As noted in \cite{swaddle_subriemannian_2017}, the above set of equations can be written as a first-order differential equation via
\begin{equation}
    \dot{\gamma}(t) = \text{proj}_\Delta(\gamma(t) \Lambda_0 \gamma(t)^\dagger)\gamma(t) \label{eqn:swadconjugation}
\end{equation}
A first-order integrator (see (\ref{eqn:firstorderintegrator})) is used to solve for $\gamma(t) = U(t)$. It is worth analysing (\ref{eqn:swadconjugation}) in light of the discussion above on conjugacy maps and their relation to time optimal geodesic paths. The $\gamma(t)$ terms in the conjugacy map:
\begin{align}
    \gamma(t) \Lambda_0 \gamma(t)^\dagger \to \Lambda_j 
\end{align}
represent the forward-solved geodesic equations \cite{earp_constructive_2005,swaddle_generating_2017} (and see section \ref{sec:geo:Geodesics and parallelism} along with discussion of KP problem solutions in sections \ref{sec:geo:KP Problems} and \ref{sec:cartan:KP problems for Lambda systems}). Given the initial condition $\Lambda_0$, $\gamma(t)$ here is the cumulative evolved operator in SU$(2^n)$ that is, for time-step $t_j$, we have:
\begin{align}
    \gamma(t_j) = \prod_{i=N}^j U_j
\end{align}
In this respect conjugating $\Lambda_0$ by the $\gamma(t_j)$ is equivalent to adopting a co-rotating basis or so-called moving frame for the Lie algebra (not dissimilar to how conjugation acts in a standard Euler decomposition such as in \cite{boozer_time-optimal_2012}). Projecting the conjugated $\Lambda_0$ back onto the horizontal space $H\M$ (i.e. $\Delta$) then defines $\Lambda_j$ as $\Lambda_0$ parallel transported along the approximate geodesic. This algorithmic approximation thus achieves (a) a way to parallel transport $\Lambda_0$ and (b) a decomposition method for generating $U_j$. 
The continuous curve $\gamma(t)$ is discretised via partitioning the parametrisation interval into $N$ temporal segments. A first-order integrator is then utilised to solve the differential equation. In continuous form, the integration equation for the unitary propagator applied over interval $\Delta t$ takes the time-dependent form (\ref{eqn:scrhodtime}):
\begin{equation}
    U(t) = \exp\left(-i \int_0^{\Delta t} \text{proj}_\Delta(\gamma(t) \Lambda_0 \gamma(t)^\dagger) dt\right)
\end{equation}
(note that in the accompanying code, the imaginary unit is incorporated into $\Delta$).
In the discrete case, the curve $\gamma(t)$ is partitioned into $N$ such segments of equal parameter-interval $h=\Delta t$, indexed by $\gamma_j$ where $j = 1,...,N$. The first-order integration resolves to:
\begin{equation}
    U_j = \exp(-ih \text{proj}_\Delta(\gamma_j \Lambda_0 \gamma_j^\dagger)) \label{eqn:firstorderintegrator}
\end{equation}
where here $U_j$ are unitaries that forward-solve the geodesic equations, represented in terms of the Euler discretisation \cite{swaddle_subriemannian_2017}:
\begin{align}
    \gamma_{j+1} &= U_j \gamma_j\\
    &=\exp(-ih \text{proj}_\Delta(\gamma_j \Lambda_0 \gamma_j^\dagger)) \gamma_j
\end{align}
where, again to reiterate, $\gamma_{j+1}$ represents the cumulative unitary propagator at time $t_{j+1}$ and $U_j$ represents the respective unitary that propagates $\gamma_j \to \gamma_{j+1}$. The Hamiltonian $H_j$ for segment $U_j$ is given by the projection onto $\Delta$:
\begin{align}
    H_j&=\text{proj}_\Delta(\gamma_j \Lambda_0 \gamma_j^\dagger) 
\end{align}
and is applied for time $h$ (though see Appendix (\ref{sec:qgml:boozerswaddlecomparison}) below for nuances regarding the interpretation of $h$ and time given the imposition of $||\text{proj}_\Delta(\Lambda_0)||=||u_0||=1$). A consequence of these formal solutions is that each $H_j$ is constrained to be generated from $\Delta$. This does not mean that only unitaries directly generated by $\Delta$ are reachable, as the action of unitaries (see (\ref{eqn:qgml:bch})) gives rise to generation of generators outside $\Delta$. It is, however, of relevance to the construction of machine learning algorithms seeking to learn and reverse-engineer geodesic approximations from target unitaries $U_T$. The consequence of this requirement is that the control functions for machine learning algorithms need only model controls for generators in $\Delta$.

%Experimental design
\section{Experimental Design} \label{sec:qgml:expdesign}
In this section, we detail our experimental design and implementation of various machine learning models that build upon and extend work in \cite{swaddle_generating_2017} applying deep learning to the problem of approximate geodesic quantum circuit synthesis. The overall objective of our experiments was to compare the performance of variety of different machine learning architectures in simulated environments in terms of generating time-optimal quantum circuits by being trained on approximate normal subRiemannian geodesic in $SU(2^n)$ Lie groups. While other methods, such as the `shooting' method \cite{wang_quantum_2015} provide alternative means of generating geodesic data,  it was shown in \cite{swaddle_generating_2017} that such methods particularly for higher-order $SU(8)$ cases led to considerable increases in runtime compared with neural network approaches. In any case, as our primary focus in this Chapter  was on investigating the utility of greybox approaches to geometric machine learning architectures, such alternative methods (for example, implementing the methods of \cite{dowling_geometry_2008, gu_quantum_2008}) of generating geodesics or approximations thereto were not canvassed. 

\subsection{Experimental objectives} \label{sec:qgml:Experimental objectives}
Synthesis of geodesics for use as training data in the various machine learning protocols utilised an adapted subRiemannian approach from \cite{swaddle_generating_2017} and \cite{boozer_time-optimal_2012}. Our overall objectives required the ability to decompose a target unitary $U_T$ in order to generate the sequence $(U_j)$ from $U_T$ and, in turn, render each $U_j$ synthesisable from a set of control amplitudes applied to generators from $\Delta$. There are a variety of classical deep learning approaches that can be adopted to solving this type of supervised learning problem, including:
\begin{enumerate}
    \item \textit{Standard neural network models}: neural network models (section \ref{sec:ml:neuralnetworkschema}) adopt variations on simply connected or other architecture that seeks to learn an optimal configuration of hidden representations in order to output (and thus generate) the desired sequence. On their own such models tend to be \textit{blackbox} models, in which algorithms are trained to learn a mapping from inputs (training data) to outputs (labels) without any necessary interpretability or clarity about the nature of the mapping or intermediate features being generated by the network;
    \item \textit{Generative models}: generative models, such as variational autoencoders (VAEs) and generative adversarial networks (GANs) and transformers seek to learn the underlying distribution of ground truth data, then use that learnt distribution to generate new similar data (distributions); and
    \item \textit{Greybox models}: greybox models, as discussed in section \ref{sec:ml:Greybox machine learning} and further on, seek to combine domain knowledge (such as laws of physics), also known as \textit{whitebox} models, together with blackbox models into a hybrid learning protocol. The particular examples we focus on are variational (parametrised) quantum circuit models.
\end{enumerate}

The practical engineering, target inputs and outputs of the various machine learning models differs depending upon metrics of success and use case. For a typical quantum control problem, the sought output of the architecture is actually the sequence of control pulses $(c_j)$ to be implemented in order to synthesise the target unitary (i.e. apply a gate in a quantum circuit). The target unitary $U_T$ is typically one of one or more inputs into the model architecture. 

The approach in \cite{swaddle_generating_2017} is blackbox in nature. In that case, the input to their model was (for their global decomposition algorithm) $U_T$ with label data the sequence $(U_j)$. The aim of their algorithm, a multi-layered Gated Recurrent Unit (GRU) RNN, was to learn a protocol for decomposing arbitrary $U_T \in SU(2^n)$ into the an estimated sequence $(\hat{U}_j)$ (sequences are indicated by parentheses). The individual $\hat{U}_j$ are then fed into a subsequent simple feed-forward fully-connected neural network whose output is an estimate sequence of controls $(\hat{c}_j)$ (where $c_j$ is used as a shorthand for each control amplitude $c_j^k$ applied to generators $\tau_k$ for segment $j$ and parentheses indicate a sequence) for generating each $\hat{U}_j$ using $\tau_k \in \Delta$. While $\hat{U}_j$ need not itself (and is unlikely to) be exactly unitary, so long as the controls $(\hat{c}_j)$ are sufficient to then input into (\ref{eqn:qgml:swaddleunitaryseq}) to generate unitary propagators, then the objective of learning the inverse mapping (\ref{eqn:qgml:inversemap}) has been achieved. No guarantees of unitarity from the learnt model are provided in \cite{swaddle_generating_2017}, instead there is a reliance upon simply finding (\ref{eqn:qgml:inversemap}) in order to provide $(\hat{c}_j)$.
As we articulate below, while this approach in theory is feasible, in practice where unitarity is required within the network itself (as per our greybox method driven by batch fidelity objective functions), a more detailed engineering framework for the networks is required. It is for this reason that we adopt a greybox approach where guarantees of unitary can be obtained via utilising a Lie-theoretic approach in which controls are learnt parameters, rather than the specific entries $(a_{ij} \in \C)$ of a unitary matrix group element.

\subsection{Models} \label{sec:qgml:Models}

\subsubsection{Geodesic deep learning architectures} \label{sec:qgml:Geodesic deep learning architectures}
Three deep learning architectures were applied to the problem of learning approximations to geodesics (definition \ref{defn:geo:Geodesic}) in SU$(2^n)$: 
\begin{enumerate}
    \item a simple multi-layer feed-forward fully-connected (FC) network (definition \ref{sec:qgml:Feed-forward neural networks1}) model implementing adaptation of (\ref{eqn:qgml:swaddleunitaryseq}) that learns controls $(c_j)$ trained against $(U_j)$  (the FC Greybox model);
    \item a greybox RNN model using GRU cells \cite{lin_geodesic_2014} in which controls $(\hat{c}_j)$ for estimated Hamiltonians $\hat{H}_j$ are learnt by being trained against $(U_j)$ (the GRU RNN Greybox model); and
    \item a fully-connected subRiemannian greybox model (the SubRiemannian model) which generates controls $(\hat{c}_j)$ by concurrently implementing (\ref{eqn:qgml:projection}) and learning the control pulses $c_{\Lambda_0}$ for the initial generator $\Lambda_0$ (that is, a model that replicates the subRiemannian normal geodesic equations while learning initial conditions for respective geodesics).
\end{enumerate}
Each model, described in more detail below, took as initial inputs the target unitary $U_T$ together with unitary sequences $(U_j)$ (as per (\ref{eqn:qgml:schrodind} above)). Each new model uses various neural network architectures to generate controls $(\hat{c}_j)$ for generators $\tau_k \in \Delta$ (where $H_j = \sum_k \hat{c}^k_j \tau_k)$ which are in turn evolved via customised layers implementing (\ref{eqn:qgml:schrodind}) in order to generate estimates $(\hat{U}_j)$. These estimates $(\hat{U}_j)$ were then compared using MSE loss using an operator fidelity metric against a vector of ones (as perfect fidelity will result in unity). A second metric of average operator fidelity was also adopted to provide a measure of how well on training and validation data (see section \ref{sec:ml:Regularisation and Hyperparameter tuning} for regularisation discussion) the networks were able to synthesise $U_j$ with respect to the estimated $\hat{U}_j$. 

Unlike the segmented neural networks for learning control pulses to generate specific $U_j$, the variable weights (and units) of the neural network were constructed with greater flexibility.
The FC Greybox, SubRiemannian and GRU RNN Greybox models tested were each tested. Note that MSE$((U_j), (\hat{U}_j))$ refers to the batch fidelity MSE described below. For each model, the inputs to the model were the target unitary $U_T$ and its corresponding sequence of subunitaries $(U_j)$. As detailed below, the penultimate layer of each model outputs an estimated sequence of subunitaries $(\hat{U}_j)$. This estimated sequence was then compared to the true sequence $(U_j)$ using operator fidelity (see (\ref{eqn:qgml:fidelity} below). This estimate of fidelity $F((U_j), (\hat{U}_j))$ was then compared using MSE against a vector of ones (i.e. ideal fidelity) which formed the label for the models. As described below, the customised nature of the models meant intermediate outputs, including estimated control amplitude sequences $(\hat{c}_j)$, Hamiltonian estimate sequences $(\hat{H}_j)$ and $(\hat{U}_j)$ were all accessible. The general architectural principles of this greybox approach are discussed in section \ref{sec:ml:Greybox machine learning}.

\subsubsection{Methods} \label{sec:qgml:Methods}
Generation of training data for each of the models tested was achieved via implementing the first-order subRiemannian geodesic equations in Python, adapting Mathematica code from \cite{swaddle_generating_2017}. A number of adaptations and modifications to the original format of the code were undertaken: (a) where in \cite{swaddle_generating_2017}, unitaries were parameterised only via their real components (to effect dimensionality reduction) (relying upon an analytic means of recovering imaginary components \cite{swaddle_subriemannian_2017}), in our approach the entire unitary was realised such that $U = X + iY$. This was adopted to improve direct generation of target unitaries of interest and to facilitate fidelity calculations. Using equation (\ref{eqn:alg:complextorealmx}) such the unitaries became expressed in terms of:
\begin{equation}
    \hat{U} = \pmat{X}{-Y}{Y}{X}
    \label{eqn:qgml:complextorealmx}
\end{equation}
where $\dim \hat{U} = \dim SU(2^{n+1})$; and (b) in certain iterations of the code for $\Lambda_0: [0,1] \to SU(2^n)$, the coefficients of the generators were derived using tanh activation functions:
\begin{align*}
\tanh(x) = \frac{e^{x} - e^{-x}}{e^{x} + e^{-x}}
\end{align*}
(with a range $-1$ to $1$ rather than the range $[0,1]$) that allowed elements of unitaries to be more accurately generated and also to test (see Appendix \ref{sec:qgml:boozerswaddlecomparison}) whether the first order integrative approach did indeed generate equivalent time-optimal holonomic paths (as in \cite{boozer_time-optimal_2012}) (see section \ref{defn:geo:holonomy}). Using tanh activation functions enabled better approximation of the relevant time-optimal control functions which give rise to the generator coefficients (for example, to reproduce the holonomic paths of \cite{boozer_time-optimal_2012}, one needs the coefficients to emulate the range of the sine and cosine control functions which characterise the time-optimal evolution in that case).

Furthermore, (c) one observation from \cite{swaddle_generating_2017} was that the training data generated unitaries relatively proximal to the identity i.e. curves that did not evolve far from their origin. This is a consequence of the time interval $\Delta t$ for each generator i.e. $\Delta t = h = 1/n_{seg}$ where $n_{seg}$ is the number of segments. The consequence of this for our results was that training and validation performance was very high for $U_T$ close to the identity (that is, similar to training sets), but declined in cases for $U_T$ further away (in terms of metric distance) from the origin. This is consistent with \cite{swaddle_generating_2017} but also consistent with the lack of generalisation performance in their model. As such, in some iterations of the experiments we scaled-up $h$ by a factor in order to seek $U_T$ which were more spread-out across the manifold. Other experiments undertaken sought to increase the extent to which training data covered manifolds by increasing the number of segments $U_j$ of the approximate geodesic while keeping $h$ fixed (between 0 and 1). We report on scale and segment number dependence of model performance below. 

In addition to these modifications, in certain experiments we also supplemented the \cite{swaddle_generating_2017} generative code with subRiemannian training data from a Python implementation of Boozer \cite{boozer_time-optimal_2012}. In this case, given the difficulty of numerically solving for arbitrary unitaries using Boozer's approach (whose solutions in the paper rely upon analytic techniques), we generated rotations about the $z$-axis by arbitrary angles $\theta$ (denoted $\eta$ in \cite{boozer_time-optimal_2012}), then rotated the entire sequence of unitaries $U_j$ by a random rotation matrix. This has the effect of generating sub-Riemannian geodesics with arbitrary initial boundary conditions and rotations about arbitrary axes, which in turn provided a richer dataset for training the various neural networks and machine learning algorithms.

For $SU(2)$, the bracket-generating set $\Delta$ can be any two of the three Pauli operators. Different combinations for $\Delta$ were explored as part of our experimental process. Our experiments focused on setting our control subset $\Delta = \{\sigma_x,\sigma_y\}$ as this allowed ease of comparison with analytic results of \cite{boozer_time-optimal_2012} and to enable assessment of how each machine learning model performed in cases where control subsets were limited, which was viewed as being more realistic in experimental contexts. Note that this corresponds to the control problem being a subRiemannian one. It aligns also with the Cartan decomposition (definition \ref{defn:alg:cartandecomposition}) of $\su(2)$ (see also Chapter \ref{chapter:Time optimal quantum geodesics using Cartan decompositions} for a comparison with analytical methods).

Test datasets for generalisation, where the trained machine learning models are tested against out of sample data, were generated using the same subRiemannian generative code above. We also sought to test, for each of $SU(2)$, $SU(4)$ and $SU(8)$, the efficacy of the models in generating sequences $(\hat{U}_j)$ that accurately evolved to randomly generated unitaries from each of those groups. The testing methodology for geodesic approximation models comprised input of the target $U_T$ of interest into the trained model with the aim of generating control pulses $(\hat{c}_j)$ from which $(\hat{U}_j)$ (and thus $\hat{U}_T$) could be generated. 

In each of the models, a customised layer generates candidate controls $(\hat{c}_j)$ in the form of variable weights which are updated during each iteration (epoch) of the model using TensorFlow's autodifferentiation architecture (which streamlines updating of variable weights). These control amplitudes are then fed into a customised Hamiltonian estimation layer which applied $(\hat{c}_j)$ to the respective generators in $\Delta$. The output of this Hamiltonian estimation layer is a sequence of control Hamiltonians $(\hat{H}_j)$ which are input into a second customised layer which implemented quantum evolution (i.e. equation (\ref{eqn:qgml:schrodind})) in order to output $(\hat{U}_j)$. A subsequent custom layer takes $(\hat{U}_j)$ and the true $(U_j)$ as inputs and calculated their fidelity i.e. it takes as inputs batches of estimates $(\hat{U}_j)$ and ground truth sequence $(U_j)$ and calculates the operator fidelity (see section \ref{sec:quant:Fidelity function}) of each $\hat{U}_j$ and $U_j$ via:
\begin{align}
    F(\hat{U}_j,U_j) & = |\Tr(\hat{U}_j^\dagger U_j)|^2/d^2
    \label{eqn:qgml:fidelity}
\end{align}
where $d = \dim U_j$. It should be noted that in this case, the unitaries are ultimately complex-valued (rather than in realised form) prior to fidelity calculations. The outputs of the fidelity layer are the ultimate output (labels) of the model (that is, the output is a batch-size length vector of fidelities). These outputs are compared to a label batch-size length vector of ones (equivalent to an objective function targeting unit fidelity). The applicable cost function used was standard MSE but applied to the difference between ideal fidelity (unity) and actual fidelity: 
\begin{align}
    C(F,1) = \frac{1}{n}\sum_{j=1}^n (1 - F(\hat{U}_j,U_j))^2
    \label{eqn:qgml:batchfidelity}
\end{align}
where here $n$ represents the chosen batch size for the models, which in most cases was 10 or a multiple thereof. It should also be noted that this approach, which we name `batch fidelity', contributed significantly to improvements in performance: previous iterations of our experiments had engineered fidelity itself as a direct loss-function using TensorFlow's low-level API, which was cumbersome,  lacking in versatility and resulting in limited improvement by comparison with batch fidelity approaches. A standard ADAM optimizer \cite{kingma_adam_2017} (with $\alpha=10^{-3}$) was used for all models.

\subsubsection{Geodesic architectures: Fully-connected Greybox} \label{sec:qgml:Geodesic architectures: Fully-connected Greybox}
\begin{figure}[h]
\begin{center}

\includegraphics[width=\linewidth]{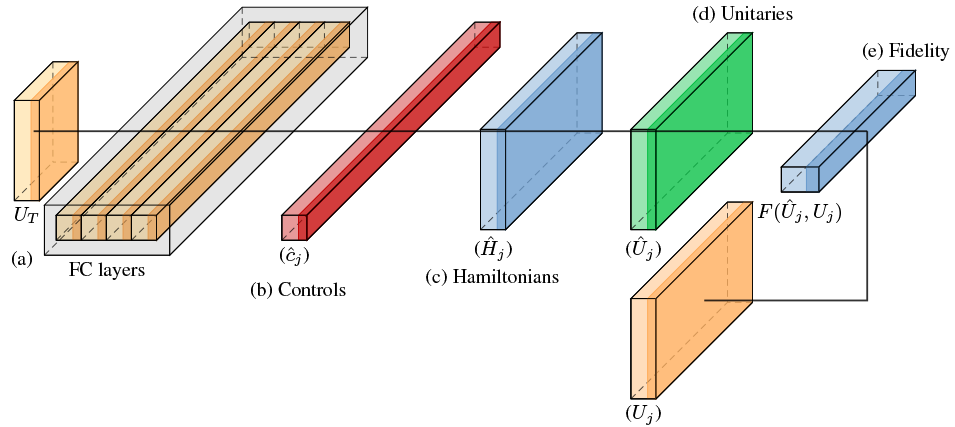}

\caption{Schema of Fully-Connected Greybox model: (a) realised $U_T$ inputs (flattened) into a stack of feed-forward fully connected layers with ReLU activations and dropout of 0.2; (b) the final dense layer in this stack outputs a sequence of controls $(\hat{c}_j)$ using tanh activation functions; (c) these are fed into a custom Hamiltonian estimation layer produce a sequence of Hamiltonians $(\hat{H}_j)$ using $\Delta$ ; (d) these in turn are fed into a custom quantum evolution layer implementing the time-independent Schr{\"o}dinger equation to produce estimated sequences of subunitaries $(\hat{U}_j)$ which are fed into (e) a final fidelity layer for comparison with the true $(U_j)$. Intermediate outputs are accessible via submodels in TensorFlow.}
\label{diagram:fcgreybox}
\end{center}

\end{figure}

To benchmark the performance of the greybox models, a blackbox model that sought to input $U_T$ and output $(\hat{c}_j)$ was constructed using a simple deep feed-forward fully-connected layer stack taking as an input $U_T$ and outputting a sequence of estimated control amplitudes $(\hat{c}_j)$. A schema of the model is shown in Figure (\ref{diagram:fcgreybox}). Subsequent customised layers construct estimates of Hamiltonians by applying $(\hat{c}_j)$ to the generators in $\Delta$, which are in turn used to generate subunitaries $\hat{U}_j$.

The stack comprised an initial fully-connected feed forward network with standard clipped ReLU activation functions (with dropout $\sim$ 0.2) that was fed $U_T$. This stack fed into a subsequent dense layer outputting $(\hat{c}_j)$ utilised tanh activation functions. Standard MSE loss against the label data $(c_j)$ was used (akin to the basic GRU in \cite{swaddle_generating_2017}). The sequence $(U_j)$ was then reconstructed using $(\hat{c}_j)$ external to the model and fidelity assessed separately. In this variation of the feed-forward fully-connected model, a basic greybox approach that instantiated the approximation (\ref{eqn:qgml:schrodind}) was adopted. 

As we discuss in Appendix \ref{chapter:Background: Classical, Quantum and Geometric Machine Learning}, greybox approaches \cite{youssry_modeling_2020, youssry_modeling_2020} represent a hybrid synthesis of `blackbox' approaches to machine learning (in which the only known data are inputs and outputs to an typical machine learning algorithm whose internal dynamics remain unknown or uninterpretable) and `whitebox' approaches, where prior knowledge of systems, such as knowledge of applicable physical laws, is engineered into algorithms. Practically, this means customising layers of neural network architecture to impose specific physical constraints and laws of quantum evolution in order to output estimated Hamiltonians and unitaries. The motivation for this approach is that it is more efficient to engineer known processes, such as the laws of quantum mechanics, into neural network architecture rather than devote computational resources to requiring the network to learn what is already known to be true (and necessary for it to function effectively) such as Schr{\"o}dinger's equation.

%========Greybox
The greybox architecture used to estimate the control pulses necessary to synthesise each $U_j$ is set-out below. This is achieved by using $\tau_i \in \Delta$ to construct estimates of Hamiltonians $\hat{H}$ and unitaries $\hat{U}$. The inputs (training data) to the network are twofold: firstly, unitaries $\hat{U}$ generated by a Hamiltonian composed of generators in $\Delta$ with uniform randomly chosen coefficients $c_j^k \in [-1,1]$, where the negative values represent, intuitively, tangent vectors pointing in the opposite direction along a Lie group manifold:
\begin{align}
    \hat{H}_j &= \sum_{k=1}^{\dim|\Delta|} \hat{c}_j^k \tau_k \qquad \hat{c}_j^k \sim U[-1,1] \\
    \hat{U}_j &=  \exp(-h\hat{H}_j)
\end{align}
(recalling $i$ is absorbed into $\tau_k$ for convenience). The coefficients $c_j$  are constructed via successive feed-forward fully-connected dense layers before being applied to the generators: they are the optimal controls being sought and represent updatable weights in the network. Secondly, a tensor of training $U_j$ (generated extrinsically from $\Delta$) is separately input into the network. 

Because TensorFlow layers require output/input as real floats, $U_j$ is separated into a real $\text{Re}(U_j)$ and imaginary $\text{Im}(U_j)$ parts which are subsequently recombined in a customised layer. The specific controls being learnt by the network were accessible using standard TensorFlow techniques that allow access to intermediate output layers (we do this by creating separate models whose output is the output of an intermediate layer). This approach allows access to intermediate outputs, such as $\hat{H}_j$ and $\hat{U}_j$. 

For training and validation of the model, we utilised the following inputs and outputs:
\begin{itemize}
    \item Inputs: $U_T$ (target unitary) and $(U_j)$ the training sequences $(U_j)$; and
    \item Outputs: Fidelity $F(\hat{U}_j, U_j) \in [0,1]$, representing the fidelities of the estimate of the sequence $(\hat{U}_j)$ from those in the training data.
\end{itemize}

In the model, $U_T$ is fed into an initial  feed-forward fully-connected layer which is in turn connected to a dense flattened layer to produce a coefficient $\hat{c}_j^k$ for each generator $\tau_k \in \Delta$ in (\ref{eqn:qgml:swaddleunitaryseq}). Thus for a model with $n_{seg}$ segments and $\dim \Delta = d$, a total of $n_{\text{seg}} \times d$ coefficients $\hat{c}^k$ are generated. 

These are then applied to the generators $\tau_k$ in a customised Hamiltonian estimation layer. The output of this layer is then input into a unitary layer which that generates each subunitary:
\begin{align}
    \hat{U}_j = \exp(h \hat{c}_j^k \tau_k)
\end{align}
with summation implied in order to generate the estimated sequence of unitaries $(\hat{U}_j)$. A subsequent custom layer calculates $F(\hat{U}_j, U_j)$. The output of this layer is a (batched) vector of fidelities which are compared against a label of ones using a standard MSE loss function and Adam optimiser. This model is the simplest of the greybox models adopted in our experiments. Pseudocode for the Fully-connected Greybox model is set-out in section \ref{sec:qgml:Algorithmic architectures} below.

%===========FC Greybox diagram

%===========LSTM GRU

\subsubsection{Geodesic architectures: GRU RNN Greybox} \label{sec:qgml:Geodesic architectures: GRU RNN Greybox}
The second category of deep learning architectures explored in our experiments were RNN algorithms \cite{goodfellow_deep_2016,hochreiter_long_1997}. LSTMs are a popular deep learning tool for the modelling of sequential datasets, such as time-series or other data in which successive data depends upon preceding data points. The interested reader is directed to a number of standard texts \cite{goodfellow_deep_2016} covering RNNs architecture in general for an overview. In short, RNNs are modular neural networks comprising `cells', self-enclosed neural networks consisting of inputs of training data, outputs and a secondary input from preceding cells. For sequential or time-series data, a sequence of modules are connected for each entry or time-step in the series, $j$. The intuition behind RNNs, such as Long-Short Term Memory networks (LSTMs), is that inputs from previous time-step cells or `memories' can be carried forward throughout the network, enabling it to more accurately learn patterns in sequential non-Markovian datasets.  The original application of GRU RNNs to solving the geodesic synthesis problem was the focus of \cite{swaddle_generating_2017}. That work utilised a relatively simple network of GRU layers, popular due to efficiencies it can provide to training regimes.

\begin{figure}[h]
\captionsetup[figure]{width=\textwidth}
\centering
\includegraphics[width=\linewidth]{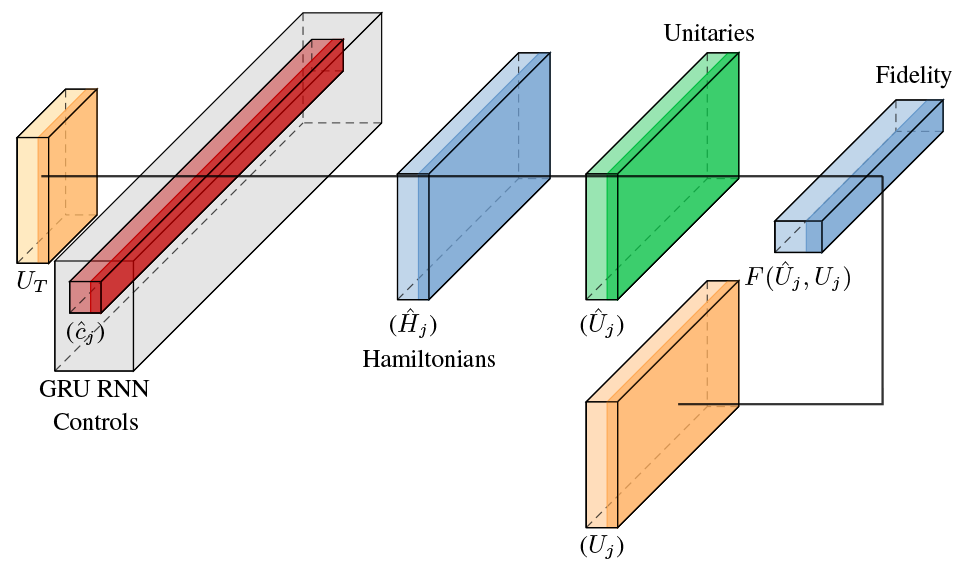}
\caption{Schema of GRU RNN Greybox model: (a) realised $U_T$ inputs (flattened) into a GRU RNN layer comprising GRU cells in which each segment $j$ plays the role of the time parameter; (b) the output of the GRU layer is a sequence of control pulses $(\hat{c}_j)$ using tanh activation functions; (c) these are fed into a custom Hamiltonian estimation layer to produce a sequence of Hamiltonians $(\hat{H}_j)$ by applying the control amplitudes to $\Delta$; (d) the Hamiltonian sequence is fed into a custom quantum evolution layer implementing the time-independent Schr{\"o}dinger equation to produce estimated sequences of subunitaries $(\hat{U}_j)$ which are fed into (e) a final fidelity layer for comparison with the true $(U_j)$. Intermediate outputs are accessible via submodels in TensorFlow.}
\label{diagram:GRU}
\end{figure}

In the present case, the aim of the GRU RNN is to generate a model that can decompose a target unitary $U_T$ into a sequence $U_j$ reachable from $I \in SU(2^n)$. The GRU RNN seeks to reverse-engineer the geodesically approximate sequence of subunitaries through a learning protocol that is itself sequential. In this model, the index $j$ of the sequence $(U_j)$ is akin to a `time slice'. At each slice $j$, the unitary $U_j$ is input into the corresponding GRU cell $G_j$ (one for each segment). A schema of the model is shown in Figure (\ref{diagram:GRU}). The cell activation functions were set to the tanh function given its range of $[-1,1]$ accorded with the range of elements of desired subunitaries. The output of the GRU cell $G_j$ then becomes, with a certain probability, an input into the successor GRU cell $G_{j+1}$ which also takes as an input the successor subunitary $U_{j+1}$ where the function tanh over operators (matrices) is understood in the usual way (see Appendix \ref{sec:qgml:GRUNN} for background).  

The output of the GRU RNN is itself a sequence of control amplitudes $(\hat{c}_j)$ from which were then applied to generators in $\Delta$ in a custom Hamiltonian estimation layer in TensorFlow in order to construct Hamiltonian estimates and quantum evolution layers to generated estimated subunitaries $\hat{U}_j$. As with other models above, the sequence $(\hat{U}_j)$ was then input into a customised batch fidelity layer for comparison against the corresponding $(U_j)$. Our variations of the basic GRU RNN differed in that rather than simply concatenating and flattening all $(U_j)$ into a long single vector for input into a single GRU cell, each $U_j$ was associated with time-slice $j$, the objective being that, a discretised output of $(\hat{U}_j)$.

%=========CHECKPOINT 2

Our main adaptation to the standard GRU RNN model was to include customised layers as described above such that the output $(\hat{U}_j)$ were themselves generated by inputting learnt coefficients $(\hat{c}_j)$ into custom Hamiltonian estimation layers (containing generators from $\Delta$), followed by quantum evolution layers (exponentiation) to generate the estimates. In this respect we followed novel approaches developed in \cite{youssry_modeling_2020, youssry_beyond_2020}, particularly around sequential Hamiltonian estimation (though we restricted ourselves throughout to square pulse forms for $(\hat{c}_j)$ only instead of also trialling Gaussian pulses). Here the aim of the GRU is to replicate the algorithmic approach in \cite{swaddle_generating_2017}, for example learning how $\Lambda_0$ is conjugated by $U_j$ in the generation of $U_{j+1}$. Again, this represents in effect a form of `whitebox' engineering in which assured knowledge, namely how unitaries approximately evolve under the cumulative action of subunitaries, is encoded into customised layers within the network (rather than having the network `deduce' this process). Pseudocode for the GRU RNN Greybox model is set-out in section \ref{sec:qgml:Geodesic architectures: GRU RNN Greybox}. 

%===GRU figure

%==========SubRiemanninan Model

\subsubsection{Geodesic architectures: SubRiemannian model}
\label{sec:qgml:Geodesic architectures: SubRiemannian model}
\begin{figure}[h]
\captionsetup[figure]{width=\textwidth}
\centering

\includegraphics[width=\textwidth]{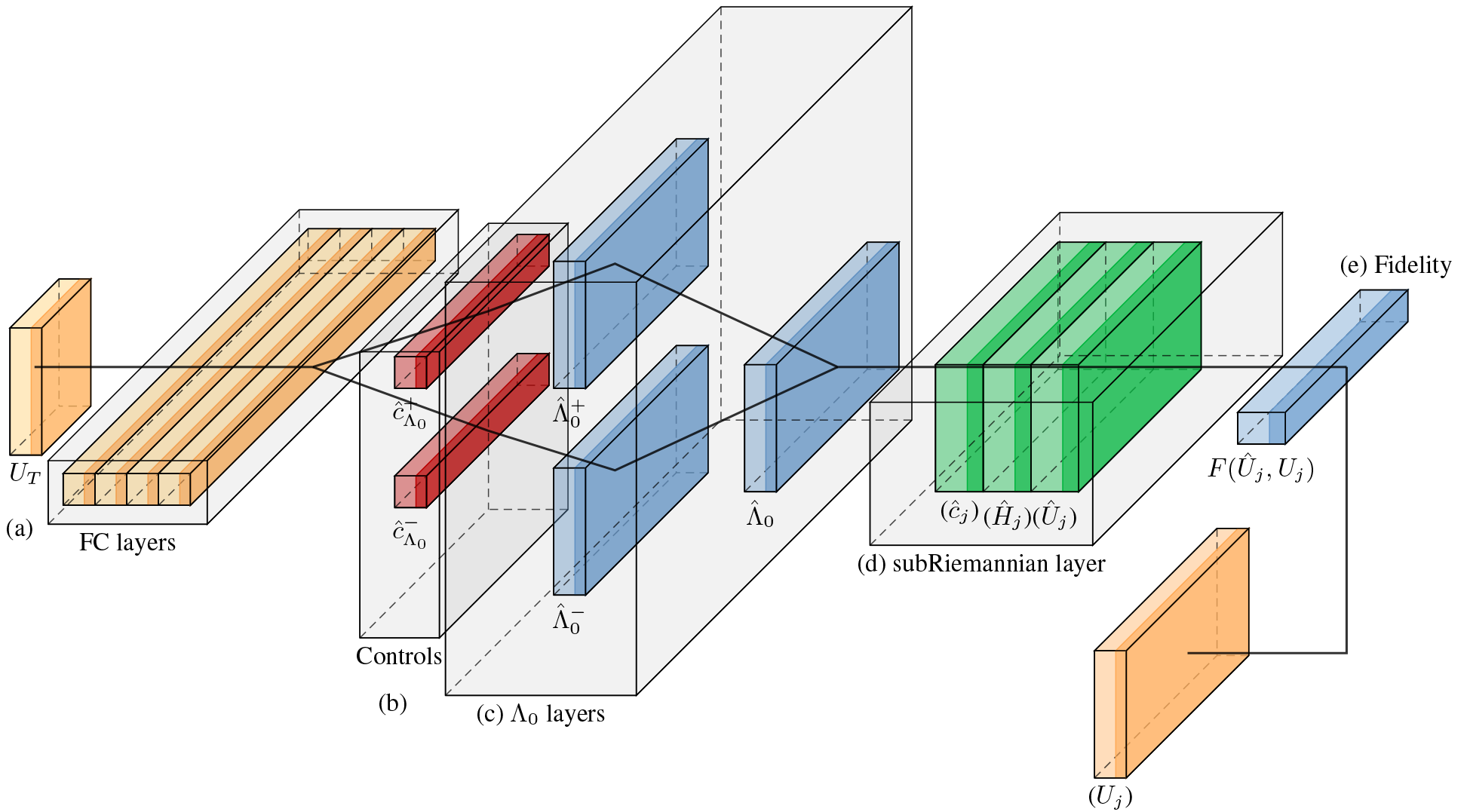}

\caption{Schema of SubRiemannian model: (a) realised $U_T$ inputs (flattened) into a set of feed-forward fully-connected dense layers (with dropout $\sim$ 0.2); (b) two layers (red) output sets of control amplitudes for estimating the positive $(\hat{c}^{+}_{\Lambda_0})$ and negative $(\hat{c}^{-}_{\Lambda_0})$ control amplitudes using tanh activation functions; (c) these are fed into two custom Hamiltonian estimation layers to produce the positive $\hat{\Lambda}^{+}_0$ and negative $\hat{\Lambda}^{-}_0$ Hamiltonians for $\Lambda_0$ using $\Delta$ or $\frak{su}(2^n)$ that are combined into a single Hamiltonian estimation $\hat{\Lambda}_0$; (d) $\hat{\Lambda}_0$ is fed into a custom subRiemannian layer which generates the control amplitudes $(\hat{c}_{j})$, Hamiltonians $(\hat{H}_{j})$ and then implements the time-independent Schr{\"o}dinger equation to produce estimated sequences of subunitaries $(\hat{U}_j)$ which are fed into (e) a final fidelity layer for comparison with the true $(U_j)$. Intermediate outputs (a) to (d) are accessible via submodels in TensorFlow. The SubRiemannian model resulted in average gate fidelity when learning representations of $(U_j)$ of over 0.99 on training and validation sets in comparison to existing GRU \& FC Blackbox models which recorded average gate fidelities of $\approx 0.70$, demonstrating the utility of greybox machine learning models in synthesising learning unitary sequences.}
\label{diagram:subR}
\end{figure}

The third model (the SubRiemannian model) architecture developed in our experiments expanded upon principles of greybox network design and subRiemannian geometry in order to generate approximations to subRiemannian geodesics. A schema of the model is shown in Figure (\ref{diagram:subR}). The choice of architecture was motivated by insights from the variational means of generating subRiemannian geodesics themselves \cite{sachkov_control_2009, frankel_geometry_2011}, namely that a machine learning model that effectively leveraged known or assumed knowledge regarding evolution of unitaries and their generation would perform better than more blackbox-oriented approaches. In essence the model algorithmically implemented the recursive method of generating approximate subRiemannian geodesics which relies only upon $\Lambda_0$ and $\Delta$ and relied upon learning the choice of initial condition $\Lambda_0$, rather than having to learn how to construct Hamiltonians or evolve according to the laws of quantum mechanics (which were instead dealt with via customised layers). 

The method in \cite{swaddle_generating_2017} assumes certain prior knowledge or information in order to generate output, including: (a) the distribution $\Delta$ i.e. the control subset in an experiment of interest; (b) the form of variational equations giving rise to normal subRiemannian geodesics; (c) hyperparameters, such as knowledge of the number of segments in each approximation and time-step $h$. The form of (\ref{eqn:qgml:projection}) provides (via the trace operation) the control amplitudes $\hat{c}_j$ for each generator for Hamiltonian $\hat{H}_j$. Once the initial generator $\Lambda_0$ is selected, given these prior assumptions, the output of geodesic approximations is predetermined. This characterisation was then used to design the network architecture: the inputs to the network were target unitaries $U_T$, together with the associated sequence $(U_j)$ and control subset $\Delta$.

The aim of the network was to, given the input $U_T$, learn the control amplitudes for generating the correct $\Lambda_0$ which, when input into the subRiemannian normal geodesic equations, generated the sequence $(\hat{U}_j)$ from which $U_T$ could be obtained (thus resulting in a global decomposition of $U_T$ into subunitaries evolved from the identity). Recall that $\Lambda_0$ is composed from $\frak{su}(2^n)$ or $\Delta$ depending on use case (the original paper \cite{swaddle_generating_2017} selects $\Lambda_0 \in \frak{su}(2^n)$). This generated $\Lambda_0$ was then input into a recursive customised layer performing the projection operation (\ref{eqn:qgml:projection}) that outputs estimated Hamiltonians, followed by a quantum layer that ultimately generated the sequence $(\hat{U}_j)$. The sequence $(\hat{U}_j)$ was then input into a batch fidelity layer for comparison against the true $(U_j)$.  Once trained, the network could then be used for prediction of $\Lambda_0$, $(U_j)$, the sequence of amplitudes $(c_i)$ and $(\hat{U}_j)$, each being accessible via the creation of sub-models that access the respective intermediate custom layer used to generate such output. Pseudocode for the SubRiemannian model is set-out in section (\ref{sec:qgml:SUB}). 

As we discuss in our results section, this architecture provided among the highest-fidelity performance which is unsurprising given that it effectively reproduces the subRiemannian generative method in its entirety. One point to note is that, while this architecture generated the best performance in terms of fidelity, in terms of the actual learning protocol (i.e. the extent to which the network learns as measured by declines in loss), it was less adaptive than other architectures. That is, while having overall lower MSE, it was initialised with a lower MSE which declined less. This is not unexpected given that, in some sense, the neural network architecture combined with the whitebox subRiemannian generative procedure overdetermines the task of learning the coefficients of a single generator $\Lambda_0$ used as an initial condition. The other point to note is that in \cite{swaddle_generating_2017}, $\Lambda_0 \in \frak{su}(2^n)$ i.e. it is drawn from the full Lie algebra, not just $\Delta$ (intuitively because it provides a random direction in the tangent space to commence evolution from). From a control perspective, however, if one only has access to $\Delta$, one cannot necessarily synthesise $\Lambda_0$, thus a second iteration of experiments where $\Lambda_0 \in \Delta$ were undertaken. The applicability of the SubRiemannian model as a means of solving the control problem is more directly related to this second case rather than the first.

% \onecolumngrid
% \begin{widetext}
%============SubRiemannian model
% \end{widetext}
%=======

%============================

% \twocolumngrid

\subsubsection{Geodesic architectures: GRU \& Fully-connected (original) model}
\label{sec:qgml:Geodesic architectures: GRU and Fully-connected (original) model}
In order to benchmark the performance of the greybox models described above, we recreated the original global and local machine learning models utilised in \cite{swaddle_generating_2017}. In that paper, the global model utilised a simple shallow-depth GRU RNN taking $U_T$ as inputs and outputting sequence estimates $(\hat{U}_j)$ (being trained on the true $(U_j)$). In this global decomposition, each element of each $U_j$ is in effect a trainable weight, with the GRU RNN returning the full $(\hat{U}_j)$ instead of only the control amplitudes as intermediate layers as in our GRU RNN Greybox model. The local model took $U_j$ as an input and output the coefficient control amplitude estimates $(\hat{c}_j)$ from which the sequence $(\hat{U}_j)$ could be reconstructed using $\Delta$. In \cite{swaddle_generating_2017}, in order to reduce parameter size of the model, the original global model was trained only on the real part of $(U_j)$ on the basis that the imaginary part could be recovered via application of the unitarity constraint (see \cite{swaddle_subriemannian_2017} for details). 

To learn the individual $U_j$ segments of the approximate geodesic unitary path, we adapted while substantially modifying the approach in \cite{swaddle_generating_2017}. In that paper, the method of learning $U_j$ segments was adopted via feeding the real part of a vectorised (i.e. flattened) unitary $U_j$ into a simple three layer feed-forward fully connected neural network. The labels for the network were the true control pulse amplitudes $c_j^k$.

Recreating these models it was found that using only the realised part of unitaries was insufficient for model performance overall, thus we included both real and imaginary parts both for model performance but also because it is unclear whether simply training alone on realised parts of unitaries affects the way in which the networks would integrate information about the imaginary parts. Furthermore, the approach in \cite{swaddle_generating_2017} did not use measures such as fidelity of more utility to quantum information practitioners, thus our model extended the original models by recreating the unitaries from the estimated controls $(\hat{c}_j)$.  

%===Results
\section{Results} \label{sec:qgml:results}
\subsection{Overview}
The motivation behind the architectures above is to develop protocols by which time-optimal quantum circuits may be implemented via sequences of control pulses applied to quantum computational systems. In this respect, the objective is for the architectures to receive a target unitary $U_T$ as input and output a sequence of control pulses $(\hat c_j)$ necessary to synthesise the estimate $\hat{U}_T$ that optimises fidelity $F(\hat{U}_T,U_T)$. Our experimental method sought to enable comparison of blackbox and greybox methods across the synthesis of unitary propagators (gates) in $SU(2^n)$ for $n = 1,2,3$ and higher order groups in order to achieve this objective. We also sought to gain insight into hyperparameters of model architecture by shedding light on, for example, the optimal number of segments, training examples and training data.  

Throughout our experiments, we observed that the selection of hyperparameters for both the training data and the models made a significant impact on performance. For example, as we discuss below, selection of different values for $h = \Delta t$ exhibited a noticeable impact on performance in terms of training/validation batch fidelity MSE and generalisation to test sets. For this reason, we extended our experiments to include progressively increasing values of $h$ from $h=1/n_{\text{seg}}$ to around $h\approx 1$.

Generalisation of models was tested via assessing the fidelities of $(\hat{U}_j)$ output by the trained models and also independently reconstructing $(\hat{U}_j)$ from the estimates of control coefficients $(\hat{c}_j)$. In this respect, our architecture benefited from the customised layering in that intermediate outputs of the models, such as control coefficients, sequences of estimated Hamiltonians $(\hat{H}_j)$, the actual unitary sequences $(U_j)$ and fidelities could all easily be extracted from the models using TensorFlow's standard Keras model functional API. As discussed in \cite{youssry_modeling_2020, youssry_beyond_2020}, one of the benefits of this type of architecture is that it allows practitioners to `open' the machine learning box, as it were, to validate at intermediate steps that the whitebox outputs of the model match expectations, which in turn is useful for model tuning and engineering.
% \onecolumngrid
\subsection{Tables and charts} \label{sec:qgml:Tables and charts}
Experimental results are set out in the tables and figures below. In Table (\ref{tab:mainresults}), each of the four models was trained and evaluated against training data from $SU(2)$, $SU(4)$ and $SU(8)$. For the greybox models, batch fidelity MSE was chosen as the relevant loss function. For the GRU \& FC Blackbox model that replicated (subject to the inclusion of imaginary parts of unitaries in training) the original machine learning models in \cite{swaddle_generating_2017}, standard MSE comparing realised unitary sequences $(U_j)$ and estimates $(\hat{U}_j)$ was used.
Average gate fidelities for training and validation data sets were also recorded, with order of magnitude of standard error provided in parentheses. Bold entries indicate the highest MSE and fidelity metrics for models trained on $SU(2)$, $SU(4)$ and $SU(8)$ training data respectively.

%========== table

\begin{table}[h!]
\resizebox{\textwidth}{!}{
\begin{tabular}{ |p{3cm}||p{1.35cm}|p{1.35cm}|p{1.65cm}|p{1.35cm}|p{1.35cm}|p{1.65cm}|p{1.35cm}|p{1.35cm}|p{1.65cm}| }
 \hline
 \multicolumn{10}{|c|}{Comparison table: training and validation $|$ $\Lambda_0 \in \frak{su}(2^n)$} \\
 
 \hline
 \multicolumn{1}{|c||}{Model} & \multicolumn{3}{c|}{SU(2)} & \multicolumn{3}{c|}{SU(4)} & \multicolumn{3}{c|}{SU(8)}  
 \\
 \hline
 Metric & MSE(T) & MSE(V) & Fidelity & MSE(T) & MSE(V) & Fidelity & MSE(T) & MSE(V) & Fidelity \\
    \hline
GRU \& FC Blackbox* & 3.693e-05
 & 3.559e-05
 & 0.6936(e-01) & 4.144e-05
 & 4.887e-05
 & 0.7170(e-02) &1.852e-04 & 4.447e-04 & 0.7231(e-02)  \\ 
    
    \hline
FC Greybox & 1.827e-05
 & 1.681e-05
 & 0.9964(e-05) & 3.924e-05
 & 4.156e-05
 & 0.9940(e-05) & 2.607e-04 & 2.450e-04 & 0.9842(e-05)  \\ 

\hline
SubRiemannian (XY) & \textbf{8.728e-09} & \textbf{3.211e-10}  & \textbf{0.9999(e-05)} & 1.521e-07 & 2.007e-07 & \textbf{0.9999(e-05)} & 1.024e-05 & 1.137e-04   & 0.9998(e-05)  \\

\hline
GRU RNN Greybox & 1.414e-07 & 1.348e-07 & 0.9998(e-05) & \textbf{9.019e-08}
 & \textbf{1.204e-07}
 & 0.9998(e-05) & \textbf{3.557e-06}
 & \textbf{1.186e-05}
 & \textbf{0.9998(e-05)}  \\

\hline

\end{tabular}}
\caption{Comparison table of batch fidelity MSE ($(U_j)$ and $(\hat{U}_j$)) for training (MSE(T)) and validation (MSE(V)) sets along with average operator fidelity (and order of standard deviation in parentheses) for four neural networks where $\Lambda_0 \in \frak{su}(2^n)$: (a) GRU \& FC Blackbox (original) (b) FC Greybox, (c) SubRiemannian model and (d) GRU RNN Greybox model. Parameters: $h = 0.1,n_{\text{seg}}=10, n_{\text{train}}=1000$; training/validation 75/25; optimizer: Adam, $\alpha\approx$1e-3. Note*: MSE for GRU \& FC Blackbox standard MSE comparing $(U_j)$ with $\hat{U}_j$. SubRiemannian and GRU RNN Greybox models outperform blackbox models on training and validation sets with lower MSE, higher average operator fidelity and lower variance.}
\label{tab:mainresults}
\end{table}

%===================

%==========Lambda_0 in \Delta table

\begin{table}[h!]
\resizebox{\textwidth}{!}{
\begin{tabular}{ |p{3cm}||p{1.35cm}|p{1.35cm}|p{1.65cm}|p{1.35cm}|p{1.35cm}|p{1.65cm}|p{1.35cm}|p{1.35cm}|p{1.65cm}| }
 \hline
 \multicolumn{10}{|c|}{Comparison table: training and validation $|$ $\Lambda_0 \in \Delta$} \\
 
 \hline
 \multicolumn{1}{|c||}{Model} & \multicolumn{3}{c|}{SU(2)} & \multicolumn{3}{c|}{SU(4)} & \multicolumn{3}{c|}{SU(8)}  
 \\
 \hline
 Metric & MSE(T) & MSE(V) & Fidelity & MSE(T) & MSE(V) & Fidelity & MSE(T) & MSE(V) & Fidelity \\
    \hline
GRU \& FC Blackbox* & 1.053e-07
 & 8.668e-08
 & 0.7180(e-02) & 1.328e-04 & 1.739e-04 & 0.9621(e-04) & 4.283e-05
 & 1.045e-04 & 0.7177(e-02)  \\

\hline
SubRiemannian (XY) & 2.616e-09
 & \textbf{9.263e-11}
  & \textbf{0.9999(e-05)} & 5.224e-08
 & 5.983e-09
 & \textbf{0.9999(e-05)} & \textbf{2.165e-07}
 & 6.874e-05
   & 0.9979(e-05)  \\

\hline
GRU RNN Greybox & \textbf{7.290e-10}
 & 7.086e-10
 & \textbf{0.9999(e-05)} & \textbf{3.478e-09}
 & \textbf{5.505e-09}
 & \textbf{0.9999(e-05)} & 2.817e-07
 & \textbf{1.092e-06}
 & \textbf{0.9994(e-05)}  \\

\hline

\end{tabular}}

\caption{Comparison table of batch fidelity MSE ($(U_j)$ v. $(\hat{U}_j$)) for training (MSE(T)) and validation (MSE(V)) sets along with average operator fidelity (and order of standard deviation in parentheses) for models where $\Lambda_0 \in \Delta$: (a) GRU \& FC Blackbox (original) (b) SubRiemannian model and (c) GRU RNN Greybox model. Parameters: $h = 0.1,n_{seg}=10, n_{train}=1000$; training/validation 75/25; optimizer: Adam, $\alpha\approx$1e-3. Note*: MSE for GRU \& FC Blackbox standard MSE comparing $(U_j)$ with $\hat{U}_j$. For this case, overall the GRU RNN Greybox model performed slightly better than the SubRiemannian model, with both outperforming the GRU \& FC Blackbox model. The FC Greybox model was not tested given its inferior performance overall.}.
\label{tab:mainresultsdelta}
\end{table}

%===================

% \twocolumngrid

\begin{figure}[h]
  \centering
  \begin{minipage}[b]{0.48\textwidth} % Adjust the minipage width to fit both images side by side
    \includegraphics[width=\textwidth]{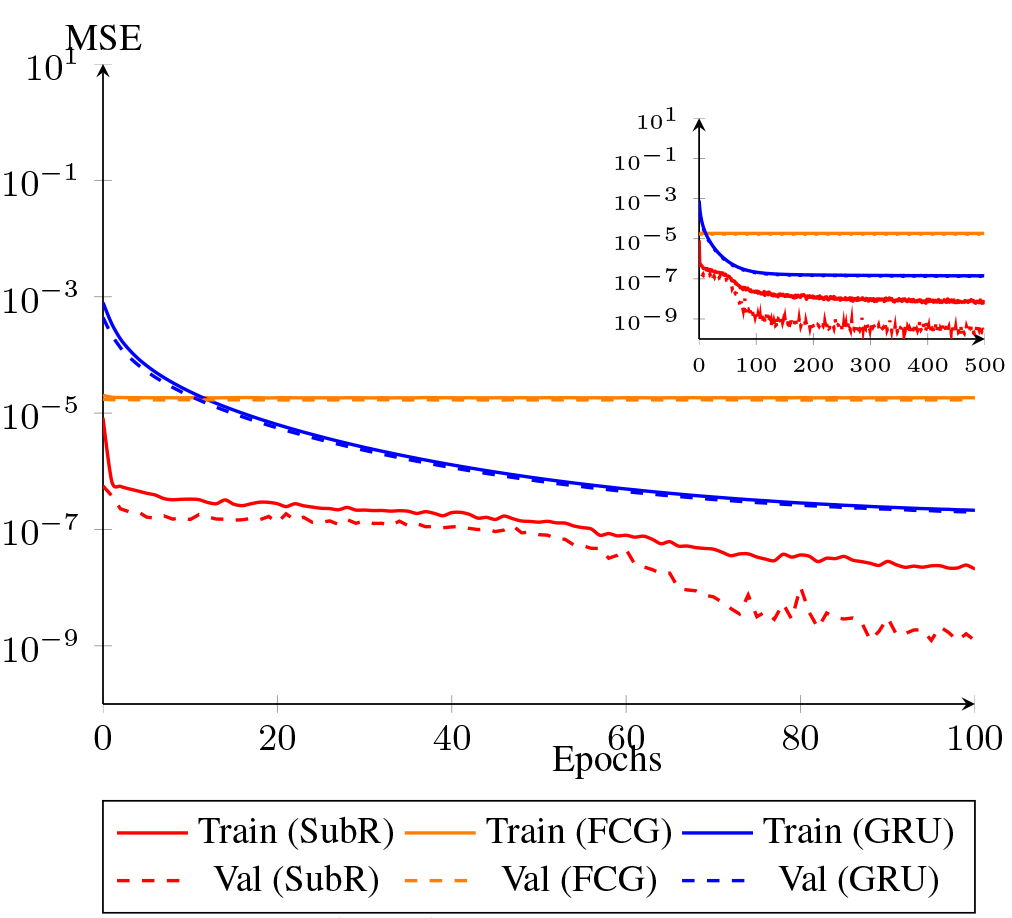}
    \caption{Training and validation loss (MSE) for SU(2).}
    \label{fig:losssu2threemodels}
  \end{minipage}
  \hfill % This will add some space between the two figures
  \begin{minipage}[b]{0.48\textwidth} % Adjust the minipage width to fit both images side by side
    \includegraphics[width=\textwidth]{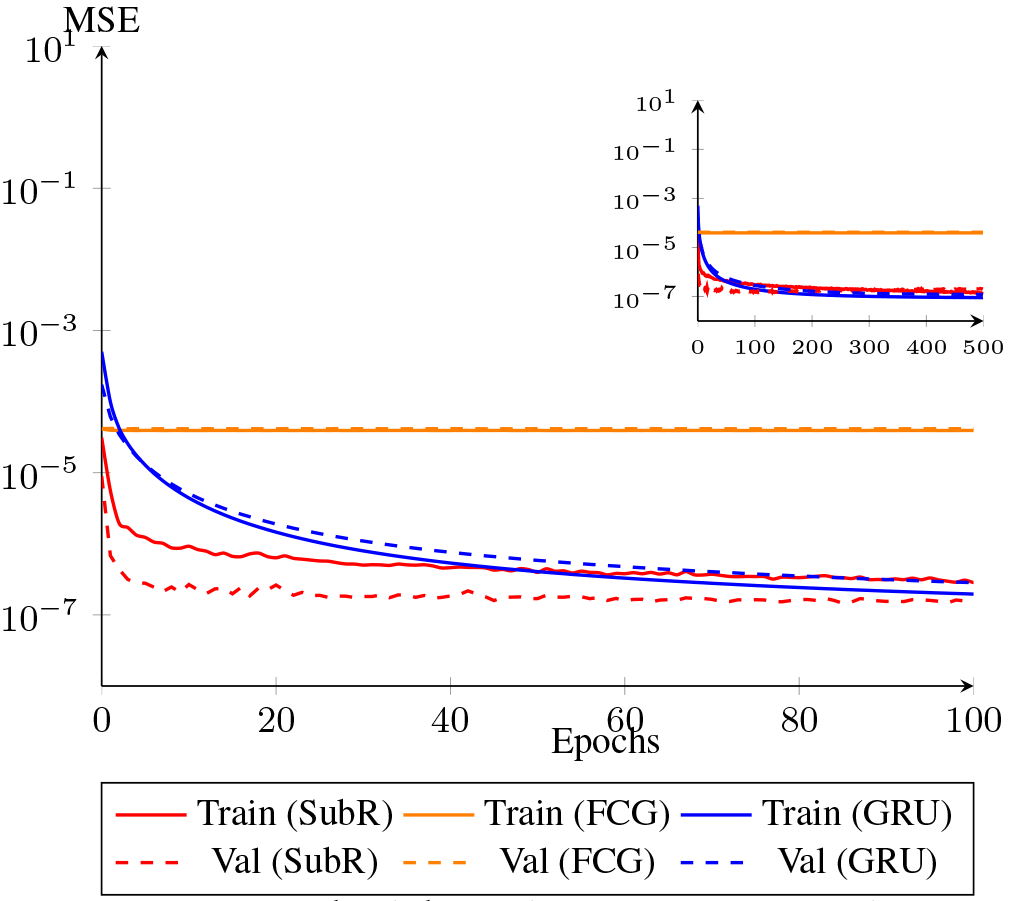}
    \caption{Training and validation loss (MSE) for SU(4).}
    \label{fig:losssu4threemodels}
  \end{minipage}
\end{figure}

%============figure
%compare h size

\begin{figure}[h]
\captionsetup[figure]{width=\textwidth}
\centering
\includegraphics[width=\linewidth]{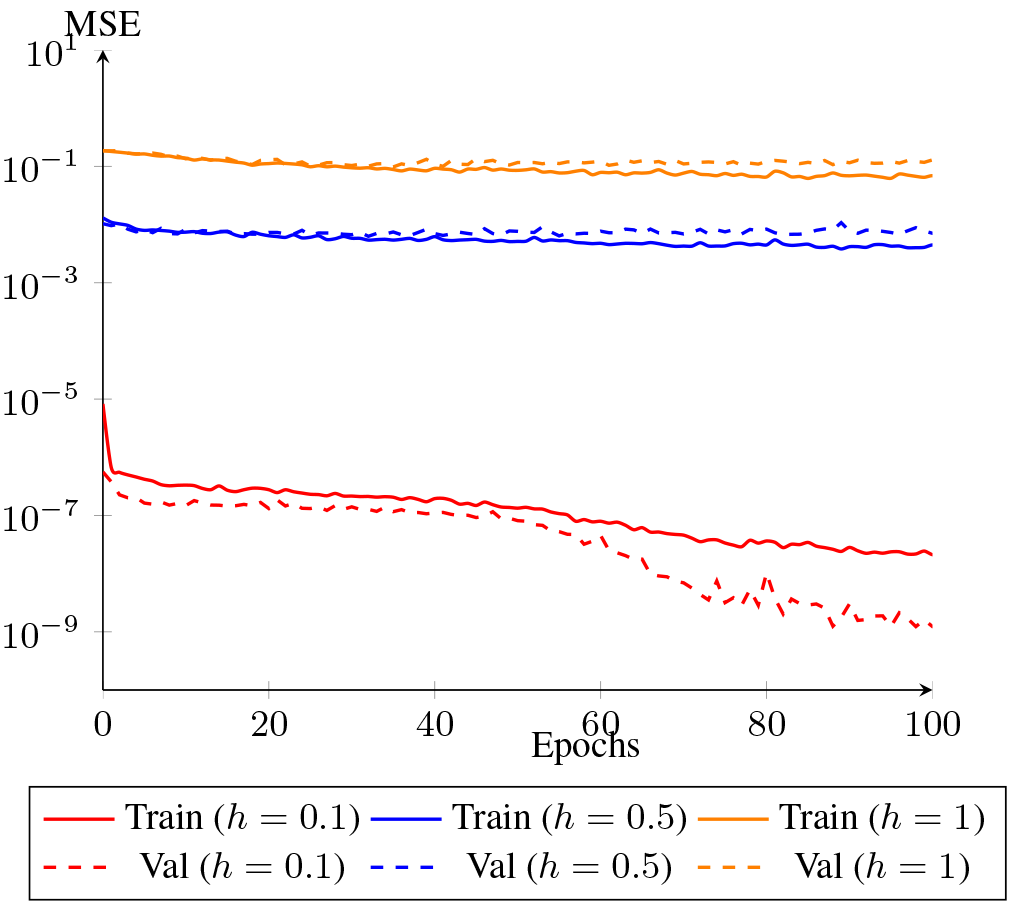}
\caption{Training and validation loss (MSE). Comparison of MSE at different time intervals for the SubRiemannian model. $h=0.1,0.5$ and $1$. $G=SU(2)$, $n_{\text{train}}=1000, n_{\text{seg}}=10$, epochs$=500$, $\Lambda_0 \in \frak{su}(2^n)$: This plot shows the differences in MSE on training and validation sets as the time-step $h=\Delta t$ varies from $0.1$ to $1$. As can be seen, larger $h$ leads to deterioration in performance (higher MSE). However, smaller $h$ can lead to insufficiently long geodesics, leading to a deterioration in generalisation. Setting $h=0.1$ (red curves) exhibits the best overall performance. Even a smaller jump up to $h=0.5$ (blue curves) exhibits an increase in MSE and decrease in performance by several orders of magnitude (and similarly for $h=1$).}
  \label{fig:lossscaledependent}
\end{figure}

%\twocolumngrid

\section{Discussion} \label{sec:qgml:discussion}

\subsection{Geodesic approximation performance} \label{sec:qgml:Geodesic approximation performance}
As can be seen from Table (\ref{tab:mainresults}), the in-sample (training/validation) performance of the models varied considerably between blackbox and greybox approaches. From a use-case and training data perspective, as can be seen from Table (\ref{tab:mainresults}), while the SubRiemannian and GRU RNN Greybox models outperformed the existing benchmark in \cite{swaddle_generating_2017} in terms of in-sample batch fidelity MSE loss, we see a decline in estimations of $U_j \in SU(2^n)$ for higher $n$. MSE overall increases with dimension $n$, which is not unexpected.  

\subsection{Greybox improvements}  \label{sec:qgml:Greybox improvements}
As can be seen from Table \ref{tab:mainresults}, the greybox models in general significantly outperformed (with fidelities around the 0.99 mark) the generic blackbox models (with fidelities in the order of 0.70) for in-sample training and validation experiments for all greybox models and all training data sets $(\Lambda \in \frak{su}(2^n)$ and $\Lambda \in \Delta$). By comparison with existing approaches in \cite{swaddle_generating_2017} and blackbox models that seek to directly synthesise control sequences $(c_j)$ or unitary sequences $(U_j)$, the SubRiemannian and GRU RNN Greybox models outperformed (batch fidelity MSE) the FC Greybox model by several orders of magnitude. This is evident most apparently in Figures (\ref{fig:losssu2threemodels}) and (\ref{fig:losssu4threemodels}). In Figure (\ref{fig:losssu2threemodels}), representing training of the models on $SU(2)$ data, the SubRiemannian model performs the best out of each model, though exhibits overfitting at around the 80 epoch level. These improvements were also accompanied by functional benefits such as the guarantees of unitarity of $U_j$. 

%======saturation

\begin{figure}[h]
\captionsetup[figure]{width=\textwidth}
\centering
\includegraphics[width=0.75\linewidth]{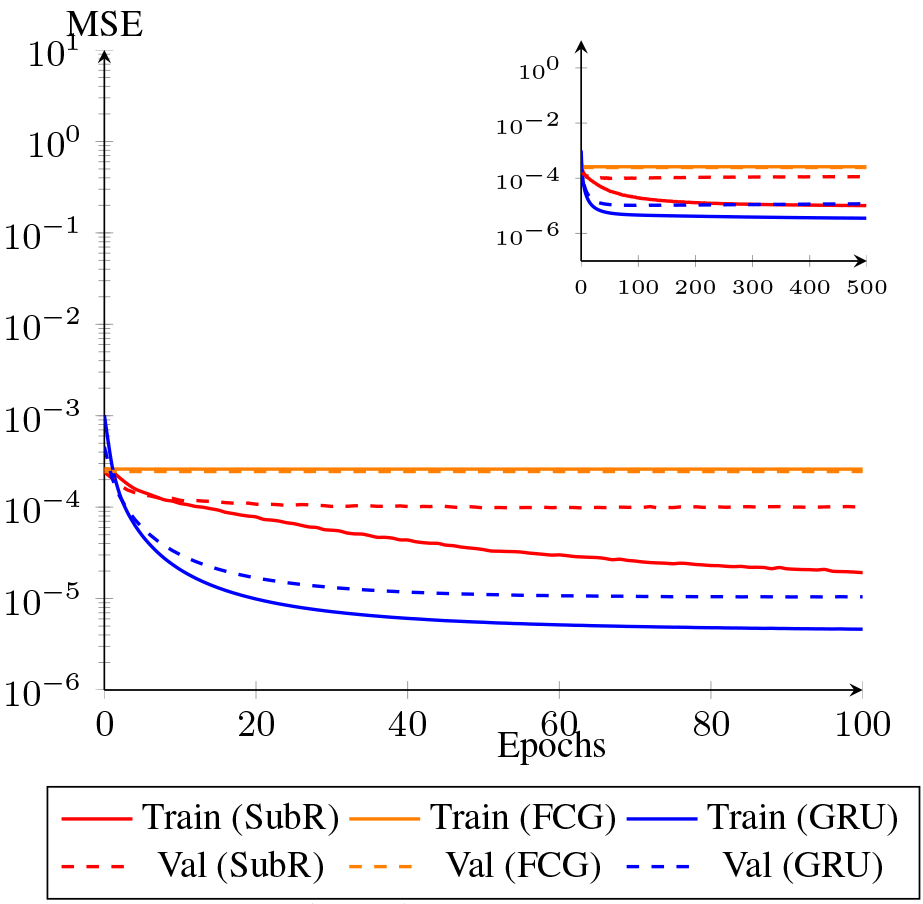}
\caption{Training and validation loss (MSE). Comparison of SubRiemannian, FC Greybox and GRU RNN Greybox models. $G=SU(8),n_{\text{train}}=1000, n_{\text{seg}}=10, h=0.1$, epochs$=500$, $\Lambda_0 \in \frak{su}(2^n)$: For $U \in SU(8)$, we see (main plot - first 100 epochs) that the GRU RNN Greybox (blue line) performs best in terms of batch fidelity MSE on training and validation sets. As shown in the inset, the GRU RNN Greybox levels out (saturates) after about 100 epochs and overall performed the best of each of the models and rendered average operator fidelities of around $0.998$. The SubRiemannian model (red) performed less-well than the GRU RNN, still recording high average operator fidelity but exhibiting overfitting as can be seen by the divergence of the validation (dashed) curve from the training (smooth) curve. The FC Greybox rapidly saturates for large $n_{\text{train}}$ and exhibits little in the way of learning. All models render high average operator fidelity $> 0.99$ and saturate after around 150 epochs (see inset).}
  \label{fig:su8multin1000nseg10}
\end{figure}

Figures (\ref{fig:losssu2threemodels}), (\ref{fig:losssu4threemodels}) and (\ref{fig:su8multin1000nseg10}) show training and validation loss for the three models for the case of $SU(2)$, $SU(4)$ and $SU(8)$ for 1000 training examples, 10 segments, $h=0.1$ and 500 epochs. All models exhibit a noticeable flatlining of the MSE loss for after a relatively short number of epochs, indicative of the models saturating (reaching capacity for learning), a phenomenon accompanied by predictable overfitting beyond such saturation points. For small $h \approx 0.1$, the batch fidelity MSE is already at very low levels of the order $\sim 10^{-5}$. Again we see these persistently low MSEs as indicative of a highly determined model in which the task of learning $\Lambda_0$ (at least for smaller dimensional SU$(2^n)$) is overdetermined from the standpoint of large hidden layers (with 640 neuron units each), together with a prescriptive subRiemannian method.  From one perspective, these highly determined architectures such as SubRiemannian model have less applicability beyond the particular use-case of learning the subRiemannian geodesic approximations specified by the method in \cite{swaddle_generating_2017}. A comparison of FC Greybox, which is a more generalisable architecture (not restricted to whitebox engineering of the subRiemannian algorithm) indicates relatively high performance measures of low MSE and high fidelity.

%=========h v performance

\begin{figure}[h]
\captionsetup[figure]{width=\textwidth}
\centering
\includegraphics[width=0.5\textwidth]{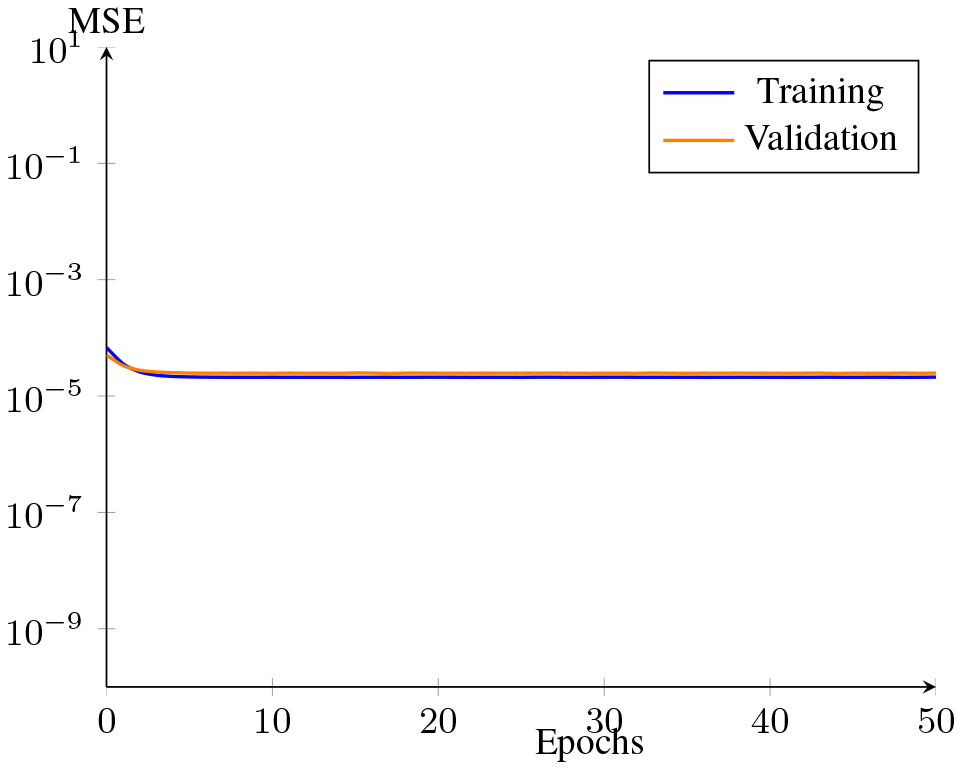}
\caption{Training and validation loss (MSE): GRU RNN Greybox. $G=SU(2),n_{\text{train}}=1000, n_{\text{seg}}=100,h=0.1$, epochs$=500$. This plot shows the MSE loss (for training and validation sets) for the GRU RNN Greybox model where the number of segments was increased from 10 to 100. As can be seen, the model saturates rapidly once segments are increased to 100 and exhibits no significant learning. Similar results were found for the SubRiemannian model. This result suggests that simply changing the number of segments is insufficient for model improvement. One solution to this problem may be to introduce variable or adaptive hyperparameter tuning into the model such that the number of segments varies dynamically.}
  \label{fig:su2n1000subR}
\end{figure}

While the SubRiemannian model performed best in the case of $\Lambda_0 \in \frak{su}(2^n)$, as is evident from Tables (\ref{tab:mainresults}
 and \ref{tab:mainresultsdelta}), the GRU RNN Greybox model performed almost as well for SU$(2)$ and moderately outperformed the SubRiemannian model for $SU(4)$ and $SU(8)$ for most cases. The GRU RNN Greybox model was noticeably faster to train than the FC Greybox model by several hours and was slightly quicker to train than the SubRiemannian model but also flatlines (saturates) relatively early as evident in Figure (\ref{fig:su2n1000subR}). This is of note considering the fact that the GRU RNN Greybox model has more parameters than the SubRiemannian model (which ostensibly needs to only learn control amplitudes for $\Lambda_0$ generation) and is consistent with the demonstrable utility of GRU neural networks for certain quantum control problems \cite{youssry_modeling_2020}. One possible reason for differences between GRU RNN Greybox and SubRiemannian models may lie in the sensitivity of $\Lambda_0$: the SubRiemannian model's only variable degrees of freedom once initiated are in the relatively few weights $c_j^k$ learnt in order to synthesise $\Lambda_0$. As the dimension of $SU(2^n)$ grows, then the coefficients of $\Lambda_0$ become increasingly sensitive, that is, small variations in $c_j^k$ have considerable consequences for shaping the evolution in higher-dimension spaces, in a sense, $\Lambda_0$ bears the entire burden of training and so becomes hypersensitive and requires ever fine-grained tuning. This is in contrast to the GRU, for example, where the availability of more coefficients $c_j^k$ means each individual coefficient $c_j^k$ need not be as sensitive (can vary more) in order to learn the appropriate sequence. \\
 
\subsection{Segment and scale dependence} \label{sec:qgml:Segment and scale dependence}
\begin{figure}[h]
\captionsetup[figure]{width=\textwidth}
\centering
\includegraphics[width=\linewidth]{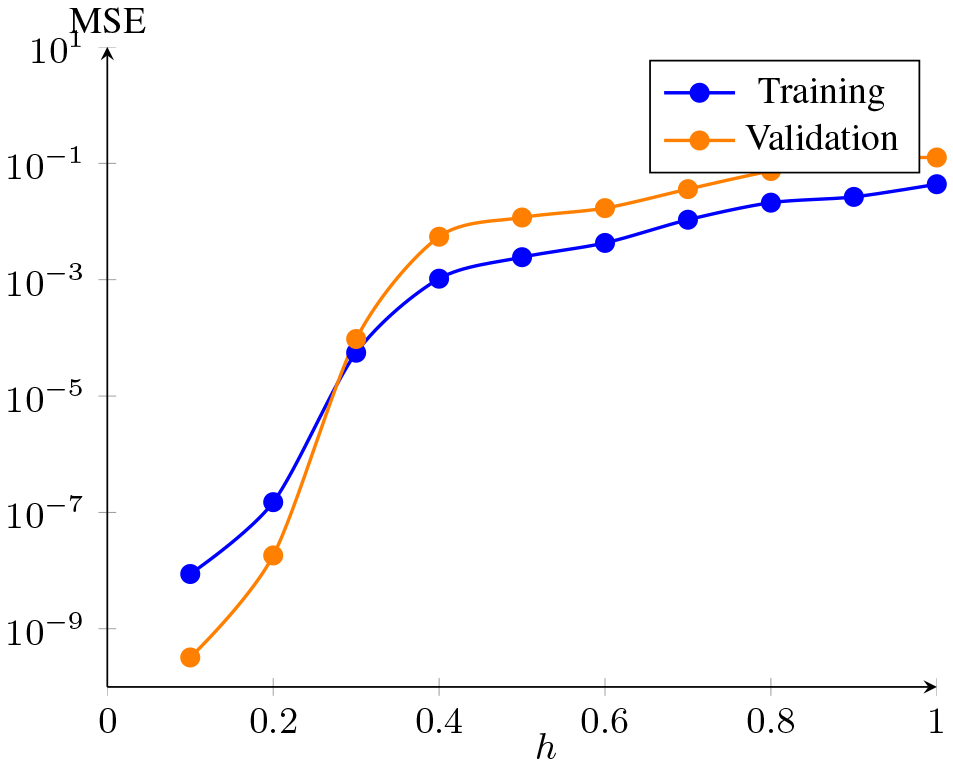}
\caption{Scale $h$ dependence (SubRiemannian model). $G=SU(2),n_{\text{train}}=1000, n_{\text{seg}}=10$, epochs$=500$. Plot demonstrates increase in batch fidelity MSE as scale $h$ ($\Delta t$) increases from 0.1 to 1, indicative of dependence of learning performance on time-interval over which subunitaries $U_j$ are evolved.}
  \label{fig:scaleplot}
\end{figure}

The experiments run across the various training sets indicated model dependence on the number of segments and scale $h$. As can be seen from Figure (\ref{fig:scaleplot}), we find that, not unexpectedly, the performance of models depends upon training data. In particular, model performance measures such as MSE and fidelity clearly depend upon time interval $h=\Delta t$: where $h$ is small, i.e. the closer the sequence of $(U_j)$ is to approximating the geodesic section, the lower the MSE and higher the fidelity. The effect on model performance is particularly evident in Figure (\ref{fig:lossscaledependent}) where increasing $h$ from 0.1 to 0.5 leads to a deterioration in loss of several orders in magnitude (particularly for $h>0.3$). As step size $h$ increases, the less approximating is the resultant curve to a geodesic. Furthermore, for larger step sizes, the conditions required for the assumption of time independence in unitary evolution (\ref{eqn:qgml:schrodind}) are less valid.

%==================generalisation
\subsection{Generalisation} \label{sec:qgml:Generalisation}
In order to test the generalisation of each model (see Appendix \ref{chapter:Background: Classical, Quantum and Geometric Machine Learning} for discussion), a number of tests were run. In the first case, a set of random target unitaries $\tilde{U}_T$ from the relevant SU$(2^n)$ group of interest were generated. These target $\tilde{U}_T$ were then input into the SubRiemannian and GRU RNN Greybox models which output the estimated approximate geodesic sequences $(\hat{U}_j)$ to propagate from the identity to $\tilde{U}_T$. An estimated endpoint target estimate $\hat{U}_T$ for the approximate geodesic was generated. This estimate was then compared against $\tilde{U}_T$ to obtain a generalised gate fidelity $F(\tilde{U}_T,\hat{U}_T)$ for each test target unitary. Second, because fidelities of test unitary targets varied considerably, in order to test the extent to which higher fidelities may be related to similarity to the underlying training set of target unitaries $\{ U_T \}_{\text{train}}$ on which the models were trained, a second fidelity calculation was undertaken. The average gate fidelity of $\tilde{U}_T$ with $\{ U_T \}_{\text{train}}$ was calculated $\bar{F}(\tilde{U}_T,\{ U_T \}_{\text{train}})$. Correlations among the two fidelities were then assessed. 

In the third case, for $SU(2)$ models trained on training data where $\Lambda_0 \in \Delta$, random test unitaries were replaced by $\tilde{U}_T$ comprising random-angle $\theta \in [-2\pi, \pi]$ $z$-rotations. The rationale was to test the extent to which a model based upon restricted control subset training and architecture could replicate unitaries generated only from $\Delta$ with high fidelity for the single qubit case of $SU(2)$ where analytic solutions to the time optimal synthesis of subRiemanninan geodesics are known \cite{boozer_time-optimal_2012}.  

%\onecolumngrid
%============figure GENERALISATION

\begin{figure}[h]
\captionsetup[figure]{width=\textwidth}
\centering
\includegraphics[width=\linewidth]{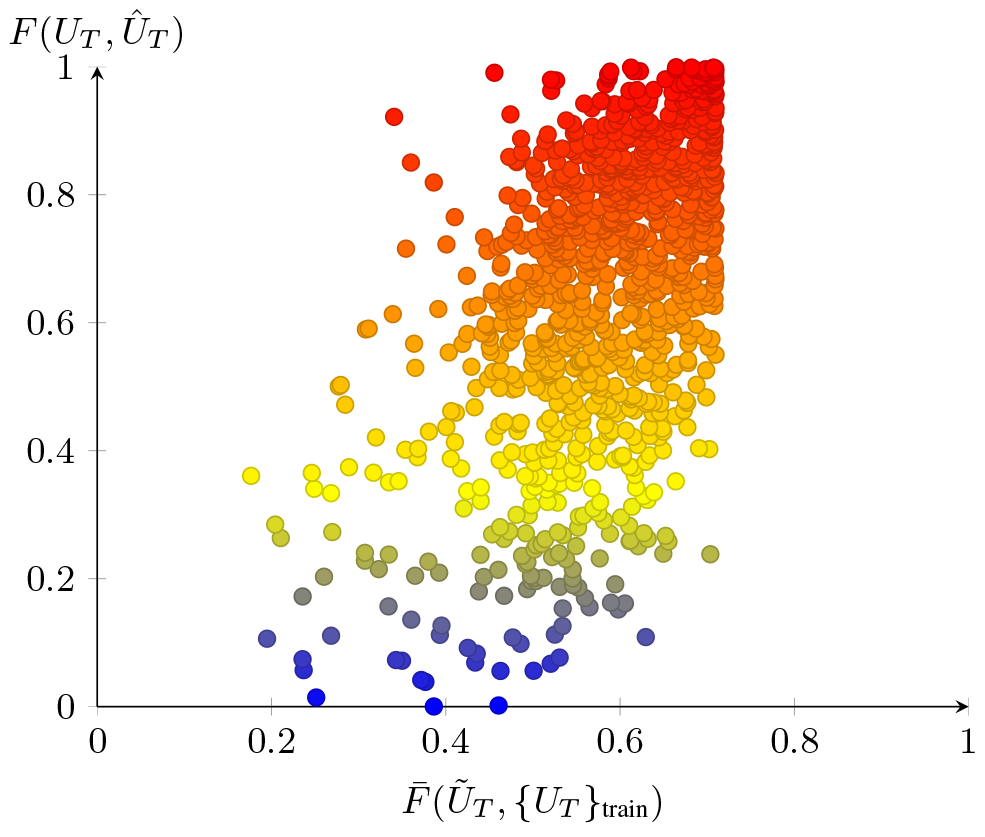}
\caption{Generalisation (SubRiemannian model). $G=SU(2),n_{\text{train}}=1000, n_{\text{seg}}=10$, epochs$=500$, $\Lambda_0 \in \frak{su}(2^n)$. Plot of generalised gate fidelity $F(\hat{U}_T,\tilde{U}_T)$  of randomly generated $\tilde{U}_T$ with the reconstructed estimate $\hat{U}_T$, versus $F(\hat{U}_T,\tilde{U}_T)$, average operator fidelity of randomly generated $U_T$ with training $\{U_T\}_{\text{train}}$ inputs to the model. The upward trend indicates an increase in operator fidelity as similarity (Pearson coefficient of $0.52$ to 95\% significance) of $U_T$ to training $\{U_T\}_{\text{train}}$ increases. Colour gradient indicates low fidelity (blue) to high fidelity (red).}
  \label{fig:su2fidvfidtrain}
\end{figure}

%=====

\begin{figure}[h]
\captionsetup[figure]{width=\textwidth}
\centering
\includegraphics[width=0.75\linewidth]{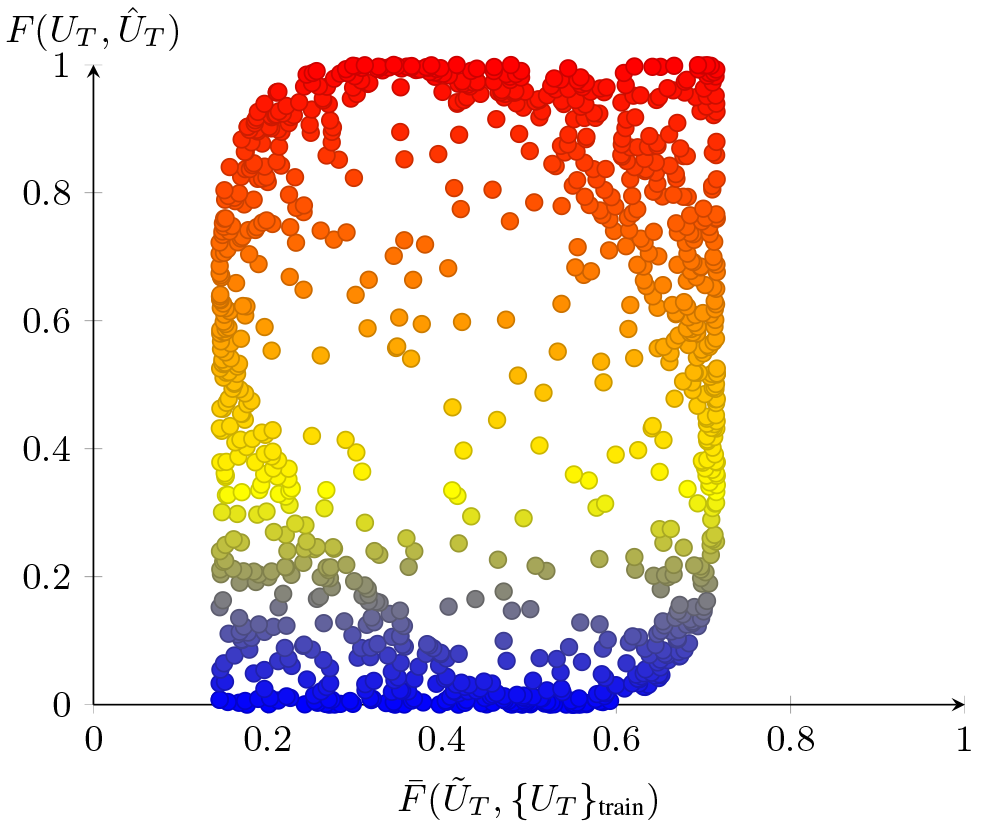}
\caption{Generalisation (SubRiemannian model). $G=SU(2),n_{\text{train}}=1000, n_{\text{seg}}=10$, epochs$=500$, $\Lambda_0 \in \Delta$. Plot of generalised gate fidelity $F(\hat{U}_T,\tilde{U}_T)$  of random-angle $\theta \in [-2\pi, \pi]$ $z$-rotations against generated $\tilde{U}_T$ with the reconstructed estimate $\hat{U}_T$, versus $F(\hat{U}_T,\tilde{U}_T)$, average operator fidelity of randomly generated $U_T$ with training $\{U_T\}_{\text{train}}$ inputs to the model. Here there is no statistically significant correlation between $U_T$ and training set $\{U_T\}_{\text{train}}$, though higher test fidelities are evident for $U_T$ bearing both high and low similarity to the training set (less dependence on similarity to training set for high fidelities).}
  \label{fig:su2fidvfidtraindelta}
\end{figure}

%====================

%========combined

\begin{figure}[!t]
  \centering
  \includegraphics[width=0.8\textwidth]{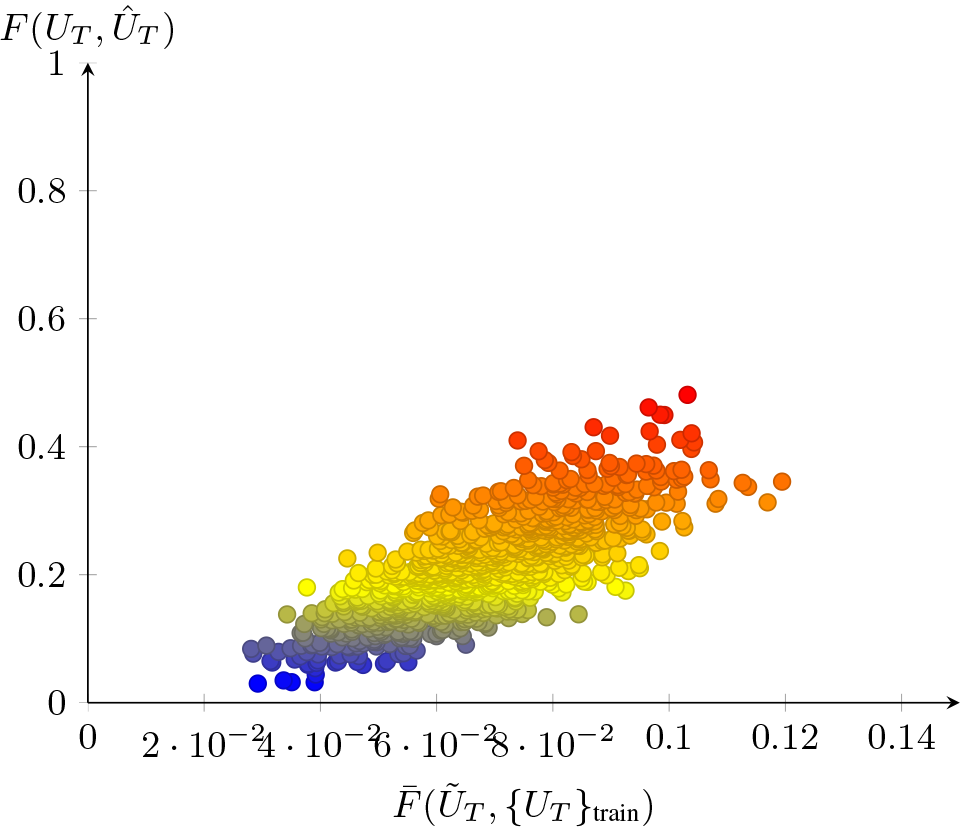}
\caption{Generalisation (SubRiemannian model).  $G=SU(8),n_{\text{train}}=1000, n_{\text{seg}}=10$, epochs$=500$, $\Lambda_0 \in \frak{su}(2^n)$. Plot of generalised gate fidelity $F(\hat{U}_T,\tilde{U}_T)$  versus $F(\hat{U}_T,\tilde{U}_T)$ (average operator fidelity against training set $\{U_T\}_{\text{train}}$). Generalisation was significantly worse for SU$(8)$, however correlation of generalised gate fidelity with similarity of $U_T$ to training sets is evident.}
  \label{fig:su8fidvfidtrain}
  \end{figure}

  % \vspace*{\floatsep}% https://tex.stackexchange.com/q/26521/5764
  \begin{figure}[!t]
  \centering
  \includegraphics[width=0.8\textwidth]{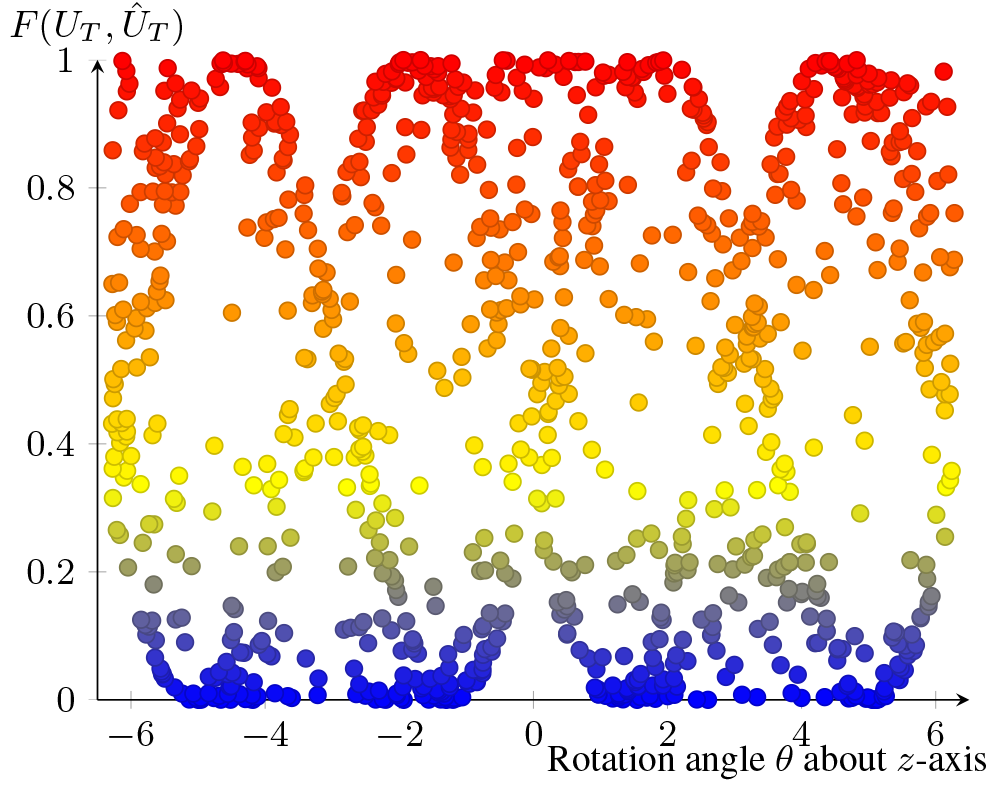}
\caption{Generalisation (SubRiemannian model). $G=SU(2),n_{\text{train}}=1000, n_{\text{seg}}=10$, epochs$=500$, $\Lambda_0 \in \Delta$. Plot of $F(\hat{U}_T,\tilde{U}_T)$  of random-angle $\theta \in [-2\pi, \pi]$ $z$-rotations $\theta$. As evident by the red high fidelities across the range $[-2\pi, \pi]$, the SubRiemannian model trained on data where $\Lambda_0 \in \Delta$ and $\Delta = \{X,Y \}$ in certain cases does generalise relatively well.}
  \label{fig:fidvanglezrotationsu2}
\end{figure}

%\twocolumngrid

Generalisation of both GRU RNN Greybox and SubRiemannian models trained on $SU(2)$ was of mixed success. Figure (\ref{fig:su2fidvfidtrain}) plots $F(\tilde{U}_T,\hat{U}_T)$ against $\bar{F}(\tilde{U}_T,\{ U_T \}_{\text{train}})$ for the SubRiemannian model (comprising only $X,Y$ generators) trained on randomly generated unitaries in $SU(2)$ where $\Lambda_0 \in \frak{su}(2^n)$ (colour gradient indicates low fidelity (blue) to high (red)). As can be seen, generalised gate fidelity varies considerably, with average generalised gate fidelity of 0.6474 with considerable uncertainty (standard deviation of 0.2288). In this case, there is a discernible relationship between fidelity and the extent to which training and test set unitaries (used in generalisation) are similar. We see this via the upward trend of fidelities as similarity of $\tilde{U}_T$ to the training set $\{ U_T \}_{\text{train}}$ increases. This is not uncommon in machine learning contexts, where the more similar data used for generalising a model are to training data, the more accurate the model. 

The model was able to generate approximations to normal subRiemannian geodesics for certain random unitaries with a fidelity of over 0.99 by comparison with the intended target $\tilde{U}_T$. A worked example illustrating the generation of specific control amplitudes using the GRU RNN Greybox model for a specific $z$-rotation is set-out in section \ref{sec:qgml:generalisation}. Identifying structural characteristics among those estimates with higher fidelity remains an open problem. 

By comparison, Figure (\ref{fig:su2fidvfidtraindelta}) plots the relationship between generalised gate fidelity and similarity to training set data for the SubRiemannian model trained on data generated where $\Lambda_0 \in \Delta$. In this case, the test unitaries $\tilde{U}_T$ were rotations by a random angle $\theta$ about the $z$-axis. No particular relationship between $\tilde{U}_T$ and the training set is apparent. Figure (\ref{fig:fidvanglezrotationsu2}) plots the same generalised gate fidelities in relation to $\theta$. Once again there is no immediately discernible pattern between the angle of the $z$-rotation and the fidelity of the estimate of $\tilde{U}_T$. We do see that high (above 0.99) fidelities are distributed across the range of $\theta$ and that there is some hollowing out of fidelities between extremes of 0 and 1.

The out of sample performance of both the SubRiemannian and GRU RNN Greybox models (in both cases limited to generators from $\Delta$) for random unitaries drawn from $SU(4)$ and $SU(8)$ was significantly worse than for $SU(2)$. Average generalised gate fidelities were below 0.5 for each of the models tested. This is not unexpected given the heightened number of parameters that the models must deploy in order to learn underlying geodesic structure increases considerably as the Lie group dimension expands. A larger training set may have some benefits, however we note the saturation of the models suggests that at least for the models deployed in the experiments described above, expanding training sets is unlikely to systematically improve the generalisation of the models. Devising strategies to address both model saturation and ways in which expanded training data could be leveraged to improve model performance remains a topic for further research.

\section{Future directions}\label{sec:qgml:conclusion}
This Chapter presents a comparative analysis of greybox models for learning and generating candidate quantum circuits from time optimally generated training data. The results from experiments above present a clear case for the benefits of greybox machine learning architecture for specific applications in quantum control. The increase in performance over blackbox models, as evidenced by training and validation average operator fidelities for synthesised quantum circuits of over 0.99 in each case, demonstrate that machine learning based methods of quantum circuit synthesis can benefit from customised architecture that engineers known or necessary information into learning protocols.
This is especially the case for quantum machine learning architectures: to the extent learning protocol resources need not be devoted to rediscovering known information or relationships, such protocols can more leverage the power of blackbox neural networks to those parts of problems which are unknown.

While the models outperformed current benchmarks on training and validation sets, they faced considerable challenges generalising well. Future work that may improve upon performance could include exploring hyperparameter learning, such as dynamically learning optimal numbers of segments, time-scale $h$ (including variable time-scale) or other metrics (such as Finslerian metrics(see section \ref{sec:qgml:Nielsen's approach} below)) for use within the subRiemannian variational algorithm itself.
The cross-disciplinary intersection of geometry, machine learning and quantum information processing provides a rich seam of emergent research directions for which the application of both geometric quantum control and greybox machine learning architectures explored in this Chapter  are potentially useful. 
It is important to note that the methods developed in this Chapter, particularly the SubRiemannian model and GRU RNN Greybox were both specifically engineered for particular objectives. While the models developed in this Chapter and experiments were tailored for the particular problem of learning subRiemannian normal geodesics for quantum circuit synthesis, the overall architectural framework in which geometric knowledge is encoded into machine learning protocols has potential for useful application in quantum and classical information processing tasks. 
Future work building upon the greybox machine learning results in this Chapter could include an exploration of ways to combine the extensive utility of symmetric space formalism, methods of Cartan and other geometric techniques with machine learning.

\section{Algorithmic architectures}
\label{sec:qgml:Algorithmic architectures} 
The section below sets-out pseudocode for the machine learning models utilised in the experiments above.
\subsection{Fully-connected Greybox model} 
\label{sec:qgml:Fully-connected Greybox model}
Pseudocode for the Fully-connected Greybox model is set-out below. Note that TensorFlow inputs required $(U_j)$ to be separated into real $\text{Re}(U_J)$ and imaginary $\text{Im}(U_J)$ parts and then recombined for input into fidelity calculations. Note the cost function $C(F,1)$ below is implicitly a function of $c_j^k$ (the sequence of which is $(c_j)$) which are the variable weights in the model. Here $\gamma$ is the learning rate for the gradient update and $\theta$ the trainable weights of the model.
\begin{algorithm}[H]
% \SetAlgoLined
 Inputs: $U_T$, $\text{Re}(U_J)$, $\text{Im}(U_J)$,$\Delta$,$h$\\
 Labels: $v = (1...1)$, $\dim v = |(U_j)|$ \\
 FC Dense Network: $U_T \to \tanh(U_T;\theta) = (\hat{c}_j)$\\
 Hamiltonian estimation: $(\hat{c}_j), \Delta \to (\hat{H}_j) = (\sum_k\hat{c}^k \tau_k)$ where $\tau_k\in \Delta$\\
 Quantum Evolution: $(\hat{H}_j) \to (\hat{U}_j) = (\exp(-h \hat{H}_j))$\\
 Fidelity: $\text{Re}(U_J),\text{Im}(U_J),(\hat{U}_j) \to F(\hat{U}_j, U_j)$\\
 MSE: $\min C(F,1) = \frac{1}{n}\sum_j^n (1 - F(\hat{U}_j,U_j))^2$\\
 Update: $\theta \to \theta - \gamma\nabla_{\theta} C(F,1)$
 \caption{Fully-connected Greybox model}
\end{algorithm}

\subsection{GRU RNN Greybox model, parameters $\theta = (w,b)$}
\label{sec:qgml:GRULSTM}
Pseudocode for the GRU RNN Greybox model is set-out below. Note that TensorFlow inputs required $(U_j)$ to be separated into real $\text{Re}(U_J)$ and imaginary $\text{Im}(U_J)$ parts and then recombined for input into fidelity calculations. Note the cost function $C(F,1)$ below is implicitly a function of $c_j^k$ (the sequence of which is $(c_j)$) which are the variable weights in the model. Here $\gamma$ is the learning rate for the gradient update and $\theta$ the trainable weights of the model.

\begin{algorithm}[H]
% \SetAlgoLined
 Inputs: $U_T$, $\text{Re}(U_J)$, $\text{Im}(U_J)$,$\Delta$, $h$\\
 Labels: $v = (1...1)$, $\dim v = |(U_j)|$ \\
 GRU RNN: $U_T \to \tanh(U_T;\theta) = (\hat{c}_j)$\\
 Hamiltonian estimation: $(\hat{c}_j), \Delta \to (\hat{H}_j) = (\sum_k\hat{c}^k \tau_k)$ where $\tau_k\in \Delta$\\
 Quantum Evolution: $(\hat{H}_j),h \to (\hat{U}_j) = (\exp(-h \hat{H}_j))$\\
 Fidelity: $\text{Re}(U_J),\text{Im}(U_J),(\hat{U}_j) \to F(\hat{U}_j, U_j)$\\
 MSE: $\min C(F,1) = \frac{1}{n}\sum_j^n (1 - F(\hat{U}_j,U_j))^2$\\
 Update: $\theta \to \theta - \gamma\nabla_{\theta} C(F,1)$
 \caption{GRU RNN Greybox model}
\end{algorithm}

\subsection{SubRiemannian model}
\label{sec:qgml:SUB}
Pseudocode for the SubRiemannian model is set-out below. Note that TensorFlow inputs required $(U_j)$ to be separated into real $\text{Re}(U_J)$ and imaginary $\text{Im}(U_J)$ parts and then recombined for input into fidelity calculations. This model learns the coefficients required to generate the initial condition $\Lambda_0$. It was found that the model performed best when positive $c_{\Lambda_0}^{+}$ and negative $c_{\Lambda_0}^{-}$ control functions were learnt separately then combined to form the coefficient $c_{\Lambda_0}$. Note the cost function $C(F,1)$ below is implicitly a function of $c_{\Lambda_0}$. Here $\gamma$ is the learning rate for the gradient update and $\theta$ the trainable weights of the model.
\begin{algorithm}[H]
% \SetAlgoLined
 Inputs: $U_T$, $\text{Re}(U_j)$, $\text{Im}(U_j)$,$A = \Delta$ or $\frak{su}(2^n)$, $U_0=I$, $h$, $n_{\text{seg}}$ \\
 Labels: $v = (1...1)$, $\dim v = |(U_j)|$ \\
 FC Dense Network: $U_T \to \tanh(U_T; \theta) = c_{\Lambda_0}^{+}, \tanh(U_T; \theta) =c_{\Lambda_0}^{-}$\\
 $\Lambda_0$ estimation:\\
 \qquad $c_{\Lambda_0}^{+}, A \to \Lambda_0^{+} = \sum_k c^{k+} \tau_k$, $\tau_k \in A$;\\
 \qquad $c_{\Lambda_0}^{-}, A \to \Lambda_0^{-} = \sum_k c^{k-} \tau_k$, $\tau_k \in A$\\
 $\Lambda_0$ layer: $\Lambda_0^{+},\Lambda_0^{-} \to \Lambda_0$ \\
 subRiemannian layer: $\Lambda_0 \to (U_j)$. Set $Y = U_0$.\\ 
 For $j$ in $n_{\text{seg}}$:\\
 \qquad $\Lambda_0 \to Y \Lambda_0 Y^\dagger = X$\\
 \qquad $X \to \hat{H}_j=\projdelta(X)$, $c_j$\\
 \qquad $\hat{H}_j \to \hat{U}_{j+1}=\exp(-h H_j)$\\
 \qquad $Y = \hat{U}_{j+1}$\\
 \qquad return $(\hat{U}_j)$\\
 Fidelity: $(\hat{U}_j),(\text{Re}(U_j)),(\text{Im}(U_J)) \to F(\hat{U}_j, U_j)$\\
 MSE: $\min C(F,1) = \frac{1}{n}\sum_j^n (1 - F(\hat{U}_j,U_j))^2$\\
 Update: $c^k \to c^k - \gamma\nabla_{c^k} C(F,1)$
 \caption{SubRiemannian model}
\end{algorithm}

\subsection{Simulation Design}
\label{sec:qgml:Simulation}
Simulation of training datasets for use in the machine learning models was undertaken in Python. The simulation was adapted from Mathematica code accompanying \cite{swaddle_generating_2017} with a number of adaptations. The code is constructed as a class with the following hyperparameters: (i) $n=\dim(SU(2^n))$ for selecting the Lie group of interest $SU(2^n)$; (ii) $n_{\text{seg}}$ the number of segments (indexed by $j$) in the global decomposition into subunitaries $(U_j)$; (iii) $n_{\text{train}}$, the number of training examples; (iv) a parameter for whether $\Lambda_0 \in \frak{su}(2^n)$ or $\Delta$; (v) a set of parameters for selecting (for $SU(2)$) which Pauli operators constituted $\Delta$; (vi) a parameter for selecting whether unitaries were to be generated in accordance with the example formulation in \cite{boozer_time-optimal_2012}, (vi) parameter for selecting $h$ (which defaults to $1/n_{\text{seg}}$ in the event of a null entry. Upon selection of parameters, the class generates an extensive selection of training data in various forms (see the relevant code repository for code with commentary), including complex and realised iterations of $U_T, (U_j), c_j, \Delta$ and other key inputs into the models. Training data was generated using a combination of standard Python numerical and scientific packages together with quantum simulation software QuTip \cite{johansson_qutip_2013}.

\subsection{Generalisation: worked example}
\label{sec:qgml:generalisation}

We include below an example of the controls generated by the GRU RNN Greybox model for $SU(2)$ when estimating an arbitrary $z$-rotation by angle $\theta$. A randomly selected angle $\theta$ between $\pm 2\pi$ was generated using Qutip, in this case $\theta=$-7.529e-01, generating the target unitary:
\begin{align*}
    U_T &= \pmat{0.930+0.368i}{0}{0}{0.930-0.368i}
\end{align*}
The generators in this case were $\sigma_x$ and $\sigma_y$. To generate an approximate geodesic sequence, we require the neural network to learn controls $ c_{x,j}$ for $X$ and $c_{y,j}$ for $Y$ for ten subsidiary Hamiltonians $H_j$ (linear compositions of $X,Y$):
\begin{align*}
    H_j &= c_{x,j}X + c_{y,j}Y
\end{align*}
The specific controls applied at time step $j$ are set-out in Table \ref{table:generalcontrols}. The resulting estimated unitary $\hat{U}_T$ generated by applying each $H_j$ for time $\Delta t$ is:
\begin{align*}
    \hat{U}_T&=\pmat{0.926+0.377i}{-0.003}{0.003}{0.926-0.377i}
    \end{align*}
with the fidelity between the target and estimate $F(\hat{U}_T,U_T) = 0.9999$.

\begin{center}
 \begin{table}[t]
 \begin{tabular}{|c|| c| c |} 
 \hline
 $k$ & $c_{x,j}$ & $c_{y,j}$ \\
 \hline
 \hline
 1 & -4.651\text{e-02} & 0.751\text{e-02} \\
 2 & -5.668\text{e-02} & 0.622\text{e-02} \\
 3 & -5.777\text{e-02} & -1.504\text{e-02} \\
 4 & -5.917\text{e-02} & 0.947\text{e-02} \\
 5 & -5.221\text{e-02} & -0.529\text{e-02} \\
 6 & -5.663\text{e-02} & 0.800\text{e-02} \\
 7 & -6.137\text{e-02} & 0.119\text{e-02} \\
 8 & -4.975\text{e-02} & -0.173\text{e-02} \\
 9 & -5.377\text{e-02} & -0.047\text{e-02} \\
 10 & -5.346\text{e-02} & 0.079\text{e-02} \\
 \hline
 \end{tabular}
 \caption{Table of control amplitudes for generation of $z$-rotation by angle $\theta = $-7.529e-01. At each time-interval $k$, controls $c_{x,j}$ are applied to $X$ generators and $c_{y,j}$ are applied to $Y$ generators to form Hamiltonian $H_j$, for time $h = \Delta t$. }
\label{table:generalcontrols}
\end{table}
 \end{center}

\section{Differential geometry and Lie groups}
\label{sec:qgml:Differential geometry and Lie groups}
\subsection{Generating subspaces for geodesics}
\label{sec:qgml:onetwobodyoperators}
\subsection{Product Operator Basis} \label{sec:qgml:Product Operator Basis}

 Our experimental results and methods focus on synthesising quantum circuits for multi-qubit systems where unitary operators are drawn from $SU(2^n)$. For such multi-qubit (qudit) systems, unitary operators $U$ belong to Lie groups $G = \sutwon$ which describe the evolution of $n$ interacting spin$-1/2$ particles. These groups are equipped with a corresponding Lie algebra of dimension $(2^n)^2-1=4^n-1$ and denoted $\g=\frak{su}(2^n)$, represented via traceless $n \times n$ skew-Hermitian ($A = -A^*$) matrices (see Appendix \ref{chapter:Background: Quantum Information Processing}). Solving time-optimal problems in such contexts often relies upon appropriate selection of a subset of generators from the Lie algebra as the control subset from which to synthesise a quantum circuit (see section \ref{sec:geo:Geometric control theory}). This is especially the case when selecting a control algebra that renders targets $U_T$ reachable in a way that approximates geodesic curves (definition \ref{defn:geo:Geodesic}) on the relevant manifold $\M$ as the choice of one set of generators over another can affect evolution time (and whether generated geodesics are indeed minimal, in cases where multiple geodesics are available such as via great circles on a 2-sphere).    
 Of importance in selecting control subsets for time-optimal synthesis of geodesics in multi-qubit systems \cite{nielsen_optimal_2006, gu_quantum_2008, wang_quantum_2015, khaneja_cartan_2001, dalessandro_lie_2008}
 is the so-called \textit{product operator basis} i.e. a basis for the Lie algebra of generalised Pauli matrices, being tensor (Kronecker) products (definition \ref{defn:quant:Tensor Product}) of elementary Pauli operators. The basis is formed by Pauli spin matrices $\{I_x, I_y, I_z\} = 1/2\{\sigma_x, \sigma_y, \sigma_z\}$ i.e. the sets of generators of rotation in two-dimensional Hilbert space (and Lie algebra basis), with usual commutation relations. A basis for $\sutwon$ comprises of many-body tensor products of these Pauli operators, i.e. for an $n$-dimensional operator, there are between $1$ and $n$ Pauli operators tensor products with identities for various indices. An orthogonal basis $\{iB\}$ (frame) for $\liesun$ is then given \cite{khaneja_cartan_2001} in closed-form via:
\begin{align*}
    B_s = 2^{q-1} \Pi_{k=1}^n (I_{k\alpha})^{a_{ks}}
\end{align*}
where $\alpha =x,y,z$ and $s$ indexes each basis element of the frame. The index $a_{ks}$ is 1 in $q$ of the indices and 0 otherwise, and is a way of representing:
\begin{align*}
    I_{k\alpha} = 1 \otimes ... \otimes I_\alpha \otimes 1
\end{align*}
where $I_\alpha$ appears only at the $k$th position with the identity appearing everywhere else. The parameter $q$ tells us how many Pauli operators are tensor-producted together e.g. $q=1$ means the basis element only has one Pauli and the rest identities; $q=2$ means we are dealing with a basis formed by tensor products of two Pauli operators and identities etc.

\subsection{One- and two-body operators} \label{sec:qgml:One- and two-body operators}
Geometric control techniques for multi-qubit systems often focus on selecting one- and two-body Pauli product operator frames (bases) for relevant control subsets \cite{dowling_geometry_2008, swaddle_subriemannian_2017} (see section \ref{sec:geo:Time optimal problems with Lie groups}). If the control subset contains only one- and two-body elements of the Lie algebra $\frak{g}$, then curves generated in the corresponding Lie group $G$ are more likely (with a number of important caveats) to be approximations to (and in the limit, as the number of gates $n \to \infty$, representations of) geodesic curves and thus time-optimal synthesis of target unitary propagators. The intuitive reason for this is that of the full Lie algebra $\g$, one- and two-body generators are less `expensive' as measure for example by a metric calculating energy (equation \ref{eqn:geo:energyhorizontalcurve}). This approach can be seen across a number of key results in the literature \cite{khaneja_cartan_2001, dowling_geometry_2008, gu_quantum_2008, wang_quantum_2015} and forms the basis for the relevant distribution used in subRiemannian variational methods in \cite{swaddle_generating_2017,swaddle_subriemannian_2017} which the protocols developed in this Chapter expand upon. The preference for one- and two-body Pauli operator frames arises in different contexts.  

For example, it is demonstrated in \cite{khaneja_cartan_2001} in the case where $G=SU(4)$ and $K = SU(2) \otimes SU(2)$ that by finding an appropriate Cartan decomposition $G=KAK$ (with associated Lie algebra decomposition $\frak{g} = \frak{p} \oplus \frak{k}$) (see sections \ref{sec:alg:Cartan decompositions} and \ref{sec:alg:Cartan algebras and Root-systems}) and maximally abelian Cartan subalgebra:
\begin{align*}
    \frak{h} = i\spn {I_xS_x, I_yS_y, I_zS_z} \subset \frak{p}
\end{align*}
(where $I_\alpha$ represent one-body terms and $S_\beta$ two-body terms), we can write $\exp(-i\frak{h}) = A$ in the $KAK$ decomposition as the exponential of a linear combination of the generators in $\frak{h}$, namely:
\begin{align*}
    U_F = K_1 \exp(-i(\alpha_1I_xS_x + \alpha_2I_yS_y + \alpha_3I_zS_z  )) K_2
\end{align*}
where $K_1,K_2 \in K = \sutwo \otimes \sutwo$. In this case, any Hamiltonian from $\frak{k}$ can be generated using the controls in $\p$ (essentially by showing they can generate the two-body terms via action of the single-body operators $I$ on $S$) and is time optimal. Because synthesis depends on the evolution of the drift Hamiltonian according to generators in $\frak{k}$ (as acted on via the adjoint action of $K$) and because this depends on the coefficients of the generators $\alpha_i$, then the minimal time is given by the coefficients of the generators in $\frak{p}$ used to steer $H_d$:
\begin{align*}
    T&=\min_{\alpha_i} \sum_{i=1}^3 |\alpha_i|
\end{align*}
hence the optimisation problem becomes relatively straight-forward. One rationale for preferring one- and two-body generators \cite{khaneja_cartan_2001} is that higher-order (i.e. more than one-body) generators are shown to have coefficients which include a scalar coupling strength $J$ between the relevant spins such that each $H_j$ has a coefficient $2\pi$ and the two-body ($I_iS_i$) term has coefficient $2\pi J$. Thus a time-optimal problem becomes a simpler optimisation problem of finding the minimal sum $\sum_i \alpha_i$ satisfying:
\begin{align*}
    U_F = Q_1 \exp(-i2\pi J(\alpha_1I_xS_x + \alpha_2I_yS_y + \alpha_3I_zS_z)) Q_2
\end{align*}
where $Q_1,Q_2 \in K$. The proof essentially relies on the fact that because synthesis of $Q_1,Q_2$ takes negligible time, then synthesis time is determined by the time to synthesise $A$ in the $KAK$ which is determined by the parameters $\alpha_i$, hence minimal time amounts to minimising the sum of $\alpha_i$. Synthesis time is thus minimal to the extent that the `fewest-body' Pauli generators are utilised in the control subset. Thus, ideally, to generate minimal (and thus time optimal) paths in $G$ to reach arbitrary target unitaries $U_T$, one should ideally choose the control subset with as few many-body terms as necessary in order to render $U_T$ reachable in a control sense. In this regard, we note recent work \cite{marvian_restrictions_2022} regarding surprising constraints on realisability and universality of unitary gate sets (in control language, reachability of circuits) for unitary transformations on composite (e.g. multi-qubit) systems generated by two-local unitaries. As noted in that work by Marvian, this may require additional algorithmic tailoring and/or the use of ancilla qubits to circumvent such restrictions. We have not in this work addressed such generic limitations, but they are an important consideration in any practical application. It is an open research question as to whether (and to what extent) machine learning techniques may also provide a means to bridge such gaps in universality arising from the tension between two-local unitaries and symmetry properties of composite systems.  

\subsection{Nielsen's approach}
\label{sec:qgml:Nielsen's approach}
Nielsen et al. \cite{nielsen_geometric_2006} also focus on adopting one- and two-body terms in their metric-based approach to characterising and generating time-optimal quantum circuits. For example, the preference for one- and two-body generators is justified by imposing  a \textit{Hamming weight} term $\text{wt}(\sigma)$ applied to the Pauli generators $\sigma$ together with a penalty function $p(\cdot)$ that penalises the control functional whenever Pauli terms of high Hamming weight are part of the control Hamiltonian. The idea is that Pauli $n$-tuples (tensor products) of anything more than one- or two -body Hamiltonians will be penalised via a higher Hamming weight as they will have many more non-identity elements, whereas one- and two-body operators have lower Hamming weight). Nielsen et al. demonstrate that selection of one- and two-body generators is optimal for calculating a lower bound on the complexity measure $m_\mathcal{G}(U)$ using Finsler metrics i.e:
\begin{equation}
    d_F(I,U) \leq m_\mathcal{G}(U)
\end{equation}
 where $\mathcal{G}$ is a universal gate set in $\sutwon$. 

The significance of restricting control subsets together with bespoke metrics when utilising geometric optimisation techniques is evident in later work \cite{nielsen_optimal_2006}. For quantum control optimisation architectures, this demonstrates the utility of Finsler metrics as a more general norm-based measure of distance (and thus optimality) together with a justification of the selection of one- and two-body generators due on the basis of Hamming weights. The use of the `penalty metric' approach (see discussion in the context of linear models in section \ref{sec:ml:Linear models} and regularisation generally in section \ref{sec:ml:Reducing empirical risk}) is explored in further work \cite{gu_quantum_2008, wang_quantum_2015} however, as noted in \cite{swaddle_subriemannian_2017}, such approaches can be convoluted without providing guarantees that optimal generators will be selected.

In \cite{nielsen_optimal_2006}, Nielsen et al. expand certain elements of the initial program combining techniques from differential geometry and variational methods to quantum circuit synthesis and quantum control. This second paper considered the difficulty of implementing a unitary operation $U$ generated by a time dependent Hamiltonian evolving to the desired $U_T$. They show that the problem of finding minimal circuits is equivalent to analogous problems in geometric control theory i.e. this Chapter has more of a focus on quantum control utilising geometric means. They select a cost function on $H(t)$ such that finding optimal control functions for synthesis of $U_T$ (evolving according to the Schrodinger equation) involves finding minimal geodesics on a Riemannian manifold $(\M,g)$ (see definition \ref{defn:geo:Riemannian Manifold}).

In this case, $H(t)$ is written in terms of a \textit{Pauli operator expansion}:
\begin{equation}
    H = \sum_\sigma^{'} h_\sigma \sigma + \sum_\sigma^{''} h_\sigma \sigma
\end{equation}
where the first summation is over one- and two-body terms, the second over all other tensor products. A cost function is constructed with a penalty term $p^2$ imposed that penalises the higher-order terms:
\begin{equation}
    F(H) = \sqrt{\sum_\sigma^{'} h_\sigma \sigma + p^2\sum_\sigma^{''} h_\sigma \sigma}
\end{equation}
with the total cost to be minimised given by:
\begin{equation}
    d(U) \equiv \int_0^T dt F(H(t))
\end{equation}
Due to parametrisation invariance, $F$ (a Finsler metric) can be rescaled such that $T = d(U)$. 
The overall effect is to demonstrate that using $O(n^2 d(I,U)^3)$ one- and two-qubit gates, it is possible to synthesise a unitary $U_A$ satisfying $||U_T - U_A|| \leq c$, where $c$ is a constant and $U_T$ is the target unitary gate. Moreover, the work demonstrates the optimality of unitary synthesis via following minimal geodesics in the Lie group manifold generated by one- and two-body generators (as we focus on below). Nielsen notes the number of one- and two-qubit terms (i.e. $\dim \Delta$) for the relevant Lie algebra is given by
\begin{equation}
    \dim \Delta=9n(n-1)/2 + 3n
\end{equation}
a relatively trivial but important feature of the machine learning code in model architectures explored above.

 Later work \cite{dowling_geometry_2008} of Nielsen and Dowling provides a more directly applicable example of how to develop analytic solutions to geodesic synthesis of unitary operations. As with the discussion above, it is worth exploring the key results from this Chapter in order to understand characteristics of relevance to any attempt to utilise geometric methods for synthesis of unitary propagators for multi-qubit systems. In the paper, they develop a method of deforming (homotopically) simple and well-understood geodesics to geodesics of metrics of interest. Intuitively, the idea is to start with a known geodesic curve between $I$ and $U_T$ and, subject to certain constraints, `bend' it homotopically (that is, via mappings which preserve topological properties) into a minimal-length curve. However, as demonstrated in \cite{dowling_geometry_2008}, a similar preference for one- and two-body terms is manifest in the applicable lifted Hamilton-Jacobi equation (this work is also important for anyone interested in geometric quantum control given its discussion of significant (and potentially intractable) complexity constraints presented by the quantum extension of the Rabarov-Rudich theorem and also extend geometric quantum computing to include ancilla qubits.

In subsequent work utilising Nielsen et al.'s approach \cite{wang_quantum_2015}, the application of penalty metrics is extended in order to show its utility in synthesising time-optimal geodesics. In that paper, it is shown that a bound on the norm of the Hamiltonian $H(t)$ is equivalent to a bound on the speed of evolution, that is, such a bound implies that minimal-time paths are minimal distance in which the norm function is used as the distance measure. Given $||H(t)||=E$, they demonstrate that for any curve connecting $I$ and $U_T$, the length of time-optimal curves is given by:
\begin{align}
    L = \int_0^T||H(t)||dt = \int_0^T Edt = ET
\end{align}
where minimising evolution time $T$ thereby minimises distance $L$. Hamiltonians of interest are confined to a control subset $\mathcal{A}$ that disjunctively partitions the Lie algebra $\mathcal{M}$ (i.e. equivalent to generators being drawn from $\frak{p}$ or $\frak{k}$ above) and cases where $||H(t)|| \leq E$ (where the Hamiltonian can rescaled i.e. reparametrised so that the norm equals $E$ at all points on the path which in effect keeps the path identical but time shorter). They introduce a slight modification, that $\Tr(H^2(t)) = E^2$ in order to introduce the \textit{quantum brachistochrone problem} (see also \cite{carlini_time-optimal_2007}), a quantum analogue of the brachistochrone (meaning `shortest time') problem from classical variational mechanics \cite{goldstein_classical_2002}.    
Their method in essence adopts the penalty-metric approach of \cite{dowling_geometry_2008} such that in the limit, the lowest-energy solution tends towards minimal time by reason of the increased cost associated with higher-order (more than one- and two-body) generalised Pauli generators. 

The approach in \cite{wang_quantum_2015} is precisely to use the penalty metric approach of Nielsen et al. to generate a subRiemannian geodesic equation in order to confine the generators of the curve on the manifold to the control subset $\mathcal{A}$. This is achieved by adopting the norm-based cost function (pseudometric) where higher-order generator terms are weighted with penalty $q$, so that minimisation will by extension favour those generators (i.e. favour generators in $\frak{p}$ not $\frak{k}$). By doing so, a sufficiently proximal initial seed for the ``shooting method'' (see \cite{wang_quantum_2015,press_numerical_2007}) is generated. This method is a generic numerical technique for solving differential equations with two-point boundary problems (where our two points are $I$ and $U_T$ on $G$) and thus generating approximate geodesics.\\
\\
One of the other connections of Nielsen et al.'s work on geometric complexity to Cartan decompositions lies in the intuitive understanding that, in a control setting, sufficiently large penalty metrics in many ways correspond to a pseudo-partitioning of bundles/spaces into those subspaces which have low energy cost to reach and those subspaces which, although in principle reachable in a control setting (definition \ref{defn:geo:Reachable set}), cannot be obtained feasibly. Thus discovering ways in which state spaces that are \textit{effectively partitioned} or which are equipped with an \textit{effective decomposition} arising from penalty metrics is of use.  

%==========
 Another motivation of restricting control subsets as in \cite{swaddle_subriemannian_2017, swaddle_generating_2017} and our experiments above) to one- and two-body terms is to be found via the  geodesic approximations via the decomposition of the Lie algebra into projective subspace operators \cite{dowling_geometry_2008,brandt_jacobi_2010,brandt_riemannian_2010}. In \cite{wang_quantum_2015}, this was achieved via setting $\mathcal{P}(H) = H_P$ for one- and two-body Pauli terms and $\mathcal{Q}(H) = H_Q$ for three- or more-body Pauli terms using the Jordan-Hahn decomposition (see equation (\ref{eqn:quant:jordanhahn})) such that
\begin{equation}
    \frak{su}(2^n) = \mathcal{P} + \mathcal{Q} \qquad H = H_P + H_Q
\end{equation}
The idea is that higher-order (three- or more-body) terms in $\{ H_Q \}$ carry a penalty parameter (weight) which is designed, when curve length is obtained via minimising the action, to penalise higher-order terms in a way that the functional (solution) to the variational problem is more likely to contain only one- and two-body terms. Thus instead of restricting the sub-algebra of controls $\frak{p}$ to only one- and two-body terms (such as is undertaken in Swaddle), they instead (as per Nielsen's original paper) begin with full access to the entire $\frak{su}(2^n)$ Lie algebra (i.e. fully controllable) and then proceed to impose constraints in order to refine this down to geodesics comprising only (or mostly) one- and two-body terms. The distinction with the subRiemannian approach adopted in \cite{swaddle_generating_2017} is that in the latter case, generators for $U$ are by design constrained to be drawn from $\mathcal{P}=\Delta$ via the projection function in equation (\ref{eqn:qgml:projection}), circumventing imposition of Finslerian penalty metrics.

\section{Comparing geodesic approximations}
\label{sec:qgml:boozerswaddlecomparison}
Generation of geodesics in a QML context relies upon the availability of ways to compare whether outputs of machine learning models do in fact closely approximate geodesic curves. Thus the availability of reliable analytic and numerical methods for the generation of geodesics for use as training, testing and validation datasets is important. 
In our work, we sought to adapt the novel algorithmic approach to geodesic synthesis from \cite{swaddle_generating_2017} to include performance metrics of relevance to quantum information processing, such as fidelity measures. By comparison, \cite{boozer_time-optimal_2012} sets out an algorithm for determining time-optimal sub-Riemannian geodesics in $SU(2)$ which can be used to benchmark the performance of different machine learning approaches to synthesising approximate geodesics. While the derivation of time optimal parameters in \cite{boozer_time-optimal_2012} relies upon complicated sequence of coordinate transformations which is not easily scalable, it does provide a useful basis for comparison with the methods in \cite{swaddle_generating_2017}. In \cite{boozer_time-optimal_2012}, it is shown that time-optimal paths, where target unitaries constitute rotations by angle $\theta$ about the $z$-axis with generators being Pauli $X$ and $Y$ operators, can be synthesised in time-optimal fashion by following `circular' or holonomic paths along which they are parallel-transported. On the Bloch sphere, this is represented as `circular' paths emanating from the north pole whose diameter increases with increasing $\theta \in [0,2\pi]$. Intuitively, the greater the angle of rotation, the greater the diameter of the holonomic path (see section \ref{defn:geo:holonomy}). 

To this end, one of the objectives of our experiments was to ascertain the reliability of a few different methods of generating geodesics using methods drawn from geometric control sources. In order to do so, we compared this variational geodesic generation \cite{sachkov_control_2009} approach to a known method for analytically determining subRiemannian geodesics in $SU(2)$ in \cite{boozer_time-optimal_2012}. By doing so, we can be confident that the variational method appropriately approximates geodesics. The challenge posed in comparing geodesic methods lies in the differing assumptions of each method: \cite{swaddle_generating_2017} constrains the norms $||\projdelta(u_0)||=||u_0||$ as a means of more efficiently generating subRiemannian geodesic approximations \cite{swaddle_subriemannian_2017}, which is in effect the time scale (or energy scale) of their method. Conversely, \cite{boozer_time-optimal_2012}, works at different scales. In practice this means the generators for unitary evolution via each method differ by a scaling related to the norm of the generators. Such different parameterisations can be understood as follows:
\begin{align*}
    &\text{Swaddle parametrisation} & \text{Boozer parametrisation}\\
    &||H_j^{(S)}||=\Omega_j &  ||H_j^{(B)}||=1\\
    &dt_j^{(S)} = h = 1/N & dt_j^{(B)} = \Omega_j h/1\\
    &t_j^{(S)} = \sum_{k=1}^j h = jh & t_j^{(B)}  = \sum_{k=1}^j\Omega_j h/1
\end{align*}
For some desired tolerance (difference) $\epsilon$, the two approximations at are identical if the cumulative norms $D(H^{(S)},H^{(B)})$ of the sum of their $j$th Hamiltonians satisfy:
\begin{align}
    D(H^{(S)},H^{(B)}) & = \sum_j \left| \left| \frac{H_j^{(S)}}{\Omega_j} - H_j^{(B)} \right| \right| < \epsilon. \label{eqn:swadboozmetric}
\end{align}
That is, we want to minimise the distance between each Hamiltonian segment. The result in \cite{boozer_time-optimal_2012} is a relatively simple control problem where the control subset consists of Pauli $\sigma_x, \sigma_y$ generators with the target a rotation about the $z$-axis by angle $\eta$, $U_T = \exp(-i\eta \sigma_z/2)$. To validate that variational subRiemannian method can reproduce the time-optimal paths from \cite{boozer_time-optimal_2012}, a transformation between the two that enables comparison of Hamiltonians at time $t_j$ respectively in each formulation must be found. Pseudocode for such a transformation (in effect, a rescaling) of Hamiltonians generated using the method in \cite{swaddle_generating_2017} by comparison with those using the method in \cite{boozer_time-optimal_2012} is set-out below (where $(S)$ indicates Hamiltonians using the method in \cite{swaddle_generating_2017} and $(B)$ the method in \cite{boozer_time-optimal_2012}).

\begin{algorithm}[H]
% \SetAlgoLined
 Generate $H_j^{(S)}$\\
 Calculate $||H_j^{(S)}|| = \Omega_j$\\
 Calculate $t_j^{(B)}  = \sum_{k=1}^j\Omega_j h$\\
 $H_0^{(B)} = \frac{1}{\Omega_j}H_1^{(B)}$\\
 Calculate $H_j^{(B)} = e^{-i\omega t_j^{(B)} \frac{\sigma_z}{2}} H_0^{(B)} e^{i\omega t_j^{(B)} \frac{\sigma_z}{2}}$
 \caption{Comparison of subRiemannian and analytic geodesic circuits in $SU(2)$}
\end{algorithm}
Here, conjugation by $\exp(-i\omega t_j^{(B)} \frac{\sigma_z}{2})$ represents the Euler decomposition of the evolution in \cite{boozer_time-optimal_2012} as if one had direct access to the generator $\sigma_z$. Alternatively, one can also compare unitaries at equivalent times via operator fidelity $F(U_j^{(B)},U_j^{(S)})$ where:
\begin{align}
    U_j^{(B)}&=\exp(-i H_j^{(B)} dt_j^{(B)})
\end{align}
where we again use the assumption:
\begin{align}
    U_F^{(B)}&\approx e^{-iH_N^{(B)} dt_N^{(B)}}... e^{-iH_1^{(B)} dt_1^{(B)}}.
\end{align}
Numerical results comparing both Hamiltonian average distance (\ref{eqn:swadboozmetric}) and fidelities for ten $U_j$ instances across $N$ segments are set-out below.

\begin{center}
 \begin{table}[t]
 \begin{tabular}{|c|| c| c |} 
 \hline
 $j$ & $D(H^{(S)},H^{(B)})$ & $F(U_j^{(B)},U_j^{(S)})$ \\
 \hline
 \hline
 1 & 0.0922 & 0.9934 \\
 2 & 0.1986 & 0.9935 \\
 3 & 0.3105 & 0.9935 \\
 4 & 0.4169 & 0.9936 \\
 5 & 0.5154 & 0.9936 \\
 6 & 0.6046 & 0.9936 \\
 7 & 0.6836 & 0.9937 \\
 8 & 0.7518 & 0.9937 \\
 9 & 0.7871 & 0.9938 \\
 10 & 0.8345 & 0.9939\\
 \hline
 \end{tabular}
 \caption{Hamiltonian distance and unitary fidelity between Swaddle and Boozer geodesic approximations.}
\label{table:1}
\end{table}
 \end{center}
Fidelity results indicate little difference between $U_j^{(S)}$ and $U_j^{(B)}$, while Hamiltonian distance increases with $j$. Overall, the results provide some measure of confidence, though not analytic certainty, that the variational subRiemannian means of geodesic approximation in \cite{swaddle_generating_2017} are useful candidates for training data.
%=====parallel transport

% \newpage
\section{Neural network and GRU architectures}\label{sec:qgml:GRUNN}
In this section, we briefly summarise aspects of the neural network architectures adopted in our experiments above. More background theory and detail can be found in Appendix \ref{chapter:Background: Classical, Quantum and Geometric Machine Learning}. 
\subsection{Feed-forward neural networks} \label{sec:qgml:Feed-forward neural networks1}
Feed-forward fully-connected neural networks (see section \ref{sec:qgml:Feed-forward neural networks1} for general discussion), such as the ones deployed in the models above, can be understood in terms of functional composition. The objective of deep feed-forward networks is to define an input-output function $z = f(a,w,b)$ where $a^l$ are inputs to the layer $l$ (setting the initial input $a^0=x$), $w^l$ is a tensor of parameters for layer $l$ to be learnt by the model and $b^l$ is a bias tensor applied to $a^l$ \cite{nielsen_neural_2015, goodfellow_deep_2016}. 
In its simplest incarnation, the feed-forward stack takes as input a flattened realised $a_0 = U_T$ (where $k$ runs over the dimension of the vector). A layer of a simple neural network consists of units or neurons activation functions $\sigma$ (in our case, the ReLU or tanh activation function) applied to the $z$ such that we have $a^l=\sigma(z^l)$, vector and bias $b$:
\begin{align}
    a^l=\sigma(z^l)=\sigma(w^l a^{l-1} + b^l)
\end{align}
where we notice that the output of the previous layer is the input vector into the subsequent layer. All final layers in the feed-forward networks used $\sigma=\tanh$ activation functions. The output of an entire layer $a^l$ is a sequence structured as a vector that then becomes the input to the next layer. Information in this compositional model flows `forward' (hence `feed-forward'). 

When the entire set of units of a preceding layer becomes an input into each unit of the subsequent layer, we say the layer is dense. The weights are updated using backpropagation and gradient descent with respect to the applicable cost functional (description from \cite{nielsen_neural_2015}
 below, here $\odot$ is the Hadamard (element-wise) product), $x$ refers to each training example (batch gradient descent example below).
 
\begin{algorithm}[H]
% \SetAlgoLined
 Input: Set $x = a^0$\\
 Feed-forward: For $m$ layers, for $l=2,...,m$ calculate: \\
 \qquad $z^{x,l} = w^l a^{x,l-1}+b^l$\\
 \qquad $a^{x,l} = \sigma(z^{x,l})$\\
 \qquad $\sigma=\tanh$ for $l=m$ \\
 Output layer $(L=m)$ error $\delta^{x,l}$: \\
 \qquad $\delta^{x,l} = ((w^{l+1})^T \delta^{x,l+1})
  \odot \sigma'(z^{x,l})$\\
  \qquad $\sigma'=\frac{\partial a_k^{x,L}}{\partial z^{x,L}_k}$\\
  \qquad $k$ runs over neurons in layer $L$\\
 Backpropagation: for layers $l=L-1,L-2,...,2$, calculate:\\
 \qquad $\delta^{x,L} = \nabla_a C_x \odot \sigma'(z^{x,L})$\\
 Gradient: cost function  gradient given by: \\
 
 \qquad $\frac{\partial c}{(\partial w_{jk}^{x,l}} = a^{x,l-1}_{k}\delta^{x,l}_j$ and $\frac{\partial C}{\partial b^{x,l}_j} = \delta^l_j$\\
 Update weights: for each layer $l=L,L-1,...,2$ update:\\
 \qquad $w^l \to
  w^l-\frac{\eta}{m} \sum_x \delta^{x,l} (a^{x,l-1})^T$\\
  \qquad $b^l \to b^l-\frac{\eta}{m}
  \sum_x \delta^{x,l}$
  \qquad 
 \caption{Stochastic gradient descent and backpropagation (batch) \cite{nielsen_neural_2015}}
\end{algorithm}

\subsection{LSTMs and GRUs} \label{sec:qgml:LSTMs and GRUs1}
Long-Short Term Memory networks and Gated Recurrent Units are a prevalent form of recurrent neural network (RNN). RNNs are networks tailored to modelling sequential data, such as time-series data, or data such as sequences of control amplitudes $(c_j)$ \cite{goodfellow_deep_2016}. For RNNs, for each time-step $t$, there is an input $x_t$ (such as $c_t$), an output $y_t$ and hidden-layer output $h_t$. The key intuitive idea behind RNNs is that $h_t$ of the network itself becomes an input into hidden layers for the immediately next time-step $t+1$.  LSTMs advance upon this concept by enabling the output of hidden layers to influence not just the immediately succeeding time-step $t+1$, but also potentially activation functions at later time steps. In this sense LSTMs allow information about previous hidden layers (or states) to function as `memory' that is carried forward.

One of the challenges regarding RNNs is the saturation of networks where new inputs to an activation function fail to contribute significantly to its output. Intuitively too much information is saturating the model, so additional information does not lead to material updates (manifest, for example in flatlining loss, as seen in some examples above). A way to overcome this problem of saturation includes to stochastically `forget' certain information in order to make room for additional information, as manifest in the forget gate of an LSTM, distinct from the update gate. GRUs \cite{cho_van_2014} by contrast seek to incorporate the output of hidden layers and updates into subsequent hidden layers as detailed below. Their popularity is often owing to their improved speedup over LSTMs for a variety of contexts.

The reset gate combines the input $x_t$ at time $t$ with the previous time-step hidden state $h_{t-1}$ to define a reset output $r_t$ \cite{cho_properties_2014}:
\begin{align*}
    r_t = \sigma(w_r x_t + u_r h_{t-1} + b_r)
\end{align*}
where $w_r, u_r$ are updatable weight matrices and $b_r$ is an applicable bias, with $\sigma$ an activation function (in our models, the tanh function to produce control amplitudes $(c_j)$ but usually the sigmoid function). The update gate remains:
\begin{align*}
    z_t = \sigma(w_z x_t + u_z h_{t-1} + b_z)
\end{align*}
This update gate is the output of the unit at time $t$. However, in order to output $h_t$, an intermediate hidden layer state is calculated:
\begin{align*}
    \tilde{h}_t = \tanh(w_h x_t + u_h(r_t \odot h_{t-1}) + b_h)
\end{align*}
where we see the $(r_t \odot h_{t-1})$ term incorporates the influence of the reset gate and previous hidden layer $h_{t-1}$ into the estimate. The final hidden layer output is then calculated by combining the Hadamard products of the update gate and previous hidden state together with the intermediate hidden state:
\begin{align*}
    h_t = z_t \odot h_{t-1} + (1-z_t) \odot \tilde{h}_t
\end{align*}
which is the ultimate output at time $t$. The incorporation of $h_{t-1}$ in this way allows influence of prior information in the sequence to influence future outputs, improving the correlation between outputs such as controls.

%======================
%======================CHAPTER: CARTAN
%======================

\chapter{Global Cartan Decompositions for $KP$ problems}
\label{chapter:Time optimal quantum geodesics using Cartan decompositions}
\section{Abstract}
Geometric methods have useful application for solving problems in a range of quantum information disciplines, including the synthesis of time-optimal unitaries in quantum control. In particular, the use of Cartan decompositions to solve problems in optimal control, especially lambda systems, has given rise to a range of techniques for solving the so-called $KP$ problem, where target unitaries belong to a semi-simple Lie group manifold $G$ whose Lie algebra admits a $\g=\k \oplus \p$ decomposition and time-optimal solutions are represented by subRiemannian geodesics synthesised via a distribution of generators in $\p$. In this Chapter, we propose a new method utilising global Cartan decompositions $G=KAK$ of symmetric spaces $G/K$ for generating time-optimal unitaries for targets $-iX \in [\frak{p},\frak{p}] \subset \frak{k}$ with controls $-iH(t) \in \frak{p}$. Target unitaries are parametrised as $U=kac$ where $k,c \in K$ and $a = e^{i\Theta}$ with $\Theta \in \frak{a}$. We show that the assumption of $d\Theta=0$ equates to the corresponding time-optimal unitary control problem being able to be solved analytically using variational techniques. We identify how such control problems correspond to the holonomies of a compact globally Riemannian symmetric space, where local translations are generated by $\p$ and local rotations are generated by $[\p,\p]$.

\section{Introduction}
Symmetry-based decompositions \cite{zeier_symmetry_2011,echeverria-enriquez_geometric_2003,sheller_symmetry_nodate,dirr_lie_2008,dalessandro_quantum_2007,jurdjevic_geometric_1997,jurdjevic_optimal_1999} are a common technique for reducing problem complexity and solving constrained optimisation problems in quantum control and unitary synthesis. Among various decompositional methods, Cartan $KAK$ decompositions (section \ref{sec:alg:Cartan decompositions}) represent a generalised procedure for decomposing certain semi-simple Lie groups\cite{helgason_differential_1979} exhibiting involutive automorphic symmetry, akin to generalised Euler or singular-value decompositions. Cartan decompositions have found specific application across a range of domains, such as synthesising time-optimal Hamiltonians for spin-qubit systems in nuclear magnetic resonance \cite{khaneja_cartan_2001,khaneja_optimal_2005,khaneja_sub-riemannian_2002}, linear optics \cite{liu_collective_2022}, general qubit subspace decomposition \cite{earp_constructive_2005}, indirectly relating to entanglement dynamics \cite{bremner_fungible_2004} and the entangling power of unitary dynamics in multi-qubit systems \cite{bullock_canonical_2004}. Other approaches in information theory \cite{drury_constructive_2008} use Cartan decompositions for quantum Shannon decompositions and quantum circuit programming \cite{tucci_introduction_2005}. More recently, their use has been proposed for efficient measurement schemes for quantum observables \cite{yen_cartan_2021}, fixed-depth Hamiltonian simulation \cite{kokcu_fixed_2021}, reducing numerical error in many-body simulations \cite{steckmann_simulating_2021} and time-reversal operators \cite{bullock_time_2005} and also measurement of quantum observables \cite{yen_cartan_2021}. Cartan decompositions have been of interest in quantum unitary synthesis due to the fact that multi-qubit systems carrying representations of $SU(2^n)$ can often be classified as type AI and AIII symmetric spaces \cite{dalessandro_introduction_2007,khaneja_cartan_2001,noauthor_cartans_2018,su_scheme_2006,tucci_introduction_2005,khaneja_sub-riemannian_2002}. Specific interest symmetric space formalism in quantum computing has largely been due to their use in synthesising time-optimal or more efficient or controllable quantum circuits \cite{khaneja_cartan_2001,earp_constructive_2005,nielsen_geometric_2006,leifer_quantum_2008,liu_collective_2022,steckmann_simulating_2021,bullock_canonical_2004,bullock_time_2005,bullock_note_2004,gokler_efficiently_2017}. 

Symmetry-based decompositions (see section \ref{sec:alg:Cartan algebras and Root-systems} and \ref{sec:ml:Symmetry-based QML}), such as Cartan decompositions, have two-fold application in quantum control problems: firstly, symmetry-decompositions can simplify unitary synthesis via reducing the computational complexity \cite{kokcu_fixed_2022}; secondly, they have specific application in quantum control settings where the control algebra (and thus set of Hamiltonians) available are only a subalgebra of the corresponding full Lie algebra $\frak{g}$ \cite{khaneja_cartan_2001,brennen_observable_2003,swaddle_subriemannian_2017}. This Chapter focuses on the second use case. For unitary targets belonging to semi-simple connected groups $U_T \in G$ amenable to Cartan decomposition as $G=KAK$ (where $A < G/K$ and $K<G$), a key challenge is identifying the Hamiltonian which will synthesise the unitary optimally. While broadly utilised, Cartan-based unitary synthesis techniques have suffered from practical limitations due to exponential circuit depth \cite{nielsen_geometric_2006,huang_explicit_2007} together with difficulty in identifying the appropriate form of Cartan decomposition \cite{kokcu_fixed_2022} and form of Hamiltonian. In this Chapter, we address such challenges by providing a generalised procedure for time-optimal unitary and Hamiltonian synthesis using a global Cartan decomposition. Specifically, we demonstrate that for parametrised unitaries with targets in $G=KAK$, by utilising what we denote the \textit{constant}-$\theta$ method, Hamiltonians composed of generators from the horizontal (antisymmetric) bracket-generating distribution of the underlying Lie algebra associated with $G$ can, in certain cases, represent time-optimal means of synthesising such unitaries.      

\section{Symmetric spaces and KP time-optimal control} \label{sec:cartan:Symmetric spaces and KP time-optimal control}
\subsection{Overview - setting the scene} \label{sec:cartan:Overview - setting the scene}
Under the postulates of quantum mechanics (axiom \ref{axiom:quant:evolution}), unitary motion is defined by the Schr\"odinger equation (definition \ref{eqn:quant:schrodingersequation}):
\begin{align}
\frac{dU}{dt} = -iHU \label{eq:cartan:2.1}
\end{align}
whose solution is often formally written as the time-ordered exponential (equation (\ref{eqn:quant:unitarysolutiontimedependentschrod})):
\begin{align}
U(T) = \mathcal{T}_+e^{-i \int_0^T H(t)dt}U_0. \label{eq:cartan:2.2}
\end{align}
Usually, these kind of equations are considered as matrix or operator equations in a given representation. However, such equations can be considered more fundamentally differential geometric in nature. This is because time-ordered exponentials can be expanded in commutators which themselves have a structure that is often universal. Universal in this context refers to the fact that a matrix representation can be one of an infinite number of otherwise inequivalent representations. Perhaps the most useful and well-known example is angular momentum, with commutators that define the group of virtual rotations. To emphasize angular momenta and rotations as universal is to appreciate that they are concepts which precede or transcend any particular matrix representations, and this universality serves as a platform for understanding details such as energy quantization.

The differential geometry of universal structures was a famous passion of Poincar\'e, who originated the ideas of ``universal covering group'' and ``universal enveloping algebra'', and Klein, for whom the ``universal properties'' of category theory are an homage (see \cite{sharpe_differential_2000, gray_felix_2005} and \cite{hawkins_emergence_2012} for historical background). In quantum physics however, although it is certainly useful to have an intuitive understanding that rotation is universal, there is usually no need to work directly with things like universal covers. Instead, the standard is to assume a Hilbert space (definition \ref{defn:quant:Hilbert Space}) and immediately consider angular momenta as acting on states to logically derive the spin quantum numbers and consequent matrix representations. Indeed, this quantum standard to act on states is why equation \eqref{eq:cartan:2.1} is often called ``Schr\"odinger.'' Followed by the basic fact that the matrix group $SU(2)$ is the fundamental representation of the universal cover of 3-dimensional rotations, one can then use this matrix group (and perhaps a little good faith) to come to understand the universal. However, if more general transformation groups are considered, calculations with matrix representations can make the universal geometry they are representing far less apparent. This article demonstrates that the so-called \textit{KP-problems}, can be solved completely with the help of some universal techniques. The $KP$ problems are a family of unitary control problems (section \ref{sec:quant:Quantum Control}) where the Hamiltonian control space $ \p $ is a Lie triple system, defined by the property $[\p, [\p, \p]] \subseteq \p$, and the target Hamiltonian is in $[\p, \p] \subseteq \k$. We expand on this below and in section \ref{sec:geo:KP Problems}.

\subsection{KP problems for Lambda systems} \label{sec:cartan:KP problems for Lambda systems}
Geometric control theory (section \ref{sec:geo:Geometric control theory}) is characterised by the study of the role of symmetries in optimal control problems. The $KP$ problem has been most extensively detailed by Jurdjevic et al. \cite{jurdjevic_geometric_1997} in the context of geometric control theory. A common application of techniques of geometric control involves so-called \textit{lambda systems}, three-level quantum systems where only two of the levels are encoded with information of interest. In such systems, the two lowest energy states of a quantum system are coupled to the highest third energy state via electromagnetic fields \cite{dalessandro_time-optimal_2020}. The typical model for the study of lambda systems in quantum control settings is the Schr\"odinger equation in the form:
\begin{align}
    \frac{dU(t)}{dt} = \hat A U(t) + \sum_j B_j U(t) \hat u_j(t) \qquad U(0) = \mathbb{I}
\label{eqn:cartan:generalschrodtimeopt}
\end{align}
where $U(t) \in SU(3)$ and $\hat A$ is a diagonal matrix comprising the energy eigenvalues of the system. The control problem becomes finding control functions $\hat u(t)$ to synthesise $U_T=U(T)$ in minimum time subject to the constraint that $||\hat u|| < C$ for constant bound $C$. The unitary targets are $U_T \in G$, a semi-simple Lie group admitting a Cartan decomposition into compact and non-compact subgroups $K$ and $P$ respectively (hence the nomenclature ``$KP$''). As discussed in \cite{albertini_symmetries_2018} and elsewhere in the literature \cite{nielsen_geometric_2006,dowling_geometry_2008}, time-optimisation can be shown to be equivalent to synthesising subRiemannian geodesics (section \ref{sec:geo:SubRiemannian Geodesics}) on a subRiemannian manifold implied by the homogeneous space $G/K$. In this picture, the minimum synthesis time equates to the arc length (equation (\ref{eqn:geo:arclengthSubRiemannian}) (measured according to the applicable subRiemannian metric) of the subRiemannian curve $\gamma$ calculated by reference to the pulses $u(t)$ (see equation (5) in \cite{albertini_symmetries_2018}). Intuitively, in this picture, the control Hamiltonian in $\p$ traces out a minimal-time along the manifold. For certain classes of $KP$ problem, Jurdjevic et al. \cite{jurdjevic_geometric_1997,jurdjevic_hamiltonian_2001,jurdjevic_optimal_1999,dalessandro_k-p_2019} have shown that geodesic solutions to equation (\ref{eqn:cartan:generalschrodtimeopt}) take a certain form (see example below).  In \cite{jurdjevic_optimal_1999} this is expressed as:
\begin{align}
    \frac{d\gamma}{dt} &=\gamma\left( e^{Kt} P e^{-At} \right)\\
    \gamma(t) &= e^{(-A + P)t}e^{-At}
    \label{eqn:cartan:jurdgeodesicsolutions}
\end{align}
with $A \in \k$ and assuming $\gamma(0)=\mathbb{I}$ for a symmetric matrix $P$ (that is in $\p$). Discussion of the case in which $e^{(-A + P)t}$ is a scalar is set out in \cite{albertini_sub-riemannian_2020}. Detailed exposition by Jurdjevic is set out in \cite{jurdjevic_geometric_1997,jurdjevic_hamiltonian_2001,jurdjevic_optimal_1999} and elsewhere. In this Chapter, we show that such results can be obtained by a global Cartan decomposition in tandem with certain constraints on chosen initial conditions $U(0)=U_0$ and the choice of generators $\Phi \in \k$ used to approximate the target unitary in $K$. 
\subsection{KAK decompositions} \label{sec:cartan:KAK decompositions}
In many quantum control and quantum algorithm settings, the target unitary $U_T$ is an element of a semi-simple Lie group $G$ such that $U_T \in G$. In certain cases, $G$ may be decomposed into a homogeneous space $G/K$ where $K$ represents an isometry (stabilizer) subgroup $K < G$. Such semi-simple Lie groups can be decomposed as $G = KAK$ where $\a \subset \p$ and $A = \exp(\a)$, a decomposition known as a Cartan decomposition \cite{dalessandro_introduction_2007,khaneja_cartan_2000,khaneja_optimal_2005,dalessandro_lie_2008,dalessandro_k-p_2019,nielsen_optimal_2006,helgason_differential_1979}. We discuss this decomposition at length in Appendix \ref{chapter:Background: Geometry, Lie Algebras and Representation Theory}, particularly in sections \ref{sec:alg:Cartan decompositions} and \ref{sec:alg:Cartan algebras and Root-systems}. The decomposition $G=KAK$ is considered a global Cartan decomposition (applying to the entire group) rendering $G/K$ a globally Riemannian symmetric space (see section \ref{sec:geo:Symmetric spaces}). The corresponding Lie algebra $\frak{g}$ may similarly be decomposed as $\frak{g} = \frak{k} \oplus \frak{p}$ where $\frak{p} = \frak{k}^\bot$. The existence of a Cartan decomposition is equivalent to the satisfaction of the following canonical Cartan commutation relations (definition \ref{defn:alg:cartandecomposition}):
\begin{align}
[\frak{k},\frak{k}] \subseteq \frak{k} \qquad [\frak{p},\frak{p}] \subseteq \frak{k} \qquad [\frak{p},\frak{k}] \subseteq \frak{p}.
    \label{eqn:cartan:cartancommrelations}
\end{align}
Given a choice of maximally non-compact Cartan subalgebra $\frak{a} \subset \frak{p}$, $G$ can be decomposed as $G = KAK$, where $A = e^{\frak{a}}$. Doing so allows unitaries in $G$ to be written as:
\begin{align}
    U = ke^{i\Theta}c
\end{align}
where $k,c \in K$ and $e^{i\Theta} \in A$ (where $\theta$ parametrises a generators in $\frak{a}$, e.g. a rotation angle). Satisfaction of Cartan commutation relations is also equivalent to the existence of an involution (see definition \ref{defn:alg:cartaninvolution}) $\chi^2=\mathbb{I}$ which partitions $G$ (and $\frak{g}$) into symmetric $\chi(\frak{k})=\frak{k}$ and antisymmetric $\chi(\frak{p})=-\frak{p}$ subalgebras. Here $K = e^\frak{k},A \subset P = e^\frak{p}$ \cite{knapp_representation_2001,helgason_differential_1979,hermann_lie_1966}. Elements of $G$ can be written in terms of the relevant group of action of subgroups $K$ and $A$, including, where relevant, unitary elements of $G$. Arbitrary targets $U_T\in G$ remain reachable, however elements of the vertical  (symmetric) subalgebra must be indirectly synthesised via application of the Lie derivative (bracket) i.e. $[\frak{p},\frak{p}] \subseteq \frak{k}$. Such decompositions are manifestly coordinate-free when represented using differential forms. Satisfying such criteria allows $G/K$ to be equivalently characterised as a Riemannian (or subRiemannian) symmetric space. \\
\\
 We propose that for certain classes of quantum control problems, namely where the antisymmetric centralizer generators parameterised by angle $\theta$ remain constant, analytic solutions for time-optimal circuit synthesis are available for non-exceptional symmetric spaces. Such cases are explicitly where control subsets are limited to cases where the Hamiltonian comprises generators from a horizontal distribution (bracket-generating \cite{helgason_differential_1979,knapp_representation_2001} and see definition \ref{defn:geo:bracketgenerating}) $\frak{p}$ with $\frak{p}\neq \frak{g}$ (where the vertical subspace (definition \ref{defn:geo:Vertical subspace}) is not null). Only access to subspace $\frak{p} \subset \frak{g}$ is directly available for control purposes. If $[\frak{p},\frak{p}] \subseteq \frak{k}$ holds, arbitrary generators in $\frak{k}$ may be indirectly synthesised (via application of Lie brackets) which in turn makes the entirety of $\frak{g}$ available and thus, in principle, arbitrary $U_T \in G$ (if $[\p,\p]=\k$) reachable (in a control sense, see definition \ref{defn:geo:Reachable set}).

 \subsection{Sketch of constant-$\theta$ method} \label{sec:cartan:sketch of constant theta method}
 Here we sketch the constant-$\theta$ method. In subsequent sections, we provide worked examples. A generalised form of the method is set out in section \ref{sec:cartan:generalmethod} below. A target unitary $U_T \in G$ may be decomposed into a Cartan $G=KAK$ coordinate system can be written in the form:
\begin{align}
    U = k e^{i\Theta}c = q e^{-ic^\inv\Theta c}
    \label{eqn:cartan:u-kak}
\end{align}
where $c,q, k \in K$ and $e^{i\Theta} \in A \subset P$. Locally forward evolution in time can be written differentially as:
\begin{align}
    dU U^\inv = -iHdt
\end{align}
where the left-hand side consists of parameters over the manifold of unitaries (or the group $G$), while the right-hand side consists of parameters that are in a geometric sense external to the manifold, in the vector field (definition \ref{defn:geo:vectorfield}) associated with the manifold.  Taking (\ref{eqn:cartan:u-kak}) as our unitary, then:
\begin{align}
    dU U^\inv = k\bigg[k^\inv dk +\cos\ad_\Theta(dc c^\inv) + i\Big(d\Theta + \sin\ad_\Theta(dc c^\inv)\Big)\bigg]k^\inv.
    \label{eqn:cartan:duuinvkak1}
\end{align}
Geometrically, we can interpret the left $K$ as a local frame (a choice of basis for the tangent space $T_p\M$ at each point $p$ of the group $G$ - see definition \ref{defn:geo:Tangent}), the right $K$ as a global azimuth (a parameter that describes the position of a point on the sphere and is the same for all points along the `longitude' (or orbits of) $K$), and the $\Theta$ as the polar geodesic connecting the two (akin to latitude). The latter two sets of parameters are the coordinates of a symmetric space. Presented in this way, the Schr\"odinger equation represents a Maurer-Cartan (differential) form \cite{helgason_differential_1979} (see definition \ref{defn:geo:Maurer-Cartan Form}), encoding the infinitesimal structure of the Lie group and satisfies the Maurer-Cartan equation (equation (\ref{eqn:geo:Maurer-Cartan equation})). For our purposes, it allows us to interpret the right-hand side in terms of the relevant principal bundle connection (see sections \ref{sec:geo:Principal fibre bundles} and \ref{sec:geo:Connections}). Intuitively, a connection provides a way to differentiate sections of the bundle and to parallel transport (definition \ref{sec:geo:Parallel transport and horizontal lifts}) along curves in the base manifold $G$. The minimal connection (see section \ref{sec:cartan:minimalconnections} for exposition) here is a geometric way of expressing evolution only by elements in $\frak{p} \in H\M$ (i.e. the horizontal subspace) where the quantum systems evolves according to generators in the horizontal control subset of the Hamiltonian given by:
\begin{align}
    i\Big(d\Theta + \sin\ad_\Theta(dc c^\inv)\Big). \label{eqn:cartan:horizontalcontrolsubalgebra}
\end{align}
The evolution of the system can be framed geometrically as tracing out a path on the manifold (a Riemannian symmetric space) according to generators and amplitudes (and controls) in the Hamiltonian. This path is a sequence of quantum unitaries. In quantum mechanics, we are interested in minimising the external time parameter. For this, we typically minimise the action such that the total time (length) of a circuit evolving is:
\begin{align}
    \Omega T = \int_\gamma\!\sqrt{(idU U^\inv, idU U^\inv)}. \label{eqn:cartan:omegatintgamma}
\end{align}
In common with geometric methods, the integral is usually parametrised by path length $s$ (so integrated between 0 and 1 - see the examples below) of the curve $\gamma$ traced out along the manifold. Here $\Omega = |H|$ and where $(X,Y)=\Tr(\text{ad}_X \text{ad}_Y)$
is the (representation independent) Killing form (see definition \ref{defn:alg:Killing form}) which, in certain cases induces a scalar form (see below and \cite{knapp_representation_2001,cahn_semi-simple_2014}).  
As is typical in differential geometry, the minimisation of evolution time of the curve $\gamma \in G$ (where $\gamma$ represents a unitary circuit parameterised by arc length (see equation (\ref{eqn:geo:arclength})) occurs over a typical geometric path length $s$ where $s \in [0,1]$ parametrises the path from beginning to end and $ds = \Omega dt$ (see the Appendix for more discussion). To find the solution for minimal time $T$, we consider paths such that $d\Theta = 0$ (and thus $d\theta=0$) which we denote the `constant-$\theta$' method. The constant-$\theta$ method implies that for the Cartan algebra parameter $\theta$, we have $d\theta = 0$.  We conjecture that upon local variation of the total time with respect to the path, we obtain Euler-Lagrange equations which show that $dq q^\inv$ must also be constant for locally time-optimal paths.
The total time for locally time-optimal constant-$\theta$ paths becomes: 
\begin{align}
    \Omega T = \min_{\theta,\phi} |\sin \text{ad}_\Theta(\Phi)| \label{eqn:cartan:OmegaT-minsinadtheta}
\end{align}
where:
\begin{align}
i\Phi = \int dc c^\inv.
\label{eqn:bigphiintdqqinv}
\end{align}
Determining the time-optimal geodesic path requires global variation over all geodesics, typically a hard or intractable problem. As shown in the literature \cite{nielsen_geometric_2006,dowling_geometry_2008,gu_quantum_2008,wang_quantum_2014}, variational methods can be used as a means of calculating synthesis time. For constant-$\theta$ paths, such calculations are significantly simplified as we demonstrate below.

\subsection{Holonomy targets} \label{sec:cartan:Holonomy targets}
The local frame has two coordinate systems that are particular, the cardinal (basis) frame ``$k$'' and the geodesic frame ``$q$.'' The cardinal frame describes the local, compact part of the group, while the geodesic frame describes the global, non-compact part of the group (see \cite{knapp_lie_1996} for detailed discussion of such concepts). We define a target in the holonomy group $K$ to be one such that: 
\begin{align}
U(T)U(0)^\inv = e^{iX} \in K. \label{eqn:cartan:u(T)u(0)-exp(iX)holonomy}
\end{align}
Holonomy group (definition \ref{defn:geo:holonomy}) refers to the set of group transformations a vector undergoes when it is parallel transported around a closed loop in a manifold. For symmetric spaces, the choice of connection gives rise to an implicit subgroup $K < G$ which acts as a holonomy group whose action traces out equivalence classes of orbit (and so in this sense $\k$ defines rotations) while $\p$ defines translations between orbits (intuitively seen in the key commutation relation $[\p,\p]\subseteq \k$). We define the connection for the geodesic frame:
\begin{align}
q^\inv dq = c^\inv\Big[(1-\cos\ad_\Theta)(dc c^\inv)\Big]c \label{eqn:cartan:qinvdq}
\end{align}
such that the Hamiltonian does not contain elements of $\frak{k}$. The intuition for a qubit is that rotations achieved via $J_z$ are by construction parallel transporting vectors via the action of $[\p,\p]\subseteq \k$. Where $J_z$ is not in the control subset, a way to parallel transport such vectors, achieving a $J_z$ rotation, but using other generators not in $K$ is needed. For constant-$\theta$ paths:
\begin{align}
\Ad_{\Phi} (\Theta) = \Theta 
%\label{eqn:cartan:adphitheta}
\hspace{50pt}
\text{and therefore}
\hspace{50pt}
X = (1-\cos\ad_\Theta)\Phi. \label{eqn:cartan:x=1-cosadtheta}
\end{align}
Note the first of these conditions is explicated as a condition that the generators comprising $\Phi \in \frak{k}$ belong to the commutant (see the general method section \ref{sec:cartan:generalmethod} for discussion). Holonomic targets may be, under certain assumptions, generated via unitaries $U \in G/K$ which in a Riemannian context are paths with zero geodesic curvature, indicating parallel translation in a geometric sense. By contrast, where the manifold is subRiemannian (i.e. where $\frak{p} < \frak{g}$) then subRiemannian geodesics may exhibit non-zero geodesic curvature by comparison with $G$ and $\frak{g}$ as a whole. 

\subsection{Symmetric space controls in $\frak{p}$} \label{sec:cartan:Symmetric space controls in p}
Summarising the above, our conjecture is that for symmetric space controls in $\frak{p}$ with holonomy targets in $\frak{k}$ that constant-$\theta$ paths are time-optimal. 
For such constant-$\theta$ paths, the objective becomes to calculate:
\begin{align}
\Omega T = \min_{\Theta,\Phi}\Big|\sin\ad_\Theta(\Phi)\Big|  \label{eqn:cartan:mintimemain}
\end{align}
under the constraints
\begin{align}
\Ad_{\Phi} (\Theta)= \Theta
\label{eqn:cartan:main-eiadPhiTheta=Theta}
\end{align}
and
\begin{align}
X = (1-\cos\ad_\Theta)\Phi.
\label{eqn:cartan:main-x=1-cosadthetabigphi}
\end{align}
The variations can be performed elegantly by the feature that $\Theta$ is in a symmetric space and $\Phi$ is in a reductive Lie group. From this point, the problem of minimising time is undertaken using variational techniques (such as Lagrange multipliers). The constant-$\theta$ assumption allows us to simplify this problem to a significant degree. We also show in our general exposition how a transformation to a restricted Cartan-Weyl basis can also assist in simplifying the often challenging global minimisation problem.

\subsection{Cartan decomposition} \label{sec:cartan:Cartan decomposition}
To explicitly connect the $G=KAK$ decomposition to the typical Schr\"odinger equation, consider the $KAK$ decomposition of a unitary $U \in G$ given by:
\begin{align}
    U = qe^a k \label{eqn:cartan:U=eqak}
\end{align}
where $q, k \in K$ and $e^a \in A$. Schr\"odinger's equation can be written consistent with the Maurer-Cartan form (see section \ref{sec:cartan:generalmethod}) as:
\begin{align}
    dU U^{-1} = -iHdt.
    \label{eqn:cartan:schrodcartandecomp}
\end{align}
Expanding out we have:
\begin{align}
    dU &= d(q e^a k) = e^a kdq + q e^a k da + qe^a dk\\
    U^{-1} &= (q e^a k)^{-1} = k^{-1}e^{-a}q^{-1}. \label{eqn:cartan:dUexpanded1}
\end{align}
Which resolves to:
\begin{align}
    dU U^\inv &= dq q^\inv + q da q^\inv + q e^a dk k^\inv e^{-a} q^\inv \\
    &= dqq^\inv + q da q^\inv + q e^{ad_a} (dk k^\inv) q^\inv. \label{eqn:cartan:dUexpanded1}
\end{align}
The adjoint term can be decomposed into symmetric and anti-symmetric parts:
\begin{align}
    e^{ad_a}(X) &= \underbrace{\cosh(\ad_a)(X)}_{\text{even powers}} + \underbrace{\sinh(\ad_a)(X)}_{\text{odd powers}} \label{eqn:cartan:adjointsymmetricantisymmetric}
\end{align}
such that:
\begin{align}
    e^{ad_a}(dk k^{-1}) = \underbrace{\cosh(\ad_a)(dk k^{-1})}_{\in \frak{k}} + \underbrace{\sinh(\ad_a)(dk k^{-1})}_{\in \frak{p}}. 
    \label{eqn:cartan:expcoshsinh}
\end{align}
As $\frak{a} \subset \frak{p}$ and given the Cartan commutation relations, we have $[\frak{a},\frak{k}] \subset \frak{p}$ while $[\frak{a},[\frak{a},\frak{k}]] \subset \frak{k}$. Thus the symmetric term in equation (\ref{eqn:cartan:expcoshsinh}) above, comprising even powers of generators is in $\frak{k}$ while the antisymmetric term, comprising odd powers of generators, will be in $\frak{p}$. Rearranging equation (\ref{eqn:cartan:schrodcartandecomp}):
\begin{align}
    dU U^\inv = q[da + \sinh(\ad_a)(dkk^\inv) + q^\inv dq + \cosh(\ad_a)(dkk^\inv)] q^\inv.
    \label{eqn:cartan:dUUsinhcosh}
\end{align}
The first two terms are in $\frak{p}$ while the latter two are in $\frak{k}$. The orthogonal partitioning from the Cartan decomposition ensures simplification such that cross-terms such as Tr$(pk)$ vanish. Thus:
\begin{align}
    \text{Tr}\left[ (dUU^\inv)^2  \right] & = \text{Tr} \left[(da + \sinh(\ad_a)(dkk^\inv))^2  \right] + \text{Tr}\left[ (q^\inv dq + \cosh(\ad_a)(dkk^\inv))^2 \right] \label{eqn:cartan:tr(duuinv)1}
\end{align}
and:
\begin{align}
    \text{Tr}\left[ (dUU^\inv)^2  \right] & = \text{Tr} \left[ da^2 \right] + \text{Tr}\left[\sinh(\ad_a)(dkk^\inv))^2  \right] + \text{Tr}\left[ (q^\inv dq + \cosh(\ad_a)(dkk^\inv))^2 \right] \label{eqn:cartan:tr(duuinv)2}
\end{align}
using $\text{Tr}(X \ad_Y(Z) = -\text{Tr}(\ad_Y(X)Z)$. 

To demonstrate the constant-$\theta$ method, we assume that our controls (generators in our Hamiltonian) are in $\frak{p}$. Doing so restricts our Hamiltonian to the horizontal subspace (definition \ref{defn:geo:Horizontal subspace}) under which the quantum state is parallel transported as the system evolves. This is equivalent to approximating a subRiemannian geodesic circuit over the relevant differentiable manifold for our chosen group $G/K$. The choice of generators that enables such parallel transport without the use of generators in $\frak{k}$ is equivalent to a Hamiltonian comprising only generators in $\frak{p}$, ensuring the quantum state undergoes parallel transport along curves $\gamma \in G/K$ such that $\nabla_{\dot{\gamma}} \dot{\gamma} = 0$ (see section \ref{sec:geo:Covariant differentiation}).  In equation (\ref{eqn:cartan:dUUsinhcosh}) this is equivalent to setting $da + \sinh(\ad_a)(dkk^\inv)=0$ or $da = -\sinh(\ad_a)(dkk^\inv)$ (in geometric parlance, setting a minimal connection see sections \ref{sec:cartan:minimalconnections} and \ref{sec:geo:Connections} generally).  
Using a change of variables we denote:
\begin{align}
    a = i\Theta \qquad k=e^{i\Phi} \qquad q=e^{i\Psi} \label{eqn:cartan:a=itheta}
\end{align}
where $dkk^\inv = id\Phi$ and $dqq^\inv = id\Psi$. Using $\cosh(\ad_{i\Theta}) = \cos(\ad_\Theta)$ and $\sinh(\ad_{i\Theta}) = \sin(\ad_\Theta)$ the connection becomes:
\begin{align}
    d\Psi + \cos(\ad_\Theta)(d\Phi) = 0. \label{eqn:cartan:connection-dpsi+cos(adtheta)dphi}
\end{align}
Assuming a constant theta $d\Theta = 0$ together with the above transformations allows our Hamiltonian in equation (\ref{eqn:cartan:dUUsinhcosh}) to be written as:
\begin{align}
    Hdt = e^{i\Psi}(-i\sin(\ad_\Theta)(d\Phi)) e^{-i\Psi}. \label{eqn:cartan:hamiltonianconnectinosettozero}
\end{align}
From this point, we must then solve the minimisation problem. 
\section{Time-optimal control examples} \label{sec:cartan:Time-optimal control examples}
We demonstrate such time-optimal analytic solutions for a few common types of symmetric space quantum control with targets in $SU(2)$ and $SU(3)$ to explicate the constant-$\theta$ method. We generalise this method in section \ref{sec:cartan:generalmethod}.
\subsection{SU(2) time-optimal control}
% \label{sec:cartan:su2timeoptimalcontrol}
Consider $G=SU(2)$ with isometry group (definition \ref{defn:quant:isometries}) $K = \text{S(U(1)} \times \text{U(1))} = \text{span}\{ e^{i \eta \sigma_z}\}$. Further define Pauli matrices $\{ \sigma_k \}$ with angular momenta $J_k = \frac{1}{2}\sigma_k$. Define the Cartan conjugation $\chi$ as:
\begin{align}
    U^\chi = e^{-iJ_z\pi} U^\dagger e^{iJ_z \pi} \label{eqn:cartan:Cartan conjugation}
\end{align}
together with a Cartan projection $\pi(U) = U^\chi U$. To define the relevant Cartan decomposition, we note the quotient group $G/K$ corresponds to the symmetric space:
\begin{align}
    S^2 \approx \frac{SU(2)}{S(U(1) \times U(1))} \equiv \text{AIII(1,1)} \label{eqn:cartan:S2}
\end{align}
where $S^2=\pi(G)=G/K$. Detail on the classification of symmetric spaces is set out in section \ref{sec:geo:Classification of symmetric spaces}). As distinct from the projection in \cite{boozer_time-optimal_2012}, the projection is representation independent. The Cartan projection above defines the relevant $G=KAK$ decomposition. The corresponding Cartan decomposition of the Lie algebra $\frak{g}$ satisfying equations (\ref{eqn:cartan:cartancommrelations}) is:
\begin{align}
    \frak{g} = \frak{su}(2) = \underbrace{\braket{-iJ_z}}_{\frak{k}} \oplus \underbrace{\braket{-iJ_x, -iJ_y}}_{\frak{p}}
    \label{eqn:frakg=k+p}
\end{align}
with Cartan subalgebra chosen to be $\frak{a} = \braket{-iJ_y}$. We can easily see the Cartan commutation relations (equation (\ref{eqn:cartan:cartancommrelations})) hold. For $S^2$ above this corresponds to the Euler decomposition:
\begin{align}
    U = e^{iJ_z \psi} e^{iJ_y \theta} e^{iJ_z \phi}. \label{eqn:cartan:eulerdecomposition}
\end{align}
Define:
\begin{align}
    k=e^{iJ_z \psi} \qquad i\Theta = iJ_y \theta \qquad c = e^{i J_z \phi} \qquad q = kc = e^{iJ_z \chi}
\label{eqn:cartan:su2:kiThetacqparams}
\end{align}
to represent the unitary as:
\begin{align}
    U = ke^{i\Theta}c = qe^{ic^\inv \Theta c}.
    \label{eqn:cartan:ukeithetac}
\end{align}
Under this change of variables, the Cartan conjugation becomes:
\begin{align}
    k^\chi = k^\inv 
\qquad
(e^{i\Theta})^\chi = e^{i\Theta}
\qquad
(UV)^\chi = V^\chi U^\chi \label{eqn:cartan: Cartan conjugation (k)}
\end{align}
with (using the adjoint action (definition \ref{defn:alg:Adjoint action})):
\begin{align}
    \pi(ke^{i\Theta}c) = e^{2c^\inv i\Theta c}. \label{eqn:cartan:pi(keiThetac)}
\end{align}
The Schr\"odinger equation then becomes:
\begin{align}
    dUU^\inv
&= k\left(k^\inv dk + \cos\ad_\Theta(dcc^\inv) + id\Theta + i \sin\ad_\Theta(dcc^\inv)\right)k^\inv\\
& = k\Big(iJ_z(d\psi+d\phi\cos(\theta)) + iJ_yd\theta - iJ_xd\phi\sin(\theta)\Big)k^\inv
\label{eqn:su2mauercartan}
\end{align}
by taking the relevant differentials and inverses in equation (\ref{eqn:cartan:su2:kiThetacqparams}), using equation (\ref{eqn:alg:econjsinhcosh}) and calculating the adjoint action on $c$. Under this Cartan decomposition, we see equation (\ref{eqn:su2mauercartan}) partitioned into symmetric $\frak{k}$ and antisymmetric $\frak{p}$ part:
\begin{align}
    dUU^\inv = k\Big(\underbrace{iJ_z(d\psi+d\phi\cos(\theta)}_{\in \frak{k}}) + \underbrace{iJ_yd\theta - iJ_xd\phi\sin(\theta)}_{\in \frak{p}}\Big)k^\inv. \label{eqn:cartan:su2schrodinger-symmantisymparts}
\end{align}
Restricting $dUU^\inv \in \frak{p}$ is equivalent to defining a minimal connection $d\psi = -d\phi \cos \theta$. That is, effectively setting the symmetric part of the Hamiltonian to zero (or in geometric parlance, restricting to the horizontal subspace corresponding to the underlying subRiemannian manifold). The Hamiltonian becomes:
\begin{align}
    H = i\frac{dU}{dt}U^\inv = k\Big(J_x\dot\phi\sin(\theta) -J_y\dot\theta \Big)k^\inv \label{eqn:cartan:su2hamiltonian1}
\end{align}
where we have defined conjugate momenta $\dot\phi,\dot\theta$ for extremisation below. To calculate the optimal time, we first define the Killing form on $\frak{g}$ for which we require normalisation of the extremised action in (such that we can define an appropriately scaled norm and metric).  For a subRiemannian space of interest in the adjoint representation, the Euclidean norm is then simply defined in terms of the Killing form as $|X| = \sqrt{(X,X)}$ such that:
\begin{align}
    |idUU^\inv|^2 = (d\psi+d\phi\cos(\theta))^2 + d\theta^2 + (d\phi\sin(\theta))^2
    \label{eqn:iduuinv^2}
\end{align}
where $|J_z|^2=\mathbb{I}$. Define the energy of the Hamiltonian as $|H| = \Omega$ such that:
\begin{align}
    \Omega t = \Omega \int_\gamma dt = \int_\gamma |idUU^\inv| =  \int_\gamma \left|i\frac{dU}{ds}U^\inv\right| ds. \label{eqn:cartan:su2omegat}
\end{align}
Here $\gamma$ defines the curve along the manifold generated by the Hamiltonian and $t$ the time elapsed. Calculating path length here is equivalent to approximating time elapsed modulo $\Omega$. Consistent with typical differential geometric methods, we parametrise by arc length $ds$\cite{do_carmo_differential_2016}. Define:
\begin{align}
    \dot t = \left|i\frac{dU}{ds}U^\inv\right| = \frac{dt}{ds} \label{eqn:cartan:parambyarclength}
\end{align}
setting optimal (minimal) time as:
\begin{align}
    T = \min_\gamma t = \min \left| \int_\gamma \dot t ds  \right|.
    \label{eqn:cartan:su2:t=mint}
\end{align}
Extremisation can be performed using the method of Lagrange multipliers and the minimal connection above. As we demonstrate below, doing so in conjunction with the constant-$\theta$ assumption simplifies the variational problem of estimating minimal time. The relevant action is given by:
\begin{align}
    S = \Omega t =  \int_\gamma\Big(\dot{t} + \lambda J_z(\dot\psi+\dot\phi\cos(\theta))\Big) ds
    \label{eqn:cartan:su2:s=omegataction}
\end{align}
noting again the role of the connection. Parametrising by arc length $s$ we have:
\begin{align}
    \left|i\frac{dU}{ds}U^\inv\right|^2 = (\dot\psi+\dot\phi\cos(\theta))^2 + \dot\theta^2 + (\dot\phi\sin(\theta))^2. \label{eqn:cartan:|du/dsu^inv}
\end{align}
Extremising the action $\delta S = 0$ resolves the canonical position and momenta via the equation above as:
\begin{align}
\Omega \frac{\delta t}{\delta \dot{\psi}} = \frac{1}{\dot{t}}\big(\dot\psi+\dot\phi\cos(\theta)\big) + \lambda
\label{eqn:cartan:lagrangedpsi}
\end{align}
\begin{align}
\Omega \frac{\delta t}{\delta \dot{\phi}} = \frac{\dot\phi}{\dot{t}} + \left(\frac{\dot\psi}{\dot{t}}+\lambda\right)\cos(\theta)
\label{eqn:cartan:lagrangedphi}
\end{align}
\begin{align}
\Omega \frac{\delta t}{\delta \dot{\theta}} = \frac{\dot\theta}{\dot{t}}
\label{eqn:cartan:lagrangedtheta}
\end{align}
\begin{align}
\Omega \frac{\delta t}{\delta \theta}
= -\left(\frac{\dot\psi}{\dot{t}}+\lambda\right)\dot\phi\sin(\theta)
\label{eqn:cartan:lagrangetheta}
\end{align}
\begin{align}
\Omega \frac{\delta t}{\delta \lambda}
=\dot\psi+\dot\phi\cos(\theta)
\label{eqn:cartan:lagrangedlambda}
\end{align}
where we have assumed vanishing quadratic infinitesimals to first order e.g. $d\phi^2=0$. We note that given $\lambda$ is constant, it does not affect the total time $T$. The choice of $\lambda$ can be considered a global gauge degree of freedom i.e. $\frac{\partial T}{\partial \lambda}=0$ (i.e. regardless of $\lambda$ minimal time, $T$ remains the same). The minimal connection constraint: 
\begin{align}
k^\inv dk = - \cos\ad_\Theta(dcc^\inv) 
\label{eqn:cartan:general:connection}
\end{align}
can be written as:
\begin{align}
    \dot\psi=-\dot\phi\cos(\theta)
    \label{eqn:cartan:minimalconnpsiphi}
\end{align} 
as we have specified $k$. The connection and equation (\ref{eqn:cartan:lagrangedlambda}) imply that $\dot\psi(s)$ becomes a local gauge degree of freedom in that it can vary from point to point along the path parameter $s$ without affecting the physics of the system (i.e. the rate of change of $\psi$ can vary from point to point without affecting the energy $\Omega$ or time $T$ of the system). That is:
\begin{align}
    \frac{\delta T}{\delta \dot\psi}=0.
\end{align}
We can simplify the equations of motion by setting a gauge fixing condition (sometimes called a gauge trajectory). Thus we select:
\begin{align}
    \dot \psi/\dot t + \lambda = 0.
    \label{eqn:cartan:gaugefixing}
\end{align}
Recalling extremisation via $dS=0$, we find equations (\ref{eqn:cartan:lagrangedphi}) and (\ref{eqn:cartan:lagrangedtheta}) become:
\begin{align}
    dS = \Omega \frac{\delta t}{\delta \dot \phi} + \Omega\frac{\delta t}{\delta \dot \theta} = \frac{\dot \phi}{\dot t} + \frac{\dot \theta}{\dot t} = 0. 
\end{align}
Thus $\dot \phi/\dot t$ and $\dot \theta/\dot t$ are constant by the constant-$\theta$ assumption i.e. $\dot\theta=0$. Minimising over constant $\dot \theta$:
\begin{align}
\Omega \frac{\partial T}{\partial \dot\theta}=(\dot\theta/\dot{t})\int_\gamma ds = 0 \label{eqn:cartan:Tindependentfromtheta}
\end{align}
confirms the independence of $T$ from $\theta$ (i.e. as $\dot t$ and the path-length $\int_\gamma ds \neq 0$). Combining the above results reduces the integrand in equation (\ref{eqn:iduuinv^2}) to dependence on the $d\phi \sin \theta$ term (as the minimal connection condition and constant $\theta$ condition cause the first two terms to vanish). 
 Such simplifications then mean optimal time is found via minimisation over initial conditions in equation (\ref{eqn:cartan:su2:t=mint}):
\begin{align}
    T = \min_{\theta,\phi} \left| \int_\gamma d\phi \sin \theta   \right|  = \min |\phi \sin \theta| \qquad \phi = \int_\gamma d\phi.
    \label{eqn:minphisintheta}
\end{align}
Note by comparison with the general form of equation (\ref{eqn:cartan:general:Tsinadthetaphi}), here the $\phi \sin \theta$ term represents $\sin \text{ad}_\Theta (dcc^\inv) = \sin \text{ad}_\Theta(\Phi)$. The above method shows how the constant-$\theta$ method simplifies the overall extremisation making the minimisation for $T$ manageable.\\
\\
Now consider holonomy targets of the form:
\begin{align}
U(T)U_0^\inv= e^{-iX} = e^{iJ_z2\eta} \in K
\end{align}
with controls only in $\frak{p} = \{iJ_x, iJ_y   \}$. By assumption $U$ is of $KAK$ form (equation (\ref{eqn:cartan:ukeithetac})):
\begin{align}
    U(T)U_0^\inv= e^{-iX} = e^{iJ_z2\eta} = ke^{i\Theta}c = qe^{ic^\inv \Theta c}.
\end{align}
Choose the initial condition as:
\begin{align}
    U_0 = e^{i\Theta}.
\end{align} 
The simplest form is when $c$ resolves to identity. This in turn requires $c\in K$ to resolve to the identity:
\begin{align}
    M \equiv \Big\{c \in K : c\Theta c^\inv =\Theta \Big\} = \{\pm \mathbb{I}\} \label{eqn:cartan:su2commutant}
\end{align}
Given we have a single element in $\k$,  this is equivalent $c = \exp(i J_z \phi)$ where:
\begin{align}
    \phi = 2\pi n
    \label{eqn:cartan:phi2pin}
\end{align}
for $n \in \mathbb{Z}$. In general we must optimise over choices of $n$, which in this case is simply $n=1$. From this choice of initial condition we have:
\begin{align}
    q(T)  &= k(T)c(T) = e^{-iX}\\
    q(T) &= e^{iJ_z\psi}e^{iJ_z\phi} = e^{iJ_z(\psi + \phi)}. \label{eqn:cartan:su2q(T)}
\end{align}
Note this is a relatively simple form of $X=(1-\cos\ad_\Theta)(\Phi)$ as $\Phi$ comprises only a single generator $J_z$. 
This condition is equivalent to:
\begin{align}
2\eta = \int_\gamma d\chi = \int_\gamma d\psi + d\phi = 2\pi n(1-\cos(\theta)). \label{eqn:cartan:su22eta}
\end{align}
Here we have again used the minimal connection constraint for substitution of variables. Thus:
\begin{align}
    \cos(\theta) = 1-\frac{\eta}{n\pi}.
    \label{eqn:cartan:costhetaetapi}
\end{align}
Substituting into equation (\ref{eqn:minphisintheta}) we have that:
\begin{align}
    \Omega T & =  \min_n 2\pi n\sqrt{\frac{2\eta}{n\pi}-\left(\frac{\eta}{n\pi}\right)^2}\\
& = 2\min_n \sqrt{\eta(2\pi n-\eta)}\\
& = 2 \sqrt{\eta(2\pi-\eta)} \label{eqn:cartan:su2OmegatT}
\end{align}
where we have used:
\begin{align}
    \sin^2\theta = \left(1-\frac{\eta}{n\pi} \right)^2 = \frac{2\eta}{n\pi}-\left(\frac{\eta}{n\pi}\right)^2
\end{align}
using trigonometric identities and setting $n=1$. Note the time optimality is consistent with \cite{boozer_time-optimal_2012}, namely: 
\begin{align}
    T = \frac{2 \sqrt{\eta(2\pi-\eta)}}{\Omega}. \label{eqn:cartan:mintimeTreBoozer}
\end{align}
We now have the optimal time in terms of the parametrised angle of rotation for $J_z$. To specify the time-optimal control Hamiltonian, recall the gauge fixing condition (equation (\ref{eqn:cartan:gaugefixing}), which can also be written $\psi/t + \lambda = 0$) such that:
\begin{align}
    \lambda & = -\dot\psi/\dot t\\
& = -\psi(T)/T\\
& = \frac{\phi}{T}\cos(\theta)\\
& = \Omega\frac{\pi-\eta}{\sqrt{\eta(2\pi-\eta)}}
\label{eqn:cartan:lambdaturningrate}
\end{align}
where we have used equations (\ref{eqn:cartan:phi2pin}) and (\ref{eqn:cartan:costhetaetapi}). Connecting the optimal time to the control pulses and Hamiltonian, note that $\lambda$ can be regarded (geometrically) as the rate of turning of the path. In particular, noting that $\lambda = -\psi(T)/T$, we can regard $\lambda$ as the infinitesimal rotation for time-step $dt$. In a control setting with a discretised Hamiltonian, we regard it as the rotation per interval $\Delta t$. Thus $-\psi(T) \to \lambda T$ and per time interval $-\psi(t) \to \lambda t$. The Hamiltonian then becomes: 
\begin{align}
H(t)= \Omega e^{-iJ_z\lambda t}e^a e^{iJ_z\lambda t} \label{eqn:cartan:su2Hamiltonian1}
\end{align}
where $a \in \frak{a}\subset \frak{p}$. Selecting $J_x$, the Hamiltonian resolves to:
\begin{align}
    H(t) &= \Omega e^{-iJ_z\lambda t/2}J_x e^{iJ_z\lambda t/2} \\
    &=\frac{\Omega}{2} \begin{pmatrix}
        e^{-i \lambda t/2} & 0 \\
        0 & e^{i\lambda t/2} 
    \end{pmatrix}
    \begin{pmatrix}
        0 & 1 \\
        1 & 0
    \end{pmatrix}
    \begin{pmatrix}
        e^{i \lambda t/2} & 0 \\
        0 & e^{-i\lambda t/2} 
    \end{pmatrix}
    \\
    &= \frac{\Omega}{2} \begin{pmatrix}
        e^{-i \lambda t/2} & 0 \\
        0 & e^{i\lambda t/2} 
    \end{pmatrix}
    \begin{pmatrix}
        0 & e^{-i \lambda t/2} \\
        e^{i\lambda t/2}  & 0
    \end{pmatrix}\\
    &= \frac{\Omega}{2} \begin{pmatrix}
        0 & e^{-i \lambda t} \\
        e^{i\lambda t}  & 0
    \end{pmatrix}\\
     &= \frac{\Omega}{2} \begin{pmatrix}
        0 & \cos(\lambda t)-i \sin(\lambda t) \\
        \cos(\lambda t)+i \sin(\lambda t)  & 0
    \end{pmatrix}\\
    &= \Omega (\cos \lambda t J_x + \sin \lambda t J_y). \label{eqn:cartan:su2Hamiltonianfinal}
\end{align}
The Hamiltonian is comprised of control subset generators $J_x, J_y \in \frak{p}$ with control amplitudes $\lambda$ given by their coefficients over time-interval $t$. 
% By normalising the Hamiltonian we obtain the equivalent result to \cite{boozer_time-optimal_2012}. 
In \cite{boozer_time-optimal_2012}, the target unitary is of the form:
\begin{align}
    U_T(t) = e^{i\frac{\eta}{2}\sigma_z} = e^{i\eta J_z}. \label{eqn:cartan:su2boozertarget}
\end{align}
With $\nu = 1 - \eta/(2\pi)$, the time-optimal solution for $\alpha(t)$ becomes $\alpha(t) = \omega t$ where:
\begin{align}
    \omega/\Omega = \frac{2\nu}{\sqrt{1-\nu^2}}
\end{align}
The minimum control time is given by:
\begin{align}
    T = \frac{\pi \sqrt{1-\nu^2}}{\Omega}.
\end{align}

\subsection{$SU(3)/S(U(1) \times U(2))$ time-optimal control} \label{sec:cartan:SU(3)}
\subsubsection{Overview}
We now consider now the constant-$\theta$ method of relevance to lambda systems, specifically for the $SU(3)/S(U(1) \times U(2))$ (AIII(3,1) type) symmetric space, distinguished by the choice of Cartan decomposition (for classification of symmetric spaces generally see section \ref{sec:geo:Symmetric spaces} and \cite{helgason_differential_1979}). The fundamental representation of $SU(3)$ generators is via the Gell-man matrices:
\begin{align}
\lambda_1 &= 
\begin{pmatrix} 
0 & 1 & 0 \\
1 & 0 & 0 \\
0 & 0 & 0 \\
\end{pmatrix} = \ketbra{0}{1} + \ketbra{1}{0} \quad
&
\lambda_2 &= 
\begin{pmatrix} 
0 & -i & 0 \\
i & 0 & 0 \\
0 & 0 & 0 \\
\end{pmatrix} = -i\ketbra{0}{1} + i\ketbra{1}{0} \\
\lambda_3 &= 
\begin{pmatrix} 
1 & 0 & 0 \\
0 & -1 & 0 \\
0 & 0 & 0 \\
\end{pmatrix} = \ketbra{0}{0} - \ketbra{1}{1} \quad
&
\lambda_4 &= 
\begin{pmatrix} 
0 & 0 & 1 \\
0 & 0 & 0 \\
1 & 0 & 0 \\
\end{pmatrix} = \ketbra{0}{2} + \ketbra{2}{0} \\
\lambda_5 &= 
\begin{pmatrix} 
0 & 0 & -i \\
0 & 0 & 0 \\
i & 0 & 0 \\
\end{pmatrix} = -i\ketbra{0}{2} + i\ketbra{2}{0} \quad
&
\lambda_6 &= 
\begin{pmatrix} 
0 & 0 & 0 \\
0 & 0 & 1 \\
0 & 1 & 0 \\
\end{pmatrix} = \ketbra{1}{2} + \ketbra{2}{1} \\
\lambda_7 &= 
\begin{pmatrix} 
0 & 0 & 0 \\
0 & 0 & -i \\
0 & i & 0 \\
\end{pmatrix} = -i\ketbra{1}{2} + i\ketbra{2}{1} \quad
&\\
\lambda_8 &= 
\frac{1}{\sqrt{3}}
\begin{pmatrix} 
1 & 0 & 0 \\
0 & 1 & 0 \\
0 & 0 & -2 \\
\end{pmatrix} = \frac{1}{\sqrt{3}}(\ketbra{0}{0} + \ketbra{1}{1} - 2\ketbra{2}{2}) \label{eqn:cartan:su3gellmanmatrices}
\end{align}
Following \cite{cahn_semi-simple_2014,byrd_geometry_1997}, we set out commutation relations for the adjoint representation of $\frak{su}(3)$, the Lie algebra of $SU(3)$ in Table (\ref{tab:su3commutationHIII}). The row label indicates the first entry in the commutator, the column indicates the second. 

\begin{landscape}
%======= Commutation Table (Main SU3)
%=========COMMUTATION TABLE (BASE)
% \begin{center}
    \begin{table}[h!]
    \centering
    \fontsize{12pt}{12pt}
    \begin{tabular}{ |c|c|c|c|c|c|c|c|c|  } 
 \hline
  & ${\color{red}-i\lambda_1}$ & ${\color{red}-i\lambda_2}$ & ${\color{red}-iH_{\text{III}}}$ & ${\color{red}-iH_{\text{III}}^\perp}$ & $-i\lambda_4 $& $-i\lambda_5 $ & $-i\lambda_6$ & $-i\lambda_7$ \\ 
 \hline
%\lambda1
 ${\color{red}-i\lambda_1}$ & \cellcolor{matteYellow}0 &\cellcolor{matteYellow} ${\color{red}i\sqrt{3}H_{\text{III}} -iH_{\text{III}}^\perp}$  & \cellcolor{matteYellow}${\color{red}-i\sqrt{3}\lambda_2}$ &\cellcolor{matteYellow} ${\color{red}i\lambda_2}$ &  $-i\lambda_7$ &  $i\lambda_6$ & $-i\lambda_5$ & $i\lambda_4$ \\
 \hline

%\lambda2
${\color{red}-i\lambda_2}$ &\cellcolor{matteYellow} ${\color{red}-i\sqrt{3}H_{\text{III}} +iH_{\text{III}}^\perp}$ &\cellcolor{matteYellow} 0  & \cellcolor{matteYellow}${\color{red}\sqrt{3}i\lambda_1}$ &  \cellcolor{matteYellow}${\color{red}-i\lambda_1}$ &  $-i\lambda_6$ &  $-i\lambda_7$ & $i\lambda_4$ & $i\lambda_5$ \\ 
\hline

%===H_{\text{III}}
 ${\color{red}-iH_{\text{III}}}$ &\cellcolor{matteYellow} ${\color{red}i\sqrt{3}\lambda_2}$ &\cellcolor{matteYellow} ${\color{red}-i\sqrt{3}\lambda_1}$ &\cellcolor{matteYellow} $0$ &\cellcolor{matteYellow} $0$ & $0$ & $0$ &  $-i\sqrt{3}\lambda_7$ &  $i\sqrt{3}\lambda_6$ \\ 
 \hline

 % ===H_{\text{III}}^\perp
${\color{red}-iH_{\text{III}}^\perp}$ &\cellcolor{matteYellow} ${\color{red}-i\lambda_2}$ &\cellcolor{matteYellow} ${\color{red}i\lambda_1}$ &\cellcolor{matteYellow} $0$ &\cellcolor{matteYellow} $0$& $-i2\lambda_5$ & $i2\lambda_4$ &  $-i\lambda_7$ & $i\lambda_6$ \\ 
 \hline

 %\lambda4
 $-i\lambda_4 $ &  $i\lambda_7$ &  $i\lambda_6$ &$0$ & $i2\lambda_5$ &\cellcolor{matteGreen}  0 & \cellcolor{matteGreen}${\color{red}-2iH_{\text{III}}^\perp}$ &\cellcolor{matteGreen} ${\color{red}-i\lambda_2}$ &\cellcolor{matteGreen} ${\color{red}-i\lambda_1}$ \\
 \hline

 %\lambda5
$-i\lambda_5 $ &  $-i\lambda_6$ &  $i\lambda_7$ &$0$&$-i2\lambda_4$ & 
 \cellcolor{matteGreen} ${\color{red}i2H_{\text{III}}^\perp}$ &\cellcolor{matteGreen} $0$ &\cellcolor{matteGreen} ${\color{red}i\lambda_1}$ &\cellcolor{matteGreen} ${\color{red}i\lambda_2}$ \\ 
 \hline

 %\lambda6
 $-i\lambda_6$ & $i\lambda_5$ & $-i\lambda_4$ &$i\sqrt{3}\lambda_7$&$i\lambda_7$ &\cellcolor{matteGreen} ${\color{red}i\lambda_2}$ &\cellcolor{matteGreen} ${\color{red}-i\lambda_1}$ &\cellcolor{matteGreen} 0 &\cellcolor{matteGreen} ${\color{red}-i\left(\sqrt{3}H_{\text{III}} + H_{\text{III}}^\perp\right)}$ \\ 
 \hline

 %\lambda7
 $-i\lambda_7$ & $-i\lambda_4$ & $-i\lambda_5$ &$-i\sqrt{3}\lambda_6$&$-i\lambda_6$ &\cellcolor{matteGreen} ${\color{red}i\lambda_1}$ &\cellcolor{matteGreen} ${\color{red}-i\lambda_2}$ &\cellcolor{matteGreen} ${\color{red}i\left(\sqrt{3}H_{\text{III}} + H_{\text{III}}^\perp\right)}$ &\cellcolor{matteGreen} 0 \\
 \hline 
\end{tabular}
    \caption{Commutation relations for generators in adjoint representation of $\frak{su}(3)$. The Cartan decomposition is $\frak{g} = \frak{k} \oplus \frak{p}$ where $\frak{k} = \spn\{-i\lambda_1,-i\lambda_2,-iH_{\text{III}},-iH_{\text{III}}^\perp\}$ (red) and $\frak{p} = \spn\{-i\lambda_4, -i\lambda_5, -i\lambda_6, -i\lambda_7\}$ (black). As can be seen visually, the decomposition satisfies the Cartan commutation relations (equation (\ref{eqn:cartan:cartancommrelations})): the yellow region indicates $[\frak{k},\k] \subset \frak{k}$, the green region that $[\frak{p},\frak{p}] \subset \frak{k}$ and the white region that $[\frak{p},\frak{k}] \subset \frak{p}$. From a control perspective, by inspection is clear that elements in $\frak{k}$ can be synthesised (is reachable) via linear compositions of the adjoint action of $\frak{p}$ upon itself (the green region) as a result of the fact that $[\p,\p]\subseteq \k$. We choose $\frak{a} = \braket{-i\lambda_5}$ with $\h = \braket{-iH_{\text{III}},-i\lambda_5}$.}
    \label{tab:su3commutationHIII}
\end{table}
% \end{center}
%=============end table
\end{landscape}

The typical Cartan decomposition of $\frak{g}=\frak{su}(3)=\k \oplus \p$ follows: 
\begin{align}
    \frak{k}&=\text{span}\{-i\lambda_1,-i\lambda_2,-i\lambda_3,-i\sqrt{3}\lambda_8\}\\
    \frak{p}&=\text{span}\{-i\lambda_4, -i\lambda_5, -i\lambda_6, -i\lambda_7\} \label{eqn:cartan:su3kpdecomposition}
\end{align}
with a maximally compact Cartan subalgebra for $\frak{g}$ often chosen to be 
\begin{align}
    \frak{h}=\text{span}\{-i\lambda_3,-i\sqrt{3}\lambda_8\}.
\end{align}
At this stage, our Cartan subalgebra is not maximally non-compact (having no intersection with $\p$), which we deal with below. Our interest is in lambda systems, that is, 3-level quantum systems where the Hamiltonian control space is a 4-dimensional space of optical transitions generated by $\p$ (where $(\lambda_1,\lambda2)$ may be substituted for $(\lambda_6,\lambda_7)$). The target space is the 4-dimensional unitary subgroup $K = S(U(1) \times U(2))$ of microwave transitions generated by $\k$. With respect to this decomposition, it is convenient to introduce a slight change of basis:
\begin{align}
    H_{\text{III}} &= -\frac{\sqrt{3}}{2}\lambda_3 +\frac{1}{2}\lambda_8 = \frac{1}{\sqrt{3}}(-\ketbra{0}{0} + 2\ketbra{1}{1} - \ketbra{2}{2}) 
\\ H_{\text{III}}^\perp &= \frac{1}{2}\lambda_3 + \frac{\sqrt{3}}{2}\lambda_8 = \ketbra{0}{0} - \ketbra{2}{2}. \label{eqn:cartan:su3HIIIHIIIperp}
\end{align}
Note for convenience:
\begin{align}
\lambda_3 &=  -\frac{\sqrt{3}}{2}H_{\text{III}} + \frac{1}{2}H_{\text{III}}^\perp \qquad \lambda_8 = \frac{\sqrt{3}}{2}H_{\text{III}}^\perp + \frac{1}{2}H_{\text{III}}\\
\lambda_3 + \sqrt{3}\lambda_8 &= 2H_{\text{III}}^\perp \qquad -\lambda_3 + \sqrt{3}\lambda_8 = \sqrt{3}H_{\text{III}} + H_{\text{III}}^\perp \label{eqn:cartan:su3lambda3lambda3+sqrt(lambda8)}
\end{align}
and:
\begin{align}
    \ad_\Theta^2 (-iH_{\text{III}}) &= 0 \qquad \ad_\Theta^2 (-iH_{\text{III}}^\perp) = (-2\theta)^{2}(-H_{\text{III}}^\perp) \label{eqn:cartan:su3adtheta2}
\end{align}
Under this change of basis:
\begin{align}
    \frak{k}&=\text{span}\{-i\lambda_1,-i\lambda_2,-iH_{\text{III}},-iH_{\text{III}}^\perp\}\\
    \frak{p}&=\text{span}\{-i\lambda_4, -i\lambda_5, -i\lambda_6, -i\lambda_7\}.
    \label{eqn:cartan:su3:kandphIIIetc}
\end{align}
Under this transformation, the maximally compact Cartan subalgebra \\$\h=\{H_{\text{III}},H_{\text{III}}^\perp\} \in k$ in (\ref{eqn:cartan:su3:kandphIIIetc}) is entirely within $\k$. In principle, to obtain a maximally noncompact Cartan subalgebra, we conjugate $\h$ via an appropriately chosen group element, thereby performing a Cayley transform (see Section 7 of Part VI of \cite{knapp_lie_1996} and for an explicit example see section \ref{sec:cartan:cayley transforms and Dynkin diagrams} below). However, in our case we can simply read off the combination $\{-iH_{\text{III}},-i\lambda_5\}$ (such that $\h \cap \p = -i\lambda_5$) from the commutation table above.
%=============optimal dicussion
The commutant of $\a = \spn\{-i\lambda_5\}$ is the subgroup:
\begin{align}
    M \equiv \big\{ k \in K : \forall i\Theta \in \a, k \Theta k^\inv \in \Theta \big\} = e^{\m} \label{eqn:cartan:su3commutant}
\end{align}
We also define:
\begin{align}
    H_{\text{III}}^{\perp''} = - \frac{\sqrt{3}H_{\text{III}} + H_{\text{III}}^\perp}{2} = \begin{pmatrix}
        0 & 0 & 0\\
        0 & 1 & 0\\
        0 & 0 & -1
    \end{pmatrix} \label{eqn:cartan:su3HIIIperp''}
\end{align}
which together with $\lambda_6$ and $\lambda_7$ define the microwave Pauli operators and the generator:
\begin{align}
    H_{\text{III}}'' = \frac{-3H_{\text{III}}^\perp - \sqrt{3 }H_{\text{III}}}{2} = \begin{pmatrix}
        -2 & 0 & 0\\
        0 & 1 & 0\\
        0 & 0 & 1
    \end{pmatrix} \label{eqn:cartan:su3HIII''}
\end{align}
which commutes with the microwave qubit. Continuing, via equations (\ref{eqn:cartan:general:optimaltime}-\ref{eqn:cartan:general:optimaltimeconstraints}) we have optimal time in the form:
\begin{align}
    \Omega T = \min_{\Theta,\Phi}|\sin \ad_\Theta (\Phi)| 
    \label{eqn:su3:optimaltimesine}
\end{align}
with initial condition $U_0$=$e^{-i\Theta} \in A$ and target unitary is given by $U_T = e^{-iX}$:
\begin{align}
     X = (1-\cos \ad_\Theta)(\Phi).
\end{align}
Here $\Phi \in \k$ and is in the commutant with respect to $\Theta$. The time-optimal circuit is generated by the Hamiltonian:
\begin{align}
H(t)= e^{-i\Lambda t}\sin\ad_{\Theta_*}(\Phi_*)e^{i\Lambda t} \qquad \Lambda = \frac{\cos \ad_{\Theta_*}(\Phi_*)}{T} \equiv \cos \ad_{\dot \Theta_*}(\Delta \Phi_*). \label{eqn:cartan:hamiltoniangeneric}
\end{align}
Here $\Theta_*,\Phi_*$ reflects a choice of parameters (e.g. $\theta^k,\phi^j$) which minimise time $T$. 
To calculate the optimal time and corresponding Hamiltonian explicitly, we note  $\ad_\Theta (\Phi) = [\Theta,\Phi] \in \frak{p}$ while $\ad_\Theta^2(\Phi) = [\Theta,[\Theta,\Phi]]  \in \frak{k}$ (see section (\ref{sec:alg:Adjoint expansions})). 

\subsubsection{General form of target unitaries in $SU(3)$} \label{sec:cartan:General form of target unitaries in SU(3)}
We set out the general form of targets using the particular Cartan decomposition and choice of basis for $SU(3)$ above. We set:
\begin{align}
    \Phi &= -i(\phi_1 \lambda_1 + \phi_2 \lambda_2 + \phi_3 H_{\text{III}}^{\perp''}   + \phi_4 \sqrt{3}H_{\text{III}}). \label{eqn:cartan:su3phi}
\end{align}
The $\cos(\ad_\Theta)$ term is proportional to the application of $\ad_\Theta^{2}$:
\begin{align}
    \cos \ad_\Theta (\Phi) &=\cos\alpha(\theta)(\Phi)\\
    &=\cos(\theta)(-i\lambda_1) + \cos(\theta)(-i\lambda_2) + \cos(2\theta)(-iH_{\text{III}}^{\perp''}) -i \sqrt{3}H_{\text{III}}
\label{eqn:su3:cosalphaexpansion}
\end{align}
using $\cos(-\theta)=\cos(\theta)$, that $\cos\alpha(\theta)=\cos(0)=1$ for $-iH_{\text{III}}$.  Our targets are in this case of the form:
\begin{align}
    X &= (1-\cos \ad_{\Theta}) (\Phi) = (1-\cos \alpha(\theta)) (\Phi)\\
    &=-i\bigg((1-\cos(\theta))\phi_1 \lambda_1 + (1-\cos(\theta))\phi_2 \lambda_2 + (1-\cos(2\theta)) \phi_3 H_{\text{III}}^{\perp''} \bigg)
    \label{eqn:cartan:SU3:X=i(phik-costheta)Hk}
\end{align}
where $1-\cos\alpha(\theta)=0$ for $H_{\text{III}}$. Note that for more general targets involving linear combinations of $H_{\text{III}}$, specific choices or transformations of $\Phi$ and transformations of $\Theta$ such that the $H_{\text{III}}$ term does not vanish may be required. The rationale for this is the subject of ongoing work and is important for proving the extent of (and any constraints upon) the constant-$\theta$ conjecture and to understanding the interplay of system symmetries with reachability of targets. As per the general method set out below, this form of $X$ can be derived as follows. We begin with the most general form of target:
\begin{align}
    X = -i(\eta_1 \lambda_1 + \eta_2 \lambda_2 + \eta_3 H_{\text{III}}^{\perp''} + \eta_4 \sqrt{3}H_{\text{III}}). \label{eqn:cartan:su3X}
\end{align}
Minimising evolution time firstly requires a choice of an initial condition. Our target can be written:
\begin{align}
    U(T)U_0^\inv=e^{-iX}=qe^{ic^\inv\Theta c} \label{eqn:cartan:su3U(T)U0inv}
\end{align}
with $U_0=e^{i\Theta}$ and the commutant condition we have that:
\begin{align}
    q(T) = k(T)c(T)=e^{\Phi'}e^{\Phi} =e^{\Phi''}= e^{-iX} \label{eqn:cartan:su3q(T)}
\end{align}
In general, we must choose $\Phi$ such that it is in the commutant:
\begin{align}
    \Phi = -i(\phi_1 \lambda_1 + \phi_2 \lambda_2 + \phi_3 \sqrt{3}H_{\text{III}} + \phi_4 H_{\text{III}}^{\perp''}) \in M \label{eqn:cartan:su3phicommutant}
\end{align}
Where a target does not comprise a generators, its coefficient may be set to zero. Gathering terms:
\begin{align}
    q(T) &= e^{-i\Phi''}\\
    \Phi'&=-i(\psi_1 \lambda_1 + \psi_2 \lambda_2 + \psi_3 H_{\text{III}}^\perp + \psi_4 \sqrt{3} H_{\text{III}})\\
    \Phi''&=-i\bigg((\psi_1 +\phi_1) \lambda_1 + (\psi_2+\phi_2) \lambda_2 \\
    &+ (\psi_3+\phi_3) H_{\text{III}}^{\perp''} + (\psi_4 + \phi_4) \sqrt{3}H_{\text{III}} \bigg)   \label{eqn:cartan:su3q(T)andphis}
\end{align}
 Using the form of minimal connection (equation (\ref{eqn:cartan:general:connection})):
\begin{align}
    \dot\psi_k = -\dot\phi_k  \cos\alpha(\theta)  \label{eqn:cartan:su3minimalconnection}
\end{align}
equate coefficients for our target Hamiltonian $X$, where, by gathering terms by generator, we note that $\eta_k = \phi_k(1-(\cos\alpha_k(\theta))$ (excluding the vanishing $-iH_{\text{III}}$ term):
\begin{align}
    \eta_1 &=\phi_1(1-\cos(\theta)) \qquad \eta_2 =\phi_2(1-\cos(\theta))\\
    \eta_3 &=\phi_3 (1-\cos(2\theta)) \qquad \eta_4. =\phi_4 (1-1)=0 e^{-iX} \label{eqn:cartan:su3etas}
\end{align}
recovering the form in equation (\ref{eqn:cartan:SU3:X=i(phik-costheta)Hk}) up to relevant generators. Continuing:
\begin{align}
   \cos(\theta)  &= 1-\frac{1}{2}\left(\frac{\eta_1}{\phi_1} +\frac{\eta_2}{\phi_2} \right) \qquad \sin(\theta) = \sqrt{1-\cos(\theta)^2}\\
   \cos(2\theta)  &= 1-\frac{\eta_3}{\phi_3} \qquad \sin(2\theta) = \sqrt{1-\cos(2\theta)^2}
   \label{eqn:su3:costhetacos2theta}
\end{align}
Note that $\eta_4$ does not contribute. Optimal time is parametrised as:
\begin{align}
    \Omega T &= \min_{\Theta,\Phi}|\sin \ad_\Theta (\Phi)| = \min_{\theta,\phi_k} |\sin \ad_{\Theta}(\Phi)|\\
    &=\min_{\theta,\phi_k} \sqrt{\left(\sum_k \phi^2_k \sin^2\alpha_k(\theta)\right)}\\
    &= \min_{\phi_k}\sqrt{ \phi_1^2\sin^2(\theta) + \phi_2^2\sin^2(\theta)+ \phi_3^2\sin^2(2\theta) } \label{eqn:cartan:su3omegaT}
\end{align}
where the minimisation depends on the choice of $\phi_k$. The choice of $\phi_i$ must be such that the commutant condition $e^{i\Phi} \in M$ is satisfied. For certain targets $X$, this means:
\begin{align}
    \Phi =-i \phi_3 \sqrt{3}H_{\text{III}}  -i2\pi k(n^x \lambda_1 + n^y \lambda_2 + n^z H_{\text{III}}^{\perp''})
\end{align}
for any $\phi_3 \in \Real, k \in \mathbb{Z}$ and unit vector $\hat n = (n^x, n^y, n^z)$. The remainders of $\p$ and $\k$ pair into root spaces of $\a$:
\begin{align}
    [i\lambda_5, i\lambda_1] &= -i\lambda_6 & [i\lambda_5, i\lambda_6] &= i\lambda_1, \\
    [i\lambda_5, i\lambda_2] &= i\lambda_7  & [i\lambda_5, i\lambda_7] &= -i\lambda_2 \label{eqn:cartan:su3rootspaces(main)}
\end{align}
and
\begin{align}
    [i\lambda_5, i\lambda_4] &= 2i\lambda_3 & [i\lambda_5, i\lambda_3] &= -2i\lambda_4 \label{eqn:cartan:su3rootspaces(commutation)}
\end{align}
In this formulation, for a target unitary $U(T) = e^{iX}$:
\begin{align}
    X &= (1-\cos \ad_\Theta)(\Phi)\\
    &=-i2\pi k((1-\cos\theta)(n^x \lambda_1 + n^y \lambda_2) + (1-\cos(2\theta))n^z H_{\text{III}}^{\perp''} \label{eqn:cartan:su3targetXwithrootspaces}
\end{align}
For simple targets, such as those dealt with in the next section, $\phi_k$ simply be an integer multiple $2\pi k$, in which case the minimisation problem becomes one of selecting the appropriate choice of $k$ that minimises $T$. Note that as discussed, this particular form of Hamiltonian cannot reach targets in $H_{\text{III}}$. Having determined $T$, the optimal time Hamiltonian is constructed as follows. Recall from equation (\ref{eqn:cartan:general:lambda=cosadthetaphiT}) that:
\begin{align}
H(t)&= e^{-i\Lambda t}\sin\ad_{\Theta_*}(\Phi_*)e^{i\Lambda t}\\
\Lambda & = \frac{\cos\ad_{\Theta_*}(\Phi_*)}{T} \label{eqn:cartan:su3lambda}
\end{align}
noting for completeness that $\Lambda \in \frak{k}$ (and $\sin \ad_\Theta(\Phi) \in \p$). We demonstrate the technique for a specific example gate in the literature below.
\subsubsection{Comparison with existing methods} \label{sec:cartan:Comparison with existing methods}
In this section, we apply the constant $\theta$ method to derive time optimal results from \cite{albertini_sub-riemannian_2020}. The $KP$ decomposition for D'Alessandro et al. in \cite{dalessandro_time-optimal_2020} and \cite{albertini_sub-riemannian_2020} is:
\begin{align}
    \frak{p}&=\frac{1}{\sqrt{2}}\text{span}\{i\lambda_1,i\lambda_2,i\lambda_4, i\lambda_5\}\\
    \frak{k}&=\frac{1}{\sqrt{2}}\text{span}\{i\lambda_3, i\lambda_6, i\lambda_7, i\lambda_8\}.
    \label{examples:cartan:dalkp}
\end{align}
As the only difference with our chosen $KP$ decomposition (up to the constant $-1/\sqrt{2}$) is swapping $-i\lambda_1,-i\lambda_2 \in \p$ and $-i\lambda_6,-i\lambda_7 \in \k$ we can use the same change of basis from $-i\lambda_3, -i\lambda_8$ to $-iH_{\text{III}},-iH_{\text{III}}^\perp$ and choice of $\a = \spn\{-i\lambda_5\}$. For convenience and continuity with \cite{albertini_sub-riemannian_2020}, we use the standard notation from that paper. The key point for our method is the choice of a Cartan subalgebra that is maximally non-compact allowing us to select $\Theta$ proportional to $-i\lambda_5$.  The form of targets in \cite{albertini_sub-riemannian_2020} $U_f$ is:
\begin{align}
    U_f &= \begin{pmatrix}
        e^{-i\phi_s} & 0 \\
        0 & \hat U_s
    \end{pmatrix}
\end{align}
where $U_s \in U(2)$ and $\det U_s = e^{i\phi_s}$. For the given $KP$ decomposition, matrices $K$ (block diagonal) and $P$ (block anti-diagonal) have the form:
\begin{align}
    K = \begin{pmatrix}
        if & 0\\
        0 & Q
    \end{pmatrix}
    \qquad 
    P = \begin{pmatrix}
        0 & \alpha & \beta \\
        -\alpha^* & 0 & 0\\
        -\beta^* & 0 & 0
    \end{pmatrix}
\end{align} 
chosen in order to eliminate the drift term. The general form of $A \in \k$ and $P =i\lambda_1 \in \p$ in that work are:
\begin{align}
    A&= \begin{pmatrix}
        a + b & 0 & 0\\
        0 & -a & -c \\
        0 &  -c & -b
    \end{pmatrix} \qquad P = i\begin{pmatrix}
        0 & 1 & 0\\
        1 & 0 & 0\\
        0 & 0 & 0
    \end{pmatrix}.
\end{align}
$P$ is expressed to be an element of a Cartan subalgebra of $\p$. However the full Cartan subalgebra is not given. In that paper, the given target is a simple Hadamard gate:
\begin{align}
    U_f &=\begin{pmatrix}
        1 & 0 & \\
        0 & \frac{1}{\sqrt{2}} & \frac{i}{\sqrt{2}}\\
        0 & \frac{i}{\sqrt{2}} & \frac{1}{\sqrt{2}}
    \end{pmatrix} = \exp(H) = \exp\left(-\frac{i \pi}{4} \lambda_6\right)
\end{align}
resulting in:
\begin{align}
A = \begin{pmatrix}
    0 & 0 & 0 \\
 0 & 0 & \pm\frac{7 i}{\sqrt{15}} \\
 0 & \pm\frac{7 i}{\sqrt{15}} & 0
\end{pmatrix} \qquad P = \begin{pmatrix}
    0 & i & 0 \\
 i & 0 & 0 \\
 0 & 0 & 0 
\end{pmatrix}.
\label{eqn:dalalbertiniAP}
\end{align}
In \cite{albertini_sub-riemannian_2020}, the solution to subRiemannian geodesics  (drawing upon results originally presented by Jurdjevic in \cite{jurdjevic_optimal_1999} (p.257) and later with more exposition in \cite{jurdjevic_hamiltonian_2001} (p.28)) relying upon, as Jurdjevic notes, the right invariance of the vector field under the action of elements of $K$) is of the form:
\begin{align}
    U_f &= \exp(A t) \exp((-A + P)t) = \exp(A t).
    \label{eqn:cartan:daljurdgeodesicsol}
\end{align}
In \cite{albertini_sub-riemannian_2020}, the results from equations (\ref{eqn:cartan:jurdgeodesicsolutions}) are assumed and expressed via the constraint $\exp((-A + P)t) = \exp(2\pi k/3) = \mathbb{I}$ (where $k \in \mathbb{Z}$) and $t = t_{\min} = 2\pi T = \sqrt{15}\pi/4$, in which case:
\begin{align}
    \exp(A \sqrt{15}\pi/4) 
    &=\exp\left(\frac{\sqrt{15}\pi}{4}\begin{pmatrix}
    0 & 0 & 0 \\
 0 & 0 & \pm\frac{7 i}{\sqrt{15}} \\
 0 & \pm\frac{7 i}{\sqrt{15}} & 0
\end{pmatrix}\right) \label{eqn:dalbertini:expAT}\\
& = \begin{pmatrix}
        1 & 0\\
        0 & \exp(- 7\pi/4 \sigma_x)  
    \end{pmatrix}
    \\
    &= \begin{pmatrix}
            1 & 0 & 0\\
            0 & \cos(-7\pi/4) & i\sin(-7\pi/4)\\
            0 & i\sin(-7\pi/4) & \cos(-7\pi/4)\\
        \end{pmatrix} = U_f
\end{align}
where we choose $c = -\frac{7 i}{\sqrt{15}}$ in the second line from equation (\ref{eqn:dalalbertiniAP}) above. If the positive value is used, a similarity transformation $K \in \exp(\k)$ of the $KAK$ decomposition $\exp(A t)P \exp(-A t)$ is required. In the general form of solution to subRiemannian geodesic equations from Jurdjevic et al., the control algebra is related to the Hamiltonian via:
\begin{align}
    \exp(A t)P \exp(-A t) = \sum_j u(t)_j B_j
    \label{eqn:dalbertini:expAtPexp-At}
\end{align}
for $B_j \in \p$ and $u_j(t) \in \Real$ (with $||u|| =M $ for $M = \Omega$ constant, by the Pontryagin `bang bang' principle).

\subsubsection{Constant-$\theta$ method} \label{sec:cartan:constantthetamethod1}
To demonstrate our constant $\theta$ method, we first obtain our target generators in terms of $\k$:
\begin{align}
    \log(U_f) = \begin{pmatrix}
        0 & 0 & 0 \\
 0 & 0 & \frac{i \pi }{4} \\
 0 & \frac{i \pi }{4} & 0 \\
    \end{pmatrix} = \frac{i\pi}{4}\lambda_6 = \left( \frac{\pi}{2\sqrt{2}} \right) \left(-\frac{i}{\sqrt{2}} \right)\lambda_6 
\end{align}
which becomes our target $X$:
\begin{align}
    X &= \left( \frac{\pi}{2\sqrt{2}} \right) \left(-\frac{i}{\sqrt{2}} \right)\lambda_6 = \frac{\pi}{4}(-i \lambda_6) = -i \eta_6 \lambda_6\\
    \Phi &=-i\phi_6 \lambda_6
\end{align}
In this relatively simple example, we can choose $\Phi$ to solely consist of $-i\lambda_6$. Setting $\Theta = -i\lambda_5$ and noting $\cos\ad_\Theta(\Phi) = \cos\alpha(\theta)(\Phi)$:
\begin{align}
    X &= (1-\cos\ad_\Theta)(\Phi)\\
    &=(1-\cos\alpha(\theta))(\Phi)\\
    &=(1-\cos(\theta))(-i\phi_6\lambda_6)\\
    &= -i\eta_6 \lambda_6
\end{align}
where we have used:
\begin{align}
    \cos \ad_\Theta(\Phi) &=  \cos \alpha(\theta) \ad^{2k}_{-i\lambda_5}(-i\lambda_6) = i\cos(\theta)\lambda_6 \qquad \cos\alpha(\theta) = \cos(\theta)\\
    \sin \ad_\Theta(\Phi) & =  \sin \alpha(\theta) \ad^{2k+1}_{-i\lambda_5}(-i\lambda_6) = i\sin(\theta)\lambda_1 \qquad \sin\alpha(\theta) = \sin(\theta)
    \label{eqn:dalcosadsinad}
\end{align}
Noting $\Ad(\exp(X))(Y)=\exp(\ad_X(Y))$ such that for $c=\exp(-i\phi_6 \lambda_6)$:
\begin{align}
    e^{-i\phi_6 \lambda_6} (-i\lambda_5) e^{i\phi_6 \lambda_6} = -i\lambda_5
    % = e^{[-i\phi_6 \lambda_6,-i\lambda_5]} = \mathbb{I} 
\end{align}
our commutant condition is satisfied for $\phi_6 = 2\pi$, then:
\begin{align}
    \frac{\pi}{4} &= 2\pi(1-\cos(\theta))\\
    \cos(\theta) &= 1 - \frac{1}{8} = \frac{7}{8}\\
    \sin(\theta) &= \sqrt{1 - \left(\frac{7}{8} \right)^2} = \frac{\sqrt{15}}{8}
\end{align}
which is ``minimum T'' in \cite{albertini_sub-riemannian_2020}. Minimal time is then:
\begin{align}
    \Omega T &= \min_{\Theta,\Phi} \big|\sin\ad_\Theta(\Phi)\big|\\
    &=\min_{\theta,\phi_k} \sqrt{\phi^2_6 \sin^2\alpha(\theta)}\\
    &=\pm 2\pi \sin(\theta)\\
    &=\pm 2\pi \frac{\sqrt{15}}{8} = \pm \frac{\sqrt{15}\pi}{4} \label{eqn:dalbertiniOmegaT}
\end{align}
which is (as time must be positive) the minimum time $t$ to reach the equivalence class of target Hamiltonians (that generate $U_f$ up to conjugation) in \cite{albertini_sub-riemannian_2020}. Note in this Chapter we denote this minimum time to reach the target as $T$. For convenience we assume $\Omega = 1$ such that $T=\sqrt{15}\pi/4$. To calculate the Hamiltonian:
\begin{align}
H(t)= e^{-i\Lambda t}\sin\ad_{\Theta_*}(\Phi_*)e^{i\Lambda t}
\end{align}
and apply our formulation:
\begin{align}
\Lambda & = \frac{\cos\ad_{\Theta_*}(\Phi_*)}{T}\\
&=\frac{2\pi\cos\theta}{T}(-i\lambda_6)\\
&= \frac{7/8}{\sqrt{15}/8}(-i\lambda_6)\\
&= \frac{7}{\sqrt{15}}(-i\lambda_6)\\
&=\begin{pmatrix}
    0 & 0 & 0\\
    0 & 0 & -\frac{7i}{\sqrt{15}}\\
    0 & -\frac{7i}{\sqrt{15}} & 0
\end{pmatrix}
\end{align}
which is $A$ in \cite{albertini_sub-riemannian_2020}. Then from equations (\ref{eqn:dalcosadsinad}) and (\ref{eqn:dalbertiniOmegaT}):
\begin{align}
    \sin\ad_\Theta(\Phi) = i2\pi\sin(\theta)\lambda_1 &=2\pi \sin(\theta)\begin{pmatrix}
        0 & i & 0 \\
        i & 0 & 0\\
        0 & 0 & 0
    \end{pmatrix} =\frac{\sqrt{15}\pi}{4}\begin{pmatrix}
        0 & i & 0 \\
        i & 0 & 0\\
        0 & 0 & 0
    \end{pmatrix}
\end{align}
which is $P$ in \cite{albertini_sub-riemannian_2020} scaled by $T$. Using our formulation, the Hamiltonian is then: 
\begin{align}
    H(t) &= e^{-i\Lambda t}\sin\ad_{\Theta_*}(\Phi_*)e^{i\Lambda t}=  e^{A t}Pe^{-A t}\\
    &= i \lambda_1 \cos\omega t \pm i\lambda_6 \sin \omega t \label{eqn:cartan:albertini:hamiltonianfinal}
\end{align}
for $\omega = 7/\sqrt{15}$. 

%==========DISCUSSION
\section{Discussion} \label{sec:cartan:discussion}
We have demonstrated that for specific categories of quantum control problems, particularly those where the antisymmetric centralizer generators parameterised by angle $\theta$ remain constant, it is possible to obtain analytic solutions for time-optimal circuit synthesis for non-exceptional symmetric spaces using a global Cartan decomposition. This is particularly true when the control subsets are restricted to cases where the Hamiltonian consists of generators from a horizontal distribution (bracket-generating \cite{helgason_differential_1979,knapp_representation_2001}) $\frak{p}$ with $\frak{p}\neq \frak{g}$ (where the vertical subspace is not null). Direct access is only available to subalgebras $\frak{p} \subset \frak{g}$. However, we have shown that if the assumptions $[\frak{p},\frak{p}] \subseteq \frak{k}$ and $d\Theta=0$ hold, arbitrary generators in $\frak{k}$ can be indirectly synthesised (via application of Lie brackets), which in turn makes the entirety of $\frak{g}$ available in optimal time. Geometrically, we have demonstrated a method for synthesis of time-optimal subRiemannian geodesics using $\frak{p}$. Note that, as mentioned in the example for $SU(2)$ above, subRiemannian geodesics may exhibit non-zero curvature with respect to the entire manifold, but this is to be expected where we are limited to a control subset. Consequently, in principle, arbitrary $U_T \in G$ (rather than just) $U_T \in G/K$ becomes reachable in a control sense.

%=====APPENDICES AS NEW SECTIONS

\section{Generalised constant-$\theta$ method}\label{sec:cartan:generalmethod}
In this section, we generalise the method detailed above. Given a $G=KAK$ decomposition, an arbitrary unitary $U_T \in G$ has a decomposition as:
\begin{align}
U = ke^{i\Theta}c = qe^{ic^\inv \Theta c}
\label{eqn:cartan:general:U=keithetac}
\end{align}
where $k,c \in K$ and $e^{i\Theta} \in A$. Define Cartan conjugation (see definition \ref{defn:alg:cartaninvolution}) as:
\begin{align}
k^\chi = k^\inv 
\qquad
(e^{i\Theta})^\chi = e^{i\Theta} 
\qquad
(UV)^\chi = V^\chi U^\chi. \label{eqn:cartan:general:(UV)chi}
\end{align}
The Cartan projection is:
\begin{align}
\pi(U) = U^\chi U \label{eqn:cartan:general:pi(U)}
\end{align}
which $U$ to an element of the subspace of $G$ that is fixed by $\chi$. Combining with 
equation (\ref{eqn:cartan:general:U=keithetac}) we have:
\begin{align}
\pi(ke^{i\Theta}c) = e^{2c^\inv i\Theta c}. \label{eqn:cartan:general:pi(keiThetac)}
\end{align}
That is, the representation of equation (\ref{eqn:cartan:general:U=keithetac}) as projected into the symmetric space $G/K$ establishing the symmetric space as a section of the $K$-bundle:
\begin{align}
\pi(G) \cong G/K. \label{eqn:cartan:general:pi(G)}
\end{align}
The existence of $\chi$ and $\pi$ are sufficient for $G/K$ to be considered globally symmetric i.e. it has an involutive symmetry at every point.  The compactness of $G$ refers to the property that $G$ is a compact Lie group, meaning it is a closed and bounded subset of the Euclidean space in which it is embedded. The symmetric space $G/K$ is equipped with a Riemannian metric, which is a smoothly varying positive-definite quadratic form on the tangent spaces of the symmetric space. Noting the Euler formula (see equation (\ref{eqn:alg:econjsinhcosh})):
\begin{align}
e^{i\Theta}Xe^{-i\Theta} = \text{Ad}_{e^{i\Theta}}(X)= e^{i\ad_\Theta}(X) = \cos\ad_\Theta(X)+i\sin\ad_\Theta(X) \label{eqn:cartan:general:eiThetaXe-iTheta}
\end{align}
where $X \in \frak{g}, e^{i\Theta} \in A \subset G$. Note here the (lower-case) adjoint action $\ad_\Theta(X)$ is the action of the Lie algebra generators on themselves (section \ref{sec:alg:Adjoint action and commutation relations}), thus takes the form of the Lie derivative (definition \ref{defn:alg:Lie algebraliederivative}) (commutator): 
\begin{align}
    \ad_\Theta(X) = [\Theta,X] \label{eqn:cartan:general:adTheta(X)}
\end{align}
whereas the (upper-case) group adjoint action $\text{Ad}_{\Theta}$ is one of conjugation of group elements (hence we exponentiate the generator $X$ implicitly). The Maurer-Cartan form (definition \ref{defn:geo:Maurer-Cartan Form}) becomes in general:
\begin{align}
dUU^\inv
&= k\left(k^\inv dk + \cos\ad_\Theta(dcc^\inv) + id\Theta + i \sin\ad_\Theta(dcc^\inv)\right)k^\inv.
\label{eqn:cartan:general:maurercartan}
\end{align}  
Recalling that Cartan conjugation is the negative of the relevant involution $\iota$, define the control space subalgebra as:
\begin{align}
\p = \{-iH : \chi(-iH) = -iH\} \label{eqn:cartan:general:piH}
\end{align}
which satisfies $[\p,\p] \subseteq \k$ and the Cartan commutation relations more generally. By restricting the Maurer-Cartan form (Hamiltonian) to the antisymmetric control subset:
\begin{align}
dUU^\inv \in \p \label{eqn:cartan:general:dUUinvinp}
\end{align}
we thereby define a minimal connection (see section \ref{sec:cartan:minimalconnections}):
\begin{align}
k^\inv dk = - \cos\ad_\Theta(dcc^\inv) 
\label{eqn:cartan:general:minconnkinvdk=-cosadthetadccinv}
\end{align}
which can as per the examples be written in its parametrised form as:
\begin{align}
\dot{\psi}^{\alpha,r}+\dot{\phi}^{\alpha,r}\cos\alpha(\Theta)=0
    \label{eqn:cartan:general:connection1}
\end{align}
here $\alpha$ is a root (functional) on $\Theta$ that selects out the relevant parameter, e.g. when $\Theta = \sum_k \theta^k H_k$ then $\alpha$ selects out the appropriate $\theta^k \in \Complex$. See section \ref{sec:alg:Abstract root systems} and Appendix \ref{chapter:Background: Geometry, Lie Algebras and Representation Theory} for more background. 

Note that if $\Theta$ comprises multiple $H_k$, then the related Hamiltonian  may also be expressed as:
\begin{align}
H = i\frac{dU}{dt}U^\inv = -k\Big(d\Theta +  \sin\ad_\Theta(dcc^\inv) \Big)k^\inv.
\label{eqn:general:H=dtheta+sinadthetadcccinv}
\end{align}\\
Given $\g$ as the Lie algebra of $G$, define the Killing form as:
\begin{align}
(X,Y)=\frac{1}{2}\text{Re}\Tr(\ad_X\ad_Y) \label{eqn:cartan:general:(X,Y)}
\end{align}
where $\ad_X$ is the adjoint representation of $X$. The Killing form (definition \ref{defn:alg:Killing form}) is used to define an inner product on $\frak{g}$ allowing measurement of lengths and angles in $\frak{g}$.
Define the inner product for weights and the Weyl vector $\rho$:
\begin{align}
(\alpha,\beta) = g^{kl} H_k H_l
\hspace{50pt}
\text{and}
\hspace{50pt}
\rho = \frac{1}{2}\sum_{\alpha\in R^+} \alpha = \sum_{k=1}^r \phi^k \label{eqn:cartan:general:weylweights}
\end{align}
where $\{H_k\}$ is a basis for the Cartan subalgebra $\frak{a}$, $R^+$ is the set of positive roots $\alpha$, $r$ is the rank, and $\{\phi^k\}$ are the fundamental weights (section \ref{sec:alg:Cartan algebras and Root-systems}). The weights $(\alpha,\beta)$ of a representation are the eigenvalues of the Cartan subalgebra $\frak{a}$, which (see below) is a maximal abelian subalgebra of $\frak{g}$. The Weyl vector $\rho$ is a special weight that is associated with the root system of $\frak{g}$ (section \ref{sec:alg:Root system properties}). It is defined as half the sum of the positive roots $\alpha$ (counted with multiplicities). The Weyl vector is used in Weyl's character formula, which gives the character of a finite-dimensional representation in terms of its highest weight \cite{hall_lie_2013} which we denote below as $\tau$. \\
\\
Define the Euclidean norm (see \ref{eqn:quant:Frobenius norm}) using the Killing form as:
\begin{align}
|X| = \sqrt{(X,X)}. \label{eqn:cartan:general:Euclidean norm}
\end{align}
We leverage the fact that the Killing form is quadratic for semi-simple Lie algebras $\frak{g}$ such that:
\begin{align}
|idUU^\inv|^2 = |ik^\inv dk + \cos\ad_\Theta(idcc^\inv)|^2 + |d\Theta|^2 + |\sin\ad_\Theta(dcc^\inv)|^2.
\label{eqn:cartan:general:maurer-cartainsquared}
\end{align}
Let the Hamiltonian have an isotropic cutoff:
\begin{align}
|H|=\Omega. \label{eqn:cartan:general:|H|}
\end{align}
By the Schrodinger equation, the time elapsed over the path $\gamma$ is given by:
\begin{align}
\Omega t = \Omega \int_\gamma dt = \int_\gamma |idUU^\inv| =  \int_\gamma \left|i\frac{dU}{ds}U^\inv\right| ds. \label{eqn:cartan:general:Omega t}
\end{align}
Define:
\begin{align}
\dot t = \left|i\frac{dU}{ds}U^\inv\right| \label{eqn:cartan:general:dot t }
\end{align}
with:
\begin{align}
T =  \min_\gamma t.
\end{align}
To perform the local minimization, we introduce a vector of Lagrange multipliers (defined below in equation (\ref{eqn:cartan:general:lagrangevector})):
\begin{align}
\Omega t =  \int_\gamma\Big(\dot{t} + \big(\Lambda,k^\inv \dot{k} + \cos\ad_\Theta(\dot{c}c^\inv)\big)\Big) ds
\label{eqn:cartan:general:Omegat=intgamma(...)}
\end{align}
note
\begin{align}
k\left(k^\inv \dot{k} + \cos\ad_\Theta(\dot{c}c^\inv)\right)k^\inv = \frac{\dot{U}U^\inv -\chi(\dot{U}U^\inv)}{2} \label{eqn:cartan:general:kkinvdotk...}
\end{align}
and
\begin{align}
k\left(i\dot\Theta + i \sin\ad_\Theta(\dot{c}c^\inv)\right)k^\inv = \frac{\dot{U}U^\inv +\chi(\dot{U}U^\inv)}{2}. \label{eqn:cartan:general:k(...)kinv}
\end{align}
We can further simplify via expanding in the restricted Cartan-Weyl basis, noting that $\alpha$ indexes the relevant roots and $s$ indexes the relevant sets of roots $r \in \Delta$. The restricted Cartan-Weyl basis allows $\frak{g}$ to be decomposed as the sum of the commutant basis vectors $H_k \in \frak{a}$ and the root vectors $E_\alpha$:
\begin{align}
    \frak{g} = H_k \bigoplus E_\alpha \label{eqn:cartan:general:g=hbigoplusealpha}
\end{align}
where there are $r$ such positive roots. The Maurer-Cartan form (equation (\ref{eqn:cartan:general:Omegat=intgamma(...)})) becomes expressed in terms of weights and weight vectors:
\begin{align}
\Theta = H_k \theta^k
\label{eqn:cartan:general:theta=hkthetak}
\end{align}
\begin{align}
i\dot{c}c^\inv = F_{\alpha,r}\dot{\phi}^{\alpha,r}
\label{eqn:cartan:general:idotccinv=Ialphadotphialphas}
\end{align}
\begin{align}
ik^\inv\dot{k} = F_{\alpha,r}\dot{\psi}^{\alpha,r}+\frak{m}
\label{eqn:cartan:general:ikinvkdot=Ialphaspsi+a}
\end{align}
and
\begin{align}
\Lambda = F_{\alpha,r}\lambda^{\alpha,r}+\frak{m}.
\label{eqn:cartan:general:lagrangevector}
\end{align}
In the above equations, the $\alpha$ are summed over restricted positive roots (of which there are $r$ many). $F_\alpha \in \k - \m$  with $\dot\phi,\dot\psi \in \Complex$ coefficients. The Lagrange multiplier vector $\Lambda$ in equation (\ref{eqn:cartan:general:lagrangevector}) is in generalised form. Note that in the $SU(2)$ case above, the simplicity of $\frak{k} = \{ J_z \}$ simplifies the multiplier term in the action equation (\ref{eqn:cartan:su2:s=omegataction}). Here the Cartan subalgebra $\h$ is given by:
\begin{align}
    \h = \{\a,\m\}. \label{eqn:cartan:general:CSAh}
\end{align}
This algebra is distinct from a maximally compact Cartan algebra in that $\h$ intersects with both $\p$ (the non-compact part of $\h$) and $\k$ (the compact part of $\h$). 
In many cases, the elements of $\h$ are themselves diagonal (see for example Hall and others \cite{hall_lie_2013,cahn_semi-simple_2014}) and entirely within $\k$. In our case, $\a \subset \p$ but $\m \subset k$. The construction of the restricted Cartan-Weyl basis is via the adjoint action of $\a$ on the Lie algebra, such that it gives rise to pairings $F_\alpha \in \k-\m, E_\alpha \in \p-\a$ conjugate under the adjoint action. 

The Cartan-Weyl basis has the property that the commutation relations between the basis elements are simplified because (i) the Cartan generators (belonging to an abelian subalgebra) commute and (ii) the commutation relations between a Cartan generator and a root vector are proportional to the root vector itself. The commutation relations between two root vectors can be more complicated, but they are determined by the structure of the root system. The set of roots are the non-zero weights of the adjoint representation of the Lie algebra. The roots form a discrete subset of the dual space to the Cartan subalgebra, and they satisfy certain symmetries and relations that are encoded in the Dynkin diagram of the Lie algebra \cite{knapp_representation_2001,helgason_differential_1979}. Transforming to the restricted Cartan-Weyl basis allows us to represent the parametrised form of equation (\ref{eqn:cartan:general:maurer-cartainsquared}) as:
\begin{align}
|\dot\Theta|^2 = g_{kl}\dot\theta^k\dot\theta^l \label{eqn:cartan:general:lagrange-dottheta2}
\end{align}
\begin{align}
| \sin\ad_\Theta(\dot{c} c^\inv)|^2 = \sum_{\alpha,r}g_{\alpha\alpha} (\dot{\phi}^{\alpha,r})^2\sin^2\alpha(\Theta) \label{eqn:cartan:general:lagrange-sinadthetaccinv}
\end{align}
\begin{align}
|k^\inv \dot{k} + \cos\ad_\Theta(\dot{c}c^\inv)|^2 = \sum_{\alpha,r}g_{\alpha\alpha} \big(\dot{\psi}^{\alpha,r}+\dot{\phi}^{\alpha,r}\cos\alpha(\Theta)\big)^2 + |\frak{m}|^2 \label{eqn:cartan:general:lagrange-kinvdotk...}
\end{align}
and using the vector of Lagrange multipliers with the connection:
\begin{align}
\big(\Lambda,k^\inv \dot{k} + \cos\ad_\Theta(\dot{c}c^\inv)\big)
= \sum_{\alpha,r}g_{\alpha\alpha} \lambda^{\alpha,r}\big(\dot{\psi}^{\alpha,r}+\dot{\phi}^{\alpha,r}\cos\alpha(\Theta)\big) + |\frak{m}|^2. \label{eqn:cartan:general:Lambda}
\end{align}
Here we have used the fact that the only non-vanishing elements of the Killing form are $g_{\alpha \alpha}$ and $g_{jk}$ (and using $g_{\alpha\alpha} = (E_\alpha,E_\alpha)$), the inner product of the basis elements with themselves, and the restricted Cartan-Weyl basis to simplify the variational equations. The nonzero functional derivatives are:
\begin{align}
\Omega \frac{\delta t}{\delta \dot{\psi}^{\alpha,r}} = g_{\alpha\alpha}\left(\frac{1}{\dot{t}}\big(\dot\psi^{\alpha,r}+\dot\phi^{\alpha,r}\cos\alpha(\Theta)\big) + \lambda^{\alpha,r}\right)
\label{eqn:cartan:general:lagrange-ELGdotphi}
\end{align}
\begin{align}
\Omega \frac{\delta t}{\delta \dot{\phi}} = g_{\alpha\alpha}\left(\frac{\dot\phi^{\alpha,r}}{\dot{t}} + \left(\frac{\dot\psi^{\alpha,r}}{\dot{t}}+\lambda^{\alpha,r}\right)\cos\alpha(\Theta)\right) \label{eqn:cartan:general:lagrange-d(dotphi)}
\end{align}
\begin{align}
\Omega \frac{\delta t}{\delta \dot{\theta}^k} = g_{kl}\frac{\dot\theta^l}{\dot{t}} \label{eqn:cartan:general:lagrange-d(dottheta)1}
\end{align}
\begin{align}
\Omega \frac{\delta t}{\delta \theta^k}
= -\sum_{\alpha,r}g_{\alpha\alpha}\left(\frac{\dot\psi^{\alpha,r}}{\dot{t}}+\lambda^{\alpha,r}\right)\dot\phi^{\alpha,r}\alpha(H_k)\sin\alpha(\Theta) \label{eqn:cartan:general:lagrange-d(dotthetak)2}
\end{align} 
\begin{align}
\Omega \frac{\delta t}{\delta \lambda^{\alpha,r}}
=\dot\psi^{\alpha,r}+\dot\phi^{\alpha,r}\cos\alpha(\theta).
\label{eqn:cartan:general:ELGlambda}
\end{align}
From the Euler-Lagrange equations (see section \ref{sec:geo:variational methods} generally) and by design from the connection we have the constraints:
\begin{align}
k^\inv \dot{k} =- \cos\ad_\Theta(\dot{c}c^\inv). \label{eqn:cartan:general:connectionkinvdotkcosadtheta}
\end{align}
We assume again the Lagrange multipliers are constant. As for the case of $SU(2)$, each Lagrange multiplier $\lambda^{\alpha,r}$ then becomes a global gauge degree of freedom in the sense that:
\begin{align}
\frac{\partial T}{\partial\lambda^{\alpha,r}}=0 \label{eqn:cartan:general:partialTpartiallambda}
\end{align}
with $\dot\psi(s)$ becoming local gauge degrees of freedom  under the constraint:
\begin{align}
\frac{\delta T}{\delta \dot\psi^{\alpha,r}}=0. \label{eqn:cartan:general:deltaTdeltadotpsi}
\end{align}
We are free to choose the gauge trajectory $\dot\psi(s)$ as:
\begin{align}
k^\inv \dot{k}/\dot{t}+\Lambda=0 \label{eqn:cartan:general:connection2}
\end{align}
With this choice, we have also by the remaining Euler-Lagrange equations:
\begin{align}
\dot{c}c^\inv/\dot{t}=\text{constant} \label{eqn:cartan:general:cotccinvdottconstant}
\end{align}
and
\begin{align}
\dot\Theta/\dot{t}=\text{constant}. \label{eqn:cartan:general:dotThetadottconstant}
\end{align}
Minimizing over the constant $\dot\theta$:
\begin{align}
\Omega \frac{\partial T}{\partial \dot\theta^k}=g_{kl}(\dot\theta^l/\dot{t})\int_\gamma ds = 0 \label{eqn:cartan:general:minimiseoverdottheta}
\end{align}
we see:
\begin{align}
\Theta = \text{constant}. \label{eqn:cartan:general:thetaconstant}
\end{align}
The above functional equations show that when the action is varied, each term in equation (\ref{eqn:cartan:general:maurercartan}) vanishes apart from the $\sin \text{ad}_\Theta(dcc^\inv)$ term. 
Calculating optimal evolution time then reduces to minimization over initial conditions:
\begin{align}
T =  \min_{\Theta,\Phi} \left|\sin\ad_\Theta(\Phi)\right|
\label{eqn:cartan:general:Tsinadthetaphi}
\end{align}
where
\begin{align}
\Phi = \int_\gamma -i dc c^\inv. 
\end{align}
We can express in terms of the Cartan-Weyl basis as follows:
\begin{align}
    \Phi&= \sum_{\alpha,r} \dot\phi^{\alpha,r} F_{\alpha,r} + \frak{m}
    \label{eqn:general:PhisumphiFalpharplusm}
\end{align}
where: 
\begin{align}
    &F_{\alpha,r} =\frac{1}{\alpha(\Theta)}\ad_\Theta (E_{\alpha,r}) \in \frak{k}-\frak{m} \label{eqn:cartan:general:Falphar}\\
    &\ad^2_\Theta (E_{\alpha,r}) = \alpha(\Theta)^2 E_{\alpha,r} \in \frak{p} - \frak{a} \label{eqn:cartan:general:ad2Ealphr}\\
    &\ad_\Theta(\frak{m}) = 0 \label{eqn:cartan:general:adthetam}
\end{align}
such that:
\begin{align}
    \sin\ad_\Theta(\dot\Phi) = \sum_{\alpha,r}\dot\phi^{\alpha,r}\sin\alpha(\Theta)E_{\alpha,r} \in \frak{p}.
    \label{eqn:cartan:general:sinadthetaPhiEalphar}
\end{align}
The minimisation can be expressed in terms of parameters noting the trace as a rank-2 tensor contraction (see section \ref{defn:geo:Tensor contractions}). The problem of finding the minimal time Hamiltonian is thereby simplified considerably. Consider targets of the form:
\begin{align}
U(T)U_0^\inv = e^{-iX} \in K.
\end{align}
Again:
\begin{align}
    U(T)U_0^\inv= e^{-iX} =   ke^{i\Theta}c = qe^{ic^\inv \Theta c}. \label{eqn:cartan:general:}
\end{align}
Explicitly we equate:
\begin{align}
    U(T) = q \qquad U(0) = e^{ic^\inv \Theta c}.  \label{eqn:cartan:general:U(T)U0}
\end{align}
As was the case for $SU(2)$, we select initial conditions as:
\begin{align}
U_0 = e^{i\Theta} \label{eqn:cartan:general:U0}
\end{align}
where we can see a quantization condition emerges by satisfying the commutant condition that $c = \exp(\Phi) \in K$ (with $\Phi \in \k$) commute with $\Theta$:
\begin{align}
\Big\{\exp(\Phi) \in K : \exp(\Phi)\Theta \exp(-\Phi) =\Theta \Big\}  = M
\label{eqn:general:csubsetidentity}
\end{align}
where $M$ can be regarded as the group elements generated by the commutant algebra $\m$, that is $M = \exp(\m)$. Such a condition is equivalent to $\Theta c \Theta^\inv = c$. In the $SU(2)$ case, because we only have a single generator in $\Phi$, the condition manifests in requiring the group elements $c \in K$ to resolve to $\pm \mathbb{I}$, which in turn imposes a requirement that their parameters $\phi_j$ be of the form $\phi_j = 2\pi n$ for $n \in \mathbb{Z}$ in order for $e^{i\phi_j k_j } = \mathbb{I}$ where $k \in \frak{k}$. In general, however, the commutant will have a nontrivial connected subgroup and it still ``quantizes'' into multiple connected components. In practice this means that $\phi_k$ may not, in general, be integer multiples of $2\pi$ and instead must be chosen to meet the commutant condition in each case. We note that:
\begin{align}
q(T)  = k(T)c(T) = e^{-iX} \label{eqn:cartan:general:q(T)}
\end{align}
or equivalently
\begin{align}
X &= \int_\gamma i dq q^\inv\\
& = \int_\gamma k\left(ik^\inv dk + idc c^\inv\right)k^\inv\\ 
& = \int_\gamma k\left(1-\cos\ad_\Theta \right)\big(idc c^\inv\big)k^\inv\\ 
& = -\left(1-\cos\ad_\Theta \right)\big(\Phi\big) \label{eqn:cartan:general:Xcetc}
\end{align}
where we have used the minimal connection in equation (\ref{eqn:cartan:general:minconnkinvdk=-cosadthetadccinv}), where the last equality comes from $k(0)=\mathbb{I}$ and the arc-length parametrisation of $\gamma$ between $s\in [0,1]$. The optimal time is:
\begin{align}
\Omega T = \min_{\Theta,\Phi} \big|\sin\ad_\Theta(\Phi)\big|
\label{eqn:cartan:general:optimaltime}
\end{align}
with minimization over constraints:
\begin{align}
e^{\Phi} \in M
\hspace{50pt}
\text{and}
\hspace{50pt}
X =  \left(1-\cos\ad_\Theta \right)(\Phi).
\label{eqn:cartan:general:optimaltimeconstraints}
\end{align}
Time-optimal control is given by Hamiltonians of the form:
\begin{align}
H(t)= e^{-i\Lambda t}\sin\ad_{\Theta_*}(\Phi_*)e^{i\Lambda t}
\label{eqn:general:timeoptimalhamiltonian}
\end{align}
where $\dot\Phi = \Phi/T$, $\Theta_*$ and $\Phi_*$ are the critical values which minimize the time,
and the multiplier is the turning rate:
\begin{align}
\Lambda & = -k^\inv \dot{k}/\dot t\\
& = -\int_\gamma k^\inv \dot{k} /T\\
& = \frac{\cos\ad_{\Theta_*}(\Phi_*)}{T}.
\label{eqn:cartan:general:lambda=cosadthetaphiT}
\end{align}
The global minimization then depends upon the choice of Cartan subalgebra $\a \ni \Theta$ (as illustrated in the examples above).

\section{Minimal connections and holonomy groups}\label{sec:cartan:minimalconnections}
A simple application of the product rule and the Euler formula:
\begin{align}
e^{i\Theta}Xe^{-i\Theta} = e^{i\ad_\Theta}(X) = \cos\ad_\Theta(X)+i\sin\ad_\Theta(X)
\end{align}
reveals:
\begin{align}
dU U^\inv = k\left(k^\inv dk + \cos\ad_\Theta(dc c^\inv)+id\Theta + i \sin\ad_\Theta(dc c^\inv)\right)k^\inv.
\end{align}
Symmetric controls:
\begin{align}
-iHdt \in \p
\end{align}
which satisfy the Lie triple property:
\begin{align}
[\p,[\p,\p]]\subseteq\p
\end{align}
define the connection:
\begin{align}
k^\inv dk = - \cos\ad_\Theta(dc c^\inv)
\end{align}
which is minimal in the sense that it minimizes the invariant line element
\begin{align}
\min_{k\in K}{|idU U^\inv|^2} =\min_{k\in K}{|ik^\inv dk + i\cos\ad_\Theta(dc c^\inv)|^2} + |d\Theta|^2+| \sin\ad_\Theta(dc c^\inv)|^2
\end{align}
where the minimization is over curves, $k(t)$.
This means that minimization with a horizontal constraint is equivalent to minimization over the entire isometry space.
Such submanifolds are said to be totally geodesic. Targets $U(T) = e^{-iX}$ with effective Hamiltonians:
\begin{align}
-iX \in \k \supseteq [\p,\p]
\end{align} 
are known as holonomies because they are generators of \textit{holonomic groups}, groups whose action on a representation (e.g. the relevant vector space) causes vectors to be parallel-transported in space i.e. the covariant derivative vanishes along such curves traced out by the group action. We discuss holonomic groups in definition \ref{defn:geo:holonomy}.  Intuitively, one can consider holonomic groups as orbits such that any transformation generated by elements of $\frak{k}$ will `transport' the relevant vector along paths represented as geodesics (definition \ref{defn:geo:Geodesic}). However, such vectors are constrained to such orbits if those generators are only drawn from $\k$ i.e. $[\k,\k] \subseteq \k$ for a chosen subalgebra $\k$. To transport vectors elsewhere in the space, one must apply generators in $\frak{p}$ which is analogous to shifting a vector to a new orbit.  Although in application, one considers $U(0)=1$, it is important to remember this variational problem is right-invariant, so one could just as well let $U(0)=U_0$ be arbitrary, as long as the target is understood to correspondingly be $U(T) = e^{-iX}U_0$. In this case, the time-optima paths are \textit{subRiemannian geodesics} (see section \ref{sec:geo:SubRiemannian Geodesics}) which differ from Riemannian geodesics to the extent that, intuitively put, evolution in the direction of $\k$ is required with the distinction between the two types of geodesics captured by the concept of geodesic curvature. 

\section{Cayley transforms and Dynkin diagrams} \label{sec:cartan:cayley transforms and Dynkin diagrams}
In this section, we give an example of computing the Cayley transform of a Cartan subalgebra $\h$ in order to increase its intersection with non-compact subalgebras of $\g$. This is of interest because the method described above in this Chapter seeks a maximally abelian subalgebra $\a \subset \h \subset \g$ from which to deduce the Cartan $KAK$ decomposition. We discuss background theory on this topic in section \ref{sec:alg:Cayley transforms} (following \cite{knapp_lie_1996}).
%==========FIGURE
%=====Cayley transform diagram(?)
\begin{figure}[h!]
    \centering
    \begin{minipage}{0.45\textwidth}
        \centering
        \begin{tikzpicture}[scale=2, >={latex}] % Scale the entire diagram
            % Circle with arrows
            \draw (0:1cm) arc (0:90:1cm);
            \draw[->] (45:1cm) arc (45:50:1cm); % Arrow for the first quarter
            \draw (90:1cm) arc (90:180:1cm);
            \draw[->] (135:1cm) arc (135:140:1cm); % Arrow for the second quarter
            \draw (180:1cm) arc (180:270:1cm);
            \draw[->] (225:1cm) arc (225:230:1cm); % Arrow for the third quarter
            \draw (270:1cm) arc (270:360:1cm);
            \draw[->] (315:1cm) arc (315:320:1cm); % Arrow for the fourth quarter
            
            % Axes
            \draw[->] (-1.2,0) -- (1.2,0); % x-axis
            \draw[->] (0,-1.2) -- (0,1.2); % y-axis
            
            % Text labels
            \node at (0,1.3) {$-iH_{\text{III}}^\perp$}; % Top of y-axis
            \node at (1.5,0) {$-i\lambda_5$}; % Right end of x-axis
            \node at (1.0,1.0) {$e^{\ad_{-i\lambda_4}\Theta}$}; % Closer to the circle arc in the top-right
        \end{tikzpicture}
        \caption{Cayley transform of $-iH_{\text{III}}^\perp$ expressed as a rotation into $-i\lambda_5$. The presence of the imaginary unit relates to the compactness here of $\g$ which reflects the boundedness and closed nature of the transformation characteristic of unitary transformations.}
        \label{fig:circle}
    \end{minipage}\hfill
    \begin{minipage}{0.45\textwidth}
        \centering
        \begin{tikzpicture}
            % Axes
            \draw[->, thick] (-3,0) -- (3,0) node[right] {$\lambda_5$};
            \draw[->, thick] (0,-3) -- (0,3) node[above] {$-iH_{\text{III}}^\perp$};
            
            % Hyperbolic curves (horizontal)
            \draw[thick, domain=-2:2] plot (\x, {sqrt(\x*\x + 1)});
            \draw[->, thick] (1.5, {sqrt(1.5*1.5 + 1)}) -- (1.6, {sqrt(1.6*1.6 + 1)}); % Arrow on top right curve
            \draw[thick, domain=-2:2] plot (\x, {-sqrt(\x*\x + 1)});
            \draw[->, thick] (-1.5, {-sqrt(1.5*1.5 + 1)}) -- (-1.6, {-sqrt(1.6*1.6 + 1)}); % Arrow on bottom left curve
            
            % Hyperbolic curves (vertical)
            \draw[thick, domain=-2:2] plot ({sqrt(\x*\x + 1)}, \x);
            \draw[->, thick] ({sqrt(1.5*1.5 + 1)}, 1.5) -- ({sqrt(1.6*1.6 + 1)}, 1.6); % Arrow on top right vertical curve
            \draw[thick, domain=-2:2] plot ({-sqrt(\x*\x + 1)}, \x);
            \draw[->, thick] ({-sqrt(1.5*1.5 + 1)}, -1.5) -- ({-sqrt(1.6*1.6 + 1)}, -1.6); % Arrow on bottom left vertical curve
            
            % Dotted lines along 45 degree axes
            \draw[dashed, domain=-2.5:2.5] plot (\x, \x);
            \draw[dashed, domain=-2.5:2.5] plot (\x, -\x);
            
            % Annotation
            \node at (2.8,2.8) {$e^{\ad_{\lambda_4}\Theta}$};
        \end{tikzpicture}
        \caption{Cayley transform of $-iH_{\text{III}}^\perp$ expressed as a rotation into $\lambda_5$. By contrast with the case to the left, the absence of the imaginary unit is indicative of non-compactness such that distances are not preserved (unlike in the unitary case where $-i$ is present).}
        \label{fig:hyperbolic}
    \end{minipage}
\end{figure}

%==========FIGURE
Recall the maximally compact Cartan subalgebra $\h=\braket{H_{\text{III}},H_{\text{III}}^\perp} \in \k$ in (\ref{eqn:cartan:su3:kandphIIIetc}) is entirely within $\k$. To obtain a maximally non-compact Cartan subalgebra, we conjugate $\h$ via an appropriately chosen group element (see Section 7 of Part VI of \cite{knapp_lie_1996}).  In principle, we could read off the combination $\braket{-iH_{\text{III}},-i\lambda_5}$ (such that $\h \cap \p = -i\lambda_5$) from the commutation table in Table \ref{tab:su3commutationHIII}. 

To demonstrate the application of a Cayley transform to obtain the requisite maximally non-compact Cartan subalgebra, we use a root (definition \ref{defn:alg:roots}) (in our case, $\gamma$ below) to construct a generator of a transformation to a new Cartan subalgebra whose intersection with $\p$ increases by one dimension. We construct a root system (section \ref{sec:alg:Root system properties}) in order to find a Cayley transformation $C$ such that:
\begin{align}
\braket{-iH_{\text{III}},-iH_{\text{III}}^\perp} \to \braket{-iH_{\text{III}},-i\lambda_5}.
\end{align}
By inspection we seek a transformation from $-iH_{\text{III}^\perp} \to -i\lambda_5$. To find the Cayley transformation, we first construct a root system via root vectors as follows:
\begin{align}
\frac{\lambda_1 + i\lambda_2}{2} &= \ketbra{0}{1}=e_\alpha &
\frac{\lambda_4 + i\lambda_5}{\sqrt{2}} &= \ketbra{0}{2} =e_\gamma & \frac{\lambda_6 + i\lambda_7}{2} &= \ketbra{1}{2} =e_{\beta}\\
\frac{\lambda_1 - i\lambda_2}{\sqrt{2}} &= e_{-\alpha} & \frac{\lambda_4 - i\lambda_5}{\sqrt{2}} &= e_{-\gamma} & \frac{\lambda_6 - i\lambda_7}{\sqrt{2}} &= e_{-\beta}. \label{eqn:cartan:cayleyrootsystem}
\end{align}
Note that the root vectors are essentially raising operators that promote the system from a lower to a higher energy state (as indicated by the ket-bra notation), while their adjoints are lowering operators. The elements of $\h$ are linear combinations of $\lambda_3, \lambda_8$ which form a basis for $\h$. To obtain the roots, we conjugate by this basis which spans our original $\h$ above:
\begin{align}
    [\lambda_3, e_{\alpha}] &= 2 e_{\alpha} & [\lambda_8, e_{\alpha}] &= 0 & \alpha &= (2,0) \\
    [\lambda_3, e_{\gamma}] &= e_{\gamma} & [\lambda_8, e_{\gamma}] &= \sqrt{3} e_{\gamma} & \gamma &= (1,\sqrt{3}) \\
    [\lambda_3, e_{\beta}] &= -e_{\beta} & [\lambda_8, e_{\beta}] &= \sqrt{3} e_{\beta} & \beta &= (-1,\sqrt{3}).
\end{align}
The roots are given by $\alpha,\gamma,\beta$. As $\gamma$ is a linear combination of $\alpha,\beta$, they are the simple roots for this system. Denote $H_{\text{III}}^\perp = H_\gamma$ then note:
%\hl{rev}[$\lambda_3 = H_\alpha],
\begin{align}
    e_\gamma + e_{-\gamma} = \ketbra{0}{2}+\ketbra{2}{0} = \lambda_4.
\end{align}
Conjugating by $H_\gamma$ we have:
\begin{align}
    [H_\gamma,e_\gamma + e_{-\gamma}] = e_\gamma - e_{-\gamma} = 2i\lambda_5
\end{align}
and:
\begin{align}
    \left[ \frac{1}{2}i\lambda_4,-iH_\gamma\right] &= -i\lambda_5 & \left[ \frac{1}{2}i\lambda_4,-iH_\text{III}\right] &= 0.
\end{align}
Cayley transformations take the form of rotations (conjugations i.e. the adjoint action) by the angle $\pi/2$. We include two diagrams in Figures \ref{fig:circle} and \ref{fig:hyperbolic} out of interest. The rotation in Fig. \ref{fig:circle} represents a transformation by a compact group element (unitary) preserving distances and angle. For contrast, a rotation where the generator lacks the imaginary unit coefficient (Fig. \ref{fig:hyperbolic}), the geometry is hyperbolic in nature, reflecting non-compact transformation. This gives an interesting geometric intuition about the role of the imaginary unit. Our generator of transformations is $i\lambda_4$, thus the Cayley transformation for $H_\gamma$ becomes:
\begin{align}
    C_\gamma &= e^{\frac{\pi}{4}\ad_{i\lambda_4}} \qquad C_\gamma(-iH_\gamma) = -i\lambda_5.
\end{align}
The maximally non-compact Cartan subalgebra $\h$ under this transformation is:
\begin{align}
    \h = \t + \a = \braket{-iH_{\text{III}},-i\lambda_5}.
    \label{eqn:su3:CSAmaxnoncompact}
\end{align}
We can visually see this via Table \ref{tab:su3commutationHIII} via the commutation relations along the $-iH_{\text{III}}$ row where such a choice of Cartan subalgebra intersects with $\p$ and $\k$. We could also have chosen $\h = \braket{-iH_{\text{III}},-i\lambda_4}$. 

\begin{figure}[ht]
\centering
\begin{tikzpicture}
  % Define the nodes
  \node (0) at (0,0) {0};
  \node (1) at (2,0) {1};
  \node (2) at (1,1.732) {2}; % The height is calculated using the Pythagorean theorem for a 1-1-sqrt(3) triangle

  % Connect the nodes to form an isosceles triangle
  \draw[->] (0) -- (1) node[midway, below] {$\alpha$};
  \draw[->] (0) -- (2) node[midway, above left] {$\gamma$};
  \draw[->] (1) -- (2) node[midway, above right] {$\beta$};

\end{tikzpicture}
\caption{A transition diagram showing the relationship between energy transitions and roots. In a quantum control context, a transition between two energy levels of an atom that can be described by a root vector in the Hamiltonian. For example, a transition between $\ket{0}\to\ket{1}$ can be described using the root vector $e_\alpha$. An electromagnetic pulse with a frequency resonant with the energy difference between these two levels can, if applied correctly, transition the system consistent with the action of $e_\alpha$.}
\label{fig:su3:rootenergytransitions}
\end{figure}
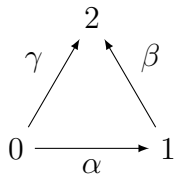

\begin{figure}[h]
\centering
\begin{tikzpicture}[scale=1.5]

% Define the radius of the circle
\def\radius{2cm}

% Draw the center node
\node[draw,circle,inner sep=2pt,fill] (center) at (0,0) {};

% Define positions of the nodes and draw the outer nodes and their opposite nodes
\node[draw,circle,inner sep=2pt,fill] (0) at (0:\radius) {};
\node[draw,circle,inner sep=2pt,fill] (180) at (180:\radius) {};
\node at (0:\radius+0.5cm) [xshift=10pt] {$\lambda_1 + i\lambda_2$};

\node[draw,circle,inner sep=2pt,fill] (60) at (60:\radius) {};
\node[draw,circle,inner sep=2pt,fill] (240) at (240:\radius) {};
\node at (60:\radius+0.5cm) {$\lambda_4 + i\lambda_5$};

\node[draw,circle,inner sep=2pt,fill] (120) at (120:\radius) {};
\node[draw,circle,inner sep=2pt,fill] (300) at (300:\radius) {};
\node at (120:\radius+0.5cm) {$\lambda_6 + i\lambda_7$};

% Draw arrows between nodes
\draw[->,thick] (center) -- (60) node[midway, above,xshift=-5pt] {$\gamma$};
\draw[->,thick] (center) -- (120) node[midway, above left,xshift=-10pt] {$\beta$};
\draw[->,thick] (center) -- (0) node[midway, above] {$\alpha$};

\end{tikzpicture}
\caption{Symmetric root system diagram for the root system described via roots in equation (\ref{eqn:cartan:cayleyrootsystem}) for the Lie algebra $\su(3)$. The roots $\alpha,\beta,\gamma$ can be seen in terms of angles between the root vectors and can be calculated using the Cartan matrix.}
\label{fig:su3:root_system_symmetric}
\end{figure}
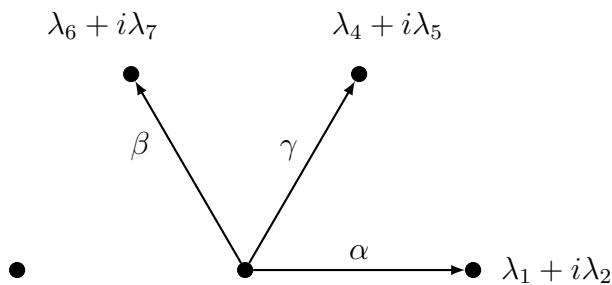

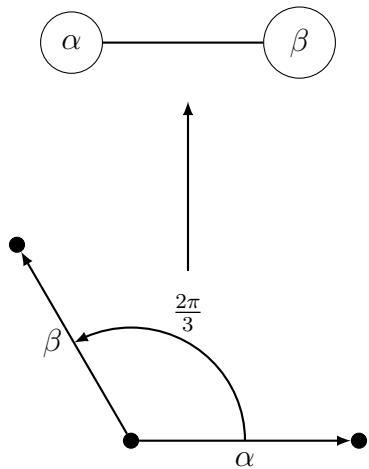
\begin{figure}[h!]
\centering

% First diagram: Dynkin diagram
\begin{tikzpicture}
  % Draw the nodes
  \node (alpha) at (0,0) [circle, draw, inner sep=5pt] {$\alpha$};
  \node (beta) at (3,0) [circle, draw, inner sep=5pt] {$\beta$};
  
  % Draw the connecting line
  \draw[thick] (alpha) -- (beta);
\end{tikzpicture}

\vspace{0.25cm} % Space between the diagrams

% Second diagram
\begin{tikzpicture}[scale=1.5]

  % Central node
  \node[draw,circle,inner sep=2pt,fill] (center) at (0,0) {};

  % Node to the right
  \node[draw,circle,inner sep=2pt,fill] (right) at (2,0) {};
  \draw[->,thick] (center) -- (right) node[midway, below] {$\alpha$};

  % Node at position (120)
  \node[draw,circle,inner sep=2pt,fill] (top) at (120:2cm) {};
  \draw[->,thick] (center) -- (top) node[midway, left] {$\beta$};

  % Arc arrow
  \draw[->,thick] (1,0) arc (0:120:1cm) node[midway, above] {$\frac{2\pi}{3}$};

  % Arrow from the arc to the Dynkin diagram
  \draw[->,thick] (0.5,1.5) -- ++(0,1.5);
% \draw[->,thick] (0,0) -- (3,1.5);
\end{tikzpicture}

\caption{Combined diagram of a Dynkin diagram and a symmetric root system with specified angles and relations.}
\label{fig:su3:CayleyDynkin}
\end{figure}

Diagrammatically, we can understand the relationship between roots and state transitions as set out in Figure \ref{fig:su3:rootenergytransitions} which shows the relationship of roots to transitions between energy states in the lambda system.
The root system can also be described as set out in Fig. \ref{fig:su3:root_system_symmetric} where the angles between root vectors (we give $\alpha$ as the example) are calculated using the Cartan matrix below.
We can represent the relevant Dynkin diagram (definition \ref{defn:alg:dynkindiagram}) for this root system. Recall the entries $A_{ij}$ of the associated Cartan matrix (definition \ref{defn:alg:Cartan matrix}) are calculated by reference to the roots as $A_{ij} = 2(\alpha \cdot \beta)/(\alpha \cdot \alpha)$ where $(\cdot)$ denotes the Euclidean inner product. The Cartan matrix encodes information about angles between simple roots and ratios of their lengths. Angles between different simple roots will be off-diagonal elements in the Cartan matrix of the form $-1,-2,...$ (diagonal entries are 2). Given $\alpha \cdot \alpha = 4$,$\beta \cdot \beta = 4$ For $(\alpha,\beta)$ and $\alpha \cdot \beta  = -2$ we have $A_{\alpha\beta} = 2(\alpha \cdot \beta)/(\alpha \cdot \alpha) = 2\cos(\theta)$ with $\theta$ being the angle between the roots. So $A_{\alpha\beta} = -1$ Noting that  and: 
\begin{align}
    \alpha \cdot \beta &= |\alpha ||\beta|\cos(\theta)\\
     -2 &= (2)(2)\cos(\theta)
\end{align}
we find $\cos(\theta)=-1/2$ such that $\theta = 2\pi/3$. 
The Dynkin diagram represents a root system for $\su(3)$. Each node represents one of the simple roots $(\alpha,\beta)$ connected by a single line. By convention, the lines connecting simple roots reflect the angle between them. For an angle of $2\pi/3$, we have one line connecting the nodes. Hence we see the connection between root systems as represented by the Dynkin diagram in Figure (\ref{fig:su3:CayleyDynkin}) (see \cite{boscain_k_2002} and references therein for a discussion on the relation of resonance to optimality). 
% %====DYNKIN DIAGRAM

\subsection{Hamiltonian}
Below is a toy model of how such roots relate to a Lambda system Hamiltonian with energy levels labeled as $\ket{0}, \ket{1}, \ket{2}$. Here the root vectors from the Lie algebra correspond to transition operators between these states and are incorporated into the Hamiltonian as:
\begin{align}
H &= \sum_{j} \omega_j H_j + \sum_{\alpha \in \Delta} g_\alpha E_\alpha \label{eqn:cartan:Hamiltonianroots}
\end{align}
where $\omega_j$ are the energy eigenvalues corresponding to the Cartan subalgebra elements $H_j$, and $g_\alpha$ are coupling constants for the transitions. The $H_j$ and $E_\alpha$ correspond to the Cartan and non-Cartan elements of $\mathfrak{g}$, respectively. The root vectors $E_\alpha$ correspond to transition operators between energy states. For example, if $E_\alpha = \ketbra{0}{1}$, it induces a transition from state $\ket{1}$ to state $\ket{0}$ when the system interacts with a resonant control field. The non-Cartan subalgebra elements, which correspond to the root vectors of the Lie algebra, drive the transitions among different energy levels. The Cartan subalgebra elements correspond to the diagonal elements in the Hamiltonian, which can be thought of as the stationary energy levels. By contrast, the non-Cartan elements are off-diagonal, represent the possible transitions or interactions between these energy levels.

\appendix
%================================
%=========================APPENDIX

% \appendix

%==========
%==========
%===QUANTUM INFORMATION PROCESSING
\chapter{Appendix (Quantum Information Processing)} \label{chapter:Background: Quantum Information Processing}
\section{Overview}
Quantum information processing is characterised by constraints upon how information is represented and processed arising from the foundational postulates of quantum mechanics. While information as the subject of enquiry within physics has a long lineage, quantum information processing as an emergent and distinct discipline arose largely through the convergence (and often synthesis) of concepts in computational and informational sciences and quantum physics, principally through the vision of quantum computing envisaged by Feynman \cite{feynman_simulating_1982} and others. The modern day field incorporates concepts from quantum physics, computer science and information and communication theory (pioneered by Shannon and others \cite{shannon_mathematical_1948}). Quantum information theory is often classified into three overlapping but distinct sub-domains of quantum computing, quantum communication and quantum sensing. Our focus in this work is in the first of these areas, quantum computing, albeit we leverage concepts drawn from communication theory such as quantum channels and registers. 

Below we provide an overview of key elements of quantum postulates \cite{nielsen_quantum_2010} and quantum information processing for classical machine learning practitioners solving optimisation problems for quantum engineering. In particular, we describe ways in which quantum data are typically represented (such as via tensors), how quantum processes are usually expressed in common programming languages. Quantum information processing draws upon formalism and conceptual framing from a range of disciplines, including analysis, information theory and complexity theory, together with algebra, geometry and other numerous branches of mathematics. All quantum information processing rests, however, upon the mathematical formalism of quantum mechanics itself. The study of such formalism usually begins with the study of quantum axioms or postulates, a set of assumptions about quantum systems and their evolution derived from observation and from which other results and properties of quantum systems are deduced. There are a variety of ways to express such axioms according to chosen formalism (and indeed debate continues within fields such as quantum foundations, mathematics and philosophy as to the appropriateness of such postulates). In the synopses below, we have followed the literature in \cite{nielsen_quantum_2011,hall_quantum_2013} and \cite{watrous_theory_2018}, making a few adjustments where convenient (such as splitting the measurement postulate in two).  We partly frame postulates information-theoretic fashion concurrently with traditional formulations. The axioms form a structured ordering of this background material, anchoring various definitions, theorems and results relevant to establishing and justifying methods and results in the substantive Chapters above. We proceed to discussing key elements of open quantum systems (relevant to, for example, the simulated quantum machine learning dataset the subject of Chapter \ref{chapter:QDataSet and Quantum Greybox Learning}) together with elements of the theory of quantum control (which we later expand upon in terms of algebraic and geometric formulations). For those familiar with quantum mechanics and quantum information processing, this section can be skipped without loss of benefit. Most of the material below is drawn from standard texts in the literature, including \cite{nielsen_quantum_2010,hall_quantum_2013,watrous_theory_2018}. Proofs are omitted for the most part (being accessible in the foregoing or other textbooks on the topic). We begin with the first of the quantum postulates (which we denote axioms following \cite{hall_quantum_2013}) of quantum mechanics, that related to quantum states.

\subsection{State space}\label{sec:chapter1:quantum:statespace}
In this section, we set out a number of standard formulations for conceptualising quantum information and quantum control problems. Advanced readers familiar with quantum formalism can progress to other sections without loss of generality. We begin with vector spaces and some results from functional analysis. In later Chapters, we will equate typical Hilbert space formalism in terms of geometric and algebraic concepts related to manifolds, fibres and connections. The first axiom concerns representations of quantum systems in terms of states.
\begin{axiom}[Quantum states]\label{axiom:quant:quantumstates}
The state of a quantum system is represented by a unit vector $\psi$ in an appropriate Hilbert space $\mathcal{H}$. Two unit vectors $\psi_1, \psi_2\in\mathcal{H}$, where $\psi_2 = c\psi_1$ for $c \in \mathbb{C}$, correspond to the same physical state. In density operator formalism, the quantum system is described by the linear operator $\rho \in \mathcal{B}(\Hilb)$ acting on $\Hilb$. For bounded $A \in \mathcal{B}(\Hilb)$, the expectation value of $A$ is $\Tr(\rho A)$. Composite quantum systems are represented by tensor products $\psi_k \otimes \psi_j$ (or equivalently in operator formalism, $\rho_k \otimes \rho_j$). 
\end{axiom}
The axiom above provides that quantum systems are represented by (unit) state vectors within a complex-valued vector (Hilbert) space $\mathcal{H}$ whose dimensionality is determined according to the physics of the problem. There are various formalisms for representing such state vectors in quantum information processing, a common one being Dirac or `bra-ket' notation. In this notation, the state vector is represented as a `ket' $\ketpsi \in \mathcal{H}$. Associated with a ket is a corresponding `bra' notation $\bra{\psi}$ which strictly speaking is a linear (one) form (or function) that acts on $\ketpsi$ such that $\braket{\psi_1|\psi_2} \in \Complex$ for two states $\ket{\psi_1},\ket{\psi_2}$. Quantum states are typically defined as elements of a Hilbert space $\Hilb$, a vector space over $\C$ equipped with an inner product. To this end, we introduce the definition of a state space.

%=====Defn: Vector Space
\begin{definition}[State space]\label{defn:quant:State space}
    A state space $V(\mathbb{K})$ is a vector space over a field $\mathbbm{K}$, typically $\mathbbm{C}$. Elements $\psi \in V$ represent possible states of a quantum system. In particular, for $a_k\in\C$ and $\psi_k \in V$, the state $\psi = \sum_k a_k \psi_k \in V$. 
\end{definition}
Evolutions of quantum systems over time are described by mappings within this state space (e.g. linear operators on $V$). Importantly, as can be seen above, any linear combination of basis states $\psi_k$ (or in braket notation, $\ket{\psi_k}$) is also a quantum state, an assumption crucial to the existence of two-level qubit systems and entanglement. The state space of a quantum system is fundamental in the description of quantum states, where the vectors are typically normalized to have a unit norm due to the probabilistic interpretation of quantum mechanics. Later, we refine our working definition of state space in terms of Hilbert spaces, operators and other concepts such as channels. Our aim is to provide sufficient depth and breadth for readers in order to connect quantum state space with concepts from algebra, representation theory, geometry and statistical learning. While we utilise standard bra-ket notation and formalism, we also incorporate a more quantum information-theoretic approach following Watrous \cite{watrous_theory_2018} who defines states in terms of \textit{registers} (alphabets or tuples thereof) as descriptions of quantum states according to an alphabet $\Sigma$. 
\begin{definition}[Classical registers and states]\label{defn:quant:Classical registers and states}
    A classical register $X$ is either (a) a simple register, being an alphabet $\Sigma$ (describing a state); or (b) a compound register, being an n-tuple of registers $X=(Y_k)$.\\
    A classical state set $\Gamma$ of $X$ is then either (i) $\Gamma=\Sigma$ or (ii) in the case of a compound register, the Cartesian product of those state sets $\Sigma = \times_k \Gamma_k$.  
\end{definition}
The formulation of classical states here maps largely to that found in computer science literature where states are constructed from alphabets etc. Classical states may be composed of subregisters which determine the overall state description. That is, a register has configurations (states) it may adopt (with subregisters determining higher-order registers). Probabilistic states are distributions over classical states the register may assume, denoted $\mathcal{P}(\Sigma)$ with states represented by probability vectors $p(a)\in \mathcal{P}(\Sigma)$. From the classical case, we can then construct the quantum information representation of a quantum register and states which are represented by density operators (defined below).
\begin{definition}[Quantum registers and states]\label{defn:quant:Quantum registers and states}
    A quantum state is defined as an element of a complex (Euclidean) space $\C^\Sigma$ for a classical state set $\Sigma$ satisfying specifications imposed by axioms of quantum mechanics (see below) as they apply to states.  
\end{definition}
In the terminology of operators (see below) a quantum state is defined as a density operator $\rho \in \mathcal{B}(\Hilb)$.
 We refer in parts to this formulation below and in later Chapters. To come to our formal definition of Hilbert spaces $\Hilb$, first we define the inner product on $V(\Km)$.

 %=====Defn: Inner Product
\begin{definition}[Inner Product]\label{defn:quant:Inner Product}
An inner product on $V(\mathbbm{K})$ is a mapping from a vector space to a field $\mathbbm{K}$ \( \langle \cdot, \cdot \rangle: V \times V \rightarrow \mathbbm{K}, (\psi,\phi) \mapsto \langle \psi,\phi \rangle \) for $\psi,\phi \in V, c \in \mathbbm{K}$ satisfying:
\begin{enumerate}[(i)]
    \item $\braket{\psi | \phi} = \braket{\phi | \psi}$.
    \item $\braket{\phi | \phi} \in \mathbb{R}^+ \text{ and } \braket{\phi | \phi} = 0 \Longleftrightarrow \phi = 0$.
    \item $\braket{c\phi | \psi} = c \braket{\phi | \psi} \text{ and } \braket{\phi | c\psi} = c^* \braket{\phi | \psi}$.
    \item $\braket{\phi + \psi | \chi} = \braket{\phi | \chi} + \braket{\psi | \chi} \text{ and } \braket{\phi | \psi + \chi} = \braket{\phi | \psi} + \braket{\phi | \chi}$.
\end{enumerate}
\end{definition}
Inner products and bilinear forms are crucial components across quantum, algebraic and geometric methods in quantum information processing. Later, we examine the relationship between differential (\textit{n}-)forms and inner products (see Appendix \ref{chapter:Background: Geometry, Lie Algebras and Representation Theory}) which are relevant to variational results in Chapter \ref{chapter:Time optimal quantum geodesics using Cartan decompositions}. There the inner product is analogous to a bilinear mapping of $V \times V^* \to \mathbb{K}$, implicit in the commonplace `braket' notation of quantum information (where kets $\ketpsi$ and $\bra{\psi}$ are duals to each other and inner products are given by $\braket{\psi|\psi}$). In the formalism below, we regard $\ket{\psi}$ as an element of a vector (Hilbert) space $V(\C)$ while the corresponding $\bra{\psi}$ as an element of the dual vector space $V^*(\C)$. Moreover, in later Chapters we adopt (and argue for the utility of) a geometric approach where inner products and norms are defined in terms of metric tensors $g$ over manifolds and vector space relationships to fibres. The inner product defined on a vector space $V$ then gives rise to a norm via the Cauchy-Schwartz inequality.
%=====Defn: Cauchy-Schwarz
\begin{definition}[Cauchy-Schwarz Inequality]\label{defn:quant:Cauchy-Schwarz Inequality}
The Cauchy-Schwarz inequality applies to inner product spaces $V(\mathbbm{K})$ such that for \( \phi, \psi \in V \):
\begin{align*}
|\langle \phi, \psi \rangle|^2 \leq \langle \phi, \phi \rangle \langle \psi, \psi \rangle.
\end{align*}
If \( ||\cdot||: V \rightarrow \mathbb{R} \) where $||\psi|| = \sqrt{\langle \psi, \psi \rangle}$, then the inner product defines a norm on $V$.
\end{definition}
%=====Defn: Norm
%====Angle
We are now equipped to define the norm on the vector spaces with specific properties as follows.
\begin{definition}[Norm]\label{defn:quant:Norm}
Define a norm on a vector space \( V \) over \( \mathbbm{K} = \mathbb{R}\) or \( \mathbb{C} \) as a mapping \( \lVert . \rVert: V \to \mathbb{R} \), \( \psi \mapsto \lVert \psi \rVert \), such that for $\psi,\phi \in V$ and $c \in \mathbbm{K}$:
\begin{align*}
    1. & \quad \lVert \psi \rVert \geq 0 \text{ and } \psi = 0 \Longleftrightarrow \lVert \psi \rVert=0 . \\
    2. & \quad \lVert c\psi \rVert = |c| \lVert \psi \rVert. \\
    3. & \quad \lVert \phi + \psi \rVert \leq \lVert \phi \rVert + \lVert \psi \rVert.
\end{align*}
\end{definition}
A norm \( \lVert . \rVert \) on \( V \) defines a distance (metric) function \( d \) on \( V \) as $d(\phi, \psi) = \lVert \psi - \phi \rVert$. As standard we can then define a generalised angle given two vectors \( \phi, \psi \in V \) in an inner product space where the cosine of the angle \( \theta \) between \( \phi \) and \( \psi \) is given by:
\begin{align}
\cos(\theta) = \frac{\langle \phi, \psi \rangle}{\| \phi \| \cdot \| \psi \|} \label{eqn:quant:cosinevector}
\end{align}
%=====Defn: Banach
A normed vector space is a \textit{Banach space}
if it is complete with respect to the associated distance function (e.g. inner product). Banach spaces are equipped with certain convergence and boundedness properties which enable among other things definitions of bounded linear functionals and dual spaces, both of which are relevant to results in later Chapters. Banach spaces allow us to define an \textit{operator norm}.
%===Operator norm
\begin{definition}[Operator norm]\label{defn:quant:Operator norm}
If $V_1(\mathbb{K})$ and $V_2(\mathbb{K})$ are normed spaces, the linear mapping $A \colon V_1(\mathbb{K}) \rightarrow V_2(\mathbb{K})$ is bounded if
\begin{align*}
\sup_{\psi \in V_1 \setminus \{0\}} \frac{\|A\psi\|}{\|\psi\|} < \infty.
\end{align*}
\end{definition}
Intuitively this tells us that linear mappings on $V(\mathbb{K})$ remain within the closure of $V(\mathbb{K})$. The set of such linear forms a complementary vector space denoted the dual space.
%=====Defn: bounded linear functional
\begin{definition}[Dual space]\label{defn:quant:Dual space}
    A bounded linear functional on a normed vector space $V(\mathbbm{K})$ is a bounded linear map $\chi: V \to \mathbbm{K}, \psi \mapsto a||\psi||$ for some (scaling) $a \in \mathbbm{K}$. The norm is given via the field norm $||.|| = |.| \in \mathbbm{K}$. The set of all such bounded linear functionals is denoted the dual space $V^*(\mathbb{K})$ to $V(\mathbbm{K})$.\\
    If $V(\mathbbm{K})$ is normed then $V^*(\mathbb{K})$ is also a Banach space. Moreover, we have that for $\psi \in V$ and $\chi \in V^*$: 
    \begin{align*}
        1.&\quad \chi(\psi) \in \mathbb{K}.\\
        2.&\quad |\chi(\psi)|= ||\chi||||\psi||.\\
        3.&\quad \forall \chi \in V^*, \chi(\psi) = 0 \Longleftrightarrow \psi = 0.
    \end{align*}
\end{definition}
Note in numbered item 2 above, duals can also be thought of as types of vectors themselves and as functions that take vectors as arguments. We discuss dual spaces and functionals in further chapters, particularly in relation to representations of Lie algebras and differential forms. With the concept of a dual space, inner product and norm, we can now define the Hilbert space.
%=====Defn: dual space
%=====Defn: Hilbert Space
\begin{definition}[Hilbert Space]\label{defn:quant:Hilbert Space}
A Hilbert space is a vector space $\mathcal{H}(\mathbbm{K})$ with an inner product \( \langle \cdot, \cdot \rangle \) complete in the norm defined by the inner product $||\psi|| = \sqrt{\langle \psi, \psi \rangle}$.
\end{definition}
The norm on $\Hilb$ satisfies:
\begin{align*}
||\psi||^2 := \sum_{j=1}^{\infty} ||\psi_j||^2 < \infty.
\end{align*} 
Often we are interested in direct sums or direct (tensor) products of Hilbert spaces such as tensor product of Hilbert spaces (crucial to multi-qubit quantum computations). We discuss these in later Chapters. In some cases, multipartite Hilbert spaces may be characterised by direct sums rather than direct (tensor) products. This is the case where a Hilbert space is decomposable into disjoint regions.

%=====Defn: Hilbert space direct sum
\begin{proposition}[Hilbert Space Direct Sum]\label{prop:quant:Hilbert Space Direct Sum} The direct sum of a sequence \( \Hilb \) of separable Hilbert spaces denoted by
\begin{align*}
H := \bigoplus_{j=1}^{\infty} H_j,
\end{align*}
has a representation as a space of sequences \( \psi = (\psi_1, \psi_2, \psi_3, \ldots) \) where \( \psi_j \in \Hilb_j \). 
\end{proposition}
A direct sum of each \( H_j \) is the set of \( \psi = (\psi_1, \psi_2, \psi_3, \ldots) \) where \( \psi_j = 0 \) for all but finitely many $\psi_j$. An inner product on the direct sum can be defined via:
\begin{align*}
\langle \phi, \psi \rangle = \sum_{j=1}^{\infty} \langle \phi_j, \psi_j \rangle,
\end{align*}
for all \( \phi_j, \psi_j \in H_j \) and $\psi,\phi \in \Hilb$. Each subspace inherits the inner product and associated completeness properties of $\Hilb$, confirming each $\Hilb_j$ as a Hilbert space. The partitioning of a Hilbert space via a Cartan decomposition of Lie algebraic generators (which form a Hilbert space as a vector space upon which an inner product is defined) $\g$ into $\g= \k \oplus \p$ is an example of such a decomposition which preserves the overall structure of the vector space within which quantum state vectors subsist \cite{schue_hilbert_1960}. Thus the essential properties of quantum state vectors are retained in the case of such decomposition. 
Hilbert spaces have a number of features of importance to quantum computational contexts. These include certain orthogonality relations which are important to their decomposition in ways that can simplify problems (such as those of unitary synthesis that we examine in later Chapters).
\begin{definition}[Orthogonality and orthonormality]\label{defn:quant:Orthogonality and orthonormality}
    A Hilbert space $\mathcal{H}$ exhibits the following relations related to orthogonality:
    \begin{enumerate}
        \item $\psi,\phi \in \mathcal{H}$ are orthogonal if $\innerproduct{\psi}{\phi}=0$.
        \item (Orthogonal space) If $\mathcal{H}_i \subset \mathcal{H}$, the orthogonal subspace $\mathcal{H}_i^\perp \subset \mathcal{H}$ of $H_i$ is:
        \begin{align*}
    \mathcal{H}_i^\perp &= \{ \phi \in \mathcal{H} | \innerproduct{\psi}{\phi} = 0 \text{ for all } \psi \in H_i \}
        \end{align*}
    \end{enumerate}
\end{definition}
In many quantum information tasks, we assume that states $\psi \in \mathcal{H}$ can be decomposed into a direct sum $\mathcal{H}_i \oplus \mathcal{H}_i^\perp = \mathcal{H}$ such that $\psi = \psi_i + \psi_i^\perp$. States are then represented in terms of an orthonormal basis for $\mathcal{H}$.
\begin{definition}[Orthonormal Basis]\label{defn:quant:Orthonormal Basis}
The set $\{e_j\}, e_j \in \mathcal{H}$, $j=1,...,\dim\mathcal{H}$ is an \textit{orthonormal} basis if $\braket{e_j, e_k}=\delta_{jk}$ (the Kronecker delta)
and $\mathcal{H}=\spn\{ae_j | a\in \C\}$. In this case:
\begin{align}
    ||\psi||^2&=\sum_j |a_j|^2 \label{defn:2.2.10sumjaj2}
\end{align}
\end{definition}
The boundedness and linearity of states means we can also consider them as (convergent) sums $\psi = \sum_j a_j e_j, a_j \in \C$ where $a_j = \braket{e_j, \psi} \in \C$ . 

\subsubsection{Tensor product states}\label{sec:quant:Tensor product states}
In many cases of interest, we have disjoint or multi-state (e.g. multi-qubit) quantum systems of interest. This is especially the case with the multi-qubit systems the subject of Chapters \ref{chapter:QDataSet and Quantum Greybox Learning} 
 and \ref{chapter:Quantum Geometric Machine Learning}. Vector spaces for multi-state systems are represented by tensor products of vector spaces and Hilbert spaces which we define below.

\begin{definition}[Tensor Product]\label{defn:quant:Tensor Product}
A tensor product of two vector spaces $V_1(\mathbb{K}), V_2(\mathbb{K})$ is itself a vector space $W(\mathbb{F})$ equipped with a bilinear map $T : V_1(\mathbb{K}) \times V_2(\mathbb{K}) \rightarrow W(\mathbb{K})$.
\end{definition}
To allow us to represent a single (equivalent) tensor product between two such vector spaces, we note the following `universal property' \cite{hall_quantum_2013} which allows elements of $W$ to be written as elements $u \otimes v$ where $u \in V_1, v, v \in V_2$. Given another vector space $Z(\mathbb{K})$ along with a bilinear mapping $\Phi : V_1 \times V_2 \rightarrow Z$, it can be shown that there exists a unique linear map $\widetilde{\Phi}: W \rightarrow Z$ such that the diagram below commutes:
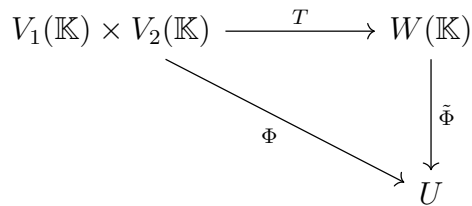
\begin{figure}[h!]
\centering
\begin{tikzcd}[row sep=huge, column sep=huge]
V_1(\mathbb{K})\times V_2(\mathbb{K}) \arrow[rd, "\Phi"'] \arrow[r, "T"] & W(\mathbb{K}) \arrow[d, "\tilde\Phi"] \\
& U
\end{tikzcd}
\caption{Commutative diagram showing the university property of tensor products discussed above.}
\end{figure}
It can then be shown that the tensor product of $V_1,V_2$ exists and is unique up to a canonical isomorphism. This allows us to represent the mapping $T$ in terms of the direct product $T:V_1 \times V_2 \to V_1 \otimes V_2, (u,v) \mapsto u\otimes v$. 

The universal property guarantees that given $\Phi: V_1 \times V_2 \to Z$, there will exist a map $\tilde\Phi: V_1 \otimes V_2 \to Z$ with $\tilde\Phi(u \otimes v = \Phi(u,v)$. Among other things this means that $\dim(V_1 \otimes V_2) = (\dim V_1)(\dim V_2)$. Under such assumptions, arbitrary $\psi \in V_1 \otimes V_2$ can then be decomposed as linear compositions of $u\otimes v$ for a given basis. It also allows the space of operators (linear maps $A_k \in End(V_k)$) to be constructed in tensor-product form by guaranteeing the existence of $A_k \otimes A_j: V_k \otimes V_j \to V_k \otimes V_j$ exhibiting (a) homomorphic structure $(A_k \otimes A_j)(u \times v) = (A_ku) \otimes (A_j v)$ and linear compositionality $(A_k \otimes A_j)(A_p \otimes A_q)=(A_kA_p) \otimes (A_jA_r)$. For tensor products of Hilbert spaces, we note that given multiple inner product spaces $V_k$ with inner products $\braket{v_p,v_q}_k$, there exists an inner product on the tensor product space $\otimes_k V_k$ given by:
\begin{align*}
    \braket{u_p \otimes v_p, u_q \otimes v_q} = \braket{u_p,u_q}_k \braket{v_p,v_q}_j
\end{align*}
where $u_p,u_q \in V_k$ and $v_p,v_q \in V_j$. In this way, a Hilbert tensor product $\otimes_k \Hilb_k$ space is equipped with sufficient structure via the inner product and bounded operators as per above (allowing, among other things, metric calculations relevant to machine learning and variational techniques to be implemented). We note for completeness that orthonormal bases $\{e_j\}_k$ for $\Hilb_k$ form a tensorial orthonormal basis for the Hilbert space $\otimes \Hilb_k$ given by $\{ \times_k e_{j,k}\}$ (tensor product of orthonormal bases).

\subsubsection{Quantum State formalism}\label{sec:quant:Quantum State formalism}
A significant proportion of quantum information research concentrates on single (or multiple) \textit{qubit} (quantum bit) systems. Qubits are two-level quantum systems (two-dimensional state spaces) comprising state vectors with orthonormal bases $\{\kz,\ko\}$ (and accompanied by corresponding operator representations below). In this formalism a \textit{qubit} can be defined as follows as a two-state quantum mechanical system:
\begin{align}
\ketpsi = a\kz + b\ko.
\label{eqn:quant:qubit}
\end{align}
Qubits are normalised such that they are unit vectors $\braket{\psi | \psi} = 1$ (that is $|a|^2 + |b|^2=1$), where $a,b \in \Complex$ are amplitudes and their squared moduli, the probabilities, for measuring outcomes of 0 or 1 (corresponding to states $\kz,\ko$ respectively). Here $\braket{\psi | \psi'}$ denotes the inner product of quantum states $\ketpsi,\ket{\psi'}$. As can be observed in equation (\ref{eqn:quant:qubit}), as distinct from a classical (deterministic or stochastic) state, quantum states may subsist ontologically in a superposition of their basis states.

The stateful characteristic of quantum systems is also manifest in the first postulate of quantum mechanics, which associates Hilbert spaces to enclosed physical systems under the assumption that $\psi \in \mathcal{H}$ provides complete information about the system \cite{nielsen_quantum_2010}. In quantum computing, qubits may be either \textit{physical} (representing the realisation of a two-level physical system conforming to the qubit criteria) or \textit{logical}, where a set of systems (e.g. a set of one or more physical qubits) behaves abstractly in the manner of a single qubits which is important in error correction. Quantum computations may also involve \textit{ancilla} qubits used to conduct irreversible logical operations \cite{nielsen_quantum_2011}.

For problems in quantum computing and control, we are interested in how quantum systems evolve, how they are measured and how they may be controlled. To this end, we include below a brief synopsis of operator formalism and evolution of quantum systems. In Appendix \ref{chapter:Background: Geometry, Lie Algebras and Representation Theory}, we characterise these formal features of quantum information processing algebraic and geometric terms. One of the consequences of definition \ref{defn:quant:State space} is the existence, unlike in the classical case, of quantum superposition states i.e. that a quantum system can exist simultaneously in multiple different states until measured. Given a quantum system in states represented by unit vectors $\ket{\psi_1}, \ket{\psi_2}, \ldots, \ket{\psi_n} \in \Hilb$, any linear combination of these states is also in $\Hilb$:
\begin{align}
\ket{\psi} = \sum_{k=1}^n c_k \ket{\psi_k} \label{eqn:quant:superposition}
\end{align}
where $c_k \in \mathbb{C}$ are complex coefficients satisfying the normalization condition $\sum_{k=1}^n |c_k|^2 = 1$. Such a superposition state $\ket{\psi}$ is a unit vector in $\mathcal{H}$. It is a valid quantum state under axiom \ref{axiom:quant:quantumstates}. The state $\ket{\psi}$ encodes the probabilities of the system being found in any of the basis states upon measurement, with the probability of finding the system in state $\ket{\psi_k}$ given by $|c_k|^2$.

\subsection{Operators and evolution}\label{sec:quant:Operators and evolution}

Often it is more convenient to work with (and consider evolutions of) functions and algebras associated with operators on $\Hilb$ rather than $\psi \in \Hilb$ directly. This operator formalism manifests in both recasting state spaces in terms of density operators and considering maps between operators in terms of quantum channels. Such mappings also serve to connect quantum formalism with the types of mappings and formalism common within machine learning and geometry. We now consider formalism associated with the second axiom of quantum mechanics, namely that physical observables correspond to real eigenvalues of (measurement) operators. 
\begin{axiom}[Observable Operators]\label{axiom:quant:observableoperators}
For every measureable physical observable quantity $m$ is defined by a real-valued function, denoted an operator, $M$ defined on classical phase space, there exists a self-adjoint (Hermitian) linear measurement operator $M$ acting on $\mathcal{H}$. The results of measurement of the state must be a real-valued eigenvalue of the measurement operator.
\end{axiom}

\subsubsection{Operator formalism}\label{sec:quant:Operator formalism}
In quantum information, operators are thought of as maps between Hilbert spaces. To understand this, we begin by considering the general class of endomorphisms on $\Hilb$ (and later describe Hilbert-Schmidt space and channels) and define $\mathcal{B}(\mathcal{H})$ as the space of bounded endomorphic linear maps of $\Hilb$:
\begin{definition}[Bounded linear operators]\label{defn:quant:Bounded linear operators}
    The space $\mathcal{B}(\mathcal{H})$ of bounded linear operators $A$ on $\Hilb$ is defined by
\begin{align*}
\mathcal{B}(\mathcal{H}) = \{ A : \mathcal{H} \rightarrow \mathcal{H} \mid A \text{ is linear and bounded} \},
\end{align*}
where $A$ is bounded if there exists a non-negative real number $K$ such that:
\begin{align*}
\| A\psi \| &\leq K\|\psi\|
\end{align*}
for all $\psi \in \Hilb$. The smallest such $K$ is the operator norm of $A$, denoted by $\| A \|$.
\end{definition}
We also note Riesz's Theorem that for any bounded linear functional $\xi:\mathcal{H} \to \C$, there exists a dual element $\chi$ such that $\xi(\psi) = \braket{\chi,\psi}$ (i.e. and so an equivalence of norms). Riesz's Theorem guarantees the existence in such cases of an adjoint operator $A^\dagger$ which crucial for Hermitian operators. 
\begin{definition}[Adjoint of an Operator]\label{defn:quant:Adjoint of an Operator}
For any $A \in \mathcal{B}(\Hilb)$, there exists a unique linear operator $A^\dagger \colon \Hilb \to \Hilb$, called the adjoint of $A$, such that
\begin{align*}
\langle \phi, A\psi \rangle = \langle A^\dagger\phi, \psi \rangle
\end{align*}
for all $\phi, \psi \in H$.
\end{definition}
The adjoint is also known as the Hermitian conjugate. In physics, sometimes the adjoint of such an operator, $A^\dagger$ is denoted $A^*$ (we reserve the $*$ notation for dual spaces). This can occasionally lead to some confusion as sometimes $A^*$ is just considered the complex (entry-wise) conjugate. We set out a few important properties of such operators below.
% [For the avoidance of doubt, in this work $A^*$ represents $\overline{A}^T$] with $\overline{A}$ representing the entry-wise conjugate.
%=====djoint Operator Properties and Hermitian Operators
\begin{proposition}[Adjoint Operator Properties and Hermitian Operators]\label{prop:quant:djoint Operator Properties and Hermitian Operators}
For all $A, B \in \mathcal{B}(H)$ and $\alpha, \beta \in \mathbb{C}$ we have
\begin{align*}
(A^\dagger)^\dagger &= A \qquad (AB)^\dagger = B^\dagger A^\dagger, \\
(\alpha A + \beta B)^\dagger &= \bar{\alpha}A^\dagger + \bar{\beta}B^\dagger \qquad
I^\dagger = I.
\end{align*}
Note too that $\|A^\dagger\| = \|A\| < \infty$. If $A^\dagger = A$, then $A$ is self-adjoint which we denote as Hermitian (or skew-self-adjoint if $A^\dagger = -A$). 
\end{proposition}
Matrix representation of Hermitian conjugation is via the (complex)-conjugate transpose $A^\dagger = (A^*)^T$. In braket notation we define $\ket{\psi}^\dagger\dot=\bra{\psi}$. The adjoint operator is also important practically as an element used when determining Cartan involutions (and thus Cartan decompositions) with respect to symmetric and anti-symmetric subspaces of $\g$. This allows us to introduce unitary operators which are crucial to quantum information processing and our results herein. Before we define unitaries, we set out a few standard important operator classes.
\begin{definition}[Positive Semidefinite Operators]\label{defn:quant:Positive Semidefinite Operators}
An operator $A \in \mathcal{B}(\mathcal{H})$ is positive semi-definite if for all $\psi \in \Hilb, \braket{\psi,A\psi} \geq 0$. This equates to $A = B^\dagger B$ for some operator $B \in \mathcal{B}(\mathcal{H})$ such that: 
\begin{align}
    \braket{\psi,A\psi} = \braket{\psi,B^\dagger B \psi} = \braket{B \psi, B\psi} = ||B \psi ||^2 \geq 0
\end{align}
Denote the set of such operators as:
\begin{align*}
\text{Pos}(\mathcal{H}) = \{B^\dagger B | B \in \mathcal{B}(\mathcal{H})\}.
\end{align*}
\end{definition}
Positive semidefinite operators are Hermitian. A \textit{positive definite operator} is a positive-semi definite operator that is also invertible i.e. $A \in Pos(\Hilb)$ such that $\det(A) \neq 0$. Positive semi-definite operators are commonly expressed via constraints on the inner product, namely that $\braket{Av,v} \geq 0$ for $v \in \Hilb$. Operators $A \in \mathcal{B}(\Hilb)$ that commute with their adjoint i.e. $A A^\dagger = A^\dagger A$ are denoted \textit{normal operators} and share the same set of eigenvectors. Using these definitions we can then specify two important classes, projection operators and isometries.

%=====Defn: projection and isometries
\begin{definition}[Projection Operators]
\label{defn:quant:projection_operators}
An operator $P \in \text{Pos}(\mathcal{H})$ is an (orthogonal) projection operator if $P^2 = P$ and, for some closed $V \in \Hilb$, $P(V)=\mathbb{I}$ and $P(V^\perp)=0$.
\end{definition}
Note that projection operators are normal operators and that the set of projection operators is denoted $\text{Proj}(\mathcal{H})$. For each $V$ there exists a $P$ such that the image of $P$ is $V$.
%=====Defn: isometries
\begin{definition}[Isometries]
\label{defn:quant:isometries} 
A bounded linear operator $A \in \mathcal{B}(\mathcal{H}, \mathcal{Y})$ (for Hilbert spaces $\Hilb,\mathcal{Y}$) is defined to be an \textit{isometry} if $\|Aw\| = \|w\|$ for all $w \in \mathcal{H}$ (i.e. it acts as a fixed point (identity) operator with respect to $\Hilb$. The statement is equivalent to $A^\dagger A = \mathbb{I}_{\mathcal{H}}$ (the identity operator on $\Hilb$). We define the set of such isometries as:
\begin{align*}
U(\mathcal{H}, \mathcal{Y}) = \{A \in \mathcal{B}(\mathcal{H}, \mathcal{Y}) \mid A^\dagger A = \mathbb{I}_{\mathcal{H}}\}
\end{align*} 
\end{definition}
In order for an isometry of the form $A \in U(\mathcal{H}, \mathcal{Y})$ to exist, it must hold that $\text{dim}(\mathcal{Y}) \geq \text{dim}(\mathcal{H})$. Every isometry preserves not only the Euclidean norm, but inner products as well: $\langle Au, Av \rangle = \langle u, v \rangle$ for all $u, v \in \mathcal{H}$. Isometries allow us to then define unitaries operators as the set of isometries $U:(\Hilb,\Hilb) \to \Hilb$ (which map $\Hilb$ to itself).

%====Defn: unitary operator
\begin{definition}[Unitary Operator]\label{defn:quant:Unitary Operator}
An operator $U \in \mathcal{B}(\Hilb)$ is unitary if:
\begin{align*}
\langle U\phi, U\psi \rangle = \langle \phi, \psi \rangle
\end{align*}
(i.e. is surjective and maintains inner products) for all $\phi, \psi \in \Hilb$. Unitary operators preserve norms $\| U\psi \| = \| \psi \|$. 
\end{definition}
Unitary operators are thus a specific type of normal operator. From this definition, we deduce that $\|U\| = 1$. $U$ is unitary if and only if $U^\dagger = U^{-1}$ such that $UU^\dagger = U^\dagger U = I$. Hermitian operators, unitary operators and positive semi-definite operators are central to quantum information processing (and the theory of quantum mechanics generally). Importantly, there is a bijection from the space of Hermitian operators on the complex space $\Hilb$ to real Euclidean spaces. 
Unitary operators represent propagators of quantum state vectors $\psi(t) \in \Hilb$ and, moreover, represent solutions to Schr\"{o}dinger's equation (which we discuss in more detail), forming a group. Not all evolution is unitary  (e.g. open quantum systems and measurement itself). However evolution of quantum systems modelled via unitary operators is usually the foundational basis for more complex models.

From the existence of normal operators we obtain the spectral theorem which states that every normal operator $X \in \mathcal{B}(\Hilb)$ can be expressed as a linear combination of projection $\Pi_k$ operators onto distinct orthogonal subspaces that allows the decomposition of such normal operators into the sum of (complex) eigenvalues and such projection operators. This fact is expressed via the spectral theorem.

\begin{theorem}[Spectral Theorem]\label{thm:quant:spectral}
Given $\Hilb$ with normal operator $X \in\mathcal{B}(\Hilb)$, there exists $\lambda_k \in \C$ ($k = 1,...,\dim\Hilb)$ and projection operators $\sum_k\Pi_k=\mathbb{I}$ in $\text{Proj}(\Hilb)$ such that:
\begin{align*}
X = \sum_{k=1}^{m} \lambda_k \Pi_k
\label{eqn:quant:spectral}
\end{align*}
\end{theorem}
This decomposition is unique in the sense that each $\lambda_k$ corresponds to an eigenvalue of $X$, and each $\Pi_k$ projects onto a subspace formed by the eigenvectors associated with $\lambda_k$. Normal operators that may be decomposed in this way have an orthonormal basis, and the same basis may be used for decomposing commuting normal operators. In particular, both Hamiltonians $H$ and measurement operators are Hermitian with real eigenvalues such that the eigenvalues in equation (\ref{eqn:quant:spectral}) above are real. Among other things, this allows the Jordan-Hahn decomposition \cite{watrous_quantum_2002} where the spectral decomposition of $H=\sum_k \lambda_k \Pi_k$ is obtained via partitioning into:
\begin{align}
    P&=\sum_k\max\{ \lambda_k,0\} \Pi_k \qquad Q = \sum_k \max\{ -\lambda_k,0\} \Pi_k
\end{align}
for $P,Q \in Pos(\Hilb), PQ=0$, allowing $H$ to be written as:
\begin{align}
    H = P-Q
    \label{eqn:quant:jordanhahn}
\end{align}
The spectral decomposition allows functions $f \in \C^m$ to be written as $f(X) = \sum_k f(\lambda_k) \Pi_k$. The spectral decomposition is also related to more general decompositions, such as singular value decompositions which hold for arbitrary $X \in \mathcal{B}(\Hilb)$, and polar decompositions (see \cite{watrous_theory_2018,hall_quantum_2013}) which are commonly used in machine learning contexts. They are also related to the later focus of this work in terms of generalised decompositions of Lie group manifolds, such as Cartan decompositions. We now define an important operator, the trace.
\begin{definition}[Trace]\label{defn:quant:Trace}
Given a non-negative self-adjoint $A \in \mathcal{B}(\Hilb)$ the trace of $A$ is given by:
\begin{align*}
\Tr(A)&=\sum_j \langle e_j, Ae_j \rangle,
\end{align*}
for an orthonormal basis $\{ e_j \}$ of $\Hilb$ (where $A$ is denoted `trace class' if $\Tr(A)$ is finite).
\end{definition}
Note that for any $A \in B(H)$, $A^\dagger$ is self-adjoint and non-negative. As a result, $\sqrt{A^\dagger A}$ is well-defined given that for such self-adjoint operators, for a bounded and measureable functional $f:\sigma(A) \to \C$ we have:
\begin{align*}
f(A)&=\int_{\sigma(A)}f(x)d\mu^A(x) \qquad A = \int_{\sigma(A)}x d\mu^A(x)
\end{align*}
where $\sigma(A)$ is a Borel $\sigma-$algebra and $\mu^A$ is a projection-valued measure on $\mu^A:\sigma(A) \to \Pi(\Hilb)$ (see section (\ref{app:Analysis and Measure Theory and Probability} for background)). The trace is independent of choice of orthonormal basis with standard properties such as cyclicity $\Tr(AB)=\Tr(BA)$ for $A,B \in \mathcal{B}(\Hilb)$. The trace operation can sometimes usefully be considered as a tensorial contraction of $A=A^i_j$ with $\delta^j_i$ i.e. $Tr(A)=A^i{}_j \delta^j_i$. For self-adjoint operators, the trace has associated eigenvalues $\lambda_j$ such that $\Tr(A)=\sum_j \lambda_j < \infty$. 

An important operation to note from a measurement and machine learning perspective is the \textit{partial trace} operation below. We can define the partial trace as follows.
\begin{definition}[Partial trace]\label{defn:quant:partialtrace}
Given a tensor product of $X_k \in \mathcal{B}(\Hilb)$, the partial trace of $X_k$ with respect to the product state is:
    \begin{align}
    (\Tr_k \otimes I)(X_{k-1} \otimes X_k \otimes X_{k+1}) = \Tr(X_k)(X_{k-1} \otimes X_{k+1}) = \Tr_{X_k}(X_{k-1} \otimes X_{k+1})
\end{align}
\end{definition}
Here $\Tr_k \otimes I$ is shorthand for the application of the trace operator to $X_k$. The partial trace has the effect of, in geometric terms (see below) contracting the relevant (tensor) product space (along the dimension associated with $X_k$) and scaling the residual tensor by the value of that trace in $\C$. In density operator formalism, the partial trace gives rise to a \textit{reduced density operator}. For $\rho_T$ representing the tensor product state $\Hilb_A \otimes \Hilb_B$, the partial tracing-out of $\Hilb_B$ in effect contracts the total tensor state up to a scalar $\trace_A$, that is:
\begin{align}
    \rho_A = \trace_B \rho_T \label{eqn:quant:reduceddensityoperator}
\end{align}

In later sections below, we note the relation of the trace to quantum measurement.

\subsection{Density operators and multi-state systems}\label{sec:quant:Density operators and multi-state systems}
Modelling multi-state and multi-qubit systems is an important feature of quantum information systems' research. It is also the primary focus of results in Chapters \ref{chapter:QDataSet and Quantum Greybox Learning} to \ref{chapter:Time optimal quantum geodesics using Cartan decompositions}. In the foregoing, typically a quantum system is described as an element of a Hilbert space $\psi \in \Hilb$ whose evolution is characterised by the operation of linear operators $\ket{\psi(t)}=U(t)\ket{\psi_0}$. However, for multi-state systems, this state formalism becomes convoluted when considering measurement and expectation outcomes of quantum systems. Hence we introduce completely positive self-adjoint operators, denoted \textit{density operators}. Such operators are akin to generalised probability distributions over expectations of observables. Thus for a distribution $\mathcal{D}$ of expectation values of observables of a measurement operator $M$, it can be shown that there exists a density operator $\rho$ where $\mathcal{D}(M) = \Tr(\rho A)$. For unit vectors in $\Hilb$, this reduces to the projection of $M$ onto $\psi$ with expectation given by $\braket{\psi,M\psi}$. First, we define a Hilbert-Schmidt operator in terms of the trace.
%=====Defn:Hilbert-Schmidt
\begin{definition}[Hilbert-Schmidt Operator]\label{defn:quant:Hilbert-Schmidt Operator}
An operator $A \in \mathcal{B}(\mathcal{H})$ is Hilbert-Schmidt if $\trace(A^\dagger A) < \infty$.
\end{definition}
All trace-class operators are Hilbert-Schmidt and the trace exhibits the usual cyclic property $\Tr(AB) =\Tr(BA)$. The inner product is given by $\braket{A,B}=\Tr(A^\dagger B)$ with the norm as per above $||A||^2 = \braket{A,A}$. With the concept of the trace we can also note a number of norms (and metrics) used further on, in particular norms (and thus metrics) for comparing quantum states (such as trace-distance related measures) and in Chapter \ref{chapter:Time optimal quantum geodesics using Cartan decompositions} for variational techniques. \textit{Schatten p-norms} are given by:
\begin{align}
    ||X||_p = \left( \Tr \left((A^\dagger A)^{\frac{1}{2}}  \right)\right)^\frac{1}{p} \label{eqn:quant:Schatten p-norms}
\end{align}
for $X \in \mathcal{B}(\Hilb)$. 
Schatten norms are important in quantum information processing in that they are preserved under unitary transformations. Schatten 2-norms (where $p=2$) are sometimes denoted the \textit{Frobenius norm} which corresponds to the Euclidean norm of an operator $X$ represented as a vector $M \in M(\C)$:
\begin{align}
    ||M||_2 =\left( \sum_{j,k}|M(j,k)|^2\right)^{1/2} \label{eqn:quant:Frobenius norm}
\end{align}
ranging over indices $j,k$. For trace distance we have as follows.
\begin{definition}[Trace distance]\label{defn:quant:Trace distance}
    A Schatten p-norm such that $p=1$ is denoted the trace distance:
    \begin{align}
        ||X||_1 = \Tr(\sqrt{X^\dagger X}).
    \end{align}
\end{definition}
This also equals the sum of singular values of $X$. The trace distance is commonly used to compare distances of quantum states via their operator representation as density operators (matrices). We now define the density operator.

%====Density matrices
\begin{definition}[Density operator]\label{defn:quant:Density operator}
A density operator is a positive semi-definite operator $\rho \in \mathcal{B}(\mathcal{H})$ that is self-adjoint and non-negative with $\Tr(\rho)=1$.
\end{definition}
Density operators tend to be the more formal representation of quantum states. The set of density operators is denoted $D(\Hilb)$ which can be shown to be a convex set. As Bengtsson et al. note \cite{bengtsson_geometry_2006} (see \S 8.7), this is crucial for the additional structure required for the set of such operators to form an algebra. For a state description given by alphabet $\Gamma$, such that $\rho_m \in D(\Hilb)$ (the density operator for the outcome described by $m \in \Sigma$ (i.e. outcomes are elements of the relevant register), which in practice depends upon measurement outcome expectations), it can be shown that:
\begin{align}
    \rho = \sum_{m\in\Gamma}p(m) \rho_m.
\label{eqn:quant:rhomixedstate}
\end{align}
Here $p$ describes the probability of observing (observations that characterise the system in state) $\rho_m$ and so in this sense represents a probability distribution over (pure) states. Equation (\ref{eqn:quant:rhomixedstate}) describes a \textit{mixed state}. Such a description is of an \textit{ensemble} of quantum states which can be construed as a function $\eta:\Gamma \to Pos(\Hilb)$ such that $\Tr(\sum_{m\in\Gamma}\eta(m))=1$. Here the trace acts as a sort of normalising measure such that $\rho_m = \eta(m)/\Tr(\eta(m))$. Density matrices characterise expectation values of observables as families of expectation values via:
\begin{align}
    \Phi_\rho(A) = \Tr(\rho A) = \Tr(A\rho) \label{eqn:quant:measurementdensitymatrixtrace}
\end{align}
where $A \in \mathcal{B}(\Hilb)$, $\Phi_\rho$ is a linear functional $\Phi_\rho:\mathcal{B}(\Hilb) \to \C$, associating operators on $\Hilb$ to scalar values in $\C$. The map $\Phi_\rho$ reflects a distribution of measurement outcomes satisfying $\Phi_\rho(I)=1$, $\Phi_\rho(A) \in \R$ for $A$ self-adjoint and non-negative if $A$ is non-negative. We denote $\rho=\ketbra{\psi}{\psi}$ with $\Tr(\ketbra{\psi}{\psi} A)=\braket{\psi,A\psi}$. Note also that relative phases $\psi_1 = e^{i\theta}\psi_2$ indicate the same states $\rho_1 = \rho_2$, reflecting the physical equivalence of states that differ only by a global phase. The trace norm above can then be used to calculate \textit{trace distance} as a measure of similarity as the Schatten 1-norm:
%=====Trace distance
\begin{definition}[Trace distance]\label{defn:quant:Tracedistance}
Given two operators $\rho,\sigma \in \mathcal{B}(\Hilb)$ representing quantum systems, the trace distance between such operators is given by:
\begin{align}
    d_T(\rho,\sigma) = \frac{1}{2}||\rho - \sigma||_1 
\end{align}
\end{definition}
Trace distance is considered a generalisation (in the quantum setting) of total variation distance between two probability distributions. The metric is a central component of the quantum machine learning objective (loss) functions used in Chapters \ref{chapter:QDataSet and Quantum Greybox Learning} and \ref{chapter:Quantum Geometric Machine Learning}. Density matrices are also convenient for distinguishing between \textit{pure} states and \textit{mixed} states which we define as follows.

\begin{definition}[Pure and Mixed States] \label{defn:quant:Pure and Mixed States}
A density operator $\rho \in \mathcal{B}(\mathcal{H})$ represents a pure state if there exists a unit vector $\psi \in \mathcal{H}$ such that $\rho$ is equal to the orthogonal projection onto $span\{\psi\}$. The density matrix $\rho$ is called a mixed state if no such unit vector $\psi$ exists.
\end{definition}
Closed-system pure states remain pure under the action generated by Hamiltonians. For pure states, $\Tr(\rho^2)=1$ while for mixed $\Tr(\rho^2) < 1$. Density matrices form a convex set such that $\lambda \rho_1 + (1-\lambda)\rho2, \lambda \in (0,1)$. Pure states are those that cannot be expressed as $\rho = \lambda \rho_1 + (1-\lambda)\rho_2$ for $\rho_1 \neq \rho_2$. Mixed and pure states can also be conceived of such that if the state of a quantum system is known exactly $\ketpsi$, i.e. where $\psi = \ketpsi\bra{\psi}$ then it is denoted as a \textit{pure state}, while where there is (epistemic) uncertainty about its state, it is a mixed state i.e. $\rho = \sum_i p_i \rho_i$ where $\trace(\rho^2)< 1$ (as all $p_i < 1$). Such properties of pure and mixed states are important tests for the effects of, for example, decoherence arising from various sources of noise (see Chapter \ref{chapter:QDataSet and Quantum Greybox Learning} for more detail).

In many applications of quantum information processing, we seek a metric to ascertain similarity between quantum states. Density matrices can be used to define a metric such as quantum relative or von Neumann entropy noted in our discussion of metrics (relevant to quantum machine learning below in subsection \ref{sec:quant:quantummetrics}):
\begin{definition}[Quantum relative (von Neumann) entropy]\label{defn:quant:quantumrelativeentropy}
    Given a density matrix $\rho \in \mathcal{B}(\Hilb)$ for a quantum state and trace operation, we define the von Neumann entropy or quantum relative entropy as:
    \begin{align*}
        S(\rho) = \Tr(-\rho \log \rho)
    \end{align*}
\end{definition}
A state $\rho$ is pure if an only if $S(\rho)=0$ and mixed otherwise. Density matrix formalism can be used to formally understand and express quantum superposition in terms of coherent (quantum) and incoherent superposition.

\begin{definition}[Coherent Superposition]
\label{defn:quant:coherent_superposition} For two quantum states $\psi_1, \psi_2 \in \mathcal{H}$ with amplitudes $c_1, c_2 \in \mathbb{C}$, the state $c_1\psi_1 + c_2\psi_2$ is called coherent if it cannot be transformed into another state with different relative phases. That is, such a state is coherent if the only case where:
\begin{align*}
    c_1e^{i\theta_1}\psi_1 + c_2e^{i\theta_2}\psi_2 = c_1\psi_1 + c_2\psi_2
\end{align*}
are physically the same state when $e^{i\theta_1} = e^{i\theta_2}$ for $\theta_1, \theta_2 \in \mathbb{R}$.
\end{definition}
Coherence is characterized by preservation of relative phases. By contrast, an \emph{incoherent superposition} of states $\psi_1, \psi_2 \in \mathcal{H}$ with probabilities $p_1, p_2 \geq 0$ such that $p_1 + p_2 = 1$ is described by the density operator $\rho = p_1 \ketbra{\psi_1}{\psi_1} + p_2 \ketbra{\psi_2}{\psi_2}$. The expectation value of an observable $A$ in this state is given by:
\begin{align*}
\Tr(\rho A) = p_1 \langle\psi_1, A\psi_1\rangle + p_2 \langle\psi_2, A\psi_2\rangle \label{eqn:quant:incoherent_superposition}
\end{align*}
which is central to measurement statistics by which states and registers are described. We now briefly describe quantum channels due to their relevance, in our context, to measurement and system evolution.

%====Quatum channels

\subsubsection{Quantum channels}\label{sec:quant:Quantum channels}
The concept of a \textit{channel} derives from information theory \cite{shannon_mathematical_1948} as a way of abstracting a medium or instrument of information transfer. The concept has since been adapted for application in quantum information theory such that a\textit{ quantum channel} is a linear map from one space of square operators to another, that satisfies the two conditions of (complete) positivity and trace preservation.

%====(Definition of a quantum channel)
\begin{definition}[Quantum Channel]\label{defn:quant:Quantum Channel}
A quantum channel  is a linear map
\begin{align*}
    \Phi \colon \mathcal{B}(\mathcal{H}_1) \to \mathcal{B}(\mathcal{H}_2)
\end{align*}
such that $\Phi$ is (a) completely positive and (b) trace-preserving. Maps satisfying these two properties are denoted CPTP maps.
\end{definition}
The set of such channels is denoted \(C(\mathcal{H}_1, \mathcal{H}_2)\) and \(C(\mathcal{H})\) for \(C(\mathcal{H}, \mathcal{H})\). Of particular importance is the concept of \textit{unitary channels}.
%====(Example of a unitary channel)
\begin{definition}[Unitary Channel]\label{defn:quant:Unitary Channel}
Given the unitary operator \(U \in \mathcal{B}(\mathcal{H})\), then the following map is a unitary channel:
\begin{align*}
    \Phi(X) = U X U^\dagger
\end{align*}
for every \(X \in \mathcal{B}(\mathcal{X})\) (e.g. $X = \rho)$ is an example of a channel, being a completely positive, trace-preserving (CPTP) map. 
\end{definition}
Such channels represent an idealised channel or memory operator, as operating on the quantum state causes no change in the state of the register \(\psi\) it acts upon (in this sense representing a fixed point operator with respect to the register state) \cite{watrous_theory_2018}. For multi-state systems, there exist product channels $C(\otimes_k\mathcal{X}_k,\otimes_k\mathcal{Y}_k)$. The trace operator can also be construed as a channel. For problems in quantum control and geometry explored below, we are mainly interested in learning the CPTP map $\Phi:U(0) \to U(T)$ satisfying some optimality condition, such as minimisation of evolution time or energy. Channels play a crucial role across quantum information processing, giving rise to considerations such as optimal representations (distinct from Lie group representations) for such channels. Importantly, the class of channels we seek to learn preserves those essential properties of quantum operators above, such as Hermiticity and trace-preservation (ensuring, for example, we remain within the closure of unitary or special unitary groups when formulating unitary sequences). Other standard results for channels, such as that for $\Hilb_1,\Hilb_2$, the set $C(\Hilb_1,\Hilb_2)$ is compact and convex are important in guaranteeing the controllability and reachability of target unitaries. We direct the reader to \cite{watrous_theory_2018} for more detailed exposition.

In the formalism of quantum information, we can also cast measurements as channels mapping to classical registers in $\R^n$. For this purpose, we introduce the notion of a quantum-classical channel.
\begin{definition}[Quantum-classical channel]\label{defn:quant:Quantum-classical channel}
    A quantum-classical channel is a CPTP map
    \begin{align*}
        \Phi \in C(\Hilb_1,\Hilb_2)
    \end{align*}
    which transforms a quantum state $\rho_1 \in \mathcal{B}(\Hilb_1)$ into a classical distribution of states represented by a diagonal density matrix $\rho_2 \in \mathcal{B}(\Hilb_2)$. 
\end{definition}
The channel is realised via a measurement comprising a dephasing channel component $\Delta$ that eliminates off-diagonal (coherent) elements i.e. it is a quantum-classical channel if $\Phi=\Delta\Phi$ where $\Delta \in C$ is a dephasing channel which has the effect of transforming off-diagonal entries in $\rho_1 \in \mathcal{B}(\Hilb_1)$ to zero while preserving diagonal entries.

Reiterating our discussion of density operators above, this equivalent to $\Phi:\rho \to \rho'$ where $\rho'$ is a diagonal density operator. As Watrous notes, these quantum-classical channels are those which can be represented as measurements of quantum registers $X$, formally described by $\rho$, with measurement outcomes $m$ stored in a register $Y$ with classical state set $\Sigma$. It can be shown that for every such channel $\Phi \in C$, there exists a unique (measurement) $\mu:\Sigma \to Pos(\Hilb)$ such that (connecting standard notation with Watrous's notation):
\begin{align}
    \Phi(X) &=\sum_{m\in\Sigma}\braket{\mu(m),X}E_{m}\\
    &=\Phi(\rho) = \sum_{m \in \Sigma} (M_m \rho M_m^\dagger) = \sum_{m \in \Sigma} p_m E_m \label{eqn:quant:measurementaschannel}
\end{align}
where $p_m = \mathrm{Tr}(M_m \rho M_m^\dagger)$ is the probability of obtaining outcome $m$, and $E_m$ represents the eigenstates corresponding to outcome $m$ projected onto a diagonal basis (i.e. diagonalised with entries corresponding to measurement outcomes equivalent to $E_{a,a}$ for $a=m$ diagonal basis operators) to represent the post-measurement state. Hence, quantum-classical channels are general measurement channels in quantum information. Other features of quantum-classical channels, such as their compactness and convexity over $\Hilb$ and the existence of product measurements (for multi-state systems), are noted in the literature (and ultimately important to both control/ reachability and the learnability. We explore their implementation in terms of machine learning techniques in later Chapters. Another concept of importance in measurement and quantum machine learning is that of partial measurements, especially the relationship between partial measurements and the partial trace operation.
\begin{definition}[Partial measurement]\label{defn:quant:Partial measurement}
    For a compound (quantum) state register $X=(Y_k), k=1,...,n$ a measurement on a single register $\mu:\Sigma \to Pos(Y_k)$ is denoted a partial measurement.
\end{definition}
Partial measurements are thus related to partial traces and, in geometric contexts (as discussed below), contraction operations over tensor fields (where the trace operation can, under certain geometric conditions relevant to quantum control, be related to tensorial contraction).

%====Quantum evolution: Hamiltonian formalism
\subsection{Quantum evolution}\label{sec:quant:Quantum evolution}
We now consider the evolution of quantum systems, beginning with the evolution axiom (adapted from the literature including \cite{hall_quantum_2013}).

\begin{axiom}[Evolution (Schr\"{o}dinger equation)]\label{axiom:quant:evolution}
Quantum state evolution is described by the following first-order differential equation, denoted the Schr\"{o}dinger equation: 
\begin{align}
    i\frac{d\psi}{dt} = H\psi \label{eqn:quant:schrodingersequation}
\end{align}
where $\psi \in \Hilb$ and  $H\in\mathcal{B}(\Hilb)$ is a fixed Hermitian operator denoted as the Hamiltonian. In the standard way we set $\hbar=1$ for convenience. 
\end{axiom}
We examine the formalism and consequences of this axiom below. In later Appendices below, we connect the evolution of quantum systems to geometric and algebraic formalism (in terms of transformations along Lie group manifolds). Before we do so, we include a short digression on the two perspectives of quantum evolution formalism.
\subsubsection{Two pictures of quantum evolution}\label{sec:quant:Two pictures of quantum evolution}
Quantum mechanics is traditionally portrayed as being describable by two somewhat different but mathematically equivalent formulations, the Schr{\"o}dinger picture, focusing on state evolution and the Heisenberg picture, focusing on operator evolution \cite{sakurai_modern_1995}.  In the Schr{\"o}dinger picture, the quantum system is represented by a wavefunction $\ket{\psi(t)}$ parametrised by time $t$ and which evolves according to equation (\ref{eqn:quant:schrodingersequation}). In this picture, the operators, representing observable, that act on $\ket{\psi(t)}$ are time-independent. Measurement probabilities of an observable (see below) are given by the inner product of $\ket{\psi(t)}$ with the observable eigenvector. By contrast, in the Heisenberg picture quantum state vectors are time-independent $\ketpsi$ and it is the operators $\hat{O}(t)$, corresponding to measureable observables, which evolve in time as:
\begin{align*}
    \hat{O}(t) = e^{iHt}\hat{O}(0)e^{-iHt}
\end{align*}
We also mention these well-known formalisms as we can connect both with a more geometric formalism for expressing quantum evolution in Chapters \ref{chapter:Quantum Geometric Machine Learning} and \ref{chapter:Time optimal quantum geodesics using Cartan decompositions}, such as the Maurer-Cartan forms and the adjoint action of Lie groups. It is also worth mentioning the hybrid \textit{interaction} (or von Neumann) picture in which both states and operators evolve, expressed (as we discuss further on) in terms of the Liouville von Neumann equation (for closed quantum systems) and the master equation (for open quantum systems \cite{wiseman_quantum_2010}):
\begin{align}
\dot{\rho} & = -i[H, \rho] + \kappa \mathcal{D}[\hat{O}]\rho \qquad \mathcal{D}[\hat{O}]\rho = \hat{O}\rho\hat{O}^\dagger  - \frac{1}{2}(\hat{O}^\dagger \hat{O}\rho + \rho \hat{O}^\dagger \hat{O}) \label{eqn:quant:masterequation}
\end{align}
where each term is parametrised by $t$ (omitted for brevity). Here $\hat{O}$ is an arbitrary operator, $\mathcal{D}$ is a superoperator term acting to transition quantum (density) operators (in this case, density operators) $\rho$ and $\kappa$ the decoherence rate.
The interaction picture is specifically relevant to our greybox open quantum systems' simulations for machine learning that is the subject of Chapter \ref{chapter:QDataSet and Quantum Greybox Learning}. We include below a number of standard definitions and results relevant to understanding the operator formalism used throughout this work. Later we connect such formalism to geometric and algebraic methods, along with formalism in quantum machine learning. 

%=====Hamiltonian formalism

\subsection{Hamiltonian formalism}\label{sec:chapter1:evolution:Hamiltonian formalism}
In density operator formalism, state evolution in Axiom (\ref{axiom:quant:evolution}) is described via:
\begin{align}
    \frac{d\rho}{dt} = -i[H,\rho] \qquad \rho(t) = e^{-iHt}\rho_0e^{iHt} \label{eqn:quant:schrodingerdensityform}
\end{align}
for $\rho_0=\rho(t=0)$. In unitary operator formalism, the above can be written:
\begin{align}
    dU U^\inv = -iHdt \label{eqn:quant:schrodingerdUUinv}
\end{align}
where $\psi(t) = U(t)\psi_0$. It is useful to unpack this formalism and define the Hamiltonian. In later Chapters, we connect the Hamiltonian to geometric, algebraic and variational principles (such as the framing of equation (\ref{eqn:quant:schrodingerdUUinv}) in terms of the Maurer-Cartan form). Hamiltonians as a formalism arise from variational techniques (see \cite{goldstein_classical_2002} for a detailed discussion) for determining equations of motion. 
\begin{definition}[Hamiltonian operator]\label{defn:quant:Hamiltonian}
    The operator $H \in \mathcal{B}(\Hilb)$ described in equation (\ref{eqn:quant:schrodingerdUUinv}) is a self-adjoint (Hermitian) denoted the Hamiltonian. The Hamiltonian operator is described as the generator of quantum evolutions. 
\end{definition}
The Hamiltonian $H(t) \in \mathcal{B}(\Hilb)$ of a system is the primary means of mathematically characterising the dynamics of quantum systems. Hamiltonians specify the data-encoding process by which information is encoded into quantum states along with how the system evolves and how the quantum computation may be controlled.  Solutions to equation (\ref{eqn:quant:schrodingerdUUinv}) are of the form:
\begin{align}
    U = \mathcal{T}_+\exp\left(-i\int_0^{\Delta t} H(t) dt\right) \label{eqn:quant:unitarysolutiontimedependentschrod}
\end{align}
where $\mathcal{T}_+$ is the time-ordering operator (described below). As can be seen, $U$ is unitary as $UU^\dagger = I$ (the adjoint action has no effect on $\mathcal{T}_+$). Being unitary, these solutions at once can be represented as linear operators on $\Hilb$ that preserve inner products (see definition \ref{defn:alg:Unitary matrix and unitary group}) and Haar measure (see definition \ref{defn:quant:Haarmeasure}). As per definition \ref{defn:quant:Unitary Channel}, these solutions can be represented as unitary channels acting on density operators via $\Phi(\rho) = U\rho U^\dagger$. Unitary evolution itself is required to preserve quantum coherence and probability measures of systems (which give rise to the enhanced computational power of quantum systems). Moreover, as we discuss in Appendix \ref{chapter:Background: Geometry, Lie Algebras and Representation Theory}, the set of unitaries forms a Lie group (see definition \ref{defn:alg:Unitary matrix and unitary group}) and so has a natural geometric interpretation in terms of a Lie group manifold $G$ whose evolution is described by a Hamiltonian composed of generators of the corresponding Lie algebra $\g$.

\subsubsection{Time-independent approximations}\label{sec:quant:Time-independent approximations}
Solving the time dependent Schr\"odinger equation given in definition \ref{eqn:quant:schrodingersequation} is often challenging or unfeasible, requiring perturbation methods or other techniques. A common approximation used in quantum information processing and quantum control is the \textit{time-independent} approximation to the Schr\"odinger equation. Recall from equation (\ref{eqn:quant:unitarysolutiontimedependentschrod}) the time-dependent solution to the close-system Schr\"odinger equation is given by:
\begin{align}
        U(T) &= \mathcal{T}_{+} e^{-i\int_0^T H(t) dt}
    \end{align}
The time-ordering reflects the fact that the generators within the time-dependent Hamiltonian do not, in general, commute at different time instants (i.e. $[H(t_i), H(t_j)]\neq 0$). In certain cases, a time-independent approximation to equation (\ref{eqn:quant:unitarysolutiontimedependentschrod}) may be adopted:
\begin{align}
    U(t) &= \mathcal{T}_{+} e^{-i\int_0^T H(t) dt}\\
    &\simeq \lim_{N\to \infty} e^{-i H(t_N) \Delta T} e^{-i H(t_{N-1}) \Delta T} \cdots  e^{-i H(t_0) \Delta T} \label{eqn:quant:timeindependentschrodunitary}
\end{align}
where $\Delta T = T/N$ and $t_j = j \Delta T$. 
The Suzuki-Trotter decomposition \cite{childs_theory_2021,suzuki_generalized_1976} and Lie-Trotter formulation:
    \begin{align}
        e^{-i(H_1+H_2)}= \lim_{n\to \infty}\left( e^{-i(H_1/n)}e^{-i(H_2/n)}\right)^n \label{eqn:quant:suzukitrotter}
    \end{align}
     under certain conditions allow the time-varying Hamiltonian to be approximated by a piece-wise constant Hamiltonian. Equation (\ref{eqn:quant:suzukitrotter}) can be seen via:
    \begin{align*}
    e^{-i(H_1+H_2)t} &= \lim_{n\to\infty} \left(1 + \frac{-i(H_1 + H_2)t}{n}\right)^n \\
    &= \lim_{n\to\infty} \left(1 + \frac{-iH_1t}{n}\right)^n\left(1 + \frac{-iH_2t}{n}\right)^n \\
    &= \lim_{n\to\infty} \left(e^{-iH_1t/n} e^{-iH_2t/n}\right)^n
\end{align*}
where we have applied equation (\ref{eqn:alg:exponentiallimit}).  In such cases, the time interval $[0,T]$ is divided into equal segments of length $\Delta t$ while the Hamiltonian is considered constant over that interval. Such results are crucial to time-independent optimal control theory where control problems are simplified by assuming that Hamiltonians may be approximated as constant over small time intervals such that the system is described by the control function $u(t)$ applied over such period. Even when only a subset of the Lie algebra $\g$ is available, we may be able to achieve universal control (that is, the ability to synthesise all generators, by repeated application of the control Hamiltonian) via commutator terms that in turn allow us to generate the full generators of the corresponding Lie algebra. This is framed geometrically in terms of the distribution $\Delta$ (a geometric framing of our set of generators) being `bracket-generating'.

The approximation is generally available under the assumption that the Hamiltonian over increments $\Delta t$ is constant as mentioned above. Of central importance to time-independent approximations it the \textit{Baker Campbell Hausdorff} formula (definition \ref{defn:alg:Baker-Campbell-Hausdorff}). For universal control in quantum systems, repeated application of the control Hamiltonian allows us to synthesize all generators of the Lie algebra corresponding to the system's dynamics. This is where the BCH formula becomes crucial, as it allows for the generation of effective Hamiltonians that include commutator terms, thus expanding the set of reachable operations. For example, the BCH formula shows us that by carefully sequencing the application of $H_1$ and $H_2$, we can effectively implement the commutator $[H_1, H_2]$ as part of the evolution:
\begin{align}
    e^{-i[H_1,H_2]\Delta t} \approx \left( e^{-iH_1\Delta t/n}e^{-iH_2\Delta t/n}e^{iH_1\Delta t/n}e^{iH_2\Delta t/n}\right)^n
\end{align}
for sufficiently large $n$ and small $\Delta t$. This approximation is powerful for designing control sequences in quantum systems with non-commuting dynamics.

\subsection{Measurement}\label{sec:quant:Measurement}
Measurement is at the centre of quantum mechanics and quantum information processing. In the section below, we set out two further axioms relating to measurement, adapted from the literature (see \cite{hall_quantum_2013,nielsen_quantum_2011}). The first concerns measurement and operators. The second concerns the characterisation of post-measurement states. Measurement formalism varies to some degree depending on how information-theoretic an approach adopted is. We incorporate elements of both the traditional and information-based formalism. We have aimed to present a useful coverage of measurement formalism specifically tailored to later Chapters, where measurement plays a critical role, for example, in both simulated quantum systems (for use in state description and other tasks) and machine learning. The first axiom defines measurement. We split this into two in terms of (a) measurement probabilities (e.g. the Born rule) and (b) the effect of measurement (wave collapse). Note here we adopt an operator formalism where measurement is represented by a set of operators $\{M \}$ the outcomes of which are measurements $m \in \Sigma$ (i.e. mapping to our register configurations) such that we can write $\{M_m | m \in \Sigma \} \subset Pos(\Hilb)$.

\begin{axiom}[Measurement]\label{axiom:quant:probabilitymeasurement}
Given $\psi \in \mathcal{H}$, the probability distribution for the measurement of an observable $m$ is such that $E[M^k] = \braket{ \psi, M^k \psi }$. Given an orthonormal basis $\{e_j\}$, we may write $\psi = \sum_j^\infty a_j e_j, a_j \in \C$ with measurement probability of observing $m$ given by $P(M=m)=|a_m|^2$ (the Born rule).
\end{axiom}
The reference to $M^k$ (following Hall \cite{hall_quantum_2013}) is to the fact that the $k$-th power of the measurement operator $M$ corresponds to the $k$-th moment of the probability distribution for measurement outcomes associated with $M$ (i.e. expectation, variance and so on). Here $M$ is a measurement operator with eigenvectors $e_m$ and eigenvalues $m$ corresponding to outcomes of measurement. 

The second measurement axiom (which can be viewed as an extension of the first) describes the effect of measurement on quantum states, namely the collapse of the wave function into eigenstates of the measurement operator. 
\begin{axiom}[Effect of Measurement]\label{axiom:quant:effectofmeasurement}
Measurement via measurement operator $M$ with measurement outcome $m \in \mathbb{R}$, causes a quantum system transition to $\psi' \in \Hilb$ such that $M\psi' = m\psi'$ where \( m \) is the eigenvalue of \( M_m \) corresponding to the observed outcome, and \( \psi' \) is the corresponding eigenstate of \( M_m \). 
\end{axiom}
This postulate is also known as the collapse of the wave-function or Copenhagen interpretation. These postulates can be expanded to encompass multiple possible measurement outcomes whereby quantum measurements are framed as sets of measurement operators $\{ M_m\}$, where $m$ indexes the outcome of a measurement (e.g. an energy level or state indicator), i.e. an observable. The probability $p(m)$ of observable $m$ upon measuring $\ketpsi$ is represented by such operators acting on the state such that measurement probability is given by:
\begin{align}
    p(m) = \braket{\psi | M_m^\dagger M_m |\psi}  = \text{tr}(M_m^\dagger M_m \rho)
\label{eqn:quant:measurementprobinnerproduct}
\end{align}
with the post-measurement state $\ket{\psi'}$ given by: 
    \begin{align}
        \ket{\psi'} &= \frac{M_m \ketpsi}{\sqrt{\braket{\psi | M_m^\dagger M_m | \psi}}}
        \label{eqn:quant:postmeasurementstate}\\
\rho'&=\frac{M_m\rho M_m^\dagger}{\braket{M_m^\dagger M_m,\rho}} \label{eqn:quant:postmeasurementstatedensity}
    \end{align}
    The set of measurement operators $\sum_m M_m^\dagger M_m = I$ reflects the probabilistic nature of measurement outcomes. Measurement is modelled as a random variable in $\Sigma$, described by the probability distribution $p \in \mathcal{P}(\Sigma)$. The act of measurement transitions $M:\rho \to \rho'$ described by equation (\ref{eqn:quant:postmeasurementstate}).  
    In quantum information theory, we further refine the concept of measurement as per below.
\begin{definition}[Measurement]\label{defn:quant:Measurementchannel}
    A measurement is defined in terms of a probability measure from the set of measurement outcomes to the set $Pos(\Hilb)$:
    \begin{align}
        \mu: \Sigma \to Pos(\Hilb) \qquad m \mapsto \mu(m) := \trace(M_m^\dagger M_m \rho) \qquad \sum_{m\in\Sigma} \mu(m)= I \label{eqn:quant:measuretoposdef}
    \end{align}
\end{definition}
Here $\Sigma$ is the set of measurement outcomes $m$ (which describe our quantum state $\rho$) following application of $M_m$ (note we include $\mu(m) := M_m$ as a bridge between the commonplace formalism of Nielsen et al. \cite{nielsen_quantum_2011} and the slightly more information theoretic form in Watrous \cite{watrous_theory_2018}. In this notation (from \cite{watrous_theory_2018}), $p(m) = \braket{\mu(m),\rho}$.
When a measurement is performed on a system described by the density operator $\rho$, the probability of obtaining outcome $m$ is given by: 
\begin{align}
p(m) = \text{Tr}(M_m \rho M_m^\dagger)
\end{align}
and the state of the system after the measurement (post-measurement state) is 
\begin{align}
\rho_m = \frac{M_m \rho M_m^\dagger}{p(m)}.
\end{align}
This measurement process can be described by a quantum channel $\mathcal{E}$ that maps the initial density operator $\rho$ to a final density operator $\rho'$ that is a mixture of the post-measurement states, weighted by their respective probabilities:
\begin{align}
\rho' = \mathcal{E}(\rho) = \sum_m p(m) \rho_m = \sum_m M_m \rho M_m^\dagger \label{eqn:quant:measurementsuperoperatordensity}
\end{align}
    Connecting the standard terminology with the more information terminology for measurement, we note that $\{M_m | m \in \Sigma \} \subset \mathcal{B}(\Hilb)$ such that $\sum_{m}M^\dagger_m M_m = I$. Measurement then corresponds $m \in \Sigma$ being selected at random (i.e. modelled as a random variable) with probability given by equation (\ref{eqn:quant:measurementprobinnerproduct}) and post-measurement state given by equation (\ref{eqn:quant:postmeasurementstatedensity}). Such measurements are described as \textit{non-destructive measurements}. 

    \subsubsection{POVMs and Kraus Operators} \label{sec:quant:POVMs and Kraus Operators}
    In more advanced treatments, positive-operator valued measure (POVM) formalism more fully describes the measurement statistics and post-measurement state of the system. For a POVM, we define a set of positive operators $\{  E_m \}=\{M^\dagger_m M_m\}$ satisfying $\sum_m E_m=\mathbb{I}$ in a way that gives us a complete set of positive operators (such formalism being more general than simply relying on projection operators). In this way the measurement operators $M$ can be regarded as \textit{Kraus operators} satisfying the requisite completeness relation \cite{kraus_states_1983,hellwig_pure_1969}. 
    As we are interested in probability distributions rather than individual probabilities from a single measurement, we calculate the probability distribution over outcomes via Born rule using the trace:
    \begin{align}
        p(E_i) = \Tr(\rho E_i)  \qquad \braket{A} = \trace(\rho A) \label{eqn:quant:measurementtracePOVM}
    \end{align} noting the expectation as per equation (\ref{eqn:quant:measurementdensitymatrixtrace}).
    Measurement thus has a representation as a quantum to classical channel $\Phi_M \in C(\X,\Y)$ (see definition \ref{defn:quant:Quantum-classical channel}) under the condition that it is completely dephasing with $\Phi_M(\rho)$ is diagonal for all $\rho \in D(\X)$. Such quantum-to-classical channels are those that can be realised as a measurement of a register $X$. As Watrous ($\S$2.3) notes, there is a correspondence between the mapping of measurement outcomes $m$ to operators $M \in Pos(\X)$. Combining Watrous and Nielsen et al.'s formalism, $\Phi(X)$ describes a classical description (distribution) of the outcome probabilities when a quantum state $\rho$ is measured:
\begin{align}
    \Phi_M(\rho) = \sum_{m \in \Sigma} \trace(M_m \rho M_m^\dagger)\ketbra{m}{m} = \sum_{m \in \Sigma} \braket{\mu(m),\rho}E_m \label{eqn:quant:measurementchannel1}
\end{align}
    \\
    \\
    Further discussion of measurement procedures is set out below (such as in Chapter \ref{chapter:QDataSet and Quantum Greybox Learning}). Note that measurements described above give rise to a set of measurement statistics and that these are used, e.g. via process or state tomography, to reconstruct or infer the state $\ket{\psi}$ or channel $U$ i.e. such that the outcome of measurements is a set of measurements $\{ m \}$ from which measurement statistics using operator $M$ are calculated. Such statistics are then used to infer or reconstruct data describing quantum states (or in operator formalism the channels or operators $U(t)$ themselves). In practice, for sequences of unitaries $(U_n(t))$ or quantum states $\ket{\psi}$, because of state collapse (\ref{axiom:quant:effectofmeasurement}), we assume that we cannot measure each unitary at each time point $t$ (albeit see discussion on post-selection measurement based quantum computing).

\subsubsection{Composite system measurement}\label{sec:quant:Composite system measurement}
    For composite systems, a measurement of the $k$-th state $\rho_k$ (or $X_k$ depending on formalism) will result in a post-measurement state conditional upon that measurement. The state $\rho_k$ will `collapse' onto one of the eigenvectors of the measurement operator $M_m$ corresponding to the observable $m$ obtained. In composite systems, because of the distributivity of scalar multiplication, the post-measurement state is then a tensor product of the original states with that eigenvector state scaled by the measurement outcome (then renormalised). This can be expressed as:
\begin{align}
    \eta&:\Sigma \to Pos(\otimes_k X_k)\\
    \eta(m)&=\Tr_{X_k}(I_{k-1} \otimes \mu(m) \otimes I_{k+1})(\rho)
\end{align}
where $X_k \in \Hilb_k$ (note sometimes the trace above is denoted $\trace_{\Hilb_k}$ to indicate tracing out of the $k$-th subsystem). The measurement $\mu(m)$ essentially maps $X_k$ (which, recall, could be a superposition state) to a classical state scaled by $m$ (and in this sense we can think of measurement as a tensorial contraction as discussed in later Chapters). In this formalism, the post-measurement state in equation (\ref{eqn:quant:postmeasurementstate}) becomes where the denominator is the normalisation scalar:
\begin{align}
    \rho'&=\frac{\eta(m)}{\Tr(\eta(m)} = \frac{\Tr_{X_k}(I_{k-1} \otimes \mu_k(m) \otimes I_{k+1})(\rho)}{\braket{\mu(m),\rho_k}}
\end{align}
We are also concerned in particular with projective measurements where each measurement is a projection operator i.e. $\mu(m) \in Proj(\Hilb)$. Each projective measurement projects the state $\rho$ into an eigenstate of the respective projection operator. The set of such operators $\{ \mu(m) | m \in \Sigma \}$ is an orthogonal set, which means there are at most $\dim \Hilb$ distinct observables $m$. It can be shown that any measurement can be framed as a projective measurement.

%=====POVM
\subsubsection{Informational completeness and POVMs}\label{sec:quant:Informational completeness and POVMs}
Positive-operator valued measure (POVM) formalism more fully describes the measurement statistics and post-measurement state of the system. For a POVM, we define a set of positive semi-definite operators $\{  E_m \}=\{M^\dagger_m M_m\}$ satisfying $\sum_m E_m=\mathbb{I}$ in a way that gives us a complete set of positive operators (such formalism being more general than simply relying on projection operators). 
%====(Definition of Positive Operator-Valued Measures (POVMs))
\begin{definition}[POVM]\label{defn:quant:POVM}
A \textit{Positive Operator-Valued Measure (POVM)} on a Hilbert space $\mathcal{H}$ is a set $\{E_m\}$ of operators that satisfies the following conditions:
\begin{align}
    &E_m \geq 0 \quad \forall m \qquad \sum_m E_m = \mathbb{I}_{\mathcal{H}}.
\end{align}
\end{definition}
Here, $E_m$ represents the effect corresponding to the measurement outcome $m$, and $\mathbb{I}_{\mathcal{H}}$ is the identity operator on $\mathcal{H}$. The operators $E_m$ are positive semi-definite and the set $\{E_m\}$ is complete in the sense that it sums to the identity, ensuring that probabilities over all outcomes sum to one.

    To estimate probability distributions, we similarly calculate the probability distribution over outcomes via Born rule using the trace $p(E_i) = \Tr(\rho E_i)$. Further discussion of measurement procedures is set out below (such as in Chapter \ref{chapter:QDataSet and Quantum Greybox Learning}). 

We are also interested, in a quantum machine learning context, to utilise informationally-complete measurements, where the set of measurements span the entire space $\mathcal{B}(\Hilb)$. This is because quantum states are uniquely determined by their measurement statistics. For problems such as state identification (tomography) and other typical tasks, understanding the underlying probability distribution for each measurement outcome enables a complete description of that state (or in Watrous's terminology, a full description of the register). 
\begin{definition}[Informationally-complete measurements]\label{defn:quant:Informationally-complete measurements}
    A measure $\mu:\Sigma \to Pos(\Hilb)$, is informationally complete if $\spn \{ \mu(m) | m \in \Sigma \} = \mathcal{B}(\Hilb)$.
\end{definition}
For diagonal states $\rho \in D(\Hilb)$, the probability vector $p \in \mathcal{P}(\Sigma)$ completely specifies $p$, which intuitively shows how density operators can be thought of as operator-analogues of probability distributions over classical states (registers). See \cite{watrous_theory_2018} for more generalised descriptions in terms of instruments.  

Finally we briefly note extremal measurements, which is a quantum-classical channel corresponding to an extreme point of all quantum-classical channels $C(\Hilb)$. A measurement $\mu$ is extremal if for all measurement choices $\mu_j,\mu_k$ with $\mu = \lambda \mu_j + (1-\lambda)\mu_k$ for $\lambda \in (0,1)$ we have $\mu_j = \mu_k$, Extremal measurements are those that cannot be expressed as non-trivial convex combinations of other POVM elements. We don't expand upon these types of measurements but note that extremal POVMs are often associated with obtaining maximum information about a quantum system. They are optimal in the sense that no other measurement can provide strictly more information about the quantum state with the fewest measurement outcomes (hence potentially providing better training data sets from a statistical learning perspective for tasks such as state discrimination). Extremal measurements are relevant to optimal design of experiments in order to assist statistical learning processes by offering data that captures the most distinct features of the quantum state's behaviour.

\subsubsection{Expectation evolution}\label{sec:quant:Expectation evolution}
The evolution of quantum states can also be characterised in terms of changes in measurement statistics over time. That is, equation (\ref{eqn:quant:schrodingerdUUinv}) can also be used to model the evolution of the measurement statistics \cite{hall_quantum_2013}. Recall that if $A \in \mathcal{B}(\Hilb)$ is Hermitian then the expectation value of $A$ for state $\psi \in \Hilb$ is:
\begin{align}
    \braket{A}_\psi = \braket{\psi,A\psi} = \trace(\rho A)\label{eqn:quant:averagevalue}
\end{align}
Equation (\ref{eqn:quant:averagevalue}) above expresses the same principle of equation (\ref{defn:quant:Expectation}) in operator formalism with the inner product. The uncertainty (variance) in measurements of $A$ is given by:
\begin{align}
    (\Delta A)^2 = \braket{A^2}_\psi - (\braket{A}_\psi)^2
\end{align}
The expectation value together with the uncertainty of measurements of $A$ are important measurement statistics of $A$ with respect to state $\psi$. We are also often interested in how measurement statistics evolve over time which can be modelled as:
\begin{align}
    \frac{d}{dt}\braket{A}_\psi = \braket{-i[A,H]}
\end{align}
where $[\cdot,\cdot]$ denotes the commutator $[A,B]=AB-BA$. The commutator (in the form of the Lie derivative) is more fully described in Appendix \ref{chapter:Background: Geometry, Lie Algebras and Representation Theory} and proposition (\ref{prop:alg:Lie bracket properties}).

\subsection{Quantum entanglement}\label{sec:quant:Quantum entanglement}
A fundamental (or in Schr\"odinger's estimation, \textit{the} fundamental) distinguishing property of quantum mechanics from its classical counterpart is the phenomenon of entanglement. Entanglement has a few different but equivalent definitions, such as being a state $\rho$ which cannot be represented as a tensor product state (definition \ref{defn:quant:Tensor Product}). More formally, entanglement can be defined as per below.
\begin{definition}[Entanglement]
Given Hilbert spaces $\mathcal{H}_A$ and $\mathcal{H}_B$ and composite system $\mathcal{H} = \mathcal{H}_A \otimes \mathcal{H}_B$. A pure state $\ket{\psi} \in \mathcal{H}$ is \textit{entangled} if it cannot be written as a product of states from each subsystem, i.e., there do not exist $\ket{\phi_A} \in \mathcal{H}_A$ and $\ket{\phi_B} \in \mathcal{H}_B$ such that:
\begin{align}
\ket{\psi} = \ket{\phi_A} \otimes \ket{\phi_B}.
\end{align}
A mixed state $\rho \in \mathcal{B}(\mathcal{H})$ is \textit{entangled} if it cannot be expressed as a convex combination of product states:
\begin{align}
\rho \neq \sum_i p_i \rho_i^A \otimes \rho_i^B,
\end{align}
where $\{p_i\}$ are probabilities, $\rho_i^A \in \mathcal{B}(\mathcal{H}_A)$, and $\rho_i^B \in \mathcal{B}(\mathcal{H}_B)$ are density matrices for the subsystems $A$ and $B$, respectively.
\end{definition}
In particular, for EPR pairs (Bell states) this means that certain states such as:
\begin{align}
    \ket{\psi} = \frac{1}{\sqrt{2}}(\ket{00} + \ket{11})
\end{align}
cannot be represented as a tensor product state. We mention entanglement only tangentially in this work but it is of course of central importance across quantum information processing.

\subsection{Quantum metrics} \label{sec:quant:quantummetrics}
Metrics play a central technical role in classical machine learning, fundamentally being the basis upon which machine learning algorithms update, via techniques such as backpropagation. Indeed metrics at their heart concern the ability to quantifiably distinguish objects in some way and to this extent have been integral to the very concept of \textit{information} as portended by Hartley in which information refers to quantifiable ability of a receiver to distinguish symbolic sequences \cite{hartley_transmission_1928}. Hartley's original measure of information prefigured advanced approaches to quantifying and describing information, such as Kolmogorov and others \cite{cover_kolmogorovs_1989}. 

Metrics for quantum information processing are related but distinct from their classical counterparts and understanding these differences is important for researchers applying classical machine learning algorithms to solve problems involving quantum data. As is commonplace within machine learning, chosen metrics will differ depending on the objectives, optimisation strategies and datasets. For a classical bit string, there are a variety of classical information distance metrics used in general \cite{nielsen_quantum_2010}. In more theoretical and advanced treatments, available metrics will depend upon the underlying structure of the problem (e.g. topology) (see \cite{watrous_theory_2018} for a comprehensive discussion). Metrics used will depend also upon whether quantum states or operators are used as the comparators, though one can relatively easily translate between operator and state metrics. We outline a number of commonly used quantum metrics below and discuss their implementation in classical contexts, such as in loss functions. Note below we take license with the term metric as certain measures below, such as quantum relative entropy, do not (as with their classical counterparts) strictly constitute metrics as such.

Being able to quantify the difference or similarity of quantum sates is fundamental to the application of any machine learning protocols in quantum information processing. This problem of state discrimination, of how accurately states can be distinguished using measurement, is examined briefly below. In particular, we focus on the key metric in later Chapters of \textit{fidelity} among target unitary operators and those generated via our algorithmic methods. We summarise briefly a few key results in the area, including the Holevo-Helstrom theorem (for providing a weighted difference between states). In machine learning, state discrimination is also central to classification methods. Indeed difference measures and metrics are central to underlying objective (loss) functions underpinning the learnability of classical and quantum machine learning algorithms. 

 \subsubsection{State discrimination}\label{sec:quant:State discrimination}
 State discrimination for both quantum and probabilistic classical states requires incorporation of stochasticity (the probabilities) together with a similarity measure. The Holevo-Helstrom theorem quantifies the probability of distinguishing between two quantum states given a single measurement $\mu$. State discrimination here is a binary classification problem. 
 \begin{definition}[Holevo-Helstrom theorem]\label{thm:quant:Holevo-Helstrom theorem}
     Given diagonal $\rho_k \in D(\Hilb)$ with $\lambda\in[0,1]$, for each choice of measurement $\mu:\{0,1 \}\to Pos(\Hilb)$, the probability of correctly ($\lambda$) or incorrectly $(1-\lambda)$ distinguishing a state is given by:
     \begin{align}
         \lambda \braket{\mu(0),\rho_0} + (1-\lambda)\braket{\mu(1),\rho_1} \leq \frac{1}{2} + \frac{1}{2}||\lambda \rho_0 + (1-\lambda)\rho_1||_1
     \end{align}
 \end{definition}
 Where, for a suitably chosen projective measurement $\mu$, the relation is one of equality. State discrimination is central to results in this work, especially in Chapters 5 and 6 where discriminating states (measuring their similarity) is the key to training quantum machine learning models. A detailed exposition of state discrimination from an information theoretic perspective can be found \cite{watrous_theory_2018}. For our primary results relating to state (and by extension operator) discrimination in later Chapters, we focus on the fidelity metric as the basis for distinguishing quantum states and operators (central to, for example, our objective functions in machine learning applications we use). We focus on fidelity below. Briefly we also mention an important feature of quantum information (especially as its relates to machine learning applications) in the form of the \textit{no cloning theorem} which provides that quantum information cannot be identically copied. That is, no procedure exists for identically replicating any arbitrary quantum state $\ket{\psi}$ on a Hilbert space $\mathcal{H}$.

 \begin{theorem}[No-Cloning Theorem]\label{thm:quant:No-Cloning Theorem}
 The no cloning theorem provides that here does not exist a completely positive, trace-preserving (CPTP) map $\Phi: \mathcal{B}(\mathcal{H}) \to \mathcal{B}(\mathcal{H} \otimes \mathcal{H})$ such that for any pure state $\ket{\psi} \in \mathcal{H}$:
\begin{align}
\Phi(\ketbra{\psi}{\psi}) = \ketbra{\psi}{\psi} \otimes \ketbra{\psi}{\psi}.
\end{align}
\end{theorem}
The no cloning theorem illustrates the fundamentally non-classical nature of quantum information. Note, however, that it does not prevent the preparation of identical quantum states (essential to machine learning protocols including those explored in Appendix \ref{chapter:Background: Classical, Quantum and Geometric Machine Learning}) i.e. identical `copies' of quantum states $\rho$ can be produced via identical preparation procedures. Rather, the no cloning theorem is a claim about operations on a state that would result in an identical copy being produced.

\subsubsection{Fidelity function}\label{sec:quant:Fidelity function}
 Of central importance to later results in Chapters \ref{chapter:QDataSet and Quantum Greybox Learning} and \ref{chapter:Quantum Geometric Machine Learning} is the fidelity function which quantifies the degree of similarity or overlap between two quantum states (or their operator representations in $Pos(\Hilb)$). Fidelity is a key metric in quantum information processing, especially in quantum control and quantum unitary synthesis problems. It is central to the quantum machine learning loss function architecture adopted in this work. As such, we provide a more in-depth assessment of the metric with regard to quantum state comparison.
\begin{definition}[Fidelity]\label{defn:quant:Fidelity}
    The fidelity between two operators $\rho,\sigma \in Pos(\Hilb)$ is given by:
    \begin{align}
        F(\rho,\sigma) = \big|\big|\sqrt{\rho}\sqrt{\sigma}  \big|\big|_1 =  \Tr\left(\sqrt{\sqrt{\sigma}\rho\sqrt{\sigma}}\right) \label{eqn:quant:fidelityfunction}
    \end{align}
\end{definition}
The fidelity function exhibits a number of properties including: (a) continuity - it is continuous at $(\rho,\sigma)$ (important for statistical learning performance); (b) symmetric $F(\rho,\sigma) = F(\sigma,\rho)$; (c) positive semi-definiteness $F(\rho,\sigma) \geq 0$ else $\rho \sigma=0$; (d) $F(\rho,\sigma) \leq \Tr(\rho)\Tr(\sigma)$ (equality if not independent); (e) preservation under conjugation by isometries (unitary channels) $V\in U(\Hilb,\Hilb)$ such that $F(\rho,\sigma)=F(V \rho V^*,V \sigma V^*)$; (f) $F(\lambda \rho,\sigma)=\sqrt{\lambda}F(\rho,\sigma)=F(\rho,\lambda \sigma)$. In terms of density operators $\rho,\sigma \in Pos(\Hilb)$, we have that  $F(\rho,\sigma) \in [0,1]$ with $F=0$ if and only if $\rho \perp \sigma$ and $F=1$ if and only if $\rho = \sigma$. The fidelity function can be shown to be jointly concave in its arguments and monotonic under the action of channels. The latter of these properties provides that for $\Phi \in C(\Hilb)$ with $\rho,\sigma \in Pos(\Hilb)$:
\begin{align}
    F(\rho,\sigma) \leq F(\Phi(\rho),\Phi(\sigma))
\label{eqn:chapter1:fidelitymonotonicity}
\end{align}
Finally, the relationship of fidelity $F$ and trace function $||\cdot ||_1$ for state $\rho,\sigma \in D(\Hilb)$ is given by Fuchs-van de Graaf inequalities, which can be expressed as:
\begin{align}
    2-2F(\rho,\sigma) \leq ||\rho - \sigma||_1 \leq 2\sqrt{1-F(\rho,\sigma)^2} \label{eqn:quant:fidelityfuchs}
\end{align}
Fidelity and trace distance are related via $D(\rho,\sigma) = \sqrt{1 - F(\rho,\sigma)^2}$. Fidelity can also be interpreted as a metric by calculating the angle $\zeta=\arccos F(\rho,\sigma)$.
Note in some texts that equation (\ref{eqn:quant:fidelityfunction}) is sometimes denoted \textit{root fidelity} while fidelity is its square (see \cite{bengtsson_geometry_2006}) and that fidelity can be related to transition probability between mixed states \cite{uhlmann_transition_1976}.  

Other metrics common in quantum machine learning literature include:
\begin{enumerate}
    \item \textit{Hamming distance}, the number of places at which two bit strings are unequal. Hamming distance is important in error-correcting contexts and quantum communication \cite{doriguello_quantum_2019}.
    \item \textit{Trace distance} or $L1$-\textit{Kolmogorov distance} described in definition \ref{defn:quant:Tracedistance}. Trace distance is a metric preserved under unitary transformations. It is therefore a widely used similarity metric in quantum information.
    \item \textit{Quantum relative entropy} (see definition \ref{defn:quant:quantumrelativeentropy}) is the quantum analogue of Shannon entropy \cite{bengtsson_geometry_2006}. It is found given by $S(\rho) = -\trace(\rho \log \rho)$ The quantum analogue of (binary) cross-entropy is in turn given by:
    \begin{align}
        S(\rho||\sigma) = \trace(\rho \log \rho) - \trace(\rho \log \sigma)
    \end{align} These measures provide a further basis for comparing for the output of algorithms to labelled data during training. 
\end{enumerate}
We discuss the use and relationship of such metrics to quantum machine learning algorithms deployed in further chapters in more detail in Appendix \ref{chapter:Background: Classical, Quantum and Geometric Machine Learning}.

%======QUANTUM CONTROL
\section{Quantum Control}\label{sec:quant:Quantum Control}
\subsection{Overview}
Quantum information tasks can be modelled in terms of a two-step experiment (a) a quantum state and measurement instrument preparation procedure (isolating the quantum system in a particular state); (ii) a measurement step where the instrument interacts with the quantum system to yield measurement statistics. Obtaining such statistics (by which the state preparation can potentially be confirmed) occurs via multiple repetitions of such experiments. While quantum systems are represented distinctly from classical systems, their state preparation, control and measurement is classically parametrised (e.g. by parameters in $\R$ or $\C$). Information can be distinguished between classically described and evolving (classical information) as distinct from being described and evolving according to quantum formalism (quantum information). In certain cases, such as with certain classes of quantum machine learning or variational algorithms \cite{verdon_quantum_2019-2,verdon_quantum_2019-3,verdon_quantum_2019-1,schuld_introduction_2015}, parameter registers may themselves be represented (or stored within) quantum registers. However, as noted above, quantum registers themselves concern distributions and evolution] over classical registers (albeit with certain non-classical features). Thus ultimately quantum information processing, and tasks involving quantum systems, involve constructions using classical information. 

As Wiseman et al. note \cite{wiseman_quantum_2010} the preparation-evolve-measure procedure of quantum parameter estimation can be framed as one where a quantum system mediates (and transforms) the classical parameters of a state preparation procedure to the classical measurement statistics. Construed as a transformation mediating the transition from classical (input) to classical (measurement/output) information, quantum systems can be seen as establishing constraints upon the accuracy of the estimation of such parameters (albeit ontological limits given the fundamental grounding of physics in quantum mechanics). For example, modelling the synthesis of a particular computation or state can be modelled as the task of quantum parameter estimation where parameters $\theta \in \C$ (such as control pulses, see below) are estimated from measurement outcomes of a state whose evolution is modelled as:
\begin{align}
    \rho_0 \to \rho_T = e^{-i Ht}\rho_0e^{i Ht} = U\rho_0 U^\dagger
\end{align}
Here the Hamiltonian is parameterised by $\theta$ as $H = \sum_\k \theta_k G_k$ where $G_k$ are generators (see below). Assuming $\rho_0$ is (sufficiently) known along with $G_k$, then ascertaining the set of controls $\theta_k$ to (optimally - say minimising time or energy) synthesise $\rho_T$ is contingent on estimates $\hat\theta$ via minimising a cost function $|\theta_k - \hat\theta_k|$. The estimates $\hat\theta$ are themselves usually constructed via comparing the measurement statistics of the estimate $\hat\rho_T$ against $\rho_T$. This is also the case when seeking to estimate, for example, state preparation parameters from measurement statistics. Thus constraints upon measurement statistics act as constraints or bounds upon how accurately quantum parameter estimation (and tasks dependent on it, such as unitary synthesis) may be performed. In this work, we primarily base our optimisations (in later Chapters) upon the fidelity measure (discussed below), assuming the existence of an (optimal) measurement protocol by which to construct our estimates of, for example, target unitaries $\hat U_T$. Below we set out a few of the key assumptions and features of quantum control formalism used in later Chapters. We revisit this formalism in Appendices below in terms of Lie theoretic and differential geometric models of quantum control.

\subsection{Evolution, Hamiltonians and control}\label{sec:quant:Evolution, Hamiltonians and control}
Our work and results in later Chapters are focused upon typical quantum control problems \cite{sachkov_control_2009, dalessandro_introduction_2007,wiseman_quantum_2010}, where it is assumed there exist a set of controls (such as pulses or voltages) with which the quantum system may be controlled or steered towards a target state (or unitary). Recall that in density matrix formalism, the Schr{\"o}dinger equation in terms of a density operator and Hamiltonian is:
    \begin{align}
        i \frac{d \rho}{dt} = [H(t), \rho(t)] \label{eqn:quant:schroddensitycontrol1}
    \end{align}
In physical systems, it corresponds to the total energy (sum of kinetic and potential energies) of the system under consideration.
In control theory, we can separate out the Hamiltonian in equation (\ref{eqn:quant:schroddensitycontrol1}) into drift and control forms.
    For a closed system (i.e. a noiseless isolated system with no interaction with the surrounding environment), it can be expressed in the general form:
    \begin{align}
        H(t) = H_0(t) \doteq H_d(t) + H_{\text{ctrl}}(t) \label{eqn:quant:hamiltoniandriftcontrol1}
    \end{align}
    $H_d(t)$ is called the drift Hamiltonian and corresponds to the natural evolution of the system in the absence of any control. The second term $H_{\text{ctrl}}(t)$ is called the control Hamiltonian and corresponds to the controlled external forces we apply to the system (such as electromagnetic pulses applied to an atom, or a magnetic field applied to an electron). This is also sometimes described as an interaction Hamiltonian (representing how the system interacts with classical controls e.g. pulses or other signals). This allows us to define a control-theoretic form of the Schr\"odinger equation.
\begin{definition}[Control (Schr\"odinger) equation]\label{defn:quant:controlequation}
    Given quantum state $\rho \in \mathcal{B}(\Hilb)$ with Hamiltonian given by $H_0(t) =  H_d(t) + H_{\text{ctrl}}(t)$ define the control form of equation (\ref{eqn:quant:schroddensitycontrol1}) as:
    \begin{align}
        i \frac{d \rho}{dt} = [ H_d(t) + H_{\text{ctrl}}(t), \rho(t)] \label{eqn:quant:schroddensitycontrol2}
    \end{align}
\end{definition}
The solution of the evolution equation at time $t=T$ is given by:
\begin{align}
    \rho(T) = U(T) \rho(0) U^{\dagger}(T),
\end{align}
where $\rho(0)$ is the initial state of the system, the unitary evolution matrix $U(t)$ is given by equation (\ref{eqn:quant:schrodingersequation}). We discuss control and its application in some detail in later Chapters, specifically relating to geometric quantum control as expressed by formal (Pontryagin) control theory where targets are unitaries $U_T \in G$ for unitary groups $G$. As D'Alessandro \cite{dalessandro_introduction_2007} notes, the typical quantum control methodology covers: (a) obtaining the Hamiltonian of the system in an appropriate form, (b) identifying the internal and interaction components of the Hamiltonian (and often calculating the energy eigenstates for a time-independent approximation) and (c) specifying a finite dimensional and bounded control system for control coefficients. \\
\\
\subsection{Control systems and strategies}\label{sec:quant:controlsystem}
We sketch some of the discussion of control theory in later Chapters here. D'Alessandro \cite{dalessandro_introduction_2007} notes that a general control system has the following form (which as we discuss is due to Pontryagin):
\begin{align}
    \dot x = f(t,x,u)    \label{eqn:quant:controlsystem1}
\end{align}
where $x$ represents the system state, $f$ is a vector field while $u=u(t)$ are real-valued time varying (or constant over small $\Delta t$ interval) control functions. The state is inaccessible in general, only via some other function or channel e.g. $g(x)$, say a measurement operation. In quantum settings, equation (\ref{eqn:quant:controlsystem1}) is described by the Schr\"odinger equation. In unitary form this is equivalent to:
\begin{align}
    \dot x \sim  \dot U \qquad f(x,t,u) \sim -iH(u(t)) U \label{eqn:quant:control-state-eqn}
\end{align}
where $H(u(t))$ is a Hamiltonian comprising controls and generators (for us, drawn from a Lie algebra $\g$). The drift Hamiltonian $H_d$ and control Hamiltonian $H_c$ combine as:
\begin{align}
    H(u) = H_d + \sum_k H_k u_k
\end{align}
where $f$ in equation (\ref{eqn:quant:control-state-eqn}) is thought of as a linear function of $x$ and an affine function of the controls $u$. For ensembles, the evolution can be described via control Hamiltonians acting upon the set of density matrices as well. Overall the generic control solution is then:
\begin{align}
    \rho(t) = U(t) \rho(0) U^\dagger(t) \qquad \dot U = -iH(u) U \qquad U(0) = \mathbb{I} \label{eqn:quant:control-math-form}
\end{align}
with solutions given by:
\begin{align}
    \dot U =-i \left(H_d + \sum_k H_k u_k \right)U \label{eqn:quant:control-unitary form}
\end{align}
Such solutions are drawn from Jurdjevic \cite{jurdjevic_geometric_1997} from geometric control theory literature. These are a major focus of our final chapter, where we show equivalent results can be obtained for certain symmetric space quantum control problems by using a global Cartan decomposition together with certain variational techniques. The form of equation (\ref{eqn:quant:control-math-form}) can be easily construed in an information-theoretic fashion, such as when logic gates (e.g. upon qubits) are sought to be engineered. 

As noted in \cite{dalessandro_introduction_2007}, an important distinction between classical and quantum control is the unavailability of classical feedback control in the quantum case as a result of the measurement axiom \ref{axiom:quant:effectofmeasurement}, namely the collapse of $\rho \to \rho'$: the modification of the state $\rho$ by measurement means that evolution must be considered as starting from a different initial condition $\rho(0)'$. Instead, we focus on \textit{open loop control schemes} which rely upon an \textit{a priori} control system. In this case, control inputs are applied without observing the current state $\rho(t)$ i.e. meaning no adjustment during the control strategy (the sequence of controls and their characteristics), such as to the sequence or amplitude of control pulses. This is distinct from \textit{feedback control} where the control strategy is dynamically adaptable based on the state (e.g. as say in reinforcement learning). 

We also note other conditions regarding reachability of targets $U_T \in G$. In particular, we rely upon the \textit{quantum recurrence theorem} \cite{liu_recurrence_2024}
\begin{definition}[Quantum Recurrence Theorem] \label{defn:quant:Quantum Recurrence Theorem}
For any $U \in G$ where $G$ represents a unitary Lie group acting on $\Hilb$ represented by $U(0)$, there exists a $\epsilon,T > 0$ such that:
\begin{align*}
\left| U(0) - U(T)U(0)(T)^\dagger \right| < \epsilon,
\end{align*}
where $|\cdot|$ denotes the operator norm. This theorem is denoted the quantum recurrence theorem.
\end{definition}
The action of the group element $U \in G$ on $\Hilb$ is the quantum analogue of the classical flow in dynamical systems. Indeed the quantum recurrence theorem is the quantum analogue of the Poincar\'e classical recurrence theorem which states that any Hamiltonian system defined on a finite phase space will eventually evolve to within arbitrarily small proximity to their initial state. The quantum recurrence theorem guarantees that quantum evolution on the action of $G$ (via conjugation) will evolve the quantum state $U$ arbitrarily close to $U_0$ after some finite time $T$, reflecting the quasi-periodic nature of the evolution in a finite-dimensional Hilbert space and follows from the spectral properties of the unitary group on such a space. Before moving to our next Chapter on Lie groups and associated algebraic concepts, we mention a few concepts from open quantum systems relevant to later Chapters.

\section{Open quantum systems}\label{sec:quant:Open quantum systems}
\subsection{Overview}
We conclude this background Appendix with a brief synopsis of open quantum systems and quantum control. This is of particular relevance to Chapter \ref{chapter:QDataSet and Quantum Greybox Learning}, where our QDataSet models the influence of noise upon one- and two-qubit systems. Moreover, it is of seminal importance in any quantum information processing task. As Wiseman and Howard note, projective measurements (see above) are inadequate for describing real measurements because, among other reasons, such measurement never measure directly the system of interest in a closed state \cite{wiseman_quantum_2010}. Rather, the system always interacts with an environment. Even if this logic is extended back to the interaction of the observer themselves, some truncation of the phenomena into system (measured via applying the measurement postulates above for projective measurement) and environment must be made (so-called Heisenberg cuts). While the focus of this work is on geometric and machine learning techniques, we briefly summarise a few relevant and key aspects of open quantum systems modelling. This is again of relevance to Chapter \ref{chapter:QDataSet and Quantum Greybox Learning} in relation to how our quantum simulations explicitly factor in a variety of noise sources. Most of our results in Chapters \ref{chapter:QDataSet and Quantum Greybox Learning} and \ref{chapter:Quantum Geometric Machine Learning} assume a simplified control problem, one of a closed quantum system, while Chapter \ref{chapter:QDataSet and Quantum Greybox Learning} demonstrates (through the $V_O$ operator) how machine learning can be used to learn those features of an open or noisy system in a way that can allow for mitigation or noise cancellation to a degree.  
    
Open quantum systems construct an overall quantum system comprising the \textit{system} (or closed quantum evolution) and the \textit{environment} (i.e. interactions with the environment). A simple Hamiltonian model for such open systems:
    \begin{align}
        H(t) = H_0(t) + H_1(t) \doteq \underbrace{H_d(t) + H_{\text{ctrl}}(t)}_{H_0(t)} + \underbrace{H_{SE}(t) + H_E(t)}_{H_1(t)}. \label{eqn:quant:openquantHamiltonianSE}
    \end{align}
    $H_0(t)$ is defined as before to encompass the drift and control parts of the Hamiltonian. The new term $H_1(t)$ now consists of two terms: $H_{SE}(t)$ represents an interaction term with the environment, while $H_E(t)$ represents the free evolution of the environment in the absence of the system. In this case, the  Hamiltonian controls the dynamics of both the system and environment combined in a highly non-trivial way. In other words, the state becomes the joint state between the system and environment. The combined system and environment then become closed. Modelling such open quantum systems is complex and challenging and is typically undertaken using a variety of stochastic master equations \cite{wiseman_quantum_2010} or sophisticated noise spectroscopy. As detailed in Chapter \ref{chapter:QDataSet and Quantum Greybox Learning}, the QDataSet contains a variety of noise realisations for one and two qubit systems together with details of a recent novel operator \cite{youssry_characterization_2020} for characterising noise in quantum systems. As such we briefly summarise a few key concepts relating to open quantum systems. Detail can be found in Wiseman and Milburn \cite{wiseman_quantum_2010} and other standard texts.

\subsection{Noise and quantum evolution}\label{sec:quant:Noise and quantum evolution}
Of importance to practical development and applications of quantum computing, especially the control of quantum systems, is understanding the role of noise and other environmental interactions via the theory of superoperators, which are defined as follows.
\begin{definition}[Superoperators]\label{defn:quant:superoperators manifold}
    A superoperator $\mathcal{S}$ is a linear map on the space of bounded linear operators on a Hilbert space $\mathcal{H}$ given as $\mathcal{S}: \mathcal{B}(\mathcal{H}) \rightarrow \mathcal{B}(\mathcal{H})$, $\hat A \mapsto \mathcal{S} \hat A$. 
\end{definition}
For a superoperator $\mathcal{S}$ to correspond to a physical process, such as a quantum channel, it must satisfy the following: (a) \textit{trace preserving or decreasing}, that is, for any state $\rho$ we have that $0 \leq \trace(\mathcal{S}\rho) \leq 1$, (b) \textit{convex linearity}, that is, for probabilities $\{p_j\}$:
    \begin{align}
        \mathcal{S} \left( \sum_j p_j \rho_j \right) = \sum_j p_j \mathcal{S}\rho_j
    \end{align}
and (c) \textit{complete positivity}, namely $\mathcal{S}$ must not only map positive operators to positive operators for the system but also when tensored with the identity operator on any auxiliary system $\mathcal{R}$:
    \begin{align}
        (\mathbb{I}_{\mathcal{R}} \otimes \mathcal{S}) (\rho_{\mathcal{R} \mathcal{S}}) \geq 0.
    \end{align}
A superoperator $\mathcal{S}$ that satisfies these properties is a CPTP map and therefore a quantum channel (by definition \ref{defn:quant:Quantum Channel}), often represented by the Kraus operators $\{E_k\}$ in the form:
\begin{align}
    \mathcal{S}(\rho) = \sum_k E_k \rho E_k^\dagger \qquad \sum_k E_k^\dagger E_k = \mathbb{I}_{\mathcal{H}} \label{eqn:quant:superoperatorkraus}
\end{align}

\subsection{Noise and decoherence}\label{sec:quant:Noise and decoherence}
Noise effects are of fundamental importance to simulating quantum systems (the subject of Chapter 5) but also the ability of machine learning algorithms to learn patterns in data. 
As shown above, open quantum systems are described by the evolution of their density operators. The formal and most common method of representing such open-state evolution is via the \textit{Lindblad master equation} which formalises non-unitary evolution arising as a result of interaction with noise sources, such as baths. The most general form of the master equation for the density operator $\rho$ of an open quantum system, preserving complete positivity and the trace of $\rho$, is given by the Lindblad equation:
\begin{definition}[Lindblad Master Equation]\label{defn:quant:lindbladmasterequation}
The mixed unitary and non-unitary time-evolution of a quantum system represented by $\rho$ interacting with its environment is given by the Lindblad master equation.
 \begin{align}
     \frac{d\rho}{dt} = -i[H, \rho] + \sum_{k} \gamma_k\left( L_k \rho L_k^\dagger - \frac{1}{2} \left\{ L_k^\dagger L_k, \rho \right\} \right) \label{eqn:quant:lindbladmasterequation}
 \end{align}   
 \end{definition}
Here $\rho$ is the density matrix of the quantum system, $H$ is the Hamiltonian of the system, dictating the unitary evolution due to the system's internal dynamics, $\sum_{k}$ indicates a sum over all possible noise channels $k$ affecting the system and $\gamma_k$ are the rates at which the system interacts with the environment through the $k^{th}$ channel, quantifying the strength of the non-unitary processes. Of importance are $L_k$, the Lindblad operators associated with each noise channel, describing how the system interacts with its environment and, in particular, how the environment transforms $\rho$ in a way that affects properties of interest, such as its coherence.

Lindblad operators $L_k$ are not superoperators themselves, but they are integral components in the definition of the \textit{Lindblad superoperator}, which describes specific physical processes (e.g., photon emission, absorption, or scattering). In contrast, the superoperator $\mathcal{L}$, represents the entire evolution of the quantum system including the dissipative dynamics. To understand this within the context of quantum trajectories, where the evolution of a system is modeled under continuous measurement, the stochastic master equation can be written as:
\begin{align}
    \frac{d\rho}{dt} = -i[H, \rho] dt + \sum_{k} \mathcal{L}[L_k]\rho dt + \sqrt{\eta}\mathcal{K}[L]\rho dW_t
\end{align} In the equations above, $L$ represents the Lindblad operator, $\rho$ is the density matrix of the system, $\eta$ is the efficiency of the measurement, and $dW_t$ is the Wiener increment representing the stochastic process of the measurement that models the infinitesimal evolution of processes that have continuous paths. The equation comprises:
\begin{enumerate}[(i)]
    \item a Hamiltonian term $-i[H, \rho]$, describing the unitary evolution of a closed system;
    \item a \textit{Lindbladian (superoperator)} term $\mathcal{L}[L]\rho$, which describes the average effects of system-environment interactions and is given by:
    \begin{align}
        \mathcal{L}[L_k]\rho = \gamma_k\left( L_k \rho L_k^\dagger - \frac{1}{2} \left\{ L_k^\dagger L_k, \rho \right\} \right)
    \end{align}
    \item a \textit{backaction (superoperator)} term $\mathcal{K}[L]\rho$, which accounts for the changes in the system due to measurement and is given by:
    \begin{align}
       \sqrt{\eta}\mathcal{K}[L]\rho = \sqrt{\eta}(L\rho + \rho L^\dagger - \rho \, \trace(L\rho + \rho L^\dagger)).
    \end{align}
    where $\eta$ denotes the influence of the backaction term.  As Wiseman and Milburn note, in the Heisenberg picture the backaction is manifest in the system operators rather than state. 
\end{enumerate}
Note for completeness that the $L_k$ terms represent the channels through which the quantum system interacts with its environment i.e. the quantum trajectories. By contrast, the backaction term seeks to model the effect of the (usually single) measurement channel (which is why we do not sum over $k$). 

\subsection{Noise and spectral density}\label{sec:quant:Noise and spectral density}
We conclude with a few remarks relevant in particular to discussions of noise spectral density in Chapter \ref{chapter:QDataSet and Quantum Greybox Learning} (section \ref{sec:qdata:Noise profile details}). In quantum control, complete knowledge of environmental noise is usually (if not always) impossible to obtain. However, in certain cases, even partial knowledge of open system dynamics can assist in achieving gate fidelities. One such example is where the control pulse is band-limited. To understand this, consider the frequency domain representation of a control pulse $u(t)$:
\begin{align}
   F(\omega) = \int^\infty_{-\infty} u(t)\exp(-i\omega t) dt
\end{align}
with frequency $\omega$ of the pulse (e.g. some signal) and $F(\omega)$ represents the amplitude of the control pulse. If $F(\omega)$ is confined within a specific frequency band, $|\omega| \leq \Omega_0$, then the quantum system's response to environmental noise can be represented by the convolution:
\begin{align}
    I(\omega) = \int_{-\infty}^\infty d\omega F(\omega) S(\omega),
\end{align}
where $S(\omega)$ is the noise power spectral density (PSD), a measure of noise intensity. In such cases, only noise components within the control's effective bandwidth, $|\omega| \leq \Omega_0$, are relevant to solving the quantum control problem. Connecting with the Lindblad master equation (see equation \ref{eqn:quant:lindbladmasterequation}), each Lindblad operator $L_k$ corresponds to a type of noise interaction between the quantum system and the environment. The decoherence rates $\gamma_k$ can be derived from correlation functions of noise operators coupling to the system as represented via the $L_k$ operators. Correlation functions capture temporal correlations of the noise operators $\beta(t)$ on the system. The (two-time or interval) correlation function $G(t)$ is defined as:
\begin{align}
    G(t) = \langle \beta(t) \beta(0)^\dagger \rangle,
\end{align}
where $\langle \cdot \rangle$ denotes the expectation with respect to the environmental state, and $\beta(t) = e^{iH_{\text{env}}t} \beta(0) e^{-iH_{\text{env}}t}$ in the Heisenberg picture with $H_{\text{env}}$ being the Hamiltonian of the environment. The PSD $S(\omega)$ is then obtained from the Fourier transform of $G(t)$:
\begin{align}
    S(\omega) = \int_{-\infty}^{\infty} G(t) e^{-i\omega t} \, dt.
\end{align}
This quantity characterizes the strength of noise or fluctuations in the environment at frequency $\omega$. As noted in the literature \cite{clerk_quantum_2008,clerk_introduction_2010}, classical PSD indicates the magnitude of noise at a particular frequency $\omega$, while quantum noise PSD indicates the magnitude of Golden rule transition rates for emission or absorption events. Dephasing rates $\gamma_k$ can thus be shown to be related to (proportional to) $S(\omega)$ (see  \cite{clerk_introduction_2010}). That is, $\gamma_k$ is determined by integrating over the PSD frequencies of interest:
\begin{align}
    \gamma_k = \eta \int_{-\Omega_0}^{\Omega_0} S(\omega) d\omega \label{eqn:quant:gammakdecoherencerateSomega}
\end{align}
where the proportionality constant $\eta$ depends upon specific interactions between the system and environment, such as coupling constants or environmental energy states.

%============

% \chapter{Appendix (Quantum)}
\section{Analysis, Measure Theory \& Probability}\label{app:Analysis and Measure Theory and Probability}
\subsection{Analysis and Measure Theory}\label{sec:quant:Analysis and Measure Theory}
In this section, we include a few concepts of relevance from analysis, measure theory and probability theory underpinning the theory of quantum information processing above. 
\begin{definition}[Compact Set]\label{defn:quant:Compact Set}
The set $A \subseteq V$ is \emph{compact} if, for every sequence $(v_j)$ in $A$, there exists a subsequence $(v_{j_k})$ that converges to a vector $v \in A$ i.e. $(v_{j_k}) \to v$. For finite spaces, a set $A \subseteq V$ is compact if and only if it is both closed and bounded by the Heine-Borel theorem.
\end{definition}
Compactness of sets is important for controllability and reachability such as via the Poincar\'{e} and quantum recurrence theorems (see definition \ref{defn:quant:Quantum Recurrence Theorem}). $A \subseteq V$ is compact if and only if it is both closed and bounded. If $A$ is compact and $f \colon A \to \mathbb{R}$ is continuous on $A$, then there exists for $f$ a maximum and minimum value on $A$ (i.e. analogous to the mean value theorem).
 For $A \subseteq V$ compact and $f \colon V \to W$ continuous on $A$, continuous maps preserve compactness i.e. the image $f(A) = \{f(v) \colon v \in A\}$ is compact.

 %===Borel sets
 Given $A \subseteq V(\C),B \subseteq W(\C)$, we set some basic properties of measurable sets relevant to various chapters.
 \begin{definition}[Borel sets]\label{defn:quant:Borel sets}
 A Borel set or subset $C \subset A$ is a set for which (a) $C$ is open in $A$, (b) $C$ is the complement of another Borel subset or (c) for a countable set $\{C_k\}, C_k \subset A$ then $C = \bigcup_k^\infty C_k$.      
 \end{definition}
The set of Borel subsets is $Borel(A)$. Functions $f:A\to B$ are Borel if $f^\infty(C) \in Borel(A)$ for all $C \in Borel(A)$ (i.e. the function preserves being Borel). Borel subsets have characteristic functions $\chi_C(u)$ which acts as a classifier (binary) where $\chi_C(u) = 1$ if $u \in C$ and 0 otherwise. Borel sets exhibit closure (algebraic) properties in that scalar multiplication of Borel functions remain Borel, while Borel functions are closed under addition. These and other properties allow us to define a (Borel) measure which is important in the underlying probabilistic quantum theories discussed above. 
 \begin{definition}[Measure]\label{defn:quant:Measure}
     A measure is a map:
     \begin{align}
         \mu: Borel(A) \to [0,\infty]
     \end{align}
     such that (a) $\mu(\null)=0$, (b) (transitivity of measure) such that:
     \begin{align*}
         \mu \bigcup_k^\infty C_k = \sum_k^\infty \mu(C_k)
     \end{align*}
 \end{definition}
 For probability we are interested in normalised measures $\mu(A) = 1$.  A Borel measure is a measure defined on the \textit{Borel} $\sigma$-\textit{algebra} of a topological space, which is the sigma-algebra generated by the open sets (or, equivalently, by the closed sets). From the measure $\mu_k$ one can also define product measures and so on. An important concept we briefly mention is that of integrable functions. 
 
 For continuous variables or continuous spectrum observables, the probability of measuring a particular outcome can be described by Borel integrable functions over the spectrum of the observable. Borel integrable functions permit integrations over Lie group manifolds in the analysis and optimization of control protocols as set out in Chapter \ref{chapter:Time optimal quantum geodesics using Cartan decompositions}
  \begin{definition}
 [Borel integrable]\label{defn:quant:Borel integrable}
     Given a measure $\mu$, define:
     \begin{align*}
         \int f(u)d\mu(u)
     \end{align*}
     An integrable (Borel) function $g:A\to\mathbb{K}$ with respect to $\mu$ is one for which there exist Borel functions $f_0,f_1:A \to [0,\infty)$ such that:
     \begin{align}
         \int g(u)d\mu(u) = \int f_0(u)d\mu(u) - \int \alpha f_1(u)d\mu(u)
     \end{align}
     where $\alpha=1$ for $\mathbb{K}=\R$ and $i$ for $\mathbb{K}=\C$. Here $g=f_1-f_0$
 \end{definition}
Integrable functions exhibit certain features such as linearity (with respect to scalars in $\mathbb{K}$).

\subsection{Probability measure}
We define a probability measure as follows, relevant in particular to both quantum information processing.

\begin{definition}[Probability measure]\label{defn:quant:Probability measure}
    For $A \in \mathcal{A} \subset \mathcal{B}(\Hilb)$, a probability measure is defined as $\mu:A \to [0,1]$ on $A\subset\mathbb{R}$ with $\mu(\mathcal{A})=1$.
\end{definition}
 In probability theory, $A$ are typically denoted \textit{events}. Being a measure, probability must exhibit countable additivity. Random variables $X$ are then defined as Borel functions $X:A\to\R$ distributed with respect to $\mu$. Standard definitions and properties of probabilities are then defined with respect to this measure theoretic definition. We then define a few standard related properties.
\begin{definition}[Expectation]\label{defn:quant:Expectation}
    The expectation value of a random variable $X$ (with respect to $\mu$) is then defined as the integral:
    \begin{align}
        E(X) = \int X(u) d\mu(u)
    \end{align}
    with:
    \begin{align}
        E(X) = \int_0^\infty Pr(X \geq \lambda)d\lambda
    \end{align}
    for $X \geq 0$.
\end{definition}
A similar formalism then applies for quantum states described via alphabets $\Sigma$ with probability vectors $p \in \mathcal{P}(\Sigma)$. The random variable $X$ takes the form of a mapping $X:\Sigma \to \R$ such that for $\Gamma \subseteq \Sigma$:
\begin{align}
    Pr(X \in \Gamma) = \sum_{a \in \Gamma}p(a)
\end{align}
with $p(a)$ the probability of the state described by a vector in $\Sigma$. Expectation becomes the familiar discretised form $E(X) = \sum_a p(a)X(a)$. It can be shown that distribution of states with respect to such probability vectors is equivalent to distribution according to Borel probability measure. A few other concepts used in this work are set out below. 
 \begin{definition}[Gaussian measure]\label{defn:quant:Gaussian measure}
     The standard Gaussian measure is a Borel measure $\gamma:Borel(\R) \to [0,1]$ where:
     \begin{align}
         \gamma(A) = \frac{1}{\sqrt{2}}\int_A \exp\left( -\frac{\alpha^2}{2} \right) d\alpha
\label{eqn:quant:prob:gaussian}
     \end{align}
 \end{definition}
The integral measure $\alpha$ in equation (\ref{eqn:quant:prob:gaussian}) is the standard Borel measure on $\R$. A standard normal random variable is such that $Pr(X \in A) = \gamma(A)$. For multidimensional systems, we note also the standard measure on $n-$dimensional systems as $\gamma_n:Borel(\R^n) \to [0,1]$ such that:
\begin{align}
         \gamma_n(A) = (2\pi)^{-n/2)}\int_A \exp\left( -\frac{||u||^2}{2} \right) dv_n(u)
\label{eqn:quant:probgaussianRn}
     \end{align}
     where $v_n$ is an $n-$fold product measure. An important feature of this measure is that it is invariant under orthogonal transformations, this includes rotations (including those generated by, for example, group elements in specific cases) e.g:
     \begin{align}
         \gamma_n(UA) = \gamma_n(A)
     \end{align}
     where $A \subseteq \R^n$ is Borel and $U \in \mathcal{B}(\R^n)$. The consequence is that for i.i.d. random variables $X_k$, the Gaussian measure projected onto a subspace is equivalent to a Gaussian measure on that subspace i.e:
     \begin{align}
         Y_k = \sum_j^n UX_j
     \end{align}

     \subsection{Haar measure}
     A Borel algebra can also be defined as the $\sigma$-algebra generated by all open subsets of the Lie topological group $G$ (being locally compact and Hausdorff). Left and right action on $S \subset G$ is defined as usual via $gS, Sg$ with left-translation (measure) invariance being defined as $\mu(gS) = \mu(S)$ and $\mu(Sg)=\mu(S)$. Using this formulation, we can define the \textit{Haar measure} \cite{collins_integration_2006} in the quantum context as a unitarily invariant Borel probability measure $\mu_H$. 
     % Definition of Haar measure with corrections
\begin{definition}[Haar measure]\label{defn:quant:Haarmeasure}
    The Haar measure on a locally compact group $G$ is a measure $\mu_H$ that is invariant under the group action. For the unitary group $U(\mathcal{X})$, it is defined as:
    \begin{align*}
        \mu_H: \text{Borel}(U(\mathcal{X})) \rightarrow [0,1]
    \end{align*}
    satisfying for all $g \in U(\mathcal{X})$ and for all Borel sets $\mathcal{A} \subseteq Borel(U(\mathcal{X}))$:
    \begin{align*}
        \mu_H(g\mathcal{A}) = \mu_H(\mathcal{A}) = \mu_H(\mathcal{A}g)
    \end{align*}
    The measure is also normalized such that $\mu_H(U(\mathcal{X})) = \mu_H(G) = 1$.
\end{definition}
The Haar measure allows definition of a uniform measure on topologically structured groups, such as locally compact unitary groups. As the measure is left- and right- invariant (see discussion of Lie groups in section (\ref{sec:alg:Lie groups}). Thus in a quantum mechanical sense ensuring unitary operators are chosen according to the Haar measure, meaning that (when selecting unitaries) each has an equal probability of being chosen (related to creating random quantum circuits and simulating quantum dynamics). One can also define the Haar measure in terms of operator-valued random variables (see \cite{watrous_theory_2018}).

%==========
%==========
%====BACKGROUND: ackground: Geometry, Lie Theory and Representation Theory
% \appendix

\chapter{Appendix (Algebra)}
\label{chapter:Background: Geometry, Lie Algebras and Representation Theory}
\section{Introduction}
\subsection{Overview}
A major focus of quantum geometric machine learning is framing quantum control problems in terms of geometric control where target quantum computations are represented as $U_T \in G$ and $G$ is a Lie group of interest. In this generic model, we are interested then in constructing time-optimal Hamiltonians composed of generators from the associated Lie algebra $\g$ (or some subspace thereof) and associated control functions $u(t)$ so as to evolve the quantum system towards the target $U_T$ in the most efficient manner possible. Doing so according to optimality criteria, such as optimal (minimum) time or energy in turn requires a deep understanding of quantum information processing concepts, but also algebraic and geometric concepts at the heart of much modern quantum and classical physics. By using methods that leverage symmetry properties of quantum systems, time-optimal solutions can often be more easily approximated to find solutions or solution forms that would otherwise be intractable or numerically infeasible to discover. This synthesis of geometry and algebra is made possible by seminal results in the history of modern mathematics which established deep connections between algebra, analysis and geometry. Connections between branches of mathematics was prefigured in much ancient meditations of natural philosophy and becoming a central motivation for so much of 19th century mathematical innovations in geometry, analysis, algebra and number theory that led to the hitherto unparalleled development of the field from Hilbert onwards. The widespread application of ideas drawn from these disciplines has been of significance within physics and other quantitative sciences, including more recently computer science and machine learning fields.

In this Appendix, we cover the key elements of the theory of Lie algebras and Lie groups relevant to the quantum geometric machine learning programme and greybox machine learning architectures discussed further on. The exegesis below is far from complete and we have assumed a level of familiarity with elementary group and representation theory. The Appendix concentrates on key principles relevant to understanding Lie groups and their associated Lie algebras, with a particular focus on matrix groups and foundational concepts such as bilinear (Killing) forms, the Baker-Campbell-Hausdorff theorem and other elementary principles. We then proceed to a short discussion of elements of representation theory, including a discussion of semi-simple Lie group classification, adjoint action, Cartan decompositions and abstract root systems. The material below is related throughout to the novel results in this work in later Chapters, as well as the other Appendices. The content below is also interconnected with the following Appendix \ref{chapter:background:Differential Geometry} on geometric principles (originally a single Appendix, it was split for readability), especially regarding differentiable manifolds, geodesics and geometric control. Most of the material is sourced from standard texts such as those by Knapp \cite{knapp_lie_1996,knapp_representation_2001}, Helgason \cite{helgason_differential_1979}, Hall \cite{hall_lie_2013} and others. As with the other Appendices, proofs are largely omitted (being accessible in the referenced texts) and content is paraphrased within exposition regarding its applicability to the main thesis. Readers with sufficient background in these topics may skip this Appendix.

% \subsubsection{Cartan's revolution}

\section{Lie theory}\label{sec:alg:Lie theory}

\subsection{Overview} 
An essential concept in physics is \emph{symmetry}, especially \emph{rotational} and \emph{Lorentz symmetry} in various physical systems. Symmetries are often described by \emph{continuous groups}—parameterized groups with a manifold structure that allows smooth operations. These are known as \emph{Lie groups}, which possess a differentiable manifold structure that supports smooth group operations. The tangent space at the identity element of a Lie group forms a \emph{Lie algebra} through a natural bracket operation, encoding the properties of the Lie group while being more tractable due to its linear structure.

In quantum mechanics, symmetry is reflected in the Hilbert space through a \emph{unitary action} of a symmetry group. This is formalized by a \emph{unitary representation}, a continuous homomorphism from a symmetry group $G$ into the group of unitary operators $U \in \mathcal{B}(\mathcal{H})$ acting on the quantum Hilbert space $\Hilb$ (axiom \ref{axiom:quant:quantumstates}). However, in reality, physical states correspond to unit vectors in $\Hilb$ differing by a phase, so we use \emph{projective unitary representations}, mapping $G$ into the quotient $U(\Hilb)/U(1)$, where $U(\Hilb)$ is the set of unitaries of $\Hilb$ and $U(1)$ is the group of phase factors, or complex numbers of unit magnitude. This captures the essence of physical state representations, as an ordinary or projective representation of a Lie group induces a corresponding representation of its Lie algebra. For example, angular momentum operators are a representation of the Lie algebra of the rotation group. Understanding the representation of Lie groups and their algebras in a way relevant to quantum information processing is therefore central to solving geometric quantum control problems, especially using machine learning. To this end, we set out background theory below. We formalise these concepts below.

\subsection{Lie groups}\label{sec:alg:Lie groups}
A topological group is a group $G$ equipped with a topology such that the group's binary operation and inverse function are continuous functions with respect to the topology. Such groups exhibit both algebraic structure and topological structure, allowing algebraic operations (due to group structure) and continuous functions on the group manifold (due to its topology). A \textit{continuous group} is one equipped with continuous group operations i.e. that $G$ is parameterised by a parameter that is continuous (and even smooth) on its domain e.g. $\theta \in I \subset \R$. If $G$ has Hausdorff ($T_2$ topology) and that parameter is differentiable, then $G$ exhibits the structure of a Lie group (and as we discuss below, that of a differentiable manifold (see definition \ref{defn:geo:Differentiable manifold})). 

\begin{definition}[Lie groups]\label{defn:alg:liegroups}
    A Lie group $G$ is a Hausdorff topological group equipped with a smooth manifold structure, namely one that supports group operations that are smooth maps.
\end{definition}
Lie groups encapsulate the concept of a \textit{continuous group} allowing the study of infinitesimal transformations (such as translations and rotations) and associated symmetry properties. The ability to analyze infinitesimal transformations is what connects Lie groups to differential geometry and physics, particularly in the context of studying continuous symmetries relevant to symmetry reduction. The corresponding Lie Groups are also equipped with product and inverse maps $\mu: G \times G \to G, \ell: G \to G$. An important property of Lie groups is left translation and the related concept of left invariance, central to the construction of invariant measures (such as the Haar measure) on groups. For a Lie group $G$, left translation by $g \in G$, denoted $L_g: G \to G, L_g(h)\mapsto gh$ for $h \in G$, is a diffeomorphism from $G$ onto itself. Right translations are similarly defined as $R_g(h) \mapsto hg$. Left invariance of a vector field $X$ is defined where $(dL_{g^\inv h})(X_g) = X_h$ (i.e. that $X$ commutes with left translations) i.e if, for all $g \in G$, we have $dL_g(X_e) = X_g$, where $e$ is the identity element in $G$ and $dL_g$ is the differential of left translation by $g$.

Given a smooth function $f:G \to \R$, we denote $f_g(h) = f(gh)$ for $g,h \in G$ the left translate. A vector field $X$ on $G$ is itself left invariant if $(Xf)_g = X(f_g)$. Such left invariant $X$ form the Lie algebra $\g$ of $G$. The vector field $X$ can then be regarded as a set of tangent vectors $X_g$ at $g \in G$. The special relationship with the identity is then further understood as follows: the mapping $X \to X_I$ (where $I$ is the identity element of $G$) is a vector space isomorphism between $\g$ to the tangent space $X_I$ (mapping vector fields to their vectors at the identity element, thus identifying the Lie algebra with the tangent space at the identity), thereby (as we discuss below and in the next Appendix) encoding the structure of the group through its tangent space at the identity. This isomorphism preserves (and thus `carries') the Lie bracket, allowing the tangent space at the identity in $G$ to be identified \textit{as} the Lie algebra $\g$ itself. Thus the entire Lie algebra, and thus group, can be studied in geometric terms of the infinitesimal translations near the identity.

\subsection{Matrix Lie groups}\label{sec:alg:Matrix Lie groups}
We begin with the general linear group $GL(n; \C)$ acting on $M_n(\C)$, the set of $\dim n$ complex matrices.
\begin{definition}[General linear group] \label{defn:alg:general linear group}
    The group of all invertible n-dimensional complex matrices is denoted the general linear group:
    \begin{align*}
        GL(n;\C) = \{ A \in M_n(\C) | \det(A) \neq 0 \}
    \end{align*}
\end{definition}
As we discuss later on, matrix representations of groups are central to many methods in quantum information processing. We note (as discussed in Chapter \ref{chapter:Quantum Geometric Machine Learning}) that $M_n(\C)$ may be identified (locally) $\C^{n^2}=\R^{2n^2}$ via representing another dimension for the imaginary terms (each complex matrix element represented by two real numbers). A \textit{matrix Lie group} is defined as a closed subgroup of $GL(n; \C)$ i.e. each such group is a real-valued submanifold of the general linear group (closed here means in the topological sense, such that limit points of sequences remain within the subgroup). To study transformations using such groups, we rely upon group homomorphisms.

\begin{definition}[Lie group homomorphism]\label{defn:alg:Lie group homomorphism}
Given matrix Lie groups $G_1$ and $G_2$, a Lie group homomorphism $\varphi:G_1\to G_2$ is a continuous group homomorphism. Such a homomorphism is an isomorphism if it is bijective and with a continuous inverse. Lie groups for which there exists such an isomorphism are equivalent (up to the isomorphism).
\end{definition}
Important groups include the group of $n-$dimensional invertible real-valued matrices $\text{GL}(n, \mathbb{R})$, and $\text{SL}(n, \mathbb{C})$ and $\text{SL}(n, \mathbb{R})$ special linear groups with determinant 1. Of particular importance is the unitary group defined below.

\begin{definition}[Unitary matrix and unitary group]\label{defn:alg:Unitary matrix and unitary group}
An operator $U \in M_n(\C)$ is unitary if $U^\dagger U = U U^\dagger = \mathbb{I}$. The set of unitary operators form the unitary group denoted $U(n)$. 
\end{definition}
This group is the group formed by the set of unitary operators in definition \ref{defn:quant:Unitary Operator}.
Unitary operators preserve inner products such that $U$ is unitary if and only if $\langle Uv, Uw \rangle = \langle v, w \rangle$ for all $v, w \in \Hilb$. The special unitary group $SU(n)$ is the set of all $U(n)$ such that $\det U(n)=1$. For the geometric study of unitary sequences (for minimisation and time-optimal problems), we characterise such sequences as paths on a Lie group manifold which is sufficiently connected to enable paths to be well defined.

\begin{proposition}[Connected Lie Group]
\label{defn:alg:connected_matrix_lie_group}
A (matrix) Lie group $G$ is \emph{connected} if for all $A, B \in G$ there exists a continuous path $\gamma : [0,1] \rightarrow M_n(\mathbb{C})$ such that $\gamma(0) = A$ and $\gamma(1) = B$ with $\gamma(t)\in G$ for all $t$. 
\end{proposition}
The connectedness of a Lie group is crucial to its representation in terms of a differentiable manifold with suitable topology for quantum information tasks.
We characterise $G$ as \emph{simply connected} if it is connected and every continuous loop (i.e. curve such that $\gamma(0)=\gamma(t)$ in $G$ can be shrunk continuously to a point $g \in G$. A matrix Lie group $G$ is considered  \emph{compact} if it is compact as a subset of $M_n(\mathbb{C}) \cong \mathbb{R}^{2n^2}$.

In particular, of relevance to later Chapters, we note that the groups $O(n), SO(n), U(n)$ and $SU(n)$ are compact. $U(n)$ is connected allowing continuous paths between $U(0) \to U(t)$. This is an important aspect of controllability (and reachability) of quantum states in later Chapters. It also impacts the learnability of unitary sequences from training data (such as discussed in Chapter \ref{chapter:Quantum Geometric Machine Learning}) i.e. non-compactness can affect the ability of a model to properly estimate or learn mathematical structure required for learning optimal controls. The group $SU(2)$, the key group for qubit-based quantum computation, is defined as:
\begin{align}
    SU(2) = \left\{ \begin{pmatrix}
\alpha & -\beta \\
\beta & \bar{\alpha}
\end{pmatrix} \, \middle| \, \alpha, \beta \in \mathbb{C}, \, |\alpha|^2 + |\beta|^2 = 1 \right\}.
\label{eqn:alg:su2group}
\end{align}
For qubits, $SU(2)$ allows for a consistent and complete description of qubit states and their transformations, which in turn is essential for the design and understanding of quantum algorithms and the behaviour of quantum systems. This is also the case for multi-qubit systems, which can be described in terms of tensor products of $SU(2)$. For multi-qubit computations (and even single-qubit ones), we are often interested in decomposition into a sequence of single-qubit gates. The simply connected nature of gates drawn from $SU(2)$ guarantees the density of universal gate sets, allowing approximation of (and controllability for) relevant target unitaries to any desired accuracy. We now set out below a few standard algebraic facts about Lie groups and Lie algebras (sourced from \cite{hall_lie_2013} and \cite{knapp_lie_1996}). In particular we emphasise concepts related to characterising quantum unitary sequences as curves along Lie group manifolds generated by the corresponding Lie algebra. First, we list a few properties of the matrix exponential of relevance.
\begin{theorem}[Matrix Exponential Properties]
\label{thm:alg:matrix_exponential_properties}
The following are properties of the matrix exponential for $X, Y \in M_n(\mathbb{C})$:
\begin{enumerate}[(i)]
    \item $e^{0} = \mathbb{I}$
    \item $e^{X^\trace} = (e^{X})^\trace$ and $e^{X^\dagger} = (e^{X})^\dagger$
    \item For $Y,X \in M_n(\C)$ invertible:
    \begin{align*}
        e^{YXY^{-1}} = Ye^{X}Y^{-1}.
    \end{align*}
    \item $\det(e^{X}) = e^{\trace(X)}$
    \item If $XY = YX$ then $[X,Y]=0$ so $e^{X+Y+\mathcal{O}[X,Y]} = e^{X}e^{Y}$ 
    \item $e^{X}$ is invertible and $(e^{X})^{-1} = e^{-X}$
    \item For if $XY \neq YX$, we have
    \begin{align*}
        e^{X+Y} = \lim_{m \to \infty} \left( e^{X/m} e^{Y/m} \right)^{m}.
    \end{align*}
\end{enumerate}
\end{theorem}

%=====Subsection: Lie algebras
\subsection{Lie algebras}\label{sec:alg:Lie algebras}
Lie algebras are a fundamental mathematical concept for describing the evolution of quantum systems, geometric structure and time-optimal control. As we expand upon below, Lie algebras provide a bridge to study symmetry groups by way of their matrix representations and transformations thereof. 
\begin{definition}[Lie algebra and Lie derivative]\label{defn:alg:Lie algebraliederivative}
     An algebra $\frak{g}$ is a vector space over a field $\Km$ equipped with a bilinear form (product) $[X,Y]$ for $X,Y \in \frak{g}$ that is linear in both variables. Such an algebra is a \textit{Lie algebra} if:
\begin{enumerate}[(a)]
    \item $[X,X]=0$ and $[X,Y]=-[Y,X]$, $\forall X,Y \in \frak{g}$; and
    \item the Jacobi identity is satisfied, namely:
    \begin{align}
        [[X,Y],Z] + [[Y,Z],X] + [[Z,X],Y]=0
    \end{align}
\end{enumerate}
\end{definition}
The product $[X,Y]$ above is denoted the \textit{Lie derivative} (sometimes the Lie bracket). In quantum information contexts, the Lie derivative is denoted the commutator. The commutator exhibits a number of properties set out below.

\begin{proposition}[Lie bracket properties]\label{prop:alg:Lie bracket properties}
    Properties of Lie bracket (commutator) include:
    \begin{enumerate}
        \item linearity in the first and antilinearity in the second arguments $[\alpha A, B + \beta C] = \alpha[A,B] + \alpha\beta^*[A,C]$ for $\alpha,\beta \in \C$; 
        \item antisymmetry $[B,A] = -[A,B]$; 
        \item derivability (Leibniz rule) $[A,BC] = [A,B]C + B[A,C]$ (related to derivations).
    \end{enumerate}
\end{proposition}
In geometric contexts, the Lie bracket (commutator) is equated with the Lie derivative, satisfying the conditions above. We use these terms interchangeably as appropriate for the context. Note that for $\g \subset \mathcal{A}$ (where $\mathcal{A}$ is an associative algebra), $\g$ can be constructed as a Lie algebra using the Lie bracket $[x,y] = xy-yx, \forall x,y \in \g$ (in such cases $\mathcal{A}$ is the universal enveloping algebra of $\g$). An important characteristic of Lie algebras is the action of a Lie algebra upon itself as represented by the adjoint action, realised in the form of the Lie derivative.

%======defn: adjoint action
\begin{definition}[Adjoint action]\label{defn:alg:Adjoint action}
    The \textit{adjoint action} of a Lie algebra upon itself $\ad: \frak{g}\to \text{End}_\Km (\frak{g})$ is given by:
\begin{align}
    \ad_X(Y) = [X,Y]
\end{align}
where $\text{End}_\Km (\frak{g})$ represents the set of $\Km$-scalar endomorphisms of $\frak{g}$ upon itself and $X,Y \in \g$. 
\end{definition}
More intuition about the relationship between the Lie derivative and differential forms is provided by showing that the adjoint action is a derivation satisfying the Jacobi identity is equivalent to:
\begin{align}
    \ad_Z[X,Y] = [X,\ad_Z(Y)] + [\ad_Z(X),Y]
\end{align}
The adjoint action is a homomorphic endomorphism of $\g$ i.e. $\ad_{[X,Y]} = \ad_X\ad_Y - \ad_Y\ad_X$ with kernel $Z_\g$ below. 

In quantum information processing, Lie algebras tend to be represented via representations of $G \subset GL(n,\C)$ or its closed subgroups (closed linear group). An important result in the theory of Lie groups is that the Lie algebra $\g$ of $G$ is canonically isomorphic with the linear Lie algebra $\frak{gl}(n,\C)$ i.e. $\mu: \g \to \frak{gl}(n,\C)$ such that $X \in \g$ can be represented in terms of real and imaginary $X_I \in \frak{gl}(n,\C)$. Thus the behaviour of $\frak{gl}(n,\C)$ may be used to study that of $\g$ and thus $G$ of interest.
\begin{definition}[Lie algebra homomorphism]\label{defn:alg: Lie algebra homomorphism}
   A linear map $\varphi:\frak{g} \to \frak{h}$ where:
\begin{align}
    \varphi([X,Y]) = [\varphi(X),\varphi(Y)]
\end{align}
where $\frak{a},\frak{b} \subset \frak{g}$:
\begin{align}
    [\frak{a},\frak{b}] = \span\{ [X,Y] | X \in \frak{a}, Y \in \frak{b} \}
\end{align}
 is defined to be a \textit{Lie algebra homomorphism} from $\frak{g}$ to $\h$.
\end{definition}
We then can define subalgebras (of critical importance to our discussion of Cartan methods below and control subsets) and ideals.
\begin{definition}[Lie subalgebras and ideals]\label{defn:alg:Lie subalgebras and ideals}
    A set $\h \subset \frak{g}$ is a \textit{Lie subalgebra} if $[\frak{h},\frak{h}] \subseteq \frak{h}$, in which case $\h$ is also a Lie algebra. A subalgebra is an \textit{ideal} if $[\h,\g] \subseteq \h$.
\end{definition}
Lie algebras can also be written as direct sums of the underlying vector spaces $\g_1,\g_2$ equipped with an (induced) Lie bracket (adjoint preserving Killing form):
\begin{align}
    [(X_1,Y_1),([X_2,Y_2])]=([X_1,Y_1],[X_2,Y_2])
\end{align}
for $X_k,Y_k \in \g_k$.

For Cartan decompositions, we require the classification of semi-simple Lie groups. This in turn requires other sundry definitions.
\begin{definition}[Centralizer and Normalizer]\label{defn:alg:Centralizer and Normalizer}
    The centralizer $Z_{\g}(\frak{s})$ of $\frak{s} \subset \g$ is a subalgebra of $\g$ comprising elements that commute with all of $\frak{s}$:
    \begin{align}
        Z_\g(\frak{s}) = \{ X \in \g | [X,Y]=0, \forall Y \in \frak{s} \}
    \end{align}
    The normalizer of $\g$ is the set of all elements of $\g$ whose commutator with $\frak{s}$ is in $\frak{s}$:
    \begin{align}
        N_\g(\frak{s}) = \{ X \in \g | [X,Y] \in \frak{s}, \forall Y \in \frak{s} \}
    \end{align}
\end{definition}
The \textit{center} $Z_\g$ is the set of $X \in G$ that commute with all elements in $\g$. To understand semi-simple Lie algebras, we can define nested commutators with a commutator series and lower central series as:
\begin{align}
    \g^0 &=\g \qquad \g^{j+1} = [\g^j,\g^j] \qquad \g &= \g^0 \supseteq \g^1 \supseteq ...\qquad \text{(commutator series)}\\
    \g^0 &=\g \qquad \g_{j+1} = [\g,\g_j] \qquad \g &= \g_0 \supseteq \g^1 \supseteq ...\qquad \text{(lower central series)}
\end{align}
From these definitions we obtain the concepts of $\g$ being \textit{solvable} if $g^j=0$ for some $j$ and \textit{nilpotent} if $g_j=0$ for some $j$ (nilpotent implies solvable). We can now define simple and semi-simple Lie algebras as follows:
%Simple and semisimple
\begin{definition}[Simple and semisimple]\label{defn:alg:Simple and semisimple}
A finite-dimensional $\g$ is simple if $\g$ is nonabelian and has no proper non-zero ideals. We say $\g$ is semisimple if $\g$ has no nonzero solvable ideals. A Lie group $G$ is said to be \textit{semisimple} if its Lie algebra $\mathfrak{g}$ is semisimple.
\end{definition}
 This last test for semisimplicity is equivalent to the radical rad $\g=0$. Simple $\g$ are semisimple by construction. They are algebras are closed under commutation $[\g,\g]=\g$ (important for control). Semisimple $\g$ have empty centers. The criterion for semi-simplicity is related together with the Killing form below to Cartan's seminal classification of Riemannian symmetric spaces (see section \ref{sec:geo:Symmetric spaces}) which is the subject of our work in Chapter \ref{chapter:Time optimal quantum geodesics using Cartan decompositions}.
 
\subsection{Killing form}\label{sec:alg:Killing form}
We now define the Killing form, a bilinear form defined on $\g$ which plays an central role in allowing algebraic concepts to be connected to geometric analogues and enabling classification of algebraic structures. In particular, as we discuss below, conditions on the Killing form allow its association as a Riemannian or subRiemannian metric (used to calculate minimal path lengths, and thus time, of curves in $G$) for symmetric spaces the subject of our final chapter. 
\begin{definition}[Killing form]\label{defn:alg:Killing form}
    The Killing form is a bilinear transformation given by:
	\begin{align}
	B(X,Y) = \Tr(\ad_X, \ad_Y)
	\end{align}
	for $X,Y \in \g$.
\end{definition}
The Killing form is an important element in Cartan's classification of semi-simple Lie algebras and, among other things, establishing appropriate metrics (and inner products) on $\g$ for variational methods discussed later on. Additionally, Cartan's criterion for semisimplicity is that the Killing form is non-degenerate, that is $\g$ is semisimple if and only if the Killing form for $\g$ is non-degenerate, namely $B(X,Y) \neq 0$ for all $X,Y \in \g$. Other concepts we briefly note are that $\g$ is \textit{reductive} if for each ideal $\a \subset \g$, there exists another ideal $\frak{b} \subset \g$ such that $\g = \a \oplus \frak{b}$ (i.e. can be written as the direct product of ideals). It can be shown this is equivalent to $\g$ being closed under the conjugate transpose (adjoint) action. Under such conditions, the Killing form can be used to define an inner product via $(X,Y) = -B(X,Y)$ and thus a metric. Moreover, the Killing form is invariant under the adjoint action of the Lie group, meaning $B(\text{Ad}_g X, \text{Ad}_g Y) = B(X,Y)$ for all $g \in G$ and $X,Y \in \g$. This invariance property of the Killing form ensures the induced metric is well-defined across the manifold $G$. We explore this in our final chapter.

\subsection{Matrix exponential}\label{sec:alg:Matrix exponential}
The matrix exponential is an important feature in the theory of Lie algebras and Lie groups (especially regarding their connection) acting as a fulcrum concept that links both operationally and semantically from algebraic theory to quantum theory. The various representations of the matrix exponential serve a variety of uses, such as in terms of infinite power series (especially relating to the use of the adjoint operator in Chapter \ref{chapter:Time optimal quantum geodesics using Cartan decompositions}) and as group elements. It is defined below.

\begin{definition}[Matrix exponential]\label{defn:alg:Matrix exponential}
The matrix exponential is defined by the following  power series: 
    \begin{align}
        e^X = \sum_{n=0}^\infty \frac{X^n}{n!}.
    \end{align}
\end{definition}
This can usefully be expressed as:
\begin{align}
     e^X = \lim_{n \to \infty} \left( \mathbb{I} + \frac{X}{n} \right)^n \label{eqn:alg:exponentiallimit}
\end{align}
where $X \in \g$ or some other vector space, such as $M_n(\C)$. More intuition can be obtained via the derivative of the matrix exponential function: 
% \hl{update}
\begin{align}
    \left. \frac{d}{dt} e^{tX} \right|_{t=0} = X. \label{eqn:alg:matrixexpderivativeatzero}
\end{align}
Equation (\ref{eqn:alg:matrixexpderivativeatzero}) provides an important way of understanding, via the representation of matrix groups, the connection between Lie algebras and Lie groups. It indicates that the derivative at the identity element of the group is precisely the Lie algebra element $X$, demonstrating how Lie algebras serve as the infinitesimal generators of Lie group actions. Moreover, the one-parameter subgroup theorem below, stating that for every element $X$ in the Lie algebra there corresponds a unique one-parameter subgroup in the Lie group with $A(t) = \exp{(tX)}$, explicitly reveals the relationship between the algebraic structure of Lie algebras and the topological and geometric structure of Lie groups. This unique mapping via the exponential allows the exploration of continuous symmetries of groups by way of their corresponding Lie algebra.
%=========
We note then the definition of a one-parameter subgroup of $G$.
\begin{definition}[One-Parameter Subgroup of \(GL(n; \mathbb{C})\)]
\label{defn:alg:one_parameter_subgroup}
A one-parameter subgroup of \(GL(n; \mathbb{C})\) is denoted \(G(\cdot)\) and is a continuous homomorphism of $\R \to GL(n; \mathbb{C})$. Equivalently, for $A \in G(\cdot)$, it corresponds to a continuous map \(A : \mathbb{R} \rightarrow GL(n; \mathbb{C})\) such that \(A(0) = \mathbb{I}\) and \(A(s + t) = A(s)A(t)\) for \(s, t \in \R\).
\end{definition}
It can then be shown that there exists a unique element of the Lie algebra (representation) used to establish the relationship between a Lie algebra and Lie group.

\begin{theorem}[One-Parameter Subgroup Exponential Representation] 
\label{thm:alg:one_parameter_subgroup_exponential}
If \(G(\cdot)\) is a one-parameter subgroup of \(GL(n; \mathbb{C})\), then there exists a unique \(X \in M_n(\mathbb{C})\) such that $A(t) = \exp{(tX)}$ parametrised by $t \in \R$.
\end{theorem}
These relations show how continuous symmetries of Lie groups may be studied via the Lie algebra as a tangent space at the identity element of the group. In quantum information contexts, the matrix exponential allows the study of system's dynamics via mapping the Hamiltonian $H$ (see definition \ref{defn:quant:Hamiltonian}) to group elements represented as unitary operators (definition \ref{defn:quant:Unitary Operator}). We now formalise this relationship.

\subsection{Lie algebra of matrix Lie group}\label{sec:alg:Lie algebra of matrix Lie group}
Lie algebras are associated with Lie groups via the exponential map.
% \hl{Update}
\begin{definition}[Lie Algebra of a Matrix Lie Group]
\label{defn:alg:lie_algebra_of_matrix_lie_group}
Given a matrix Lie group $G \subseteq GL(n; \mathbb{C})$, the Lie algebra $\mathfrak{g}$ for $G$ is defined as follows:
\begin{align*}
    \mathfrak{g} = \{ X \in M_n(\mathbb{C}) \mid e^{tX} \in G \text{ for all } t \in \mathbb{R} \}.
\end{align*}
\end{definition}
Note that $X \in \mathfrak{g}$ if and only if the one-parameter subgroup is in closure of $G$. For $X \in \mathfrak{g}$, it is sufficient that $e^{tX} \in G$ for all $t \in \R$.
% \hl{Paraphrased}
For any matrix Lie group $G$, the Lie algebra $\mathfrak{g}$ of $G$ has the following properties. The zero matrix $0$ belongs to $\mathfrak{g}$. For all $X$ in $\mathfrak{g}$, $tX$ belongs to $\mathfrak{g}$ for all real numbers $t$. For all $X$ and $Y$ in $\mathfrak{g}$, $X + Y$ belongs to $\mathfrak{g}$. For all $A \in G$ and $X \in \mathfrak{g}$ we have $AXA^{-1} \in \mathfrak{g}$. For all $X$ and $Y$ in $\mathfrak{g}$, the commutator $[X, Y]$, defined by
\begin{align}
[X, Y] := XY - YX = \frac{d}{dt}e^{tX}Ye^{-tX} \bigg|_{t=0}
\label{eqn:alg:liealgderivatidentity}
\end{align}
belongs to $\mathfrak{g}$. The first three properties of $\mathfrak{g}$ say that $\mathfrak{g}$ is a real vector space. Since $M_n(\mathbb{C})$ is an associative algebra under the operation of matrix multiplication, the last property of $\mathfrak{g}$ shows that $\mathfrak{g}$ is a real Lie algebra. Equation (\ref{eqn:alg:liealgderivatidentity}) also indicates that the Lie algebra $\g$ is the set of derivatives at $t=0$ of all smooth curves $\gamma(t) \in G$ where such curves equal the identity at zero. It thus provides a bridge between typical algebraic formulations of commutators and geometric representations of tangents. In the next section we provide more detail on the relationship between Lie theory and differential geometric concepts such as curves. Note that we can define the Lie algebras of $\u(n)$ and $\frak{su}(n)$:
\begin{align}
    \u(n)&=\{ X \in M_n(\C) | X^\dagger = -X \}\\
    \frak{su}(n) &= \{X \in \u(n) | \Tr(X)=0  \}
    \label{eqn:alg:xderivofetx}
\end{align}
The significance of the exponential map can be further understood as follows. Group actions $\Phi: G \to G$ have a corresponding map in the Lie algebra $\phi:\g \to g$ such that: 
\begin{align}
    \Phi(e^{tX})=e^{t\phi X}
    \label{eqn:alg:groupalgebrahomocorro}
\end{align}
where $\phi(X)=\frac{d}{dt}\Phi(e^{tX})|_{t=0}$ (implying smoothness for each $X$). Here $\phi$ is a linear homomorphism such that:
\begin{align}
    \phi([X,Y])&=[\phi(X),\phi(Y)] \qquad \phi(AXA^\inv) = \Phi(A)\phi(X)\Phi(A)^\inv
\end{align}
where $A\in G$ and $X \in \g$. For a group $G$ and corresponding algebra $\g$, $X \in \g$ if and only if $e^X \in G$.

A key feature in quantum control contexts is framing control of the evolution of quantum systems in terms of paths over unitary Lie group manifolds. For this to be the case, in essence we require that Lie group homomorphisms be characterised or determined by Lie algebra homomorphisms - that is, we essentially want guarantees that arbitrary paths along the manifold can be generated by applying controls to generators in $\g$. This is particularly the case when the Lie groups of interest exhibit topological properties of being simply connected (meaning any close curve in $G$ can be shrunk to an arbitrary element of $G$ itself - or that is, there are no `holes'). In a control context, the simply connected nature of the underlying group manifold is important for reachability of targets. If a group $G$ is connected and simply connected, then it can be shown that the Lie algebra and Lie group are related such that there is unique mapping $\Phi$ such that the group-algebra homomorphism correspondence set out in equation (\ref{eqn:alg:groupalgebrahomocorro}) holds. 
From a geometric perspective, the following relationship between Lie algebras and tangent spaces is set out.
\begin{theorem}[Lie algebra tangent space correspondence]\label{thm:alg:Lie algebra tangent space correspondence}
    Each Lie algebra $\g$ of $G$ is equivalent to the tangent space to $G$ at the identity. The algebra $\g$ is equivalent to the set of $X \in M_n(\C)$ such that there exists a smooth curve $\gamma:\R \to M_n(\C) \subseteq G$ with $\gamma(0)=I,\gamma'(0)=X$.
\end{theorem}
This can be seen given equation (\ref{eqn:alg:xderivofetx}) in concert with $\gamma(t) = e^{X(t)}$ (such that $\gamma'(0) \in \g$). Moreover, for a matrix Lie group that is connected, then it can be shown that for $\gamma \in G$, there is a finite set of elements in $\g$ that generate $\gamma$. These facts are important to the characterisation and control of quantum unitary sequences (and circuits) in terms of paths along manifolds in Chapters \ref{chapter:Quantum Geometric Machine Learning} and \ref{chapter:Time optimal quantum geodesics using Cartan decompositions}. The relationship between Lie algebra and group homomorphisms is not simply one-for-one, as can be seen the fact that $\su(2) \simeq \frak{so}(3)$ while $SU(2)$ and $SO(3)$ are not isomorphic (see literature \cite{knapp_lie_1996,knapp_representation_2001} etc for more detail). We note finally for completeness that on occasions one is interested in the universal cover $\tilde G$ of a group $G$ which effectively is a simply connected matrix Lie group which `covers' the space of interest (in the sense that there is a group homomorphism between the groups and isomorphism among their algebras that allows the study or manipulation of group $G$, which may not be simply connected, via $\tilde G$ and $\tilde \g$).

\subsection{Homomorphisms}\label{sec:alg:Homomorphisms}
The relationship between Lie algebras and Lie groups via the exponential map is considered a lifting of homomorphisms of Lie algebras to homomorphisms of (analytic) groups. That is, for $G,H$ analytic subgroups, $G$ simply connected and the Lie algebra homomorphism $\varphi: \g \to \h$, then there exists a smooth homomorphism $\Phi: G \to H$ such that $d\Phi = \varphi$. As \cite{knapp_lie_1996} notes, there are two equivalent ways to express such a lifting: either relying on lifting homomorphisms to simply connected analytic groups, or relying on existence theorems for sets of solutions to differential equations. The first approach involves defining curves $\gamma: \R \to G$ with $\gamma(t) \mapsto \exp(tX)$. One then defines $d/dt$ and $\tilde X$ as left invariant fields on $\R,G$ such that:
\begin{align*}
    d\gamma(t)\left(\frac{d}{dt}\right)f = \frac{d}{dt}f(\gamma(t)) = \frac{d}{dt}f(\exp(tX))= \tilde X f(\exp(tX))
\end{align*}
hence we see the important explicit relationship between the left-invariant vector fields $\tilde X$ and $d/dt$. Among other things, this shows the mapping $X \to \exp(X)$ is smooth (given the smoothness of $d/dt$). Expressed in local coordinate charts, the equation above represents a system of differential equations satisfied by curves $\gamma(t)$ represented as exponentials. Among other aspects, this relation enables the machinery of analysis to be brought to bear in group theoretic problems in quantum information contexts. For example, it enables the application of Taylor's theorem such that:
\begin{align*}
    (\tilde X^n)(g \exp (tX)) = \frac{d^n}{dt^n}(f(g\exp (tX))
\end{align*}
for $g\in G, f \in C^\infty$. Noting also that:
\begin{align}
    \tilde X f(g) = \frac{d}{dt}f(g \exp t X) \big|_{t=0}
\end{align}
shows again how the operation of left-invariant vector fields $\tilde X$ is equated with differential operators. 

\subsection{Baker-Campbell-Hausdorff theorem}\label{sec:alg:Baker-Campbell-Hausdorff theorem}
As Hall \cite{hall_lie_2013} notes, an important feature of the correspondence between Lie algebras and Lie groups is to show that the Lie group homomorphism $\Phi: G \to H$ (for Lie groups $G$ and $H$) defined by $\Phi(\exp(X)) = \exp(\phi(X))$ where $\phi: \g \to \h$ is a Lie algebra homomorphism. A consequence of this homomorphism is the \textit{Baker-Campbell-Hausdorff} (BCH) theorem which allows a direction relationship between the properties of the exponential map and Lie algebraic operations to become apparent.  
\begin{definition}[Baker-Campbell-Hausdorff]\label{defn:alg:Baker-Campbell-Hausdorff}
The Baker-Campbell-Hausdorff formula shows that the map $\Phi$ is identified from $U_e \in G$ into $H$ (where $e$ is the identity element of $G$) is a local homomorphism by the Lie group-Lie algebra homomorphism, such that for sufficiently small $X,Y, Z \in \g$:
\begin{align}
    \exp(X)\exp(Y) = \exp\left(X + Y + \frac{1}{2}[X,Y] + \frac{1}{12}[X,[X,Y]] ...  \right) \label{eqn:alg:Baker-Campbell-Hausdorff}
\end{align}
\end{definition}
(see Hall \cite{hall_lie_2013} $\S 5$ for proofs including the Poincar\'e integral form). The BCH formula is important in quantum control settings in particular as we discuss in other parts of this work. Moreover, equation (\ref{eqn:alg:Baker-Campbell-Hausdorff}) fundamentally elucidates the important role of adjoint action (the commutator) in shaping quantum state evolution via its effect upon exponentials (as diffeomorphic maps on $\M$) which arises from the non-commutativity of the Lie bracket. 

%===========ROOT SYSTEMS/DIAGRAMS (MOVE)

%==========REPRESENTATION THEORY

\section{Representation theory}\label{sec:alg:Representation theory}
\subsection{Overview}
In this section, we briefly cover elements of the representation theory of semi-simple groups. This is relevant to geometric treatments involving Lie algebras. Representations are abstractions (homomorphisms) of groups (in terms of actions on generalised linear (invertible) vector spaces, i.e. on  $GL(n; \C)$ the group of invertible linear transformations of a finite-dimensional vector space $V$ for which the associated Lie algebra is $\frak{gl}(n,\C)$.

\subsection{Representations}\label{sec:alg:Representations}
We begin with the definition of representations.
\begin{definition}[Finite-Dimensional Representation of a Lie Group]\label{defn:alg:Finite-Dimensional Representation of a Lie Group}
\label{defn:alg:finite_dim_representation_of_lie_group}

A finite-dimensional representation of $G$ is a continuous homomorphism of $\pi:G \to GL(V)$. A representation satisfies:
\begin{align}
    \pi([X,Y]) = [\pi(X),\pi(Y)]
\end{align}
for $X,Y \in G$ and equivalently for the lie algebra $\pi: \g \to \frak{gl}(V)$.
\end{definition}
An \textit{invariant subspace} for a representation $\varphi:\g \to GL(V)$ on a subspace $U \subset GL(V)$ is one such that $\varphi(X)U \subseteq U$ for $X \in \g$ (i.e. if the only $G$-invariant subspaces of $V$ under $\pi$ are the trivial subspace $\{0\}$ and $V$ itself). A representation on a non-zero space $GL(V)$ is \textit{irreducible} if the only invariant subspaces are 0 and $GL(V)$. That is, given $\pi: G \to GL(V)$, a subspace $U \subset V$ is invariant if $\pi(g)w \in W$, for all $g \in G, w \in W$. 
% \begin{definition}[Irreducible representations]
% \label{defn:alg:irreducible_representation}
% \end{definition} 
Two representations of $G$ are isomorphic if and only if the Lie algebra representations are isomorphic. This result allows us in particular to move between the fundamental and adjoint representation with implications, for example, with respect to optimal parametrisation of invariant neural networks whose parameters are correlated to the choice of representation \cite{larocca_group-invariant_2022,larocca_theory_2023}.
Representations $\varphi,\varphi'$ are equivalent if there exists an isomorphism between the two vector spaces $\chi:V\to V'$ such that $\chi \varphi(X) = \varphi'(X)E$. A \textit{completely reducible} subspace is then one which can be decomposed into the direct sum of irreducible representations i.e. such that $V = \oplus_i U_i$ for $U \subset V, i = 1,...,\dim V$.

\subsection{Complex Semi-simple Lie Algebras}\label{sec:alg:Complex Semi-simple Lie Algebras}
Of central importance to our result in Chapter \ref{chapter:Time optimal quantum geodesics using Cartan decompositions} are what are known as Cartan decompositions of symmetric space Lie groups as a technique for solving certain classes of time-optimal quantum control problems. In this section, we provide a brief summary of Cartan decompositions, their relationship to Lie groups and algebras together with root systems and Dynkin diagrams. The section largely summarises standard work in the literature \cite{knapp_lie_1996,hall_beyond_2007,helgason_differential_1962,helgason_differential_1979}. Quantum information processing problems in this work (and much of the field) are often related to semisimple Lie groups and Lie algebras. In our case, we are interested in Cartan subalgebras $\h \subset \g$ that are abelian and certain subclasses of maximally abelian subalgebras. The Cartan subalgebra forms the generators the adjoint action of which on $\g$ enable its root-space decomposition which in turn gives rise to an abstracted root system. The root system can then in turn be related (given a fixed ordering or choice of basis) to simple roots. These roots can be used to form a basis of $\g$, from which a Cartan matrix and Dynkin diagram may be constructed and, as we show in Chapter \ref{chapter:Time optimal quantum geodesics using Cartan decompositions}, is related to analytic forms of time-optimal unitary synthesis. The relevance of root systems, Cartan matrices and such diagrams in quantum control settings is explored primarily in the final Chapter of this work.

 %=========Adjoint action
\subsection{Adjoint action and commutation relations} \label{sec:alg:Adjoint action and commutation relations}
% \subsubsection{Adjoint action}
An important representation used throughout this work is the \textit{adjoint representation}. We set out its definition below.
\begin{definition}[Adjoint representation]\label{defn:alg:adjointrepresentation}
    Given a Lie group $G$, Lie algebra $\g$ and $X \in \g$ with smooth isomorphism $\Ad_g(h) = g h g^\inv$ where $g,h \in G$, there exists a corresponding Lie algebra isomorphism $\ad_X: \g \to \g$ such that:
    \begin{align*}
        \exp((\ad_\g)(X)) = g(\exp(X))g^\inv
    \end{align*}
    i.e. $\exp(\ad_\g(X))=\Ad_G(\exp(X))$.
\end{definition}
In the above, the former being the Lie algebra adjoint, the latter the group adjoint action, the two concepts being related by the exponential map. We denote $\ad_g(X) = g X g^\inv$. We note that each Lie group exhibits the structure of a manifold that is real and analytic (under multiplication and inversion and for the exponential map). Moreover, as noted in \cite{knapp_lie_1996}, it can be shown that each real Lie algebra has a one-to-one finite dimensional representation on complex vector spaces $V(\C)$.

\subsection{Adjoint expansions} \label{sec:alg:Adjoint expansions}
We note for completeness a few additional concepts. Firstly \textit{automorphisms} of a Lie algebra $\g$ are invertible linear maps $L$ such that $[L(X),L(Y)]=L([X,Y])$ for $X,Y \in \g$. In addition we have that $\ad_{L(X)} = L(\ad_X)L^\inv$. The Killing form is invariant under such automorphisms $B(L(X),L(Y))=B(X,Y)$ for $X,Y \in G$.

Note that the adjoint action of group elements can be expressed via conjugation as $\text{Ad}_h(g) = hgh^\inv$where $g,h \in G$. The adjoint action of a Lie algebra upon itself, which we focus on, is via the commutator namely $\ad_X(Y) = [X,Y]$ where $X,Y \in \frak{g}$. Thus terms such as $\ad_\Theta(\Phi)$ involve calculation of commutation relations among generators. We also note the following relation (utilised in Chapter \ref{chapter:Time optimal quantum geodesics using Cartan decompositions}):
\begin{align}
    e^{i\Theta}Xe^{-i\Theta} = e^{i\ad_\Theta}(X) = \cos\ad_\Theta(X)+i\sin\ad_\Theta(X).
    \label{eqn:alg:econjsinhcosh}
\end{align}
The $\cos \ad_\Theta (X)$ term can be understood in terms of the cosine expansion:
\begin{align}
    \cos \ad_\Theta (X) = \sum_{n=0}^\infty \frac{(-1)^n}{(2n)!} (\ad_\Theta))^{2n} (X).
\label{eqn:alg:cosadthetaXexpansion}
\end{align}
Thus each term is a multiple of $\ad_\theta^2$. For certain choices of generators $\Theta$ and $X \in \frak{k}$ this effectively means that the adjoint action acts as an involution up to scalar coefficients given by the relevant root functional 
$\alpha(X)$ (see below). Thus in the $SU(2)$ case, $\ad^{2n}_{J_y}(J_z) \propto J_z$ (with $n=0,1,2,...$), each application of the adjoint action acquires a $\theta^2$ term (from $\Theta = i\theta J_y$), such that the series can be written:
\begin{align}
    \cos \ad_{\Theta} (-iJ_z) = \sum_{n=0}^\infty \frac{(-1)^n}{(2n)!} (\theta))^{2n} (J_z) = \cos(\theta)(-iJ_z)
\label{eqn:alg:su2cosadthetaXexpansion}
\end{align}
More generally, the Cartan commutation relations exhibit such an involutive property given that for $\Theta \in \frak{k}$:
\begin{align} \ad_\Theta(\frak{k})=[\Theta,\frak{k}] \subset \frak{p} \qquad \ad^2_ \Theta(\frak{k})=[\Theta,[\Theta,\frak{k}]] \subset \frak{k}
\end{align}
In the general case, assuming an appropriately chosen representation, each application of the adjoint action by even $n$ or odd $2n+1$ powers (and thus our $\cos \theta$ and $\sin \theta$ terms) will be scaled by an eigenvalue $\alpha$, such eigenvalue being the root (see below for an example). 
Thus we have $\ad_{\Theta}^{2n}(\frak{k}) \subset \frak{k}$ and:
\begin{align}
\cos(\ad_\Theta) \Phi = \cos\alpha(\theta)\Phi 
\label{eqn:alg:cosadthetaphi=costhetaphi}
\end{align}
(for $\Phi \in \k$) such as in equation (\ref{eqn:cartan:main-x=1-cosadthetabigphi}) and similarly for the sine terms. For the $\sin\ad_\Theta(X)$ terms, by contrast, the adjoint action is of odd order: 
\begin{align}
    \sin \ad_\Theta (X) = \sum_{n=0}^\infty \frac{(-1)^{n}}{(2n+1)!} (\ad_\Theta))^{2n+1} (X)
\label{eqn:alg:cosadthetaXexpansion}
\end{align}
such that $\ad_{\Theta}^{2n+1}(\frak{k}) \in \frak{p}$, hence the sine term arising in our Hamiltonian form equation (\ref{eqn:general:timeoptimalhamiltonian}) which, when conjugated with $\Lambda$ results in a Hamiltonian given in terms of control generators, namely:
\begin{align}
    [\sin \ad_\Theta(\frak{k}),\Lambda] \subset \frak{p}
\end{align}
By convention, sometimes one sees in the literature $KAK$ decompositions written in the form $e^{k_1}\frak{a}e^{k_2}$ where $k_1,k_2 \in \frak{k},e^{k_i}\in K$. Here the two exponentials are group elements while $\a \subset \p$. Thus in equation (\ref{eqn:alg:econjsinhcosh}) we write the adjoint action as conjugation given it involves the Lie group (rather than Lie algebraic) elements, while for the action of a Lie algebra on itself, we would use the commutation relation. 
% For completeness \hl{\textbf{FIX}}: 
% \begin{align}
%     e^{k_1} e^{a} \approx e
% \end{align}
The choice is usually one of convenience.

%===========Real forms and Root systems
\subsection{Real and complex forms}\label{sec:alg:Real and complex forms}
 Complex Lie algebras $\g$ (over $\C$) and real Lie algebras $\g_0$ (over $\R$) are related via complexification and the related concept of the real form of a complex Lie algebra. Such complexification of Lie algebras is relevant in particular to formulating Cartan decompositions of Lie algebras and their associated groups (see definition \ref{defn:alg:cartandecomposition}). We begin with a definition of real forms.
\begin{definition}[Real form]\label{defn:alg:realforms}
    We say that a real vector space $V(\R)$ is the real form of a complex vector space $W(\C)$ when the two are related as:
    \begin{align*}
        W^\R = V \oplus i V
    \end{align*}
    Where we denote $W$ the complexification of $V$.
\end{definition}
This equivalence is a statement that a real vector space can be complexified in the manner above. For Lie algebras, we often denote a real Lie algebra (as a vector space over $\R$) via a 0 subscript $\g_0$. In such terms, we may denote:
\begin{align}
    \g = \g_0^\C = \g_0 + i\g_0 \label{eqn:alg:liealgebracomplexification}
\end{align}
as the complexification of $\g_0$ (and equivalently indicating that $\g_0$ is the real form of the complex Lie algebra $\g$). Doing so allows the retention of certain properties of $\g_0$ while allowing complex coefficients yet retaining structure of interest from $\g_0$ (including decompositional structure). We can describe the real forms of $\g$ via $\g=\g_0 + i\g_0$, analogous to splitting $\C^n$ into $\R^{2n}$ (parametrising the imaginary part). In particular for $M\in M_n(\C)$ we have under this transformation (used in equation (\ref{eqn:qgml:complextorealmx}) below):
\begin{align}
    Z(M) &= \begin{pmatrix}
        \text{Re} M & -\text{Im} M\\
        \text{Im} M & \text{Re} M
        \label{eqn:alg:complextorealmx}
    \end{pmatrix}
\end{align}

\section{Cartan algebras and Root-systems}\label{sec:alg:Cartan algebras and Root-systems}
\subsection{Complex Semi-simple Lie Algebras}\label{sec:alg:Complex Semi-simple Lie Algebras}
Cartan subalgebras are of central importance across an expansive variety of physical and informational problems in physics and quantum computation. The algebras derive from Cartan's study of symmetry groups and provide a means of characterising the underlying or fundamental properties of such groups, such as their associated root systems. In turn, as we discuss and show throughout this work, those symmetry-related properties, such as roots and weights, have significant implications for symmetry reduction and characterisation of a wide range of physical and mathematical phenomena. This includes, in particular, weight systems (or set of weights) associated with representations of a Lie algebra or Lie group. Weights are elements of the dual space (see definition \ref{defn:quant:Dual space}) of a Cartan subalgebra and label the eigenspaces of the representation with respect to the Cartan subalgebra.
Root systems are primarily used in the classification and structure theory of Lie algebras, while weight systems are used in the study and classification of their representations. As noted in the literature, the process involves (a) identifying the (complex semi-simple Lie algebra) $\g$, (b) selecting a Cartan subalgebra $\h$, (c) constructing an abstract reduced root system, (d) selecting a choice of ordering and (e) then constructing a Cartan matrix.

\subsection{Roots}\label{sec:alg:Roots}

We define root systems as per below \cite{knapp_lie_1996,hall_lie_2013}. Given a subalgebra $\h \subset \g$ with a (diagonal) basis $H$ and a dual space $\h^*$ with basis elements $e_j \in \h^*$, we recall that the duals are a vector space of functionals $e_j: V \to \C$. For a matrix Lie algebra, We can construct a basis of $\g$ given by $\{\h,E_{ij}\}$ where $E_{ij}$ is 1 in the $(i,j)$ location and zero elsewhere. We are interested in the adjoint action of elements of $H \in \h$ on each such eigenvector as:
\begin{align*}
    \ad_H(E_{ij}) = [H,E_{ij}] = (e_i(H)-e_j(H))E_{ij} = \alpha E_{ij}
\end{align*}
where $e_j$ selects out the $(j,j)^{th}$ element of a matrix (recalling duals $e_j(v)=v_j \delta_{ij}, v_j \in \C$ form a vector space of linear functionals). That is $\alpha = e_i - e_j$ is a linear functional on $\h$ such that $\alpha:\h \to \C$. Such functionals $\alpha$ are denoted roots. Thus $\g$ can be decomposed as:
\begin{align}
    \g = \h \oplus_{i\neq j} \C E_{ij} = \h \oplus_{\alpha \in \Delta} \g_\alpha \qquad \g_\alpha = \{X \in \g | \ad_H(X) = \alpha X, H \in \h \} \label{eqn:alg:rootdecompositionLiealgebra}
\end{align}
This is the root space decomposition of $\g$ with $X \in \g_\alpha$, denoted the \textit{Cartan-Weyl basis}. As we see below, such decomposition is related to the existence of Cartan subalgebras and eventually Cartan decompositions, relevant to results in above Chapters. The commutation relations are given by:
\begin{align}
[\mathfrak{g}_{\alpha}, \mathfrak{g}_{\beta}] := \begin{cases}
\mathfrak{g}_{\alpha + \beta} & \text{if } \alpha + \beta \text{ is a root} \\
0 & \text{if } \alpha + \beta \text{ is not a root or } 0 \\
\subseteq \mathfrak{h} & \text{if } \alpha + \beta = 0.
\end{cases} \label{eqn:alg:rootcommutationrelations}
\end{align}
with $[E_{ij}, E_{ji}] = E_{ii} - E_{jj} \in \h$. Roots are then defined as positive or negative (with $-$ and may be ordered as a sequence. Each root is a  nongeneralised weight of $\ad_\h(\g)$ (a root of $\g$ with respect to $\h$). The set of roots is denoted $\Delta(\g,\h)$. 

\subsection{Cartan subalgebras}\label{sec:alg:Cartan subalgebras}
Given a Lie algebra representation, we can construct a system analogous to a root system, via a weight space and weight vectors, as follows.
\begin{definition}[Weights]\label{defn:alg:weights}
    Given a representation $\pi: \h \to V(\C)$ and roots $\alpha \in \h^*$, define the generalised weight space as:
    \begin{align*}
        V_\alpha = \{  v \in V | (\pi(H)-\alpha(H)1)^nv=0, \forall H \in \h \}
    \end{align*}
    for $V_\alpha\neq 0$.
\end{definition}
The elements of $v \in V_\alpha$ are generalised weight vectors with $\alpha$ the weights. If this condition holds, we say $\h$ is a nilpotent Lie algebra, such that there are finitely many generalised weights and $V = \oplus_\alpha V_\alpha$ (weight space decomposition). Weights needn't be linearly independent and are dependent upon roots. For a Lie algebra, we denote $V_\alpha$ as $\g_\alpha$. It can be shown that for such a nilpotent $\h$ relative to $\ad_g \h$ (the action on $\h$), the generalised weight vectors have the following properties: (a) $\g = \oplus \g_\alpha$, (b) $\h \subseteq \g_0$ and (c) $[\g_\alpha, \g_\beta] \subseteq \g_{\alpha + \beta}$ (zero if $\alpha + \beta$ is not a weight etc). Note that
\begin{align*}
    \g_\alpha &= \{  X \in \g | (\ad_H-\alpha(H)1)^nv=0, \forall H \in \h \}\\
    \g &= \h \oplus_{\alpha \in \Delta} \g_\alpha \label{eqn:alg:rootspace decompositiongalpha}
\end{align*}
i.e. $\pi$ is the adjoint action of $\h$ on $X \in \g$. Here $\g_0$ (not to be confused with the real $\g_0$ we discuss below) is then the weight space subalgebra associated with the zero weight under this adjoint action, it denotes the weight space corresponding to the zero weight in the decomposition of the Lie algebra $\g$ (the centralizer of $\h \subset \g$). We come now to the definition of a Cartan subalgebra.
\begin{definition}[Cartan subalgebra]\label{defn:alg:cartansubalgebra}
  A Cartan subalgebra $\h \subset \g$ is a nilpotent Lie algebra of a finite-dimensional complex Lie algebra $\g$ such that $\h = \g_0$. Cartan subalgebras are maximally abelian subalgebras of $\mathfrak{g}$. 
\end{definition}
The algebra $\h$ is a Cartan subalgebra if it is the normalizer with respect to $\g$, that is if $N_\g(\h) = \{ X\in\g | [X,\h] \subseteq \h \}$. It can be shown that all finite complex Lie algebras $\g$ have Cartan subalgebras. For complex semisimple Lie algebras, Cartan algebras $\h$ are abelian ($B([H_k,H_j])=0$ for Killing-form $B$ and $H_k,H_j \in \h$). A Cartan subalgebra $\h$ is the maximally abelian subalgebra such that the elements of $\ad_\g(\h)$ are simultaneously diagonalisable. Moreover, all such Cartan subalgebras of $\g$ are conjugate via an automorphism of $\g$ in the form of an adjoint action. With this understanding of Cartan subalgebras we can now generalise the concepts of roots.

\subsection{Root system properties} \label{sec:alg:Root system properties}
Roots exhibit a number of properties \cite{knapp_lie_1996}. We set out a number of these relevant to results in later Chapters and in particular our methods in Chapter \ref{chapter:Time optimal quantum geodesics using Cartan decompositions}. 
\begin{definition}[Roots]\label{defn:alg:roots}
    Non-zero generalised weights of $\ad_\h(\g)$ are denoted the \textit{roots} of $\g$ with respect to the Cartan subalgebra. Denote the set of roots $\alpha$ is $\Delta(\g,\h)$
\end{definition}
Given a Killing form $B(\cdot,\cdot)$, a few standard results hold: (a) $B(\g_\alpha,\g_\beta)=0$ for $\alpha,\beta \in \Delta$ (including $0$); (b) $\alpha \in \Delta \implies -\alpha \in \Delta$, (c) each root $\alpha$ is associated with a $H_\alpha \in \h$ such that we can identify a correspondence between the root functional and Killing form given by:
\begin{align*}
    \alpha(H) = B(H,H_\alpha) \qquad \forall H \in \h
\end{align*}
noting that $\Delta$ spans $\h^*$. Importantly we have that for each eigenvector $E_\alpha \in \g$ (non-zero) that:
\begin{align*}
    [H,E_\alpha] &= \alpha(H)E_\alpha
\end{align*}
(see \cite{knapp_lie_1996} for standard results e.g Lemma 2.18). Other results include that $\ad_\h(\g)$ is simultaneously diagonalisable, and that the Killing form has an expression in terms of the functional:
\begin{align*}
    B(H,H') = \sum_\alpha \alpha(H)\alpha(H') \qquad B: \h \times \h \to \C
\end{align*}
together with normalisability (such that $B(E_\alpha,E_{-\alpha})=1)$ and standard commutation relations:
\begin{align}
    [H_\alpha,E_\alpha] &= \alpha(H_\alpha)E_\alpha & [H_\alpha,E_{-\alpha}] &= -\alpha(H_{-\alpha})E_{-\alpha} & [E_\alpha,E_{-\alpha}] &= H_\alpha.
\end{align}
Finally in this section, we note concepts of relevance to Cartan matrices applied in the final chapter. We denote \textit{root strings}, or $\alpha$-strings as follows: given $\alpha \in \Delta, \beta \in \Delta \cup \emptyset$, the $\alpha$-string that contains $\beta$ is the set of all elements $\beta + n\alpha \in \Delta \cup \emptyset$ for $n \in \Z$. Essentially denoting a (symmetric) string of roots $\beta + n\alpha$ for $-p \leq n \leq q$ ($p,q \geq 0)$ with:
\begin{align*}
    p-q = \frac{2\braket{\beta,\alpha}}{\braket{\alpha,\alpha}}.
\end{align*}

%====Abstract root systems
\subsection{Abstract root systems}\label{sec:alg:Abstract root systems}
Abstract root systems are an important way of capturing the essence of the symmetries within a Lie algebra via enabling the construction of relations among roots that may be interpreted using the formalism and language of symmetry. 
\begin{definition}[Root System]\label{defn:alg:Root System}
A root system $(V, \Delta)$ is a finite-dimensional $\Delta \in V(\R)$ equipped with an inner product $\langle \cdot , \cdot \rangle$ and norm $|\cdot |^2$ with a subset $\Delta = \{\alpha \in V | \alpha \neq 0 \}$ satisfying (a) $V = \text{span}(\Delta)$, (b) $\alpha \in \Delta$ and $c \in \mathbb{R} \implies c\alpha \in \Delta$ only if $c = \pm 1$, (c) given the linear transformation $s_{\alpha}$:
\begin{align*}
    s_{\alpha} \cdot \beta = \beta - 2 \frac{\langle \beta, \alpha \rangle}{\langle \alpha, \alpha \rangle} \alpha, \quad \beta \in E
\end{align*}
for $\alpha,\beta \in \Delta$ then $s_{\alpha} \cdot \beta$ and (d):
\begin{align*}
    \frac{\langle \beta, \alpha \rangle}{2 \langle \alpha, \alpha \rangle} \in \Z.
\end{align*}
\end{definition}
The rank of $V$ is $\dim(V)$. The elements $\alpha \in \Delta$ are roots. A root system is reduced if it satisfies (a) to (d) above, or non-reduced if it doesn't satisfy (b). Symmetric spaces root systems of interest in this work are of the non-reduced type. $s_\alpha$ is a reflection about the hyperplane with respect to $\alpha$ (an orthogonal transformation with determinant $-1$). Such reflections are said to generate an orthogonal group of $V$, the Weyl group $W$, generated by such reflections. This general formulation allows root systems to be associated to all complex semisimple Lie algebras. 

\subsection{Reduced abstract root systems}\label{sec:alg:Reduced abstract root systems}
A reduced abstract root system in $\h_0^*$ is one where $\alpha \in \Delta$ but $2\alpha \notin \Delta$. Familiar Lie algebraic concepts apply to root systems such as the classification of \textit{reducible} root systems (orthogonal decompositions $V = \oplus V_k$ such that $\alpha$ belongs to a $V_k$) and corresponding irreducible root systems. An important element in the geometric characterisation of root systems is set out below. Given $\alpha,\beta$ (linearly independent) and $\braket{\alpha,\alpha} \geq \braket{\beta,\beta}$ it can be shown that there are limited allowed angles $\theta$ and length ratios among the roots via the proposition that: (a) $\braket{\alpha,\beta}=0$, (b) $\braket{\alpha,\alpha} = \braket{\beta,\beta}$ with $\theta=\pi/3$ or $2\pi/3$, (c) $\braket{\alpha,\alpha} = 2\braket{\beta,\beta}$ (for $\theta=\pi/4,3\pi/4)$ or (d) $\braket{\alpha,\alpha} = 3\braket{\beta,\beta}$ (for $\theta=\pi/6,5\pi/6)$. If $\theta \in (\pi/2,\pi)$ (obtuse) then $\alpha+\beta$ is a root or $\theta \in (0,\pi/2)$ then $\pm(\alpha-\beta)$ are roots. These angles can be used to plot the relevant root system (see Chapter \ref{chapter:Time optimal quantum geodesics using Cartan decompositions}). In more general settings, the geometric representation of roots, under certain assumptions, then exhibits certain permutation symmetries and the applicable Weyl group is defined in terms of these. 

\subsection{Ordering of root systems}\label{sec:alg:Ordering of root systems}
One final concept before coming to the Cartan matrix is that of fixing a lexicographic ordering related to the concept of positivity of roots. Given such a set of roots:
\begin{align*}
    \Delta = \{\alpha_1, \alpha_2, \ldots, \alpha_l\}
\end{align*}
we introduce an ordering as follows. A root $\beta = \sum_{i=1}^l b_i\alpha_i$ is positive with respect to the lexicographic ordering if the first non-zero coefficient $b_j$ in its expansion with respect to the basis $\Delta$ is positive. Conversely, the root $\beta$ is negative if this first non-zero coefficient is negative. The lexicographic order is defined by comparing roots: for two distinct roots $\beta = \sum_{i=1}^l b_i\alpha_i$ and $\gamma = \sum_{i=1}^l c_i\alpha_i$, we say that $\beta > \gamma$ if, at the first index $j$ where $b_j$ and $c_j$ differ, we have $b_j > c_j$. This comparison allows partitioning the root system into two disjoint sets $\Delta^+$ and $\Delta^-$ of positive and negative roots.  

The highest root in a root system is the maximal element with respect to this lexicographic ordering and is always a positive root. This root plays a pivotal role in the structure and representation theory of semisimple Lie algebras.

As we articulate in Chapter \ref{chapter:Time optimal quantum geodesics using Cartan decompositions}, the construction of the abstract root system for Lie algebras (related to quantum control problems of interest) involves determining root systems. We note that $\alpha$ is \textit{simple} if $\alpha > 0$ and it is linearly independent of other roots. It can be shown that any positive root can be decomposed as $\alpha = \sum_i n_i \alpha_i$ (with $\alpha_i$ simple and $n_i \geq 0$). The sum of $n_i$ is an integer denoted the \textit{level} of $\alpha$ with respect to the set of simple roots (important as the set of simple roots generates the entire root system through integer linear combinations).

\subsection{Cartan matrices}\label{sec:alg:Cartan matrices}
We now define the Cartan matrix used in Chapter \ref{chapter:Time optimal quantum geodesics using Cartan decompositions}. To do so we fix a root system $\Delta$, assume each $\alpha \in \Delta$ is reduced and fix an ordering as per above. Data from Cartan matrices may be represented in diagrammatic form via Dynkin diagrams.
\begin{definition}[Cartan matrix]\label{defn:alg:Cartan matrix}
    Given $\Delta \in V$ and simple root system $\Pi = \{\alpha_k\}$, $k=1,...,n=\dim V$, the Cartan matrix of $\Pi$ is an $n \times n$ matrix $A = (A_{ij})$ where:
    \begin{align}
        A_{ij} = 2\frac{\braket{\alpha_i,\alpha_j}}{|\alpha_i|^2}
    \end{align} 
\end{definition}
A variety of other standard properties apply to abstract root systems $\Delta$ constraining the roots and their ratios (e.g. $2\braket{\beta,\alpha}/|\alpha^2|=0,\pm 1,\pm 2, \pm 3, \pm 4$ and $\pm 1$ if $\alpha,\beta$ are non-proportional with $|\alpha| \leq |\beta|$). These and other constraints limit the ratios and relation of roots in ways that allow their representation in terms of Cartan matrices and diagrammatic representations. Different orderings (enumerations) of $\Pi$ lead to different $A_{ij}$, however they are all conjugate to each other. Cartan matrices satisfy certain properties e.g. $A_{ij} \in \Z, A_{ii}=2, A_{ij}\leq 0 (i\neq j), A_{ij}=0 \leftrightarrow A_{ji}=0$ and the existence of a diagonal matrix $D$
that renders $A$ symmetric positive definite under conjugation. The (multiple) block diagonality of a Cartan matrix implies its reducibility, otherwise the matrix is irreducible. From the Cartan matrix, we can construct a Dynkin diagram (as distinct from a root diagram) as follows. 

%===Dynkin diagram

\begin{figure}[h]
\centering
\begin{tikzpicture}
  % Define the style for the nodes
  \tikzset{dynkinnode/.style={circle, draw, minimum size=25pt, inner sep=0pt}}
  
  % Draw the first node
  \node[dynkinnode] (first) {};
  
  % Label above the first node
  \node[above=1mm of first] {1};
  
  % Label below the first node
  \node[below=1mm of first] {$\alpha_1$};

  % Draw the second node
  \node[dynkinnode] (second) [right=2cm of first] {};

  \node[below=1mm of second] {$\alpha_2$};

  % Draw the third node
  \node[dynkinnode] (third) [right=2cm of second] {};

  % Label below the third node
  \node[below=1mm of third] {$\alpha_n$};
  
  % Connect the nodes
  \draw[-] (first) -- (second);
  \draw[dashed] (second) -- (third);
\end{tikzpicture}
\caption{Expanded Dynkin diagram of type \( A_n \) with labeled vertices and edges. The numbers above the nodes indicate the length of the roots relative to each other. $A_{ij}A_{ji}$ determines the number of lines (or the type of connection) between vertices $i$ and $j$. This product can be 0 (no connection), 1 (single line), 2 (double line), or 3 (triple line), representing the angle between the corresponding roots. Additionally, when $A_{ij}A_{ji} > 1$, an arrow is drawn pointing from the longer root to the shorter root. }
\label{fig:alg:dynkin_diagram_an_expanded}
\end{figure}
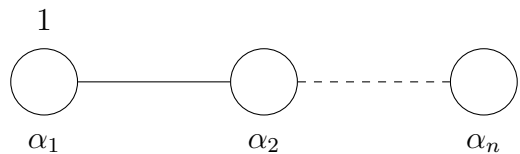

\begin{definition}[Dynkin diagram]\label{defn:alg:dynkindiagram}
    A Dynkin diagram is a graph diagram of a set of simple roots $\Pi$. Let each simple root $\alpha_i$ be a vertex of the graph, with a weight proportional to $\braket{\alpha_i,\alpha_i} = |\alpha_i|^2$. The graph construction rules are as follows. Two simple roots  $\alpha_i,\alpha_j$ are connected by $A_{ij}A_{ji}$ edges (which may be zero).
\end{definition}
While Dynkin diagrams are not focus for use in this work, we do relate them to quantum control results in the Chapter \ref{chapter:Time optimal quantum geodesics using Cartan decompositions}, hence here we mention a few sundry properties. For an $\l \times l$ Cartan matrix $A$, the associated Dynkin diagram has the following structure: (a) there are at most $l$ pairs of vertices $i<j$ having one edge (at least) between; (b) the diagram has no loops; and (c) at most three edges (triple points) connect to any node. The Cartan matrix may be used to determine a Dynkin diagram (up to scaling factors) with a description set out in Fig. \ref{fig:alg:dynkin_diagram_an_expanded} (see \cite{knapp_lie_1996} for more detailed exposition). 

%=====Dynkin diagrams 

By constructing an abstract reduced root system, one can then show that for two complex semisimple Lie algebras with isomorphic abstract reduced root systems, the associated Cartan matrices are isomorphic.  If two complex semisimple Lie algebras have the same Dynkin diagram, they are isomorphic, thereby sharing the same algebraic structure. Operations on Cartan matrices correspond to operations on Dynkin diagrams, allowing visualisation of transformations. It can be shown that the choice of Cartan matrix is independent of the choice of positive ordering up to a permutation of index and that the Cartan matrix determines the reduced root system. The Weyl group, as the set of orthogonal transformations , can be used to determine $\Delta^+$ and $\Pi$.

\section{Cartan decompositions}\label{sec:alg:Cartan decompositions}
An important set of theorems for the results in Chapters \ref{chapter:Quantum Geometric Machine Learning} and \ref{chapter:Time optimal quantum geodesics using Cartan decompositions} are those related to the Cartan decomposition of semisimple Lie groups and their associated algebras. The Cartan decomposition allows decomposition of a classical matrix Lie algebra into Hermitian and skew-Hermitian components. For a given a complexified semisimple Lie algebra $\g = \g_0^\C$, a corresponding Cartan involution $\theta$ is associated with the decomposition $\g = \k \oplus \p$. In turn, this allows a global decomposition of $G=KAK$ where $K = e^\k$ and $A = e^\a$ where $\a \subset \p$ (a maximal abelian subspace of $\p$), which can be viewed analogously in terms of the polar decompositions of matrices. Of importance to our result in Chapter \ref{chapter:Time optimal quantum geodesics using Cartan decompositions} is that Cayley transforms (see below) may be used to transform between any Cartan subalgebras up to conjugacy. We use this result to define our appropriate control decomposition in Chapter \ref{chapter:Time optimal quantum geodesics using Cartan decompositions}. We briefly set out relevant background theory below, before moving to geometric concepts and interpretation.  

\subsection{Compact Real Forms} \label{sec:alg:Compact Real Forms}
For classical semisimple groups, the matrix Lie algebra $\g_0$ over $\R,\C$ is closed under conjugate transposition, in which case $\g_0$ is the direct sum of its skew-symmetric $\k_0$ and symmetric $\p_0$ members. Recall from equation (\ref{eqn:alg:liealgebracomplexification}) that complexification of $\g_0$ is denoted $\g = \g_0^\C = \g_0 + i\g_0$. The real Lie algebra $\g_0$ has a decomposition as the direct sum $\g_0 = \k_0 \oplus \p_0$. We can construct a real vector space of complex matrices (that is, where coefficients of matrices are real) as $\u_0 = \k_0 + i\p_0$ as a subalgebra. As Knapp notes \cite{knapp_lie_1996}, there are certain requirements with respect to $\g_0$ and $\u_0$ such as $\k_0 = \g_0 \cap \u_0$ and $\p_0 = \g_0 \cap i\u_0$ which allow us to decompose $\g_0 = \k_0 \oplus \p_0$ in a way that in turn allows the complexification:
\begin{align}
    \g = \k \oplus \p
\end{align}
As the focus of this work is geometric and machine learning techniques for Lie groups and algebras of relevance to quantum control, we omit specific standard steps showing the complexification of $\g = \k_0 \oplus \p_0$ from this work, see \cite{knapp_lie_1996,knapp_representation_2001,hall_lie_2013} for more detail. Note that we can define the mapping $\theta(X) = -X^\dagger$ (negative of complex conjugate transpose) as an involution (that squares to identity) with $\theta[X,Y] = [\theta(X),\theta(Y)]$. We then define a Cartan involution as follows.
\begin{definition}[Cartan involution]\label{defn:alg:cartaninvolution}
    Given a Killing form $B$, $X,Y\in\g$ and an involution $\theta$, a Cartan involution is a positive definite symmetric bilinear form:
    \begin{align}
        B_\theta(X,Y) = -B(X,\theta Y)
    \end{align}
\end{definition}
It can be shown (see below) that involutions on $\g_0$ induce Cartan involutions on $\g$ (for complex semisimple $\g$, the only Cartan involutions of $\g^\R$ are conjugations with respect to real forms of $\g$). 
Cartan involutions are conjugate up to inner automorphisms Int $\g$. This result is related to the existence statement that $\g$ is equipped with a Cartan involution which is so conjugate (the proofs are usually via $\g_0$ which is then complexified). From the existence of this involution, we infer the existence of the relevant Cartan decomposition.
%===========
\begin{definition}[Cartan decomposition]\label{defn:alg:cartandecomposition}
    Given a complex semisimple Lie algebra $\g$ with Cartan involution $\theta$, $\g$ can be decomposed as:
    \begin{align}
        \g = \k \oplus \p
    \end{align}
    with $\k$ the +1 symmetric eigenspace $\theta(\k)=\k$ and $\p$ the -1 anti(skew)symmetric eigenspace $\theta(\p)=-\p$ satisfying the following commutation relations:
    \begin{align}
        &[\k,\k] \subseteq \k &[\k,\p] &\subseteq \p & [\p,\p] \subseteq \k \label{eqn:alg:cartancommutationrelationsmain}
    \end{align}
\end{definition}
Here $\k,\p$ are orthogonal with respect to the Killing form in that $B_\g(X,Y)=0=B_\theta(X,Y)$ i.e. for $X \in \k$ and $Y \in \p$ we have $B(X,Y) = 0$. For the corresponding Lie group $G$, the existence of the Cartan decomposition of $\g$ (where $K \subset G$ is analytic with Lie algebra $\k$) gives rise to a global Cartan decomposition of $G$. The global Cartan decomposition $G = K \exp(\p)$ together with the fact that $\p = \bigcup_{k \in K}\Ad_k\a$ means that the global Cartan decomposition is expressible as $G=KAK$ which we define below. 
\begin{definition}[KAK Decomposition]\label{defn:alg:KAKdecomposition}
    Given $G$ with a Cartan decomposition as in definition (\ref{defn:alg:cartandecomposition}), $G$ can be decomposed as $G=KAK$ where $K = \exp(\k), A=\exp(\a)$. Every element $g \in G$ thus has a decomposition as $g=k_1 a k_2$ where $k_1,k_2 \in K, a \in A$.
\end{definition}
Where such a Cartan decomposition is obtainable, we denote (as discussed further on) $G/K$ as a global Riemannian symmetric space. Note that in this case we can decompose an arbitrary $U \in G$ as:
\begin{align}
    U = kac \qquad U = pk \label{eqn:alg:cartankakandpk}
\end{align}
where $k,c \in \exp(\k) = K$ and $p \in \exp(\p)=P$ and $a \in \exp(\frak{a}) \in A \subset P$. 

The Cartan decompositions of semi-simple Lie algebras and groups is central to results in our Chapter on global Cartan decompositions for time optimal control. It is also important to results relevant to Chapter \ref{chapter:Quantum Geometric Machine Learning} given the role of Cartan decompositions in quantum control formalism \cite{khaneja_cartan_2001,khaneja_optimal_2005,brockett_sub-riemannian_nodate,dalessandro_introduction_2007}. In particular, we rely upon the $KAK$ decomposition and the assumption that $a \in \exp(\a)$ is parametrised by a constant parameter (the `constant-$\theta$' ansatz) which, as we demonstrate, simplifies the minimisation problem for finding certain time (and energy) optimal sequences of unitaries.

\subsection{Compact and non-compact subalgebras}\label{sec:alg:Compact and non-compact subalgebras}
For our results in Chapter \ref{chapter:Time optimal quantum geodesics using Cartan decompositions}, we require transformation of the relevant Cartan subalgebra (e.g. for $\su(3)$) to one which is maximally non-compact in order to prove our results relating to time-optimal control. This is because we require our maximally abelian subalgebra $\a \subset \h$ to be an element of our control subset, i.e. we require $\a \subset \p$. Because Cartan decompositions $\g = \k \oplus \p$ with associated involutions and $\h \subset \g$ are conjugate up inner automorphisms of $\g$ (denoted $\g$, namely those which have a representation as conjugation by an element $g \in G$, that is $\h' = \text{Int}_g(\h) = g \h g^\inv$) we can apply a generator in $\k$ to transform to a new subalgebra that is maximally non-compact (see below). In the literature, it can be shown that in particular any such $\h$ is thereby conjugate to a $\theta-$stable Cartan subalgebra (for involution $\theta$). This means that we can decompose the Cartan subalgebra as $\h = \t \oplus \a$ where $\t \subset \k, \a \subset \p$. As Knapp notes, the roots of $(\g,\h)$ are real on $\a_0 \oplus i\t_0$. The subalgebras $\t$ and $\a$ are compact and non-compact subalgebras of $\h$ respectively, with $\dim \t$ being the compact dimension and $\dim \a$ being the compact dimension. The Cartan subalgebra $\h = \t \oplus \a$ is maximally compact if its compact subalgebra is as large as possible and maximally non-compact in the converse case. As $\a \subset \p$, for a given Cartan decomposition, $\h$ is maximally compact if and only if $\a$ is a maximally abelian subspace of $\p$. 

%====Cayley transforms
\subsection{Cayley transforms}\label{sec:alg:Cayley transforms}
As we discuss in Chapter \ref{chapter:Time optimal quantum geodesics using Cartan decompositions}, our Cartan decomposition for quantum control problems related to symmetric spaces requires $\a \subset \h$ which is a maximally non-compact Cartan subalgebra. Cartan subalgebras $\h$ are conjugate to other Cartan subalgebras via a transformation of the form $h' = K\h K^\inv$. In certain cases, such as the typical Cartan decomposition of $SU(3)$ that we explicitly explore in our final chapter, $\h$ is not maximally non-compact (e.g. for $\su(3)$ choosing $\h = \{ \lambda_3, \lambda_8\}$ Gell-man generators renders $\h \subset \k$ while $\h \cap \p = \emptyset$). To remedy this, we apply Cayley transforms of which there are two types: (a) one constructed from an imaginary non-compact root $\beta$ (denoted by Knapp as $c_\beta$) whose application to $\h$ causes the intersection $\h \cap \p$ to increase by one dimension; and (b) another constructed from a real root $\alpha$ (denoted by Knapp as $d_\alpha$) whose application to $\h$ causes the intersection $\h \cap \p$ to decrease by one dimension. Our interest is in the first of these, $c_\beta$. An example of the form of such a transformation for $\sl(2,\C)$ is given by:
\begin{align}
    c_\beta &= \Ad\left(\exp\left(i\frac{\pi}{4}X\right)\right) \qquad X = \overline{E}_\beta + E_\beta \in \g_\beta \subset \p
\end{align}
where $E_\beta \in \g_\beta, \overline{E}_\beta \in \g_{-\beta}$. In our example, we choose $\gamma = (1,\sqrt{3})$ such that:
\begin{align*}
    E_\gamma + E_{-\gamma} = \lambda_4
\end{align*}
where $\lambda_4$ is the corresponding Gell-man matrix (see equation (\ref{eqn:cartan:su3gellmanmatrices}). We include a worked example of a Cayley transformation in Chapter \ref{chapter:Time optimal quantum geodesics using Cartan decompositions}. 
Having covered key elements of algebra and representation theory relevant to our results' Chapters, we now cover relevant background in differential geometry and then geometric control theory. Our focus is largely on the synthesis of geometric and algebraic formalism, where specific Lie groups are represented in terms of differentiable manifolds, Lie algebras in terms of fibre bundles (the Lie algebra of a Lie group can be associated with the structure group of a principal fibre bundle, and Lie algebra-valued differential forms play a crucial role in defining connections on these bundles) and where transformations are given in terms of connections (such as Levi-Civita and others). We also relate basic elements of the theory of Riemannian and subRiemannian manifolds, geodesic theory and the application of variational methods to the theory of Lie algebras and groups articulated above in a quantum context. Finally, we integrate our algebraic and geometric exegesis into the context of geometric control theory, with a specific focus on quantum control relevant to the main results of this work.

%========DIFFERENTIAL GEOMETRY
\chapter{Appendix (Differential Geometry)}\label{chapter:background:Differential Geometry}
The study of geometric methods usually begins with the definition and study of the properties of differentiable manifolds. One begins with the simplest concept of a set of elements, characterised only by their distinctiveness (non-identity), akin to pure spacetime points $\M$. The question becomes how to impose structure upon $\M$ in a useful or illuminating manner for explaining or modelling phenomena. To do so, in practice one assumes that $\M$ is a topological space, following which additional structural complexity, necessary for certain operations, such as analytical operations (differentiation etc) or measures (for integration), is added. The material on differential geometry in this Appendix is drawn from standard texts in the field \cite{isham_modern_1999,frankel_geometry_2011,montgomery_tour_2002,do_carmo_differential_2016,knapp_lie_1996,helgason_differential_1979,goldstein_classical_2002,kobayashi_foundations_1963}. Most results are presented without proofs which can be found in the collection of resources noted above. The Chapter begins with an outline of basic conceptualisation of manifolds, tangent planes, pushforwards and bundles. It them moves onto vector (and tensor) fields, $n$-forms, connections and parallel transport. The second section of this Appendix concentrates on geometric control theory, drawn primarily from the work of Jurdjevic \cite{jurdjevic_geometric_1997,jurdjevic_non-euclidean_1995,jurdjevic_control_1981,jurdjevic_optimal_1999,jurdjevic_hamiltonian_2001}, Boscain \cite{boscain_introduction_2021,boscain_invariant_2008}, D'Alessandro \cite{dalessandro_introduction_2007} and others. It focuses primarily on classical geometric control theory, but the results, especially where controls are constructed from semi-simple Lie groups and algebras, carry over to the quantum realm.

\section{Differential geometry}
\subsection{Manifolds and charts}\label{sec:geo:Manifolds and charts}
We begin with the concept of a \textit{topological manifold} $\M$, being a connected Hausdorff ($T_2$) space (a topological space such that for any two distinct points there exist disjoint neighbourhoods). From this point, we defined coordinate charts and atlases as follows.
\begin{definition}[Coordinate charts and atlases]\label{defn:geo:Coordinate charts and atlases}
    A \textit{coordinate chart} is a pair $(U,\phi)$ on $\M$ where $U \subset \M$ (an open subset) and $\phi:U \to \Km^m$ with $p \mapsto \phi(p)$ (usually taking $\Km = \R $ or $\C$) is a homeomorphism. The charts $\phi,\phi^\inv$ together form a family denoted as an \textit{atlas}.
\end{definition}
An atlas constitutes the assignment of a collection of open charts to the underlying manifold, which in turn allows the imposition of differential structure on the manifold. 

\begin{definition}[Differentiable manifold]\label{defn:geo:Differentiable manifold}
    A differentiable manifold is a Hausdorff topological space $\M$ together with a global differentiable structure.
\end{definition}
The coordinates of a point $p$ are maps from $p\in U \subset  \mathcal{M}$ to the Euclidean space $\K^m$ i.e. $(\phi^1(p),...,\phi^m(p))$. The \textit{coordinate functions} are each of these individual functions $\phi^\mu: U \to \K$. We have $m$ of them so that's how we get the map from $U$ to $\K^m$. These functions are defined in terms of a chain, or rather, composition that takes information from $p$, to $\K^m$ to $\K$. An example is Euclidean space $\Real^m$ which can be equipped with a differential structure via imposing a globally-defined coordinate chart, where the coordinates of vectors $v \in \Real^m$ have components $v=(x^1,...,x^m)$. Any finite dimensional vector space $V$ can be regarded as a differentiable manifold by choosing a basis set of vectors and using this to map $V$ isomorphically to $\R^m$ e.g. if $\{e_1,...,e_m \}$ is a basis of $V$, then the vector $v = \sum_i^m v^i e_i$ which becomes mapped to the $n$-tuple $(v^1,...,v^m)\in \R^m$. Again, the key point is that $\{e_i\}$ is the basis of $V$, whereas the coordinate functions (or set thereof) are the elements parametrised by $\R^m$. For certain manifolds, such as the sphere $S^1$ or the torus, there is no way of parametrising the space with a single coordinate function, hence we may require an atlas with different coordinate functions. In this case a mapping between coordinate functions that preserves smoothness is required. Of crucial importance for our analysis is the fact that a Lie group $G$ (see definition \ref{defn:alg:liegroups}) is a group that is also a smooth manifold via the smoothness of group operations.
%Lie group manifold
\begin{proposition}[Lie Group (Manifold)]\label{prop:geo:Lie Group (Manifold)} A Lie group $G$ equipped with a differentiable structure is a smooth differentiable manifold via the smoothness of group operations (arising from the continuity properties of $G$ i.e. being a topological group) given by multiplication $m: G \times G \to G$, defined by $m(g, h) = gh$, and inversion $i: G \to G$, defined by $i(g) = g^{-1}$.
\end{proposition}
Functions defined on $p \in \M$ that overlap multiple charts are required to be differentiable. Functions that preserve such structure are played by the role of $C^r$ functions, those which are differentiable up to order $r$. In particular we are interested in local representatives, being functions between manifolds $f: \M \to \N$ which have a representation (in a geometric sense) as a function between the coordinate charts $(U,\phi)_{\mathcal{M}} \to (V,\psi)_\mathcal{N}$. The idea is that we can say the function $f$ between manifolds is a $C^r$ function if all local representatives between the two coverings are themselves $C^r$ functions i.e. we can designate $f$ as a $C^r$ function if for all coverings (basically atlases) of $\mathcal{M},\mathcal{N}$ by charts (or rather coordinate neighbourhoods), the local representative functions are $C^r$. Functions that are differentiable require $r \geq 1$, while smooth functions require $C^\infty$.
\\
\\
Usually we are working with $C^\infty$ spaces. The complete atlas (a $C^\infty$ atlas) provides the differential structure required for differential calculus and analytic approximations thereby allowing a differentiable structure to be imposed on $\M$ (where $\Km = \C$ one also imposes holomorphic constraints upon functions defined on $\M$). For brevity, we denote a manifold as differentiable whether complex or real. In differential geometry, the structure (topology) preserving map is fulfilled by a $C^\infty$ (or technically $C^r$ where $r$ may or may not be finite) function such that the function $f:\M \to \N$ is a $C^\infty$ diffeomorphism assuming $f$ is a bijection and $f,f^\inv$ are both $C^\infty$. Two diffeomorphic manifolds can be regarded as `equivalent' or `copies' of a single abstract manifold. 
\\
\\
We denote $F$ as the set of smooth differentiable functions $C^\infty(\M)$ (with which we shall mainly be concerned with). Such functions compose differentiably and remain homeomorphic. The set of such functions $C^\infty$ forms an algebraic derivation (see below). It can be shown under typical criteria for $A \in V_{open} \in M$, there exists a $\varphi \in C^m$ such that $\varphi(C)=1,\varphi(\Bar{V})=0$ i.e. functioning as an indicator or classifier function (something of relevance to statistical learning of, for example, classifiers). Of relevance to product spaces in quantum information, product spaces of two manifolds $M \times N$ are represented by products in the vector space representation (see paper), so that the product is itself a manifold (topologically).We leave as understood usual features such as definitions of covering, paracompactness and normality of $\M$. With these concepts, we now move onto concepts of curves and tangent spaces. 

\subsubsection{Tangent spaces}\label{sec:geo:Tangent spaces}
Tangent spaces are particularly important in algebraic geometry, geometric control and geometric approaches to machine learning. The most basic example of a tangent space for a point in some vector space $x \in R^n$ with the tangent space typically denoted as $T\M$, representing the tangent space to the manifold $\M$. It is often illustrated by reference to the real vector algebra as:
\begin{align}
        T_x R^n:= \{ v \in \R^{n+1} | x \cdot v = 0 \}\label{eqn:geo:tangentspaceTxSn}
    \end{align}
to indicate the changed dimensionality (intuitively being constructed from derivatives at $p \in \M$) hence being orthogonal. More constructively for our purposes, tangent spaces can be thought of in a more algebraic way connected with the local differentiable properties of functions on the manifold such that tangent spaces, like differentiable manifolds, make no reference to \textit{ambient spaces} in their definition e.g. higher-dimensional vector spaces. As Isham \cite{isham_modern_1999} notes ``[t]he crucial question, therefore, is to understand what should replace the intuitive idea of a tangent vector as something that is tangent to a surface in the usual sense and which, in particular, `sticks out' into the containing space'' (p.72). To answer to this question is to construe the tangent as a property of curves in the manifold. We define a tangent vector in terms of properties of curves $\gamma \in \M$ which are in turn usually parametrised in a particular way (using affine parameters or arc length). Moreover, because there are many curves that are tangent, the concept of a tangent vector as an equivalence class of curves is used. 

\begin{definition}[Curves on manifold]\label{defn:geo:Curves on manifold}
    A curve $\gamma$ on a manifold $\M$ is a smooth (i.e. $C^\infty$) map:
    \begin{align*}
        \gamma: &\R \supset (-\epsilon,\epsilon) \to \mathcal{M} \qquad t \mapsto \gamma(t)
    \end{align*}
     (i.e. from an interval in $\R$ crossing 0 into $\mathcal{M}$), such that, $\gamma(0)=p$.
\end{definition}

As $t$ varies `infinitesimally' within $\epsint$, the curve moves infinitesimally away from $p \in \M$ but that it stays within the neighbourhood $U_p$. Next, we define a tangent as a property of curves.
\begin{definition}[Tangent]\label{defn:geo:Tangent}
    Curves are tangent if the image of two curves (i.e. two maps) at $t=0, t\in (-\epsilon,\epsilon)$ and their derivatives are identical. Formally, two curves $\gamma_1, \gamma_2$ are \textit{tangent} at a point $p \in \mathcal{M}$ if:
    \begin{enumerate}[(i)]
        \item $\gamma_1(0) = \gamma_2(0) = p$; and
        \item The derivatives of two curves in a local coordinate system are the same i.e. in some local coordinate system $(x^1,...,x^m)$ around the point $p$, the two curves are `tangent' in the usual sense as curves in $\R^m$ i.e:
        \begin{align*}
            \left.\frac{dx^i}{dt}(\gamma_1(t))\right|_0 = \left.\frac{dx^i}{dt}(\gamma_2(t))\right|_0 \qquad i=1,...,m
        \end{align*}
    \end{enumerate}
\end{definition}
If $\gamma_1,\gamma_2$ are tangent in one coordinate system, then they are tangent in any other coordinate system that covers $p \in \mathcal{M}$. Hence the definition of tangent is independent of the coordinate system chosen. The equivalence class of curves satisfying this condition at $p \in \M$ is sometimes denoted $[\gamma]$ which can be shown to form a vector space, hence we can reasonably denote tangents at $p$ as vectors. Note that curves may not be tangent at another point $q \in \M$, hence tangent vectors are thought of as local (around $p$) equivalence relations, hence the importance of connections (see below) that, intuitively speaking, tell us how to map between tangent spaces across a manifold. The idea here is that the equivalence relation among curves arises from the tangent relation (curves are equivalent where at $p$ (for $t=0$) and their differentials with respect to coordinate functions are identical. From this construction of the tangent we obtain a number of important definitions:
\begin{enumerate}[(i)]
    \item \textit{Tangent space}: the tangent space $T_p\mathcal{M}$ to $\mathcal{M}$ at $p \in \mathcal{M}$ is the set of all tangent vectors at $p$.
    \item \textit{Tangent bundle}: is defined as the union of tangent spaces i.e. $T\mathcal{M}:= \bigcup_{p \in \mathcal{M}} T_p \mathcal{M}$.
    \item \textit{Projection map}: a map $\pi: T \mathcal{M} \to \mathcal{M}$ from the tangent bundle to $\mathcal{M}$ associated with each point $p \in \mathcal{M}$. This map is `natural' in the sense that it associates each tangent plane (set of tangent vectors) with the point $p\in\mathcal{M}$.
\end{enumerate}
The tangent vector $v \in T_p\mathcal{M}$ can be also be construed as a \textit{directional derivative} on functions that are defined on the manifold, $f$ on $\mathcal{M}$ by defining their `action' on the function $f$ (which we denote $v(f)$ in terms of a differential operator on $f$ evaluated at the identity of the parametrisation of the curve itself (i.e. 0 point of interval)):
    \begin{align*}
        v(f) = \frac{df(\gamma(t))}{dt}\bigg|_0
\label{eqn:geo:directionalderivative}
    \end{align*}
The directional derivative points in the direction of steepest ascent. From this, we can define the \textit{gradient} as follows.
\begin{definition}[Gradient]\label{defn:geo:gradient}
     Given a smooth function $f: \mathcal{M} \rightarrow \mathbb{R}$ on a differentiable manifold $\mathcal{M}$, the gradient of $f$ at a point $p \in \mathcal{M}$, denoted $\nabla f(p)$, is the unique vector in the tangent space $\nabla f(p) \in T_p\mathcal{M}$ that satisfies:
    \begin{align*}
        v(f) = \langle \nabla f(p), v \rangle 
    \end{align*}
    for all $v \in T_p\mathcal{M}$ where $v(f)$ denotes the directional derivative of $f$ in the direction of $v$ (as above), and $\langle \cdot , \cdot \rangle$ denotes the applicable metric (see below) on $\mathcal{M}$.  
\end{definition}
In a coordinate frame $(U, \phi)$ around $p$, with $\phi = (x^1, \dots, x^n)$, the gradient of $f$ can be expressed as:
    \begin{align*}
        \nabla f(p) = \sum_{j=1}^{n} \left( \frac{\partial f}{\partial x^j} \circ \phi^{-1} \right)(\phi(p)) \frac{\partial}{\partial x^j}\bigg|_{\phi(p)}
    \end{align*}
    where $\frac{\partial}{\partial x^j}\big|_{\phi(p)}$ are the basis vectors of the tangent space $T_p\mathcal{M}$ in the coordinate chart, and $\frac{\partial f}{\partial x^j}$ are the partial derivatives of $f$ with respect to the coordinates $x^j$. The inverse $\phi^{-1}$ reverses this mapping, translating coordinates in $\mathbb{R}^n$ (or $\mathbb{C}^n$) back to points on the manifold $\mathcal{M}$. We can understand how the $\nabla$ `points' in the direction of steepest ascent by noting that that:
    \begin{align}
        \nabla(f(p)) \cdot v = |\nabla(f(p))||v|\cos\theta
    \end{align}
    where $\cos\theta$ is given by equation (\ref{eqn:quant:cosinevector}). Then $|\nabla(f(p))|$ is maximal for $\cos(\theta)=1$ which occurs when $v$ and $\nabla(f(p))$ are parallel. Hence $\nabla(f(p))$ as a vector points in the direction of maximal increase i.e. steepest ascent. To obtain the steepest gradient descent, we utilise $-\nabla$.
Although in this work we only touch upon the geometric machine learning, this type of framing of tangents is important for geometric framing of statistical learning methods, such as gradient descent (see section \ref{sec:ml:Optimisation and Gradient Descent}), specifically in relation to stochastic gradient descent optimisation (definition \ref{defn:ml:Gradient descent optimisation}). 

As is shown in the literature, tangent spaces $T_p\M$ at $p$ can be considered as real vector spaces. The intuition for this is that two curves $\gamma_i,\gamma_j$ cannot be added directly because $\M$ lacks the structure to do so. However, the coordinate space to which we have a mapping $\Real^m$ is a vector space, thus one considered how each curve $\gamma_i,\gamma_j$ is represented in $\Real^m$ such that $t \mapsto \phi \circ \sigma_1(t) +  \phi \circ \sigma_2(t)$ which we can think of as mapping from $\R$ into $\mathcal{M}$ then into $\R^m$. The object $\phi \circ \sigma_1(t) +  \phi \circ \sigma_2(t)$ is considered a curve in $\R^m$ which itself passes through the origin $\R^m$ when $t=0$. This additive object is then considered a curve in $\mathcal{M}$ that passes through $p$ when $t=0$. We then define vector addition and scalar multiplication.  From this we can show that $T_p \mathcal{M}$ is a real vector space and that the definitions above that allow assertion of the existence of a vector space of tangent vectors independent of the choice of chart and distinct from curves $\gamma_i$. Because of this vector structure, the tangent bundle is a \textit{vector bundle}.

\subsubsection{Push-forwards}
An important map to mention in geometry is that of a \textit{push-forward}. Recall that tangent spaces are akin to a local linearisation of the manifold. That is, a map $h: \M \to \N$ can be linearised via the architecture of tangent spaces and their vector space structure. Given this map between manifolds $\M,\N$, we are interested in maps between their corresponding tangent spaces $T\M,T\N$. This map is denoted the push-forward $h_*$ and is defined as follows.
%Definition:pushforwards
\begin{definition}[Push-forward]\label{defn:geo:Push-forward}
Given the mapping $h: \mathcal{M} \to \mathcal{N}$ and $v \in T_p \mathcal{M}$, then a \textit{push-forward} is a function taking a vector in the tangent space associated with $p \in \mathcal{M}$, namely $v \in T_p \mathcal{M}$ and represented as $h_*(v)$ to the tangent space associated with the element of $\mathcal{N}$ we denote via the map $h(p) \in \mathcal{N}$:
     \begin{align}
         h_*:& T\M \to T\N\\
         v &\mapsto h_*(v)\\
         h_*(v)&:=h \circ v \label{eqn:geo:pushforward}
     \end{align}
     where $v = [\gamma]$.
\end{definition}
Pushforwards are \textit{compositional} mappings between tangent spaces which compose linearly. It can be shown that certain commutativity diagrams hold with respect to pushforwards such that, as maps between tangent spaces, they are independent of intermediate tangent spaces through which they transition.

%Tensor fields and tangent spaces
\subsection{Tangent spaces and derivations}\label{sec:geo:Tensor fields and tangent spaces}
To understand the connection between geometry and Lie algebras, and ultimately symmetry groups relevant to quantum machine learning, we focus on tangent spaces, vector fields and later tensor fields which exhibit the structure of a derivation. Recall that a derivation is a function on an algebra which generalizes certain features of the derivative operator. For an algebra $A$ over a field $K$ (also a ring), a \textit{derivation} of $A$ is a map $D: A \to A$ such that:
    \begin{align}
        D(\alpha f + \beta g) &= \alpha Df + \beta Dg \qquad \text{(scalar distributivity)}\\
        D(fg) &= f D(g) + gD(f) \qquad \text{(product rule aka Leibniz's law)}
    \end{align}
    where $\alpha,\beta \in \Km; f,g \in A$. Intuitively, we can think of derivations as typical linear operators that allow differentiation along a direction. $C^\infty$ forms an algebra such that (this also relates to the condition of being a \textit{derivation}):
    \begin{align}
        (\lambda f)(p) &= \lambda f(p) \qquad \text{(Scalar multiplicative distributivity)}\\
        (f + g) (p) &= f(p) + g(p) \qquad \text{(Additive distributivity)}\\
        (fg) (p) &= f(p) g(p) \qquad \text{(Multiplicative distributivity)}
    \end{align}
    for $\lambda \in \mathbb{F}$ (the field), $p \in M$ and $f,g \in C^\infty$. Importantly, tangents vectors can be construed as operators or linear maps acting on functions $f \in \cinfm$ by noting they satisfy the requirements of a derivation. This is a useful way to build intuition around tangent vectors as differential operators upon a space and can be seen via the definition of tangent vectors themselves i.e:
     \begin{align*}
        v(f) = \frac{df(\gamma(t))}{dt}|_0
    \end{align*}
    where $v=[\gamma]$, the equivalence class of curves tangent at $p \in \mathcal{M}$ (sometimes tangent vectors are described as a derivation map from such a ring of functions on $\cinfm$ into $\Real$). The space of derivations $D_p\M$ (i.e. $D$ at $p \in \M$) satisfies the requirements of being a real vector space if we define:
    \begin{align*}
        (v_1 + v_2)f = v_1(f) + v_2(f), (rv)f = rv(f)\qquad v_1,v_2,v\in D_p(\M), r \in \R, f \in C^\infty(\M).
    \end{align*}
    As shown in the literature, it can be shown that the vectors in such a space of derivations have a representation as vectors in the space constituted by partial derivatives as vectors, allowing the equating of derivations at $p$ with the tangent space at $p$. In this representation, the partial derivatives represent the basis vectors of the vector space with coefficients (component or coordinate functions) as:
    \begin{align}
      v \in D_p(\M) \implies v = \sum_{\mu = 1}^m v(x^\mu)\pardiv = \sum_{\mu = 1}^m v^\mu\pardiv \label{eqn:geo:tangentpartialderivsum}
    \end{align}
    where the set of real numbers $\{ v^1,...,v^m \}$ or $\{ v^\mu \}$ are the components of the vector in the `direction' of $\partial_{x^\mu}$. That is, it can then be shown that there exists an isomorphism between $T_p\M$ and $D_p\M$ allowing, in particular expression of important terms such as the Jacobian in terms of local representatives of push-forward maps between tangent planes.

    %Vector fields and n-forms
    \subsubsection{Vector fields}\label{sec:geo:Vector fields and n-forms}
    Of particular importance to quantum information and algebraic transformations in this work is the concept of a \textit{vector field} as an assignment of vectors in tangent planes to each point in $\M$.

\begin{definition}[Vector field]\label{defn:geo:vectorfield}
A vector field $X$ on a $C^\infty$ manifold $\M$ is a \textit{derivation} of the algebra $C^\infty(M)$ by way of an assignment of $X_p \in T_p\M, \forall p \in \M$. Given $f \in C^\infty(\M)$ and vector fields $X,Y \in D(\M)$, then $Xf$ and $X + Y$ are also vector fields as follows:
    \begin{align}
        X_p f: \M \to \Real & \qquad p \mapsto (Xf)(p) \\         X + Y&: p \to Xp + Yp
    \end{align}
    for $p \in \M$.
\end{definition}
To each $p \in \M$ is associated a vector field. We think of vector fields as operators on functions $f$ which are themselves defined on $\M$. Note in the set of vector fields,  Helgason \cite{helgason_differential_1979} denotes $X,Y \in D(M)$ as $\mathcal{D}^1(\M)$. If the above conditions are met, then $D(\M)$ is a module over the ring of functions $F$ on $C^\infty(M)$. If both vector fields are in this module, i.e. if $X,Y \in D(\M)$, then we have that $XY-YX$ is also a derivation of $C^\infty(M)$ and is itself a vector field. This vector field (again, a mapping of the algebra over the field, in this case over $C^\infty(M)=A$ in the above definition of derivation) is denoted:
    \begin{align}
        [X,Y] = XY-YX
    \end{align}
and aligns as the Lie derivative (commutator) (see definition (\ref{defn:alg:Lie algebraliederivative})). That is, for the map $X:C^\infty(\M) \to C^\infty(\M)$, the map above from $\M \to \R$ defined by $(Xf)(p) = X_p F$ defines a geometric characterisation of the Lie derivative of the function $f$ along the vector field (assignment of tangent vectors which are operators) $X$. The set of all vector fields $X$ on $\M$ is sometimes denoted $VFld(\M)$ \cite{isham_modern_1999} or $\mathfrak{X}(\mathcal{M})$ (or $\Gamma(\M)$) and carries the structure of a real vector space.
\\
\\
Note that a vector field can be regarded as a first-order differential operator: as the tangent vector $X_p \in T_p\M$ is differential operator $C^\infty$ functions defined on $\M$. A vector field, as an assignment of such tangent vectors, can be considered first-order differential operator on $\M$. Thus for a neighbourhood around $p$ given by $U\in\M$, the vector field related to $X$ can be defined as:
    \begin{align*}
        X = \sum_{\mu =1}^m (X^\mu)\frac{\partial}{\partial x^\mu}
    \end{align*}
    where $X^\mu$ are the coefficient functions and the partial derivatives the basis. The functions $X^\mu$ are the \textit{components} of the vector field $X$ with respect to the coordinate system associated with the chart $(U,\phi)$. The components are sometimes denoted $X^\mu$ (though notation may vary) i.e. they are the coordinate functions that tell us the components of the vector at some point $p\in\M$ given a coordinate system. The form of the components depends upon the coordinate chart in use. Given two coordinate charges $(U,\phi),(U',\phi')$ with $U \bigcup U' \neq \emptyset$, the tangent vector should be defined in a way that it is equivalent across two coordinate systems, that is:
    \begin{align*}
        X = \sum_\mu X^\mu \partial_{x^\mu} = \sum_\nu X^{\nu'} \partial_{x^{\nu'}}
    \end{align*}
We may then express the coefficient functions in one coordinate chart (or frame) in terms of another:
    \begin{align*}
        X^{\nu'} = \sum_\mu X^\mu \frac{\partial x^{\nu'}}{\partial x^\mu}
    \end{align*}
\subsubsection{Vector fields and commutators}\label{sec:geo:Vector fields and commutators}
We explicate a number of relevant features of vector fields and commutators, given their importance in both quantum and geometric methods adopted in the substantive Chapters above. Commutators arise in this context by considering composition of vector fields: naively multiplying two vector fields does not obtain a third. $X,Y$ can be viewed as linear maps from $C^\infty(\M) \to \cinfm$. So we can define the composition $X \circ Y: \cinfm \to \cinfm$ as $X \circ Y(f) = X(Y(f))$ however this is not a vector field because the Leibnizian derivation property (`product rule') isn't satisfied. Instead, it can be shown that $X \circ Y - Y \circ X$ satisfies the derivation criteria (see text) and so is a vector field:
    \begin{align*}
        (X \circ Y - Y \circ X)(fg) = g(X \circ Y - Y \circ X)f + f(X \circ Y - Y \circ X)g
    \end{align*}
    As discussed above, this composed vector field is the \textit{commutator} of vector fields $X,Y$ and is denoted $[X,Y]$. In component form the commutator is represented as:
     \begin{align}
        [X,Y]^\mu = \sum_\nu (X^\nu Y^\mu_{\nu} - Y^\nu X^\mu_{\nu}) \label{eqn:geo:commutatorcomponentform}
    \end{align}
The commutator is both antisymmetric $[X,Y]=-[Y,X]$ and importantly satisfies the Jacobi identity. The relation to Lie algebras is apparent. In general a Lie algebra defined as a vector space $\g$ together with a bilinear map $[\cdot,\cdot]:\g \times \g \to \g,(A,B) \mapsto [A,B]$ satisfies (a) antisymmetry and (b) the Jacobi identity for elements $A,B \in \g$. This bilinear mapping is not associative (ordering matters), forming a non-associative algebra. Note that the Lie algebra `bracket' is defined as compositions which satisfy antisymmetry and Jacobi, rather than directly as the commutator. It turns out that for certain vector spaces (those with which we are concerned in this work), the commutator will satisfy the requirements for being a Lie algebra. One such example is the set of $M(n,\C)$ where the Lie bracket is defined to be the commutator $[A,B]:=AB-BA$ as discussed above. For our purposes, the results relating to vector antisymmetry and satisfaction of the Jacobi equation imply that the set of vector fields $X$ on $\M$ (using the same notation for a set as an individual field) have the structure of a real Lie algebra. Note also that, if $h: \M \to \N$ and the associated pushforward satisfies $h_*:T_p\M \to T_{h(p)}\N$ then each manifold is considered $h$-related.

\subsubsection{Integral curves and Hamiltonian flow} \label{sec:geo:Integral curves and Hamiltonian flow}
We now connect the theory of vector fields with integral curves and Hamiltonian flow, on our way towards understanding the theory of Riemannian and sub-Riemannian geodesic curves. Vector fields can be regarded as generators of an `infinitesimal' diffeomorphism of a manifold e.g. in the form $\delta(x^\mu) = \epsilon X^\mu(x)$. Recall that tangent vectors can be regarded as (a) equivalence classes of curves or (b) derivations defined at a point in the manifold. The question is: given a vector field $X$ on $\M$, is it possible to `fill' $\M$ with a family of curves in such a way that the tangent vector to the curve that passes through any particular point $p \in \M$ is just the vector field $X$ evaluated at that point, i.e. the derivation $X_p$? This idea is important in general canonical theory of classical mechanics, where the state space of a system is represented by a certain (even-dimensional e.g. symplectic phase-space) manifold $\mathcal{M}$ and physical quantities are represented by real-valued differentiable functions on $\mathcal{M}$. The manifold $\mathcal{M}$ is equipped with a structure that associates to each such function (representing physical quantities) $f:\mathcal{M} \to \R$ a vector field $X_f$ on $\mathcal{M}$. The family of curves that `fit' $X_f$ play an important role. In particular for quantum systems, curves associated with the vector field $X_H$ where $H: \mathcal{M} \to \R$ is the \textit{energy function}, are dynamical trajectories of the system i.e. the Hamiltonian flow. Of particular importance to our search for time-optimal curves (geodesics) in later chapters are integral curves. We want to find a single curve that (i) passes through $p\in\M$ and (ii) is such that the tangent vector at each point along the curve agrees with the vector field at that point. For this we use the definition of an \textit{integral curve}.
%
%definition integral curves
\begin{definition}[Integral curves]\label{defn:geo:Integral curves}
Given vector field $X$ on $\M$, the integral curve of $X$ passing through $p \in \M$ is a curve $\gamma: (-\epsilon,\epsilon) \to \M, t \mapsto \gamma(t) $ such that we have $\gamma(0) = p$ and the push-forward satisfies:
    \begin{align*}
        \gamma_*\left( \frac{d}{dt} \right)_t = X_{\gamma(t)} \qquad \forall t \in (-\epsilon,\epsilon) \subset \R
    \end{align*}
\end{definition}
The components of $X^\mu$ of the vector field $X$ determine the form taken by the integral curve $t \mapsto \gamma(t)$. Remember, the vector field (and tangent vectors) should be thought of as differential operators, so we have:
    \begin{align*}
        X^\mu(\gamma(t)) = \frac{d}{dt}x^\mu (\gamma(t))
    \end{align*}
    with the boundary condition that $x^\mu(\gamma(0)) = x^\mu (p)$. Certain other properties are also sought on integral curves, such as completeness properties such that the curves are defined for all $t$ (associated with assumptions regarding the compactness of the manifold).  
\\
\\
We are also interested in the extent to which a vector field can be regarded as the generator of infinitesimal translations on the manifold $\M$. To this end we define a one-parameter group of local diffeomorphisms (see definition \ref{defn:alg:one_parameter_subgroup}) at a point $p \in \M$ consisting of the following (a triple with certain properties): (i) an open neighbourhood $U(p)$, (ii) a real constant $\epsilon > 0$ and (iii) a family of diffeomorphisms of $U$, given by $\{ | \phi_t | |t|>\epsilon\}$ onto the open set $\phi_t(U) \subset \M$, i.e. $\phi_t: \epsint \times U \to \M$ and $(t,q) \mapsto \phi_t(q)$. The group has the following properties: (a) maps from the parameter and neighbourhood to $\M$ are smooth, i.e. $\epsint \times U \to \M, (t,q) \mapsto \phi_t(q)$ is smooth on both domains; (b) compositions follow $\phi_s(\phi_t(q)) = \phi_{s + t}(q)$ and $\M$, i.e. $\phi_0 (q) = q$. The set of local diffeomorphisms is denoted Diff($\M$). The term \textit{one-parameter subgroup} refers to the composition property (b) above and that the map $t \to \phi_t$ can be thought of as a representation (vector space with map) of part of the additive group of the real line. Families of local diffeomorphisms can be thought of as \textit{inducing} vector fields. Through each point $q\in\M$, there passes a local curve $t \mapsto \phi_t(q)$. Because of the existence of this curve, we can construct a vector field on $U$ by taking tangents to this family of curves at $q$. This vector field is denoted $X^\phi$, the vector field induced by the diffeomorphism $\phi$, more formally:
        \begin{align*}
            X_q^\phi (f):=\frac{d}{dt}f(\phi_t(q))|_0 \qquad \forall q \in U \subset \M.
        \end{align*}
 It can then be shown that $\forall q \in U$ the curve $t \mapsto \phi_t(q)$ is an integral curve of $X^\phi$ for $|t|<\epsilon$.

 \subsubsection{Local flows}\label{sec:geo:Local flows}
 We now briefly mention local flows and its connection with Diff$(\M)$. A \textit{local flow} of a vector field is a one-parameter local diffeomorphic group such that the vector fields induced by the group are the vector field itself. More formally, for a vector field $X$ defined on an open subset $U \subset \M$, the local flow of $X$ at $p$ is a one-parameter group of local diffeomorphisms defined on some open subset $V \subset U$ such that $p \in V \subset U$ and such that the vector field induced by this family equals $X$. In a local coordinate system, local flow - the  family of diffeomorphisms $\phi_t^X$ - is expressed in the following way:
    \begin{align*}
        X^\mu(\phi_t^X(q)) = \frac{d}{dt}x^\mu (\phi_t^X(q))
    \end{align*}
    around $q \in U \subset \M$. We then Taylor expand the coordinates (which are in a vector space) $x^\mu(\phi_t^X(q))$ of the transformed point $\phi_t^X(q)$ near $t=0$, resulting in:
    \begin{align*}
        x^\mu(\phi_t^X(q)) = x^\mu(q) + tX^\mu (q) + O(t^2)
    \end{align*}
  Recalling that $x^\mu(\phi_t^X(q)) \in \Real^m$, the expansion shows that the coordinates $x^\mu$ of the transformed of $q \to \phi_t^X(q)$ are approximated by the original coordinates $x^\mu(q)$ translated to first order by $tX^\mu(q)$. As $X^\mu$ belongs to the vector field, this is what allows us to say that the vector field generates `infinitesimal' transformations on the manifold because it is responsible (up to approximation) for `shifting' from $q$ to $\phi_t^X(q)$ in terms of a local coordinate system on the manifold. We discuss the relationship of integral curves and local flows to time-optimal geodesics in the context of geometric control theory further on in section \ref{sec:geo:Symplectic manifolds and Hamiltonian flow} on Hamiltonian flow, providing a connection between geometric formalism and quantum information processing. 

%======SECTION: Cotangent vectors and dual spaces
\subsection{Cotangent vectors and dual spaces}
In this section, we discuss important concepts of one-forms, cotangent vectors, leading to discussion of pullbacks, tensors and $n$-forms along with corresponding definitions of covariance and contravariance. 

\subsubsection{Dual spaces}\label{sec:geo:Dual spaces and one forms}
Recall from above the definition (\ref{defn:quant:Dual space}) that for any vector space $V = V(\mathbb{K})$ we define the associated dual space $V^*$ as the space of all bounded linear maps $L:V \to \mathbb{K}$ (see definition \ref{defn:quant:Bounded linear operators}). Here we focus on $\mathbb{K}=\R$ for exposition. In this mapping formulation, the \textit{dual} of a real vector space $V$ is the collection $V^*$ of all linear maps $L:V \to \R$. The action of $L \in V^*$ on $v \in V$ is usually denoted $\braket{L,v}$ or sometimes with the subscript $V$ as well. Note that here $\braket{\cdot,\cdot}$ represents the evaluation of the functional at a vector in the sense of assigning a scalar to each vector $v \in V$. Note that by the Riesz representation theorem (for finite dimensional $V$), the isomorphism between $V$ and $V^*$ means that we can associate to each vector $v$ a unique such functional $L_v$, thus allowing us to view the evaluation of functionals in this way as a generalisation of the inner product (hence the notational similarity). The dual itself can be given the structure of a real vector space by leveraging the vector-space properties of $\R$ into which $L$ maps i.e.:
    \begin{enumerate}[(i)]
        \item $\braket{L_1 + L_2,v} := \braket{L_1,v} + \braket{L_2,v}$ ;
        \item $\braket{rL,v}:= r\braket{L,v}, r \in \Real$.
    \end{enumerate}
If $V$ with $\dim V < \infty$ has a basis $\{e_1,...,e_n\}$, then the \textit{dual basis} for $V^*$ is a collection of vectors $\{f^1,...,f^n \}$ (they are vectors because they are linear maps $L \in V^*$ which is a vector space). This set $\{f^i \}$ is uniquely specified by the criterion that:
    \begin{align}
        \braket{f^i, e_j} = \delta^i_j         \label{eqn:geo:dualdelta}
        \end{align}
Note equation (\ref{eqn:geo:dualdelta}) may also be written as $f^i(e_j) = \delta^i_j$ where duals take (or act upon) vectors as their arguments, which is in essence an application of the Riesz representation theorem. Because of the isomorphism between a vector and its dual, one can `invert' dual and vector spaces, so the dual is the argument of a vector map to a scalar, but for clarity one usually maintains the formalism above. For maps between different vector spaces $L: V \to W$, then $L$ induces a dual mapping $L^*: W^* \to V^*$ on $k \in W^*$ by:
    \begin{align*}
        \braket{L^*k,v}_V:=\braket{k,Lv}_W
    \end{align*}
    which says that the dual-map $L^*$ acting on $k$ in $W^*$ (itself a dual space) generates a map $L^*k$ which lives in $V^*$. This map then acts on $v \in V$ (taking it to $\R$), acting as a pullback (discussed below): $L^*$ is pulling back the linear functional $k$ from $W^*$ to a linear functional in $V^*$. To explicate the relationship between $V$ and $V^*$, it can be shown that there exists a canonical map $\chi:V \to (V^*)^*$ where $\braket{\chi(V),\ell}_{V^*} = \braket{\ell,v}_V$ where $V \in V, \ell \in V^*$ which is an isomorphism if $\dim V < \infty$. This concept captures the spirit of what is sometimes described as $V$ being equivalent to the `dual of its dual'. With these concepts to hand, we can now define cotangent structures.

    \begin{definition}[Cotangent vectors]\label{defn:geo:Cotangent vectors}
        A cotangent vector is a map from tangent space to $\R$. At a point $p\in\M$, it is defined as a \textit{real linear map}: 
    \begin{align*}
        k:T_p\M \to \R
    \end{align*}
    The value (which is in $\R$) of $k$ when acting on $v\in \tpn$ is denoted $\braket{k,v}$ or $\braket{k,v}_p$ (analogous to using inner-product symbolism for mapping to scalar or contraction).
    \end{definition}
The \textit{cotangent space} is the space of all cotangent vectors i.e. at $p \in \M$ it is the set $T^*_p\M$ of all such linear maps constituting cotangent vectors. It is a vector space and is the dual of the vector space $\tpn$. Similarly, the \textit{cotangent bundle} $T^*\M$ is the collection (a vector bundle) of all such cotangent for all $p \in \M$ i.e. $T^*\M = \bigcup_{p\in\M} T^*_p\M$. 

\subsubsection{One forms}\label{sec:geo:One-Forms}
We now come to the important definition of a one-form. In the field of geometric optimal control, one-forms are essential for formulating and solving control problems. One-forms can represent constraints and objectives in control problems, particularly in the context of Hamiltonian systems, where the dynamics can be expressed in terms of symplectic geometry (which we discuss in section \ref{sec:geo:Symplectic manifolds and Hamiltonian flow}).
%one-form definition
\begin{definition}[One-form]\label{defn:geo:One-form}
A one form $\omega$ defined on $\M$ is an assignment of a cotangent vector (being a smooth linear map) $\omega_p$ to every point $p \in \M$. We can consider a one-form as a map from a vector field (as a vector space) to $\R$ i.e. $\omega: X \to \R$ for $X \in \mathfrak{X}(\M)$ i.e. a one-form is a field of cotangent vectors, meaning it assigns a cotangent vector to each point in a smoothly varying manner.
\end{definition}
One-forms are useful and important for understanding tensorial contractions and, under certain conditions, relate to inner products. Noting that $\braket{\omega,X}(p):=\braket{\omega_p,X_p}_p$, we observe that the one-form as an assignment of a cotangent vector to each point $p \in \M$, in turn induces a map from the tangent plane at $p$ to $\R$. In this sense the representation $\braket{\omega_p,X_p}_p$ specifies a map $\omega$ from the vector field at $p$ to $\R$ and that this mapping is smooth i.e. infinitely-differentiable. Connecting with the definition of the inner product (equation (\ref{defn:quant:Inner Product})) above, we can consider an inner product in terms of tangent spaces:
\begin{align}
    \braket{\cdot,\cdot}: T_p\M \times T_p\M \to \R \label{eqn:geo:oneforminner product}
    \end{align}
and one forms. Consider a metric tensor $g$ which is a type $(0,2)$-tensor meaning it takes two vectors as an argument and returns a scalar. Choosing an inner product $\braket{\cdot,\cdot}$ on a vector (tangent) space $T_p\M$ gives rise to an isomorphism:
    \begin{align*}
        \varphi:& T_p\M \to T^*_p\M \qquad
        v \mapsto \braket{v,\cdot}.
    \end{align*}
The function $\varphi(u)$ takes $u \in \tpn$ and returns its dual such that that $\varphi(u)(v) = \braket{u,v}$ i.e we have a function taking a vector, obtaining its dual then taking the inner product with $v$ and returning a real number. In this sense, we can identify the inner product as the action of a covector on a vector i.e:
    \begin{align*}
        \braket{u,v} = \varphi(u) v.
    \end{align*}
   In terms of components of the metric tensor, this is equivalent to:
    \begin{align*}
        \braket{u,v} = g_{ij}u^i v^j = u_jv^j
        \label{eqn:tensordualinnerproductmetric}
    \end{align*}
    but noting that there is no canonical identification of $V$ with $V^*$. Such concepts are important in quantum and geometric methods, especially Hamiltonian dynamics where, if a system has a configuration space $Q$ then the space of classical states is defined to be the cotangent bundle i.e. set of all linear maps from the tangent plane into $\R$ (see \cite{isham_modern_1999}). We briefly note the expression of dual vectors in terms of components and basis elements.\\
    \\
As discussed above, selecting a local coordinate system around $p \in \M$ gives rise to a basis set associated with $p \in \M$:
        \begin{align*}
            \bigg\{ \pardivx \bigg\}.
        \end{align*}
        This is a basis for the set of derivations and thus of the tangent space at $p$ by the isomorphism above. There is also an associated basis about $p$ for the corresponding dual $\tpndual$. It is denoted by what via differential operators:
        \begin{align*}
            (dx^1)_p,...,(dx^m)_p.
        \end{align*}
        As with the basis of the dual space above, the basis for the dual space $\tpndual$ is given via a set of linear operators in the dual space which when composed with the basis vectors of $\tpn$ (i.e. the $\pardivx$ operators), leads to deltas. That is:
        \begin{align}
            \bigg\langle{(dx^\mu)_p,\left( \frac{\partial}{\partial x^\nu} \right)\bigg\rangle}_p := \delta^\mu_\nu. 
            % \label{eqn:geo:innerproductvectordualdelta}
        \end{align}
        In this way, we can expand dual vectors $k \in \tpndual$ in terms of component functions and this basis:
        \begin{align*}
            k = \sum_\mu^m k_\mu (dx^\mu)_p
        \end{align*}
        where the components $k_\mu \in \R$ of the vector for the coordinate system and thus the basis vectors $\pardivxdual$ are given via the bilinear map (e.g. inner product in certain contexts) of the dual vector $k$ with the basis element of the tangent vector:
        \begin{align*}
            k_\mu = \bigg\langle{k,\left( \pardivx \right)_p\bigg\rangle}_p.
        \end{align*}
        Remembering that $k$ is a map acting on $v \in \tpn$, then we can write the bilinear map pairing $\braket{k,v}$ similarly as:
        \begin{align*}
            \braket{k,v} = \sum_\mu^m k_\mu v^\mu
        \end{align*}
        recalling $v^\mu = v(x^\mu)$ and the resemblance here to inner products.
        Using this formalism, we can choose a local representation for one forms:
        \begin{align}
            \omega = \sum_\mu \omega_\mu dx^\mu
        \label{eqn:geo:dual-omegasumomegamudxmu}
        \end{align}
        i.e. we can express a one-form (cotangent vector) in terms of the basis set of the dual space where the above is a shortened expression for:
        \begin{align*}
            \omega_p = \sum_\mu \omega_\mu(p) (dx^\mu)_p
        \end{align*}
        for $p \in U \subset\M$. The components $\omega_\mu$ of $\omega$ are \textit{functions} on $U$:
        \begin{align*}
            \omega_\mu(q) := \bigg\langle{\omega, \pardivx\bigg\rangle}_p
        \end{align*}
        where $q \in U \subset \M$.
Such a construction shows how a one-forms can be construed as smooth cross-sections of $T^*\M$ analogous to vector fields defined as smooth sections of $T\M$ (see \cite{isham_modern_1999}).
\subsubsection{Pullbacks of one-forms}
 We now describe the pullback property of one-forms.  While it isn't true in general that a map $h: \M \to \N $ (between manifolds) can be used to push-forward a vector field on $\M$ i.e. we cannot always use some commuting differential form to act as our pushforward in the general case, this map $h$ \textit{can} be used to \textit{pull-back} a one-form on $\N$. This feature is connected with the way in which the global topological structure of a manifold is reflected in \textit{DeRham cohomology groups} of $\M$ defined using differential forms and to the fact that fibre bundles also pull-back (not push-forward) (see Frankel \cite{frankel_geometry_2011} $\S13$ for detailed discussion).
%pullback definition
\begin{definition}[Pull-back]\label{defn:geo:Pull-back}
    A pullback is the dual of the map between tangent spaces across manifolds. If $h: \M \to \N$ and we have the linear map (the pushforward) $h_*: \tpn \to T_{h(p)}\N$, then then \textit{pullback map} is defined as:
    \begin{align}
        h^*:T_{h(p)}^*\N \to T_{p}^*\M \label{eqn:geo:pullback}
    \end{align}
    and is considered under certain conditions the dual of $h_*$. This means that for all maps (which are also vectors in the associated vector space) $T_{h(p)}^*\N$ and $v \in \tpn$ we have :
    \begin{align*}
        \braket{h^*k,v}_p := \braket{k,h_*v}_{h(p)}
    \end{align*}
\end{definition}
To unpack the formalism, note that $T^*_{h(p)}\N$ is a reference to the dual of the tangent space for $p\in\M$ mapped to $h(p)\in\N$, so we can see $h^*$ as `pulling back' to the dual space of $p\in\M$. The term $\braket{h^*k,v}_p$ refers to $h^*$ (the pullback) acting on $k$ i.e. it `pulls back' the map $k$ living in $T_{h(p)}^*\N$ to a map living in $T_{h(p)}^*\M$. Thus $h^*k$ is a map living in $T_{h(p)}^*\M$ which is itself a linear map that acts on the tangent space $\tpn$, i.e. $v \in \tpn$. The definition says that this pullback is equivalent (diagrammatically) to $k$ acting on $h_*$ having been pushed forward to $T_{h(p)}\N$ which means it cans be acted on by $k \in T_{h(p)}^*\N$. In both cases, we have a mapping to $\R$.
\\
\\
We can also extend pullbacks to one-forms. Recall, a one-form is an assignment of a cotangent vector $\omega_p$ to $p\in\M$ and this cotangent vector, which is a mapping from $\tpn$ to $\R$, acts on the tangent vector at $p$, i.e. $v \in \tpn$ to $\R$. We define a pullback of one-forms as follows.
\begin{definition}[Pull-back (one-forms)]\label{defn:geo:Pull-back (one-forms)}
    If $\omega$ is a one-form on $\N$, then the pullback of $\omega$ is itself a one-form $h^*\omega$ on $\M$ expressed as:
    \begin{align*}
        \braket{h^*\omega,v}_p := \braket{\omega,h_* v}_{h(p)}
    \end{align*}
    for all $p \in \M$ and $v \in \tpn$. 
\end{definition}

It can be shown (see \cite{isham_modern_1999} for specifics exposition) that the pullback $h^*$ of $\omega$ (again, in the dual space of $\N$) evaluated at $p \in \M$ is given by the pullback acting on that one-form $\omega$, i.e. $h^* \omega$ when then acts on the \textit{basis} elements of $\tpn$, namely the $\pardivx$ terms, mapping them to $\R$ via the bilinear form. This is definitionally equated with the one-form $\omega$ acting on the basis element $\pardivx$ of $\tpn$ once it has been `pushed forward' to $\tpnn$ by the pushforward map $h_*$, which also is equivalent to an evaluation of the one-form at $h(p)$ (as we are assuming the existence of $h: \M \to \N$ here). The point of the pullback is to allow the structure of maps defined on or in relation to one manifold, say $\N$, to be expressed or represented in terms of structure on or related to another manifold $\M$.

    \subsubsection{Lie derivatives and pullbacks}\label{sec:geo:Lie derivatives and pullbacks}
    We now specify the relationship of Lie derivatives (and thus commutators) to pullbacks. Using definition (\ref{defn:geo:Pull-back (one-forms)}), the pull-back of a one-form $\omega$ on $\mathcal{M}$ can be related to a vector field $X$ with an associated one-parameter group of local diffeomorphisms $t \to \phi_t^X$. For each $t$, there exists an associated pull-back $\phi_t^{X^*}$. This one-parameter family of pulled-back forms describes how the one-form $\omega$ changes along the flow lines (integral curves) of $X$. In this formalism we can connect Lie derivatives to one forms and pullbacks. The \textit{Lie derivative} of $\omega$ associated with vector field $X$, denoted $L_X \omega$, is then the \textit{rate of change} of $\omega$ along the flow-lines (integral curves) of the one-parameter group $\phi_t^X$ of diffeomorphisms associated with the vector field $X$. That is:
    \begin{align}
        L_{X}\omega := \left.\frac{d}{dt}\right|_{t=0} \phi_t^{X^*}\omega
    \label{eqn:geo:liederivativepullback}
    \end{align}
    Note by comparison that the Lie derivative can be expressed as:
\begin{align}
    \frac{d}{dt}\bigg|_{t=0} \phi^X_{-t_*}(Y) = [X,Y] = L_XY
\end{align}
Recall that that $X$ in $\phi_t^X$ refers to the group of \textit{local} diffeomorphisms that induce the vector fields $X$. Then $\phi_t^{X^*}$ is the family of diffeomorphisms that generate the \textit{dual} cotangent vector field $X^*$. Equation (\ref{eqn:geo:liederivativepullback}) can be understood as follows. Recall that $L_Xf = Xf$ and $L_XY = [X,Y]$ (as naive multiplication of vector fields lacks the derivation quality). It can be shown that $L_X\braket{\omega,Y} = \braket{L_X\omega,Y} + \braket{\omega,L_XY}$. 
    Recall $\phi^X_t$ representing the flow of $X$ on $\mathcal{M}$ (a one-parameter group of diffeomorphisms). This says that each value of $t$ is associated with a diffeomorphism from $\M$ to itself. The flow moves points along the integral curves of the vector field $X$. Associated with such flow is an inverse flow, indicated by the negative sign in $\phi^X_{-t}$, in essence a rewinding of the flow (which recall is possible as a diffeomorphism). The operation $\phi^X_{-t}$ moves points $p\in\M$ along integral curves in the opposite direction to $\phi^X_t$. In this way, we can see how the commutator (Lie derivative) $[X,Y]$ expresses the rate of change of the vector field $Y$ along integral curves of $X$. Moreover, $df$ can be used to define the \textit{exterior derivative}.
 \begin{definition}[Exterior derivative]\label{eqn:geo:Exterior derivative}
     The exterior derivative of $f \in \cinfm$ is given by:
     \begin{align*}
        \braket{df,X}:=Xf = L_Xf
    \end{align*}
for vector fields $X$ on $\M$.
 \end{definition}
   In local coordinates, this has a representation as:
    \begin{align*}
        (df)_p = \sum_\mu^m \left( \frac{\partial}{\partial x^\mu} \right)_p f (dx^\mu)_p
    \end{align*}
    i.e. we represent the function in the dual basis $dx^\mu$ of $X^*$, then take the derivative with respect to the coordinates in $X$, i.e. $\partial_\mu$. Note also the commutation of the pullback $h^*$ with $f$, $h^*(df) = d(f \circ h)$. 

\subsection{General tensors and $n$-forms} \label{sec:geo:General tensors and n-forms}
We can now express tensors using the geometric formalism above, which we utilise for our discussion of time-optimal geometric control. Generalising tangent and cotangent vectors requires tensors and $n$-forms. To do this, we adopt the idea of the \textit{tensor product} of two vector spaces $V$ and $W$, denoted (as per definition (\ref{defn:quant:Tensor Product})) by $V \otimes W$ with a geometric framing. The tensors examined above can then be represented as compositions of tangent and cotangent spaces.

\begin{definition}[Tensor types (geometric)]\label{defn:geo:Tensor types (geometric)}
    Tensors are vector products of tangent spaces and/or cotangent spaces. A \textit{tensor of type} $(r,s) \in T_p^{r,s}\M$ at a point $p\in\M$ belongs to the tensor product space:
    \begin{align}
        T_p^{r,s}\M := \Bigg[\otimes^r T_p\M \Bigg] \otimes \Bigg[\otimes^s T^*_p \M\Bigg]\label{eqn:geo:Trstensortype}
    \end{align}
     i.e. $r$ tensor products of the tangent space tensor-producted with $s$ tensor products of the dual (cotangent) space. 
\end{definition}
The basis of $T^{r,s}_p\M$ can be thought of as composed of the tensor products of $r$ basis elements of the tangent space (partial derivatives) and $s$ basis elements of the dual (cotangent) space (differentials):
        \begin{align*}
    T_p^{r,s}\M=\Bigg[ \otimes_{i=1}^r \left( \frac{\partial}{\partial x^{\mu_i}} \right) \Bigg] \otimes \Bigg[ \otimes_{j=1}^s (dx^{v_j}) \Bigg].
        \end{align*}
In this formulation, with vectors and their duals, tensors as linear maps becomes apparent. We can represent the $T_p^{r,s}$ type tensor in terms of a multilinear mapping of the Cartesian product of a vector space and its dual to $\R$ which, in functional notation (Cartesian products describing the function of tensors rather the direct product) is given by:
    \begin{align}
        \bigg(\times^r \tpndual\bigg) \times \bigg(\times^s \tpn\bigg) \to \R \label{eqn:geo:tensorasmapping}
    \end{align}
where a multilinear map is viewed as a function that takes $r$ tangent vectors as input and $s$ cotangent vectors as input and maps them to $\R$ (the difference being notational, $r$ always indexes the tangent vectors, $s$ always indexes the cotangent vectors). Tensors take products of vectors and covectors as their arguments, thus tensor maps in effect take tensors as their arguments. The upper index $r$ indices the number of contravariant components of the tensor associated with $\tpn$. These are akin to `directional' components transforming in the same way as coordinate differentials $dx^i$. Contravariant components are typically denoted via superscripts corresponding to directions the tensor acts along with respect to covectors. The subscript index $s$ denotes the number of covariant components, consisting of linear functionals in $\tpndual$ (recall for duals and vectors we have components and bases). This formulation is usefully understood in terms of equation (\ref{eqn:geo:dualdelta}) i.e. $\braket{f^i, e_j} = \delta^i_j $. Each basis element of the vector and its dual either annihilate or lead to unity (scaled by any coefficients $a,b$), thus if we denote component coefficients $a,b \in \Real$ for vectors and duals $a f^i,b e_j$, then:
\begin{align}
    \braket{a f^i, b e_j} = ab\braket{f^i, e_j} = ab\delta^i_j
\end{align}
Thus the definition of vectors and duals, as bases of $\tpn$ and $\tpndual$ shows how the tensorial map effectively acts as contraction (see below) which reduces the dimensionality of the input vectors and covectors (the input tensor) and then multiplies the remainder by the corresponding scalar product of the real coefficients of the basis vectors for the vector and its dual so annihilated. As we see below, the usual inner product on vector spaces can be framed in the language of metric tensors (where the metric tensor acts as a bilinear form resulting in a scalar product). Framing tensors as multilinear maps connects to concepts further on of, for example, \textit{metric tensors} which act on $\tpn$ and $\tpndual$ to contract tensor products to scalars, forming the basis for measuring the time, energy or length of paths given by Hamiltonian evolution in our quest for time-optimal geodesics. In quantum information, the eponymous tensor networks represent an important application of tensor formalism for a variety of use cases, such as error correction and algorithm design \cite{wood_tensor_2015,orus_practical_2014}. We note for completeness a few types of special $(r,s)$ tensors. Note the relation between the tangent planes and cotangent planes:
    \begin{enumerate}[(i)]
        \item $T_p^{0,1}\M = \tpndual$ is the cotangent space at $p$ (space of covectors or one-forms). An example at $p \in \M$ would be a vector in $\tpn$ with one contravariant component;
        \item $T_p^{1,0}\M = \tpn = (\tpndual)^*$ (dual of dual returns the tangent plane). An example would be a covector or one-form in $\tpndual$ consisting of one covariant component;
        \item $T_p^{r,0}\M$ is the space of \textit{r-contravariant} tensors (multilinear functions of covectors);
        \item $T_p^{0,s}\M$ is the space of \textit{s-covariant} tensors (multilinear functions of vectors); 
        \item $T_p^{r,s}\M$ a mixed tensor comprising both contravariant (acting on vectors) and covariant (acting on covectors) elements. Such higher-rank tensors are used to represent more complex objects, such as curvature tensors or stress-energy tensors.
    \end{enumerate}
\textit{Covariant} tensors are multi-linear maps that take vectors (from $\tpn$) as inputs and return scalars, while \textit{contravariant} tensors take covectors (from $\tpndual$) as inputs. From this we obtain a \textit{tensor field} on $\M$ (a generalisation of the concept of a vector field), which is a smooth association of a tensor of type $(r,s)$ for $p \in \M$. It is instructive to observe how $\tpn$ is associated with contravariant transformations while $\tpndual$ is associated with covariant transformations. This is intuitively related to the fact that dual spaces (cotangent spaces) consist of linear maps that act on vectors in the tangent space, hence they transform covariantly (i.e. they co-vary with vectors). On the other hand, structures that transform `opposite to the way tangent vectors do', i.e. according to the basis changes in the cotangent space, are said to transform contravariantly. Since tensor spaces can involve multiple layers of dual spaces, we consider their transformation properties in terms of how they relate to the transformations of the basis in the tangent and cotangent spaces. Tensors can be defined as the unique bilinear map $V^* \times W^* \to \R$ which evaluates to $\braket{k,v}\braket{l,w} \in \R$ on $(k,l) \in V^* \times W^*$. We also can now usefully define a tensorial contraction as a generalisation of equations (\ref{eqn:geo:dualdelta}). 
Contraction is an operation that reduces the order of a tensor by pairing covariant vectors, indicated by (lower) indices with  contravariant vectors, indicated by (upper) indices and summing over the dimension of the manifold, effectively treating the contraction over indices as if they are in an inner product, akin to a trace operation. Recalling that a tensor is a mapping (a function), then we can define the contraction as a function, being a tensor, that takes as its argument another tensor. In general, the contracting tensor can be of any order, but in practice contraction is often performed using metric tensors, tensors which, as maps to $\R$, satisfy the conditions of being metric (we discuss this below). The most general form of a contraction is set out below. Following \cite{helgason_differential_1979,isham_modern_1999} first recall that vectors and their duals are related via:
\begin{align*}
\langle X_{i} \otimes \cdots \otimes X_{r}, \omega_{s'} \otimes \ldots \otimes \omega'_{r} \rangle &= \prod_{i,j} \omega_{j}(X'_{i})\omega'_{i}(X_{j}),
\end{align*}
for $X_{i}, X'_{j} \in T^r\M$ and $\omega_{j}, \omega'_{i} \in T^*\M$. We then define tensor contractions as follows.
%tensor contraction definition
\begin{definition}[Tensor contractions]\label{defn:geo:Tensor contractions}
A $(1,1)$-contraction is a linear mapping $C^i_j: T^r_s \M \to T^{r-1}_{s-1}\M$ with $(C^i_j(T^r_{s}))_p = C^i_j(T^r_{s,p})$ for $T \in T^r_s \M, p \in \M$. The mapping $C^i_j$ is the contraction of the $i$th contravariant index and $j$th covariant index:
\begin{align}
C^i_j\underbrace{(\otimes_{j=1}^r X_j \otimes_{i=1}^s \omega_i )}_{(r,s)} &= \langle X_i, \omega_j \rangle (X_1 \otimes ... \otimes \hat{X}_i ... \otimes X_r \otimes \omega_1 \otimes \cdots \otimes \hat{\omega}_j ... \otimes \omega_s)\\
&= \delta^j_i (X_1 \otimes ... \otimes \hat{X}_i ... \otimes X_r \otimes \omega_1 \otimes \cdots \otimes \hat{\omega}_j ... \otimes \omega_s)\\
&= \underbrace{(X_1 \otimes ... \otimes \cancel{X_j} ... \otimes X_r \otimes \omega_1 \otimes \cdots \otimes \cancel{\omega_j} ... \otimes \omega_s)}_{(r-1,s-1)} \label{eqn:geo:Tensor contractions}
\end{align}
Where the $\hat{X_i}$ symbol denotes removal of the $i^{th}$ element (following \cite{helgason_differential_1979}). 
\end{definition}
Thus contraction reduces the rank of a tensor by one in both the covariant and contravariant indices. Note that we can in principle construct a more general contraction mapping the product of individual contractions. In particular we examine metric tensors, which provide a way to compute inner products between tangent vectors, thereby inducing a geometry and way of measuring distance (and thus time optimality) on $\M$.

\subsubsection{Metric tensor}\label{sec:geo:Metric tensor}
We now move to the definition of the \textit{metric tensor}, of fundamental importance to time optimal and machine learning approaches adopted in this work.
\begin{definition}[Metric tensor]\label{def:geo:metric_tensor}
    The metric tensor is a [0,2]-form tensor field mapping $g_p: T_p\M \times T_p\M \to \R$ given by:
    \begin{align}
        g := g_{ij} dx^i \otimes dx^j \qquad g_{ij} := \braket{e_i,e_j} \qquad g^{ij:}=\braket{dx^i,dx^j} \label{eqn:geo:metric_tensor}
    \end{align}
    where $e_i=\partial_i=\partial_{x_i}$ are basis elements of $T_p\M$ and $dx_i$ are the corresponding dual basis elements for the inverse metric tensor $g^{ij}$.
\end{definition}
Recall in this notation that $v = v^j e_j$ so that and similarly for $v$ expressed in the dual basis i.e. $v=v_j dx^j$. The metric tensor can be described in terms of Metric tensors induce an inner product space on $T_p\M$ allowing vector both lengths (magnitudes) and angles between vectors to be calculated. The metric tensor can be used for raising and lowering indices in effect converting between vectors and their covectors acting on the vector \(v\) to produce its covariant components \(v_i\) through index lowering. This process is described by:
\begin{equation}
    v_i = g_{ij} v^j.
\end{equation}
which can be seen, assuming an orthonormal basis $\{e_i\}$ as:
\begin{align*}
\braket{v,e_j}=\braket{v^ie_i,e_j}=v^i\braket{e_i,e_j}=v^i\delta_{ij}=v_j
\end{align*}
where $g_{ij}=\delta_{ij}$ in the orthonormal case (see \cite{frankel_geometry_2011} p.lix).

\subsubsection{$n$-forms and exterior products} \label{sec:geo:n-forms and exterior products}
We conclude this section introducing $n$-forms and exterior products. Consider the following elementary $n$-forms, assuming $0 \leq n \leq \dim\mathcal{M}$ with $n \in \mathbb{N}$. A $0$-form is a function in $C^\infty(\mathcal{M})$. A one-form ($1$-form) at each point $p \in \mathcal{M}$,  assigns a linear functional that maps vectors in $T_p\mathcal{M}$ to $\mathbb{R}$. It is a section of the cotangent bundle $T^*\mathcal{M}$ defined as follows.
%n-form definition
\begin{definition}[$n$-form]\label{defn:geo:nform}
    An $n$-form is form of a \textit{tensor field}, represented by $\omega$, of type $(0,n)$ that is \textit{totally skew-symmetric}. For any permutation $P$ of the indices $1,2,\ldots,n$:
    \begin{align*}
   \omega(X_1,\ldots,X_n) = (-1)^{\deg(P)}\omega(X_{P(1)},\ldots,X_{P(n)})
    \end{align*}
\end{definition}
where $X_i$ are arbitrary vector fields on $\mathcal{M}$, and $\text{deg}(P)$ is the degree of the permutation $P$ ($+1$ even, $-1$ odd). The set of all $n$-forms on $\mathcal{M}$ is denoted $A^n(\mathcal{M})$. An $n$-form is therefore a particular way of assigning $n$ cotangent vectors to $p \in \M$. We can now define an important concept, the \textit{wedge} or \textit{exterior product} in terms of a tensor product of $n$-forms.

\begin{definition}[Exterior product]\label{defn:geo:Exterior product}
    For $\omega_1 \in A^{n_1}(\M),\omega_2 \in A^{n_2}(\M)$ the \textit{wedge or exterior product} of n-forms $\omega_1,\omega_2$ is the $(n_1 + n_2)$-form, denoted $\omega_1 \wedge \omega_2$ and defined as:
    \begin{align}
        \omega_1 \wedge \omega_2 = \frac{1}{n_1! n_2!} \sum_{\sigma(P)}(-1)^{\deg(P)}(\omega_1 \otimes \omega_2)^P \label{eqn:geo:Exterior product}
    \end{align}
    where $\sigma(P)$ denotes permutations over $P$.
\end{definition}
The permutations $\sigma(P)$ are understood as follows. For a tensor field $\omega$ of type $(0,n)$, the \textit{permuted} tensor field $\omega^P$ is defined to be the permutation map applied to the permutation index i.e:
    \begin{align*}
        \omega^P(X_1,...,X_n) = \omega(X_{P(1)},...,X_{P(n)})
    \end{align*}
    for all vector fields $X_1,...,X_n$ on the manifold $\M$. The factor $1/(n_1!n_2!)$ is a normalisation factor given the ways of permuting $\omega_1,\omega_2$, while the sum over permutations $\sigma(P)$ ensures all orderings of vectors are considered and that the result is antisymmetric with respect to exchange of vectors. When applied to vector fields, the result is a real number interpreted as an oriented volume spanned by those vectors associated with $p \in \M$. Intuitively one can think of how in two dimensions the wedge (or cross) product leads to a volume (area) which can be given an orientation based on the direction of its vectors. While not a focus of this work, we note that the generalised pullback commutes with the wedge product, noting for $h:\M \to \N$ and differential forms $\alpha,\beta$, the pullback is a homomorphism $h^*(\alpha \wedge \beta) = (h^*\alpha) \wedge (h^*\beta)$. Wedge products turn vectors into a graded algebra, important in various algebraic geometry techniques. The basis of $n$-forms at $p \in \M$ is given by the wedge product of differentials:
    \begin{align*}
        (dx^{\mu_1})_p \wedge ... \wedge (dx^{\mu_n})_p
    \end{align*}
    Indeed it can be shown that $d\omega(X_i)$ at $p \in \M$ only depends upon vector fields $X_i$ at $p$, a fact related to the property of the exterior derivative being $f$-linear:
    \begin{align*}
    d\omega(X_1,...,fX_i,...,X_{n+1}) &= fd\omega(X_1,...,X_i,...,X_{n+1})\\
    d\omega(fX,Y) &= fd\omega(X,Y)
\end{align*}
Such formulations then form the basis for important theorems related to DeRham cohomology (see Frankel \cite{frankel_geometry_2011}). In this context the \textit{exterior derivative} can be defined as follows.
%==
\begin{definition}[Exterior Derivative]\label{defn:geo:exterior_derivative}
Given $A(\M) := T^*\M$, the \textit{exterior derivative} is a map $d: A^n(\mathcal{M}) \to A^{n+1}(\mathcal{M})$, which takes an $n$-form to an $(n+1)$-form. For an $n$-form $\omega \in A^n(\mathcal{M})$ given locally by
\[
\omega = \sum_{i_1, \ldots, i_n} \omega_{i_1 \ldots i_n} dx^{i_1} \wedge \ldots \wedge dx^{i_n},
\]
the exterior derivative is defined as
\begin{align}
d\omega &= \sum_{i_1, \ldots, i_n} d\omega_{i_1 \ldots i_n} \wedge dx^{i_1} \wedge \ldots \wedge dx^{i_n}
\end{align}
where $\quad d\omega_{i_1 \ldots i_n} = \sum_{j} \frac{\partial \omega_{i_1 \ldots i_n}}{\partial x^j} dx^j$.
\end{definition}
The exterior derivative has the property that $d(d\omega) = 0$ for any form $\omega$, and it is linear over the smooth functions on $\mathcal{M}$, i.e., for a smooth function $f$ and a form $\alpha$, we have $d(f\alpha) = df \wedge \alpha + f d\alpha$. It is relevant in particular to the Maurer-Cartan form and Cartan structure equations discussed elsewhere.

\subsection{Tangent planes and Lie algebras}\label{sec:geo:Tangent planes and Lie algebras}
In this section, we recount a few properties of Lie groups and Lie algebras from earlier sections, connecting them explicitly to geometric formulations used in later chapters. Recall from proposition (\ref{prop:geo:Lie Group (Manifold)}) that a Lie group is a group equipped with the structure of a differentiable manifold such that the group operations are smooth i.e. where the map in parameter space that takes us from $g_1$ (parametrised by $\theta_1$) to $g_2$ in $G$ is a differentiable map. \\
\\
A fundamental aspect of Lie group theory is the isomorphism between the set of all left-invariant vector fields on a Lie group $G$, denoted $L(G)$, and the tangent space $T_eG$ at the identity element $e$ of $G$. This isomorphism implies that one can understand the behaviour of left-invariant vector fields by examining transformations within $T_eG$. In particular, this concept is foundational when considering the exponential map. The tangent space $T_eG$ is, effectively, the Lie algebra of the Lie group $G$, and mappings between different Lie groups $G \to H$ can be studied by examining the corresponding maps between their Lie algebras $T_eG \to T_eH$ where here $e$ denotes the identity element for $G$. In addition, given the completeness of left-invariant vector fields $G$, we can extend integral curves using the group structure even if $G$ is not compact. Moreover, we can see how the exponential map is constructed by considering the unique integral curve of a left-invariant vector field at $t=0$, mapping $t \mapsto \exp(tA)$, where $A$ is an element of the Lie algebra $T_eG$. The map is formally defined as $\exp: T_eG \to G$, such that $\exp(A) = \exp(tA)$ evaluated at $t=1$. It can be demonstrated that the exponential map is a local diffeomorphism near the identity $e$, mapping a neighborhood in $T_eG$ smoothly onto a neighborhood in $G$. This elucidates that the exponential map generates a one-parameter subgroup of $G$ (see definition \ref{defn:alg:one_parameter_subgroup}), and, in fact, every one-parameter subgroup of $G$ can be expressed in the form $t \mapsto \exp(tA)$, a fact related to the bijective correspondence between one-parameter subgroups of the Lie group $G$ and elements of its Lie algebra discussed earlier. Thus the neighborhood of $e$ in $G$, which is diffeomorphically mapped by $\exp: T_eG \to G$, is filled with the images of these subgroup maps, allowing investigations into these subgroups by examining the local structure around the identity element of the Lie algebra.
\subsubsection{Left and right translation} \label{sec:geo:Left and right translation}
Of importance to connecting Lie groups and algebras to geometric constructs is the existence of left and right translations of $G$ onto itself, which can be used to map the local tangent bundle around the entire group. Recall \textit{left translations} and \textit{right translations} of $G$ are diffeomorphisms given by the group actions:
    \begin{align*}
        r_g:& G \to G, \qquad &l_g: G \to G\\
        &g'\mapsto g'g \qquad &g' \mapsto gg'
    \end{align*}

A vector field $X$ on a Lie group $G$ is \textit{left-invariant} if it is $l_g$-related to itself for all $g \in G$ i.e.:
    \begin{align}
        l_{g^*}X = X, \forall g \in G \label{eqn:geo:leftinvarianttranslation}
    \end{align}
    or equivalently:
    \begin{align*}
        l_{g^*}(X_{g'}) = X_{gg'}, \forall g,g' \in G
    \end{align*}
    with a similar notion of \textit{right invariance}. Here $l_{g^*}$ denotes the pushforward, being the derivative map of the left action $L_g$ i.e. $g$ maps between manifolds, while $g^*$ is the corresponding push-forward mapping between their respective tangent spaces. Connecting with the notation above, given an isomorphism between the tangent space at identity and $L(G)$ given by $\xi: T_e G  \to L(G)$ with $\xi(A) = L^A$ for $A \in T_eG$ then we define $L_g^A = l_{g^*}A$ for all $g \in G$. The set of left invariant vector fields is denoted $L(G)$. Indeed one of the important properties of $X$ being left-invariance is that it is complete (Isham \cite{isham_modern_1999} $\S 4.2$). Similarly, local compactness is important in quantum contexts as it relates to the existence, for example, of the Haar measure (discussed in Appendix A above) where the measure exists on $G$ if $G$ is (quasi)-invariant under left and right translations.  For vector fields $X_1, X_2$ on $\M$ that are by a pushforward $h_*$ mapped to vector fields $Y_1,Y_2$ on a manifold $\N$ where $h: \M \to \N$, the \textit{commutator} $[X_1,X_2]$ is $h$-related to the commutator $[Y_1,Y_2]$. For left-invariant vector fields $X_1,X_2$, we then have that the commutator is invariant under the left-invariant actions (translations) of $G$, namely:
    \begin{align}
        l_{g^*}[X_1,X_2] = [l_{g^*}X_1,l_{g^*}X_2] = [X_1,X_2] \label{sec:geo:leftinvariancecommutator}
    \end{align}
    which shows that $[X_1,X_2] \in L(G)$.  It can be shown that the set $L(G)$ is a `sub-Lie algebra' of the infinite-dimensional Lie algebra of all vector fields on the manifold $G$. This is equivalent to the \textit{Lie algebra} of $G$ and represents a way of construing the Lie algebra in terms of important invariant properties of vector fields. It can also be shown that there is an isomorphism between the tangent space at the identity $e \in G$ i.e. $T_eG$ and $L(G)$, allowing $L(G)$ (and diffeomorphisms of $G$) to be explored via actions upon $T_eG$ (at the identity). The commutator $[A,B] \in T_eG$ (for $A,B \in T_eG$) is then defined to be the unique element in $T_eG$ satisfying:
\begin{align*}
        L^{[A,B]} = [L^A,L^B].
    \end{align*}
    Expanding on the above notation, here $L^A := L^A_g = l_{g_*}A, \forall g \in G$ i.e. $L^A$ is the vector field on $G$ defined by such left translation and $L^{[AB]} = l_{g_*}[A,B]$ representing the left invariant vector field associated with the Lie bracket $[A,B] \in T_eG$. We note also the structure constants of a Lie algebra $L(G) \simeq T_eG$ arising from commutators of its basis elements $E_\alpha$ that is:
    \begin{align*}
        [E_\alpha,E_\beta] = \sum_{\gamma = 1}^n C_{\alpha\beta}^\gamma E_\gamma
    \end{align*}
    for $C_{\alpha\beta}^\gamma \in \Real$ where $C_{\alpha\beta}^\gamma$ denotes the relevant structure constant (which play roles in quantum mechanics, geometry and elsewhere). The adjoint map also has an expression in terms of actions on $T_eG$ and that left-invariant fields $X$ such that integral curves of $X$ can be extended for all $t \in \R$.

\subsubsection{Right and left invariance and Schr\"odinger's equation}\label{sec:geo:Right and left invariance and Schrodinger's equation}
Because of its relevance in particular to our Chapter \ref{chapter:Time optimal quantum geodesics using Cartan decompositions} results, we set out the significance of left and right invariance in relation to quantum control. Our problem in that Chapter is to determine the optimal time to synthesise a unitary $U_T \in G$ using only controls in $\p$, the antisymmetric subspace corresponding to the Cartan decomposition $\g = \k \oplus \p$, where $\g$ is the Lie algebra of $G$. Now consider the Schr\"odinger equation:
\begin{align}
    \frac{dU}{dt} = -iH(t)U.
    \label{eqn:geo:leftrightinvarschrod}
\end{align}
Presented in this way, the Hamiltonian on the left of $U$, is the generator of (infinitesimal) translations (to second order) $U \to U + dU$ over increments $dt$. $U$ is our unitary at $t$ while $U + dU$ is our unitary at $t + dt$. The Hamiltonian is applied for $dt$ i.e. $dU/U = -iH(t)dt$. Thus time in equation (\ref{eqn:geo:leftrightinvarschrod}) moves from right to left. Intuitively this is because $U$ on the right-hand side represents the quantum state at time $t$, while $dU$ on the left-hand side represents it at the later time $t + dt$. When we say right or left translation by $g$, we can think of right translation as following the flow of time once its direction is chosen, and left translation in this case as time flowing `backwards' relative to that choice. Thus right and left action by $h \in G$ is expressed as follows:
\begin{align}
    \frac{dU}{dt}h & = -iH(t)Uh & &\text{(right action)}
\end{align}
whereas left action is given by:
\begin{align}
    h\frac{dU}{dt} & = -ihH(t)U = -i(hH(t)h^\inv) hU &\text{(left action)}.
\end{align}
Thus we can see how, once an implicit direction is chosen, as distinct from right action, left action acts to conjugate (thus transform by an effective rotation) the Hamiltonian $H(t) \to hH(t)g^\inv$ (recalling here that the action of $h$ on $\g$ is by conjugation, expressed as the commutation $\g \to [X_h,\g]$). Thus while the representation of left or right invariance can be chosen initially in an arbitrary fashion, \textit{once chosen} in certain circumstances they are not equivalent. 

\subsubsection{Exponentials, integral curves and tangent spaces}\label{sec:geo:Exponentials, integral curves and tangent spaces}
We briefly note a few connections between the exponential map (discussed in Appendix \ref{chapter:Background: Geometry, Lie Algebras and Representation Theory}) and the geometric formalism above. The exponential map can be related to integral curves, i.e. as the unique integral curve satisfying:
    \begin{align}
        t &\to \gamma^{L^A}(t) \qquad       A=\gamma_*^{L^A} \left( \frac{d}{dt} \right)_0 \label{eqn:geo:integralcurveexpleftinvariant}
    \end{align}
    of the left invariant vector field $L^A$ associated with the identity in $\M$, that is $\gamma^{L^A}(0)=e$ and which is defined for all $t \in \Real$. The notation $\gamma^{L^A}$ refers to the integral curve generated by the left-invariant vector field $L^A$ originating from the identity element $e \in G$. For any $g \in G$, the left translation map $L_g:G \to G, L_g(h) = gh, h \in G$ does not alter $L^A$. Here $A$ is the tangent vector to the curve $\gamma^{L^A}$ at $e$ where $\gamma_*^{L^A}$ represents the pushforward of the curve's tangent vector at $t=0$. The integral curve is then written as a mapping from the affine parameter $t \mapsto \exp(t A)$ with $A \in T_eG$. The map $\exp: T_eG \to G$ is defined via $\exp A = \exp(tA)\big|_{t=1}$, reflecting the convention that one moves from the identity $e \in G$ along the integral curve generated by $A$ to $\exp(A) \in G$ with time $t \in [0,1]$. This reflects the idea that evolution of curves in $G$ can be framed in terms of evolution from the identity element and therefore studied in terms of the canonical Lie algebra associated with $T_eG$. \\
    \\
    Other results mentioned in Appendix \ref{chapter:Background: Geometry, Lie Algebras and Representation Theory} also apply, such as the exponential map being local diffeomorphism from $T_eG$ to $e \in G$ and that $t \mapsto \exp(At)$ is considered a unique one-parameter subgroup of $G$ (indeed that all such one-parameter subgroups are of that form for $A \in T_eG \simeq L(G)$). This reiterates the one-to-one association between one parameter subgroups of $G$ and the exponential map. Left-invariance also applies to $n$-forms and that Lie algebras can be said to have associated with them dual Lie algebras (see literature for standard discussion).

\subsubsection{Maurer-Cartan Form} \label{sec:geo:Maurer-Cartan Form} Differential forms are an important tool for exhibiting how algebraic properties of $\g$ manifest within or affect geometric and topological properties of a system. Of note in this regard is the Maurer-Cartan form which is a $\g$-valued one-form on $G$. The form can be understood as relating structure constants and wedge products to the derivative of a one-form, providing a means of expressing $v \in T_p\M$ in terms of $\g$. Given the Maurer-Cartan equation:
\begin{align}
    d\omega^\alpha + \frac{1}{2}\sum_{\beta,\gamma=1}^n C_{\beta \gamma}^\alpha \omega^\beta \wedge \omega^\gamma=0 \label{eqn:geo:Maurer-Cartan equation}
\end{align}
we observe that the commutator of one-forms is a two-form which can be obtained via exterior differentiation. The term $d\omega^\alpha$ is the exterior derivative of the form, and the wedge product $\omega^\beta \wedge \omega^\gamma$ is summed over the structure constants $C_{\beta \gamma}^\alpha$ of the Lie algebra, weighted by the Lie bracket.

\begin{definition}[Maurer-Cartan Form] \label{defn:geo:Maurer-Cartan Form}
Let $G$ be a Lie group. The Maurer-Cartan form is a $\mathfrak{g}$-valued one-form $\omega$ on $G$, where $\mathfrak{g}$ is the Lie algebra of $G$. For each $g \in G$, $\omega_g: T_gG \rightarrow \mathfrak{g}$ is defined by
\begin{align*}
\omega_g(v) = l_{g^{-1}_*}(v)
\end{align*}
where $l_{g^{-1}_*}$ denotes the pullback of the left multiplication by $g^{-1}$ in $G$, and $v \in T_gG$ is a tangent vector at $g$. 
\end{definition}
Here $l_{g^{-1}_*}: T_p\M \to \g$. The one-form $\omega_g$ relates the tangent space at any point $g \in G$ back to the tangent space at the identity and therefore represents a map from $T_p\M \to \mathfrak{g}$. As Sharpe \cite{sharpe_differential_2000} notes, the equation is of profound use in classifying properties of spaces. In particular, Cartan's structural equations (see Theorem \ref{thm:geo:Cartanstructuralequations} below) can be expressed as:
\begin{align}
    d\omega_g = \frac{1}{2}[\omega_g,\omega_g]  \qquad [\omega_g,\omega_g] := \frac{1}{2}[\omega_g,\omega_g](u,v) = [\omega_g(u),\omega_g(v)] \label{eqn:geo:cartancurvatureequation}
\end{align}
which is also described as the \textit{Cartan curvature} equation as it describes local curvature (where $d\omega_g =0$ equates to flatness).  Note we can derive Cartan curvature equation (\ref{eqn:geo:cartancurvatureequation}) from equation (\ref{eqn:geo:Maurer-Cartan equation}).

Connecting the concepts above to quantum information problems in this work, the form takes a tangent vector $v \in T_gG$ and returns a left-invariant vector field associated with $v$, meaning the field remains constant along the flow. Here $l_{g^{-1}}$ translates $v$ back to the identity then pushes it forward to $g'$ using left translation. 

As we discuss, the Maurer-Cartan form assists in identifying left-invariant vector fields invariant under the group action, simplifying equations of motion in optimal control contexts. For optimal control, we are also interested in transformation groups. For example, a group $G$ acting on $\M$ does so if there exists a homomorphism such that $\gamma: G \to \sigma(\M)$ with $g \mapsto \gamma_g$ where $\sigma(G)$ is the permutation group of $\M$ (with an equivalent anti-homomorphism for right action). Left action on a differentiable manifold can then be considered as a homomorphism from the group (of diffeomorphisms on $\M$) to curves $\gamma$, that is $g \to \gamma_g$ where $\gamma: \text{Diff}(\M) \to \M$. This gives rise to the notion of \textit{equivariant} mappings, being structure preserving maps between pairs of group actions of $G$ on manifolds $\M,\N$ where $g \mapsto \gamma_g, g\mapsto \gamma_g'$ on $\M$ and $\N$ respectively where $\gamma_g' \circ f = f \circ \gamma_g$. Equivariance has recently become of interest in quantum machine learning via the construction of equivariant neural networks where the network architecture is constructed to be equivariant \cite{meyer_exploiting_2022-1,cohen_equivariant_2021,mernyei_equivariant_2022}. 
\\
\\
We can connect different types of group actions discussed in earlier parts of this work to their geometric equivalents:
\begin{enumerate}[(i)]
    \item \textit{Kernel}: kernel of a $G$-action is the subgroup of $G$ defined by:
    \begin{align*}
        K = \{  g \in G | gp = p, \forall p \in \M \}.
    \end{align*}
    \item \textit{Effective group action}: is where the kernel equals the identity, that is $K = \{ e \}$.
    \item \textit{Free group action}: if for all $p \in \M$ it is the case that $\{   g \in G | gp = p \} = \{e \}$ i.e. $\M$ is moved away from itself except with the exception of the unit element $e$ i.e. $e=p$ is the only element such that $gp=p$.
    \item \textit{Transitive group action}: G-action is \textit{transitive} if any pair of points $p,q\in\M$ can be \textit{connected} by an element of the group, i.e. there exists a $g \in G$ such that $p = gq$. In transitive action, the whole of $\M$ can be probed by the $G$-action. This is complementary to the idea that in an effective action the whole of $G$ can be proved by its action on the manifold $\M$.
    \item \textit{Orbit}: the orbit $O_p$ of the G-action through $p$ is the set of all points in $\M$ that can be reached from $p$, that is:
    \begin{align}
        O_p = \{ q \in \M | \exists g \in G, q = gp  \}. \label{eqn:geo:orbit}
    \end{align}
    When a group $G$ acts on a set $X$ (group action on $X$) it \textit{permutes} the element of $X$. The path of an element of $X$ moves around a \textit{fixed path} is its orbit. An example is the \textit{sphere} $S^2$ acted on by the \textit{circle group} $S^1$ via rotations around the $z$ axis, this leaves the elements of the set invariant. Such action is analogous to the action of the compact subgroup $K$ under a Cartan  decomposition $G = K \oplus P$, the intuitive idea being that each $K$ defines an orbit about which $k \in K$ transform, but that to reach different orbits, one requires the action of some element in $P$.
    \item \textit{Stabiliser (isotropy) groups}: if G action on $\M$ is transitive with $G_p$ of the action is given by:
        \begin{align}
            G_p = \{ g \in G | gp = p  \}. \label{eqn:geo:stabiliser}
        \end{align}
        The stabilisers of an element $p \in \M$ can be thought of as group actions (or permutations of $\M$ which form a group homomorphically mapped to $G$) which leave $p$ unchanged, i.e. the set of group action identity elements with respect to $p \in \M$ where we can think of a set of permutations which leave $p$ unchanged or `in the same position'. Stabiliser groups are fundamental to quantum information (especially surface codes). In our final chapter, the stabiliser group $K$ of $G$ is of fundamental importance, noting, for example, that stabiliser groups along an orbit are conjugate i.e. if $p,q$ belong to the same orbit of a $G$-action, they can be reached via some $g \in G$. 
\end{enumerate}

\subsubsection{Infinitesimal transformations and adjoint action}\label{sec:geo:Infinitesimal transformations and adjoint action}
An important action throughout this work is the \textit{adjoint action} of $G$ on itself (discussed in Chapter 2): 
        \begin{align*}
            Ad_g(g') = gg'g^{-1}.
        \end{align*}
        The kernel of this action is the centre $C(G)$ because the centre is the set of elements of $G$ that commute with every element in G, but then for an $g' \in C(G)$ we have $Ad_g(g')= gg'g^{-1}=g'gg^{-1}=g'$ i.e. the (effective) kernel of this action of conjugating by $g$ is the set of elements in $g$ that result in this action being equivalent to the identity. The relation between the commutator of a pair of Lie algebra elements (crucial for all quantum control problems) and the push-forward $Ad_{g*}$ of the adjoint action of $G$ can be expressed as follows. Let $A,B \in T_eG$ with Lie bracket $[A,B]$, then the bracket has a representation as differentiated adjoint:
\begin{align*}
    [A,B] = \frac{d}{dt}Ad_{\exp(tA)*}(B)|_{t=0}.
\end{align*}
Recalling that the exponential is given such that $\exp(Ad_{g*}(B)) = g(\exp(B))g^{-1}$, it can be shown that the relation between the commutator can be thought of in terms of this conjugacy:
\begin{align*}
    \exp(tA)\exp(B)\exp(-tA) = \exp(t[A,B] + O(t^2))
\end{align*}
and $X^{Ad_{g*}(A)} = \delta_{g^{-1}*}(X^A)$ where $\delta_{g^{-1}*}$ is the differential (pushforward) of the map $g^\inv$ acting on $X^A$. That is, the vector field associated with the adjointly transformed Lie algebra element $A$ is equivalent to the transformed vector field $A$. 

We conclude this section via reiterating in a geometric context the key results related to Lie algebra homomorphisms and connecting these to vector fields. This is an important result for further on when we introduce the \textit{connection} between vector (fibre) bundles that serves a pivotal role in our final chapter. In the theory of infinitesimal transformations, we can represent the Lie algebra $\g$ of $G$ via vector fields on $\M$ on which $G$ acts. This mapping is a homomorphism of Lie algebras and associates $X^A$ with $A \in T_eG$. Assume $G$ right-acts on $\M$. Given a mapping
    \begin{align*}
        A \mapsto X^A
    \end{align*}
    that assigns to each $A \in T_eG$ the vector field $X^A$ (denoted the \textit{induced vector field} on $\M$) is a homomorphism from the Lie algebra $L(G) \simeq T_eG$ 
    into the infinite-dimensional Lie algebra of all vector fields on $\M$. This is denoted:
    \begin{align*}
        [X^A, X^B] = X^{[A,B]}
    \end{align*}
    for all $A,B \in T_eG \simeq L(G)$. For left actions, the map $A \mapsto X^A$ is an anti-homomorphism $[X^A, X^B] = X^{[A,B]}$ from $T_eG \simeq L(G)$. If the action is effective on $\M$, then the map $A \mapsto X^A$ is an isomorphism from $T_eG \simeq L(G)$ to Lie algebras on $\M$.

\subsection{Fibre bundles}\label{sec:geo:Fibre bundles}
In many cases of physical importance, the vector space $V$ arises as a representation space of some \textit{internal symmetry group}, especially for example in symmetry reduction techniques used in geometric control theory. Fibre bundles allow for the study of situations where the structure of interest over each point of a base space is more complex than just a vector space, allowing, for example, gauge symmetries to be identified (e.g. where gauge symmetries are not completely represented as transformations on vector spaces). They are of particular relevance in gauge theories providing a means of describing fields and their connections over manifolds.  
In quantum information problems involving systems with a geometric or topological character, quantum state space can be considered as sections of a fibre bundle where the base space represents a space of parameters or configurations, and the fibres represent the Hilbert spaces associated with each configuration. In geometric control, we often look to find ways of modelling the control space modeling control inputs and state spaces as fibres. Fibre bundles are an abstraction of vector bundles and are utilised for when, for example, mappings lack vector space properties. 
We begin with the concept of a \textit{bundle}, defined as a triple $(E, \pi, \mathcal{M})$, where $E$ and $\mathcal{M}$ are linked via a continuous projection map $\pi: E \to \mathcal{M}$. 
We briefly sketch out the theory of fibre bundles and principal bundles as these are touched upon in other sections.  Define a \textit{bundle space} as $E = \bigcup_{p \in \mathcal{M}} F_p$, with $F_p$ being fibres. A fibre is then defined as follows.
\begin{definition}[Fibre]\label{defn:geo:fibre}
    The fibre $F_p$ over $p$ is the inverse image of $p$ under $\pi$. Geometrically, it arises via the map $\pi^{-1}:\mathcal{M} \to T\mathcal{M}$, an example of a fibre bundle associating $p \in \mathcal{M}$ with the tangent space $T_p \mathcal{M}$. Formally, we define the fibre $F_p$ as follows. The projection $\pi$ associates each fibre $F_p$ with a point $p \in \mathcal{M}$, where:
\begin{align*}
    F_p = \pi^{-1}(\{p\}),
\end{align*}
defines $F_p$ as the preimage of $p$ under $\pi$ (that is the set of all points in $E$ mapped to $p \in \M$).
\end{definition}
 Certain bundles have the special property that the fibres $\pi^{-1}(\{p\}), p \in \M$ are all \textit{homeomorphic} (\textit{diffeomorphic} for manifolds) to $F$. In such cases, $F$ is known as the \textit{fibre} of the bundle and the bundle is said to be a \textit{fibre bundle} (we explore this more via \textit{associated bundles} below). For vectors, this is the set of all vectors that are tangent to the manifold at the point $p$. The fibre bundle is sometimes visualised in diagrammatic form (see Isham \cite{isham_modern_1999} $\S 5.1$):
    \[ \begin{tikzcd}
F \arrow{r} & E \arrow{d}{\pi} \\%
& \M
\end{tikzcd}
\]
Related to fibre bundles is the idea of a \textit{section}. The section of a bundle $(E,\pi,\M)$ is a \textit{map} from the base space (manifold) to the total space:
    \begin{align*}
        s: \M \to E
    \end{align*}
    such that the image of each point $p \in \M$ lies in the fibre $\pi^{-1}(\{ p\})$ over $p$ such that:
    \begin{align*}
        \pi \circ s = id_\M
    \end{align*}
    So the section is the map from the manifold such that we take $p$, send it up to the total space $E$, then that image of $p$, when subject to $\pi$, takes us back to $p \in M$, hence $\pi \circ s$ is equivalent to the identity on $\M$. One can also define maps between bundles and pullbacks among fibres analogous to the case for vector bundles.

\subsubsection{Principal fibre bundles}\label{sec:geo:Principal fibre bundles}
An important canonical fibre bundle is the principal fibre bundle whose fibre acts as a Lie group in a particular way. For our purposes, they allow for the definition of \textit{connections} (see below) which describe how fibres (or vector spaces) are connected over different points in $\M$. Connections are of fundamental importance to results in our final Chapter and also to definitions of vertical and horizontal subspaces in subRiemannian control problems further on. A connection on a principal bundle defines a notion of horizontal and vertical subspaces within the tangent space of the total space. This distinction is crucial for defining parallel transport and curvature, concepts that are central to understanding the dynamics and control of systems with symmetry. Firstly, we define a principal fibre bundle as follows.

\begin{definition}[Principal fibre bundle]\label{defn:geo:Principal fibre bundle}
    A principal fibre a bundle $(E,\pi,\M)$ is a \textit{G-bundle} (a \textit{fibre bundle}) if the total space $E$ is a \textit{right-G space} and if the bundle $(E,\pi,\M)$ is isomorphic in the following way:
    \begin{align*}
        (E,\pi,\M) \simeq (E,\rho, E/G)
    \end{align*}
    where $E/G$ is the \textit{orbit space} of the $G$-action on $E$ and $\rho$ is here the usual projection map such that we have the diagram:
    \[ \begin{tikzcd}
E \arrow{d}{\pi} \arrow{r}{u} & E \arrow{d}{\rho}\\%
\M \arrow{r}{\simeq} & E/G
\end{tikzcd}
\]
\end{definition}
A principal fibre bundle has a typical fibre that is a Lie group $G$, and the action of $G$ on the fibres is by right multiplication, which is free and transitive. The fibres of the bundle are the orbits of the $G$-action on $E$ and hence are not generally homeomorphic to each other. All non-principal bundles are associated with the principal bundle. \textit{Principal} $G$ \textit{bundles} are when $G$ acts freely on $E$ (that is, \textit{free action} is when the only element of $G$ that acts as an identity element on $x \in E$ is the identity in $G$ itself). Given a closed subgroup $H \subset G$, then the quotient space $G/H$ is also an orbit space with fibre $H$. When $G$ is the fibre itself and the action on $G$ is both free and transitive, we use the notation $P$ to denote the total space (i.e. $E$). The notation $P$ is used to emphasise that each fibre is isomorphic to $G$ itself and that all fibres in the bundle are homogeneous i.e. they are all structurally isomorphic (so we can utilise a single representation for each). A few other concepts to note include:
\begin{enumerate}[(i)]
    \item \textit{Principal total space (P)}. For principal $G$ bundles, the total space is often denoted as a \textit{principal total space} where the principal $G$ bundle is then indicated by the triple $(P,\pi,\M)$ where $P$ is the principal total space. There exist \textit{principal maps} between two such bundles i.e. $u: P \to P'$ for $(P,\pi,\M)$ and $(P',\pi',\M')$. The mapping is $G$-equivariant as $u(pg) = u(p)g$. Here $G$ acts on $P$ and $P$ projects onto $\M$ via $\pi$.
    \item \textit{Triviality}. The \textit{triviality} (or \textit{trivialisation}) of a principal $G$-bundle relates to whether the bundle expressly respects the product structure of the base space with $G$. A principal $G$-bundle is trivial if it is isomorphic to the product bundle $(\M \times G, pr_1,\M)$ where $pr_1:\M \times G \to \M$ is a projection onto $\M$. Triviality requires (i) \textit{compatibility} with projection maps so that diffeomorphisms $f$ respect the fibration structure of $P$ by mapping fibres in $P$ over $p \in \M$ to $\{p\} \times G$ and (ii) \textit{equivariance} with respect to $G$ such that $f(pg) = f(p)g$, where $f(p) \in \M \times G$ (such that $f$ preserves the group action structure of the bundle).
\end{enumerate}
For the avoidance of doubt, we have so far (with slight abuse of notation for convenience) been equating $G \equiv \M$ (as distinct from explicitly notationally indicating the group acting on the manifold $\M \times G)$, where $T_gG$ for $g \in G$ captures the infinitesimal directions in which $G$ can evolve, mapping directly to tangent spaces $T_p\M$ on the manifold. Usually the principal bundle $(P,\pi,\M)$ is introduced as a more abstract formulation to cater for where $G$ acts on a different manifold (e.g. where $G \neq \M$). The formulation of total space $P$ allows consideration of how the tangent spaces $T_gG$ (that is for our formulation, $T_p\M$) are related across the entirely of $\M$. Thus in our treatment, we are interested in how $G$ acts upon itself, which in the language of quantum operators is how operators act upon themselves e.g. $U_1(t)U_2(t)=U'(t)$ can be regarded as group elements acting on themselves. 

\subsubsection{Associated bundles}
While principal fibre bundles above allow us to abstractly associate the group action of $G$ to a manifold, the fibres remain abstract. In practice we want an association between fibre bundles and more familiar structures from a geometric control and quantum information processing perspective, e.g. we want our fibres to have the structure of vector spaces, tangent space or Lie algebras. For this we turn to the concept of \textit{associated fibre bundles} which enable the construction of bundles with fibres that are not necessarily groups but can be any space on which the group acts. This is particularly relevant to our use of geometric methods where fibres are vector spaces, such as Lie algebras or tangent spaces. The idea \cite{isham_modern_1999} is that an associated bundle can be constructed where $G$ acts as a group of transformations on $F$. 

\begin{definition}[Associated bundle]\label{defn:geo:Associated bundle}
    Given a principal fibre bundle $(P, \pi, \mathcal{M})$ and a left $G$-space $F$, an associated fibre bundle is a fibre bundle $(P_F, \pi_F, \mathcal{M})$ where the principal total space $P_F$ is defined as the quotient $(P \times F)/G$ (also represented as $P \times_G F$ being the space formed by the Cartesian product of $P$ and $F$ modulo the group action). 
\end{definition}
The projection $\pi_F: P_F \to \mathcal{M}$ is induced by the projection $\pi: P \to \mathcal{M}$ of the principal bundle. For each $p \in \M$, the space $\pi^\inv_F(p)$ is homeomorphic to $F$, in essence allowing us to select structures, such as vector spaces or tangent spaces, which respect the fibre bundle structure of the principal $G$-bundle. In this way, the principal bundle structure can be extended over the chosen fibres e.g. vector spaces in a way that varies smoothly from point to point over $\mathcal{M}$. This is in turn central to the concept of a connection, which gives a covariant derivative along the fibres of the bundle.

The primary form of associated bundle we are interested in are\textit{ vector bundles}, which are important due to their being equipped with a natural vector space structure. For convenience we retain $E$ instead of $P_F$.
\begin{definition}[Vector bundle]\label{defn:geo:Vector bundle}
    An $n$-dimensional complex or real vector bundle $(E,\pi,\M)$ is a type of fibre bundle in which each fibre exhibits the structure of an $n$-dimensional vector space. For each $p \in \M$ there exists a neighbourhood $U$ and way of locally expressing  trivialisation of the bundle via $h: U \times \R^n \to \pi^{-1}(U)$ such that for all  $ y \in U, h: \{ y \} \times \R^n \to \pi^{-1}(\{ y\})$ is linear.
\end{definition}
Vector bundles are the primary type of (associated) fibre bundle with which we are concerned in this work. The concept has broad application, e.g. in machine learning by allowing the representation of data points that are sensitive to inherent symmetries in the data. In quantum settings, unitary operations and quantum channels, which describe the evolution of quantum states, can be seen as bundle maps that act on the sections of vector bundles. Unitary operations, representing reversible quantum evolutions, can be modeled as isomorphisms of Hilbert spaces $\Hilb$ (definition \ref{defn:quant:Hilbert Space}) as fibres which preserving their structure. Quantum channels, which describe more general (possibly irreversible if decohering) quantum evolutions, can be represented as linear maps between fibres (see \cite{bengtsson_geometry_2006} for more detail).

\subsection{Connections}\label{sec:geo:Connections}
Connections are a fundamentally important concept in geometry, encapsulating the concept of differentiation of vector fields along curves on manifolds. A connection on a manifold $\mathcal{M}$, or more specifically on a tangent bundle $T\mathcal{M}$, provides a systematic way to parallel transport vectors along paths, allowing the comparison of tangent spaces at different points on $\mathcal{M}$, effectively facilitating the extension of the notion of directional derivatives to curved spaces. Intuitively the idea of a connection is a means of associating vectors between infinitesimally adjacent tangent planes $T_p\M \to T_{p+dp}\M$ as one progresses from $\gamma(t)=p$ to $\gamma(t+dt)=p+dt$ on $\M$. Another way to think about them is that they provide a way of identifying how $T_p\M$ transforms along the curve $\gamma(t)$, giving conditions to be satisfied, in the form of transformation rules for such vectors, for tangent vectors $\dot\gamma(t)$ to remain parallel as they are transported along a curve (parallel transport or the vanishing of the covariant derivative, which we discuss below). 

Connections are also a fundamental means of distinguishing between Riemannian and subRiemannian manifolds via the decomposition of fibres (and Lie algebras) into horizontal and vertical subspaces. They are thus central to quantum and classical control problems, such as the $KP$ control problem we study extensively in the final Chapter. In Riemannian geometry, for example, Christoffel symbols represent coefficients of connection on the bundle of frames (set of bases) for $\M$. Connections, as we explain below, are related to notions of parallel transport and covariant differentiation. The underlying idea of parallel transport and covariant differentiation requires that one compares points in neighbouring fibres (or vector spaces, in the case of a vector bundle) in a way that isn't dependent upon a particular local coordinate system (or trivialisation). Thus a concept of directionality is needed such that vector fields point from one fibre to another. Vector fields arising from Lie algebras lack this intrinsic orientation of directional pointing. The connection provides a concept of directionality by partitioning the fibre into horizontal and vertical subspaces as discussed below.

Consider a principal $G$-bundle $(P, \pi, \mathcal{M})$ above. A connection on $P$ provides a smooth splitting of the tangent space $T_pP$ at each point $p \in P$ into vertical and horizontal subspaces, $T_pP = V_pP \oplus H_pP$, where $V_pP$ is tangent to the fibre and $H_pP$ is isomorphic to $T_{\pi(p)}\mathcal{M}$. To understand this formalism, we expand upon important concepts of vertical and horizontal subspaces. These are fundamental in later chapters, where synthesis of time-optimal (approximate) geodesics arise by way of choosing Hamiltonians comprising generators from horizontal subspaces (Lie subalgebras). Evolution according to horizontal subspaces in Riemannian manifolds relates to vanishing (zero) covariant derivatives.

\begin{definition}[Vertical subspace]\label{defn:geo:Vertical subspace}
   The vertical subspace $V_pP$ at $p \in P$ is defined as the kernel of the differential of the projection map $\pi: P \rightarrow \mathcal{M}$, i.e., 
    \begin{align*}
    V_pP = \ker(\pi_*|_p) = \{ v \in T_pP \mid \pi_*(v) = 0 \}.
    \end{align*}
\end{definition}
where $\pi_*$ is the associated mapping i.e. $\pi_*: T_pP \to T_{\pi(p)}\M$. Connections can be associated with $L(G)$ valued one-forms $\omega$ (which is the assignment of dual spaces for each $p \in \M)$ where $\omega_p(X^A)=A$ for $p \in \M$ and $A \in L(G)$. Intuitively, $V_pP$ consists of tangent vectors to $P$ at $p$ that are ``vertical'' in the sense that they point along the fibre $\pi^{-1}(\pi(p))$. These vectors represent infinitesimal movements within the fibre itself, without leading to any displacement in the base manifold $\mathcal{M}$. Vector fields $X^A_p$ belong to the subspace $V_pP \subset T_pP$  intuitively they point `along' the fibres. This vertical subspace is defined by:
    \begin{align*}
        V_pP = \{ \tau \in T_pP | \pi_* \tau = 0 \}
    \end{align*}
By contrast, horizontal vectors represent transformations between fibres. We formalise the definition of a horizontal subspace below.   
\begin{definition}[Horizontal subspace] \label{defn:geo:Horizontal subspace}
The horizontal subspace $H_pP$ at $p \in P$, on the other hand, is selected by the connection and complements the vertical subspace within $T_pP$. It is formally defined such that $\tau \in H_pP$ if and only if $\omega_p(\tau)=0$ (i.e. the kernel of the one-form) and can be related as:
\begin{align}
    H_p = \{ \tau \in T_pP | \pi_* \tau \neq 0\}.
\end{align}
\end{definition}
Specifically, a connection on $P$ provides a smooth assignment of a horizontal subspace to each point in $P$ such that the tangent space at any point $p$ decomposes as a direct sum:
    \begin{align*}
    T_pP = V_pP \oplus H_pP.
    \end{align*} 
The horizontal subspace consists of those vectors that are orthogonal (under a chosen Riemannian or pseudo-Riemannian metric on the bundle) to the vertical space with respect to a chosen connection. Vectors in $H_pP$ are ``horizontal'' in the sense that they correspond to displacements that lead to movement in the base manifold $\mathcal{M}$ when considered under parallel transport defined by the connection. This can be understood diagramatically. The projection map $\pi: P \to \M$ induces a (push-forward) map $\pi_*: T_pP \to T_{\pi(p)}\M$:
    \begin{figure}
        \centering
        \[ \begin{tikzcd}
    & TP \arrow{d} \arrow{r}{\pi_*} & T\M \arrow{ddl}\\
G \arrow{r} & P \arrow{d}{\pi}  & \\%
& \M  & 
\end{tikzcd}
\]
        \caption{Commutative diagram showing the relationship of the connection (map from $G \to P \to \M$), the projection map $\pi: P \to \M$ and induced horizontal map (the pushforward) $\pi^*$}
\label{fig:geo:connectionfibresetc}
    \end{figure}
In this structure, $A \mapsto X^A$ is an isomorphism of $L(G)$ onto $V_pP$. The vertical subspace thus comprises those vectors that are in the kernel of the map $\pi_*$, i.e. they map to the nullspace of $T\M$ and such vectors do not generate evolutions in the base manifold $\M$. Analogously, these vectors `point up' so do not specify a direction to move on $\M$, akin to a vector pointing vertically out of a surface. The horizontal subspace, by contrast, comprises vectors that generate translations on the base manifold. This is intuitively equivalent to those vectors `pointing' in some direction of translation on the manifold from say $p \to p' \in \M$. Because $p'$ has its own fibre (e.g. vector bundle), then those horizontal subspace vectors `point' horizontally towards other vector bundles.  
\\
\\
It is useful to dwell on the notion of orthogonality between vertical and horizontal subspaces implied by the connection. Indeed in many ways it is more illuminating to begin with this concept of orthogonality to build intuition around the terms vertical and horizontal. This provides means of distinguishing between vertical and horizontal movements in $P$. The choice of horizontal subspace $H_pP$ at $p \in P$ is not arbitrary. Rather, it is subject to conditions ensuring consistency with the group action and smooth variation across $P$:
\begin{align*}
    \delta_{g*}(H_pP) = H_{pg}P, \quad \forall g \in G, p \in P,
\end{align*}
where $\delta_g(p) = pg$ denotes the right action of $G$ on $P$. This condition ensures that vectors being deemed horizontal is consistent with the geometric structure of the bundle as defined by the group action.

Intuitively, vertical subspaces manifest the internal symmetry of the bundle encoded by the Lie group $G$, while horizontal subspaces encapsulate the geometry of how the bundle expands over the base manifold. This geometric structure, facilitated by the connection, is crucial for defining parallel transport,  curvature and ultimately geodesics (and their approximations) that characterise time-optimality.
We can thus formally define a connection \cite{isham_modern_1999,frankel_geometry_2011} and \cite{helgason_differential_1979} in these terms.
%Connection definition
\begin{definition}[Connection]\label{defn:geo:Connection} A \textit{connection} on a principal $G$-bundle given as:
\begin{align*}
    G \to P \to \M
\end{align*}
is a smooth assignment of horizontal subspaces $H_pP \subset T_pP$, to each point $p$ in the total space $\in P$ such that:
\begin{enumerate}[(i)]
    \item $T_pP \simeq V_pP \oplus H_p P, \forall p \in P$ (i.e. decomposable into direct sum of horizontal/vertical)
    \item $\delta_{g*}(H_pP) = H_{pg}P, \forall g \in G, p \in P$ (subject to right action) where $\delta_g(p) = pg$ denoting right action.
\end{enumerate}
\end{definition}
An example of a connection is the covariant derivative, which describes how vectors transform across the fibre (vector) bundle.The commutative diagram in Figure (\ref{fig:geo:connectionfibresetc}) illustrates how the projection map $\pi: P \to \M$ gives rise to an \textit{induced horizontal map} from the tangent space to the total space:
\begin{align*}
    \pi_*: T_pP \to T_{\pi(p)}\M.
\end{align*}
By construction, the kernel of this map is the vertical subspace. Connections can also be understood usefully in terms of one forms.  Connections can be associated with certain $L(G)$-valued one-forms $\omega$ on $P$. Recall that the map $\ell: L(G) \to VFlds(P), A \mapsto X^A$, so $\ell^{-1}$ maps back to the Lie algebra $L(G)$. If $\tau \in T_pP$, then:
\begin{align*}
   \omega_p: T_pP \to P, \omega_p(\tau) = \ell^{-1}(V_pP(\tau))
\end{align*}
which maps back to $V_pP$. As this map takes us back to $L(G)$, then $\ell$ is the isomorphism of $L(G)$ with $V_pP$, so we can associate one-forms with maps between the tangent space and the Lie group as follows:
\begin{enumerate}[(i)]
    \item $\omega_p(X^A) = A, \forall p \in P, A \in L(G)$;
    \item $\delta_g^* = Ad_{g^{-1}}\omega$, with $\delta_g^* \omega (\tau) = Ad_{g^{-1}}\omega(\tau)$;
    \item $\tau \in H_pP \iff \omega_p(\tau) = 0 $.
\end{enumerate}
As touched upon above, the horizontal subspace of the tangent space $TP$ can thus be thought of as the kernel of the one-form that maps to the left-invariant Lie group ($L(G)$) (see section \ref{sec:geo:leftinvariancecommutator}):
\begin{align*}
    H_pP = \{ \tau \in T_pP | \omega_p(\tau) = 0, \omega_p(\tau) = \ell^{-1}(V_pP(\tau))  \}
\end{align*}
In this formulation, the connection can be expressed as an $L(G)$-valued one form on $P$ satisfying (i) to (iii) above. The map $\ell: L(G) \to VFlds(P)$ maps Lie algebra elements to vector fields on $P$ such that $\ell^{-1}(V_pP(\tau))$ would map this vertical component back to an element of the Lie algebra.

\subsubsection{Relation to Cartan decompositions}\label{sec:geo:Relation to Cartan decompositions}
The decomposition into vertical and horizontal subspaces discussed above is particularly relevant to time optimal control problems. In essence, given the Cartan decomposition (\ref{defn:alg:cartandecomposition}) $\g = \k \oplus \p$ we associate $\k$ as the vertical and $\p$ as the horizontal subspace. We can then understand the symmetry relations expressed by the commutators related to this vertical and horizontal sense of directionality: i.e. given $[\k,\k]\subset \k, [\k,\p]\subset \p$ and $[\p,\p]\subset \k$ we can see that the horizontal generators under the adjoint action shift from $p \in G/K$ to $p' \in G/K$, while the vertical generators in k do not translate those points in $G/K$. We introduce two definitions to elucidate this further, the \textit{Cartan connection}.
\begin{definition}[Cartan Connection]\label{defn:geo:CartanConnection}
A Cartan connection on a principal $G$-bundle $(P,\pi,\M)$ (with group $G$, subgroup $H$ (and subalgebra $\h$) and homogeneous space $G/H$) consists of an atlas (see definition \ref{defn:geo:Coordinate charts and atlases}) $U \in \M$ and a $\mathfrak{g}$-valued one-form $\theta_U: TU \to \g$ defined on each chart such that (a) $\mathfrak{h}$: $\theta_U \mod \mathfrak{h}$ is a linear isomorphism for every $u \in U$ and (b) for charts $U,V$, then $h: U \cap V \rightarrow H$:
\begin{align}
    \theta_V = \text{Ad}(h^{-1})\theta_U + h^*\omega_H \label{eqn:geo:thetaUoneformcartanmaurer}  
\end{align}
    where $\omega_H$ is the Maurer-Cartan form of $H$.
\end{definition}
Note that Cartan curvature (equation (\ref{eqn:geo:cartancurvatureequation})) can be expressed as:
\begin{align}
\Omega_U = d\theta_U + \frac{1}{2}[\theta_U, \theta_U] \label{eqn:geo:cartancurvatureequation2}
\end{align}
We can understand the commutation relations in terms of principal bundles and horizontal/vertical subspaces as follows: 
\begin{enumerate}
    \item \textit{Vertical subspace $\k$}. The first relation $[\mathfrak{k}, \mathfrak{k}] \subseteq \mathfrak{k}$ provides that the Lie bracket of two vertical generators remains within the vertical subspace. Geometrically, this means that transformations generated by elements of $\mathfrak{k}$ remain within the fibres of the principal bundle $G \rightarrow G/K$. Such vertical movements do not evolve to other points in the base space $G/K$, reflecting an intrinsic, self-contained dynamics within each fibre, akin to internal symmetries. Geometrically, translations by elements of $\mathfrak{k}$ keep points within the same orbit in $G/K$.
    \item \textit{Horizontal to vertical}. The second commutation relation $[\mathfrak{p}, \mathfrak{p}] \subset \mathfrak{k}$, indicates that the commutator of two horizontal elements results in a vertical element. This property is particularly relevant as it relates to the curvature of the connection in $G/K$. The horizontal generators under the adjoint action can lead to a shift that results in a vertical movement, suggesting that purely horizontal displacements can induce curvature in the space, effectively ``bending'' into the vertical subspace. This phenomenon is a hallmark of the non-trivial geometry in subRiemannian spaces and is crucial for understanding the underlying geometric structure of $G/K$. Elements of $\mathfrak{p}$ correspond to horizontal directions that can be projected onto the base space $G/K$, effectively causing translations from one orbit to another. This curvature is a manifestation of how horizontal translations alter the geometric structure of the space, moving points across different orbits in $G/K$ (see below).
    \item \textit{Vertical acting on horizontal}. The relation $[\mathfrak{k}, \mathfrak{p}] \subset \mathfrak{p}$ implies that the action of vertical elements on horizontal ones results in horizontal displacements. This can be interpreted as the influence of the group's internal symmetries on the directionality of movement within $G/K$. In other words, the vertical generators, through their adjoint action, can modify the direction of horizontal generators without leaving the horizontal plane, thereby affecting the trajectory of points in $G/K$ without transitioning to vertical movement.
\end{enumerate}
Recall that in the full Lie group $G$, orbits are generated by the action of its subgroups, including $K$, on points within $G$. For a subgroup $K$ and an element $g \in G$, the orbit of $g$ under the action of $K$ is defined as $\mathcal{O}_g = {k \cdot g \mid k \in K}$. These orbits represent continuous trajectories or paths within $G$ that are traced out by the action of $K$, reflecting the inherent symmetry structure imposed by $K$ on $G$. When considering the quotient space $G/K$, which represents the space of left cosets of $K$ in $G$, the action of $K$ on $G$ translates differently. Here, each point in $G/K$ corresponds to an orbit of $K$ in $G$, and the quotient space essentially collapses these orbits to single points, with the projection map $\pi: G \rightarrow G/K$ sending elements of $G$ to their corresponding orbits in $G/K$. 

\subsubsection{Fixed points and orbits}\label{sec:geo:Fixed points and orbits}
Translations by elements of $\mathfrak{k}$ preserve the orbit of a point in $G/K$ (reflecting the isotropy subgroup $K$'s action where $K$ is the stabiliser (see equation (\ref{eqn:geo:stabiliser})) of a point in $G/K$) while translations by elements of $\mathfrak{p}$ have the capacity to navigate across different orbits, revealing the transitive but stratified nature of the action of $G$ on $G/K$. This can be related to the fixed-point operation of $K$. In the quotient space $G/K$, the orbits of $K$ in $G$ appear as if $K$ acts as a fixed-point operator. This perspective arises because, within $G/K$, each orbit $\mathcal{O}_g$ is identified with a single point, and the distinct actions of elements of $K$ on points in $G$ that would have moved them along their orbits are now seen as leaving the corresponding points in $G/K$ invariant. That is:
\begin{align}
    k(gK) = (kg)K = gK
\end{align}
for all $k \in K$ and $g \in G$, where $gK$ denotes the coset (or point in $G/K$) corresponding to $g$. The action of $K$ on any $g \in G$ thus translates to the invariance of the coset $gK$ in $G/K$, effectively rendering $K$ as acting by fixed points in the quotient space. To this end, each fibre of $G/K$ can be thus regarded as an equivalence class of orbits. For any Lie group $G$ and a subgroup $K$, the quotient space $G/K$ is constructed by partitioning $G$ into equivalence classes under the equivalence relation:
\begin{align*}
    g \sim g' \longleftrightarrow g' = gk
\end{align*}
for $k \in K$. This relation groups elements of $G$ into sets where each set, or equivalence class, contains elements that can be transformed into one another by the right action of elements of $K$. Mathematically, an equivalence class of $g \in G$ can be denoted by the coset $gK = \{gk \mid k \in K\}$, which represents an orbit of $g$ under the action of $K$.

In the quotient topology of $G/K$, each point corresponds to one such equivalence class or orbit. The projection map $\pi: G \rightarrow G/K$ maps each element $g \in G$ to the equivalence class $gK$ it belongs to. The preimage $\pi^{-1}(\pi(g)) = gK$ under this map is a fibre over the point $\pi(g)$ in $G/K$. Therefore, each fibre of $G/K$ is representative of an orbit in $G$ under the action of $K$, signifying that the entire set of elements in $G$ that are related through multiplication by elements of $K$ are collapsed to a single point in $G/K$. This illustrates how the quotient space encapsulates the idea of moving from the specific (individual group elements in $G$) to the general (equivalence classes or orbits in $G/K$) by abstracting away the internal symmetries represented by $K$.

This explicitly builds intuition for the symmetry reducing properties of $KP$ decompositions and the fundamental role of connections in defining such symmetries on manifolds. Moreover, as we discuss below, the synthesis of connections in terms of vertical and horizontal spaces, related to the homogeneous space $G/K$ allows for a detailed understanding of Riemannian and subRiemannian geometry and, in turn, geometric control theory.

\subsection{Parallel transport and horizontal lifts}\label{sec:geo:Parallel transport and horizontal lifts}
We come now to important concepts related to geodesics and time optimality. We are interested in the concept of parallel transport in a principal bundle $\xi = (P,\pi,\M)$ with a connection given (via the one-form) $\omega$. We are interested in \textit{horizontal vector fields} whose flow lines move from one fibre to another, intuitively constituting translation across a manifold via generators in the horizontal subspace.  Parallel transport shifts vector fields vectors along integral curves such that they are parallel according to a specified connection. The concept of horizontal lifts is central to this process by enabling (i) preservation of a concept of direction and (ii) preserving `horizontality' such that the notion of straightness or parallelism of Euclidean space can be extended to curved manifolds, essential for the study of geodesics. Note that while all geodesics are generated by generators in the horizontal subspace $H_p\M$, not all curves generated by horizontal generators are geodesics.

\begin{definition}[Horizontal lift]\label{defn:geo:Horizontal lift}
    Given a vector field $X$ on $\mathcal{M}$ and connection expressed in terms of $\omega$, its horizontal lift $X^{\uparrow}$ is a vector field on $P$ that is everywhere horizontal with respect to the one-form $\omega$ such that given $\pi_*: H_p P \to T_{\pi(p)}\M$ is an isomorphism, for all $X$ on $\M$ there exists a \textit{unique} vector field $X^{\uparrow}$ satisfying the following for all $p \in P$:
    \begin{enumerate}[(i)]
        \item $\pi_*(X^{\uparrow}_p) =  X_{\pi(p)}$
        \item $V_p(X^{\uparrow}_p)=0$
    \end{enumerate}
The vector field $(X^{\uparrow}_p)$ is called the \textit{horizontal lift} of $X$ as it `lifts' up the vector field $X$ on $\M$ into the \textit{horizontal subspace} of $TP$.
\end{definition}
The requirement of $V_p(X^{\uparrow}_p) = 0$, indicates that that $X^{\uparrow}$ lies entirely in the horizontal subspace $H_p\M$, encapsulates the essence of parallel transport as maintaining the direction of $X$ through the fibres of $P$. For a smooth curve $\gamma$ in $\mathcal{M}$, a horizontal lift $\gamma^\uparrow: [a,b] \to P$ is a curve whose tangent vectors are in the horizontal subspaces $H_{\gamma^\uparrow(t)}P$ as:
\textit{horizontal} in that:
    \begin{align*}
        &V_p[\gamma^\uparrow] = 0\\
        & \pi(\gamma^\uparrow(t)) = \gamma(t), \forall t \in [a,b].
    \end{align*}
Given a point $p \in P$ above $\gamma(a)$, there exists a unique horizontal lift $\gamma^\uparrow$ starting at $p$. This uniqueness underscores the connection's role in determining a specific way to transport points along curves in $\mathcal{M}$, embodying the geometric intuition behind parallel transport.
\begin{definition}[Parallel vector fields]\label{def:geo:parallelvectorfields}
   For a curve $\gamma(t) \in \M$, the vector field $X$ is parallel along $\gamma(t)$ if the covariant derivative vanishes along the curve, that is if:
   \begin{align}
       \nabla_{\dot \gamma(t)}X(t) = 0 \qquad \forall t \in [a,b]
\end{align}
where $\nabla$ is the covariant derivative defined below.
\end{definition}
Note often we use the notation for $\dot \gamma(t) \in T_p\M$ the terms $\nabla_{\dot \gamma(t)}:= \nabla_{\gamma(t)}$ where $\gamma(t)$ is any curve that belongs to the equivalence class of $[\dot \gamma(t)]$ (Isham $\S 6.7$). In this work we generally use the notation $\nabla_{\dot \gamma(t)}$ to emphasise the covariant derivative is with respect to $\dot \gamma(t)$. Parallel transport of $v \in T_{\gamma(a)}\M$ along $\gamma(t)$ for $t \in [a,b]$ is then defined by the following map $\tau$ which takes $v$ to the corresponding vector field $X(b) \in T_{\gamma(b)\M}$. The vector field $X$ is the unique vector field along $\gamma$ such that $X(a)=v$ and $\nabla_{\dot \gamma(t)}X(t) = 0$ is satisfied. Parallelism is expressible in terms of horizontal lifts as follows \cite{isham_modern_1999,frankel_geometry_2011}.
%parallel transport
\begin{definition}[Parallel transport]\label{def:geo:Paralleltransport} 
For $\gamma: [a,b] \to\M$, the notion of \textit{parallel transport} along $\gamma$ from $\gamma(a)$ to $\gamma(b)$ is defined as a map $\tau: T_{\gamma(a)}\M \to T_{\gamma(b)}\M$ satisfying the following:
    \begin{align*}
        \tau: \pi^{-1}(\{ \gamma(a) \}) \to \pi^{-1}(\{ \gamma(b) \})
    \end{align*}
\end{definition}
The map represents assigning, to each $\pi^{-1}(\{ \gamma(a) \})$, a point in the horizontal lift of $\gamma(b)$. That is, assigning $\gamma^\uparrow(b) \in \pi^{-1}(\{ \gamma(b) \})$ to $\pi^{-1}(\{ \gamma(a) \})$ where $\gamma^\uparrow$ is the \textit{horizontal lift} of the curve $\gamma$ that passes $p$ at $t=a$.

Parallel transport is defined across fibres, where horizontal curves are mapped into horizontal curves by the right action ($\delta_g$) of $G$ on $P$, for all $g \in G$ that commute with $\tau$ i.e. $\tau \circ \delta_g = \delta_g \circ \tau$. It can be shown that if $p_1 = \tau(p)$ then $p_1g = \tau(p)g = \tau(pg)$ with the result that $\tau$ is a bijection among fibres. The concept of parallel transport is then extended to fibres via the holonomy group. This is important in Chapter \ref{chapter:Time optimal quantum geodesics using Cartan decompositions}, where we are interested in holonomic paths for time-optimal synthesis. The holonomy group $\text{Hol}_p(\omega)$ consists of all elements of $G$ that can be realized as the result of parallel transporting a reference fibre along closed curves (loops) starting and ending at $p \in \M$.
\begin{definition}[Holonomy group]\label{defn:geo:holonomy}
For a loop $\gamma: [0,1] \rightarrow \mathcal{M}$ with $\gamma(0) = \gamma(1) = p$ and a horizontal lift $\gamma^{\uparrow}$ of $\gamma$ starting at a point $x \in P$ above $p$, the endpoint of $\gamma^{\uparrow}$ is related to $x$ by an element of $G$. The set of all such elements, for all possible loops $\gamma$ based at $p$, forms the holonomy group $\text{Hol}_p(\omega)$.
\end{definition}
For a vector field $X$ to be parallel along $\gamma$ (for all $t \in [a,b]$), the requirement $\nabla_{\dot \gamma(t)}X(t) = 0$ says that the rate of change of the vector field and the connection's adjustment of that rate of change sums to zero. Holonomy reflects how connections on $\M$ influence the paths taken by curves on the manifold, where curves traced out by the action of the holonomy group constitute (time-optimal) subRiemannian or Riemannian geodesics (discussed further on in this work).

\subsection{Covariant differentiation}\label{sec:geo:Covariant differentiation}
We now come the important concept of covariant differentiation. As noted in the literature, the challenge of identifying the derivative of a section lies in how to compare neighbouring points in two different fibres in a way that does not depend on the local coordinate frame. For bundles equipped with a connection that allows a pullback points one fibre to those in another, this comparison can be undertaken. Consider the map $\psi: \mathcal{M} \rightarrow P_V$ where $\psi$ is a section, being a smooth map that assigns a vector in the fibre each point in $\mathcal{M}$ (here $P_V$ is the total space of the vector bundle where each fibre is a vector space isomorphic to $V$ (following the notation of \cite{isham_modern_1999})). We wish to understand how $\psi$ varies along curves in the manifold. This concept is expressed via the \textit{covariant derivative} as a generalised gradient (definition \ref{defn:geo:gradient}). Recall that $\mathfrak{X}(\mathcal{M})$ denotes the set of smooth vector fields on $\M$. 
%covarient derivative
\begin{definition}[Covariant derivative]\label{defn:geo:covariantderivative}
    Denote a principal $G$-bundle $\xi = (P, \pi, \mathcal{M})$ with $V$ a vector space linear representation of $G$. Consider the curve $\gamma : [0, \epsilon] \to \mathcal{M}$ such that $\gamma(0) = p_0 \in \mathcal{M}$, and let $\psi : \mathcal{M} \to P_V$ be a section of the associated vector bundle. Then the covariant derivative of $\psi$ in the direction $\gamma$ at $p_0$, is
\begin{align*}
\nabla_{\gamma}\psi := \lim_{t \to 0} \frac{\tau_{t}^{-1} \psi(\gamma(t)) - \psi(\gamma(0))}{t}
\label{eqn:geo:covariantderiv1}
\end{align*}
where $\tau_{t}^{-1}$ is the parallel transport operator from the vector space $\tau_{t}^{-1}(\{\gamma(t)\})$ to the vector space $\tau_{t}^{-1}(\{x_0\})$.
\end{definition}
Note we use a slight abuse of notation where $\nabla_\gamma$ refers more properly to $\nabla_{\dot\gamma}$ where $\dot\gamma(t)$ is the tangent vector to the curve $\gamma$ at the point $\gamma(t)$. Here $\tau_{t}^{-1}$ transports vectors back along $\gamma(t) \to \gamma(0)$ while preserving parallelism. We interchange with the notation $\nabla_X$ which typically refers to the covariant derivative along a vector field $X \in \mathfrak{X}(\M)$ on a manifold $\mathcal{M}$. Here, $X$ is a smooth section of the tangent bundle $T\mathcal{M}$, and $\nabla_X$ is an operator that acts on a vector field or section of a vector bundle. If two curves $\gamma_1,\gamma_2$ are tangent at $p\in\M$ then we have $\nabla_{\gamma_1}\psi = \nabla_{\gamma_2}\psi$. Moreover, connecting with the notation:
\begin{enumerate}[(a)]
    \item For $v \in T_x\M$, the covariant derivative of $\psi$ of $P_V$ along $v$ is represented as $\nabla_v \psi = \nabla_\gamma \psi$ where $\gamma$ here is any curve in the equivalence class $v=[\gamma]$.
    \item For the vector field $X$ on $\M$, the covariant derivative along $X$ is the linear operator: 
    \begin{align*}
        \nabla_X: \Gamma(P_V) \to \Gamma(P_V)
    \end{align*}
    on the set of sections $\Gamma(P_V)$ of the vector bundle $P_V$ associated with $p \in M$, that is:
    \begin{align*}
        \nabla_X\psi(p) = \nabla_{X_p}\psi.
    \end{align*}
\end{enumerate}
Consider now affine connections, noting $\mathfrak{X}(\M)$ is also denoted $\Gamma(P_V)$. We note that $\nabla_X$ on $\mathfrak{X}(\M)$ exhibits the structure of derivation:
\begin{align}
    \nabla_X(f\psi) = f\nabla_X\psi + X(f)\psi
\end{align}
for $f \in \cinfm$. The $\nabla_X$ operator is also linear in $\mathfrak{X}(\M)$ (which can also be regarded as a module over $\cinfm$). These properties are expressed by considering $\nabla_X$ as an \textit{affine connection}.
%affine connection
\begin{definition}[Affine connection]\label{defn:geo:Affine connection}
    An affine connection on $\M$ is an operator $\nabla: \mathfrak{X}(\M) \times \mathfrak{X}(\M) \to \mathfrak{X}(\M)$ which associates with $X \in \mathfrak{X}(\M)$ a linear mapping $\nabla_X$ of $\mathfrak{X}(\M)$ into itself i.e. $(X,Y) \mapsto \nabla(X,Y)=\nabla_XY$ satisfying the following two conditions:
\begin{align*}
(\nabla_1) \quad &\nabla_{fX+gY} = f\nabla_X + g\nabla_Y; \\
(\nabla_2) \quad &\nabla_X(fY) = f\nabla_X(Y) + (Xf) Y
\end{align*}
for $f, g \in C^\infty(M), X, Y \in \mathfrak{X}(\M)$.  
\end{definition}
The affine connection is one way to define the covariant derivative $\nabla_X$ on a vector bundle $(P_V,\pi_V,\M)$, which in turn can be related to the connection one-form on the principal bundle $\xi$ (see above). The affine connection together with a local coordinate chart $(U,p)$ can be defined via:
\begin{align*}
    \nabla_{\partial_i} = \Gamma^k_{ij} \partial_k
\end{align*}
where $\Gamma^k_{ij} = (\nabla_{\partial_i} \partial_j)^k$ are \textit{Christoffel symbols of the second kind} and $(\cdot)^k$ indicates the $k$-th component. For the covariant derivative of $Y$ with respect to $X$, such $k$-th coordinate is given by:
\begin{align*}
\left(\nabla_X Y\right)^k = X^i (\nabla_i Y)^k = X^i \left(\frac{\partial Y^k}{\partial x^i} + \Gamma^k_{ij} Y^j\right).
\end{align*} 

\subsection{Geodesics and parallelism}\label{sec:geo:Geodesics and parallelism}
We can use the above concepts to define a notion of parallelism leading to a definition of geodesics which are curves that locally minimise distance and are solutions to the geodesic equation derived from a chosen connection. Denote $\gamma: I \to \M, t \mapsto \gamma(t)$ for an interval $I \subset \Real$ (which we generally without loss of generality specify as $I=[0,1]$) with an associated tangent vector $\dot\gamma(t)$. Here $\gamma$ is regular. Two vector fields $X,Y$ are parallel along $\gamma(t)$ if:
\begin{align*}
    \nabla_XY = 0 \qquad \forall t \in I.
\end{align*}
This definition of parallelism involves values of $X,Y$ only on the curve $\gamma$. In local coordinates $\{x_i\}$, it can be shown that their exist coordinate functions $X^i,Y^j$ for $i,j \leq m$ on the coordinate neighbourhood $U$ as per above such that:
\begin{align*}
    X = \sum_i X^i \left(\frac{\partial}{\partial_{x_i}} \right) \qquad Y = \sum_j Y^j \left(\frac{\partial}{\partial_{x_j}} \right)
\end{align*}
\begin{align*}
\nabla_X(Y) &= \sum_k \left( \sum_i X^i \frac{\partial Y^k}{\partial x_i} + \sum_{i,j} X^i Y^j \Gamma^k_{i,j} \right) \frac{\partial}{\partial x_k} \quad \text{on } U
\end{align*}
with the result:
\begin{align}
\frac{dY^k}{dt} + \sum_{i,j} \Gamma^k_{i,j} \frac{dx_i}{dt} Y^j &= 0 \quad (t \in J) \label{eqn:geo:geodesicchristoffel1}
\end{align}
which can be shown \cite{helgason_differential_1979} to represent equation (\ref{eqn:geo:covariantderiv1}) in the limit as $t \to 0$. Equation (\ref{eqn:geo:geodesicchristoffel1})  indicates that the component form of the covariant derivative along $\gamma(t)$ is zero (hence indicative of parallel transport). The covariant derivative operator $\nabla_X$ has a number of properties, including that (i) it is a derivation (an $(r,s)$-tensor being drawn from the derivation space of tangent and cotangent vectors respectively), (ii) it preserves tensors (i.e. it maps $(r,s)$ tensors to $(r,s)$ tensors) $\nabla_X: T^{r,s}\M \to T^{r,s}\M$ and (iii) commutes with all contractions $C^i_j$ (see above). From this we come to the first iteration of a \textit{geodesic}.
%geodesic definition
\begin{definition}[Geodesic]\label{defn:geo:Geodesic}
    A curve $\gamma: I \to \M, t \mapsto \gamma(t)$in $\M$ is denoted a geodesic if the set of tangent vectors $\{\dot\gamma(t)\} = T_{\gamma(t)}\M$ is parallel with respect to $\gamma$, corresponding to the condition that $\nabla_{\gamma}\dot\gamma=0$, which we denote the geodesic equation. 
\end{definition}
In a coordinate fame the geodesic equation is expressed as:
\begin{align*}
    \frac{d^2u^\gamma }{ds^2} +   \Gamma^\gamma_{\alpha \beta} \frac{du^\alpha }{ds} \frac{du^\beta }{ds} = 0 \label{eqn:geo:geodesiccoordinateframe}
\end{align*}
where:
\begin{align*}
\Gamma^\gamma_{\alpha \beta} = \frac{1}{2} g^{\gamma \mu} \left( \frac{\partial g_{\mu \alpha}}{\partial u^\beta} + \frac{\partial g_{\mu \beta}}{\partial u^\alpha} - \frac{\partial g_{\alpha \beta}}{\partial u^\mu} \right)
\end{align*}
are the Christoffel symbols of the second kind (essentially connection coefficients) with $g^{\gamma \mu}$ the inverse of the metric tensor and $ds$ usually indicates parametrisation by arc length. It can be shown that given an affine connection above, for any $p \in \M$ and $X_p$, there exists a unique maximal geodesic $t \mapsto \gamma(t)$ such that $\gamma(0)=p$ and $\dot\gamma(0)=X_p$, i.e. it cannot be extended to a larger interval. As Nielsen et al. note \cite{nielsen_optimal_2006,dowling_geometry_2008}, the geodesic equation is solved either (a) as an initial value problem by specifying $\gamma(0)=p,\dot\gamma(0)=v$ for $v \in T_p\M$, or (b) a boundary value problem for $\gamma(0)=p$ and $\gamma(1)=q$ using variational methods. For Lie group manifolds, all geodesics are generated by generators from the horizontal subspace $H_p\M$, but not all curves generated from the horizontal subspace are geodesics. Note (for clarity) with regards to $H_pP$ and $H_p\M$ that we can denote the image of \(H_pP\) under \(\pi_*\) as \(H_p\mathcal{M}\), representing bundle's horizontal structure within the tangent space at \(\pi(p)\) in \(\mathcal{M}\), thereby elucidating how the geometry of the bundle expands over the base manifold via the connection.

\section{Riemannian manifolds}\label{sec:geo:Riemannian manifolds}
The covariant derivative / affine connection $\nabla$ on $\M$ allows us to define concepts of \textit{torsion} and \textit{curvature} tensor fields. First, we define the celebrated Riemann curvature tensor. The tensor intuitively represents 

\begin{definition}[Riemann curvature tensor]\label{defn:geo:Riemann curvature tensor}
Given an $(\M,g)$ Riemannian manifold (see below) and set of vector fields $\mathfrak{X}(\M)$, define the following $(1,3)$ tensor field:
\begin{align}
    &\mathfrak{X}(\M) \times \mathfrak{X}(\M) \times \mathfrak{X}(\M) \to \mathfrak{X}(\M) \\
    &(X,Y,Z) \mapsto R(X,Y)Z &= (\nabla_X\nabla_Y - \nabla_Y\nabla_X - \nabla_{[X,Y]})Z \\
    & &= \nabla_X\nabla_Y Z - \nabla_Y\nabla_X Z - \nabla_{[X,Y]}Z
\end{align} 
for any vector fields $X, Y, Z \in \mathfrak{X}(\mathcal{M})$, where $\nabla$ denotes the Levi-Civita connection associated with the metric $g$.
\end{definition}
The Riemann curvature tensor measures the extent to which the affine connection (covariant derivative) is not commutative i.e. the failure of the covariant derivatives to commute and therefore the extent of intrinsic curvature of the manifold. In this way, it corresponds to the non-holonomy of $\M$. Recall $\text{Hol}_p(\omega)$ for a loop $\gamma$ reflects parallel transportation of a vector $v \in T_p\M$ to and from $p \in \M$ along $\gamma$. The deviation of $v$ from its initial position can be given by the action of the Riemann curvature tensor along the loop expressed by the Ambrose-Singer Holonomy Theorem \cite{ambrose1953theorem} which states that the Lie algebra of $\text{Hol}_p(\omega)$ at $p \in \M$ is generated by the values of the curvature $\Omega$ over all horizontal two-planes in $T_pP$ where $p$ is the fibre over $x$ (see also section \ref{sec:geo:sectionalcurvature} below). The properties of $\text{Hol}_p(\omega)$ that describe how parallel transport around closed loops $\gamma$ distort vectors are related to the curvature experienced by these vectors in plans spanned by tangent vectors at $T_x\M$. Integrating $R_{\mu \nu}$ over all such two-dimensional subspaces identifies the set of possible curvatures that parallel transport can induce i.e. each element in $\text{Hol}_p(\omega)$ corresponds to a unique way of parallel transporting and the relation to curvature quantifies the amount of twisting or distortion of the vector ($R_{\mu \nu}$ being responsible for both the curvature experienced and the Lie algebra of $\text{Hol}_p(\omega)$).

Given on Riemannian manifolds the affine connection is given by the Levi-Civita connection that is metric compatible and torsion free, the Riemann curvature tensor has an expression in terms of the second covariant derivative:
\begin{align*}
    R(X,Y) = \nabla^2_{X,Y} - \nabla^2_{Y,X}
\end{align*}
indicating how the non-commutativity of the second covariant derivatives of vector fields $X$ and $Y$ represents the intrinsic curvature of $\M$. Given the relationship between second derivatives and curvature, one can see how the Riemann curvature tensor gives rise to a measure of curvature on manifolds. In coordinate notation, it is denoted as:
\begin{align*}
    R^{\alpha}_{\beta\gamma\delta} = \Gamma^{\alpha}_{\beta\delta,\gamma} - \Gamma^{\alpha}_{\beta\gamma,\delta} + \Gamma^{\mu}_{\beta\delta} \Gamma^{\alpha}_{\mu\gamma} - \Gamma^{\mu}_{\beta\gamma} \Gamma^{\alpha}_{\mu\delta}.
\end{align*}
To obtain measures of curvature, one can then perform contractions with the metric tensor $g$ to obtain the \textit{Ricci tensor} which is obtained by performing a tensor contraction over the first and third indices i.e. $R_{\mu\nu} = R^\lambda_{\mu\lambda\nu}$. Scalar curvature is then obtained via contraction with the inverse metric tensor $R = g^{\mu\nu}R_{\mu\nu}$, which is in effect a trace operation.
Given the definitions
\begin{align*}
T(X, Y) &= \nabla_X(Y) - \nabla_Y(X) - [X, Y] \qquad (\text{torsion tensor}) \\
R(X, Y) &= \nabla_X\nabla_Y - \nabla_Y\nabla_X - \nabla_{[X,Y]} \qquad (\text{curvature tensor}),
\end{align*}
With an appropriate one-form, we can then define a \textit{torsion tensor field} as the argument for the one-form mapping $\omega: T*\M \times T\M \times T\M  \to \R$ with $(\omega,X,Y) \mapsto\omega(T(X,Y))$ where $T(X,Y) \in T^1_2\M$ (i.e. a type (1,2) tensor). We can define the \textit{curvature tensor field} $\omega: T*\M \times T\M \times T\M  \times T\M \to \R$ with $(\omega,X,Y,Z) \mapsto\omega(R(X,Y) \cdot Z)$ where $R(X,Y) \cdot Z \in T^1_3\M$. It can be shown that the relevant one-forms in turn determine the structure of the covariant derivative on $\M$, while it can be shown (via Cartan's structure equations) that the forms $\omega$ are themselves described by the torsion and curvature tensor fields as per below (adapted from Theorem 8.1 in \cite{helgason_differential_1979}, $\S$8). The relation of curvature and torsion to differential forms is shown explicitly in Cartan's structural equations which we set out below.

\begin{theorem}[Cartan structural equations]\label{thm:geo:Cartanstructuralequations}
Given a principal bundle connection (definition \ref{defn:geo:Connection}), Cartan's structural equations are given by:
\begin{align}
    d\omega^i &= - \sum_{p} \omega^i_p \wedge \omega^p + \frac{1}{2} \sum_{j,k} T^i_{jk} \omega^j \wedge \omega^k \\
    d\omega^i_l &= - \sum_{p} \omega^i_p \wedge \omega^p_l + \frac{1}{2} \sum_{j,k} R^i_{ljkp} \omega^j \wedge \omega^k
\end{align}
\end{theorem}
Here for completeness, $d\omega^i$ represents the exterior derivative (how $\omega^i$ changes in all directions), $\omega^i_p$ is a one form representing connection coefficients, encoding information about non-linear behaviour, the $\wedge$ product captures the idea of an oriented area spanned by the one-forms $\omega^i$. Cartan's structural equations provide an important tool for using the local behaviour of a connection to understand global properties of $\M$, such as curvature and torsion. 

\subsection{Riemannian Manifolds and Metrics}\label{sec:geo:Riemannian Manifolds and Metrics}
We now come to identifying how to calculate lengths of curves on manifolds relevant to our substantive Chapters. Firstly, we define a (pseudo)-Riemannian structure in terms of a tensor field. In doing so we gain further insight into metrics in terms of tensor fields of type (0,2). A Riemannian manifold is the tuple $(\M,g)$ (i.e a manifold $\M$ with a metric $g$) where to each $p \in \M$ is assigned a positive definite map $g_p: T_p\M \times T_p\M \to \R$ (described usually in terms of being an inner product) and an associated norm $||X_p||: T_p\M \to \R$. We integrate these concepts in the formalism above as follows.
\begin{definition}[Riemannian structure]\label{defn:geo:(Pseudo)-Riemannian structure}
    Define a pseudo-Riemannian structure on $\M$ as a type-(0,2) tensor field $g$ such that (a) $g(X,Y) = g(Y,X)$ (symmetric) and (b) for $p \in \M$, $g$ is a nondegenerate bilinear form $g_p: T_p\M \times T_p\M \to \R$.
\end{definition}
If we assume $g_p$ is positive definite for all $p \in \M$ then we speak of Riemannian structures and manifolds as follows.
\begin{definition}[Riemannian Manifold]\label{defn:geo:Riemannian Manifold}
    A Riemannian manifold is a connected $\cinfm$ manifold with a Riemannian structure such that there exists a unique affine connections satisfying the following:
    \begin{enumerate}[(i)]
        \item $\nabla_X Y - \nabla_Y X = [X,Y]$ (namely the torsion tensor $T=0$); and 
        \item $\nabla_Z g = 0$ (parallel transport preserves the inner product on tangent spaces).
    \end{enumerate}
\end{definition}
By applying $\nabla_Z$ to the tensor field formed from $X \otimes Y \otimes g$, can specify the Riemann connection. We understand this as follows. An affine connection $\nabla$ is metric compatible with a metric $g$ when: 
\begin{align*}
X\innerproduct{Y}{Z} = \innerproduct{\nabla_X Y}{Z} +  \innerproduct{Y}{\nabla_X Z}
\end{align*}
which can be expressed using the metric $g$ as:
\begin{align*}
    Xg({Y},{Z}) = g({\nabla_X Y},{Z}) +  g({Y},{\nabla_X Z}).
\end{align*}
Metric-compatible connections preserve angles, orthogonality and lengths of vectors in $T_p\M$ when transported along $\gamma(t)$. The fundamental theory of Riemannian geometry is an existence and uniqueness theorem which asserts the existence of a torsion-free metric compatible connection, denoted the \textit{Levi-Civita connection}, which can be expressed via the Koszul formula as per below.
\begin{align}
    2g(X, \nabla_Z Y) = Z g(X,Y) + g(Z,[X,Y]) + Y g(X,Z) + g(Y,[X,Z]) - Xg(Y,Z) - g(X,[Y,Z]).
\end{align}
An example of a connection where torsion is non-zero yet remains metric compatible is the Weitzenb\"ock connection. The metric $g$ determines the local structure but not global structure. We now define the Riemannian metric.
\begin{definition}[Riemannian metric]\label{defn:geo:Riemannian metric}
    A Riemannian metric is an assignment to each $p \in \M$ of a positive-definite inner product:
    \begin{align*}
        g_p: T_p\M \times T_p\M \to \R
    \end{align*}
    with an induced norm:
    \begin{align*}
        ||\cdot ||_p&: T_p\M \to \Real\\
        (v,w)&\mapsto \sqrt{g_p(v,w)}.
    \end{align*}
\end{definition}
The Riemannian metric is a $(0,2)$-form thus a tensor, aligning definition (\ref{def:geo:metric_tensor}). Recall $(0,2)$-forms at $p \in \M$ have a representation as the tensor product of two cotangent vectors at $p$. Given a coordinate basis for $T^*_pM$ at $p$, denoted $\{dx^1, dx^2, \ldots, dx^n\}$. In the case of a $(0,2)$-form $g$, we can express it in this basis as:
% \begin{align*}
% \omega &= \sum_{i,j} \omega_{ij} \, dx^i \otimes dx^j,
% \end{align*}
\begin{align*}
g &= \sum_{i,j} g_{ij} \, dx^i \otimes dx^j,
\end{align*}
where the $g_{ij}$ are the components of the form with respect to the basis, and they can be functions on the manifold. We can then identify an induced norm on $\M$ using this metric. For $X \in T_pM$, denote:
\begin{align}
    \left(g_p(X,X)\right)^{1/2} \dot = || X ||.
\end{align}
We may now define the important concept of arc (curve) length which is fundamental to later chapters. Given the curve (segment) $t \mapsto \gamma(t)$ with $t \in [\alpha,\beta]$ and metric $g_\gamma = g$, define the \textit{arc length} of $\gamma$ as follows:
\begin{definition}[Arc length]\label{defn:geo:arclength}
    Given a curve $\gamma(t) \in \M$ with $t \in [\alpha,\beta]$ and metric $g$, the arc length of the curve from $\gamma(0)$ to $\gamma(T)$ is given by:
    \begin{align}
    \ell(\gamma) = \int_0^T \left( g(\dot\gamma(t),\dot\gamma(t)) \right)^{1/2} dt. 
    \label{eqn:geo:arclength}
\end{align}
\end{definition}
Given any normal neighbourhood $N_0$ of $0 \in \M$ and defining $N_p = \exp(N_0)$, then for any $q \in N_p$, denote the unique geodesic in $N_p$ joining $p$ to $q$ as $\gamma_{pq}$. In this case it can be shown that $\ell(\gamma_{pq}) < \ell(\gamma)$ for each curve segment joining $p$ and $q$. Under this assumption, we can impose a metric upon the Riemannian manifold $\M$ as follows. Assuming $\M$ is simply connected, then all $p,q \in \M$ can be joined via a curve segment. The \textit{distance} of $p,q$ is then defined by the infimum of the shortest curve measured according to the equation (\ref{eqn:geo:arclength}) above and is defined as:
\begin{align}
    d(p,q) = \inf_\gamma L(\gamma).
    \label{eqn:geo:riemannmetric}
\end{align}
Here $d(p,q)$ satisfies the usual axioms of metrics, namely (a) symmetry $d(p,q) = d(q,p)$, (b) triangle inequality $d(p,q) \leq d(p,r) + d(r,q)$ and (c) positive definite $d(p,q) = 0$ if and only if $p=q$.
It can then be shown that for such a Riemannian manifold $\M$ with metric $d$ given by equation (\ref{eqn:geo:riemannmetric}) together with a curve $\gamma_{pq}$ joining $p,q \in \M$, then if $\ell(\gamma_{pq}) = d(p,q)$ then $\gamma_{pq}$ is a geodesic. Then consider an open ball around $p$ with radius $r$, $B_r(p)$ and associated sphere around $p$ denoted $S_r(p)$. Assume an open ball $V_r(0) = \{X \in T_p \M | 0 \leq ||X|| \leq r \}$ is a normal neighbourhood of $0 \in T_p\M$. In this case it can be shown that $B_r(p) = \exp(V_r(0))$. For complete Riemann manifolds (where every sequence in $\M$ is convergent), each pair of points $p,q \in \M$ can be joined by a geodesic of length $d(p,q)$. 

The preservation of distances can then be understood in terms of total geodesicity of manifolds. 
A sub-manifold $S$ of a Riemannian manifold $\M$ is \textit{geodesic at} $p$ if each geodesic tangent to $S$ at $\p$ is also a curve in $S$. The submanifold $S$ is \textit{totally geodesic} if it is geodesic for all $p \in S$. It can then be shown that if $S$ is totally geodesic, then parallel translation along $\gamma \in S$ always transports tangents to tangents, that is $\tau: S_p\M \to S_p\M$. Summarising the above points: 
\begin{itemize}
    \item \textit{Riemann Connection}: 
This refers to the affine connection on a Riemannian manifold $\M$, representing differentiation of vector fields $\mathfrak{X}(\M)$ along curves $\gamma$ and for calculations related to parallel transport.
\item \textit{Riemann metric}: This provides a notion of distance on $\M$, providing a means of measuring lengths of curves and angles between vectors, thereby defining the manifold geometry. 
\item \textit{Riemann curvature tensor}: This tensor $R_{\mu\nu}$ quantifies curvature on $\M$ and derives from the   Riemann connection. Importantly, it is  independent of any coordinate embedding.
\item \textit{Levi-Civita Connection}: Given all connections on a Riemannian manifold with metric $(\M,g)$, the Levi-Civita connection is the unique connection that is both torsion-free and metric-compatible, i.e. preserving the Riemannian metric under parallel transport (under which the covariant derivative of the metric tensor vanishes).
\end{itemize}

\subsection{Fundamental forms}\label{sec:geo:Fundamental forms}
In this section we briefly set out a few key theorems and definitions from standard differential geometry relating to fundamental forms and curvature which are familiar for differential geometry on Euclidean spaces. In particular we connect standard results to the coordinate-free formalism adopted above.
\begin{enumerate}[(i)]
    \item \textit{First fundamental form}. The first fundamental form is an expression of the metric tensor $g$ on $\M$ and is denoted:
    \begin{align}
        I_p(X,Y) = g_p(X,Y) \qquad X,Y \in T_p\M.
    \end{align}
    Given a coordinate chart $\phi:\R^2 \to \M$ where $\phi(X,Y) \to p\in\M$, define as usual $\phi_X = \partial_X \phi$ and similarly for $v$ and dual $X'=dX/dt$. Then (recalling $g(X,V)=\braket{X,Y}$):
    \begin{align}
        E &= g(\phi_X,\phi_X) \qquad F = g(\phi_X,\phi_Y)\qquad G = g(\phi_Y,\phi_Y)\\
        g_{ij} &= \pmat{E}{F}{F}{G}\\
        I_p &= E (u')^2 + 2Fu'v' + G(v')^2 = g_{ij}\frac{dx^i}{dt}\frac{dx^j}{dt} = ds^2{dt}
    \end{align}
    with equivalents for higher-dimensional manifolds. The first fundamental form thus represents a quadratic form associated with the metric tensor \cite{frankel_geometry_2011}.
    \item \textit{Second fundamental form}. The second fundamental form describes how surfaces curve in ambient space i.e. by reference to a normal to surface. In general, define a level set function $f: \M \to \R$ as the set of points $p \in \M$ such that $f(p)=c \in \R$:
    \begin{align}
        S_c = \{ p \in \M | f(p) = c\}.
    \end{align}
    $S_c \subset \M$ will be a submanifold if $\Delta f \neq 0, \forall p \in S_c$. Recall that $\nabla f$ points in the direction of the steepest incease in $f$. For $\nabla f(p) \neq 0$, $f$ is non-constant except along directions tangent to $S_c$. Thus $S_c$ resembles a hypersurface in $\M$ (where $\dim S_c = \dim \M-1$), which is a generalisation of $\M$ being, for example, $\R^3$ with $S_c = \R^2$ and $\nabla_f$ being the normal vector `pointing out' of the surface. Thus we can denote a generalised unit normal vector via $N = \nabla f/||\nabla f||$. The second fundamental form is then defined as:
    \begin{align*}
        II_p(X,Y) = g(\nabla_X N,Y) = \braket{S(X),Y}
    \end{align*}
    for $X,Y \in T_p\M$. Here $S(X)=\nabla_X N$, the shape operator which encodes information about a manifold's (or submanifold's) curvature by describing how the normal vector field $N$ changes along the vector field $X$ for $X \in T_p\M$ (i.e. as one moves tangent to the surface). We then have in the coordinate frame:
        \begin{align*}
        e&=-\braket{N_u, \phi_u} \qquad        f=-\braket{N_u, \phi_v} = -\braket{N_v, \phi_u}\qquad 
        g=\braket{N_v, \phi_v}\\
       II_p&=e (u')^2 + 2fu'v' + g(v')^2
    \end{align*}
    \item \textit{Gauss and Codazzi equations}. The Gauss equation provides that the Gauss curvature is a measure of intrinsic curvature of the manifold. In a coordinate frame, given the first and second fundamental forms in a coordinate basis with components $\phi_u,\phi_v$ with principal curvatures given by $dN = \text{diag}(k_1,k_2)$. Gaussian curvature $K$ is then:
    \begin{align}
        K & = \frac{\det (h_{ij})}{\det (g_{ij})} = \frac{eg -f^2}{EG - F^2}   k_1 k_2 \\
        H& = \frac{1}{2} (k_1 - k_2) = \frac{1}{2}\frac{eE - 2fF + gG}{EG-F^2}
    \end{align}
    where $H$ is mean curvature (the average of principal curvatures $k_1,k_2$. Gaussian curvature is extrinsic to the manifold, while mean curvature depends on how it is embedded in ambient space. In coordinate free form, the Gauss equation can be expressed as:
    \begin{align}
        [\nabla_X,\nabla_Y]Z = -K \det(X,Y)JZ
    \end{align}
    where $\det(X,Y)$ is the \textit{area form} of a Riemannian manifold and $J$ is an endomorphism (mapping) $J:\M \to \M$ such that $\braket{JX,Y}=\det(X,Y)$ and is denoted an \textit{almost complex structure} $J^2=-I$ (and so intuitively acts to rotate by $\pi/2$ in a way that transitions for example $JX\to Y$). These results provide that the Gaussian curvature $K$ is invariant under local isometry, a statement of Gauss's celebrated \textit{Theorem Egregium} (see \cite{frankel_geometry_2011}). Additionally, the \textit{Codazzi equation} provides that the covariant derivative of $II_p$ is symmetric, that is:
    \begin{align}
       [\nabla_X,\nabla_Y]N =  (\nabla_X \nabla_Y - \nabla_Y \nabla_X)N = \nabla_{[X,Y]}N \label{eqn:geo:Codazzi equation}
    \end{align}
    which, when satisfied, ensures that the local shape of the surface given by normal curvature $\nabla_X N$ is consistent with the global properties of the manifold (thereby providing an integrability condition). 
\end{enumerate}

\subsection{Curvature and forms}\label{sec:geo:curvatureandforms}
\subsubsection{Overview}
Curvature is a concept of fundamental importance in geometric methods across quantum information, algebra, machine learning and control. Curvature also plays a profound role in the classification of symmetric space of interest in quantum information, where in many cases computations may be expressed in terms of the evolution of curves over unitary Lie group manifolds generated by Hamiltonians drawn from corresponding Lie algebras. In this section, we contextualise curvature in the language of differential forms, beginning with the concepts of sectional curvature and revisiting curvature tensor fields on the way to defining the celebrated Riemann curvature tensor.

\subsubsection{Sectional curvature}\label{sec:geo:sectionalcurvature}
We define sectional curvature on the way to defining the formalism of Riemannian symmetric spaces, the subject in particular of our final chapter. Sectional curvature $K(\sigma_p)$ provides a means of classifying curvature on Riemannian manifolds via a two-dimensional sub-bundle $\sigma_p \subset T_p\M$. Geometrically it corresponds to the Gaussian curvature with plane $\sigma_p$ constructed from tangents to geodesics $\gamma$ in the direction of $\sigma_p$.  Recall the definition of the section of a fibre bundle.
\begin{definition}[Section of a Fibre Bundle]\label{defn:geo:Section of a Fibre Bundle}
Let $ \pi: E \to \M $ be a fibre bundle with total space $ E $, base space $ \M $, and fibre $ F $. A section of this fibre bundle is a continuous map $ s: \M \to E $ such that $ \pi \circ s = \text{id}_M $, the identity map on $\M $. In other words, for each point $ p \in \M $, $ s(p) $ is a point in the fibre over $ p $, i.e., 
\begin{align}
\pi(s(p)) = p, \quad \forall p \in \M.
\end{align}
\end{definition}
This condition ensures that the section lifts each point in the base manifold back to a specific point in the total space, intersecting each fibre exactly once. For a \textit{vector bundle section} then the section is a map assigning to each $p \in \M$ a vector in the corresponding fibre. We now define the sectional curvature as taking a two-dimensional sub-bundle of the section in order to calculate curvature. The intuitive idea is that sectional curvature allows one to observe how a plane embedded in the manifold bends or curves as it moves through the space.
\begin{definition}[Sectional curvature]\label{def:geo:sectionalcurvature}
    For a Riemannian manifold $\M$ and point $p \in \M$ with a two-dimensional sub-bundle $\sigma_p \subset T_p\M$ with $X,Y \in \sigma_p$, define the sectional curvature $K: T_p\M \times T_p\M \to \R, (X,Y) \to K(X,Y)$ as:
    \begin{align*}
K(X, Y) &= \frac{g(R(X, Y)Y, X)}{g(X,X)g(Y,Y) - g(X,Y)^2}\\
&=\frac{\langle R(X, Y)Y, X\rangle}{\langle X, X\rangle \langle Y, Y\rangle - \langle X, Y\rangle^2}
\end{align*}
where $R$ is the Riemann curvature tensor, $g$ is the Riemannian metric tensor and $g(X,Y)=0$ (orthogonal).
\end{definition}
Sectional curvature $K$ thus measures how the Riemann curvature tensor, when acting on $X,Y$ relates to the area spanned by $X$ and $Y$. This spanning of an area is also described by a ``2-Grassmanian bundle'', which intuitively speaking is the set of  
Riemannian manifold then has constant curvature $c$ if $K(H) = c$. Indeed it can be shown (see \cite{helgason_differential_1979}, Prop. 12.3) that sectional curvature determines the Riemannian curvature such that if for two bundles $T_p\M,T_{q}\M$, $K_p = K_{q}$, then $R_p=R_q$. For a two-dimensional manifold, the sectional curvature of the plane spanned by $(X,Y)$ is the Gaussian curvature at that point. Given Gaussian curvature is restricted to two-dimensions, sectional curvature in one sense allows a generalisation of Gaussian curvature measurement to higher-dimension manifolds.

\section{Symmetric spaces}\label{sec:geo:Symmetric spaces}
\subsection{Overview}
Symmetric spaces were originally defined as Riemannian manifolds whose curvature tensor is invariant under all parallel translation. Cartan allowed the classification of all symmetric spaces in terms of classical and exceptional semi-simple Lie groups. Cartan's classification of symmetric spaces (see Chapters IV and IX of \cite{helgason_differential_1979}) is of seminal importance. Locally they are manifolds of the form $\R^n \times G/K$ where $G$ is a semi-simple lie group with an involutive automorphism whose fixed point set is the compact group $K$ while $G/K$, as a homogeneous space, is provided by a $G$-invariant structure.  

As Helgason \cite{helgason_differential_1979} notes, Cartan adopted two methods for the classification problem. The first, based on holonomy groups provides that for $p \in \M$ Riemannian, then the holonomy group (definition \ref{defn:geo:holonomy}) of $\M$ is the group of all linear transformations of the tangent space $T_p\M$ corresponding to parallel transportation of closed curves $\gamma \in \M$. Parallel translation is equated with the action of the holonomy group and leaves the Riemannian metric invariant, leading to a method of classification. The more direct second method involves demonstrating the invariance of the curvature tensor under parallel transport is equivalent geodesic symmetry being a local isometry for all $p \in \M$ through which the geodesic cure passes. Such spaces equipped with geodesic symmetry and global isometry are equipped with a transitive group of isometries $K$ such that space is represented as a coset space $G/K$ with an involutive isomorphism whose fixed point is the isometry group $K$.  The study of symmetric spaces then becomes a question of studying specific involutive automorphisms of semi-simple Lie algebras, thus connecting to the classification of semi-simple Lie groups. 
\\
\\
\textit{Geodesic symmetry} is defined in terms of diffeomorphisms $\varphi$ of $\M$ which fix $p \in \M$ and reverse geodesics through that point (i.e. when acting the map $\varphi: \gamma(t=1) = q \to \gamma(t=-1)=q'$ for $\gamma(0)=p$, such that $d\varphi_p: T_p\M \to T_p\M, X \mapsto d\varphi(X) = -X$ (i.e. $d\varphi_p \equiv -I$ where $I$ denotes the identity in $T_p\M$). Here $d\varphi$ can be thought of as the effect of $\varphi$ on the tangent space e.g. in Lie algebraic terms $\varphi(\gamma(t))=\varphi(\exp(X))=\exp(d\varphi(X))=\exp(-X)$ for $X \in \g$. This means manifold $\M$ is locally symmetric around $p \in \M$. That is, the manifold $\M$ is \textit{locally (Riemannian) symmetric} if each geodesic symmetries is isometric, i.e. if there is at least one geodesic symmetry about $p$ which is an isometry (and globally so if this applies for all $p \in \M$). As noted in the literature, this is equivalent to the vanishing of the covariant derivative of the curvature tensor along the geodesic. A globally symmetric space is one where geodesic symmetries are isometries for all $\M$.
A manifold being \textit{affine locally symmetric} is one which, given $\nabla$, the torsion tensor $T$ and curvature tensor $R$, $T=0$ and $\nabla_Z R=0$ for all $Z \in T\M$. 
%Riemannian local symmetric space
\begin{definition}[Riemannian local symmetric space]\label{defn:geo:Riemannian local symmetric space}
    A Riemannian manifold $\M$ if for all $p \in \M$ there exists a normal neighbourhood $N_p$ where the geodesic symmetry for $p$ is an isometry. In other words, a Riemannian local symmetric space is a manifold $\M$ where the Riemann curvature tensor is invariant under all parallel translations:
    \begin{align*}
        \nabla_X R = 0.
    \end{align*}
\end{definition}
It can be shown that this is equivalent to the sectional curvature being invariant under parallel translations. In the formalism of Helgason, given a Riemannian manifold $\M$ with Riemannian structure $Q$, an analytic Riemannian manifold is one where both $\M$ and $Q$ are analytic. From this we obtain the global form i.e. a \textit{Riemannian globally symmetric} space if all $p \in \M$ are fixed points of an involutive symmetry $\theta$ such that $\theta^2 = \mathbb{I}$. In this case $\theta$ is also a geodesic symmetry in that $\theta(p)$ induces $\dot\gamma_p \to -\dot\gamma_p$ i.e. a reflection symmetry along the geodesic. As Helgason, following Cartan, notes, in such cases certain facts follow from being a Riemannian globally symmetric space $G/K$ (in which case $(G,K)$ is sometimes denoted a Riemannian symmetric pair), with a Cartan decomposition as discussed above and an association of $H_p\M \sim \p$ and $V_p\M \sim \k$. 
It can be shown then that the Riemann curvature tensor for the symmetric space (homogeneous) $G/K$ with a Riemannian metric allows for curvature, via $R$:
% , to be related to the Lie triple property, namely:
\begin{align}
    R_p(X,Y)Z = -[[X,Y],Z] \qquad X,Y,Z \in \p.
\end{align}
The curvature tensor is independent of the group action of $h \in G$, that is, the curvature tensor remains invariant. In geometric terms, this is reflected by the fact that the pullback of the metric tensor by $h$ corresponds to the original metric tensor itself. This $G$-invariance property of symmetric spaces $G/K$ means that the parallel transport of vectors remains the same. 

Curvature plays an important role in the classification of symmetric spaces. Three classes of symmetric space can be classified according to their sectional curvature as follows. Given a Lie algebra $\g$ equipped with an involutive automorphism $\theta^2=\mathbb{I}$, with corresponding group $G$ with $G/K$ as above, we have a $G$-invariant structure on the Riemannian metric i.e. $g(X,Y) = g(hX,hY)$ for $h\in G/K$. Then the three types of symmetric space are:
\begin{enumerate}[(i)]
    \item $G/K$ compact, then $K(X,Y) > 0$;
    \item $G/K$ non-compact, then $K(X,Y) < 0$; and
    \item $G/K$ Euclidean, then $K(X,Y)=0$.
\end{enumerate}
The non-compactness or compactness can be calculated via the rank of $\M=G/K$, being the maximum dimension of a subspace of $T\M$ where $K=0$.

\subsection{Classification of symmetric spaces}\label{sec:geo:Classification of symmetric spaces}
The classification of Riemannian symmetric spaces by Cartan (see Knapp \cite{knapp_lie_1996} and Helgason \cite{helgason_differential_1979} for detailed discussion, especially Helgason Chapter IX) terms of simple Lie groups is set out below in Table \ref{tab:classification}. Type I spaces are non-compact and are associated with non-compact simple Lie groups, while Type II spaces are their compact counterparts. Type III spaces are distinct in that they are Riemannian duals of themselves. Type IV spaces involve complex structures and are related to complexifications of simple Lie algebras. The families of simple Lie groups are denoted by $A_n$, $B_n$, $C_n$, and $D_n$, where $A_n$ corresponds to the special linear group $\text{SL}(n+1, \mathbb{C})$, $B_n$ to the special orthogonal group $\text{SO}(2n+1)$, $C_n$ to the symplectic group $\text{Sp}(n)$, and $D_n$ to the special orthogonal group $\text{SO}(2n)$. Additionally, the exceptional Lie groups are denoted as $E_6$, $E_7$, $E_8$, $F_4$, and $G_2$, each signifying a unique algebraic structure that induces specific geometric properties on the associated symmetric spaces.

\begin{table}[h!]
\footnotesize
\caption{\label{tab:classification}Classification of Riemannian Globally Symmetric Spaces (adapted from Helgason \cite{helgason_differential_1979} as compiled in Wikipedia)}
% \begin{ruledtabular}
\begin{tabular}{llll}
Type & Lie Algebra Class & Symmetric Space & Alternative Description \\
\hline
I (II) & AI($n$) & $SU(n)/SO(n)$ & - \\
I (II) & AII($2n$) & $SU(2n)/Sp(2n)$ & $SU(2n)^*/Sp(2n)$ \\
I (II) & AIII($n,r$) & $U(n)/[U(r)\times U(n-r)]$ & $U(r,n-r)/[U(r)\times U(n-r)]$ \\
I (II) & BDI($n,r$) & $SO(n)/[SO(r)\times SO(n-r)]$ & $SO(r,n-r)/[SO(r)\times SO(n-r)]$ \\
I (II) & DIII($n$) & $SO(2n)/U(n)$ & $SO(2n)^*/U(n)$ \\
I (II) & CI($n$) & $Sp(2n)/U(n)$ & $Sp(2n,\mathbb{R})/U(n)$ \\
I (II) & CII($n,r$) & $Sp(2n)/[Sp(2r)\times Sp(2n-2r)]$ & $Sp(2r,2n-2r)/[Sp(2r)\times Sp(2n-2r)]$ \\
III (IV) & A($n$) & $SU(n) \times SU(n) / SU(n)$ & - \\
III (IV) & BD($n$) & $SO(n) \times SO(n) / SO(n)$ & $SO(n,\mathbb{C}) / SO(n)$ \\
III (IV) & C($n$) & $Sp(2n) \times Sp(2n) / Sp(2n)$ & - \\
\end{tabular}
% \end{ruledtabular}
\end{table}

\section{SubRiemannian geometry}\label{sec:geo:SubRiemannian geometry}
We now discuss subRiemannian geometry which is fundamental to Chapters \ref{chapter:Quantum Geometric Machine Learning} and \ref{chapter:Time optimal quantum geodesics using Cartan decompositions} of this work where the problem at hand is to develop techniques for time optimal synthesis in relation to subRiemannian manifolds. Most of the material is drawn from Montgomery's seminal work on subRiemannian geometry \cite{montgomery_tour_2002} together with other standard source material on topic. That is, subRiemannian geometry involves a manifold $\M$ together with a distribution upon which an inner product is defined.  Distribution in this context refers to a linear sub-bundle of the tangent bundle of $\M$ and corresponds to the horizontal subspace of $T\M$ discussed above and where the vertical subspace is non-null. SubRiemannian manifolds are characterised where the distribution is not the entire bundle, or in the language of Lie algebra, where for a decomposition $\g = \k \oplus \p$, we have for our accessible (or control) subspace $\p \subset \g$ rather than $\p = \g$. This means that the generators or vector fields $\mathfrak{X}(\M)$ are constrained to limited directions. A transformation not in the horizontal distribution may be possible (as we discuss for certain subspaces where $[\p,\p]\subseteq \k$), but such that the geodesic paths connecting the start and end points will be longer than for a Riemannian geometry on $\M$. Formally, we define a subRiemannian manifold as follows.

\begin{definition}[SubRiemannian manifold and distributions]\label{defn:geo:SubRiemannian manifold}
A subRiemannian manifold (geometry) on $\M$ consists of a distribution $\Delta$, being a vector sub-bundle $H_p\M \subset T\M$ together with a fibre inner product on $H_p\M$. The sub-bundle corresponds to the \textit{horizontal distribution}, having the meaning ascribed to horizontal subspace $H_p\M$ above.
\end{definition}
A curve on $\M$ is a \textit{horizontal curve} if it is tangent to $H_p\M$. Similar to Riemannian manifolds, \textit{subRiemannian length} $\ell = \ell(\gamma)$ (for $\gamma$ smooth and horizontal) is then defined similarly as:
\begin{align}
    \ell(\gamma) = \int || \dot \gamma(t) || dt \label{eqn:geo:arclengthSubRiemannian}
\end{align}
where $||\dot \gamma(t)|| = \sqrt{\braket{\dot\gamma(t),\dot\gamma(t)}}$ computed over the horizontal subspace $H\M$. \textit{SubRiemannian distance} $d_S$ is defined as the infimum of all such lengths, namely $d_S(A,B) = \inf \ell(\gamma)$ of all curves connecting $A,B \in \M$. To guarantee that a horizontal curve has a minimal or geodesic distance, we require that the curve $\gamma$ be absolutely continuous which is to say that its derivative $\dot\gamma(t) \in H_p\M$ for all $t$. In this case we can define a subRiemannian \textit{minimising geodesic} as that absolutely continuous horizontal path that realises the distance between two points in $\M$. Sometimes we define the \textit{energy} of a horizontal curve as:
\begin{align}
    E(\gamma) = \int_\gamma \frac{1}{2}||\dot\gamma(t)||^2 \label{eqn:geo:energyhorizontalcurve}
\end{align}
as it can be more convenient to minimise this quantity rather than $\ell(\gamma)$. A horizontal curve $\gamma$ that minimises $E$ among all curves also minimises length and can be parametrised by constant speed (analogous to parametrisation by constant arc length for minimal geodesics) via $d(q_0,q_1)/T$ for $\gamma(0)=q_0, \gamma(T)=q_1 \in \M$.

\subsection{SubRiemannian Geodesics}\label{sec:geo:SubRiemannian Geodesics}
As Montgomery notes, while a Riemannian metric is defined by a covariant $(0,2)$ form. In subRiemannian geometry, the metric is defined only on a subset of the tangent space (the horizontal distribution), unlike in Riemannian geometry where it is defined over all $T_\M$. Instead, it is usual for a \textit{subRiemannian metric} to be encoded as a contravariant symmetric two-tensor as a section of the bundle. We denote this the \textit{cometric} as follows:
\begin{definition}[SubRiemannian (co)metric]\label{defn:geo:SubRiemannian (co)metric}
    The subRiemannian metric, or cometric, is a section of the bundle $T\M \otimes T\M$ of symmetric bilinear forms on the cotangent bundle $T^*\M$. This gives rise to a fibre-bilinear form:
    \begin{align*}
        \braket{\cdot,\cdot}: T^*\M \otimes T^*\M \to \R.
    \end{align*}
\end{definition}
The form  defines a symmetric bundle that maps $\beta: T^*\M \to T
\M, \alpha(\beta_p(\mu)) = ((\alpha,\mu))_q$ where $\alpha,\mu \in T^*_p\M$ and $p \in \M$. Symmetric here refers to the fact that $\beta$ equals its adjoint $\beta^*$. The cometric then satisfies the conditions that $\text{Im}(\beta_p)=H_p\M$ and it considered a subRiemannian inner product $\alpha(v) = \braket{\beta_p(\alpha),v}_p$ for $v \in H_p\M$, $\alpha \in T^*\M$. Using the subRiemannian metric we specify a system of Hamilton-Jacobi equations on $T^*\M$ the solution to which is a subRiemannian geodesic.

This formalism can be used to define a \textit{subRiemannian Hamiltonian} given by $H(p,\alpha) = \frac{1}{2}(\alpha,\alpha)_p$ where $\alpha \in T^*\M$ and $(\cdot,\cdot)_p$ is the cometric. As Montgomery notes, $\beta$ uniquely determines the subRiemannian structure. The formalism also has its own form of Hamiltonian differential equations (see \cite{montgomery_tour_2002} Appendix A) given by $\dot x^i = \partial H\partial p_i, \dot\partial p_i = -\partial H/\partial x^i$, denoted the \textit{normal geodesic equations}. Here $x^i$ are the coordinates and $p_i$ are momenta functions for coordinate vector fields. 

In this sense, Riemannian geometry is a special case of subRiemannian geometry where the distribution $\Delta = \g$ or the entire tangent bundle and with the cometric $g^{\mu\nu}$. As with the Riemannian case, there are equivalent existence theorems regarding minimal geodesics.
\begin{theorem}[Normal subRiemannian geodesics]\label{thm:geo:Normal subRiemannian geodesics}
    Given a solution to the subRiemannian equations on $T^*\M$ given by $(\gamma(t),p(t))$ with $\gamma(t) \in \M$, then every sufficiently short curve segment $\gamma(t)$ is a minimising subRiemannian geodesic and $\gamma$ is the unique minimising geodesic between endpoints $\gamma(0)=q,\gamma(T)=p$. 
\end{theorem}
Here $\gamma(t)$ is a projected curve, i.e. a curve constituted via projection onto the subRiemannian manifold. As noted in the literature, there are subRiemannian geometries for which minimal geodesics exist that are not solutions to subRiemannian Hamiltonian equations, so-called `singular' geodesics. The existence of subRiemannian geodesics and importantly the ability for curves generated by distributions $\Delta$ to effectively cover the entirety of $\M$ (thus, in a quantum information context, making any $U(t)$ in principle reachable via only the subset $\p$), relies upon bracket-generating properties of $\Delta$ and Chow's theorem.
%Bracket generating
\begin{definition}[Bracket generating (distribution)]\label{defn:geo:bracketgenerating}
    A distribution $\Delta$ (as a collection of vector fields $\{X_i\}$) is called bracket generating if the application of the Lie derivative among all $X_i$ spans the entire tangent bundle. Equivalently, a distribution (in our case defined by a control subset of the Lie algebra) $\Delta = \h \subset \g$ for a Lie algebra $\g$ is bracket generating if repeated application of the Lie derivative (commutator) (definition \ref{defn:alg:Lie algebraliederivative}) obtains the entirely of $\g$.
\end{definition}
By a theorem of Chow and Raschevskii (see \cite{montgomery_tour_2002}), it can be shown that if $\M$ is connected with a bracket generating distribution and complete (with respect to the subRiemannian metric), then there exists a minimising geodesic connecting any two points of $\M$. For a decomposition $\g = \k \oplus \p$, then we are interested in bracket generating control subsets $\p$ such that $[\p,\p]\subseteq\k$ such that we have access to the entirety of $\g$ (recall $\p$ is our horizontal subspace (distribution) $H_p \M$).

A subRiemannian structure on $\M$ is, as noted above, given by the sub-bundle distribution $\Delta \subseteq H_p\M$. Noting the canonical projection $\pi_\Delta: \Delta \to \M$ being the canonical projection of the bundle with fibres at $p \in \M$ given by $\Delta_p = \pi^\inv_\Delta \subseteq H_p\M$. Assume $\dim \Delta_p = m$, independent from the dimension of $p \in \M$. Such a setup corresponds to, in traditional control contexts, the existence of $m$ smooth vector fields $\mathfrak{X}_p(\M) = \{ X_1,...,X_m \}$ such that all points $p \in \M$ share the same distribution i.e. $\Delta_p$. We assume that $\mathfrak{X}_p$ is bracket generating such that it is the smallest $\p$ which can generate the entire $\g$. With the assumption that $\Delta_p$ is the same for all $p\in\M$, then we have that for all $p \in \M$, $\g = T_p\M$. The Riemannian metric restricted to $\Delta_p \in T_p\M$ provides a smooth, positive definite inner product for generators in $\Delta_p$ (also, for convenience, we assume the usual orthonormality of vectors and covectors).

\section{Geometric control theory}\label{sec:geo:Geometric control theory}
\subsection{Overview}
The primary problem we are concerned with in our final two chapters is solving time optimal control problems for certain classes of Riemannian symmetric space. In our case, time optimal control is equivalent to finding the time-minimising subRiemannian geodesics on a manifold $\M$ corresponding to the homogeneous symmetric space $G/K$. Our particular focus is the $KP$ problem, where $G$ admits a Cartan $KAK$ decomposition where $\g = \k \oplus \p$, with the control subset (Hamiltonian) comprised of generators in $\p$. In particular such spaces exhibit the Lie triple property $[[\p,\p],\p] \subseteq \k$ given  $[\p,\p]\subseteq \k$. Here, $\g$ remains in principle reachable, but where minimal time paths constitute subRiemannian geodesics. Such methods rely upon \textit{symmetry reduction} \cite{zeier_symmetry_2011,echeverria-enriquez_geometric_2003}. As D'Alessandro notes \cite{dalessandro_introduction_2007}, the primary problem in quantum control involving Lie groups and their Lie algebras is whether the set of reachable states $\mathcal{R}$ (defined below) for a system is the connected Lie group $G$ generated by $\g = \spn\{-H(u(t))\}$ for $H \in \g$ (or some subalgebra $\h \subset \g$) and $u \in U$ (our control set, see below). This is manifest then in the requirement that $\mathcal{R} = \exp(\g)$. In control theory $\g$ is designated the \textit{dynamical Lie algebra} and is generated by the Lie bracket (derivative) operation among generators in $H$. The dynamical adage is a reference to the time-varying control functions $u(t)$.
\subsection{Geometric control preliminaries}\label{sec:geo:Geometric control preliminaries}
Recall from section \ref{sec:quant:Quantum Control} that we define a control system (definition \ref{sec:quant:controlsystem}) as time-dependent vector fields defined over a differentiable manifold $\M$ parametrised by control functions $u(t)$ which give rise to a set of solutions to the differential state equation (\ref{eqn:quant:controlsystem1}) in the form of admissible control trajectories $\gamma(t)$. Optimal control aims to minimise a functional (an objective) over such control trajectories subject to constraints. In control theory \cite{schattler_geometric_2012} ($\S 2$ and $\S 4$) we define a \textit{control system} as follows.

\begin{definition}[Control system] \label{defn:geo:controlsystem}
    A control system is a 4-tuple $\Sigma = (\M,U,f,\mathcal{U})$ where $\M$ is the state space represented as a differentiable manifold, $U \subset \R^m$ the set of controls for $u_k(t) \in U$, $f$ are the dynamics and $\mathcal{U}$ is the set of admissible controls.
\end{definition}
In the general case, $\M$ is a $C^r$ manifold (see section \ref{sec:geo:Manifolds and charts}) and $\mathcal{U}$ are Lebesgue integrable bounded functions in $U$ (Lesbegue measurability is required as restricting to piecewise continuous controls does not guarantee optimality). The dynamics where $\dot \gamma(t) = f(t,\gamma,u)$ (i.e. equation (\ref{eqn:quant:controlsystem1}) are defined such that $f: \R \times \M \times U \to \M$. The values of $t \in I$ over which equation (\ref{eqn:quant:controlsystem1}) has a unique solution is denoted the \textit{admissible control trajectory} which we discuss below in the following subsections.

Returning to horizontal curves above, we assume that such curves $\gamma: \Real \supset [0,T] \to \M$ satisfy certain usual Lipschitz continuity requirements (almost everywhere differentiable as per above). We assume that $||\dot\gamma(t)||$ is essentially bounded as follows.
\begin{definition}[Essentially bounded]\label{defn:geo:Essentially bounded}
A tangent $\dot\gamma(t)$ is essentially bounded where there exists a constant $N$ and mapping $H: [0,T] \to T\M, t \mapsto H(t) = \dot \gamma$ such $\braket{H(t),H(t)}_S \leq N$ for all $t \in [0,T]$.     
\end{definition}
This condition ensures that $H(t) = \dot\gamma(t)$ almost everywhere along the curve (which we assume to be regular $\dot\gamma \neq 0$). As discussed above, horizontal curves are then those whose tangents (generators) are in the horizontal distribution for such a curve, that is where $\dot\gamma(t) \in \Delta_{\gamma(t)}$. Recalling our vector fields $X_j$ are differential operators on $\gamma(t)$, we can then write the curve in terms of control functions and generators.
%curve control functions and generators
\begin{definition}[Horizontal control curves]\label{defn:geo:Horizontal control curves}
    Given $\gamma(t) \in \M$ with $\dot\gamma(t) \in \Delta_{\gamma} \subseteq H_p\M$, we can define horizontal control curves as:
    \begin{align}
        \dot\gamma(t) = \sum_j^m u_j(t) X_j(\gamma(t))
\label{eqn:geo:horizontalcurves}
    \end{align}
    where $u_j$ are the control functions given by:
    \begin{align}
        u_j(t) = \braket{X_j(\gamma(t)),\dot\gamma(t)}. \label{eqn:geo:controlshorizontal}
    \end{align}
\end{definition}
The boundedness of $\dot\gamma(t)$ sets an overall bound on $u_j(t)$. Recall that the length of a horizontal curve is given by:
\begin{align}
    \ell(\gamma) = \int_0^T ||\dot \gamma(t)|| dt =  \int_0^T \sqrt{\braket{\dot\gamma(t),\dot\gamma(t)}} dt = \int_0^T \sqrt{\sum_j^m u^2_j(t)} dt. \label{eqn:geo:arclengthparamby}
\end{align}
When $[0,T]$ is normalised this is equivalent to parametrisation by arclength. As noted in \cite{albertini_symmetries_2018}, the consequence of the Chow-Raschevskii theorem, $\M$ being connected and $\mathfrak{X}(\M)$ being bracket generating renders $(M,d)$ a subRiemannian metric space where $d_S$ is the subRiemannian distance function.

\subsection{Time optimal control}\label{sec:geo:Time optimal control}
For subRiemannian and Riemannian manifolds, the problem addressed in our final two chapters in particular is, for a given target unitary $U_T \in G$, how to identify the minimal time for and Hamiltonian to generate a minimal geodesic curve such that $\gamma(0)=U_0$ and $\gamma(T)=U_T$ for $\gamma:[0,T] \to \M$. This is described \cite{albertini_symmetries_2018} as the \textit{minimum time optimal control problem} \cite{jurdjevic_geometric_1997}. In principle, the problem is based upon the two equivalent statements (a) $\gamma$ is a minimal subRiemannian (normal) geodesic between $U_0$ and $U_T$ parametrised by constant speed (arc length); and (b) $\gamma$ is a minimum time trajectory subject to $||\vec u|| \leq L$ almost everywhere (where $u$ stands in for the set of controls via control vector $\vec u$). SubRiemannian minimising geodesics starting from $q_0$ to $q_1$ (our $U_T$) subject to bounded speed $L$ describe \textit{optimal synthesis} on $\M$. There are two loci of geodesics related to their optimality.
\begin{enumerate}[(i)]
    \item \textit{Critical locus} ($CR(\M)$) being the set of points where minimising geodesics are not optimal. This is defined such that for $p \in CR(\M)$, the horizontal curve $\gamma$ is a minimising geodesic for all points $t \in [0,T]$ but not for infinitesimally extended interval $[0,T+\epsilon]$, the set of such points being \textit{critical points}; and
    \item \textit{Cut locus} $(CL(\M))$ being the set of points $p \in \M$ reached by more than one minimising geodesic $(\gamma_i,\gamma_j)$ over $t \in [0,T]$, where at least one is optimal (such points $p \in \M$ then being \textit{cut points}).
\end{enumerate}
When minimising $\gamma$ are analytic functions of $t \in [0,\infty)$ i.e. $CL(\M) \subseteq CR(\M)$, then cut points are critical points.  In terms of optimal control, the critical locus indicates points at which a geodesic ceases to be time optimal beyond a marginal extension of the parameter interval as per above. Conversely, the cut locus represents points where multiple minimal geodesics intersect, delineating the farthest extents of unique geodesic paths within the reachable set, thereby affecting optimality by introducing alternative minimal paths. An important concept with a geometric framing and one drawn from control theory is that of a reachable set.
%reachable set definition
\begin{definition}[Reachable set]\label{defn:geo:Reachable set}
    The set of all points $p \in \M$ such that, for $\gamma(t): [0,T] \to \M$ there exists a bounded function $\vec u, ||\vec u||\leq L$ where $\gamma(0)=q_0$ and $\gamma(T) = p$ is called the reachable set and is denoted $\mathcal{R}(T)$.
\end{definition}
Note for reachable sets under usual assumptions we have $T_1 \leq T_2$ implies $\mathcal{R}(T_1) \subseteq \mathcal{R}(T_2)$. For optimal trajectories, $p = \gamma(T)$ belong on the boundary of $\mathcal{R}(T)$. In general, targets are in the space of equivalence class of orbits of $[p]$. For our particular case, we assume that $G=\M$. For convenience and notational efficiency we denote by $g \in G$ the relevant group diffeomorphisms of $\M$. We also make reasonable assumptions about the existence of a minimal orbit type and that for points on the same orbit $q_2 = g q_1$, then one can transition between minimising geodesics parametrised by constant speed $L$ via the group action $\gamma_1 = g \gamma_2$, which effectively pushes $\gamma$ around the orbit while satisfying time optimality.

\subsection{Variational methods} \label{sec:geo:variational methods}
Of fundamental importance to optimal control solutions (in geometric and other cases) is the application of variational calculus in terms of Euler-Lagrange, Hamiltonian and control optimisation formalism represented by the Pontryagin Maximum Principle below. In the working below, we follow Jurdjevic's geometric framing of the Pontryagin maximum principle (PMP) (see \cite{jurdjevic_geometric_1997}, Chapter 11). The maximum principle represents an extension of techniques developed by Weierstrass and others, with a particular focus on optimality conditions and the distinction between `weak' (Euler-Lagrangian) and `strong' (PMP) minima. The PMP is a cornerstone of optimal control theory. It sets out the necessary and sufficient conditions for optimal control of curves $\gamma(t)$ for dynamical systems defined on a differentiable manifold $\mathcal{M} \sim \mathbb{R}^n$. The principle assumes the existence of functions $(f_1(\gamma(t), u(t)), \ldots, f_m(\gamma(t), u(t)))$ on $\mathcal{M}$ with $u(t)$ control variable $u \in U \subset \mathbb{R}^m$. Here $U$ is the control set which is parameterized alongside $\gamma$ as $u = u(t), \gamma = \gamma(t)$ for $t \in I = [t_0, t_1]$ (under usual assumptions of boundedness and measurability of $t$). The evolution of $\gamma(t)$ in $\mathcal{M}$ is driven by the interaction of the state and control according to the differential state equation:
\begin{align}
\dot{\gamma}_i(t) = f_i(\gamma(t), u(t))
\end{align}
for almost all $t \in I$. Solution curves $\gamma(t)$ are typically unique under these conditions. The PMP approach employs a variational equation to analyze how small variations in the initial conditions affect the system's evolution. This principle is crucial for determining optimal trajectories that satisfy both the dynamical constraints imposed by the system and the objective of minimizing a specific cost function.

\subsubsection{Pontryagin Maximum Principle}\label{sec:geo:Pontryagin Maximum Principle}
Consider as our differentiable manifold $\M \sim \R^n$ on which we define $\cinfm$ functions $(f_1(\gamma,u),...,f_m(\gamma,u))$ where our coordinate charts $\phi \in S \subset \R^n$ parametrise curves $\gamma(t) \in \M$. We have as our control variable $u \in U \subset \R^m$ where $U$ is our control set. Both are parametrised as $u=u(t),\gamma=\gamma(t)$ for $t \in I = [t_0,t_1]$ (assuming boundedness and measurability). The evolution of $\gamma \in \M$ is determined by the state and the control, according to the differential equation. We have the \textit{state equation}:
\begin{align}
    \dot\gamma_i(t) = f_i(\gamma(t),u(t)) \label{eqn:geo:pontrydotgamma}
\end{align}
for almost all $t \in I$. Solution curves $\gamma(t)$ can in typical circumstances be regarded as unique.  The classical PMP approach then posits a \textit{variational equation} such that for the integral curve $\gamma(t)$, which expresses how small variations in the initial conditions of a dynamical system affect its evolution. For a solution curve $\gamma(t)$, the matrix:
\begin{align}
    A_{ij}(t) = \frac{\partial f_i}{\partial \gamma_j}(\gamma(t),u(t))
\end{align}
expresses how the vector field (whose components are $f_i)$ changes around $\gamma(t)$. The \textit{variational system} along the curve $\gamma(t), u(t)$ is then:
\begin{align}
    \frac{dv_i}{dt}=\sum_j^n A_{ij}(t)v_j(t)
\end{align}
a set of differential equations describing the evolution of a small perturbation (i.e. a variation) of $\gamma(t)$ by $v(t)$. The solution to $v(t)$ describes how sensitive the system is to such perturbations over time and thus how stable it is. Related to the variational system is the \textit{adjoint system} a set of differential equations for optimality with respect to $p(t)$:
\begin{align}
    \frac{dp_i}{dt} = -\sum_j^n p_j A_{ji}(t) \label{eqn:geo:pontryadjointsystem}
\end{align}
where $p_i$ are \textit{costate variables} (see below, essentially akin to Lagrange multiplier terms). Solutions to equation (\ref{eqn:geo:pontryadjointsystem}) allow construction of the time optimal Hamiltonian. Solutions $\gamma(t),v(t)$ satisfy:
\begin{align}
\sum_{i=1}^{n} p_i(t) v_i(t) = c,
\end{align}
for some constant $c$, such that the solutions have a constant inner product over time, constituting level sets in their applicable phase space. The systems can be expressed via the Hamiltonian as a function of $\gamma, p$ and $u$ as follows:
\begin{align}
    H(\gamma, p, u) &= p_0 f_0(\gamma, u) + \sum_{i=1}^{n} p_i f_i(\gamma, u) \label{eqn:geo:PMPHamiltonian}
\end{align}
with:
\begin{align}
\frac{d\gamma_i}{dt} &= \frac{\partial H}{\partial p_i}(\gamma(t), p(t), u(t)) = f_i(\gamma(t),u(t)) \\
\frac{dp_i}{dt} &= -\frac{\partial H}{\partial \gamma_i}(\gamma(t), p(t), u(t)) = \sum_j^n p_j \frac{\partial f_i}{\partial x_i}(\gamma(t),u(t)). \label{eqn:geo:pontryHamiltonianequations}
\end{align}
Note that $f_0(\gamma(t),u(t))$ is the function we seek to minimise (hence $p_0 \leq 0$ below) while costates $p_i$ act as in effect Lagrange multiplier terms such that $f_i$ represent the rate of cost for exerting control $u(t)$ given state $\gamma(t)$. Optimisation is then expressed in terms of a \text {cost functional} which quantifies the evolution along the curve from points $p(t_0)$ to $p(t_1)$ in $S \in \M$:
\begin{align*}
    C(u) = \int_{t_0}^{t_1} f_0(\gamma(t), u(t)) dt \label{eqn:geo:costfunctional}
\end{align*}
where $(\gamma(t),u(t))$ is denoted a \textit{trajectory} and $C(u)$ the cost function. A trajectory is \textit{optimal} relative to $\gamma(t_0)=a,\gamma(t_1)=b$ if it is the minimal cost among all trajectory costs from $a$ to $b$. The optimal control problem is then constituted by finding an optimal trajectory. The PMP then specifies necessary conditions that any optimal trajectory must satisfy. From this, we can obtain the concept of the \textit{maximal Hamiltonian} $H_M$ associated with an integral curve $(\gamma(t),p(t),u(t))$:
\begin{align}
    H_M(\gamma(t),p(t)) = \sup_{u\in U}H(\gamma(t),p(t),u(t)).
\end{align}
The PMP transforms the problem of minimising $C(u)$ into one of Hamiltonian maximisation. It does so by the introduction of the costate variables $p(t)$ which are analogous to Lagrange multipliers, enabling inclusion of system dynamics in the optimisation. Thus the Hamiltonian is defined to include the dynamics and cost function with co-state variables. In essence it assumes an optimal minimum length given by the cost $p_0f_0(\gamma,u)$ term which is negative and then adds to the Hamiltonian the $\sum_i p_i f_i(\gamma,u)$ representing the additional energy or cost from the controls. The PMP is formally given below. 
\begin{definition}[Pontryagin maximum principle]\label{defn:geo:Pontryagin maximum principle}
    The maximum principle for solving the optimal control problem provides that for a trajectory $(\gamma(t),u(t))$ evolving from $a \to b$ over interval $t \in I=[0,T]$, there exists a non-zero absolutely continuous curve $p(t)$ on $I$ satisfying:
    \begin{enumerate}[(i)]
        \item $(\gamma(t),p(t),u(t))$ is a solution curve to the Hamiltonian equations (\ref{eqn:geo:pontryHamiltonianequations}).
        \item $H(x(t), p(t), u(t)) = H_M(x(t), p(t))$ for almost all $t \in [0, T]$; and
    \item $p_0(T) \leq 0$ and $H_M(x(T), p(T)) = 0$.
    \end{enumerate}
\end{definition}

%PMP and quantum control
\subsubsection{PMP and quantum control}\label{sec:geo:PMP and quantum control}
Following \cite{dalessandro_introduction_2007,jurdjevic_geometric_1997}, typical quantum control problems are fundamentally  take one of three forms: (a) the problem of Mayer (where $C(u) = C(\gamma(T),T)$, (b) Lagrange where $C(u) = \int_0^T L(\gamma(t), u(t), t), dt$ where $C_T$ denotes some cost functional for time $T$, combined into a more general Bolza form:
\begin{align}
    C(u) = C_T(\gamma(T),T) + \int_0^T L(\gamma(t), u(t), t) dt \label{eqn:geo:PMPBolzaloss}
\end{align}
subject to the constraint imposed by the Schr\"odinger equation (\ref{eqn:quant:schrodingersequation}). Both the Hamiltonian and unitary target $U$ are complex-valued, however they can be represented in real values via the mapping $\C^n \to \R^{n^2}$. For the optimal solution, one assumes the existence of a set of optimal controls $u_j(t)$ which may be unknown and replaces them with a control approximating $u$, given by $u^\epsilon$. The cost function is then:
\begin{align}
    C(u^\epsilon) - C(u) \geq 0.
\end{align}
Following this, there are then two types of variation condition, known as \textit{strong variation} and \textit{weak variation} describing how varied the choice of $u^\epsilon$ is from the true $u$, which can be used to optimise. The PMP problem above is based upon a \textit{strong variation} then involves assumptions about how much $u^\epsilon$ varies over intervals. One assumes that for $\tau \in (0,T]$:
\begin{align}u^\epsilon =
      \begin{cases}
      u & \text{if }t \in [0,\tau - \epsilon] \quad \text{and} \quad  (\tau-\epsilon,\tau]\\
      v & t \in (\tau,T]    
      \end{cases}\label{eqn:geo:PMPtimes}
\end{align}
where $v$ is any value in the admissible control set $U \in \R^m$. In this context, the PMP is then expressed as that assuming $u(t)$ as the optimal control and $\gamma(t)$ as a solution equation (\ref{eqn:geo:pontrydotgamma}) in the form of Schr\"odinger's equation exist, there exists a non-zero (vector) solution to the adjoint equations and terminal equation at $T$:
\begin{align}
    \dot p^T = -p^T f_i(\gamma(t),u(t)) \qquad p^T(T)=-\phi_\gamma(\gamma(T))
\end{align}
where for all $\tau \in (0,T]$:
\begin{align}
    p^T(\tau)f(\gamma(\tau),u(\tau) \geq p^T(\tau)f(\gamma(\tau),u(\tau).
\end{align}
Practically speaking, to apply the PMP in quantum contexts, the following procedure is used. First we define the Hamiltonian (equation (\ref{eqn:geo:pontryHamiltonianequations})). For each state $\gamma$ and costate $p$, we maximise the Hamiltonian over $u \in U$ such that:
\begin{align}
    H(p,\gamma,u) \geq H(p,\gamma,v) \label{eqn:geo:PMPHgeqH}
\end{align}
recalling $v$ is determined as per equation (\ref{eqn:geo:PMPtimes}). The solutions to this maximisation problem enable us to write $u = u(\gamma,p)$ i.e. as a function of the state and costate. We can convert the general loss (equation (\ref{eqn:geo:PMPBolzaloss})) into a `Mayer form' via expressing $\phi$ as a function of $\gamma$ and the loss Lagrangian $L$. In this case, the differential equations (\ref{eqn:geo:pontryHamiltonianequations}) then become:
\begin{align}
p_i^\T = -p_i^T f_i(\gamma_i,u(\gamma_i,p_i) - \mu L \qquad \dot \mu = 0
\end{align}
with boundary conditions that $\gamma(0)=\gamma_0$ (and that $p^T(T)=-\phi_\gamma(\gamma(T))$. Controls that satisfy these conditions are denoted \textit{extremal}. The Hamiltonian then is written as follows as a function of the controls:
\begin{align}
    H(p,\gamma,u) &= p^T f(\gamma,u) - L(\gamma,u)\\
    &=p^T H(u)\gamma - L(\gamma,u)\\
    &=H(u)p + L(\gamma,u) \qquad \text{(by skew-symmetry)}
\end{align}
with adjoint equations:
\begin{align}
    \dot p^T = -p^T H(u) _ L^T \qquad p(T) = -\phi^T(\gamma(T))
\end{align}
for a typical qubit control problem where $H \in \g = \su(2)$ for final state target $U_f \in G= SU(2)$, with Pauli generators $\sigma_j$ where $H=\sum_j u_j(t) \sigma_j$. The cost function is then of the form:
\begin{align}
    J(u_j) = Re(U^\dagger(T)U_f) + \eta \int_0^T \sum_j u_j(t)\sigma_j dt
\end{align}
where $\eta$ is a parameter controlling the contribution of the second term to the cost functional. It can be shown (see \cite{dalessandro_introduction_2007} $\S 6$) that the optimal controls are then of the form:
\begin{align}
    u_j(t) = A \cos(\omega t + B)
\end{align}
where $A,\omega$ and $B$ are free parameters which can be determined through a minimisation procedure. As noted in the literature, the solutions to the PMP equations are extremal curves such that the trajectories $\gamma(t)$ can be said to reside on the boundary of the reachable set $\mathcal{R}$. Solving the PMP can be challenging. One class of tractable problems is when the targets $U_T \in G$ belong to a Riemannian symmetric space where the symmetry properties of the related Cartan decomposition can, in certain cases, allow the control solutions $u_j$ to be found in closed form. This is the main focus of Chapter \ref{chapter:Time optimal quantum geodesics using Cartan decompositions}. In Chapter \ref{chapter:Quantum Geometric Machine Learning}, we also discuss and examine the work of Nielsen et al. and others on geometric quantum control \cite{gu_quantum_2008,dowling_geometry_2008,nielsen_geometric_2006} where minimisation of paths on corresponding Lie group SU($2^n$) manifolds is related to obtaining minimum bounds on circuit complexity (a research programme built upon in later work by Brown and Susskind \cite{brown_complexity_2019,lin_complexity_2020}).

%=========CONTROL AND LIE GROUPS
\subsection{Time optimal problems with Lie groups}\label{sec:geo:Time optimal problems with Lie groups}
In this section, we examine the general variational problem above framed in terms of the geometric and algebraic formalism of Lie groups. One of the motivations for the geometric turn in much of control theory (where applicable) is that geometric tools and techniques have been able to provide an efficient framework through which to satisfy the optimisation requirements of the PMP. Recall the two overarching principles for optimal control are (a) establishing the \textit{existence} of controllable trajectories (paths) and thus reachable states; and (b) to show that the chosen path (or equivalence class of paths) is \textit{unique} by showing it meets a minimisation (thus optimisation) criteria. For certain classes of control problem, one can leverage geometric and symmetry properties to satisfy these requirements. 

When the target manifold $\M$ is a Lie group $G$ with a Lie algebra $\g$, then for any element in $G$ to be reachable is equivalent to $G \sim \exp(\g)$ by the properties of the exponential homomorphism (section \ref{sec:geo:Exponentials, integral curves and tangent spaces}). This will be the case when Hamiltonians $H$ are constructed from a distribution $\Delta \subseteq \g$ (with controls applied to generators in $\Delta$) that is a bracket-generating set (definition \ref{defn:geo:bracketgenerating}) such that the application of the Baker-Campbell-Hausdorff formula (definition \ref{defn:alg:Baker-Campbell-Hausdorff}) generates arbitrary targets in $G$. 
Thus satisfies the existence requirement for PMP solutions. The uniqueness requirement is then satisfied via results from geometry that guarantee (a) the metric structure that allows calculation of arc-length or energy and (b) that provides for there to be a solution (or class of solutions) which are unique by reason of minimising time or energy as calculated by way of the said metric, that is, unique geodesics. See standard presentations in \cite{sachkov_control_2009,jurdjevic_geometric_1997} for more technical exposition. Thus geometric control involving Lie groups can, in certain cases, offer a way to considerably simplify complex PMP optimisation problems. We elucidate specific applications of the PMP in the quantum case below.

\subsubsection{Dynamical Lie algebras}\label{sec:geo:Dynamical Lie algebras}
Certain approaches to geometric control use the language of \textit{dynamical Lie algebras} defined by the criterion that:
\begin{align}
    \mathcal{R} = \exp(\h) \label{eqn:geo:liealgebracontrollable}
\end{align}
for some Lie algebra $\h$ where $\h = \spn\{ -iH(u(t))\}$. `Dynamical' here refers to the fact that $\h$ is composed of generators multiplied by control functions $u(t)$ which are time-dependent and thus dynamic. Such a condition is equivalent to complete controllability, namely that every $U_T \in G$ is obtainable via a particular sequence of controls $u(t)$. The condition relies upon the bracket-generating characteristics of $\h$ (see definition \ref{defn:geo:bracketgenerating}) to recover the full Lie algebra $\g$. 

In some literature \cite{dalessandro_introduction_2007}, the number of applications of the commutator in order to obtain $\g$ is denoted the \textit{depth} of the bracket. Moreover, the condition in equation (\ref{eqn:geo:liealgebracontrollable}) relies upon the fact that $\exp(\h) \subset G$ and that $G$ is compact. This is an important point: in our final Chapter (and in subRiemannian geometric control problems generally) admitting a Cartan decomposition $\g = \k \oplus \p$, the isometry group $\exp(\k)$ is not compact with respect to $G$ (as $[\k,\k] \subseteq \k$). One cannot reach arbitrary states (limits) in $\p$. However, symmetric spaces admitting such a decomposition exhibit the Lie triple property with respect to $\p$, namely that $[\p,[\p,\p]] \subseteq \p $ where $[\p,\p] \subseteq \k$, which is a necessary requirement which can be expressed as: 
\begin{align}
    [[\p,\p],\p] \subseteq \p \label{eqn:geo:lietripleproperty}
\end{align}

Controllability is sometimes segmented into three types: (a) \textit{pure state controllable}, where for $U(0),U(T) \in G$, there exist controls $\{ u_j(t) \}$ rendering both reachable; (b) \textit{equivalent state controllable}, where $\{ u_j(t) \}$ exist to reach $U_1(T)$ up to some phase factor $\phi$ such that $U(T) = \exp(i\phi)U_1(T)$; and (c) \textit{density-matrix controllable} such that there exist controls $\{ u_j(t) \}$ enabling $\rho_i \to \rho_k$ for all $\rho_i.\rho_k \in \mathcal{B}(\Hilb)$ (i.e. any state is reachable from any other). 
\\
\\
\subsubsection{Symplectic manifolds and Hamiltonian flow}\label{sec:geo:Symplectic manifolds and Hamiltonian flow}
We can express the above in terms of symplectic manifolds and Hamiltonians which elucidates the deep connection between Hamiltonian dynamics and the geometric formalism above. Given a control system defined on $\M$, the Hamiltonian can be represented as a map $H:T^*\M \to \R$ which generates a flow (see definition \ref{sec:geo:Local flows}) on the cotangent bundle which represents the phase space of the quantum system (as we discuss below). As discussed below (definition \ref{defn:geo:Hamiltonian vector field}), one forms define a Hamiltonian vector field associated with the Hamiltonian. Hamiltonian flow is the local flow generated by a Hamiltonian vector field $X_H$ in a symplectic manifold. That is, the flow $\phi_t^{X_H}$ (to connect with local flow (section \ref{sec:geo:Local flows})) constitutes evolution in phase space along integral curves consistent with symplectic structure. The flow is given by Hamilton's equations and in a quantum context by Schr\"odinger's equations. 

For more context, as Hall \cite{hall_quantum_2013} notes, intuitively a symplectic manifold $\M$ is one with sufficient structure to define the Poisson bracket of two functions $f_1,f_2 \in \cinfm$ on it.
%symplectic manifolds
\begin{definition}[Symplectic manifold]
A symplectic manifold is a smooth differentiable manifold equipped with a non-degenerate $(0,2)$-form $\omega$ on $\M$.  
\end{definition}
Symplectic manifolds are even dimensional as odd-dimensional bundles cannot be equipped with non-degenerate skew bilinear forms. Thus we assume $\M$ is of dimension $2n$ for convenience. It can be shown that the cotangent bundle $T^*\M$ of any manifold has the structure of a symplectic manifold. Given a coordinate system $x_i$ then, as per our discussion above, we can represent an element in $T_p^*\M$ as $\phi = X^j dx_j$ with $p_j$ the coordinate functions in $\R$. Recall that $\pi^\inv: M \to T^*M$. 

Define a one-form $\theta(X) = \phi(\pi^*(X))$ sometimes denoted the \textit{tautological one-form} (so called because it returns the value of $\theta$ acting on tangent vectors at $p$). Define a (0,2)-form via the exterior product $\omega = d\theta = dp_j \wedge dx_j$. As discussed above, the non-degeneracy of $\omega$ allows the identification of $T_p\M$ with $T^*_p\M$. The two-form $\omega$ also has a corresponding dual defined on $T^*\M$ (taking duals as its arguments) given by $\omega^\inv: T^*\M \times T^*\M \to \R$. Then, given $f,g \in \cinfm$, we can define the Poisson bracket in differential formalism as:
\begin{align}
    \{f,g\} = -\omega^\inv(df,dg) \label{eqn:geo:symplecticomegapoisson}
\end{align}
which anti-commutes $\{g,f\} + \{f,g\}=0$ with $\{f,gh\} = \{f,g\}h+ g\{f,h\}$. The term $\omega^\inv$ is taken from Hall \cite{hall_quantum_2013} an denotes the bilinear form on $T^*\M$ that arises by way of the canonical identification of $T^*\M$ with $T\M$ if $\omega$ is non-degenerate. With these identifications and in particular the isomorphism described above, we can define the \textit{Hamiltonian vector field} as follows.
\begin{definition}[Hamiltonian vector field]\label{defn:geo:Hamiltonian vector field}
    For $f \in \cinfm$, we denote $X_f \in T\M$ the Hamiltonian vector field associated to $f$ if:
    \begin{align}
        df = w(X_f,\cdot) \label{eqn:geo:Hamiltonian vector field}
    \end{align}
    where $X_f$ corresponds to $df$ under such isomorphism given by $\omega$.
\end{definition}
In this case, we can relate such Hamiltonian vector field to the Poisson bracket via:
\begin{align}
    X_f(g) = \{f,g\} = -X_g(f) \qquad \omega(X_f,X_g) = -\{f,g\}
\end{align}
for $f,g \in \cinfm$. Thus conceptually we  connect vector fields to Poisson brackets in Hamiltonian formalism. We then define the \textit{Hamiltonian flow} generated by $f$, as a local flow $\varphi^f$ being the local flow generated by $-X_f$, such flow preserving $\omega$, expressed via $L_{X_f}\omega = 0$. 

We then have that $f,g,h \in \cinfm$ form a Lie algebra under the Poisson bracket satisfying the Jacobi equation, which in turn allows us to transition from the Poisson bracket to the commutator via:
\begin{align}
    [X_f,X_g] = X_{\{f,g\}} \label{eqn:geo:commutatorpoisson}
\end{align}
For the flow $\varphi$ generated by $-X$ on $\M$, if $\varphi$ preserves $\omega$ then we can represent $X = X_f$. In this case, we can assert that the flow is locally (or, if applicable to all $\M$, globally) Hamiltonian. We then have the concept of a \textit{Hamiltonian generator} where $\varphi = \varphi^f$. As Hall notes \cite{hall_quantum_2013}, any smooth function on a symplectic manifold $\M$ generates a Hamiltonian flow. We can then define the familiar Hamiltonian $H$ defined on $T*\M$. We then obtain the definition of a Hamiltonian and Hamiltonian system in such geometric terms.
%Hamiltonian systems and Hamiltonians
\begin{definition}[Hamiltonian systems and Hamiltonians]\label{eqn:geo:Hamiltonian systemsandHamiltonians}
    A Hamiltonian system on a symplectic manifold $\M=T^*\M$ is a tuple $(\M, \Phi^H)$ where $\Phi^H$ represents the Hamiltonian flow generated by the Hamiltonian $H$ on $\M$. Conserved quantities for the Hamiltonian system are represented by functions $f$ such that $f(\Phi^H(v))$ is independent of $t$ for each $v \in \M$ 
\begin{align}
\frac{d}{dt} f(\Phi^H_t(z)) = \{ f, H \}(\Phi^H_t(z)),
\end{align}
for all $z \in N$, or, more concisely,
\begin{align}
\frac{df}{dt} = \{ f, H \}.
\end{align}
The functions $f$ are conserved quantities if and only if $\{ f, H \} = 0$.
\end{definition}
The Hamiltonian $H$ dictates the flow or vector field on this manifold by determining the direction and rate of change of state variables in $\mathcal{M}$ (i.e. curves $\gamma(t)$ represented as sequences of unitaries $U(t)$).  

We briefly connect to the formulation of Hamiltonians, as the generator of time translations, in terms of controls $u_j(t)$ and generators in $\g$. The Hamiltonian determines a Hamiltonian vector field on the symplectic manifold $T^*\M$ (the cotangent bundle of $\M$). The $(0,2)$-symplectic form $\omega$ encodes information about the evolution of $\gamma(t) \in \M$ as shown in equation (\ref{eqn:geo:symplecticomegapoisson}). Given a Hamiltonian function $H(q,p)$ on a symplectic manifold $\M$ equipped with $\omega$, the Hamiltonian vector field $X_f$ is defined as per equation (\ref{eqn:geo:Hamiltonian vector field}) above, recalling $dH$ is the differential of $H$. In local phase space coordinates $(p,q)$ we have $\omega = \sum_j^n dp_j \wedge dq^j$ where $n$ denotes the degrees of freedom of the system. The vector field is then defined using Hamilton's equations:
\begin{align}
    X_f = \sum_j^n \left(\frac{\partial H}{\partial p_j}\frac{\partial}{\partial q^j} -  \frac{\partial H}{\partial q^j}\frac{\partial}{\partial p_j}  \right)
\end{align}
In canonical position and momenta coordinates $(q^j, p_j)$:
\begin{align}
    \dot q^j &= \frac{\partial H}{\partial p_j} \qquad 
    \dot p_j &= -\frac{\partial H}{\partial q^j}.
\end{align}
Hence we see how the Hamiltonian \( H \) determines the rates of change (i.e., the velocities) of the coordinates \( q^j \) and \( p_j \), establishing the dynamics of the system on the symplectic manifold. The Hamiltonian vector fields and the corresponding flows not only preserve this symplectic structure but also facilitate the analysis of dynamical properties, such as the conservation laws dictated by the Poisson brackets, which are essential in the study of classical Hamiltonian systems. While quantisation methods allow transition from classical to quantum formalism, there are a number of subtleties between geometric phase spaces in classical and quantum case (see \cite{de_silva_barbosa_2019} which also discusses the significance of the Bell–Kochen–Specker theorem for translating between classical and quantum phase space formalism).

\subsubsection{Geodesics and Hamiltonian Flow}
Finally in this section, we mention connections with the symplectic Hamiltonian flow formalism above and optimal curves on $\M$. We can relate a geodesic $\gamma(t)$ on a Riemannian (or subRiemannian) manifold $\M$ to the Hamiltonian dynamics above by considering the cotangent bundle $T^*\M$ and a form of (kinetic) Hamiltonian given by $H(x,p) = \frac{1}{2} g^{ij}(x) p_i p_j$, where $g^{ij}$ is the inverse the metric tensor (see section \ref{sec:geo:Metric tensor}) i.e. we have a Hamiltonian quadratic in momenta. In this case, it can be shown that Hamilton's equations are equivalent to the geodesic equations, providing a phase space formulation of geodesic flows. The Hamiltonian system is characterised by geodesic flow on $T\mathcal{M}$, and the geodesics represent integral curve projections of this Hamiltonian flow onto $\mathcal{M}$. The Hamiltonian vector fields above preserve the symplectic form $\omega$ on $T^*\mathcal{M}$, implying that the energy $H$ is conserved along these flows, thereby defining a Hamiltonian flow where $dH = 0$ along the trajectories. Thus we see how symplectic representations of state evolution can provide conceptual maps between Hamiltonians in a quantum information context and time-optimal geodesic curves $\gamma(t)$ from a geometric perspective (see Terry Tao's blog \cite{tao_terence_euler-arnold_2010}, noting also that often one reframes the problem of geodesic flow in terms of energy rather than geodesic paths due to degeneracy among such paths i.e. where paths belong to a cut locus above such that there is more than one optimal geodesic).

\subsection{KP Problems}\label{sec:geo:KP Problems}
A particular type of subRiemannian optimal control problem with which we are concerned in Chapter \ref{chapter:Time optimal quantum geodesics using Cartan decompositions} is the $KP$ problem. In control theory settings, the problem was articulated in particular via Jurdjevic's extensive work on geometric control \cite{jurdjevic_abstract_1970,jurdjevic_control_1972,jurdjevic_control_1981,jurdjevic_controllability_1978}, drawing on the work of \cite{brockett_sub-riemannian_nodate} as particularly set out in \cite{jurdjevic_geometric_1997} and \cite{jurdjevic_hamiltonian_2001}. Later work building on Jurdjevic's contribution includes that of Boscain \cite{boscain_introduction_2021,boscain_k_2002,boscain_invariant_2008}, D'Alessandro \cite{dalessandro_introduction_2007,dalessandro_k-p_2019} and others. $KP$ problems are also the focus of a range of classical and control problems focused on the application of Cartan decompositions of target manifolds $G=K \oplus P$, such as in nuclear magnetic resonance \cite{khaneja_time_2001} and others (see our final chapter). The essence of the $KP$ problem is where $\M$ corresponds to a semi-simple Lie group $G$ together with right-invariant vector fields $\mathfrak{X}(\M)$ equivalent to the corresponding Lie algebra $\g$. In this formulation, the Lie group and Lie algebra can be decomposed according to a Cartan decomposition (definition \ref{defn:alg:cartandecomposition}) which is, recalling equation:
\begin{align*}
    \g = \k \oplus \p
\end{align*}
together with the Cartan commutation relations (Lie triple):
\begin{align*}
    [\k,\k] \subseteq \k \qquad [\k,\p] \subseteq \p \qquad [\p,\p] \subseteq \k
\end{align*}
and equipped with a Killing form (definition \ref{defn:alg:Killing form}) which defines an implicit positive definite bilinear form $(X,Y)$ which in turn allows us to define a Riemannian metric restricted to $G/K$ in terms of the Killing form. Thus for our purposes, we understand the $KP$ problem as a minimum time problem for a subRiemannian symmetric space $G/K$. 

\subsection{SubRiemannian control and symmetric spaces}\label{sec:geo:SubRiemannian control and symmetric spaces}
Assume our target groups are connected matrix Lie groups (definition \ref{defn:alg:connected_matrix_lie_group}). Recall equation (\ref{eqn:geo:horizontalcurves}) can be expressed as:
\begin{align}
    \dot\gamma(t) = \sum_j X_j(\gamma) u_j(t)
\end{align}
where $X_j \in \Delta = \p$, our control subset. For the $KP$ problem, we can situate $\gamma(0) = \mathbb{I} \in \M$ (at the identity) $||\hat u|| \leq L$, in turn specifying a reachable set $\mathcal{R}(T)$.  As D'Alessandro et al. \cite{albertini_symmetries_2018,dalessandro_time-optimal_2020} note, reachable sets for $KP$ problems admit reachable sets for a larger class of problems. Connecting with the language of control, we can frame equation (\ref{eqn:geo:horizontalcurves}) in terms of \textit{drift} and \textit{control} parts with:
\begin{align}
    \dot\gamma(t) = A \gamma(t) + \sum_j X_j(\gamma_j) u_j(t) \label{eqn:geo:horizontalcontroSchrod}
\end{align}
where $A\gamma(t)$ represents a drift term for $A \in \k$. Our unitary target in $G$ can be expressed as:
\begin{align}
    \dot \gamma(t) = \sum_j \exp(-At)X_j\exp(A t) \gamma(t) u_j
    \label{eqn:geo:dotU=sumjKAKdal}
\end{align}
for bounded $||A_p|| = L$. For the $KP$ problem, the PMP equations are integrable. One of Jurdjevic's many contributions was to show that in such $KP$ problem contexts, optimal control for $\vec u$ is related to the fact that there exists $A_k \in \k$ and $A_p \in \p$ such that:
\begin{align}
    \sum_j^m X_j u_j(t) = \exp(At)X_j\exp(-A t).
\end{align}
Following Jurdjevic's solution  \cite{jurdjevic_hamiltonian_2001} (see also \cite{albertini_symmetries_2018}), optimal pathways are given by:
\begin{align}
    \dot\gamma(t) &= \exp(A_kt)A_p\exp(-A_k t) \gamma(t) \qquad \gamma(0)=1 \\
    \gamma(t) &= \exp(-A_k t) \exp((A_k + A_p)t)
\end{align}
resulting in analytic curves. As we explore in our final Chapter and as noted in the literature, one can select $A_p \in \a \subset \p$ where $\a$ is the non-compact part of a maximally abelian Cartan subalgebra in $\p$. In this regard, we see conjugation of a Cartan subalgebra element by elements of $K$, reminiscent of the $KAK$ decomposition itself. It is also worth noting that $\a$ being a maximal subalgebra means that equation (\ref{eqn:geo:dotU=sumjKAKdal}) is invariant under the action of $K \in \k$, reflected in the commutation relation $[\k,\p] \subseteq \p$. Albertini et al. \cite{albertini_symmetries_2018} note that for $\gamma(t) \notin CL(\M)$ with $H$ the isotropy group of $\gamma$ then the tuple $(A_k,A_p)$ minimising the geodesic is invariant under the action of $h \in H$. This gives rise to a general method for time optimal $KP$ control problems set out in \cite{albertini_symmetries_2018}: (i) identify the symmetry group of the problem, (ii) specify $G/K$, (iii) find boundaries of $\mathcal{R}$ (which may require numerical solution), (iv) find the first value of $t$ such that $\pi(\gamma(t)) \in \pi(\mathcal{R})$ and (v) identifying the orbit space within which the optimal control exists and then moving within an orbit to the final target $U_T = \gamma(T)$. We draw the reader's attention to work of Jurdjevic \cite{jurdjevic_optimal_1999,jurdjevic_geometric_1997} for discussion and in particular D'Alessandro \cite{dalessandro_introduction_2007} ($\S6.4.2$) for detailed exposition of this Cartan-based solution to the $KP$ problem. In our final chapter, we revisit this method demonstrating how our novel use of a global Cartan decomposition and variational methods give rise to time optimal synthesis results consistent with this literature.

%=========BACKGROUND: QUANTUM MACHINE LEARNING CHAPTER

\chapter{Appendix (Quantum Machine Learning)}
\label{chapter:Background: Classical, Quantum and Geometric Machine Learning}

\chapter*{ABC}

\section{Introduction}
In this Appendix, we survey literature from classical and quantum machine learning relevant to later chapters. We include a high-level review of key concepts from statistical learning theory for both context and utility. We specifically focus on deep learning architecture adopted in Chapters \ref{chapter:QDataSet and Quantum Greybox Learning} and \ref{chapter:Quantum Geometric Machine Learning}, building up towards an exegesis on neural networks by showing how they are a non-linear extension of generalised linear models. We focus on specific aspects of neural network architecture and optimisation procedures, specifically stochastic gradient descent and its implementation via backpropagation equations. We then provide a short overview the burgeoning field of quantum machine learning. We track how quantum analogues of classical machine and classical statistical learning have developed, noting key similarities and differences. In particular, in order to contextualise the learning protocols adopted in Chapters \ref{chapter:QDataSet and Quantum Greybox Learning} and \ref{chapter:Quantum Geometric Machine Learning}, we provide a comparative analysis of learning in a quantum context, with a specific comparative analysis between quantum and classical backpropagation techniques. We also focus on the relationship between learning (in both classical and quantum contexts \cite{scott_aaronson_learnability_2007}) and measurement, emphasising the hybrid nature of QML learning protocols (in the main) arising from dependence upon quantum-classical measurement channels. We examine how techniques from geometry, algebra and representation theory have been specifically (and relatively recently, in some cases) integrated into both classical and quantum machine learning strategies, such as in the form of equivariant and dynamical Lie-algebraic methods \cite{villar_scalars_2021,skolik_equivariant_2022}. In doing so, we provide a survey overview of geometric machine learning, together with the use of representation theory and differential geometry in relevant aspects of classical machine learning, briefly summarising recent results in invariant and geometric QML \cite{cerezo_building_2019,cerezo_cost_2021,cerezo_variational_2021,cerezo_challenges_2022}. Finally, we map out key architectural characteristics of hybrid quantum-classical methods denoted as greybox machine learning used in our work above.

% \section{Classical machine learning}

% \section{Deep learning}
\section{The Nature of Learning}\label{sec:ml:The Nature of Learning}

\subsection{Taxonomies of learning}\label{sec:ml:Taxonomies of learning}
In this section, we set out a brief synopsis of key principles of machine learning, including foundational elements of supervised and unsupervised models, architectural principles regarding data, models and loss functions. We cover principles behind training, optimisation, regularisation and generalisation, together with the primary classes of algorithm used in this work. Our treatment focuses on both classical and quantum forms of machine learning.

A common taxonomy for approaching learning problems involves classification in terms of concepts of data, modelling, risk/loss and algorithm type (supervised or unsupervised learning) \cite{schuld_machine_2021}. The concept of \textit{learning} is then manifest in the ability of a model to accurately predict or reproduce known information (such as labels from training data) and to generalise to out-of-sample data well. Other adjuncts for characterising learning of algorithms include algorithmic expressibility or complexity measures \cite{aaronson_quantum_2013}. A \textit{model} is then represented as an algorithm which achieves these objectives, as measured by figures of merit, such as empirical risk, training and generalisation error, algorithmic complexity and others. We discuss these in more detail with our summary of statistical learning theory concepts below. In many cases, the models in question, both classical and quantum, are extensions of well-understood statistical modelling techniques, such as generalised linear models \cite{mccullagh_generalized_1989,hastie_elements_2013}, graphical models, kernel methods \cite{cristianini_introduction_2000} or deep-learning algorithms \cite{goodfellow_deep_2016}. Quantum machine learning adapts concepts from modern machine learning (and statistical learning) theory to develop learning protocols for quantum data, or using quantum algorithms themselves including generative models \cite{wolf_transformers_2020,vaswani_attention_2017} have been adapted in quantum settings. The landscape of QML is already vast thus for the purposes of our work it will assist to situate our results within the useful schema set out in \cite{schuld_machine_2021} below in Table (\ref{table:quantumclassical}). Our results in Chapters \ref{chapter:QDataSet and Quantum Greybox Learning} and \ref{chapter:Quantum Geometric Machine Learning} fall somewhere between the second (classical machine learning using quantum data) and fourth (quantum machine learning for quantum data) categories, whereby we leverage classical machine learning and an adapted bespoke architecture equivalent to a parametrised quantum circuit \cite{benedetti_parameterized_2019} (discussed below). 

%===QUANTUM CLASSICAL TABLE
\begin{center}
\begin{table}
\begin{tabular}{ |p{3cm}||p{3cm}|p{3cm}|p{3cm}|  }
 \hline
 \multicolumn{4}{|c|}{QML Taxonomy} \\
 \hline
 \textbf{QML Division}  & \textbf{Inputs} & \textbf{Outputs} & \textbf{Process} \\
 \hline
 Classical ML   & Classical    & Classical &   Classical \\
 \hline
 Applied classical ML &   Quantum (Classical)  & Classical (Quantum)   & Classical\\
 \hline
 Quantum algorithms for classical problems &Classical & Classical&  Quantum\\
 \hline
 Quantum algorithms for quantum problems    &Quantum & Quantum&  Quantum\\
 \hline
\end{tabular}
\caption{Quantum and classical machine learning table. Quantum machine learning covers four quadrants which differ depending on whether the inputs, outputs or process is classical or quantum.}
\label{table:quantumclassical}
\end{table}
\end{center}

\section{Statistical Learning Theory}\label{sec:ml:Statistical Learning Theory}
\subsection{Classical statistical learning theory} \label{sec:ml:Classical statistical learning theory}
The formal theory of learning in computational science is classical statistical learning theory \cite{vapnik_nature_1995,james_introduction_2013,vapnik_overview_1999,hastie_elements_2013} which sets out theoretical conditions for function estimation from data. In the most general sense, one is interested in estimating $Y$ from as some function of $X$, or $Y = f(X) + \epsilon$ (where $\epsilon$ indicates random error independent of $X$). Statistical learning problems are typically framed as \textit{function estimation models} \cite{vapnik_nature_1995} involving procedures for finding optimal functions $f: \mathcal{X} \to \mathcal{Y}, x \mapsto f(x) = y$ (where $\X,\Y$ are typically vector spaces) where it is usually assumed that $X,Y \sim \Prb_{XY}$ i.e. random variables drawn from a joint distribution with realisations $x,y$ respectively. Formally, this model of learning involves (a) a generator of the random variable $X \sim \Prb_X$, (a) a supervisor function that assigns $Y_i$ for each $X_i$ such that $(X,Y) \sim \Prb_{XY}$ and (c) an algorithm for learning a set of functions (the hypothesis space) $f(X,\Theta)$ where $\theta \in \Theta$ parametrises such functions.

Typically $X$ are denoted the features and $Y$ are denoted the labels. Feature and label data is usually classified into \textit{discrete} or \textit{continuous} data. The two primary types of learning problems are supervised and unsupervised learning (defined below) where the aim is to learn a \textit{model family} $\{f\}$ or distribution $\Prb$. 
\begin{definition}[Supervised learning] \label{defn:ml:Supervised learning}
    Given a sample $D_n = \{ X_i, Y_i\}$ comprising tuples $(X_i, Y_i) \in \mathcal{X} \times \mathcal{Y}$ (where for simplicity it is assumed $D_n$ is i.i.d) where $X,Y \sim \Prb_{XY}$, a supervised learning task consists of learning the mapping $f: \X \to \Y$ for both in-sample and out-of-sample $(X_i,Y_i)$.
\end{definition}
\textit{Unsupervised learning} is defined as for $D = \{X_i\}, X \sim \Pr_X$ unsupervised learning does not utilize corresponding output labels $Y_i$ for each input $X_i$. Instead, the objective is to discover underlying patterns, structures, or distributions within the dataset $D_n$, such as clusters or classifications. \textit{Semi-supervised} learning is where some but not all $X_i$ are labelled. \textit{Self-supervised learning} is where a model $f$ is trained on data in $\X$ itself (e.g. where traditionally unlabelled data is used in effect as a label, such as where $\X$ is a sequence of data and each element in the sequence is predicted using the previous items in the sequence). The primary subject of this work is supervised learning of unitary operators as elements of Lie groups of relevance to quantum information processing, with a focus on $U(t) \in G$ for $G=SU(2)$ or $SU(3)$. 

Finding optimal functions intuitively means minimising the error $\epsilon = |f(x)-y|$ for a given function. This is described generically as \textit{risk minimisation} where, as we discuss below, there are various measures of risk (or uncertainty) which typically are expectations (averages) of loss. The objective of supervised learning then becomes one of \textit{generalisation}, which is intuitively the minimisation of risk across both in-sample and out-of-sample data. More formally, given a loss function $L:Y \times y \to R, (f(x),y) \mapsto L(f(x),y)$ and a joint probability distribution over $X \times Y$ denoted by $(X,Y) \sim \Prb_{XY}$ and associated density $p_{XY}$ (with usual assumptions regarding applicable measure, Lebesgue for continuous, counting for discrete), then an expectation operator can be defined as:
\begin{align}
    E[f(X,Y)] = \int f(x,y)d \Prb_{XY}(x,y) = \int f(x,y) p_{XY}(x,y). \label{eqn:ml:expectation operator}
\end{align}
The \textit{statistical risk} or \textit{true risk} of $f$ is then given by:
\begin{align}
    R(f) = E[L(f(X),Y) | f]. \label{eqn:ml:statlearning true risk}
\end{align}
Statistical risk is an estimator, representing the average (expected) loss of $f$ with respect to $L$. The minimum risk (\textit{Bayes risk}) is the infimum of $R(f)$ over all such measureable functions (learning rules) $f$ denoted $R^* = \underset{f \in \mathcal{F}^*}{\inf} R(f)$ where $\mathcal{F}^*$ represents the set of measureable functions $f$ such that $\mathcal{F} \subset \mathcal{F}^*$. Of course the challenge is that $\Prb_{XY}$ is usually not known such that the estimation of $f$ relies upon random samples of training data comprising $n$ examples of input $x$ and label $y$ data, denoted $D_n = \{ X_i, Y_i\}^n$ (where for simplicity it is assumed $D_n$ is i.i.d and $x,y$ represent realisations of random variables $X,Y$). The learning task then becomes framed as finding an estimate $\hat{f}$ of $f$ (conditional upon $D_n$) among a set of functions (prediction rules) $\mathcal{F}$. This functional estimator conditioned on the sample is denoted $\hat{f}_n(x) = f(x; D_n)$. The learning objective then aims to minimise $R(\hat{f}_n)$ with high probability over the distribution of data, i.e. we want to minimise the \textit{expected risk} given by $E[R(\hat{f}_n)]$. Intuitively this means the average error (risk) of the prediction rules is minimised over the distribution (of samples) $D_n$ i.e. $\hat{f}_n$ performs well on average across $D_n$. Thus the objective becomes one of optimising the \textit{selection rule} of $f \in \mathcal{F}$ given training data $D_n$ (i.e for the specific algorithm selecting $f$ given $D_n$) rather than to minimise a specific $f$.

\subsubsection{Empirical risk}\label{sec:ml:Empirical risk}
The concept of \textit{empirical risk} enables an estimation of statistical risk on out of sample (unseen) data using $D_n$ \cite{vapnik_overview_1999}, a sample of $n$ i.i.d. samples of training data i.e. $D_n$ allows us to estimate.
\begin{definition}[Empirical risk]\label{defn:ml:empiricalrisk}
    Given a sample $D_n = \{X_i,Y_i\}^n$, loss function $L$ and a function (hypothesis) $f \in F$, the empirical risk is denoted:
    \begin{align}
    \hat{R}_n(f) = \frac{1}{n}\sum_i^n L(f(X_i),Y_i)  
    \label{eqn:ml:empiricalrisk}
\end{align}
and represents an estimate of the average loss over the training set comprising $n$ samples of $X_i,Y_i$.
\end{definition}
The objective then becomes learning an algorithm (rule) that minimises empirical risk, thereby obtaining an optimal estimator across sampled and out-of-sample data, namely:
\begin{align}
    \hat{f}_n = \arg \underset{f \in \mathcal{F}}{\min} \hat{R}_n(f)
    \label{eqn:ml:fhatopt}
\end{align}
i.e. the $f$ that minimises $\hat{R}$. As the number of samples $n \to \infty$, then by assumption (of i.i.d. $D_n$), $\hat{R}(f) \to R(f)$. Equation (\ref{eqn:ml:empiricalrisk}) is typically reflected in (batch) gradient descent methods that seek to solve for the optimisation task (\ref{eqn:ml:fhatopt}). Typical gradient descent rules adopt an \textit{update} rule, which can usefully (for the purposes of comparison with quantum) be conceived of as a state transition rule. An important (and ubiquitous) class of optimisation algorithm is gradient descent (and backpropagation) which requires constraining estimators $\hat{f}$ that enable $\hat{R}_n(f)$ to be \textit{smooth} (continuous, differentiable), a requirement of $f \in C^k$, the class of all $k$-differentiable functions (ideally $C^\infty)$. Moreover, it is typical that $f$ is chosen to be a \textit{parameterised} function $f = f(\theta)$ such that the requisite analytic structure (parametric smoothness) is provided for by the parametrisation, typically where parameters $\theta \in \mathbb{R}^m$. The parameters $\theta$ are often denoted the \textit{weights} of a neural network, for example where $\hat{R}_n(f(\theta))$ is often just denoted as a function of the parameters (given the data is fixed). \\
\\
Under such assumptions, it can be shown (assuming the analyticity of $L$) that:
\begin{align}
     \hat{R}_n(f) = \frac{1}{n}\sum_i^n L(F(X_i(\theta)),Y_i) \label{eqn:ml:statisticalrisklossfunction}
\end{align}
is smooth in $\theta$, which implies the existence of a gradient $\nabla_\theta \hat{R}_n(f(\theta))$ which is a key requirement of backpropagation. The update rule is then a transition rule for $\theta$ that maps at each iteration (epoch) $\theta_{i+1} = \theta_i - \gamma(n) \nabla_\theta \hat{R}_n(f(\theta))$. Here $\gamma(n) \geq 0$ is the \textit{step-size} whose value may depend upon the iteration (epoch) $n$ (with $\sum^\infty \gamma(n) = \infty$ and $ \sum^\infty \gamma^2(n) < \infty$, the square-integrable condition with respect to the Lebesgue measure \cite{vapnik_overview_1999}). 

\subsubsection{Common Loss Functions}\label{sec:ml:Common Loss Functions}
A variety of common loss functions are used as empirical risk estimators. Two popular choices across statistics and also machine learning (both classical and quantum) are (a) mean-squared error (MSE) and (b) root-mean squared error (RMSE). Given data $D_n(X,Y)$ with $(X_i, Y_i) \sim \Prb_{XY}$ and a function estimator $f_\theta$ as per above, MSE is defined as follows.
\begin{definition}[Mean Squared Error] \label{defn:ml:Mean Squared Error}
    The MSE for a function $f$ parameterized by $\theta$ over a dataset $D_n$ is:
    \begin{align}
        \text{MSE}(f_\theta) = \frac{1}{n}\sum_{i=1}^n \left(\hat f_\theta(X_i(\theta)) - Y_i\right)^2 \label{eqn:ml:MSE}
    \end{align}
\end{definition}
i.e. $L_2$ loss. MSE calculates the average of the squares of the differences between predicted and label values. RMSE is defined as $\sqrt{\text{MSE}(f_\theta)}$. Other common loss functions include (i) cross-entropy loss e.g. Kullback-Leibler Divergence (see \cite{hastie_elements_2013} $\S14)$ for classification tasks and comparing distributions (see section \ref{sec:quant:quantummetrics} for quantum analogues), (ii) mean absolute error loss and (iii) hinge loss. The choice of loss functions has statistical implications regarding model performance and complexity, including bias-variance trade-offs.

\subsubsection{Model complexity and tradeoffs} \label{sec:ml:Model complexity and tradeoffs}
As we discuss below, there is a trade-off between the size of $\mathcal{F}$ and empirical risk performance in and out of sample. Such risk $\hat{R}_n(f)$ can always be minimised by specifying a large class of prediction rules. The most extreme example being $f(x) = Y_i$ when $x=X_i$ and zero otherwise (in effect, $\mathcal{F}$ containing a trivial mapping of $X_i$ to $Y_i$ for all $X_i$). In this case, $\hat{R}_n(f) \to 0$, but performs poorly on out of sample data. Such an example also provides more context on what it means to learn, namely that learning is properly characterised as selecting $\mathcal{F}$ that minimises $\hat{R}_n(f)$ across in-sample and out-of-sample data. The size and complexity of prediction rules $\mathcal{F}$ illustrates a tradeoff between approximation and estimation (known as `bias-variance' tradeoff for squared loss functions). This tradeoff is formally represented by \textit{excess risk}, the difference between expected empirical risk and Bayes risk: 
\begin{align}
    E[R_n(\hat{f}_n)] - R^* = \underbrace{E[\hat{R}_n(\hat{f}_n)] - \underset{f \in \mathcal{F}}{\inf} R(f)}_{\text{estimation error}} + \underbrace{\underset{f \in \mathcal{F}}{\inf} R(f) - R^*}_{\text{approximation error}} \label{eqn:ml:excessrisk}
\end{align}
% \hl{\textbf{DO WE NEED OPTIMISATION ERROR?}}
Recalling $E[\hat{R}_n(\hat{f}_n)]$ is contingent upon $\hat{f}$ estimated from $D_n(X,Y)$ (the data samples) and that the optimal $f$ (one that minimises statistical risk) is represented by $\underset{f \in \mathcal{F}}{\inf} R(f)$, estimation error captures how well $\hat{f}$, which is learnt from data, performs against all possible choices in $\mathcal{F}$. By contrast, as approximation error indicates the best choice of $f$ to minimise statistical risk, then the approximation error indicates the deterioration in performance arising from restricting $\mathcal{F}$ to subsets of all possible $f \in \mathcal{F}^*$ (i.e. all $f$ measurable). The tradeoff is characterised by the fact that if $\mathcal{F}$ becomes large (more prediction rules), estimation error tends to zero (i.e. there is little that must be learned from the data), but the approximation error increases. Recalling that by increasing the size of $\mathcal{F}$, the size of $\hat{R}_n(\hat{f})$ can be minimised (intuitively, we have a greater selection of $f$ to choose from according to which $\hat{R}_n(\hat{f})$ may be made small). But the chosen $f$ that renders $\hat{R}_n(\hat{f})$ minimal, namely $\hat{f}_n$, will overfit the data because $\hat{R}_n(\hat{f})$  is not a non-optimal estimate of the true risk $\R_n(\hat{f}_n)$.

\subsection{Reducing empirical risk}\label{sec:ml:Reducing empirical risk}
Two empirical risk-based minimisation methods for handling overfitting are (a) to limit the size of $\mathcal{F}$ in order to control estimation error. However, given the estimation/approximation error tradeoff, placing upper bounds on estimation error also places lower bounds on approximation error; and (b) including a penalty metric (the basis of regularisation) that penalises model complexity. In this latter case, we express the inclusion of the penalty (cost) term $C_P(f)$ as:
\begin{align}
    \hat{f}_n = \arg\min_{f \in \mathcal{F}} \{ \hat{R}_n(f) + C_P(f)   \}. \label{eqn:ml:penaltycostregularisation}
\end{align}
In practice, $C_P(f)$ may be proportional to the degree of $f$ (e.g. for a polynomial) or the norm of the derivative of $f$. Often $C_P(f)$ is proportional to parameter norms. Limiting training data size $n$ is one method of limiting the size of $\mathcal{F}$ (such as the method of sieves) on the assumption that $\mathcal{F}$ increases monotonically with $n$. In Bayesian contexts, where empirical risk reflects a log likelihood function, $C(f)$ is interpretable as incorporating prior knowledge about the likelihood of models. In this case, setting $\exp(-C_P(f))$ is a prior probability distribution on $\mathcal{F}$ (i.e. the probability for $f$), then $f$ being probable equates to small $C_P(f)$ and vice versa. Other models, such as `description length methods', estimate model complexity via the number of bits required to represent the model. Other techniques include hold-out methods involving splitting training data $D$ into training sets $D_T$ and test (or validation) sets $D_V$, where $D_T$ is used to select $\hat{f}_n$ and assessed (via a separate risk measure) against $D_V$. Knowing how to partition the training data can be difficult. A common approach in machine learning is to use $k$-fold cross validation which randomly splits data into training and test sets, removing entry $k$ from $D$ and minimising the regularised empirical risk \cite{hastie_elements_2013,goodfellow_deep_2016}. Theoretically demonstrating guarantees for improved performance can remain challenging, however (see \cite{hastie_elements_2013} $\S7.10$ for a general discussion). 

Regularisation and penalty-based methods are an important means of addressing empirical risk and form the basis of regularisation to deal with overfitting. We examine these in particular when reviewing the work of Nielsen et al. \cite{nielsen_geometric_2006,gu_quantum_2008,nielsen_optimal_2006} and penalty metrics for geometric quantum control.  

\subsection{No-Free Lunch Theorems} \label{sec:ml:No-Free Lunch Theorems}
Statistical learning theory (both classical and quantum) also has a number of \textit{no free lunch theorems} which place bounds upon the generalisation performance of general-purpose optimisation algorithms. The classical no free lunch theorem \cite{wolpert_no_1997} asserts that for any algorithm $A$, when averaged over all possible problems, the expected performance of $A$ is equivalent to any other algorithm $A'$. Mathematically, this can be represented in terms of risk across different types of learning problems.

Given the set $\mathcal{F}$ of all possible functions $f:\mathcal{X} \to \mathcal{Y}$ and the set of all distributions $\mathcal{P}$, for any two algorithms $f_A$ and $f_{A'}$ the \textit{no free lunch} theorem \cite{wolpert_no_1997} provides that:
\begin{align}
\int_{\mathcal{P}} R_{f_{A}}(L(f(X),Y)) \, d\mathcal{P} = \int_{\mathcal{P}}R_{f_{A'}}(L(f(X),Y)) \, d\mathcal{P} \label{eqn:ml:nofreelunch}
\end{align}
where $R_f(L(f(X),Y))$ is the true risk (equation \ref{eqn:ml:statlearning true risk}) associated with algorithm $A$ for a particular distribution $\Prb_{XY}$ in predicting function $f$, and $f \in \mathcal{F}$. The theorem indicates that no algorithm can universally minimise true (statistical) risk across all distributions. In particular, the theorem sets bounds upon how well models can generalise. Below we discuss the quantum analogue of the no free lunch theorem.  

\subsection{Statistical performance measures} \label{sec:ml:Statistical performance measures}
Before moving onto consideration of deep learning algorithms, we mention a few common statistical measures (including those used in Chapter \ref{chapter:QDataSet and Quantum Greybox Learning}) in the context of statistical learning theory. Much machine learning and statistical modelling, especially classification algorithms, utilise measures such as true positive (TP), true negatives (TN), false positives (FP) and false negatives (FN) by reference to an functional binary classifier estimator $\hat f(X): \X \to \Y$ where $\Y = \{0,1\}$. These can be related in statistical learning terms such as accuracy, a metric used in Chapter \ref{chapter:Quantum Geometric Machine Learning} (see results' tables and discussion in section \ref{sec:qgml:Tables and charts}) to compare model performance among candidate quantum machine learning architectures.
Accuracy measures the proportion of correct predictions among the total number of cases examined and can be linked with the expectation of the indicator loss:
\begin{align}
\text{Accuracy} &= \frac{\text{TP} + \text{TN}}{\text{TP} + \text{FP} + \text{TN} + \text{FN}} \\
&= 1 - E[L(f(X), Y)] =1 - E[\mathbb{I}(f(X_i)=Y_i))]. \label{eqn:ml:measures-accuracy}
\end{align}
In terms of statistical learning measures, empirical risk with an indicator function $\mathbb{I}(f(X_i)=Y_i))$ which is 1 if $f(X_i)=Y_i$ or 0 otherwise can be written as:
\begin{align}
    \hat R_n(f) &= \frac{1}{n}\sum_{i=1}^n (1-\mathbb{I}(f(X_i)-Y_i))\\
    &=1-\frac{1}{n}\sum_{i=1}^n (\mathbb{I}(f(X_i)-Y_i))
\end{align}
where $\frac{1}{n}\sum_{i=1}^n (\mathbb{I}(f(X_i)-Y_i))$ is the proportion of correct predictions. Accuracy can be written as:
\begin{align}
    \text{Accuracy}(f) = \frac{1}{n}\sum_{i=1}^n \mathbb{I}(f(X_i)=Y_i))
\end{align}
in which case empirical risk is one minus the accuracy. Given that $E[L(f(X),Y)] = \lim_{n\to \infty} \hat R_n (f)$, then we can understand accuracy as per above in equation (\ref{eqn:ml:measures-accuracy}). Thus we see the inverse relationship between minimising empirical risk and maximising accuracy.

\section{Deep Learning}\label{sec:ml:Deep Learning}
Chapters \ref{chapter:QDataSet and Quantum Greybox Learning} and \ref{chapter:Quantum Geometric Machine Learning} of this work concern the construction of a range of deep learning architectures to solve optimisation problems in quantum control. In this section, we cover the key elements of the machine learning architectures adopted in those chapters. We focus on neural network formalism, with a particular emphasis on `greybox' methods whereby neural network architectures are encoded with \textit{a priori} information relevant to the problem at hand, such as via encoding specific laws, rules or transformations (such as, in our case, the time-independent approximation of Schr\"odinger's equation) which assist the learning process. 

\subsection{Linear models}\label{sec:ml:Linear models}
Neural networks used in this work derive ultimately from an adaptation of statistical methods related to generalised linear models. In this section, we briefly survey the key features of generalised linear models as they relate to neural networks. We subsequently define neural network architecture in terms of non-linear functional composition of a variety of (usually) linear models, via layers in a deep neural network having a representation as a directed graph. Tracking the construction of neural networks by building up intuition through linear models and then non-linear networks is a useful way of both demystifying much of the jargon around neural networks, but also serves to connect the architecture to concepts across our previous chapters. We begin with basic linear models.
\begin{definition}[Linear models]\label{defn:ml:Linear models}
    Given $X \in \R^m$ and $Y \in \R$ where samples $(X,Y) \sim \Prb_{XY}$ are identically and independently distributed, we define a basic linear model for estimating $Y$ (assuming it as a scalar for simplicity) as:
    \begin{align}
        \hat Y = \omega_0 + \sum_{j=1}^m X_j^T \omega_j + \epsilon
    \end{align}
    where $\omega_0 \in \R$  is the estimate of bias while $\omega \in \R^m$ is a vector of coefficients (weights) with $\epsilon$ uncorrelated (random) errors.
\end{definition}
Sometimes $\omega_0$ (being constant) is absorbed such that $\hat Y = X^T \omega$ where it is understood that the zeroth index is identity (so we just have the $\omega_0$ term). We drop the transpose symbol on $X$ and other tensors from hereon in, it being understood. The coefficients $\hat \omega$ are determined according to a chosen minimum estimate of empirical risk, usually least squares. Where $\omega_j \in \R^m$ satisfies smoothness and other (mathematically) nice properties, then we can regard $\omega_j$ as being drawn from a differentiable manifold $\M \simeq \R^m$ (homeomorphically) where the basis of $\omega_j$ can be considered to be (tangent) vectors, each of which has directions $j=1,...,m$. An extension to forms of generalised additive model can then be undertaken by considering \textit{projection pursuit regression} (used by Hastie et al. \cite{hastie_elements_2013} as we see below to connect with neural network architecture). In generalised linear model theory, minimising empirical risk is often undertaken by imposing penalty metrics in objective/loss functions in order to steer models away from high-variance (and thus overfitting of) classes of estimates $\hat f$. Such methods are sometimes denoted `shrinkage methods' in the vernacular of statistical learning as they aim to minimise coefficient variance while retaining predictive power. \textit{Ridge regression}, common across the sciences, is one such model utilising ridge functions, being functions that are a combination of univariate functions with an affine transformation. Ridge coefficients are then those $\hat\omega$ which minimise a penalised loss function, parametrised by some parameter $\lambda \in \R$ (so higher values of $\|\omega\|_2^2$ incur a higher penalty): 
\begin{align}
    % General form of ridge regression
\hat{\omega}_{\text{ridge}} = \arg\min_{\omega} \left\{ L(Y, f(X)) + \lambda \|\omega\|_2^2 \right\} \label{eqn:ml:ridgeregression}
\end{align}
where here $C_P(f)=\lambda \|\omega\|_2^2$ connecting with the discussion of penalty terms above. As we can see (via the loss function, for simplicity we assume $L \in \R$), $f:\R^m \to \R$, while we can regard the addition of the $\lambda \|\omega\|_2^2 $ term (denoted a regularisation term) as one that penalises larger weights $\omega$. The regularisation term thereby aims to address overfitting by reducing the magnitude of $\omega$ and thus the variance of the model. A useful way of connecting generalised linear models with neural networks is to via projection pursuit models which represent linear combinations of non-linear ridge functions. We do so in order to elucidate the geometric and directional nature of statistical learning in this way, in turn connecting with intuition for quantum geometric machine learning.

A ridge function defined as $f(X) = g\braket{X, e}$ for $X \in \R^m$, $g: \R \to \R$ being a univariate function (a function on the loss function in effect), $e \in \R^m$ with the inner product as defined previously (definition \ref{defn:quant:Inner Product}). The ridge function is constant along directions orthogonal to $e_m$. By construction, vectors in hyperspaces of $R^m$, denoted $e_1,...,e_{m-1}$ are orthogonal to $e_m$ i.e. $\braket{e_j,e_m}=0$ for $j=1,...,m-1$. Thus:
\begin{align}
f\left(X + \sum_{j=1}^{m-1} b_j e_j\right) = g\left(\braket{X, e_m} + \sum_{j=1}^{m-1} b_j \braket{e_j , a_m}\right) = g(\braket{X,e}) = f(X) \label{eqn:ml:ridgefunction}
\end{align}
which elucidates the invariance under the affine transformation. Ridge functions then form the basis for what in statistical learning is denoted project pursuit regression \cite{hastie_elements_2013} with the general form \cite{friedman_projection_1981}:
\begin{align}
    f(X) = \sum_n^N g_n(\omega_m X) \label{eqn:ml:projectpursuitregression}
\end{align}
for dimension $N$. In this formulation, $\omega_m$ are $m$ unit vectors in the $m$-directions of the manifold $\M$. The key element is that the estimator $\hat Y=f(X)$ is a function of derived features $V_m = \omega_m X$ rather than $X$ directly. Note for consistency that our general weight tensor $\omega$ is a scaled version of $\omega_m$. The ridge functions $g_n(\omega_m X)$ vary only in the directions defined by $\omega_m$ where the feature $V_m = \omega_m X$ can be regarded as the projection of $X$ onto the unit vector $\omega_m$. In general $g_n$ there is a wide variety of non-linear functions to choose from. As Hastie et al. \cite{hastie_elements_2013} note, if $N$ is sufficiently large, then the model can approximate \textit{any} continuous function in $\R^m$ (see \cite{goodfellow_deep_2016} $\S6.4.1$). Thus projection pursuit regression models can be regarded as \textit{universal approximators}, a key characteristic of the success of neural networks.

\subsection{Neural networks}\label{sec:ml:Neural networks}
 The formal definition of neural networks by Fiesler \cite{fiesler_neural_1994} provides a taxonomy to understand variable elements.
\begin{definition}[Neural network] \label{def:ml:neuralnetworkgeneral}
    A neural network can be formally defined as a nested 4-tuple $NN=(T,C,S(0),\Phi)$ where $T$ is network topology, $C$ is connectivity, $S(0)$ denotes initial conditions and $\Phi$ denotes transition functions. 
\end{definition}
The formal definition of a neural network has influenced later literature and technical work in the field. We discuss briefly some of the elements below:
\begin{enumerate}
    \item \textit{Topology}. A directed graph $G=(V,E)$ of functional composition, with vertices $V$ representing neurons and $E$ edges between neurons comprising. The topology comprises (a) a framework of $L_n$ layers each with $N_n$ neurons; and (b) a set of connectivity relations between source neurons in $L_0$ exhibiting certain properties such as (i) intraconnectivity, self-connectivity or supraconectivity, (ii) symmetry or asymmetry and (iii) order;
    \item \textit{Constraints}. Bounds upon (a) weight value range $||\omega_{ij}|| \leq K_1 \in \R$, (b) local thresholds (biases of offsets) and (c) activation range i.e. $||\sigma|| \leq K_2 \in \R$;
    \item \textit{Initial state}. covering initialisation of (a) weights, (b) local thresholds and (c) activation values; and
    \item \textit{Transition functions}. These are transition functions for (a) neuron functions (specifying output of neuron via activation function), (c) learning rules (updating weights and offsets), (c) clamping functions that keep certain neurons constant and (d) ontogenic functions, those that change network topology.
\end{enumerate}
Modern formulations generally track, with some differences, this taxonomy, with each category and subcategory being a widely studied subdiscipline of machine learning research, such as those concerned with optimal network topology, pretraining and tuning of hyperparameters (see \cite{goodfellow_deep_2016}). Of note is that while statistical learning theory and computer science can provide some theoretical bases for network design, in general there are few or limited theoretical guarantees relating network architectural features (such as network topology or choice of initialisation). As such, tuning such architectural characteristics (which are often represented as hyperparameters of a model or architecture) tends to be experimental (or in reality trial-and-error) and empirical in nature. Moreover, as we show in particular in Chapter \ref{chapter:Quantum Geometric Machine Learning}, network architecture and connectivity is often a key and important element in both the performance of networks but also their description via visual or other means.\\
\\
\subsubsection{Neural network components}\label{sec:ml:Neural network components}
An elementary neural network can be understood as a non-linear extension of the generalised linear models discussed above within the network taxonomy. In the examples below, neural network layers $a^{(l)}$ are indexed by $l=0,...,L$ where $l=0$ is the first layer and $l=L$ the output layer. We consider a simple fully-connected feed-forward network comprising an input layer $a^{(0)}$ with $m$ inputs i.e. $X \in \R^m$ (whose activation function, as we discuss below, can be considered identity neuron functions), a set of hidden layers $a^{(1)}$ to $a^{(l-1)}$ and a final layer $a^{(L)}$ that provides predicted outputs. Each layer comprises $N$ neurons so that $a^{(l)}=(a^{(l)}_1,...,a^{(l)}_{n_l})$ where $n_l$ is the number of neurons in layer $l$. That is, each neuron has its own activation function though in practice these are considered by convention the same for each neuron (for simplicity we keep the number of neurons in each layer constant). Each neuron in each layer takes the inputs of (all or some) neurons in the previous layer (or the initial data $X$) as an argument for its own activation function (which is usually non-linear) comprising a weight tensor (which for consistency with linear models we generally denote $\omega$) and a bias term $\omega_0$ which we usually absorb into $\omega$. Here we adopt the general case of $\omega \in \R^{n \times m}$ to allow for subsequent layers that may have $n$ neurons. Activation functions for the basis for the non-linearity of neural networks. We define activation functions below.
%========
\begin{definition}[Activation function]\label{defn:ml:Activation function}
    Given an input vector $X \in \R^m$, weight tensor $\omega \in \R^{n \times m}$ and bias term $\omega_0 \in \R^n$ we define the affine transformation $z = \omega X + \omega_0 \in \R^n$. An activation function is then defined as:
    \begin{align*}
        \sigma&:\R^m \to \R^n \qquad \sigma_k: \R \to \R \qquad k=1,...,n\\
        \sigma(z) &= \sigma(\omega X + \omega_0) = \begin{pmatrix}
            \sigma_1(z), ...,\sigma_n(z)
        \end{pmatrix}
    \end{align*}
    where the activation function $\sigma(z)$ is (usually) applied element-wise.
\end{definition}
Each neuron in each layer is constituted by an activation function $\sigma_k$ and together they form an $n$-dimensional layer (or more precisely each layer has its own number of neurons $n_l$ but we leave this understood). Activation functions $\sigma$ are generally classified into one of the following types: (i) \textit{ridge functions} (as described above) where $f(X) = g(\omega X + \omega_0)$ for a non-linear ridge function $g:\R \to \R$, (ii) \textit{radial functions} where $f(X) = g(||X - c||)$ where $c$ is the centre of the radial basis function (e.g. for a radius $r = ||X-c||$), $g:\R \to \R$ is a non-linear function with $||\cdot||$ a norm (usually Euclidean) or (iii) \textit{fold functions} such as the softmax function. Fold functions, such as softmax, are particularly important as they are often interpreted formally or heuristically as probabilities relating to classification. 
Let $ \sigma: \mathbb{R}^m \rightarrow S $ be a function, where $ S $ is the standard $(m-1)$-simplex defined as:
\begin{align}
S = \left\{ p\in \mathbb{R}^m \mid p_i \geq 0, \forall i \in \{1, \dots, m\}, \text{ and } \sum_{i=1}^m p_i = 1 \right\}. \label{eqn:ml:simplex}
\end{align}
Each point within the simplex $ S $ represents a possible probability distribution over $m$ discrete outcomes, where $p_i$ is the probability of the $i$-th outcome.
The function $\sigma$ transforms a vector $z \in \mathbb{R}^m$ into a vector $p \in S$ such that each component of $p$, denoted as $p_i$, can be interpreted as the probability of the $i$-th outcome. Framed in this way, the activation satisfies requirements for interpretation as a probability distribution, namely (i) $p_i \geq 0$ (non-negativity) and (ii) $\sum_{i=1}^m p_i = 1$. An example of an activation function often adopted as a proxy for probability is the \textit{softmax function} \cite{martins_softmax_2016}, being a type of fold function:
\begin{align}
p_i = \frac{e^{z_i}}{\sum_{j=1}^m e^{z_j}} \label{eqn:ml:activationsoftmax}
\end{align}
which effectively converts a vector of real numbers into a probability distribution of possible outcomes.

\subsubsection{Layers in neural networks}\label{sec:ml:Layers in neural networks}
As discussed above, clustering or layering of neurons is a hallmark of neural network architecture. While fully-connected feed-forward networks (where each neuron in each layer is forward-connected to each other) exhibit relatively symmetric structure, not all networks are as simply described given concurrent parallel networks, or common practices such as \textit{residual connections} that may connect initial layers to layers further down the hierarchy directly rather than through intermediate layers. Nevertheless, for exposition we consider in essence a fully-connected feedforward network. To do so, we describe a canonical multi-layer perceptron model of a feed-forward network (directly related to one of the candidate models used in Chapter 6, namely the fully-connected model). In the following we denote $n_l$ the number of neurons for layer $l$.
\begin{definition}[Feed-forward neural network] \label{defn:ml:Feed-forward neural network}
    A feed-forward neural network (also known as a multi-layer perceptron network) consists of multiple layers $a^l$ of neurons $a^l_i$ such that each neuron in each layer (other than the first input layer) is a (composition) function of each neuron in the preceding layer. Formally:
    \begin{align}
    a_i^{(l)} = \sigma^{l}_i\left(\sum_{j=1}^{n_{l-1}} \omega_{ij}^{(l)} a_j^{(l-1)} + \omega_{i0}^{(l)}\right) = \sigma^{l}_i\left(\sum_{j=0}^{n_{l-1}} \omega_{ij}^{(l)} a_j^{(l-1)}\right)\label{eqn:ml:feedforwardactivation-al}
\end{align}
where $l$ indexes each layer, $i$ each neuron in that layer $a^l_i$ valued by an activation function $\sigma^{(l)}_i$ for that neuron in that layer (definition \ref{defn:ml:Activation function}) which is itself a function of the sum of weight matrices $\omega_{ij}^{(l)}$ applied to the output of each neuron of the previous layer $a_j^{(l-1)}$ together with a bias term $\omega_{i0}^{(l)}$.
\end{definition}
Sometimes the literature just uses $a^{(l)}_i$ or $\sigma^{(l)}_i$ for neurons, but the distinction is usually maintained to emphasise the activation function as a function and $a^{(l)}_i$ as a neuron. Neurons are also occasionally referred to as `units'. Note that in the right-most term above we have absorbed the bias $\omega_{i0}$ into the summation for convenience (which can be achieved by considering the zeroth neuron in the layer $a^{l-1}_0$ as the identity). Following the classification of networks above, a fully-connected neural network with layers indexed from $l=0,...,L$ can then be formally described below. For convenience in the following, we denote the linear argument of activation functions as follows:
\begin{align}
    z_i^{(l)} = \sum_{j=0}^{n_{l-1}} \omega_{ij}^{(l)} a_j^{(l-1)} \label{eqn:ml:backprop-zi(l)}
\end{align}
which indicates how the $j$th neuron in the previous layer $l-1$ feeds into the $i$th neuron of the $l$th layer, weighted by the corresponding weight $\omega_{ij}^{(l)}$. We include a diagram depicting the fully-connected feed forward neural network in Fig. \ref{fig:ml:fullyconnectedNNwithmatrix}.
%Neural network schema
\subsubsection{Neural network schema}\label{sec:ml:neuralnetworkschema}
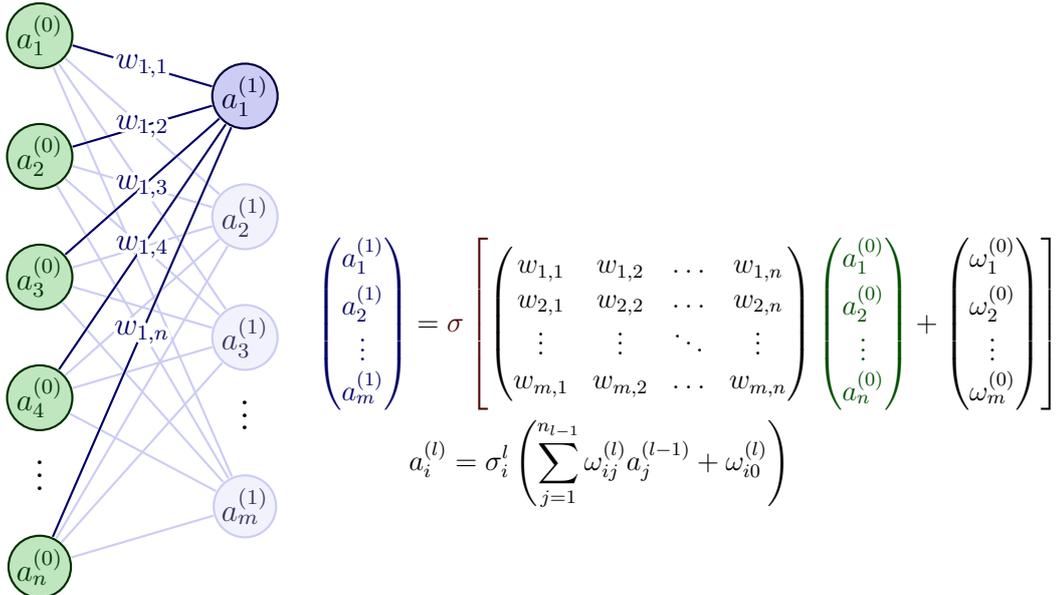
\begin{figure}
    \centering
\begin{tikzpicture}[x=2.7cm,y=1.6cm]
  \message{^^JNeural network activation}
  \def\NI{5} % number of nodes in input layers
  \def\NO{4} % number of nodes in output layers
  \def\yshift{0.4} % shift last node for dots
  
  % INPUT LAYER
  \foreach \i [evaluate={\c=int(\i==\NI); \y=\NI/2-\i-\c*\yshift; \index=(\i<\NI?int(\i):"n");}]
              in {1,...,\NI}{ % loop over nodes
    \node[node in,outer sep=0.6] (NI-\i) at (0,\y) {$a_{\index}^{(0)}$};
  }
  
  % OUTPUT LAYER
  \foreach \i [evaluate={\c=int(\i==\NO); \y=\NO/2-\i-\c*\yshift; \index=(\i<\NO?int(\i):"m");}]
    in {\NO,...,1}{ % loop over nodes
    \ifnum\i=1 % high-lighted node
      \node[node hidden]
        (NO-\i) at (1,\y) {$a_{\index}^{(1)}$};
      \foreach \j [evaluate={\index=(\j<\NI?int(\j):"n");}] in {1,...,\NI}{ % loop over nodes in previous layer
        \draw[connect,white,line width=1.2] (NI-\j) -- (NO-\i);
        \draw[connect] (NI-\j) -- (NO-\i)
          node[pos=0.50] {\contour{white}{$w_{1,\index}$}};
      }
    \else % other light-colored nodes
      \node[node,blue!20!black!80,draw=myblue!20,fill=myblue!5]
        (NO-\i) at (1,\y) {$a_{\index}^{(1)}$};
      \foreach \j in {1,...,\NI}{ % loop over nodes in previous layer
        %\draw[connect,white,line width=1.2] (NI-\j) -- (NO-\i);
        \draw[connect,myblue!20] (NI-\j) -- (NO-\i);
      }
    \fi
  }
  
  % DOTS
  \path (NI-\NI) --++ (0,1+\yshift) node[midway,scale=1.2] {$\vdots$};
  \path (NO-\NO) --++ (0,1+\yshift) node[midway,scale=1.2] {$\vdots$};
  
  % EQUATIONS
  \def\agr#1{{\color{mydarkgreen}a_{#1}^{(0)}}} % green a_i^j
  \node[below=16,right=11,mydarkblue,scale=0.95] at (NO-1)
    {$\begin{aligned} %\underset{\text{bias}}{b_1}
       &= \color{mydarkred}\sigma\left( \color{black}
            w_{1,1}\agr{1} + w_{1,2}\agr{2} + \ldots + w_{1,n}\agr{n} + b_1^{(0)}
          \color{mydarkred}\right)\\
       &= \color{mydarkred}\sigma\left( \color{black}
            \sum_{i=1}^{n} w_{1,i}\agr{i} + b_1^{(0)}
           \color{mydarkred}\right)
     \end{aligned}$};
  \node[right,scale=0.9] at (1.3,-1.3)
    {$\begin{aligned}
      {\color{mydarkblue}
      \begin{pmatrix}
        a_{1}^{(1)} \\[0.3em]
        a_{2}^{(1)} \\
        \vdots \\
        a_{m}^{(1)}
      \end{pmatrix}}
      &=
      \color{mydarkred}\sigma\left[ \color{black}
      \begin{pmatrix}
        w_{1,1} & w_{1,2} & \ldots & w_{1,n} \\
        w_{2,1} & w_{2,2} & \ldots & w_{2,n} \\
        \vdots  & \vdots  & \ddots & \vdots  \\
        w_{m,1} & w_{m,2} & \ldots & w_{m,n}
      \end{pmatrix}
      {\color{mydarkgreen}
      \begin{pmatrix}
        a_{1}^{(0)} \\[0.3em]
        a_{2}^{(0)} \\
        \vdots \\
        a_{n}^{(0)}
      \end{pmatrix}}
      +
      \begin{pmatrix}
        \omega_{1}^{(0)} \\[0.3em]
        \omega_{2}^{(0)} \\
        \vdots \\
        \omega_{m}^{(0)}
      \end{pmatrix}\right]\\
      & a_i^{(l)} = \sigma^{l}_i\left(\sum_{j=1}^{n_{l-1}} \omega_{ij}^{(l)} a_j^{(l-1)} + \omega_{i0}^{(l)}\right)
    \end{aligned}$};
  
\end{tikzpicture}
    \caption{Schema of the first two layers of a fully-connected feed-forward neural network (definition \ref{defn:ml:Feed-forward neural network}) together with associated matrix and algebraic representation. Here $a_i^{(l)}$ are the input layer neurons, $\omega_{lj}^{(l)}$ are the weights (absorbing bias terms) for neuron $a_j^{(l)}$ (diagram adapted from \cite{neutelings_izaak_tikznet_nodate}).}
    \label{fig:ml:fullyconnectedNNwithmatrix}
\end{figure}

A schema of the toy model is set out in Figure \ref{fig:ml:fullyconnectedNNwithmatrix}:
\begin{enumerate}
    \item \textit{Input layer}. The input layer is represented by feature data $X=(x_1,...,x_m) \in \R^m$. For the input layer ($l=0$), we can just regard the activation function as the identity $a_j^{(0)} = x_j$, the $j$-th (for $j=1,...,n_0$).
    \item \textit{Hidden layer}. Here the layers indexed by $l$ comprise neurons indexed by $i$ represented by their activation function $a_i^{(l)}$. A layer $l$ comprises $n_l$ neurons. For a fully-connected feed-forward layer, this means all neurons in the immediately preceding layer are inputs into each activation function (neuron) of the immediately subsequent layer:
    \begin{align}
        a^{(l)} = (a^{(l)}_1,...,a^{(l)}_{n_l}) 
        % = \sigma(\alpha_0 + \alpha_m^T X) \qquad 
    \end{align}
    % where $\alpha_j:=w_j$ represent the weights (including bias for $0$) of the hidden layer. 
    using equation (\ref{eqn:ml:feedforwardactivation-al}) above. 
    
\item \textit{Output layer}. The output layer $a^L$ is then chosen to accord with the problem at hand, so for classification problems it may be either a binary classification or link-function (such as a logit function) giving an outcome $\sigma_l \in [0,1]$ interpretable as a probability (e.g. $\sigma^{(l)}_i$ where for example to classify $K$ objects we would have $i=1,...,K$). The overall output of the network is the estimator $\hat Y$, denoted as $\hat Y = (\hat Y_1, \hat Y_2, ..., \hat Y_{n_L})$, where $n_L$ is the number of neurons in the layer:
\begin{align}
   f(X) = \hat Y = a^{(L)} = \left( a_1^{(L)}, a_2^{(L)}, ..., a_{n_L}^{(L)} \right). \label{eqn:ml:neuralnetworkfinallayer}
\end{align}
\end{enumerate}
Sometimes in the literature the output layer may itself be subject to an additional transformation (see for example \cite{hastie_elements_2013} $\S 11.6$) but such transformations can always just be cast as a final layer whose activation function is simply whatever the transformation is. 

As Hastie et al. \cite{hastie_elements_2013} note, the non-linearity of the activation functions $\sigma$ expand the size of the class of candidate functions $\mathcal{F}$ that can be learnt. Functionally, the network can be regarded as a functional composition, with each layer representing a function comprising linear inputs into nonlinear activations $\sigma^l$ applied sequentially. We can regard the neural network as a functional composition among activation functions with inputs $X$ and outputs $a^{(L)}$.
While not the focus here, in geometric contexts doing so can allow us to frame the network as maps among differentiable manifolds. We also note that multilayer feed-forward networks are of particular significance due to the universal approximation theorem \cite{hornik_multilayer_1989} for learning arbitrary functions (and other universal and function approximation results) with a quantum analogue that functions can be represented to arbitrary error at exponential depth (which is practically infeasible).  

%========diagram

%========diagram

\section{Optimisation methods}\label{sec:ml:Optimisation methods}
\subsection{Optimisation and Gradient Descent}\label{sec:ml:Optimisation and Gradient Descent}
The \textit{learning} component of machine learning is an optimisation procedure designed to reduce excess risk via minimising empirical risk. 
 The main optimisation algorithm used across most machine learning involves some sort of \textit{stochastic gradient descent} whereby parameters $\theta$ of estimated functions $\hat f(\theta)$ are updated according to a generalised directional derivative. In keeping with the geometric framing of much of this work, we present a definition below. Stochastic gradient descent is a type of gradient descent where only a randomly sampled subset of $D_n(X,Y)$ is used to train the model (in practice this involves a batch of randomly sampled training data). 
% \hl{REDO}
% Let $M$ be a smooth manifold representing the parameter space of a model, with a Riemannian metric $g$. Consider a loss function $L: M \to \mathbb{R}$ which is a smooth function on $M$. In the context of machine learning, stochastic gradient descent is an optimization technique to find a local minimum of $L$. \hl{EDIT}\\
% \\
Neural networks are parametrised by a set of unknown weights which are updatable (learnable) by means of updating according to an optimisation procedure. We now discuss probably the most important optimisation technique in quantum machine learning (as with classical), namely backpropagation based upon stochastic gradient descent.
\subsection{Backpropagation}\label{sec:ml:Backpropagation}
Backpropagation \cite{rumelhart_learning_1986,werbos_backpropagation_1990,lecun_optimal_1990,lecun_gradient-based_1998} is a stochastic gradient descent-based method for optimising model performance across neural network layers. For neural networks, the most common approach is to use the method of \textit{gradient descent} to update parameters, with the particular method by which this update is calculated being via backpropagation equations. It represents, subject to a number of conditions, an efficient way of performing gradient descent calculations in order to update parameters $\omega$ of models $f_\theta$ to achieve some objective, such as minimisation empirical risk (equation (\ref{defn:ml:empiricalrisk})) (loss) by, having calculated an estimate $\hat f_\theta(X)$, propagating the error $\delta \sim |\hat f_\theta - Y|$ through the network. Propagation here refers to the use of the calculated error $\delta$ in updating parameters $\omega$ across network layers. Backpropagation consists of two phases: (a) a \textit{forward pass}, this involves calculating each layer's activation function $a_i^{(l)}$ (see equation (\ref{eqn:ml:feedforwardactivation-al})); and (b) a \textit{backward pass} where the backpropagation updates are calculated. From equation (\ref{eqn:ml:statisticalrisklossfunction}), assume a loss function denoted generically by $\hat R_n(\omega)$ (in Chapters \ref{chapter:QDataSet and Quantum Greybox Learning} and \ref{chapter:Quantum Geometric Machine Learning}, we use variations of mean square error). Firstly, for gradient descent recall the directional derivative \ref{eqn:geo:directionalderivative}) in differential form for Riemannian and subRiemannian manifolds and gradient (definition \ref{defn:geo:gradient}).
Recall our network is a composition of (activation) functions parametrised by weight tensors $\omega$. For completeness, in this section: $\omega_{ij}^{(l)}$ is the weight vector for neuron $i$ in layer $l$ that weights neuron $j$ in layer $l-1$, $a^{(l)}$ is layer $l$, $a_i^{(l)}$ is the $ith$ neuron in layer $l$ and $n_l$ is the number of neurons (units) in layer $l$. We define gradient descent for optimisation using these definitions as follows. 
\begin{definition}[Gradient descent optimisation]\label{defn:ml:Gradient descent optimisation}
    Optimisation by gradient descent is defined as a mapping:
    \begin{align}
\omega_{ij}^{(l+1)} &= \omega_{ij}^{(l)} - \gamma_l \sum_{k=1}^{N} \frac{\partial \hat R_k}{\partial \omega_{ij}^{(l)}} = \omega_{ij}^{(l)} - \gamma_l \sum_{k=1}^{N} \nabla_{\omega_{ij}^{(l)}} \hat R_k     
\label{eqn:ml:backpropupdatemain}
\end{align}
where (a) $\hat R_k$ are loss/error values for the $k$th data point in the training set $D_n(X,Y)$, (b) $\omega_{ij}^{(l)}$ denote the weights of the $i$th neuron for the $l$th layer weighting the $j$th neuron of the $l-1$th layer and (c) $\gamma_l$ the learning rate for that layer (which is usually constant across layers and often networks). 
\end{definition}
Equation (\ref{eqn:ml:backpropupdatemain}) updates each weight $\omega_{ij}^{(l)}$ by reference to each example $(X_i,Y_i)$, however in practice a subsample of $D_n$ is used denoted a \textit{batch}, such that $\hat R_k$ is an average over the size of the batch size $N_B$ i.e. $(1/N_B)\sum_k^{N_B} \hat R_k$ (we omit the summation below for brevity) (note we set $k=i$ for consistency with the choice of each data point $(X_i,Y_i)$). Calculating $\nabla_{\omega_{ij}^{(l)}} \hat R_i$ relies upon the chain rule. First, consider how $\hat R_i$ varies in the linear case of $z^{(l)}_i$ (without applying the non-linear activation function $\sigma$):
\begin{align}
\frac{\partial R_i}{\partial \omega_{ij}^{(l)}} = \frac{\partial R_i}{\partial z_{i}^{(l)}}\frac{\partial z_{i}^{(l)}}{\partial \omega_{ij}^{(l)}}  
% \qquad 
% \frac{\partial R_i}{\partial \alpha_{m\ell}} &= s_{mi} x_{i\ell} 
\label{eqn:ml:gradientdescent-errors}
\end{align}
The first of these terns is denoted the error while the second term is shown to be equivalent to the preceding layer:
\begin{align}
    \delta^{(l)}_i &= \frac{\partial R_i}{\partial z_{i}^{(l)}} \label{eqn:ml:backpropRz}\\
    \frac{\partial z_{i}^{(l)}}{\partial \omega_{ij}^{(l)}} &= \frac{\partial }{\partial \omega_{ij}^{(l)}} \left(\sum_{\mu=0}^{n_{l-1}} \omega_{i\mu}^{(l)} a_\mu^{(l-1)} \right) =  a_j^{(l-1)} \label{eqn:ml:backpropzomega}
\end{align}
where the partial derivatives vanish in equation (\ref{eqn:ml:backpropzomega}) for all but $\mu=j$. The $\delta^{(l)}_i$ term in equation (\ref{eqn:ml:backpropRz}) is denoted the error. As we show below, $\delta^{(l)}_i$ is dependent upon errors in the $l+1$th layer, hence the error terms propagate `backwards', giving the name backpropagation.

For the output layer, we note that $\hat R_i=\hat R_i(\hat Y_i,Y_i) = \hat R_i (\sigma^{(L)}_i(z^{(L)}_i),Y)$ thus by the chain rule:
\begin{align}
     \delta^{(L)}_i &= \frac{\partial \hat R_i}{\partial z_{i}^{(L)}} = \frac{\partial \hat R_i}{\partial \sigma_{i}^{(L)}} \frac{\partial \sigma_{i}^{(L)}}{\partial z_{i}^{(L)}} = \frac{\partial \hat R_i}{\partial \sigma_{i}^{(L)}}  \sigma_{i}^{'(L)}(z_{i}^{(L)})
\end{align}
and:
\begin{align}
   \frac{\partial R_i}{\partial \omega_{ij}^{(l)}} = \delta^{(l)}_i a^{(l-1)}_j.
\end{align}
For hidden layers $l<L$ we have:
\begin{align}
    \delta^{(l)}_i &= \frac{\partial \hat R_i}{\partial z_{i}^{(l)}} = \sum_{\mu=1}^{n_{l+1}} \frac{\partial \hat R_i }{\partial z_{\mu}^{(l+1)}} \frac{\partial z_{\mu}^{(l+1)}}{\partial z_{i}^{(l)}}\\
    &=\sum_{\mu=1}^{n_{l+1}}  \delta^{(l+1)}_\mu \frac{\partial z_{\mu}^{(l+1)}}{\partial z_{i}^{(l)}}.
    % \label{eqn:ml:backpropRz}
\end{align}
Noting (equation (\ref{eqn:ml:backprop-zi(l)})) then:
\begin{align}
   z_\mu^{(l+1)} &= \sum_{i=0}^{n_{l}} \omega_{i\mu}^{(l+1)} \sigma^{l}_i(z_i^{(l)})\\
   \frac{\partial z_{\mu}^{(l+1)}}{\partial z_{i}^{(l)}} &=\omega_{i\mu}^{(l+1)} \sigma^{l'}_i(z_i^{(l)})
\end{align}
resolves to the \textit{backpropagation formula}:
\begin{align}
    \delta^{(l)}_i &= \sigma^{l'}_i(z_i^{(l)}) \sum_{\mu=1}^{n_{l+1}} \omega_{i\mu}^{(l+1)}  \delta_\mu^{(l+1)}  \label{eqn:ml:backpropagationequationssdelta} 
\end{align}
with:
\begin{align}
    \frac{\partial \hat R_i}{\partial \omega_{ij}^{(l)}} &= \nabla_{\omega_{ij}^{(l)}} \hat R_i = \sigma^{l'}_i(z_i^{(l)}) a^{(l-1)}_j \sum_{\mu=1}^{n_{l+1}} \omega_{i\mu}^{(l+1)}  \delta_\mu^{(l+1)} \label{eqn:ml:backpropfinal}
\end{align}
We can see from equation (\ref{eqn:ml:backpropfinal}) that the error $ \delta^{(l)}_i$ for layer $l$ is dependent on errors in the $l+1$th layer i.e. $\delta^{(l+1)}_\mu$. In this sense, errors propagate backwards from the final output layer to the first layer. Backpropagation thus relies firstly on the forward pass, the estimates $\hat f_k(X)$ are computed which allows computation of first $\delta_k^L$ (errors based on outputs and labels) (the forward phase) following which they are `back-propagated' throughout the network via the backpropagation equations (\ref{eqn:ml:backpropagationequationssdelta}) (the `backward phase'). Error terms for layer $l-1$ are calculated by weighting via $\omega^{(l+1)}_{i\mu}$ the error terms for the next layer $\delta^{(l+1)}_\mu$ and then scaling via $\sigma^{l'}_i(z_i^{(l)})$. Doing so allows computation of the gradients in equation (\ref{eqn:ml:gradientdescent-errors}). A \textit{training epoch} is then one full round of forward and backward passes. Note that equation (\ref{eqn:ml:backpropupdatemain}) sums over the training set, referred to as \textit{batch learning} (the most common case). Alternatively, one can perform backpropagation point-wise based on single observations $(X_i,Y_i) \in D_n$. 

\textit{Quantum backpropagation} also brings with it a number of subtleties arising because of the effect of measurement on quantum states. Thus the classical analogue of backpropagation does not carry over one-for-one to the quantum case. In many cases, models learn classically parametrised quantum circuit features via implementing offline classical backpropagation (as per above) conditional upon measurement statistics. Examples of quantum-related backpropagation proposals include, for example, a series of papers by Verdon et al. \cite{verdon_learning_2019,verdon_quantum_2017,verdon_quantum_2019-1,verdon_quantum_2019-2,verdon_quantum_2019-3,verdon_universal_2018} where continuous parameters are quantised and encoded in a superposition of quantum registers enabling a backpropagation-style algorithm. Other examples, including quantum neural network proposals \cite{beer_quantum_2021,beer_quantum_2022,beals_quantum_2001} provide automatic differentiation as a means of propagating a form of automatic measurement that is then propagated through a quantum analogue of a fully-connected network. In each case, one must remember that the optimisation strategy relies upon \textit{measurement} which (as per definition \ref{defn:quant:Measurementchannel}) is an inherently quantum-to-classical channel.

\subsection{Natural gradients}\label{sec:ml:Natural gradients}
We briefly mention the concept also of \textit{natural gradients} here, drawn primarily from \textit{geometric information theory} \cite{nielsen_geometric_2015,nielsen_elementary_2020} as a means of connecting with the formalism of Appendix \ref{chapter:Background: Geometry, Lie Algebras and Representation Theory}. As noted in the literature \cite{nielsen_elementary_2020,amari_information_2016}, local gradient descent based optimisation depends on the choice of local parameters. Given (in the scalar case) two parameterisations $\eta$ and $\theta$, with initialisations $\theta_0=\theta(\eta_0)$ and $\eta_0 = \eta(\theta_0)$, we will usually have $\eta(\theta) \neq \theta(\eta)$ such the path and stationary endpoint of gradient descent in parameter space may be different (using $L$ as our loss function to avoid confusion with $\mathcal{R}$ below):
\begin{align}
    \theta_{t+1} &= \theta_t - \alpha_t \nabla_\theta L_\theta(\theta_t) \neq 
    \eta_{t+1} = \eta_t - \alpha_t \nabla_\eta L_\eta(\eta_t) \label{eqn:ml:naturalgradient-problemstatement}
\end{align}
referring to a generic smooth loss function $L \in \cinfm$. Here $\alpha_t$ is step-size in the gradient descent function. The solution to this problem is to select a \textit{natural gradient} \cite{amari_natural_1998} which provides a directional derivative in the intrinsically steepest direction with respect to the (Riemann) metric tensor. In information theory, parameter space can be modelled in terms of a differentiable manifold where updating parameters corresponds to routes or curves on the manifold. In this case, points on the manifold $p \in \M$ have a representation, for example, as a set of parameters (say a vector) associated with that point i.e. $\theta_p$. Stochastic gradient descent on Riemannian manifolds ($\M,g$) \cite{bonnabel_stochastic_2013}, denoted by $\nabla_{\M}$, utilises the exponential map (definition \ref{defn:alg:Matrix exponential}) $\exp:T_p\M \to \M$ for its update protocol such that:
\begin{align}
    p_{t+1} = \exp_{p_t}(-\alpha_t \nabla_{\M}L(p_t)) \qquad \nabla_{\M}L(p) = \nabla_v(L(\exp_p(v)))\big|_{v=0} \label{eqn:geo:naturalSGD-Riemanngradient}
\end{align}
where $\nabla_v$ is the directional derivative given in equation \ref{eqn:geo:directionalderivative} and $\alpha$ is step size (see section \ref{sec:ml:Regularisation and Hyperparameter tuning}). Because calculating the exponential term in equation (\ref{eqn:geo:naturalSGD-Riemanngradient}) can be difficult, instead mapping to the parameter bundle occurs via a first-order Taylor expansion of the exponential \textit{Euclidean retraction} $\mathcal{R}$ (adapted from \cite{nielsen_elementary_2020}):
\begin{align}
    \theta_{t+1} =  \mathcal{R}_{\theta_t}\left(-\alpha_t \nabla_{\theta} L_\theta(\theta_t)\right)
\end{align} 
which shifts $p$ by some infinitesimal amount. Natural gradient descent is then given by:
\begin{align}
    \theta_{t+1}= \theta_t-\alpha_t g^{-1}_\theta(\theta_t) \nabla_{\theta} L_\theta(\theta_t) \label{eqn:ml:Natural gradient descent-main}
\end{align}
where the \textit{natural gradient} $\nabla_N$ is defined as:
\begin{align}
    \nabla_N L_\theta(\theta) := g^{-1}_\theta(\theta) \nabla_\theta L_\theta(\theta).  \label{eqn:ml:Natural gradient descent-nabla}
\end{align}
Here $g_\theta^\inv(\theta)$ is the dual Riemannian metric and is invariant under invertible smooth changes of parametrisation. Note the presence of $g^\inv$ indicates that the gradient, which transforms contravariantly. The benefit of the natural gradient is that it is invariant under coordinate transformations, though there are some subtleties regarding convergence discussed in the literature \cite{bonnabel_stochastic_2013,nielsen_geometric_2015,amari_natural_1998}. The quantum analogue of natural gradient descent is briefly touched upon in Appendix \ref{chapter:Background: Classical, Quantum and Geometric Machine Learning} as a technique for optimisation.

\subsection{Regularisation and Hyperparameter tuning}\label{sec:ml:Regularisation and Hyperparameter tuning}
As the abstract definition of neural networks in section \ref{def:ml:neuralnetworkgeneral} above indicates, there is a wide combinatorial landscape of parameters for the design of neural network architectures, including topology, choices of activation functions, loss functions, regularisation, dropout and so on. Hyperparameters are settings of a neural network model which are not adapted or updated by the learning protocol of the model in a direct sense (though model performance may inform ancillary hyperparametric models). In effect changing the hyperparameter changes the neural network model itself. We set out briefly a few such parameters, techniques and hyperparameters used in the results detailed in later chapters.
\begin{enumerate}[(i)]
    \item \textit{Learning rate}. The learning rate parameter $\gamma_r$ (or \textit{gain} in adaptive learning) in equations (\ref{eqn:ml:backpropupdatemain}) represent the step size at each iteration. Mathematically it represents a scaling of the (directional derivative) $\nabla_\omega R$ and is among the most important hyperparameters for network performance.
    \item \textit{Epochs}. The number of epochs over which training (a full forward- and backward-pass of the backpropagation algorithm in equation (\ref{eqn:ml:backpropagationequationssdelta})) is a tunable parameter of neural networks. In principle the number of epochs can affect the sensitivity of weights to training data and can risk overparametrisation such that models do not generalise well. Conversely, insufficient epochs may not sufficiently minimise empirical risk (definition \ref{defn:ml:empiricalrisk}).
    \item \textit{Regularisation}. As flagged above, regularisation is a means of seeking to minimise variance, and thus overfitting, of models trained on in-sample training data where inference (i.e. predicting / inferring label outputs $Y$) is undertaken on out-of-sample feature data $X$.
    \item \textit{Generalisation}. Generalisation refers to a statistical measure of risk that captures the out-of-sample performance of a model (see discussion above). 
\end{enumerate}
Finally, we briefly note that while statistical learning theory provides a generalised framework for analysing model performance, deep learning networks often exhibit singularities i.e. they are singular models. The fact that over-parametrised deep learning models can generalise well runs counter somewhat to the implied trade-off between model complexity and performance at the heart of statistical learning theories \cite{zhang2021understanding}.  In such cases, alternative approaches such as singular learning theory provide means of estimating appropriate statistical learning figures of merit for models (see seminal work by Watanabe \cite{watanabe_algebraic_2009} generally and \cite{wei2022deep} for more detail). 

\section{Quantum Machine Learning} \label{sec:ml:Quantum Machine Learning}
In this section we consider how the principles of classical statistical learning and classical machine learning algorithms are related to the hybrid quantum-classical models explored later on in this work. In line with our taxonomy focusing on data, models and metrics (measures), we start with a brief synopsis of how data (quantum or classical) is actually encoded for use in quantum machine learning scenarios. In this section, we cover a number of important differences between classical and quantum machine learning. We only sketch a few germane points related to the objectives of our work. As we discuss in Chapter \ref{chapter:QDataSet and Quantum Greybox Learning}, quantum machine learning is now a vast and variegated discipline combining elements of quantum computation and information processing, statistical learning theory, communication theory and other areas. The unifying factor behind quantum machine learning is the use of quantum data or quantum information processes in ways that enable, constitute, or rely upon some type of learning protocol as measured or indicated by relevant metrics, such as in-sample performance, variance, accuracy and so on. The spectrum of quantum machine learning algorithms is expansive, covering quantum neural networks, parametrised (variational) quantum circuits, quantum support vector machines and other formulations \cite{cerezo_challenges_2022}. Similarly, the application of deep learning principles to quantum algorithm design, such as via quantum neural networks (see \cite{schuld_machine_2021,beer_quantum_2022} for an overview), quantum deep learning architectures (including quantum analogues of graph neural networks, convolutional neural networks and others \cite{verdon_learning_2019,verdon_universal_2018,verdon_quantum_2019-1,verdon_quantum_2017}) speaks to the diversity of approaches now studied in the field. Coupled with these approaches, emerging trends in the use of symmetry techniques, such as dynamical Lie algebraic neural networks, symmetry-preserving learning protocols and geometric quantum machine learning approaches (discussed in particular in section \ref{sec:ml:Geometric QML}) offer new methods for improving on problem-solving in the quantum and classical domain. Much literature is also devoted to understanding the differences, some subtle, some less-so, between classical and quantum machine learning both at the theoretical level of statistical learning, down to the practical implementation of algorithms in each case (examples include literature in quantum algorithm design and complexity theory \cite{aaronson_how_2022}). In this section, we provide a short high-level summary of some of the key features of quantum machine learning relevant to Chapters \ref{chapter:QDataSet and Quantum Greybox Learning} and \ref{chapter:Quantum Geometric Machine Learning}, with a particular focus on variational or parametrised quantum machine learning circuits. First-off, we consider a few key principle differences between classical and quantum approaches in machine learning, starting with the fundamental distinction between the need for quantum channels to preserve unitarity (definition \ref{defn:quant:Unitary Channel}) versus the dissipative nature of classical neural networks \cite{schuld_quest_2014}.

\subsection{Neural networks and quantum systems} \label{sec:ml:Neural networks and quantum systems}
As Schuld et al. note \cite{schuld_quantum_2014}, the challenge for adapting classical neural network architecture to the quantum realm in part lies in structural distinctions between classical and quantum computing. Classical neural network architectures are fundamentally based upon non-linear and often dissipative dynamics, as distinct from the linear unitary dynamics of quantum computing. Classical neural networks, operations such as activation functions and normalisation can lead to a loss of information (e.g. applying a ReLU activation \cite{boob_complexity_2020}) can set negative values to zero, constituting information loss. Classical neural networks also are often irreversible (such as pooling or other convolutional network operations) such the network is not, functionally speaking, locally bijective between layers. Quantum information processing, by contrast, is constituted by unitary operations on $\Hilb$ which preserve the axiomatic linearity of quantum state space (i.e. that a quantum system remains a linear combination (superposition) of basis states $\sum_j a_j \ket{\psi}_j$ following unitary evolution). Furthermore, unitary evolution is information-preserving in that all unitary operations are in principle reversible (ensuring, among other things, conservation of total probability). Thus from the onset, there are key distinctions to be resolved and bridged for neural network architecture and quantum information processing to be synthesised.

Early progenitors of quantum machine learning architectures adapting principles form neural networks include \textit{Hopfield networks} \cite{hopfield_neural_1982}, being a (directed) graph network where weights are indexed by $i$ and the neurons by $j$. Hopfield networks usefully illustrate the differences between classical and quantum networks. Hopfield networks exhibit \textit{associative (content-addressable) memory}, which allows retrieval of the network state from $P$ stored network states 
\begin{align*}
    X^P = \{ (x^1_1,...,x_N^1),...,(x_1^P,...,x_1^P)\}.
\end{align*}
Associative memory therefore allows computation using incomplete input information instead of the pattern's exact storage address in RAM (basis of human memory and learning). Modern day neural networks represent functional compositions of non-linear equations. Early forms of neurons included \textit{perceptrons} (sometimes denoted binary neurons or linear threshold functions) such as the McCulloch-Pitt neuron (see \cite{schuld_introduction_2015,schuld_quest_2014} for a review of the early history of the field). Fundamentally, as noted in the literature, if quantum neural networks or machine learning circuits merely induced measurements in order to update the parameters, then they would proceed classically and in effect destroy (due to measurement collapse (axiom \ref{axiom:quant:effectofmeasurement}) \cite{schuld_quest_2014}) the underlying quantum properties of the data. Early techniques for addressing this implicit challenge included leveraging quantum measurement to select system eigenstates \cite{kak_quantum_1995}, the use of dissipative quantum systems (being revisited now in light of thermodynamic hybrid quantum computing), encoding parameter evolution in quantum terms \cite{perus_neural_2000}, use of entanglement \cite{behrman_quantum_1996}, teleportation and back-action based approaches \cite{abrams_quantum_1999}.

\subsection{Quantum statistical learning} \label{sec:ml:Quantum statistical learning}
Quantum learning tasks are similarly contingent upon the particular functional forms to be learnt. For example, \cite{huang_information-theoretic_2021} consider a class of learning problems concerned with learning the function:
\begin{align}
    f(x) = \text{tr}(O \mathcal{E}(\ket{x}\bra{x}))
    \label{eqn:ml:preskillfn}
\end{align}
where $O$ represents a typical known observable (operator) and $\mathcal{E}$ represents a completely positive trace preserving map i.e. a channel representing physical (possibly quantum) state evolution. Equation (\ref{eqn:ml:preskillfn}) is generic, representing classical inputs mapped to real numbers (e.g. results of Hermitian measurement operators $O$). Observation (measurement) constitutes an interaction between an experiment and the physical (quantum) system $\mathcal{E}$, which can be thought of as querying $\mathcal{E}$ by way of $O$.  As with classical, the learning problem is to estimate a function $h(x) \approx f(x)$ while minimising statistical and empirical risk, but with the added constraint of minimising the number of queries on $\mathcal{E}$. 

The model in \cite{huang_information-theoretic_2021} is in many ways generic and reflective of many QML architectures, reflecting the fact that information extraction for updating protocols in effect represents a quantum-classical channel (definition \ref{defn:quant:Quantum-classical channel}) via measurement or in-built trace operations. A comparison with quantum control is again useful in this regard. For hybrid classical-quantum structures, empirical risk minimisation is with respect to a \textit{classical} representation $f(x)$ (e.g. contingent upon measurement outcomes $m$ given measurement operator $M$) and typically a classical loss function (e.g. fidelity measures, quantum Shannon entropy or some other function of input and output state representations). Physical intuition is useful here: classical gradient descent then enters into Hamiltonian $H$ governing the evolution (and transition) of the quantum system via controls (e.g. coefficients $c_i$ that affect generator amplitudes). Other results in quantum-related statistical learning include bounds on the expressibility of variational quantum circuits \cite{du_learnability_2020,raghu_expressive_2017,chen_expressibility_2021,sim_expressibility_2019,nakaji_expressibility_2021}. 
\\ 
\\
The limits of quantum machine learning algorithms from a statistical learning perspective have been examined throughout the literature \cite{ciliberto_statistical_2020}. For example, it is shown in \cite{ciliberto_statistical_2020} that challenges persist for quantum machine learning algorithms with polyalgorithmic time complexity in input dimensions. In this situation, the statistical error of the QML algorithm is polynomially dependent upon the number of samples in the training set, such that the best error rate $\varepsilon$ is achievable only if statistical and approximation error scale equivalently, which in turn requires approximation error to scale polynomially with the number of samples. For logarithmic dependency on sampling, it is shown that the additive sampling error from measurement further introduces a polynomial dependency on the number of samples. Such constraints affect the viability of whether certain popular or mooted QML algorithms may in fact provide any advantage over classical analogues. 
Thus in many cases proposed QML algorithms face challenges posed by barren plateaus, entanglement growth and statistical learning barriers. Thus there is, as with classical machine learning, a need to consider how algorithms may (if at all) be architected in order to solve such scaling challenges. Doing so is one of the motivations for this work and its exploration of techniques from geometry, topology and symmetry algebras. As discussed in the introduction, in a machine learning context, the ability to encode prior or known information, offers one route to such efficiency, hence our synthesis of what we denote (and further articular) as \textit{greybox} QML together with geometric QML. 

\subsection{Quantum measurement and machine learning} \label{sec:ml:Quantum measurement and machine learning}
As discussed in Appendix \ref{chapter:Background: Quantum Information Processing}, quantum systems are constructed on the basis of underlying measurement instruments and their accompanying theory. In quantum information processing, measurement is an integral part of the evolution and characterisation of systems as distinct from classical theory. Recall from section \ref{sec:quant:Measurement} that measurement can be considered as a quantum-classical channel (definition \ref{defn:quant:Quantum-classical channel}) and equation (\ref{eqn:quant:measurementaschannel}) that stochastically maps from states $\rho$ to a classical register (of measurement outcomes) $\Sigma$. As per the measurement axioms of quantum mechanics, the effect of measurement is to collapse a state $\rho$ into an eigenvalue of the relevant measurement operator $M_m$ (axiom \ref{axiom:quant:effectofmeasurement}). 
 
Thus the question arises as to how machine learning models of quantum phenomena, trained on stochastic quantum measurement data (such as reconstructions of unitaries from measurement statistics) ought to integrate such stochasticity and measurement practices into their algorithms. To begin with, it helps to identify key structural differences between the classical and quantum measurement and how these differences may impact learning protocols to be adopted:
\begin{enumerate}[(i)]
    \item \textit{Probabilistic outcomes}. Quantum measurements are inherently probabilistic (axiom \ref{axiom:quant:probabilitymeasurement}) such that measurements (of outputs) will differ given identical inputs, but moreover it may be unclear what intermediate quantum evolution has occurred. Classical machine learning is no stranger to handling uncertainty or probabilistic characteristics of computing, however quantum computing presents distinct challenges given the nature of quantum measurement. 
    \item \textit{Wave function collapse}. Unlike the classical case, quantum measurement decoheres the state into eigenstates of the measurement operator $M_m$ and unlike classical measurement, the decohering process of quantum measurement channels thus leads to information loss, thus the dynamic updating becomes problematic (which is also the reason online control is problematic).
    \item \textit{No cloning and non-repeatability}. No cloning theorems applicable in quantum information (theorem \ref{thm:quant:No-Cloning Theorem}) mean that classical machine learning methods relying upon copying data (especially during training versus simple initial state preparation) may be unavailable.
\end{enumerate}

Our models of quantum simulation in Chapter \ref{chapter:QDataSet and Quantum Greybox Learning} specifically incorporate measurement data into the QDataSet in Chapter \ref{chapter:QDataSet and Quantum Greybox Learning}. By contrast, in Chapter \ref{chapter:Quantum Geometric Machine Learning}, we train a greybox neural network to generate approximate geodesics. The network is trained on geodesic sequences generated from horizontal distributions $\Delta$. Our target (label) data in that case are unitary targets $U_T$, but the construction of these ultimately depends upon measurement, which we assume has been undertaken (we do this in order to focus upon the key objectives of modelling time optimal paths rather than tomographical reconstruction). We discuss a few of these issues below in the context of parameter shift rules.

\subsection{Barren Plateaus}\label{sec:ml:Barren Plateaus}
Barren plateaus are a phenomena related to training of QML algorithms using gradient-based methods. Early papers on quantum algorithm-related barren plateaus \cite{mcclean_barren_2018} identified phenomena where the variance of a gradient expectation was shown to decrease exponentially in the number of qubits. In this sense they are somewhat analogous to the classical \textit{vanishing gradient problem} \cite{pascanu_difficulty_2013,goodfellow_deep_2016} albeit with specific differences (see \cite{larocca_thanasilp_cerezo_2024} for a comparison). The vanishing variance of the gradient is known as a barren plateau, reflecting the geometric intuition where the process of gradient descent seeking (local or global minima) finds itself in sparse of `flat' (zero-manifold curvature) region. This in line with the principle underlying the directional derivative (see definition \ref{eqn:geo:directionalderivative} and gradients \ref{defn:geo:gradient}) i.e. that generally, zero curvature implies a form of (pseudo)-isotropy where there is no necessary preferred direction of steepest descent, thus confounding the gradient descent protocol. Since the original paper, a plethora of studies \cite{holmes_barren_2021,holmes_connecting_2022,liu_presence_2021,patti_entanglement_2021} have examined conditions under which barren plateaus may or may not arise, together with strategies for preprocessing or tuning initial conditions (such as parameter initialisation strategies $\theta$ \cite{marrero_entanglement-induced_2021} or Bayesian methods \cite{rad_surviving_2022}) to mitigate or even side-step such issues. Other examples using control paradigms include \cite{larocca_diagnosing_2022} which leverage Lie algebras encoded within certain ansatzes (including popular Hamiltonian variational ansatzes \cite{mele_avoiding_2022}) to show guarantees for avoiding barren plateaus using such ansatzes may not exist, or showing noise-induced barren plateaus for variational quantum algorithms \cite{wang_noise-induced_2021}. Conversely, for certain classes of QML algorithms such as quantum convolutional neural networks, barren plateaus may be absent assuming the architecture is appropriately chosen \cite{pesah_absence_2021}. Examples or strategies to address barren plateaus include using adaptive quantum natural gradients for greater stability and optimised descent \cite{haug_optimal_2021}, the use of parameter initialisation \cite{liu_parameter_2021} (a state efficient ansatz) ground state preparation, the use of Gaussian parameter initialisations \cite{zhang_gaussian_2022} or even category-theoretic methods \cite{zhao_analyzing_2021}. Experimentally, we found, when tuning our quantum deep learning architectures in Chapters \ref{chapter:QDataSet and Quantum Greybox Learning} and \ref{chapter:Quantum Geometric Machine Learning}, initialising parameters as per \cite{pascanu_difficulty_2013} worked effectively to avoid such issues. Barren plateaus are particularly relevant for researchers designing QML algorithms.

%==========
\subsection{Encoding data in quantum systems} \label{sec:ml:Encoding data in quantum systems}
In this section, we examine protocols for encoding information in quantum algorithms used in quantum machine learning. Most quantum information optimisation problems involve information encoded in quantum systems, either by construction in an experiment involving quantum systems themselves, or via encoding exogenous or classical information into quantum systems (such as qubits) in order to leverage the benefits of quantum computation. Both approaches involve the input into quantum states in a process known as \textit{state preparation}. The way in which data is encoded in quantum systems affects the performance and expressiveness many quantum algorithms \cite{schuld_machine_2021}. Information is usually encoded using one of four standard encoding methods including: (a) basis encoding, (b) amplitude encoding, (c) qsample encoding and (d) dynamic encoding. The first of these, \textit{basis encoding}, is a technique that encodes classical information into quantum basis states. Usually, such procedures involve transformation of data into classical binary bit-strings $(x_1,...,x_d), x_i \in \{0,1\}$ then mapping each bit string to the quantum basis state of a set of qubits of a composite system. For example, for $x \in \Real^N$, say a set of decimals, is converted into a $d$-dimensional bit string (e.g. $0.1\to 00001..., -0.6 \to 11001..$) suitably normalised such that $x = \sum_k^d (1/2^k)x_k$. The sequence $x$ is given a representation via $\ketpsi \propto \ket{000001\,11001}$ (see \cite{schuld_supervised_2018}). \textit{Amplitude encoding} associates normalised classical information e.g. for an $n$-qubit system (with $2^n$ different possible (basis) states $\ket{j}$), a normalised classical sequence $x\in \Complex^{2^n}, \sum_k |x_k|^2=1$ (possibly with only real parts) with quantum amplitudes $x=(x_1,...,x_{2^n})$ can be encoded as $\ket{\psi_x}=\sum_j^{2^n} x_j \ket{j}$. Other examples of sample-based encoding (e.g. \textit{Qsample} and \textit{dynamic} encoding are also relevant but not addressed here). From a classical machine learning perspective, such encoding regimes also enable both features and labels to be encoded into quantum systems.

%===POSSIBLY MOVE

%====Basis orthonormal

In Chapters \ref{chapter:QDataSet and Quantum Greybox Learning} and \ref{chapter:Quantum Geometric Machine Learning}, we assume the standard orthonormal computational basis $\{\kz,\ko\}$ such that $\braket{1 | 0}=\braket{0 | 1}=0$ and $\braket{1|1}=\braket{0|0}=1$. Quantum states encode information of interest and use in optimisation problems. They are not directly observable, but rather their structure must be reconstructed from known information about the system. In machine learning contexts, quantum states may be used as inputs, constituent elements in intermediate computations or label (output) data. For example, in the QDataset (explored in Chapter \ref{chapter:QDataSet and Quantum Greybox Learning}), intermediate quantum states at any time step may be reconstructed using the intermediate Hamiltonians and unitaries for each example. The code repository for the QDataSet simulation provides further detail on how quantum state representations are used to generate the QDataSet \cite{perrier_qdataset_2021}. Depending on machine learning architecture, quantum states will usually be represented as matrices or tensors and may be used as inputs (for example, flattened), label data or as an intermediate input, such as in intermediate layers within a hybrid classical-quantum neural network (see \cite{perrier_quantum_2020, youssry_modeling_2020}). For example, consider the matrix representation of eigenstates of a Pauli $\sigma_z$ operator below:
\begin{align}
\sigma_z = \begin{pmatrix} 
1 & 0 \\
0 & -1
\end{pmatrix} 
\end{align}
In the computational basis, this operator has two eigenstates $\kz,\ko$ for eigenvalues $\lambda=1,-1$:
\begin{align}
\kz =
\begin{pmatrix} 
1 \\
0
\end{pmatrix} \qquad \text{for} \qquad \lambda = 1 \qquad
\ko =
\begin{pmatrix} 
0 \\ 
1
\end{pmatrix} \qquad \text{for} \qquad  \lambda=-1 
\end{align}
where we have adopted the formalism that the $\lambda=1$ eigenstate is represented by $\kz$ and the $\lambda=-1$ eigenstate is represented by $\ko$ (our choice is consistent with Qutip - practitioners should check platforms they are using for the choice of representation). These eigenstates have a density operator representation as:
\begin{align}
\rho_{\lambda=1}=\ketbra{0}{0} \qquad  \rho_{\lambda=-1} = \ketbra{1}{1}
\end{align}
with matrix representations:
\begin{align}
\ketbra{0}{0} = \begin{pmatrix} 
1 & 0 \\
0 & 0
\end{pmatrix}\qquad  \ketbra{1}{1} = \begin{pmatrix} 
0 & 0 \\
0 & 1
\end{pmatrix}.
\end{align}
Each element $\ketbra{a}{b}=E_{a,b}$ can be considered as an operator basis representation for the relevant density operators (not to be confused with root system vectors in representation theory below). 
For machine learning practitioners, one way to think about density operators (see definition \ref{defn:quant:Density operator}) is by associating rows and columns with bra and ket vector representations:
\begin{align}
\rho = a\ketbra{0}{0} + b\ketbra{0}{1} + c\ketbra{1}{0} + d\ketbra{1}{1} \dot{= }
\begin{blockarray}{ccc}
& \bz & \bo \\
\begin{block}{c (cc)}
\kz \qquad & a & b  \\
\ko \qquad & c & d  \\
\end{block}
\end{blockarray}
\end{align}
where $a,b,c,d \in \Complex$ are the complex-values amplitudes respective. Given $\rho = \sum_{p_i} \rho_i$, the diagonal elements $a_{n}$ of the density matrix describe the probability $p_n$ of the system residing in state $\rho_n$, that is 
\begin{align}
\rho_{nn} = a_{n}a_n^* = p_n \geq 0
\end{align}
For pure states, the diagonal along the density matrix will only have one non-zero element (i.e. it will be 1) so that $\rho = \rho_i$. A mixed state will have multiple entries along the diagonal such that $ 0 \leq a_n < 1 $. For example, the $\sigma_z$ eigenvectors have the representation:
   \begin{align} 
   \ketbra{0}{0} \dot{=} 
\begin{blockarray}{ccc}
 & \bz & \bo \\
\begin{block}{c (cc)}
  \kz \qquad & 1 & 0  \\
  \ko \qquad & 0 & 0  \\
\end{block}
\end{blockarray}
\qquad
 \ketbra{1}{1} \dot{=} 
\begin{blockarray}{ccc}
 & \bz & \bo \\
\begin{block}{c (cc)}
  \kz \qquad & 0 & 0  \\
  \ko \qquad & 0 & 1  \\
\end{block}
\end{blockarray}
 \end{align}
Sometimes the density matrix representation of a state will be equivalent to the outer product of the state, but caution should be applied as this is not generally the case. Translating between the nomenclature and symbolism of quantum information to a more familiar matrix representation used in machine learning assists machine learning researchers to develop their algorithmic architecture. For example, the QDataSet simulation code utilises state space representations of data and operations thereon in order to generate the output contained in the datasets themselves. To recover a quantum state $\ket{\psi(t_j)} = U(t_j)\ket{\psi_0} \approx \prod_i U_i^j \ket{\psi_0}$, one can apply the sequence $U_i$ up to $i=j$ (note the order of application is such that $U_j...U_0 \ketpsi$).

\section{Variational quantum circuits}\label{sec:ml:Variational quantum circuits}
In this section, we sketch the general use of variational quantum algorithms adopted in Chapters \ref{chapter:QDataSet and Quantum Greybox Learning} and \ref{chapter:Quantum Geometric Machine Learning}. As Schuld et al. \cite{schuld_machine_2021} note, variational quantum circuits can be interpreted as both deterministic and probabilistic machine learning models for functions $f_\theta: \X \to \Y$ (or learning a probability distribution $\Prb_\theta$). As noted in the literature, quantum circuits parametrised by $\theta$ which are updated according to some objective (cost) function as part of an optimisation process can in some sense naturally be regarded as akin to machine learning models \cite{benedetti_parameterized_2019,cerezo_variational_2022}. Firstly, we define a \textit{parametrised quantum circuit} in terms of a unitary $U(\theta)$ dependent upon parameters $\theta \in \R^m$.
\begin{definition}[Parametrised quantum circuits]\label{defn:ml:Parametrised quantum circuits}
    A parametrised unitary operator is a unitary $U(\theta)(t)$ operator acting on a Hilbert space $\Hilb$ parametrised by a set of parameters $\theta \in \R^m$. A parametrised quantum circuit is then a sequence of unitary operations:
    \begin{align}
        U(\theta,t) = \mathcal{T}_+ \exp\left(-i\int_0^T H(\theta,t')dt'\right).\label{eqn:ml:parametrised circuit-timedependent}
    \end{align}
\end{definition}
In the time-independent approximation, we have
\begin{align}
    U(\theta,t)\bigg|_{t=T} = U_{T-1}(\theta_{T-1})(\Delta t)...U_1(\Delta t)(\theta) \label{eqn:ml:parametrised circuit-time-independent}
\end{align}
i.e. a sequence of unitaries acting on the initial state $\ket{\psi(0)}$ which we can represent via $U_0(t)$. We assume left-action on states $\ket{\psi}$ where dependence upon $t$ is understood. Recall each $U_i(\theta_i(t))$ above is the solution to the time-dependent Schr\"odinger equation (equation \ref{eqn:quant:schrodingersequation}) as a sequence of integrals. For simplicity, we assume the time-independent approximation (equation (\ref{eqn:quant:timeindependentschrodunitary})). The aim of a variational quantum algorithm (VQA) is to find an optimal set of parameters minimising a cost functional on the parameter space given by $C(\theta): \R^\m \to \R$ such that:
\begin{align*}
    \theta_* = \arg\min_\theta C(\theta).
\end{align*}
The term variational in this quantum context \cite{peruzzo_variational_2014} derives from the underlying principles of variational calculus of finding optimal solutions, such as (originally speaking) computing, for example, lowest energy states of a system. As Schuld et al. note, the term \textit{variational quantum circuit} is sometimes used interchangeably with parametrised quantum circuit (see Benedetti et al. \cite{benedetti_parameterized_2019} for a still-salient review). In this work, we adopt this association between the two. We set out the deterministic and probabilistic form below. For a deterministic model, we let $f_\theta:\X \to \Y$ and denote $U(X,\theta)$ (a quantum circuit) for initial state $X \in \X$ and parameters $\theta \in \R^m$. Let $\{O_k\}$ represent the set of (Hermitian) observable (measurement) operators. Then the following function represents a deterministic model:
    \begin{align*}
        f_\theta(X) = \trace(O \rho(X,\theta)) \label{eqn:ml:qml-VQC-ftheta}
    \end{align*}
    where $\rho(X,\theta)=U(X,\theta)\rho_0 U(X,\theta)^\dagger$.
% \begin{definition}[Deterministic quantum model]
% \end{definition}
Common approaches to the form of $U(X,\theta)$ include, in analogy with classical machine learning, $U(X,\theta)=W(\theta)S(X)$ where $W(\theta)$ represents a parametrised weight tensor with $S(X)$ representing the encoding of initial features. In this work, we adopt geometric quantum control-style architectures where the unitary channel layers in each network constructed from Hamiltonian layers comprising dynamical Lie algebra generators $X_j \in \g$ (see section \ref{sec:geo:Dynamical Lie algebras}) and associated time-dependent control functions $u_j(t)$ (sometimes denoted $c_j(t)$)). 
In terms of measurement statistics, then for $U(\theta)\ket{\psi(0)} = \ket{\psi(t)}$, then the class of functions in equation (\ref{eqn:ml:qml-VQC-ftheta}) we seek to learn are represented by measurement statistics, i.e. a mapping from inputs (e.g. quantum data $x$) to $\R$ as:
\begin{align}
    f_\theta(x) = \sum_m m p(m)
\end{align}
ranging over measurements $\{m \}$. In practice because of post-measurement state collapse, one takes the expectation to estimate $f_\theta$ as $\hat f_\theta(x) = (1/N)\sum_i^N m^{(s)}$ where $m^{(s)} \in \{ m \}$ i.e. the outcome of a random measurement from the set of possible outcomes. Reducing error $f_\theta - \hat f_\theta \leq \epsilon$ requires $\mathcal{O}(\epsilon^{-2})$ measurements, growing quickly with precision. Thus one must repeat the experiment $N$ times depending on $\epsilon$ sought. In general we are interested in learning distributions of measurement statistics (from which to reconstruct states or operators) given inputs $X \in \X$ and conditional upon labels (outputs) $\Y$. In measurement terms, we regard $Y_i \in \Y$ as measurement outcomes so we interested in a \textit{supervised probabilistic quantum model} for a conditional distribution given by $\Prb_\theta(m|X)$, not just $\Prb_\theta(X)$ itself. Because probabilistic quantum models generate samples, they are often considered \textit{generative models}. \\
\\
\\
The typical supervised QML task for unitary synthesis involves a data space $D_n(\X,\Y)$ where feature data is drawn from quantum state data for a Hilbert space $\X \sim \Hilb$ and $\Y \sim \Km$ (which can be real $\R$ or complex $\C$ -valued) labels. In many cases, including our two machine learning-based chapters, $\Hilb$ is a tensor product for qubits $\Hilb_\otimes = \otimes_n \Hilb$ where $\dim \Hilb_\otimes = 2^n$. The form of $D_n(\X,\Y)=\{ \rho_i, y_i \}_i$ where $D_n$ is sampled from an underlying population. In Chapter \ref{chapter:Quantum Geometric Machine Learning} for example, the training data generated is somewhat unusual in that label data (that which we want to accurately predict) is the sequence $n$ unitaries $(U_n)$ (i.e. so $(U_n) \in \Y$) where the target (final state) unitary $U_T$ is actually fed in via an input layer (i.e. $U_T \in \X$). In that case, as we discuss, what we are interested in is conditioned on $U_T$, what set of controls $c_j(\Delta t)$ can be generated such that the resulting sequence $(U_n(\Delta t_n))$ leads to $U_T$. Conversely, for the QDataSet in Chapter \ref{chapter:QDataSet and Quantum Greybox Learning}, the training data may vary, such as being classical measurement statistics arising from $m$ Pauli measurements (in which case $\Y \subset \R^m$). As noted in our discussion on measurement (section \ref{sec:quant:Measurement}), for our final two chapters we assume access to a measurement process that allows us data about (of) $U$.
\\
\\
%Parameter manifolds
Finally, we include a few comments regarding geometric formulations of parametrisations above. Recall that unitary elements in our approach are parametrised by control functions $u(t)$. However these control functions are themselves parametrised by $\theta$, i.e. $u(t) = u(\theta(t))$. In this work, we adopt the time-independent approximation assuming $H(t)$ is constant for over small $\Delta t$ then effectively we can treat the controls as parametrised by $\theta$ and in this way our unitaries $U=U(\theta(t))$. In many cases, our parameter e.g. $\theta \in \R^m$ can be framed as the fibre of an underlying manifold $\K$ such the commutative diagram in Figure \ref{fig:ml:parametermanifolds} holds. While not the subject of this work, understanding the relationship between parameter manifolds and target (label) data manifolds (often described geometrically in terms of \textit{embedding}) is an important research direction utilising techniques from geometry and machine learning, with specific applications to solving problems in quantum computing.
\begin{figure}[h!]
    \centering
    \begin{tikzcd}
&\R \arrow[r] &\mathbb{R}^m \arrow[r,"h_*"'] \arrow[d,  "\pi_{\mathcal{K}}"']
& \mathbb{R}^{2n} \arrow[d,  "\pi_{\mathcal{M}}"'] \\
& &\mathcal{K} \arrow[r, "h"']
& \mathcal{M} \cong G
\end{tikzcd}
    \caption{Manifold mapping between parameter manifold $\mathcal{K}$ and its fibre bundle $\R^m$ and target manifold $\M$ with associated fibre bundle spaces $\R^{2n}$. Optimisation across the parameter manifold can be construed in certain cases as an embedding of the parameter manifold within the target (label) manifold $\M$, where such embeddings may often be complex and give rise to non-standard topologies.}
    \label{fig:ml:parametermanifolds}
\end{figure}

\subsection{QML Optimisation}\label{sec:ml:QML Optimisation}
The quantum supervised learning problem above becomes to learn the distribution $\Prb_\theta(Y|X)$. The most common technique for parametrised quantum circuits is \textit{automatic differentiation} \cite{bartholomew-biggs_automatic_2000} via which one can implement stochastic gradient descent and backpropagation algorithms.
As with the classical case, a loss function is chosen, typically the cost functional $C(f_\theta,\hat f_\theta) = C(\theta)$. Thus for example a loss function based on the \textit{fidelity} metric (definition \ref{defn:quant:Fidelity}) adopted in equation (\ref{eqn:ml:batchfidelityMSE}) in Chapter \ref{chapter:Quantum Geometric Machine Learning}, denoted \textit{batch fidelity} takes the mean squared error (MSE, see equation (\ref{eqn:ml:MSE}) above) of the loss (difference) between fidelity of our unitary estimator and the target unitary for empirical risk as:
\begin{align}
    C(F(\theta),1) = \frac{1}{n}\sum_{j=1}^n (1 - F(\hat{U(\theta)}_j,U(\theta)_j))^2. \label{eqn:ml:batchfidelityMSE}
\end{align}
This equation implicitly relies upon assumptions regarding measurement protocols that allow us to specify $\hat U_j$ and $U_j$. Because we are using channels $U_j$ for states $\rho_j \in \mathcal{B}(\Hilb)$ we could equivalently specify equation (\ref{eqn:ml:batchfidelityMSE}) in terms of the measurement statistics required to (tomographically) reconstruct $U_j$ (something we touch upon in Chapter \ref{chapter:QDataSet and Quantum Greybox Learning}), in which case it could be written in terms of informationally complete measurement statistics (see section \ref{sec:quant:Informational completeness and POVMs}) with loss functions given in terms of expectation statistics (equation \ref{eqn:quant:measurementdensitymatrixtrace}) continent upon parameters (i.e. controls) $\theta$. Before moving onto a sketch of our greybox variational quantum architecture adopted in Chapters \ref{chapter:QDataSet and Quantum Greybox Learning} and \ref{chapter:Quantum Geometric Machine Learning}, it is worth saying a few words about differences in implementation of gradient descent in the classical and quantum case and parameter shift rules in particular. 

\subsection{QML Gradients} \label{sec:ml:QML Gradients}
The discussion of gradient-based optimisation methods above is classically focused, relying on mathematically nice properties of differentiable manifolds $\M$ (such as configuration space or registers parametrising quantum states or operators). Equivalent and analogous methods of calculating gradients in quantum information processing have been studied in the context of a variety of quantum and hybrid quantum-classical machine learning architectures. Early examples include utilising quantum encoding for faster numerical gradient estimation \cite{jordan_fast_2005} including in the multivariate case \cite{gilyen_optimizing_2017} There are a number of candidate optimisation strategies in the literature that effectively adapt classical stochastic gradient descent to the quantum case, including the following: 
\begin{enumerate}[(i)]
\item \textit{Parameter shift rules}. Parameter shifts \cite{mitarai_quantum_2018,wierichs_general_2022} in the context of gradient descent represent a method of calculating gradients by shifts of parameters. By evaluating the cost function $C(\theta)$ at the value of two different parameters $\theta, \theta + \epsilon_\theta$, the rescaled difference $\nabla_\theta - \nabla_{\theta + \epsilon}$ forms an unbiased estimate of $\nabla_\theta$, usually restricted to gates with two different eigenvalues \cite{wierichs_general_2022}. The reason this matters is because in certain cases classical stochastic gradient descent is inappropriate or unable to obtain a well-defined gradient. For a cost functional $C(\theta)$ we calculate the gradient as:
    \begin{align}
    \nabla_\theta C = \frac{C(\theta + \epsilon) - C(\theta - \epsilon)}{2}
    \end{align}
    ($\epsilon$ is usually set to $\pi/2$ for operators with $\pm 1$ eigenvalues). An example would include where a quantum circuit includes a rotation gate $R_x(\theta)=\exp(-i\sigma_x \theta/2)$. Connecting with section \ref{sec:quant:Expectation evolution}, the gradient of the expectation of an observable $A$ with respect to a parametrised gate $U(\theta)$ in simple form is:
    \begin{align}
        \nabla_\theta \braket{A}_{U(\theta)} = \frac{\braket{A}_{U(\theta+\epsilon)} - \braket{A}_{U(\theta-\epsilon)}}{2}
    \end{align}
    providing a means of understanding how measurement statistics vary as $U(\theta)$ evolves.  
    \item \textit{Quantum natural gradients}. the quantum natural gradient method \cite{stokes_quantum_2020} related to the Fisher information metric and Fubini-Study metric $F_Q$ provides a candidate quantum analogue for quantum-specific natural gradient descent (see section \ref{sec:ml:Natural gradients}):
    \begin{align}
    \theta_{t+1} = \theta_{t} - \eta g^+(\theta_t)\nabla \mathcal{L}
    \end{align}
    where $\eta$ is the learning rate, loss function $\mathcal{L}$ and $g^+$ is related to the pseudo-inverse of the Fubini-Study metric tensor. This metric tensor is sometimes denoted the `quantum geometric tensor' which effectively involves inner-product contractions between states and their derivatives as a means of calculating the directional derivatives.
    
    \item \textit{Quantum autodifferentiation and backpropagation}. In a series of papers, Beer et al. \cite{beer_quantum_2021,beer_quantum_2022,beer_training_2020} propose a quantum analogue of autodifferentiation and backpropagation for (fully connected for the most part) quantum neural network architectures. Quantum autodifferentiation extends the concept of computational graphs to quantum circuits, enabling the calculation of gradients via a quantum version of the backpropagation algorithm. The proposal is based upon a dissipative quantum neural network architecture leveraging perceptron unitaries acting on $m+n$ qubits with $(2^{m+n})^2-1$ parameters. The circuit comprises $L+2$ layers of qubits (+2 being one for input, one for output, the rest $L$ being hidden). Each perception is represented by a unitary with information being propagated essentially via partial trace (definition \ref{defn:quant:partialtrace}) operations. More specifically (see $\S 4$ therein) each layer of qubits $\rho^{l-1}$ is tensored with the subsequent layer of qubits $\rho^l$ initialised in $\ketbra{0}{0}$. A set of unitaries are applied to both layers with the $l-1$-th layer being traced out. This mapping of layers $l-1$ to $l$ is denoted via the map $\mathcal{E}_t^l$. The backpropagation step works similarly but by applying the adjoint of this layer-to-layer map $\mathcal{F}_t^l$ first to the tensor product of layer $l$ and $l-1$ then tracing out layer $l$. The loss function $\mathcal{L}$ is a function of the fidelity of the labelled (target) states and the final output layer of the network. In particular, for fidelity-based loss functions with a quantum neural network architecture where each layer comprises qubits acted upon by unitaries, errors are backpropagated using a unitary operation on qubits that encodes the error given by:
    \begin{align*}
        U_j^l(t+\epsilon) = \exp(i\epsilon K_j^l (t))U^l_j(t)
    \end{align*}
    for $\epsilon$ training step size where:
    \begin{align*}
        K_j^l(t) = \frac{\eta 2^{m-l-1}i}{S} \sum_x \trace_{\text{rest}} \left( M_j^l(x,t) \right)
    \end{align*}
    where $M_j^l$ encodes the errors calculated via a commutator term which essentially propagates from the input layer to the $l$th layer for one term and from the label state to the $l+1$-th state, the difference between the unitaries between any two states then will show up as a non-zero commutator which is what $M_j^l$ encodes. This it is shown is equivalent to calculating the derivative of $\mathcal{L}$ for gradient update purposes via the equations above, with the benefit that only two layers need to be updated at any one time rather than needing to update all layers to do so.   
    \item \textit{Backaction-based gradients}. In \cite{verdon_universal_2018}, Verdon et al.  introduce a quantum-specific analogue of backpropagation, \textit{baqprop} (backward propagation of phase errors) based on a back-action arising from interaction of the quantum system with an external apparatus (e.g. measurement). The model adopted seeks to calculate the gradient of an oracle function $f$ whose input is a continuous register (see definitions \ref{defn:quant:Classical registers and states} and \ref{defn:quant:Quantum registers and states}). Given position $\Phi_j$ and momentum $\Pi_j$ operators for a multi-(two) qubit register, a von Neumann measurement $e^{-i f(\hat{\phi}_1)\hat{\Pi}_2} : |\phi_1, \phi_2\rangle \rightarrow |\phi_1, \phi_2 + f(\phi_1)\rangle$ representing a controlled displacement of the second register which computes the function $f$ on the first register and stores the result in the second. While in the position basis it appears the first register is unchanged, in the momentum it can be shown that the momentum of the first register is affected. This can be seen via expressing the operation via the adjoint action (definition \ref{defn:alg:Adjoint action}) in the Heisenberg frame this is $Ad_{e^{i f(\hat{\phi}_1)\hat{\Pi}_2}}(\hat{\Pi}_2) = \hat{\Pi}_2 + f'(\hat{\phi}_1)$ which can be shown via $ \hat{\Pi}_1 = Ad_{e^{i f(\hat{\phi}_1)\hat{\Pi}_2}}(\hat{\Pi}_1) + f'(\hat{\phi}_1)\hat{\Pi}_2$. Information about the gradient is then obtained by various measurement protocols on $\hat \Pi_1$ i.e. the momentum of the first register is shifted in proportion to the gradient $f'$. The actual optimisation procedures applied involve quantum-tunnelling style propagation of errors and hybrid style momentum measurement gradient descent.
\end{enumerate}

\section{Symmetry-based QML}\label{sec:ml:Symmetry-based QML}
In this final section, we sketch an outline of the general structure of greybox variational quantum circuits that are the subject of our next two chapters. We connect the general architecture to three related fields in quantum and classical machine learning (a) Lie algebraic methods (such as Lie group machine learning \cite{moskalev_liegg_2023,lu_survey_2020}), (b) geometric machine learning \cite{bronstein_geometric_2021} and (c) geometric quantum machine learning \cite{nguyen_theory_2022,larocca_group-invariant_2022}. 

\subsection{QML Optimal Control} \label{sec:ml:QML Optimal Control}
In terms of the discussion of optimal control examined in the previous chapter, the backpropagation algorithms are designed to learn function estimates $f_\theta$ that map from time-optimal geodesic data to sequences of controls. An effective learning protocol would thereby, in principle, learn controls (and thus unitary sequences) satisfying the requisite PMP conditions. These approaches seek to learn parameterised functions $u_j(t)$ in the case of control theory and weights $\omega$ in the case of neural networks to optimise an objective function (maximising the PMP Hamiltonian or minimising the loss/cost function $L$ in backpropagation). In essence the conceptual connection between controls acting on generators, updated to satisfy the constraints of the optimal control problem, with updatable weights acting on input features in order to satisfy the objective function of minimising loss provides a bridge between classical control theory and machine learning. Indeed the deep connections between the two are very much manifest across the literature, especially in the work of Bellman \cite{bellman_dynamic_1956,bellman_introduction_2016}.

In Chapter \ref{chapter:Quantum Geometric Machine Learning}, we assume that the sequence of unitaries (generated from a theoretically justified means of obtaining subRiemannian geodesics) is time-optimal and thus bounded by total time (path-length) $L$, thus adopt a learning protocol that seeks to learn the sequence of controls $u(t) \in [0,1]$ or $u(t) \in [-1,1]$ (using a tanh activation function) which minimises the error (fidelity) of $\hat U_j(t)$ versus $U_j(t)$ for a sequence of unitaries $j=1,...,n$. Put another way, we assume by reason of the assumption that training data consists of geodesic paths (represented by sequences of unitaries) whose length is therefore minimal and unique. From equation \ref{eqn:geo:arclength}, because the controls set the length $\ell(\gamma)$, the sequence of controls obtaining $\hat U_j(t) \simeq U_j(t)$ will be minimal (up to some error tolerance).

\subsection{Geometric information theory} \label{sec:ml:Geometric information theory}
Geometric information theory emerged in the second-half of the twentieth century via the application of techniques from differential geometry to information theory. In particular this saw results mapping concepts of Riemannian manifolds and metrics across to information manifolds and figures of merit common in statistics, such as Fisher information. The fundamental organising principle is to regard information-theoretic problems of interest in terms of geometric concepts. Thus information is presented as inhering within a differentiable manifold e.g. Riemannian or Kahler manifold. Information theory based upon specific measures of information \cite{nielsen_geometric_2015}. As Amari \cite{amari_information_2016,amari_natural_1998} (credited with first equating Fisher information and the Riemannian metric) notes, information geometry is a means of exploring the world using modern (differential) geometry, as a discipline arising from the use of such techniques to study symmetry and invariance properties involved in statistical inference. In the interests of time and space, we do not in this work delve into information geometry in any significant way, however note its use in machine learning in a variety of areas, such as clustering, support vector machines, belief propagation and boosting. The underlying idea of information geometry is to cast information and probability distributions in terms of geometric structures such as differentiable manifolds. Other areas of machine learning exploring geometric and algebraic techniques include Lie group machine learning \cite{li_lie_2019,moskalev_liegg_2023} where, similar to the use of Lie groups in quantum contexts, machine learning networks and algorithms are constructed in order to respect and leverage symmetries underpinning groups and their corresponding Lie algebras (see \cite{lu_survey_2020,li_theory_2007} comprehensive reviews and theory). 

\subsection{Geometric QML}\label{sec:ml:Geometric QML}
In recent years, interest in applying geometric techniques to assist in symmetry reduction and symmetry-enhanced quantum machine learning network design have gained prominent. We summarise some of the key results relevant to our complementary work below. This emerging sub-discipline, sometimes named geometric quantum machine learning \cite{cerezo_challenges_2022,larocca_theory_2021} seeks to incorporate symmetry properties into the design of machine learning architecture as a means of improving figures of merit, such as runtime complexity, metrics (such as fidelity) or generalisation and overparametrisation. This is often achieved via seeking to build parametrised quantum circuits or quantum neural network architectures which, as functional compositions, respect underlying symmetries of input data, or which lead to outputs which respect symmetry properties, such as outputs in terms of unitaries or parametrisations thereof. For example, \cite{schatzki_theoretical_2022} design quantum neural network layers $\mathcal{F}$ that permutation symmetry under the action of the permutation group $S_n$. In this case, $\sigma\mathcal{F(\sigma)} = \mathcal{F}(\sigma(\psi))$ for $\sigma \in S_n$.\\
\\
Central to the GQML programme is the use of symmetry-respective architectures. This is usually in the form of neural network layers consistent of unitaries generated by generators of dynamical Lie groups (see section \ref{sec:geo:Dynamical Lie algebras}). The use of symmetry reduction is shown in a number of works to, for example, reduce the effective parametrisation required in order to meet certain performance thresholds \cite{nguyen_theory_2022}. Relatedly, the reduction in parametrisation owing to such embedded symmetries has been shown to improve other architectural features, such as the risk of \textit{barren plateaus} \cite{arjovsky_invariant_2019}. Geometric QML is the closest extant discipline related to the present work. The distinction between this work and various geometric QML literature \cite{larocca_group-invariant_2022,larocca_diagnosing_2021,larocca_theory_2021,larocca_exploiting_2020,nguyen_theory_2022,cerezo_challenges_2022,cerezo_variational_2020,ragone_representation_2023,caro_generalization_2022} in this emerging field is that they are usually concerned with constructing quantum machine learning architectures (e.g. variational quantum circuits) which respect compositionally certain symmetries by encoding Lie algebraic generators within VQC layers, whereas the work in Chapter \ref{chapter:Quantum Geometric Machine Learning} for example does not seek an entire network that respects symmetry properties, but rather only a sub-network of layers whose output (e.g. in the form of unitary operators $U(t)$) respects symmetries inherent in unitary groups. Moreover, our work in that Chapter is concerned with design choices to enable the network to learn geodesic approximations via simplifying the problem down to one of learning optimal control functions.

\section{Greybox machine learning} \label{sec:ml:Greybox machine learning}
Our approach in our Chapters above represents a related, but distinct, stream of research leveraging algebraic and geometric techniques for solving time-optimal unitary synthesis problems. Our general approach to architecture, focusing on quantum geometric machine learning, is set out below. The method is a hybrid of quantum and classical architectures which we denote \textit{greybox architecture} or greybox learning which leverages variational parametrised quantum circuits (equation \ref{eqn:ml:parametrised circuit-time-independent}) that embed symmetries of interest (via Lie algebra generators and unitary activation functions) in order to generate unitary targets respecting symmetry structures. The underlying idea is that known a priori information is encoded within a classical neural network stack, obviating the need for the network to learn such rules (such as requirements regarding unitarity or Hamiltonian construction) and thereby affording guarantees (such as unitarity of outputs) in a way that blackbox learning only at best approximates. Although we do not formally prove it, our focus is on the idea that by doing so we are in effect minimising empirical risk by reducing the variance in the models that would otherwise be required to for example learn and implement the Schr\"odinger equation from scratch. We can analogise between directionality of gradient descent updates to estimate $\hat Y$ and using machine learning methods to solve time-optimal problems. Sketching this out, in the previous Appendices, we articulated how $U_T \in G \simeq \M$ (for Lie group $G$) are generated (via Hamiltonian flow) by Hamiltonians $H$ from the corresponding Lie algebra $\g$. By the correspondence between tangent vectors and Lie algebra basis elements, we can then regard learning optimal control functions $u_j(t)$ (for directions $j$ where $j = 1,...,\dim \g$) as equivalent optimisation in a machine learning context. The controls $u_j(t)$ in effect parametrise our system consistent with the generalised Pontryagin method, with the adjustment of the control functions to minimise our loss functions (explicated below in terms of fidelity $F(\hat U_T,U_T)$) geometrically represented as transformations in one or more of the directions of elements of $H_p\M$ (our distribution $\Delta$). The \textit{geometric} features we are seeking to leverage are the symmetry properties of subRiemannian symmetric spaces, in particular discovering strategies for learning sequences of controls for generating time-optimal geodesic paths along the manifold of interest. We describe this relationship below.
\begin{enumerate}[(i)]
    \item \textit{Objective}. Firstly, our problem is one of using machine learning to synthesise time-optimal unitaries relevant to quantum computational contexts. Specifically, this means using machine learning architecture to solve certain control systems expressed through the Pontryagin Maximum Principle (definition \ref{defn:geo:Pontryagin maximum principle}). Our approach focused on quantum control lends itself to variational quantum algorithms from a design-choice perspective.
    \item \textit{Control optimisation}. Our aim is to therefore generate a sequence of control pulses that may be applied to corresponding Lie algebra generators (in our case corresponding to single and multi-qubit systems) in order to generate sequences of time-optimal (or as approximately time-optimal as we can get) unitary operators. given the objectives and constraints required, including (a) ensuring outputs respect unitary constraints and (b) controls (and by extension path-length) respects minimisation (and geodesic) constraints.
    \item \textit{Inputs}. The inputs layers $a^{(0)}$ of the network take as their feature data unitaries $U(t) \in G$ or in the case of multi-qubit systems, $G \otimes G$ for Chapters \ref{chapter:QDataSet and Quantum Greybox Learning} and \ref{chapter:Quantum Geometric Machine Learning}, $G=\text{SU}(2)$. The unitaries have a representation by way of the usual representation of $SU(2)$ in terms of the matrix group $M_2(\C)$ (see definition \ref{defn:alg:connected_matrix_lie_group}). Denoting such matrices $X \in M_2(\C)$ (or $X_1 \otimes X_2 \in M_2(\C) \otimes M_2(\C)$ (by the single $X$ for convenience), the initial layers tend to transform (e.g. flatten) such matrices into a realised form. The activation function of these input layers tends to then just be the identity:
\begin{align}
    a^{(0)} = I(X)
\end{align}
consistent with the generic neural network schema in section \ref{sec:ml:neuralnetworkschema}.
\item \textit{Feed-forward layers}. The input layers are then fed into a sequence of typical classical neural network layer, in our case feed-forward layers (see definition \ref{defn:ml:Feed-forward neural network}):
\begin{align}
    a^{(1)}&=\sigma(\omega a^{(0)} + b).
\end{align}
The feed-forward layers tend to be, at least in our examples, fully-connected and dense. 
\item \textit{Control pulse layers}.  As we sketch below, parameter network layers (parametrised by $\theta$) feed into layers comprising control pulses $u_j(t)$. The neurons within these layers are characterised by activation functions such that $u_j(t) \in [0,1]$, where we rely implicitly on the fact that for the bounded control problem, we can arbitrarily scale (normalise) controls to fit within minimum length $L$ (see definition \ref{defn:geo:arclength}).
\item \textit{Hamiltonian and Unitary layers}.  The control pulse layers then feed into bespoke layers to (i) construct the Hamiltonian from fixed Lie algebra generators in $\g$ which is then (ii) input into a layer comprising an exponential activation that constructs time-independent approximation to time-dependent unitary operator. In all cases, our target of interest is the optimal unitary for time $t_j$. In Chapter \ref{chapter:Quantum Geometric Machine Learning}, we are interested in the sequence of controls $(u_i(t))$ that generate a sequence of estimated unitary operators $(\hat U_j(t))$ which, under the assumption of time-independence (see definition \ref{eqn:quant:timeindependentschrodunitary}), allows us to generate our target $U(T) \in G$. Each node generating a unitary has $k$ generators and control functions $u_i(t), i=1,...k$ where $k=\dim\g$ (or whatever the control subset is). Our labels and our function estimates are \textit{sequences} of unitaries and, as we note below, the target $U_T$ is actually one of the inputs to the models.
\item \textit{Optimisation}. The optimisation occurs by reference to a cost function based on the fidelity metric (definition \ref{defn:quant:Fidelity}). The fidelity of each unitary estimate in the sequence $\hat U_j(t)$ is estimated. 
From a control theory perspective, the optimisation is classical offline control, where an exogenous protocol is used to optimise and then adjust the parameters $\theta$ accordingly which ultimately feed into adjusting the controls $u_i(t)$. Several classes of network are used, including a vanilla (basic) fully-connected feedforward network, an LSTM and the greybox approach which we detail here. The networks are trained on data $D_n(X,Y)$ generated from distribution $\Delta \in \su(2)$, such that the resulting Hamiltonians (and thus curves on $\M$) are time-optimal, these are subRiemannian geodesics.
\item \textit{Measurement}. In Chapter \ref{chapter:QDataSet and Quantum Greybox Learning}, measurement statistics for each time slice are captured in the data (from which one can reconstruct unitaries or Hamiltonians for such time slice), whereas for Chapter \ref{chapter:Quantum Geometric Machine Learning}, the existence of an external process that allows for $\hat U(t)$ and $\hat U(t)$ to be constructed is assumed.
\end{enumerate}

\backmatter
\bibliographystyle{IEEEtran}
% \bibliography{IEEEabrv, IEEEfull, references}
\bibliography{IEEEabrv, IEEEfull, referencesfinal3}
\newpage\null\thispagestyle{empty}\newpage
\end{document}